# Improved Numerical Generalization of Bethe- Weizsacker Mass Formula

**Strachimir Cht. Mavrodiev**

**Institute for Nuclear Research and Nuclear Energy7, Sofia, BAS, Bulgaria**

## ABSTRACT

In this paper is presented explicit improved numerical generalization of Bethe-Weizsacker mass formulae which describes the values of measured 2654 nuclear mass in AME2012 nuclear database with accuracy less than 2.2 MeV, starting from the number of protons $Z = 1$ and number of neutrons $N = 1$.

In the obtained generazation of the Bethe-Weizsacker formula the influence of magic numbers and boundaries of their influence between them is defined for nine proton (2, 8, 14, 20, 28, 50, 82, 108, 124) and ten neutron (2, 8, 14, 20, 28, 50, 82, 124, 152, 202) magic numbers.

## INTRODUCTION

The history and development of Bethe-Weizsacker mass formulae was presented in details in paper [1] – accuracy for nuclear masess bether or equale to 2.6 MeV .

The purpose of the present work is to obtain the improved, compared with [1] explicit form of of BW formulae as function of Z and N, which describes the values of nuclear masses from most recent evaluation database AME2012 (December 2012 - [2, 3]). The masses extrapolated from systematics and marked with the symbol # in the error column [4] are not taken into account here.

These aim have been reached using Alexandrov dynamic autoregularization method (FORTRAN code REGN-Dubna [5-17]) for solving the overdetermined algebraic systems of equationswhich is constructive development of Tikhonov regularization method [18-20]. One have to note that the use of procedure LCH (developed by Alexandrov and Mavrodiev) permits to dicover the explicit form of unknown function [21-26].

The basis for the classic BW mass formula is sketched in Sec.1 of this paper. The explicit form of numerical generalization of BW mass formula is described in Sec.2. The behavior of different part of generalized BW mass formula like functions of variables A, Z and N is presented in Sec.3. The results and graphical presentation of residuals Res = Expt – Th like functions of variables A, Z and N is presented in Sec. 4. The Fortran source code of the generalized BW mass formula is gived in Appendix A.The description of the experimental nuclear mass values from AME2012 database is given in Table 4, Appendix B. In Appendix C in Table 7 are presented the predicted values of the binding energy, nuclear mass, atomic mass and mass excess for supper havy nuclei analyzed in paper [37] . In **Appendix D** is presented the accuracy of calculation the total $Q_a$ and kinetic $E_a$ energies of alpha decay using the improved BW formulae.

## 1. THE BETHE-WEIZSACKER MASS FORMULA AND BINDING ENERGY OF THE NUCLEUS

The nuclear masses can be calculated from the formula:

$$M_{Nucl}(A,Z) = Zm_p + Nm_n - AE_B(A,Z) \qquad (1)$$

where $Z$ and $N$ are the numbers of protons and neutrons, $m_p$ and $m_n$ are their masses correspondingly, A = Z + N and $E_B(A,Z)$ is the binding energy per nuclei.

In the theory of the liquid drop model, proposed by George Gamow [27, the BW formulae for binding energy per nuclon is given by:

$$E_B(A,Z) = Volume - Surface\frac{1}{A^{\frac{1}{3}}} - Charge\frac{Z(Z-1)^2}{A^{\frac{4}{3}}} - Symmetry\frac{(N-Z)^2}{A^2} + Pairing\ \frac{\boldsymbol{\delta(A,Z)}}{A^{\frac{3}{2}}}, \qquad (2)$$

where

$$\delta(A,Z) = \begin{cases} +1, \text{for even } N,Z \\ -1, \text{for odd } N,Z \\ 0, \quad for \text{ odd } A = Z+N \end{cases} \qquad (3)$$

The improving of formula (1) had been proposed in many papers: [20 – 28].

For performing the digital generalization of BW mass formula we accept that the Volume, Surface, Charge, Symmetry, Pairing and powers of A- 1/3, 4/3, 2, and 3/2 from formulae (2) are s unknown functions of A, Z and unknown parameters a = ($a_i$, i=1,…N).

If we accept the notation $Vol = Volume,\ Cha = Charge,\ Sym = Symmetry,$ $Wig = Pairing$ , magic numbers correction function $K_{MN}(A,Z,a)$ and for powers $P_1(A,Z,a)$, $P_2(A,Z,a)$, $P_3(A,Z,a)$, $P_4(A,Z,a)$ the formula of the binding energy will has a form

$$E_B(A,Z,a) = Vol(A,Z,a) - Sur(A,Z,a)\frac{1}{A^{(P_1(A,Z,a)}} - Cha(A,Z,a)\frac{Z(Z-1)}{A^{P_2(A,Z,a)}} - Sym(A,Z,a)\frac{(N-Z)^2}{A^{P_3(A,Z,a)}} + Wig(A,Z,a)\frac{\delta(A,Z)}{A^{P_4(A,Z,a)}} + K_{MN}(A,Z,a) \qquad (4)$$

For more convenient start of iteration procedure the inicial values of new unknown functions are choosen near to the values of constants from papers [4], [27 – 37] as well as the values of powers in formulae (2). The explicit form of this ten unknown functions will be discovered by the solution of inverse problems, defined from overdetermined systems of nonlinear equations for binding energy

$$E_B^{Expt}(A_j,Z_j) = E_B^{Th}(A_j,Z_j,a), \qquad (5)$$

nuclear mass

$$M_{Nucl}^{Expt}(A_j,Z_j) = M_{Nucl}^{Th}(A_j,Z_j,a), \qquad (6)$$

atomic masses

$$M_{At}^{Expt}(A_j,Z_j) = M_{At}^{Th}(A_j,Z_j,a) \qquad (7)$$

and mass exccess

$$M_{Exc}^{Expt}(A_j,Z_j) = M_{Exc}^{Th}(A_j,Z_j,a), \qquad (8)$$

where $j = 1, \dots, 2564$ and a is a set unknown digital parameters.

The relations between the values of nucleimass $M_{Nucl}(A,Z)$, atomic mass$M_{At}(A,Z)$, mass excess $M_{Exc}(A,Z)$, hydrogen atom mass$m_H$, proton mass $m_P$ and the neutron mass $m_N$ [2-4] are:

$$M_{At}(A,Z) = Zm_H + Nm_N + A\,E_B(A,Z), \qquad (9)$$

$$M_{Nucl}(A,Z) = M_{At}(A,Z) - (Z\,m_e + A_{el}Z^{2.39} + B_{el}Z^{5.35}) \qquad (10)$$

and

$$M_{Exc}(A,Z), \quad = M_{At}(A,Z) - Au, \qquad (11)$$

where $A_{el} = 1.44381 \times 10^{-5}\ MeV$ and $B_{el} = 1.55468 \times 10^{-12}\ MeV$, $a_{N-1}$=2.39, $a_N$=5.35 [19, 30], the mass of the Hydrogen atom$m_H = 938.782303(0.084)\ MeV$, $\mathrm{m_n} = 939.56538(4.56)$ MeV,$\mathrm{m_P} = 938.272\,046(21)$ MeV,$\mathrm{m_{el}} = 0.510\,998\,928(11)$ MeVand $\mathrm{u} = 931.494\,061(21)$ MeV. The $1-\sigma$ uncertainties in the last digits of the above values are given in parentheses after the values.

The error analysis of the digital parameters $\{a_i\}$ need more power computer facilities and work time. So, the correlations between parameters and their exclusion from unknown function will be done in the next research.

The using the LCH procedure, realized in the REGN program, permits us to choose the "better" function out of two functions with the same $\chi^2$.

## 2. The explicit form of the numerical generalization of BW mass formula

The linearly independent arguments of numerical generalization $\{v\} = \{v_i,\ i=1,\dots,9\}$ can be choosen as follow:

$$v_1 = \frac{Z}{A},\ v_2 = \frac{N}{A},\quad v_3 = \frac{N-Z}{A},\quad v_4 = \frac{Z}{N+1},\quad v_5 = log(A+1),\quad v_6 = \frac{1}{log(A+1)}, \qquad (12)$$

where *Z* and *N* are the numbers of protons and neutrons in nuclei, *A* = *Z* + *N* $and$

$$v_7 = \begin{cases} 0, \text{for odd } A \\ 1, \text{for even } A \end{cases},\ v_8 = \begin{cases} 0, \text{for odd } Z \\ 1, \text{for even Z} \end{cases},\ v_9 = \begin{cases} 0, \text{for odd } N \\ 1, \text{for even N} \end{cases} \qquad (13)$$

Solving the ouverdetemined nonlinear algebraic systems (5 -8) with condition Expt=Theory for binding energy, nuclear mass, atomic mass and mass excess, using step by step different models for unknown functions $Vol(A,Z,a)$, $Sur(A,Z,a)$, $Cha(A,Z,a)$, $Sym(A,Z,a)$, $Wig(A,Z,a)$, $P_1(A,Z,a)$, $P_2(A,Z,a)$, $P_3(A,Z,a)$, $P_4(A,Z,a)$ and $K_{MN}(A,Z,a)$, using the possibilities of FORTRAN code REGN [5 – 17] for to choose the "better" function (LCH procedure –[21 - 26], we receive their explicit forms as follow:

$$Vol(A,Z,a) = \exp\left(a_1 + P(v,a,I_s)\right) + C(v,a,N_0), \qquad (14.1)$$

$$Sur(A,Z,a) = \exp\left(a_2 + P(v,a,I_s + Np)\right) + C(v,a,N_0 + N_i), \qquad (14.2)$$

$$Cha(A,Z,a) = \exp\left(a_3 + P(v,a,I_s + 2.Np)\right) + C(v,a,N_0 + 2.N_i), \qquad (14.3)$$

$$Sym(A,Z,a) = \exp\left(a_4 + P(v,a,I_s + 3.Np)\right) + C(v,a,N_0 + 3.N_i), \qquad (14.4)$$

$$Wig(A,Z,a) = \exp\left(a_5 + P(v,a,I_s + 4.Np)\right) + C(v,a,N_0 + 4.)N_i \qquad (14.5).$$

The parametrized powers that were implemented in Eqs.7 have been obtained using the same LHC procedure and defined as:

$$P_1(A,Z,a) = \exp\left(a_6 + P(v,a,I_s + 5.Np)\right) + C(v,a,N_0 + 5.N_i), \qquad (14.6)$$

$$P_2(A,Z,a) = \exp\left(a_7 + P(v,a,I_s + 6.Np)\right) + C(v,a,N_0 + 6.N_i), \qquad (14.7)$$

$$P_3(A,Z,a) = \exp\left(a_8 + P(v,a,I_s + 7.Np)\right) + C(v,a,N_0 + 7.N_i), \qquad (14.8)$$

$$P_4(A,Z,a) = \exp\left(a_9 + P(v,a,I_s + 8.Np)\right) + C(v,a,N_0 + 8.N_i), \qquad (14.9)$$

where

$$P(v,a,i) = \exp\left(-\left(\sum_{j=1}^{3}\sum_{l=1}^{4} a_{i+l+4(j-1)} v_l^j + a_{i+13} v_6 + a_{i+14} v_5\right)^2\right) \qquad (15)$$

and

$$C(v,a,i) = \exp(a_{i+1} v_7 + a_{i+2} v_8 + a_{i+2} v_9) \qquad (16).$$

The different values of the proton and neutron magic numbers were considered in different inverce problems and the dependence on the magic numbers and the bounaries between them were determined.

For obtaining the dependencies from magic numbers and boundaries between them, which define their influence, there was formulated different inverse problems with different values of proton and neutron magic numbers.

The LCH analysis of solutions of different inverse problems gives the explicit form of function $K_{MN}(A,Z,a,i)$ for nine proton and ten neutron magic numbers (see Appendics A) as follow:

$$K_{Mn}(A,Z,a) = \Big(\left[w_Z\left(1 + C(v,a,N_0 + 9.\ N_i)\right) + G(v,a,N_1)\right] \frac{\exp\left(\frac{-(Z-Z_{MN})^2}{2G(v,a,N_1 + 2.N_w)}\right)}{(Z-Z_{MN})^2 + G(v,a,N_1 + 2.N_w)} +$$

$$\left[w_N\left(1 + C(v,a,N_0 + 10.\ N_i)\right) + G(v,a,N_1 + N_w)\right] \frac{\exp\left(\frac{-(N-N_{MN})^2}{2G(v,a,N_1 + 3.N_w)}\right)}{(N-N_{MN})^2 + G(v,a,N_1 + 3.N_w)}\Big)/A^N \qquad (17)$$

where $Z_{MN}$ and $N_{MN}$ are the nearest to *Z* and *N* = *A* - *Z* magic numbers, correspondingly, $w_Z$ and $w_N$ are the half sum of corresponding magic numbers between which are *Z* and *N*.

The explicit form of the function $G(v,a,i)$ is

$$G(v,a,i) = \exp\left(a_{i+15} - \left(\sum_{j=1}^{3}\sum_{l=1}^{4} a_{i+l+4(j-1)} v_l^j + a_{i+13} v_6 + a_{i+14} v_5\right)^2\right) \qquad (18)$$

The integer numbers in formulaes (14 - 18) have the values as follow: $I_s = 9$, $Np = 15 = N_w$, $N_i = 4$, $N_0 = Is.(1 + Np)$, $N_1 = N_0 + N_d$, $N_d = 44$, $\mathrm{N} = N_0 + N_d + 4.N_w + 1 = 249$.

The investigation of parameter's errors and correlations between them will be presented in next work. Our estimation is that after using LCH procedure the number of prarameters will be less in times with the same $\chi^2$.

**3.The behavior of the different part of generalized BW mass formulae like functions of variables A, Z and N**

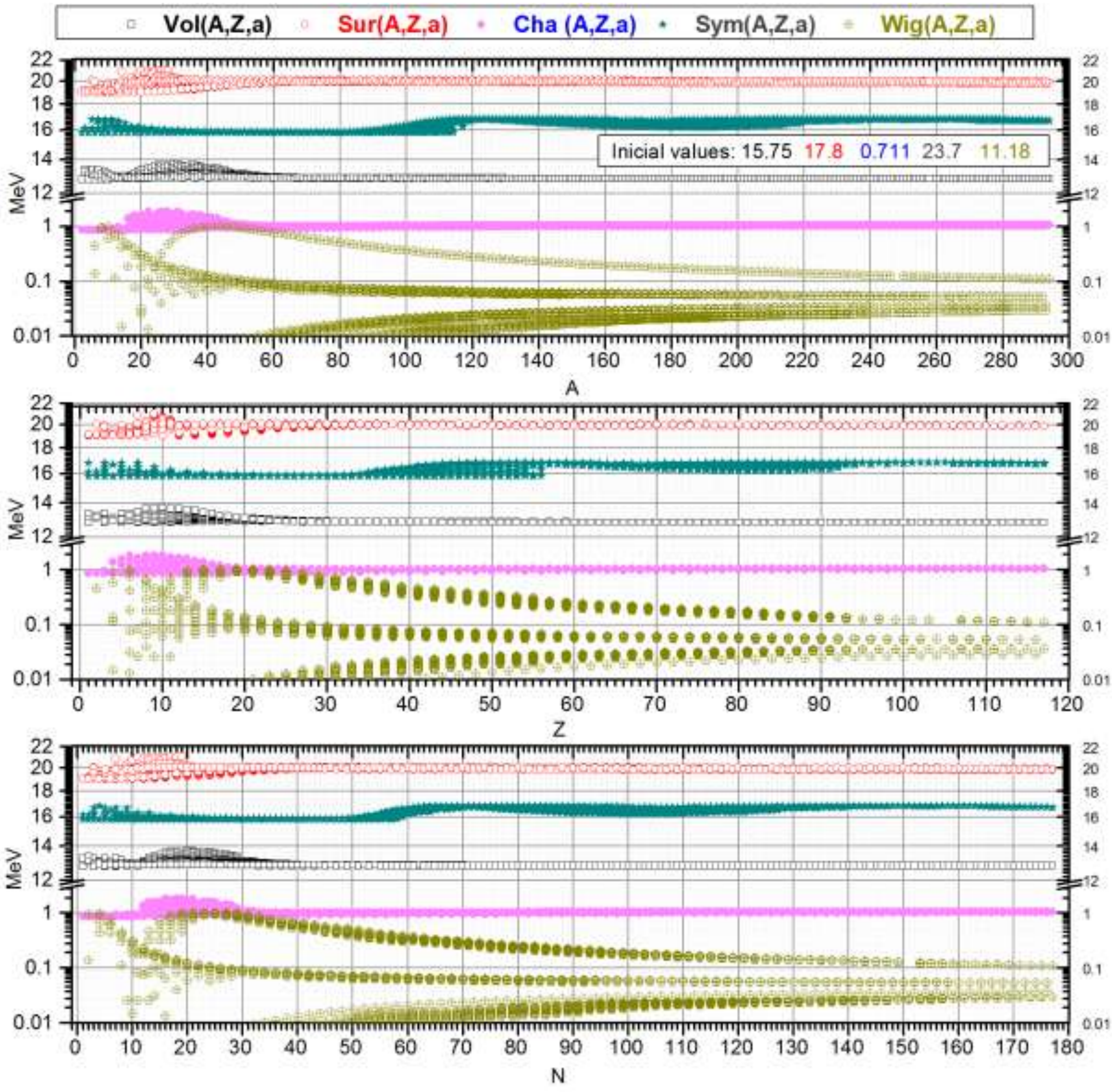

**Fig.1** The behavior of the structures $Vol(A, Z, a)$, $Sur(A, Z, a)$, $Cha(A, Z, a)$, $Sym(A, Z, a)$ *and* $Wig(A, Z, a)$, *see* $Eqs. 14.1 - 5$, as functions of A, Z and N.

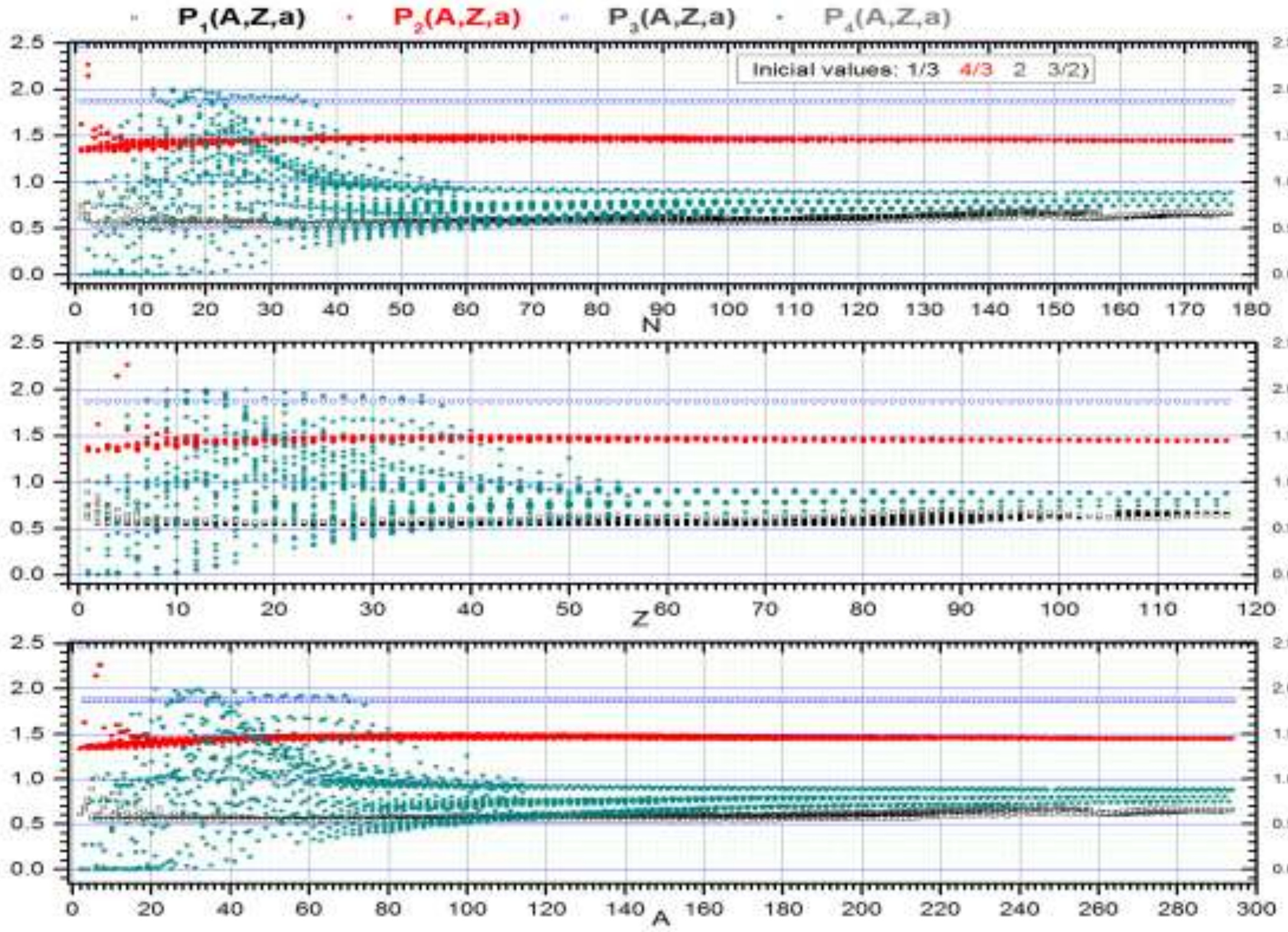

**Fig.2** The behavior of power factors $P_1(A, Z, a)$, $P_2(A, Z, a)$, $P_3(A, Z, a)$, $P_4(A, Z, a)$, see Eqs. 14.6 – 9, as functions of A, Z and N, see Eqs.17.

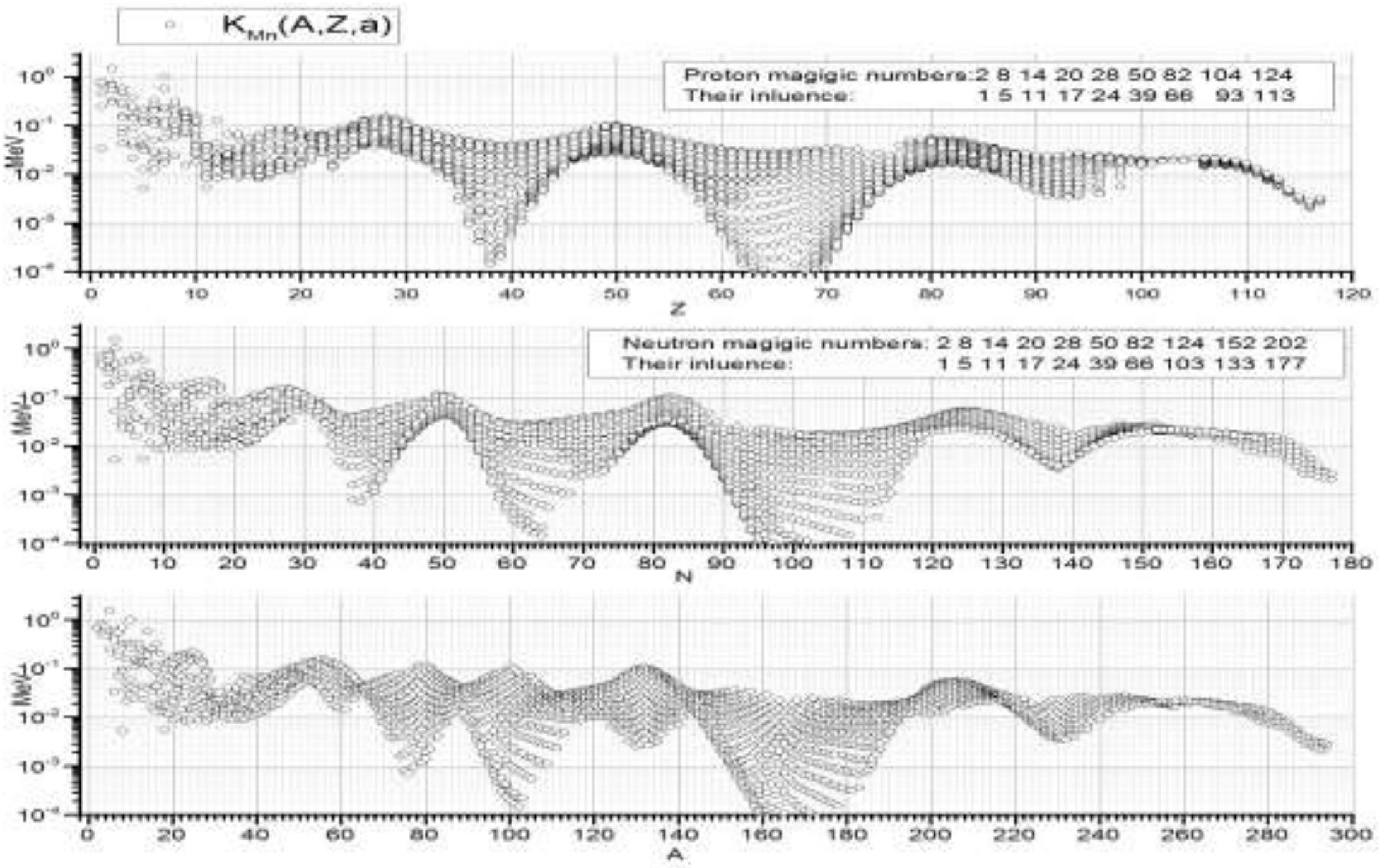

**Fig.3** The magic numbers correction energy function $K_{MN}(A, Z, a, i)$ of the magic numbers, see Eqs.17, as function of the proton $Z$, neutron $N$ and atomic mass number $A$ respectively

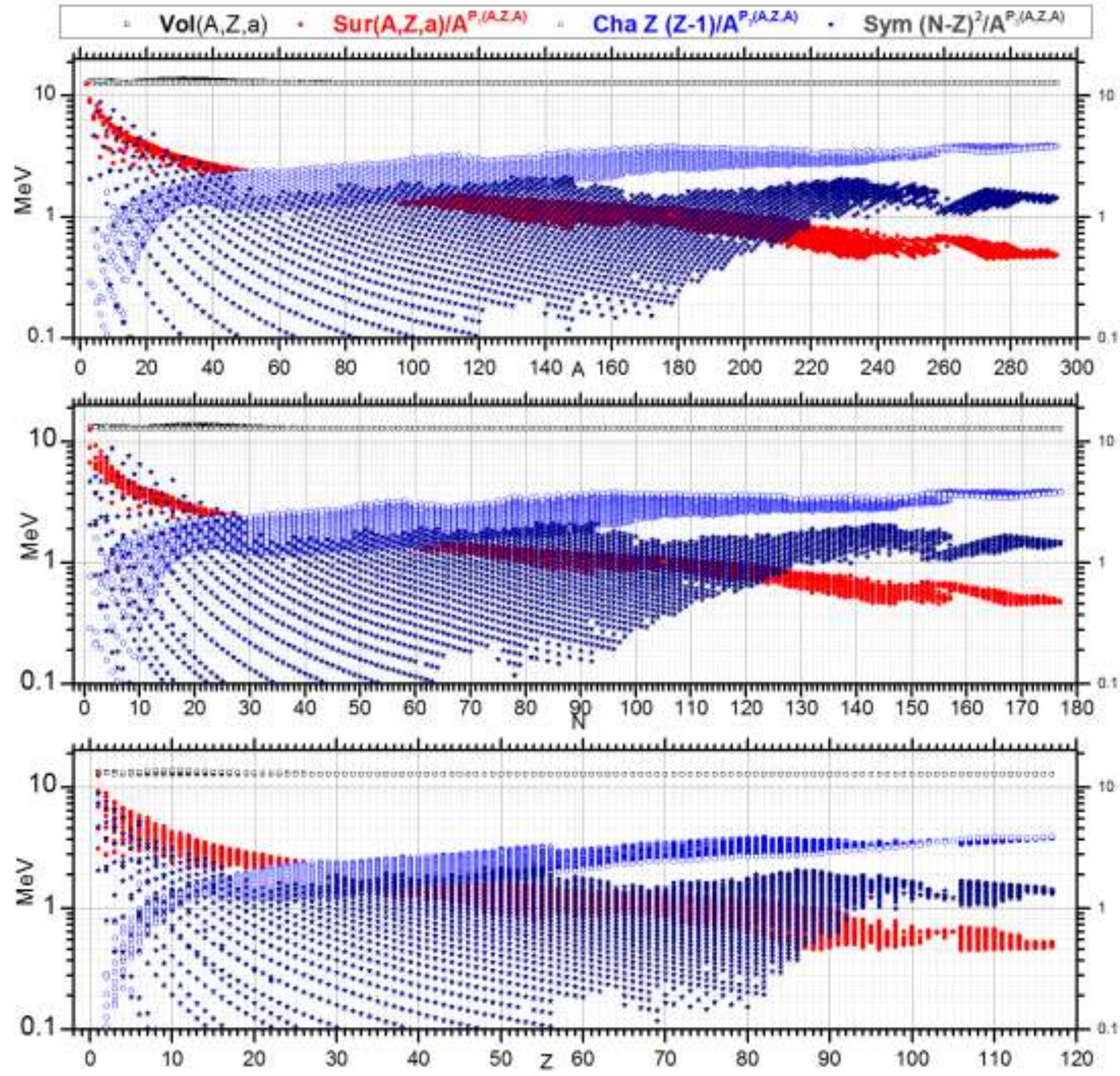

**Fig.4** The behavior of structures $Vol(A,Z,a)$, $Sur(A,Z,a)\frac{1}{A^{P_1(A,Z,a)}}$, $Cha(A,Z,a)\frac{Z(Z-1)}{A^{P_2(A,Z,a)}}$, $Sym(A,Z,a)\frac{(N-Z)^2}{A^{P_3(A,Z,a)}}$ as functions of A, Z and N, see Eqs. 4 and 17.

## 4. Results

The challenge of low energy nuclear physics to describe the dependence of the binding energy, nuclear and atomic mass, mass excess as functions of the number of protons and neutrons is presented.

This result was established by using the ecperimental data from AME2012 [2,3] database, the inverse problem method for discovering the explicit form of unknown theoretical function (model) and the values, based on the REGN (L. Alekasandrov-Regularized Gauss-Newton iteration method) [5 - 17] for solving the overedeterminednonlinear system of equations.One have to note that the LCH-weighting procedure [21-26] of the REGN program permits to choose the better function out of two functions with the same $\chi^2$.

The essential advantage of the Alexandrov method [5 - 17] from other similar methods is extremely effective ideology regularization of inverse problem solution, which on each iteration step controls not only the actual decision, but, very importantly, uncertainty of the solution. At the same time, the transition from the mathematical theory of the autoregularizated iterative processes, which is based on meaningful theorems of convergence L. Aleksandrov [6] to Fortran codes (REGN-Dubna [8], FXY-Sofia-Dubna [17]) is very complicated, but technically clear work.

In the next figures are presented the experimental and theoretical values and its residuals for binding energy, mass excess, nuclear and atomic mass as function of mass number A, proton number Z and neutron number N.

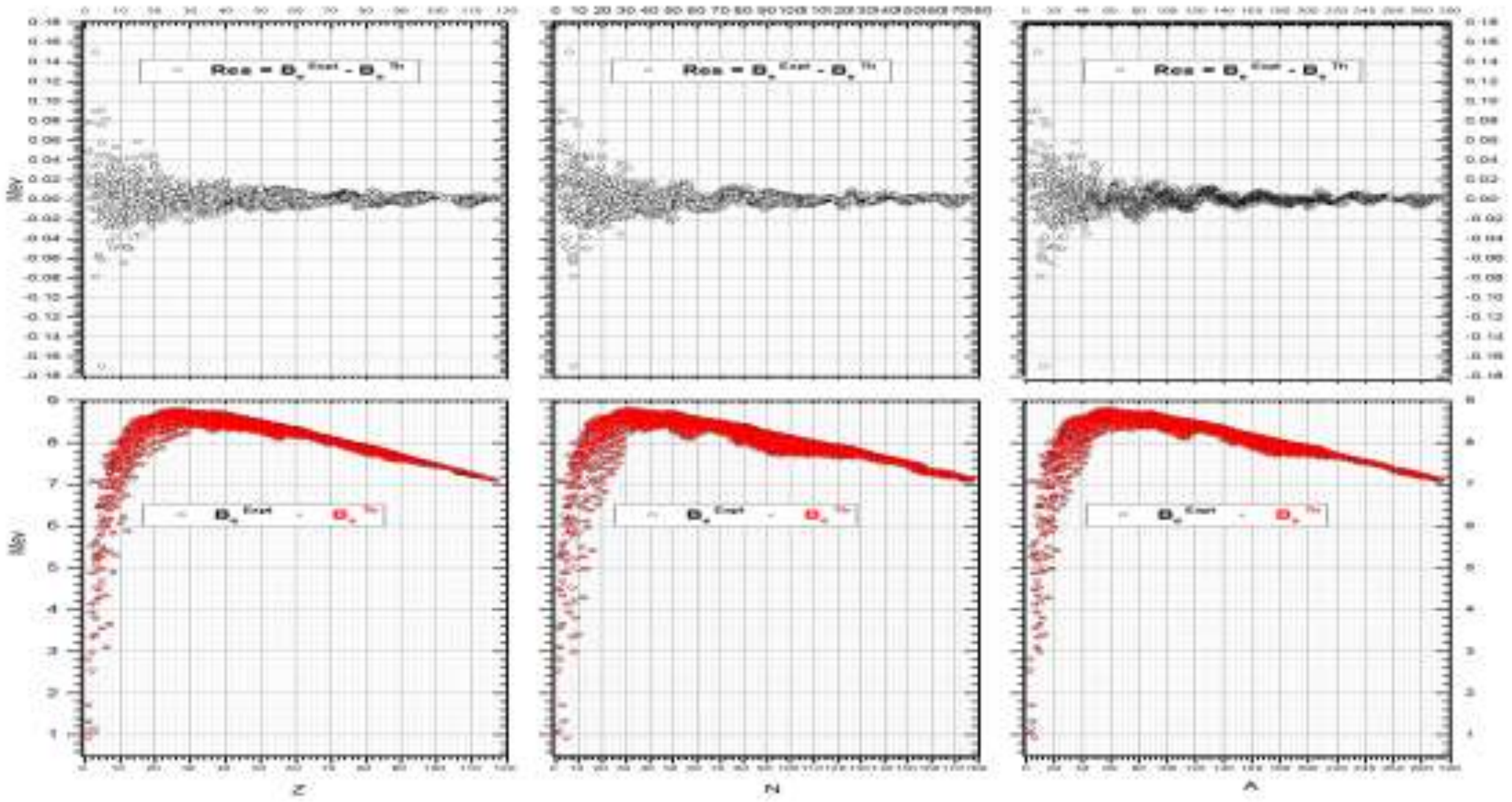

Fig.5 Experiment and theory of the Binding energy and its residuals as function of A, Z and N

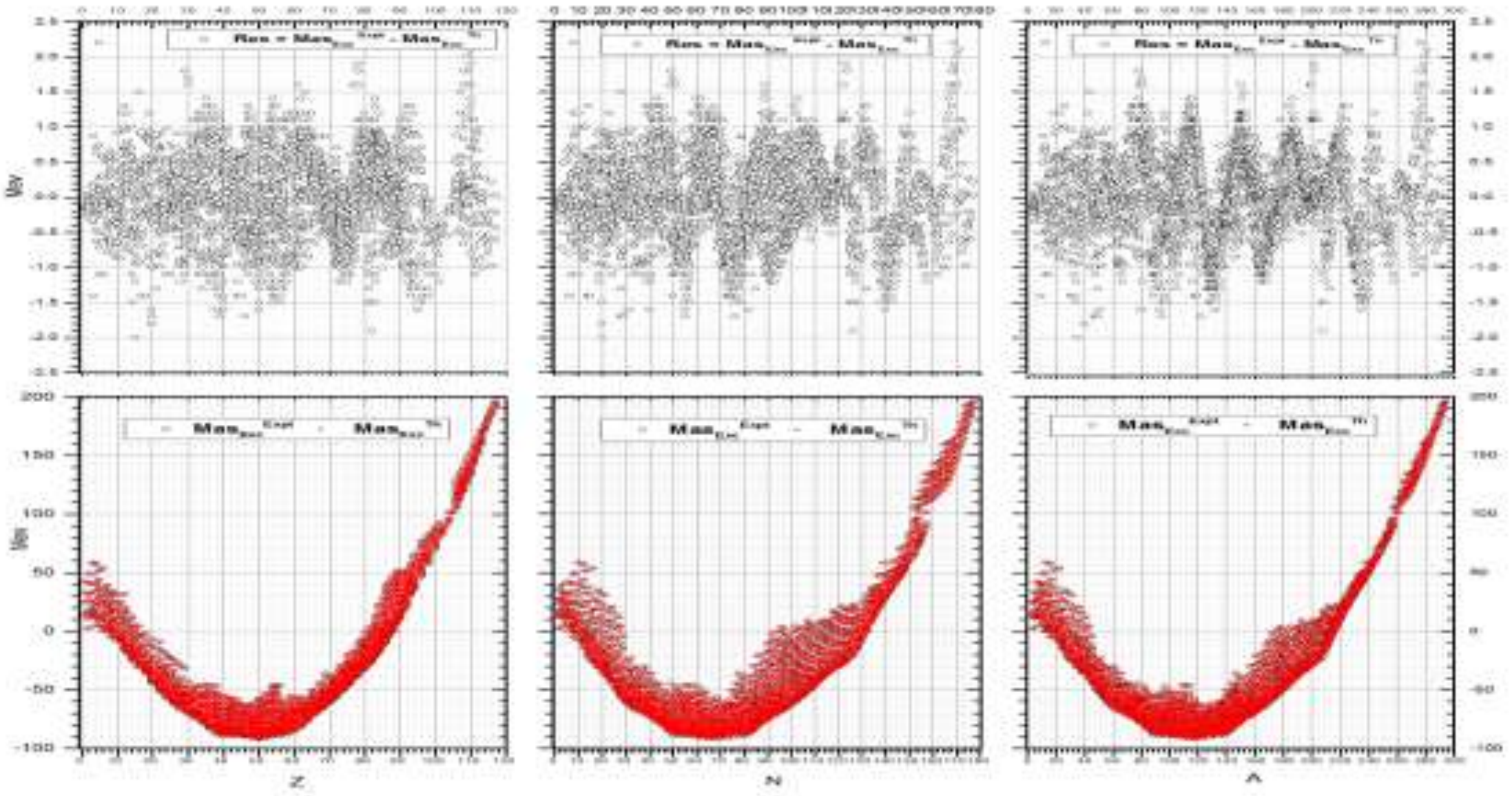

**Fig.6** Experiment and theory of Mass excess and its residuals as function of A, Z and N

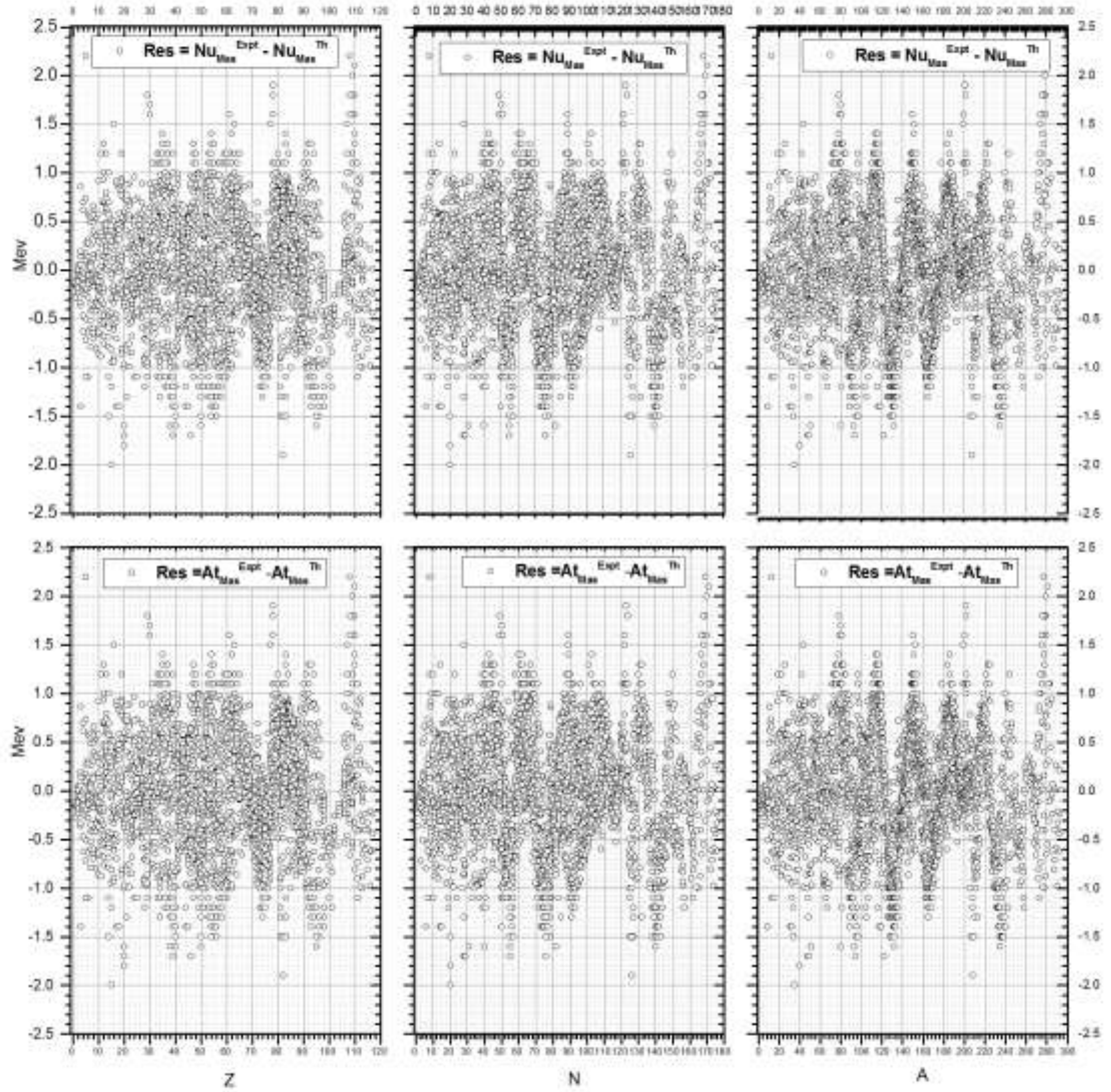

**Fig.7** Residuals of the Nuclear and Atomic masses as functions of mass number A, Z and N

The explicit form for fuctions $E_B^{Th}(A, Z, a)$, $M_{Nucl}^{Th}(A, Z, a)$, $M_{At}^{Th}(A, Z, a)$, $M_{Exc}^{Th}(A, Z, a)$ and the solution for the values of digital parameters {a} (see Application 1) describe 2564 nuclei, atomic masses, mass excess and binding energies starting from A = 2 ( Z = 1, N = 1) with relative error$\bar{\epsilon}_{\rm r}$

$$\bar{\epsilon}_{\rm r} = \frac{1}{N} \sum_{i=1}^{N} \frac{Expt\,(A_i, Z_i) - Th\,(A_i, Z_i, a)}{Expt\,(A_i, Z_i)} \,, \qquad (19)$$

The resuduals, which are the difference between experiment and model values

*Residual = Expt – Th*

bilongs to the interval (-2.0- 2.20) Mev for nuclear, atomic mass and mass excess and to (-0.17, 0.15) for binding energy.

The $\chi^2$ test (estimation of describing accuracy) has been done to determine what significance there is with this value of $\chi^2$ using the formula (see Eq.6 in [28]):

$$\chi^2 = \sum_{k=1}^{M} \left(\frac{Expt\,(A_k,Z_k) - Th(A_k,Z_k,a)}{\sigma(A_k,Z_k)}\right)^2 \qquad (20)$$

where

$$\sigma(A_k,Z_k) = C \times \sigma_{Stat}\,(A_k,Z_k) + Percent\ \times Expt(A_k,Z_k) \qquad (21)$$

Here $\sigma_{Stat}\,(A_k,Z_k)$ is the uncertainty of a nuclei as it has been reported in AME2012, $C$ and $Percent$ are the nuisance parameters, where C=1 and $Percent$ is percentage of the given experimental value. The Table 1 illustrates the quality of descriptions of the binding energy, the nuclear, atomic masses and the mass excess assuming different hypothesis for the $Percent, \bar{\epsilon}_{\mathrm{r}}$ , $\chi^2$ $and$ $\chi_n$,

$$\chi_n = \sqrt{\frac{\chi^2}{M-N}}, \qquad (22)$$

where $\boldsymbol{M-N}$ is the number of degrees of freedom.

Note, that some masses of nuclei are measured with very high precision, which can noticed from mass excess column in AME2012, but due to artificial cut off of the significant digits the uncertainties for these nuclei are given as zero uncertainty. Since we do not know the exact numbers we treat uncertainties for these nuclei as 1% of the given experimental value.

Table 1

| | Percent | $\bar{\epsilon}_{\mathrm{r}}$ | $\chi^2$ | $\chi_n$ |
|---|---|---|---|---|
| $B_e$ | 0.290E-02 | 0.417E+00 | 2564. | 1.052 |
| $M_{Nucl}$ | 0.118E-04 | -0.672E-03 | 2468. | 1.033 |
| $M_{At}$ | 0.118E-04 | -0.672E-03 | 2467. | 1.032 |
| $M_{Exc}$ | 0.217E+00 | 0.241E+02 | 2464. | 1.032 |

The nuclear drip lines are the boundary delimiting the zone of Z, N in which atomic nuclei lose stability due to the transmutation of neutrons (down one) as well as because of Coulomb repulsion of protons (up). To find where these drip lines are on the nuclear landscape we need to know the values of Z and N, where the separation energy is changed it sign. The coordinates, where separation energies change the sign can be calculated by using the explicit form for binding energy $\mathrm{E_B(A,Z,a)}$ (Eq.(4)).

The formulae for two neutrons and two proton separation energies are

$$S_{2p}(Z,N,a) = (Z+N)E_B^{Th}(Z,N,a) - (Z-2+N)E_B^{Th}(Z-2,N,a) \qquad (23)$$

and

$$S_{2n}(Z,N,a) = (Z+N)E_B^{Th}(Z,N,a) - (Z+N-2)E_B^{Th}(Z,N-2,a), \qquad (24)$$

as well as simple algoritm for calculation of coordinates Z, N.

The next two figures illustrate the behavior of calculated two proton and neutron drip- lines and their asymptotic.

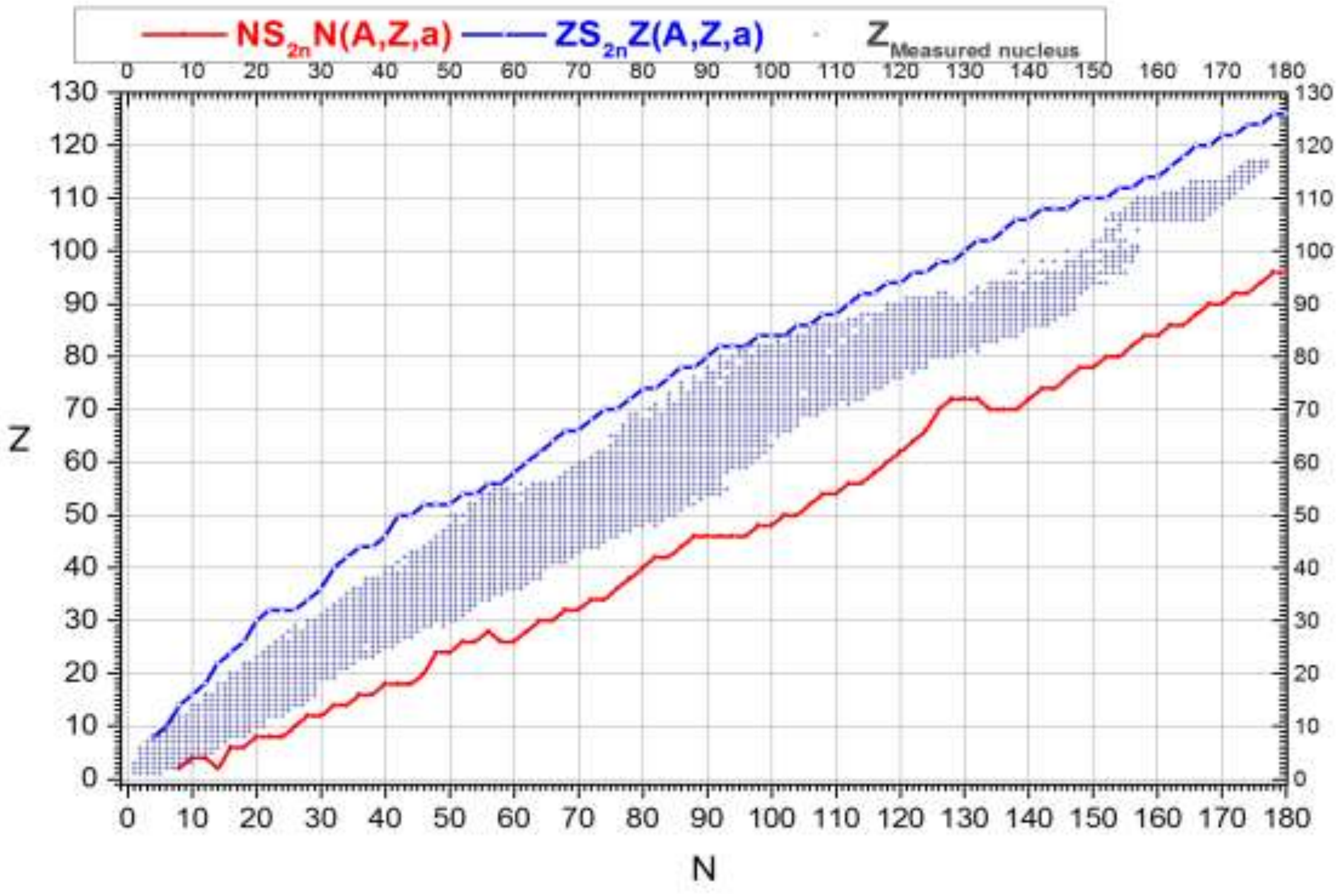

**Fig.8** The behavior of calculated proton and neutron drip- lines compared with Z,N coordinates of 2564 measured nucleus.

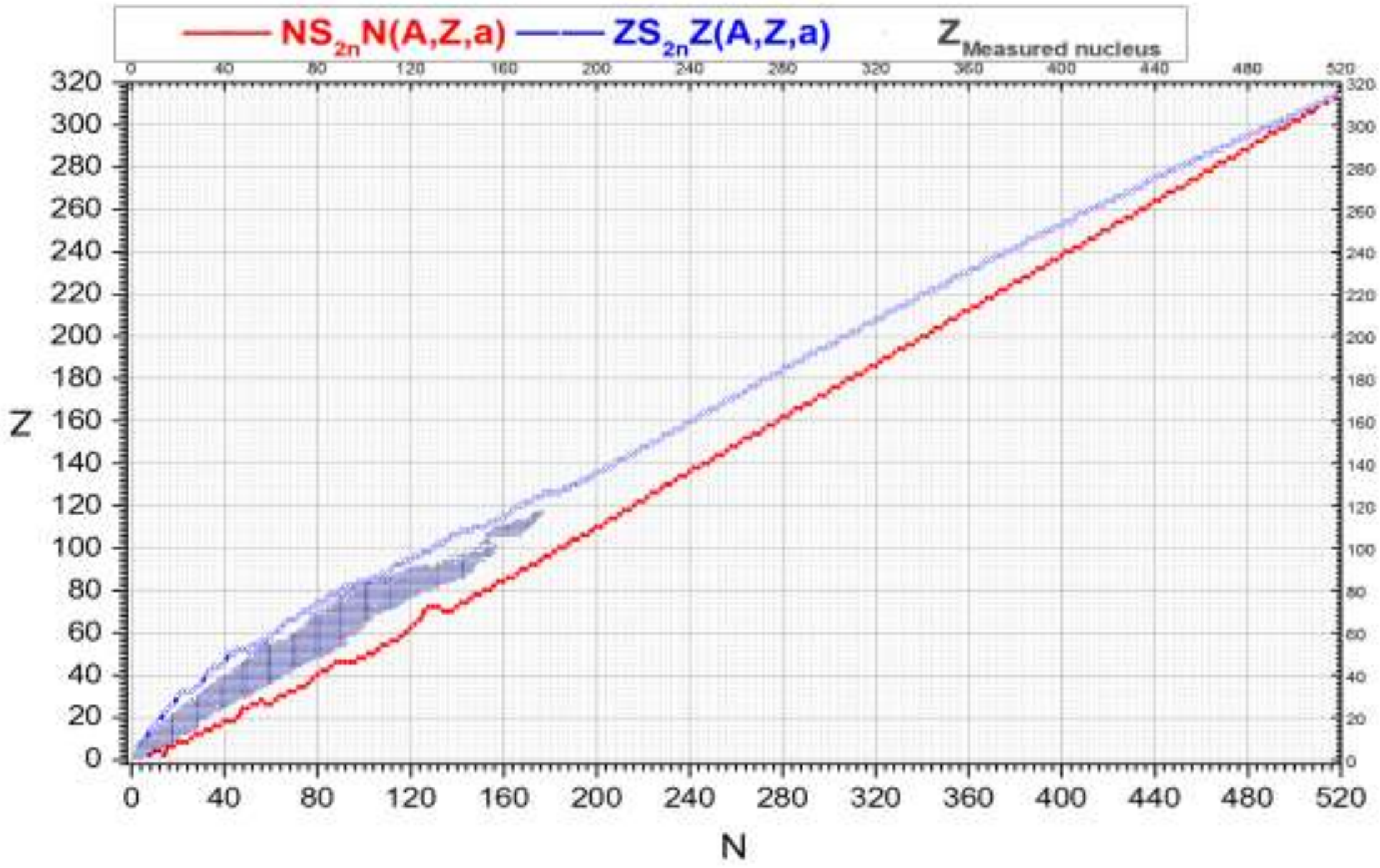

**Fig.9** The asimptotocbehavior of calculated proton and neutron $S_{2n}$ drip-lines.

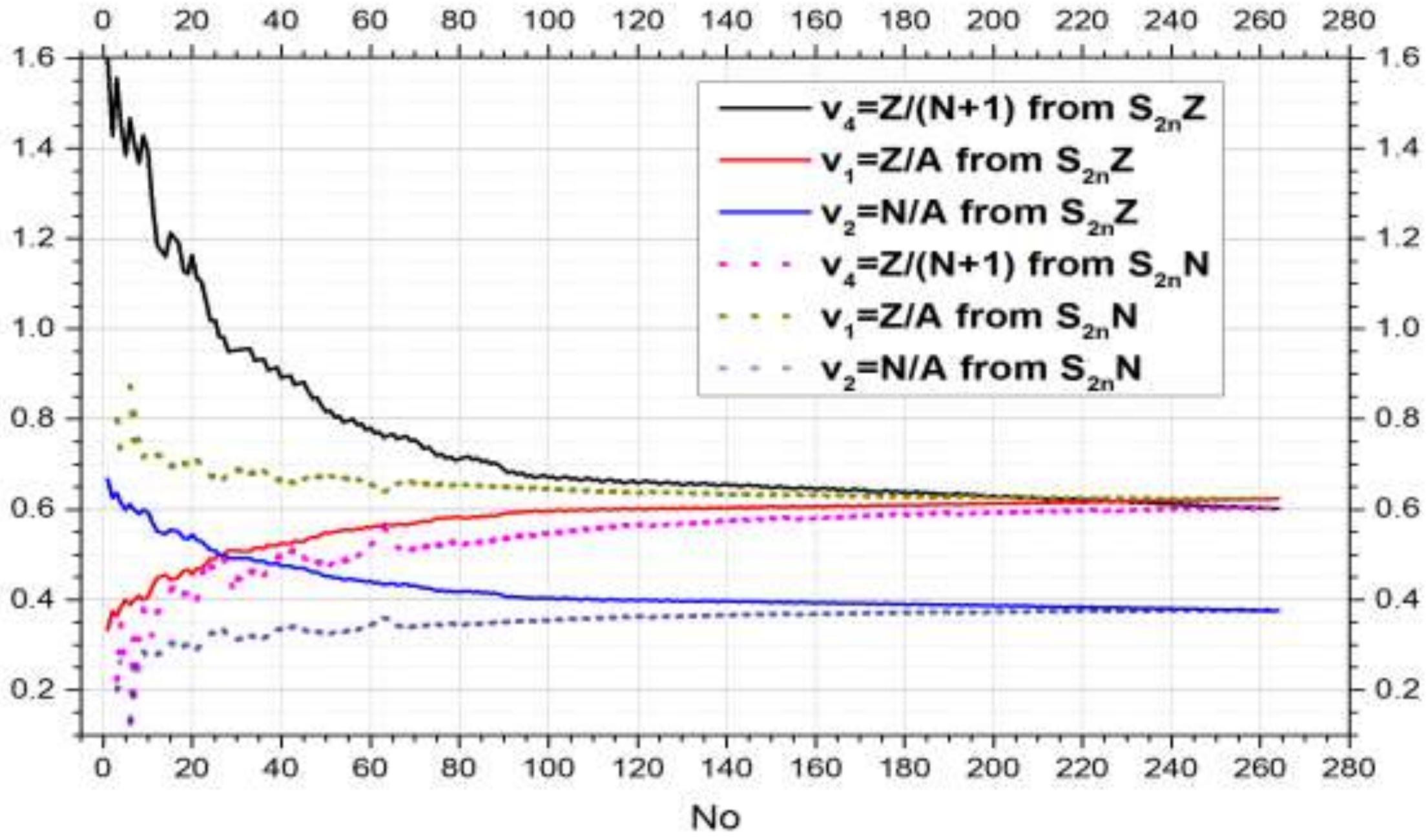

**Fig.10** Grafical explanations of th asymptotic behavior presented in **Fig. 9,** see Eqs.(12) and (23-24)

## Conclusion

In this paper is presented an improved numerical generalization of Bethe- Weizsacker mass formulae which describes the values of measured 2654 nuclear mass in AME2012 nuclear database with residduals from -2.00 to 2.20 MeV for nuclear, atomic mass, mass excess and in interval -0.17 - 0.15 MeV for binding energy.

The rediscovered in paper [1] well known proton and neutron magic numbers as well the discovered (108, 124 for Protons and 152, 202 for Neutrons) new one

| Proton magic numbers | Neutron magic numbers |
|---|---|
| $1 \leq 2 < 5$ | $1 \leq 2 < 5$ |
| $5 \leq 8 < 11$ | $5 \leq 8 < 11$ |
| $11 \leq 14 < 17$ | $11 \leq 14 < 17$ |
| $17 \leq 20 < 24$ | $17 \leq 20 < 24$ |
| $24 \leq 28 < 39$ | $24 \leq 28 < 39$ |
| $39 \leq 50 < 66$ | $39 \leq 50 < 66$ |
| $66 \leq 82 < 95$ | $66 \leq 82 < 103$ |
| $95 \leq 108 < 116$ | $103 \leq 124 < 138$ |
| $116 \leq 124 <$ unknown | $138 \leq 152 < 177$ |
| | $177 \leq 202 <$ unknown |

Table 2: The range of influence of proton and neutron magics numbers obtained in this work.

was confirmed.

The first interesting application of the proposed explicit form of improved numerical generalization of Bethe-Weizsacker mass formulae seems to be the calculation of two proton and neutron drip-lines and their intercept in Z=320, N=530, A=850.

Proposed BW formulae can be used for calculation of not known nuclear mass, the total and kinetc energy of proton, alpha, cluster decays and spontaneously fissions- see Appendix C.

The used in this paper approach for generalization of BW mass formulae can be applied for the actualization the half- life models (see for example [37] which describes the increezing volume of experimental data in NuDat database. Such actualization can be used for reasurching the problems in super havy nuclei stability islands.

**ACKNOWLEDGMENTS: The outhor is thankful to Svetla Drenska nad Maksim Deliyergiyev for fruitfull discutions.**

## REFERENCES

--------------------------------------------

[1] S. CHT. MAVRODIEV AND M.A. DELIYERGIYEV, NUMERICAL GENERALIZATION OF THE BETHE-WEIZSACKER MASS FORMULA, HTTP://ARXIV.ORG/ABS/1602.06777, [NUCL-TH], 22 FEB 2016.

[2] G. AUDI, M. WANG, A. WAPSTRA, F. KONDEV, M. MACCORMICK, X. XU, B. PFEIFFER, THE AME2012 ATOMIC MASS EVALUATION, CHINESE PHYSICS C 36 (12) 1287, 2012.

[3] G. AUDI, F. KONDEV, M. WANG, B. PFEIFFER, X. SUN, J. BLACHOT, M. MACCORMICK, THE NUBASE2012 EVALUATION OF NUCLEAR PROPERTIES, CHINESE PHYSICS C 36 (12) 1157, 2012.

[4] M. W. KIRSON, MUTUAL INFLUENCE OF TERMS IN A SEMI-EMPIRICAL MASS FORMULA, NUCLEAR PHYSICS A 798

[5] L. ALEKSANDROV, REGULARIZED COMPUTATIONAL PROCESS FOR THE ANALYSIS OF EXPONENTIAL PROBLEMS, J. COMP. MATH. AND MATH. PHYS. 10, VOL. 5 ,1285-1287,1970 .

[6] L. ALEKSANDROV, REGULARIZED INTERATIVE PROCESSES FOR THE SOLUTION OF NONLINEAR OPERATIONAL EQUATIONS, COMMUNICATIONS JINR P5-5137, DUBNA, 1970.

[7] L. ALEKSANDROV, AUTOREGULARIZED ITERATION PROCESSES OF NEWTON-CANTAROVICH TYPE, COMMUNICATIONS JINR P5-5515, DUBNA, 1970.

[8] L. ALEKSANDROV, REGULARIZED COMPUTATIONAL PROCESS OF NEWTON CANTAROVICH TYPE, J. COMP. MATH. AND MATH. PHYS., 11, VOL. 1,36-43,1971.

[9] L. ALEKSANDROV, ON THE REGULARIZATION OF ITERATION PROCESSES OF NEWTON TYPE, COMMUNICATIONS JINR P5-7258, DUBNA, 1973.

[10] L. ALEKSANDROV, A PROGRAM FOR THE SOLUTION OF NONLINEAR SYSTEMS OF EQUATIONS BY MEANS OF REGULARIZED ITERATION PROCESSES OF GAUSS-NEWTON TYPE, COMMUNICATIONS JINR P5-7259, DUBNA, 1973.

[11] L. ALEKSANDROV, REGULARIZED PROCESSES OF NEWTON TYPE FOR THE SOLUTION OF NONLINEAR SYSTEMS OF EQUATIONS ON COMPUTER, JINR, B1-5-9869, DUBNA, 1976.

[12] L. ALEKSANDROV, ON NUMERICAL SOLUTION ON COMPUTER OF THE NONLINEAR NONCORRECT PROBLEMS, COMMUNICATIONS JINR P5-10366,DUBNA, 1976.

[13]L. ALEXANDROV, THE PROGRAM REGN (REGULARIZED GAUSS-NEWTON ITERATION METHOD) FOR SOLVING NONLINEAR SYSTEMS OF EQUATIONS, USSR COMPUT. MATH. AND MATH. PHYS. 11 (1) (1970) 36–43, [IN RUSSIAN] ZH. VYCHISL. MAT. MAT. FIZ., JINR DUBNA PREPRINTS: P5-7258, P5-7259, DUBNA, 1973.

[14] L. ALEKSANDROV, PROGRAM SYSTEM REGN FOR SOLUTION OF NONLINEAR ( SYSTEMS OF EQUATIONS, RSIC CODE PACKAGE PSR-165[OAK RIDGE NATIONAL LABORATORY COMPUTER CODE NUMBER: P00165].

[15] L. ALEKSANDROV, THE NEWTON-KANTOROVICH REGULARIZED COMPUTING PROCESSES, USSR COMPUT. MATH. AND MATH. PHYS. 11 (1) (1971) 46 – 57. DOI:HTTP://DX.DOI.ORG/10.1016/0041-5553(71)90098-X.

[16] L. ALEKSANDROV, M. DRENSKA, D. KARADJOV, PROGRAM CODE REGN (CODE SYSTEM FOR SOLVING NONLINEAR SYSTEMS OF EQUATIONS VIA THE GAUSS-NEWTON METHOD), RSIC-RSIC-31JINR 61-11-82-767, DUBNA, 1982.

[17] L. ALEKSANDROV, PROGRAM AFXY (ANALYZE FX=Y) FOR INVESTIGATION OF NONLINEAR SYSTEMS, PRIVATE COMMU NICATIONS, 2005.
[18] A. N. TIKHONOV, V. Y. ARSENIN, METHODS OF SOLUTION OF ILL-POSED PROBLEMS, NAUKA, LENINGRAD, 1986, KLUWER ACADEMIC PUBL., DORDRECHT, 1995.
[19] A. N. TIKHONOV, A. S. LEONOV, A. G. YAGOLA, NONLINEAR ILL-POSED PROBLEMS, NAUKA, LENINGRAD, CHAPMAN AND HALL, 1995.
[20] A. N. TIKHONOV, A. V. GONCHARSKII, V. V. STEPANOV, A. G. YAGOLA, REGULARIZING ALGORITHMS AND A PRIORI INFORMATION, NAUKA, MOSCOW, [IN RUSSIAN ], 1983.
[21] L. ALEKSANDROV, S. C. MAVRODIEV, S.DEPENDENCE OF HIGH-ENERGY HADRON-HADRON TOTAL CROSS-SECTIONS ON QUANTUM NUMBERS, COMM. JINR E2-9936 ,20 P, 1976.
[22] L. ALEKSANDROV, S. DRENSKA, S. CHT. MAVRODIEV, THE DEPENDENCE ON THE QUANTUM NUMBERS OF THE EFFECTIVE RADIUS OF HADRON INTERACTIONS, PHYSICS OF ATOMIC NUCLEI 32 (2). DOI:10.1134/S1063778815060150.
[23] S. DRENSKA, S.CHT. MAVRODIEV, A.N. SISAKIAN, ON THE BEHAVIOR OF INCLUSIVE HADRON CROSS-SECTIONS AT LARGE TRANSFERRED MOMENTA AND CERN ISR AND SPS COLLIDER ENERGIES, JINR-E2-83-587, AUG 1983. 14PP., INT. SEMINAR 'HIGH ENERGY PHYSICS AND QUANTUM FIELD THEORY', PROTVINO, USSR, 1983.
[24 S. DRENSKA, S.CHT. MAVRODIEV, EFFECTIVE RADIUS AND TOTAL CROSS-SECTIONS OF HADRON - HADRON INTERACTIONS OF ELEMENTARY PARTICLES. (IN RUSSIAN), JINR-P2-11159, DEC 1977. 21PP., SOV.J.NUCL.PHYS., 28:385,1978, YAD.FIZ.28:749-760,1978.
**[25]** L. ALEKSANDROV, S. CHT. MAVRODIEV, A.N. SISSAKIAN, ON THE INVESTIGATION OF SOME NONLINEAR PROBLEMS IN HIGH ENERGY PARICLE AND COSMIC RAY PHYSICS, PHYSICS AT THE FUTURE COLLIDERS, OCTOBER 16-21, TBILISI, GEORGIA, 2005
[26] S. DRENSKA, A. GEORGIEVA, V. GUEORGIEV , R. ROUSEV , P. RAYCHEV , UNIFIED DESCRIPTION OF THE LOW LYING STATES OF THE GROUND BANDS OF EVEN-EVEN NUCLEI*,* PHYS*.* REV. C32, 1833, 1976.
[27] G. GAMOW, CONSTITUTION OF ATOMIC NUCLEI AND RADIOACTIVITY, 1931.
[28] J. FRENCH, E. HALBERT, J. MCGRORY, S. WONG, COMPLEX SPECTROSCOPY, ADVANCES IN NUCLEAR PHYSICS 3193–258, 1969.
[29] S. GORIELY, F. TONDEUR, J. PEARSON, A HARTREE-FOCK NUCLEAR MASS TABLE, ATOMIC DATA AND NUCLEAR DATA TABLES 77 (2) , 311 – 381. DOI:HTTP://DX.DOI.ORG/10.1006/ADND.2000.0857, 2001.
[30] G. T. GARVEY, I. KELSON, NEW NUCLIDIC MASS RELATIONSHIP, PHYS. REV. LETT. 16, 197–200, 1966, DOI:10.1103/PHYSREVLETT.16.197.
[31] J. JANECKE AND P.J. MASSON, MASS PREDICTIONS FROM THE GARVEY-KELSON MASS RELATIONS, ATOMIC DATA AND NUCLEAR DATA TABLES 39 (2), 265 – 271. DOI:HTTP://DX.DOI.ORG/10.1016/0092-640X(88) 90028-9, 1988.
[32] S. LIRAN, N. ZELDES, A SEMIEMPIRICAL SHELL-MODEL FORMULA, ATOMIC DATA AND NUCLEAR DATA TABLES , DOI:HTTP://DX. DOI.ORG/10.1016/0092-640X(76)90033-4.
[33] R. NAYAK, L. SATPATHY, MASS PREDICTIONS IN THE INFINITE NUCLEAR MATTER MODEL, ATOMIC DATA AND NUCLEAR DATA TABLES 73 (2) (1999) 213 – 291. DOI:HTTP://DX.DOI.ORG/10.1006/ADND.1999.0819.
[34] J. DUFLO, A. ZUKER, MICROSCOPIC MASS FORMULAS, PHYS. REV. C 52 (1995) R23–R27. DOI:10.1103/PHYSREVC.52.R23. URL HTTP://LINK.APS.ORG/DOI/10.1103/PHYSREVC.52.R23
[35] J. DUFLO, A. P. ZUKER, THE NUCLEAR MONOPOLE HAMILTONIAN, PHYS. REV. C59 (1999) R2347–R2350. DOI:10.1103/PHYSREVC.59.R2347. URL HTTP://LINK.APS.ORG/DOI/10.1103/PHYSREVC.59.R2347
[36] G. T. GARVEY, W. J. GERACE, R. L. JAFFE, I. TALMI, I. KELSON, SET OF NUCLEAR-MASS RELATIONS AND A RESULTANT MASS TABLE, REV. MOD. PHYS. 41S1–S80, DOI:10.1103/REVMODPHYS.41.S, 1969
[37] YU. TS. OGANESSIAN, V.K. UTYONKOV, SUPERHEAVY NUCLEI FROM $^{48}$CA INDUCED REACTIONS, NUCLEAR PHYSICS A 944, 62–98, HTTP://DX.DOI.ORG/10.1016/J.NUCLPHYSA.2015.07.003, 2015

## APENDIGS A

```
      FUNCTION aNuclMass(Proton,aNeutron,BiEnTh,AtMassTh,aMassExcTh)
     IMPLICIT DOUBLE PRECISION(A-H,O-Z)
     common/nhelp/lexpt,iSP,nPow,nBWp,MnZ,MnN,N0,N1,Nqn
    common/Structure/c1,c2,c3,c4,c5,pow1,pow2,pow3,pow4
     common/bwigner/BWZ,BWN,CorMn,BetheWeizsacker
     common/help1/Vol,Sur,Cha,Sym,Wig,CorectionMN
     common/variables/AA,a1,a12,a13,a2,a22,a23,a3,a32,a33,a4,a42,a43,a5,a6,AAA,AAZ,AAN
     DIMENSION A(249)
A(  1)=  0.25507416191372E+01; A(  2)=  0.29480322227196E+01; A(  3)= -0.15240794353594E+00; A(  4)=  0.27613067152223E+01; A(  5)= -0.14608531349138E+02;
A(  6)= -0.60910715941880E+00; A(  7)=  0.28938286067961E+00; A(  8)=  0.62797867413768E+00; A(  9)= -0.64541092218686E+01; A( 10)=  0.60617738340954E+01;
A( 11)= -0.25239057703281E+02; A( 12)= -0.24802772686869E+02; A( 13)= -0.34894373553149E+01; A( 14)=  0.42265262354888E+01; A( 15)= -0.16867830544139E+02;
A( 16)= -0.25978663300929E+02; A( 17)= -0.55805059209406E+02; A( 18)= -0.14296562680995E+02; A( 19)= -0.90728241772492E+01; A( 20)=  0.10517549079459E+03;
A( 21)=  0.31007164406804E+02; A( 22)=  0.92081832141815E+02; A( 23)=  0.85523056628911E+01; A( 24)= -0.12478033498314E+02; A( 25)=  0.34119508067253E+02;
A( 26)= -0.17548718215732E+02; A( 27)= -0.51996525481234E+02; A( 28)=  0.87901786730791E+02; A( 29)= -0.10579631642206E+02; A( 30)= -0.61909390001168E+02;
A( 31)= -0.16218654332653E+03; A( 32)= -0.34647679384139E+03; A( 33)= -0.86047356196937E+02; A( 34)= -0.31932453217226E+02; A( 35)=  0.37051397812016E+03;
A( 36)=  0.21349622755167E+03; A( 37)=  0.34990103635321E+02; A( 38)=  0.14562907221657E+02; A( 39)=  0.17628284980770E+02; A( 40)=  0.26065110339320E+02;
A( 41)= -0.41349886019565E+02; A( 42)= -0.60198346817141E+02; A( 43)= -0.62825496305597E+02; A( 44)=  0.44854668929078E+02; A( 45)= -0.17097388296446E+02;
A( 46)=  0.61639048303003E+02; A( 47)= -0.29061194467403E+02; A( 48)=  0.46590734525005E+02; A( 49)=  0.25828489181034E+00; A( 50)= -0.83717845271728E+01;
A( 51)=  0.15511019804264E+02; A( 52)=  0.12058757755147E+03; A( 53)=  0.12128631061721E+02; A( 54)= -0.83023410732454E+01; A( 55)=  0.34060108010142E+03;
A( 56)= -0.97789821427346E+02; A( 57)= -0.41078512289741E+03; A( 58)= -0.32449245336032E+03; A( 59)=  0.30197302386850E+03; A( 60)= -0.12128440149168E+03;
A( 61)=  0.13021094330472E+03; A( 62)= -0.45648431792406E+03; A( 63)=  0.17802293740510E+03; A( 64)= -0.29545546426174E+02; A( 65)=  0.45456572777452E+03;
A( 66)=  0.27238788912592E+03; A( 67)=  0.21195485374410E+03; A( 68)=  0.11421368624124E+02; A( 69)=  0.22053384705547E+03; A( 70)=  0.66456987007168E+00;
A( 71)=  0.67819483341362E+00; A( 72)=  0.51238275124587E+00; A( 73)=  0.66054280105701E+00; A( 74)=  0.58068506990505E+00; A( 75)=  0.59375099958725E+00;
A( 76)=  0.50424792921496E+00; A( 77)=  0.57529607704856E+00; A( 78)=  0.53989062081345E+00; A( 79)=  0.54962466883348E+00; A( 80)=  0.50144795131888E+00;
A( 81)=  0.53434149547078E+00; A( 82)=  0.80970461792498E+00; A( 83)=  0.88282671796841E+00; A( 84)=  0.84339731582067E+00; A( 85)=  0.14263448222761E+02;
A( 86)=  0.53127623197596E+01; A( 87)= -0.25403261437180E+01; A( 88)= -0.14485345328915E+02; A( 89)=  0.17884433735423E+01; A( 90)=  0.13398429976175E+02;
A( 91)= -0.17680225859764E+02; A( 92)=  0.15389274808763E+02; A( 93)=  0.13961246614744E+01; A( 94)=  0.33202560640427E+01; A( 95)=  0.57632220940582E+01;
A( 96)= -0.43395216772207E+01; A( 97)= -0.42965950934656E+01; A( 98)= -0.56597615406018E+00; A( 99)= -0.40561452494548E+01; A(100)= -0.35631155909735E+02;
A(101)= -0.34488620740778E+01; A(102)=  0.25896052223211E+02; A(103)=  0.10264705887015E+03; A(104)= -0.30273306463950E+02; A(105)=  0.70619973908207E+01;
A(106)=  0.11763471867012E+02; A(107)= -0.52471943889610E+02; A(108)= -0.10842640148347E+02; A(109)= -0.30819723720464E+01; A(110)=  0.78252695301474E+01;
A(111)=  0.11481257309520E+02; A(112)=  0.52832505413751E+02; A(113)=  0.20575510612006E+01; A(114)= -0.52970902082589E+02; A(115)=  0.58404470169100E+00;
A(116)=  0.64420482386302E+00; A(117)=  0.56473703487318E+00; A(118)=  0.58662107374489E+00; A(119)=  0.52805205987106E+00; A(120)=  0.59122058494864E+00;
A(121)=  0.50879639761575E+00; A(122)=  0.53778680328172E+00; A(123)=  0.50673919266588E+00; A(124)=  0.55333944906857E+00; A(125)=  0.49570651697715E+00;
A(126)=  0.51450347209055E+00; A(127)=  0.66852146352280E+00; A(128)=  0.82381955336915E+00; A(129)=  0.73407745570732E+00; A(130)=  0.79493640747106E+02;
A(131)= -0.64656654640051E+02; A(132)= -0.14478571283225E+03; A(133)=  0.88846627193616E+02; A(134)=  0.12650245770061E+03; A(135)= -0.18013553004714E+02;
A(136)=  0.20031436632192E+03; A(137)= -0.23663412582440E+03; A(138)=  0.10274048786932E+03; A(139)=  0.53657166944923E+02; A(140)=  0.26151116610173E+03;
A(141)=  0.69694010795576E+02; A(142)= -0.29538622533278E+02; A(143)=  0.40574328628229E+01; A(144)=  0.91139403091438E+01; A(145)= -0.53312251965351E+01;
A(146)=  0.19259225826202E+01; A(147)=  0.32261517791472E+01; A(148)= -0.18541010014966E+01; A(149)=  0.13249503481173E+02; A(150)=  0.11072302249996E+02;
A(151)=  0.11947708229257E+02; A(152)= -0.53481404958122E+00; A(153)=  0.10114248073711E+02; A(154)=  0.46261487743981E+01; A(155)=  0.52411582077914E+01;
```

```
A(156)=  0.12314068853845E+01; A(157)= -0.39891848295882E+02; A(158)=  0.58311290117070E+01; A(159)=  0.87722866609269E+01; A(160)= -0.18564150326085E+01;
A(161)=  0.72831841603052E+02; A(162)= -0.16864086111513E+02; A(163)= -0.14351540852334E+02; A(164)= -0.17316214465805E+01; A(165)=  0.64875543459977E+01;
A(166)= -0.42814963482145E+01; A(167)= -0.37001151004671E+02; A(168)=  0.14062320111665E+02; A(169)=  0.27377053500450E+01; A(170)=  0.12516306376308E+01;
A(171)=  0.14369002462184E+01; A(172)=  0.15132899786070E+01; A(173)= -0.15823312209869E+02; A(174)=  0.50508829992826E+01; A(175)= -0.94591972708125E+01;
A(176)=  0.86317001284259E+01; A(177)= -0.13224567908162E+02; A(178)=  0.16379618258142E+02; A(179)= -0.46128808198233E+01; A(180)=  0.39333294025268E+00;
A(181)= -0.13926376077819E+02; A(182)=  0.55197943596927E+01; A(183)=  0.24225525144603E+02; A(184)= -0.30647578529774E+00; A(185)=  0.11003794043948E+01;
A(186)=  0.81263804986962E+00; A(187)=  0.17881132239828E+01; A(188)= -0.64177701407203E+01; A(189)= -0.12863314586693E+03; A(190)= -0.10949706797996E+03;
A(191)=  0.18671543496175E+02; A(192)=  0.50279187172749E+02; A(193)= -0.48028359133688E+02; A(194)= -0.29243567627517E+02; A(195)=  0.82942062819614E+02;
A(196)=  0.15649466746047E+02; A(197)=  0.24781043254308E+01; A(198)=  0.11295507986138E+00; A(199)= -0.66128324530029E+02; A(200)= -0.17316945208955E+02;
A(201)=  0.12249257915447E+03; A(202)=  0.17859910624676E+02; A(203)=  0.39599907514277E+01; A(204)= -0.48170297315741E+03; A(205)= -0.25756661283479E+03;
A(206)=  0.20415736040862E+03; A(207)=  0.63459661450534E+03; A(208)= -0.12487662352516E+03; A(209)=  0.84935976818059E+02; A(210)=  0.67467416089113E+03;
A(211)= -0.14758224587331E+03; A(212)=  0.24284896249588E+03; A(213)=  0.71281171889176E+02; A(214)= -0.13650305819098E+04; A(215)= -0.56130536604560E+02;
A(216)= -0.15164293882114E+03; A(217)= -0.11368543801350E+02; A(218)=  0.45663967463640E+01; A(219)=  0.71545263865989E+02; A(220)=  0.40014679350369E+02;
A(221)= -0.21712487341746E+02; A(222)= -0.57271891081000E+01; A(223)=  0.35548815228447E+02; A(224)= -0.95472233995775E+01; A(225)= -0.66624503468226E+02;
A(226)= -0.82568814476865E+02; A(227)= -0.26357468887255E+01; A(228)= -0.44256155134240E+00; A(229)=  0.13231975233085E+03; A(230)=  0.46232430670616E+02;
A(231)= -0.53793834109989E+02; A(232)= -0.21307033916800E+01; A(233)=  0.39416453400446E+01; A(234)= -0.78518977765971E+02; A(235)= -0.41158373404377E+02;
A(236)= -0.73097467533245E+01; A(237)=  0.15090415062851E+03; A(238)= -0.64256071927857E+02; A(239)= -0.67787890697522E+02; A(240)= -0.59251645617283E+02;
A(241)= -0.16833414874648E+03; A(242)= -0.89606531543556E+02; A(243)= -0.12143451488117E+02; A(244)=  0.29648125749522E+03; A(245)=  0.69031158138327E+02;
A(246)=  0.12122745041242E+03; A(247)=  0.54884238069675E+01; A(248)=  0.10085869536863E+02; A(249)=  0.59777558299538E+00

    pi2 = 2.d0*0.3141592653589793D1
   HAtomMass = 938.782303d0     ! 2.072E-20 [MeV]
   aNeutronMass = 939.56538d0   !2.072*10^(-14) [MeV]
   ProtonMass = 938.272046d0
   ElectronMass = 0.510998928d0
   u = 931.494061D0
   Ael = 1.44381E-05
   Bel = 1.55468E-12

  iStr = 5; iPow = 4;  iSP= iStr + iPow ! number in inicials
  nPow = 15;   nBWp= 15; nBW = 4*nBWp
  MnZ = 9; MnN = 10; Ndop = 44
  N0 = isp*(1+nPow);  N1= N0+Ndop; Nqn = N1+nBW
   N = Nqn  + 1 !; print *,N; pause

  AA = Proton + aNeutron
   a1 = Proton/AA; a12 = a1**2; a13 = a1**3
    a2 = aNeutron/AA; a22 = a2**2; a23 = a2**3
     a3 = (aNeutron-Proton)/AA; a32 = a3**2; a33 = a3**3
```

```
        a4 = Proton/(aNeutron+1); a42 = a4**2; a43 = a4**3
         a5 = dlog(AA+1.d0); a6 = 1.d0/a5
       Z0 =  FunZ(N,a, Proton,WZ)
      aN0 =  FunN(N,A,aNeutron,WN )
 !         print *,AA,Proton,aNeutron,Z0,aN0; pause
       AAA = 1.D0
          if( int(AA/2)*2.ne.AA) AAA = 0.D0
       ZZZ = 1.D0
          if( int(Proton/2)*2.ne.Proton) ZZZ = 0.D0
       ANN = 1.D0
          if( int(aNeutron/2)*2.ne.aNeutron) ANN = 0.D0
    AAA = AAA/AA; AAZ = ZZZ/Proton; AAN = ANN/(aNeutron+1.d0)
           CorMN =  BrWig(Proton,Z0,WZ,aNeutron,aN0,WN,a,n,N1)
!   N0 = isp(1+nPow)  N1= N0+nBW; Nqn = N1+15 +38 qn parameters
    CorectionMN = CorMN
     c1 = eexp( a(1)) + CorPow(a,n,isp)             +  CorS(a,n,N0)          ! (15.75d0)
      c2=  eexp( a(2)) + CorPow(a,n,isp + nPow )     +  CorS(a,n,N0+4)       ! (17.8d0)
       c3 = eexp( a(3)) + CorPow(a,n,isp + 2*nPow )   +   CorS(a,n,N0+8)     ! ( 0.711d0)
        c4 = eexp( a(4)) + CorPow(a,n,isp + 3*nPow )   +   CorS(a,n,N0+12)    ! (23.7d0)
         c5 = eexp( a(5)) + CorPow(a,n,isp + 4*nPow )   +    CorS(a,n,N0+16) ! (11.18d0)
      pow1 =  eexp( a(6)) + CorPow(a,n,isp + 5*nPow )    +    CorS(a,n,N0+20)    ! 1.d0/3
       pow2 =  eexp( a(7)) + CorPow(a,n,isp + 6*nPow )    +    CorS(a,n,N0+24)   ! 4.d0/3
        pow3 =  eexp( a(8)) + CorPow(a,n,isp + 7*nPow )    +    CorS(a,n,N0+28)  ! 2.d0
         pow4 =  eexp( a(9)) + CorPow(a,n,isp + 8*nPow )    +    CorS(a,n,N0+32) ! 3.d0/2
!       Print *,'N0=',N0,'N1=',N1,'Nqn=',Nqn, ' N =', N; pause
      Vol =               c1                                              !1
       Sur =              c2/AA**Pow1                                                                  !2  1/3
        Cha =             c3*Proton*(Proton-1.d0)/AA**Pow2                                      !3  4/3 ZZ(ZZ-1.d0)
         Sym =  c4*(aNeutron-Proton)**2/AA**Pow3                               !4  2 2 (AN-ZZ)**2
       if( int(AA/2)*2.ne.AA) WigE=0.d0 !  A odd
       if( int(Proton/2)*2.ne.Proton.and. int(aNeutron/2)*2.ne.aNeutron) WigE =-1.d0      !odd odd
       if( int(Proton/2)*2.eq.Proton.and. int(aNeutron/2)*2.eq.aNeutron) WigE = 1.d0      !even even
      Wig = c5*WigE/AA**Pow4       !5 +-0 3/2
          BetheWeizsacker = Vol - Sur - Cha - Sym + Wig
     BiEnTh = BetheWeizsacker + CorMN
      AtMassTh = Proton*HAtomMass + aNeutron*aNeutronMass - AA*BiEnTh
       aNuclMass = AtMassTh - (ElectronMass*Proton + Ael*Proton**2.39D0 + Bel*Proton**5.35D0)
```

```
        aMassExcTh = AtMassTh - AA*u
       RETURN;  END
!******************************************************************************************************************
      Function CorS(a,n,i)
      IMPLICIT DOUBLE PRECISION(A-H,O-Z)
      common/variables/AA,a1,a12,a13,a2,a22,a23,a3,a32,a33,a4,a42,a43,a5,a6,AAA,AAZ,AAN
      dimension a(n)
      CorS =  eexp(- (a(i+1)*AAA + a(i+2)*AAZ + a(i+3)*AAN + a(i+4))**2 )
      return; end
!******************************************************************************************************************C
      Function CorPow(a,n,i)
      IMPLICIT DOUBLE PRECISION(A-H,O-Z)
      common/variables/AA,a1,a12,a13,a2,a22,a23,a3,a32,a33,a4,a42,a43,a5,a6,AAA,AAZ,AAN
      dimension a(n)
          c1 = a(i+1)*a1 + a(i+2)*a2 + a(i+3)*a3 + a(i+4)*a4
           c2 = a(i+5)*a12 + a(i+6)*a22 + a(i+7)*a32 + a(i+8)*a42
            c3 = a(i+9)*a13 + a(i+10)*a23 + a(i+11)*a33 + a(i+12)*a43
             c4 =  a(i+13)*a6   + a(i+14)*a5 + a(i+15)
      CorPow =  eexp(- ( c1 + c2  + c3 + c4  )**2 )
      return; end
!******************************************************************************************************************C
      Function BrWig(Z,Z0,WZ,aN,aN0,WN,a,n,i)
      IMPLICIT DOUBLE PRECISION(A-H,O-Z)
      common/variables/AA,a1,a12,a13,a2,a22,a23,a3,a32,a33,a4,a42,a43,a5,a6,AAA,AAZ,AAN
     common/bwigner/BWZ,BWN,CorMn,BetheWeizsacker
      common/nhelp/lexpt,iSP,nPow,nBWp,MnZ,MnN,N0,N1,Nqn
      dimension a(n)
      Z02 = (Z-Z0)**2; aN02 = (aN-aN0)**2; A02 = (AA-AA0)**2
        AmpZ =  CorAmp(a,n,i) + WZ*(1.d0 + CorS(a,n,N0+36))
         AmpN =   CorAmp(a,n,i +  nBWp) + WN*(1.d0 + CorS(a,n,N0+40))   !
        GamZ =   CorGam(a,n,i + 2*nBWp ) + WZ
         GamN =   CorGam(a,n,i + 3*nBWp ) + WN
        BWZ  =  AmpZ*eexp(-Z02/GamZ)/(Z02+GamZ)
         BWN  =  AmpN*eexp(-aN02/GamN)/(aN02+GamN)
      BrWig =   (BWZ  +  BWN)/AA**a(N)
!       print '(2f5.0,3f6.0,6e12.4,2e14.6)',Z,aN,wZ,wN,WA,AmpZ,AmpN,AmpA,BwZ,BwN,BwA,a(i+15),a(i+15+nBWp)  ; pause
!              Print *,'N0=',N0,'N1=',N1,'Nqn=',Nqn, ' N =', N
```

```
! pause
      return; end
!****************************************************************************************************************************C
      Function CorAmp(a,n,i)
      IMPLICIT DOUBLE PRECISION(A-H,O-Z)
      common/variables/AA,a1,a12,a13,a2,a22,a23,a3,a32,a33,a4,a42,a43,a5,a6,AAA,AAZ,AAN
      dimension a(n)
          c1 = a(i+1)*a1 + a(i+2)*a2 + a(i+3)*a3 + a(i+4)*a4
           c2 = a(i+5)*a12 + a(i+6)*a22 + a(i+7)*a32 + a(i+8)*a42
            c3 = a(i+9)*a13  + a(i+10)*a23  + a(i+11)*a33  + a(i+12)*a43
              c4 =  a(i+13)*a6   + a(i+14)*a5
      CorAmp =   eexp(a(i+15) - ( c1 + c2 + c3  + c4  )**2)
      return; end
!****************************************************************************************************************************C
      Function CorGam(a,n,i)
      IMPLICIT DOUBLE PRECISION(A-H,O-Z)
      common/variables/AA,a1,a12,a13,a2,a22,a23,a3,a32,a33,a4,a42,a43,a5,a6,AAA,AAZ,AAN
      dimension a(n)
          c1 = a(i+1)*a1 + a(i+2)*a2 + a(i+3)*a3 + a(i+4)*a4
           c2 = a(i+5)*a12 + a(i+6)*a22 + a(i+7)*a32 + a(i+8)*a42
            c3 = a(i+9)*a13 + a(i+10)*a23 + a(i+11)*a33 + a(i+12)*a43
             c4 =  a(i+13)*a6   + a(i+14)*a5
      CorGam =  eexp( a(i+15) - ( c1 + c2 + c3 + c4  )**2)
      return; end
!****************************************************************************************************************************C
      Function  FunZ(N,A,x,WZ)
      IMPLICIT DOUBLE PRECISION (A-H,O-Z)
!      Proton Magic numbers 2,8,14,20,28,50,82,108,124
     common/nhelp/lexpt,iSP,nPow,nBWp,MnZ,MnN,N0,N1,Nqn
      dimension aMn(9) ,aB(9),A(N)
     data aMn/2.d0,8.d0,14.d0,20.d0,28.d0,50.d0,82.d0,108.d0,124.d0/
     data aB/1.d0,5.d0,11.d0,17.d0,24.d0,39.d0,66.d0,95.d0,116.d0/
     !   do i = 1, MnZ; aMn(i) =  (a(Nqn+i));  aB(i) =   (a(Nqn + MnZ + i))
     !   enddo
        do i = 1,MnZ - 1
         if(x.ge.aB(i).and.x.lt.aB(i+1))    then
          FunZ = int(aMn(i))
```

```
          Wz   = (aMn(i+1) - aMn(i))/2
         return; endif
        enddo
         if(x.ge.aB(MnZ))        FunZ = int(aMn(MnZ))
          if(x.ge.aB(MnZ))    WZ  = (aMn(MnZ) - aMn(MnZ-1))/2
!               Print *,'N0=',N0,'N1=',N1,'Nqn=',Nqn, ' N =', N
!  pause
      return; end
  !************************************************************************************************************C
      Function  FunN(N,A,x,WN)
      IMPLICIT DOUBLE PRECISION (A-H,O-Z)
!     Neutron Magic numbers 2,8,14,20,28,50,82,124,152,202
     common/nhelp/lexpt,iSP,nPow,nBWp,MnZ,MnN,N0,N1,Nqn
      dimension aMn(10),aB(10),A(N)
     data aMn/2.d0,8.d0,14.d0,20.d0,28.d0,50.d0,82.d0,124.d0,152.d0,202.d0/
    data aB/1.d0,5.d0,11.d0,17.d0,24.d0,39.d0,66.d0,103.d0,138.d0,178.d0/
!       do i = 1, MnN; aMn(i) =  (a(Nqn + 2*MnZ + i));  aB(i) =   (a(Nqn + 2*MnZ + MnN + i));
!       enddo
        do i=1,MnN-1
         if(x.ge.aB(i).and.x.lt.aB(i+1))   Then
          FunN = int(aMn(i))
           WN   = (aMn(i+1) - aMn(i))/2
         return; endif
        enddo
         if(x.ge.aB(MnN))      FunN = int(aMn(MnN))
          if(x.ge.aB(MnN))      WN   = (aMn(MnN) - aMn(MnN-1))/2
!                Print *,'N0=',N0,'N1=',N1,'Nqn=',Nqn, ' N =', N
! pause
      return; end
```

# APPENDIX B

| No | El | A | Z | N | N-Z | $B_e^{Expt}$ | $B_e^{Th}$ | $ResB_e$ | $Nu_{Mass}^{Expt}$ | $Nu_{Mass}^{Th}$ | $ResNu_{Mass}$ | $At_{Mass}^{Expt}$ | $At_{Mass}^{Th}$ | $ResAt^{Mass}$ | $Mas_{Exc}^{Expt}$ | $Mas_{Exc}^{Th}$ | $ResMas_{Exc}$ |
|---|---|---|---|---|---|---|---|---|---|---|---|---|---|---|---|---|---|
| 1 | H | 2 | 1 | 1 | 0 | 1.112283 | 1.033632 | 7.9E-02 | 1875.613 | 1875.769 | -1.6E-01 | 1876.124 | 1876.280 | -1.6E-01 | 13.136 | 13.292 | -1.6E-01 |
| 2 | H | 3 | 1 | 2 | 1 | 2.827266 | 2.777549 | 5.0E-02 | 2808.921 | 2809.069 | -1.5E-01 | 2809.432 | 2809.580 | -1.5E-01 | 14.950 | 15.098 | -1.5E-01 |
| 3 | He | 3 | 2 | 1 | -1 | 2.572681 | 2.494993 | 7.8E-02 | 2808.391 | 2808.623 | -2.3E-01 | 2809.413 | 2809.645 | -2.3E-01 | 14.931 | 15.163 | -2.3E-01 |
| 4 | H | 4 | 1 | 3 | 2 | 1.720450 | 1.685368 | 3.5E-02 | 3750.086 | 3750.226 | -1.4E-01 | 3750.597 | 3750.737 | -1.4E-01 | 24.621 | 24.761 | -1.4E-01 |
| 5 | He | 4 | 2 | 2 | 0 | 7.073915 | 7.062302 | 1.2E-02 | 3727.379 | 3727.424 | -4.5E-02 | 3728.401 | 3728.446 | -4.5E-02 | 2.425 | 2.470 | -4.5E-02 |
| 6 | Li | 4 | 3 | 1 | -2 | 1.153761 | 1.064487 | 8.9E-02 | 3749.766 | 3750.121 | -3.6E-01 | 3751.299 | 3751.654 | -3.6E-01 | 25.323 | 25.678 | -3.5E-01 |
| 7 | H | 5 | 1 | 4 | 3 | 1.336360 | 1.312550 | 2.4E-02 | 4689.851 | 4689.970 | -1.2E-01 | 4690.363 | 4690.481 | -1.2E-01 | 32.892 | 33.011 | -1.2E-01 |
| 8 | He | 5 | 2 | 3 | 1 | 5.512132 | 5.521208 | -9.1E-03 | 4667.679 | 4667.633 | 4.7E-02 | 4668.701 | 4668.655 | 4.7E-02 | 11.231 | 11.184 | 4.7E-02 |
| 9 | Li | 5 | 3 | 2 | -1 | 5.266132 | 5.250901 | 1.5E-02 | 4667.616 | 4667.690 | -7.4E-02 | 4669.149 | 4669.223 | -7.4E-02 | 11.679 | 11.753 | -7.4E-02 |
| 10 | H | 6 | 1 | 5 | 4 | 0.961640 | 0.914511 | 4.7E-02 | 5630.329 | 5630.611 | -2.8E-01 | 5630.840 | 5631.122 | -2.8E-01 | 41.876 | 42.158 | -2.8E-01 |
| 11 | He | 6 | 2 | 4 | 2 | 4.878519 | 4.861478 | 1.7E-02 | 5605.534 | 5605.635 | -1.0E-01 | 5606.556 | 5606.657 | -1.0E-01 | 17.592 | 17.693 | -1.0E-01 |
| 12 | Li | 6 | 3 | 3 | 0 | 5.332331 | 5.330097 | 2.2E-03 | 5601.518 | 5601.529 | -1.2E-02 | 5603.051 | 5603.062 | -1.2E-02 | 14.087 | 14.098 | -1.1E-02 |
| 13 | Be | 6 | 4 | 2 | -2 | 4.487247 | 4.470562 | 1.7E-02 | 5605.295 | 5605.392 | -9.7E-02 | 5607.339 | 5607.437 | -9.7E-02 | 18.375 | 18.472 | -9.7E-02 |
| 14 | He | 7 | 2 | 5 | 3 | 4.123057 | 4.122498 | 5.6E-04 | 6545.509 | 6545.512 | -2.8E-03 | 6546.531 | 6546.534 | -2.8E-03 | 26.073 | 26.076 | -2.5E-03 |
| 15 | Li | 7 | 3 | 4 | 1 | 5.606439 | 5.613890 | -7.5E-03 | 6533.832 | 6533.778 | 5.4E-02 | 6535.365 | 6535.311 | 5.4E-02 | 14.907 | 14.853 | 5.4E-02 |
| 16 | Be | 7 | 4 | 3 | -1 | 5.371548 | 5.354030 | 1.8E-02 | 6534.183 | 6534.303 | -1.2E-01 | 6536.227 | 6536.347 | -1.2E-01 | 15.769 | 15.889 | -1.2E-01 |
| 17 | B | 7 | 5 | 2 | -3 | 3.558705 | 3.536763 | 2.2E-02 | 6545.579 | 6545.729 | -1.5E-01 | 6548.135 | 6548.285 | -1.5E-01 | 27.677 | 27.827 | -1.5E-01 |
| 18 | He | 8 | 2 | 6 | 4 | 3.924520 | 3.924666 | -1.5E-04 | 7482.540 | 7482.537 | 2.2E-03 | 7483.562 | 7483.560 | 2.2E-03 | 31.610 | 31.607 | 2.6E-03 |
| 19 | Li | 8 | 3 | 5 | 2 | 5.159712 | 5.183193 | -2.3E-02 | 7471.365 | 7471.175 | 1.9E-01 | 7472.898 | 7472.708 | 1.9E-01 | 20.946 | 20.756 | 1.9E-01 |
| 20 | Be | 8 | 4 | 4 | 0 | 7.062435 | 7.043381 | 1.9E-02 | 7454.849 | 7454.999 | -1.5E-01 | 7456.894 | 7457.044 | -1.5E-01 | 4.942 | 5.091 | -1.5E-01 |
| 21 | B | 8 | 5 | 3 | -2 | 4.717153 | 4.626653 | 9.1E-02 | 7472.318 | 7473.039 | -7.2E-01 | 7474.874 | 7475.594 | -7.2E-01 | 22.922 | 23.642 | -7.2E-01 |
| 22 | C | 8 | 6 | 2 | -4 | 3.101524 | 3.086318 | 1.5E-02 | 7483.949 | 7484.067 | -1.2E-01 | 7487.016 | 7487.134 | -1.2E-01 | 35.064 | 35.182 | -1.2E-01 |
| 23 | He | 9 | 2 | 7 | 5 | 3.349029 | 3.335285 | 1.4E-02 | 8423.360 | 8423.483 | -1.2E-01 | 8424.382 | 8424.505 | -1.2E-01 | 40.936 | 41.058 | -1.2E-01 |
| 24 | Li | 9 | 3 | 6 | 3 | 5.037768 | 4.887475 | 1.5E-01 | 8406.868 | 8408.219 | -1.4E+00 | 8408.401 | 8409.752 | -1.4E+00 | 24.955 | 26.305 | -1.4E+00 |
| 25 | Be | 9 | 4 | 5 | 1 | 6.462668 | 6.437614 | 2.5E-02 | 8392.750 | 8392.973 | -2.2E-01 | 8394.795 | 8395.018 | -2.2E-01 | 11.348 | 11.571 | -2.2E-01 |
| 26 | B | 9 | 5 | 4 | -1 | 6.257070 | 6.199978 | 5.7E-02 | 8393.307 | 8393.818 | -5.1E-01 | 8395.863 | 8396.373 | -5.1E-01 | 12.416 | 12.927 | -5.1E-01 |
| 27 | C | 9 | 6 | 3 | -3 | 4.337423 | 4.317493 | 2.0E-02 | 8409.290 | 8409.465 | -1.8E-01 | 8412.357 | 8412.533 | -1.8E-01 | 28.911 | 29.086 | -1.8E-01 |
| 28 | He | 10 | 2 | 8 | 6 | 2.997616 | 2.979404 | 1.8E-02 | 9363.090 | 9363.272 | -1.8E-01 | 9364.112 | 9364.294 | -1.8E-01 | 49.172 | 49.353 | -1.8E-01 |
| 29 | Li | 10 | 3 | 7 | 4 | 4.531351 | 4.497823 | 3.4E-02 | 9346.460 | 9346.793 | -3.3E-01 | 9347.993 | 9348.326 | -3.3E-01 | 33.053 | 33.386 | -3.3E-01 |
| 30 | Be | 10 | 4 | 6 | 2 | 6.497630 | 6.473836 | 2.4E-02 | 9325.503 | 9325.739 | -2.4E-01 | 9327.548 | 9327.783 | -2.4E-01 | 12.607 | 12.843 | -2.4E-01 |
| 31 | B | 10 | 5 | 5 | 0 | 6.475075 | 6.491397 | -1.6E-02 | 9324.435 | 9324.269 | 1.7E-01 | 9326.991 | 9326.824 | 1.7E-01 | 12.051 | 11.884 | 1.7E-01 |
| 32 | C | 10 | 6 | 4 | -2 | 6.032034 | 6.027298 | 4.7E-03 | 9327.572 | 9327.615 | -4.3E-02 | 9330.639 | 9330.682 | -4.3E-02 | 15.699 | 15.742 | -4.3E-02 |
| 33 | N | 10 | 7 | 3 | -4 | 3.643664 | 3.651246 | -7.6E-03 | 9350.162 | 9350.081 | 8.0E-02 | 9353.740 | 9353.660 | 8.0E-02 | 38.800 | 38.719 | 8.1E-02 |
| 34 | Li | 11 | 3 | 8 | 5 | 4.155381 | 4.233863 | -7.8E-02 | 10285.629 | 10284.764 | 8.6E-01 | 10287.162 | 10286.298 | 8.6E-01 | 40.728 | 39.863 | 8.7E-01 |
| 35 | Be | 11 | 4 | 7 | 3 | 5.952540 | 6.008582 | -5.6E-02 | 10264.567 | 10263.948 | 6.2E-01 | 10266.611 | 10265.993 | 6.2E-01 | 20.177 | 19.558 | 6.2E-01 |
| 36 | B | 11 | 5 | 6 | 1 | 6.927716 | 6.939113 | -1.1E-02 | 10252.546 | 10252.418 | 1.3E-01 | 10255.102 | 10254.974 | 1.3E-01 | 8.668 | 8.539 | 1.3E-01 |
| 37 | C | 11 | 6 | 5 | -1 | 6.676374 | 6.687441 | -1.1E-02 | 10254.017 | 10253.892 | 1.3E-01 | 10257.084 | 10256.959 | 1.3E-01 | 10.650 | 10.524 | 1.3E-01 |
| 38 | N | 11 | 7 | 4 | -3 | 5.364038 | 5.408193 | -4.4E-02 | 10267.159 | 10266.669 | 4.9E-01 | 10270.738 | 10270.248 | 4.9E-01 | 24.304 | 23.813 | 4.9E-01 |
| 39 | Li | 12 | 3 | 9 | 6 | 3.799100 | 3.810905 | -1.2E-02 | 11225.315 | 11225.171 | 1.4E-01 | 11226.848 | 11226.705 | 1.4E-01 | 48.920 | 48.776 | 1.4E-01 |
| 40 | Be | 12 | 4 | 8 | 4 | 5.720722 | 5.779237 | -5.9E-02 | 11200.962 | 11200.257 | 7.0E-01 | 11203.006 | 11202.301 | 7.0E-01 | 25.078 | 24.373 | 7.1E-01 |
| 41 | B | 12 | 5 | 7 | 2 | 6.631221 | 6.692904 | -6.2E-02 | 11188.742 | 11187.999 | 7.4E-01 | 11191.298 | 11190.554 | 7.4E-01 | 13.369 | 12.626 | 7.4E-01 |
| 42 | C | 12 | 6 | 6 | 0 | 7.680144 | 7.683105 | -3.0E-03 | 11174.861 | 11174.822 | 3.9E-02 | 11177.928 | 11177.889 | 3.9E-02 | 0.000 | -0.040 | 4.0E-02 |
| 43 | N | 12 | 7 | 5 | -2 | 6.170109 | 6.174902 | -4.8E-03 | 11191.688 | 11191.626 | 6.2E-02 | 11195.266 | 11195.204 | 6.2E-02 | 17.338 | 17.275 | 6.3E-02 |
| 44 | O | 12 | 8 | 4 | -4 | 4.890195 | 4.901927 | -1.2E-02 | 11205.753 | 11205.607 | 1.5E-01 | 11209.843 | 11209.697 | 1.5E-01 | 31.915 | 31.768 | 1.5E-01 |
| 45 | Li | 13 | 3 | 10 | 7 | 3.403015 | 3.397946 | 5.1E-03 | 12166.230 | 12166.294 | -6.4E-02 | 12167.763 | 12167.827 | -6.4E-02 | 58.341 | 58.405 | -6.4E-02 |
| 46 | Be | 13 | 4 | 9 | 5 | 5.241435 | 5.205479 | 3.6E-02 | 12141.037 | 12141.502 | -4.7E-01 | 12143.081 | 12143.546 | -4.7E-01 | 33.659 | 34.124 | -4.6E-01 |
| 47 | B | 13 | 5 | 8 | 3 | 6.496406 | 6.664271 | -1.7E-01 | 12123.429 | 12121.243 | 2.2E+00 | 12125.984 | 12123.799 | 2.2E+00 | 16.562 | 14.376 | 2.2E+00 |
| 48 | C | 13 | 6 | 7 | 1 | 7.469849 | 7.387465 | 8.2E-02 | 12109.480 | 12110.547 | -1.1E+00 | 12112.547 | 12113.614 | -1.1E+00 | 3.125 | 4.192 | -1.1E+00 |
| 49 | N | 13 | 7 | 6 | -1 | 7.238863 | 7.235055 | 3.8E-03 | 12111.189 | 12111.234 | -4.5E-02 | 12114.768 | 12114.813 | -4.5E-02 | 5.345 | 5.390 | -4.4E-02 |
| 50 | O | 13 | 8 | 5 | -3 | 5.811762 | 5.861812 | -5.0E-02 | 12128.448 | 12127.792 | 6.6E-01 | 12132.538 | 12131.882 | 6.6E-01 | 23.115 | 22.459 | 6.6E-01 |
| 51 | Be | 14 | 4 | 10 | 6 | 4.993897 | 4.948539 | 4.5E-02 | 13078.826 | 13079.459 | -6.3E-01 | 13080.871 | 13081.504 | -6.3E-01 | 39.954 | 40.587 | -6.3E-01 |
| 52 | B | 14 | 5 | 9 | 4 | 6.101644 | 6.120582 | -1.9E-02 | 13062.024 | 13061.756 | 2.7E-01 | 13064.580 | 13064.312 | 2.7E-01 | 23.664 | 23.395 | 2.7E-01 |
| 53 | C | 14 | 6 | 8 | 2 | 7.520319 | 7.474314 | 4.6E-02 | 13040.869 | 13041.509 | -6.4E-01 | 13043.936 | 13044.577 | -6.4E-01 | 3.020 | 3.660 | -6.4E-01 |
| 54 | N | 14 | 7 | 7 | 0 | 7.475614 | 7.493891 | -1.8E-02 | 13040.201 | 13039.941 | 2.6E-01 | 13043.780 | 13043.519 | 2.6E-01 | 2.863 | 2.602 | 2.6E-01 |
| 55 | O | 14 | 8 | 6 | -2 | 7.052301 | 7.025989 | 2.6E-02 | 13044.834 | 13045.197 | -3.6E-01 | 13048.924 | 13049.287 | -3.6E-01 | 8.007 | 8.370 | -3.6E-01 |
| 56 | F | 14 | 9 | 5 | -4 | 5.285208 | 5.323110 | -3.8E-02 | 13068.279 | 13067.742 | 5.4E-01 | 13072.881 | 13072.344 | 5.4E-01 | 31.964 | 31.427 | 5.4E-01 |
| 57 | B | 15 | 5 | 10 | 5 | 5.880002 | 5.804882 | 7.5E-02 | 13998.813 | 13999.936 | -1.1E+00 | 14001.368 | 14002.492 | -1.1E+00 | 28.958 | 30.081 | -1.1E+00 |
| 58 | C | 15 | 6 | 9 | 3 | 7.100169 | 7.120343 | -2.0E-02 | 13979.216 | 13978.910 | 3.1E-01 | 13982.283 | 13981.977 | 3.1E-01 | 9.873 | 9.566 | 3.1E-01 |
| 59 | N | 15 | 7 | 8 | 1 | 7.699460 | 7.664374 | 3.5E-02 | 13968.933 | 13969.455 | -5.2E-01 | 13972.512 | 13973.034 | -5.2E-01 | 0.101 | 0.623 | -5.2E-01 |
| 60 | O | 15 | 8 | 7 | -1 | 7.463692 | 7.458360 | 5.3E-03 | 13971.176 | 13971.251 | -7.5E-02 | 13975.266 | 13975.341 | -7.5E-02 | 2.856 | 2.930 | -7.4E-02 |
| 61 | F | 15 | 9 | 6 | -3 | 6.481455 | 6.427923 | 5.4E-02 | 13984.615 | 13985.412 | -8.0E-01 | 13989.217 | 13990.014 | -8.0E-01 | 16.807 | 17.603 | -8.0E-01 |
| 62 | Be | 16 | 4 | 12 | 8 | 4.285285 | 4.279805 | 5.5E-03 | 14959.307 | 14959.393 | -8.6E-02 | 14961.351 | 14961.437 | -8.6E-02 | 57.447 | 57.532 | -8.5E-02 |
| 63 | B | 16 | 5 | 11 | 6 | 5.507317 | 5.490636 | 1.7E-02 | 14938.461 | 14938.725 | -2.6E-01 | 14941.017 | 14941.281 | -2.6E-01 | 37.112 | 37.376 | -2.6E-01 |
| 64 | C | 16 | 6 | 10 | 4 | 6.922054 | 6.909779 | 1.2E-02 | 14914.531 | 14914.724 | -1.9E-01 | 14917.598 | 14917.791 | -1.9E-01 | 13.694 | 13.886 | -1.9E-01 |
| 65 | N | 16 | 7 | 9 | 2 | 7.373796 | 7.347244 | 2.7E-02 | 14906.010 | 14906.430 | -4.2E-01 | 14909.588 | 14910.009 | -4.2E-01 | 5.684 | 6.104 | -4.2E-01 |
| 66 | O | 16 | 8 | 8 | 0 | 7.976206 | 7.998708 | -2.3E-02 | 14895.077 | 14894.712 | 3.7E-01 | 14899.167 | 14898.802 | 3.7E-01 | -4.737 | -5.103 | 3.7E-01 |
| 67 | F | 16 | 9 | 7 | -2 | 6.963731 | 6.924706 | 3.9E-02 | 14909.983 | 14910.601 | -6.2E-01 | 14914.585 | 14915.203 | -6.2E-01 | 10.680 | 11.298 | -6.2E-01 |
| 68 | Ne | 16 | 10 | 6 | -4 | 6.083216 | 6.070303 | 1.3E-02 | 14922.777 | 14922.977 | -2.0E-01 | 14927.890 | 14928.091 | -2.0E-01 | 23.986 | 24.185 | -2.0E-01 |
| 69 | B | 17 | 5 | 12 | 7 | 5.266461 | 5.254961 | 1.1E-02 | 15876.613 | 15876.806 | -1.9E-01 | 15879.169 | 15879.362 | -1.9E-01 | 43.771 | 43.963 | -1.9E-01 |
| 70 | C | 17 | 6 | 11 | 5 | 6.558090 | 6.541670 | 1.6E-02 | 15853.362 | 15853.638 | -2.8E-01 | 15856.429 | 15856.705 | -2.8E-01 | 21.031 | 21.306 | -2.7E-01 |
| 71 | N | 17 | 7 | 10 | 3 | 7.286229 | 7.265348 | 2.1E-02 | 15839.690 | 15840.041 | -3.5E-01 | 15843.268 | 15843.619 | -3.5E-01 | 7.870 | 8.220 | -3.5E-01 |
| 72 | O | 17 | 8 | 9 | 1 | 7.750728 | 7.709013 | 4.2E-02 | 15830.499 | 15831.204 | -7.0E-01 | 15834.590 | 15835.294 | -7.0E-01 | -0.809 | -0.105 | -7.0E-01 |
| 73 | F | 17 | 9 | 8 | -1 | 7.542328 | 7.522755 | 2.0E-02 | 15832.748 | 15833.075 | -3.3E-01 | 15837.350 | 15837.677 | -3.3E-01 | 1.952 | 2.278 | -3.3E-01 |
| 74 | Ne | 17 | 10 | 7 | -3 | 6.640499 | 6.659721 | -1.9E-02 | 15846.785 | 15846.452 | 3.3E-01 | 15851.899 | 15851.565 | 3.3E-01 | 16.500 | 16.166 | 3.3E-01 |
| 75 | B | 18 | 5 | 13 | 8 | 4.973602 | 4.979167 | -5.6E-03 | 16816.184 | 16816.081 | 1.0E-01 | 16818.739 | 16818.636 | 1.0E-01 | 51.847 | 51.743 | 1.0E-01 |
| 76 | C | 18 | 6 | 12 | 6 | 6.426195 | 6.441310 | -1.5E-02 | 16788.743 | 16788.468 | 2.8E-01 | 16791.810 | 16791.535 | 2.8E-01 | 24.918 | 24.642 | 2.8E-01 |
| 77 | N | 18 | 7 | 11 | 4 | 7.038562 | 6.994438 | 4.4E-02 | 16776.427 | 16777.217 | -7.9E-01 | 16780.005 | 16780.795 | -7.9E-01 | 13.113 | 13.902 | -7.9E-01 |
| 78 | O | 18 | 8 | 10 | 2 | 7.767097 | 7.788771 | -2.2E-02 | 16762.019 | 16761.624 | 4.0E-01 | 16766.109 | 16765.714 | 4.0E-01 | -0.783 | -1.179 | 4.0E-01 |
| 79 | F | 18 | 9 | 9 | 0 | 7.631638 | 7.628299 | 3.3E-03 | 16763.164 | 16763.218 | -5.4E-02 | 16767.765 | 16767.820 | -5.4E-02 | 0.873 | 0.927 | -5.4E-02 |
| 80 | Ne | 18 | 10 | 8 | -2 | 7.341257 | 7.348282 | -7.0E-03 | 16767.096 | 16766.964 | 1.3E-01 | 16772.210 | 16772.077 | 1.3E-01 | 5.318 | 5.184 | 1.3E-01 |
| 81 | Na | 18 | 11 | 7 | -4 | 6.202276 | 6.216829 | -1.5E-02 | 16786.304 | 16786.035 | 2.7E-01 | 16791.929 | 16791.660 | 2.7E-01 | 25.037 | 24.767 | 2.7E-01 |
| 82 | C | 19 | 6 | 13 | 7 | 6.118333 | 6.110877 | 7.5E-03 | 17727.732 | 17727.870 | -1.4E-01 | 17730.799 | 17730.937 | -1.4E-01 | 32.413 | 32.550 | -1.4E-01 |
| 83 | N | 19 | 7 | 12 | 5 | 6.948583 | 6.917005 | 3.2E-02 | 17710.663 | 17711.259 | -6.0E-01 | 17714.242 | 17714.838 | -6.0E-01 | 15.856 | 16.450 | -5.9E-01 |
| 84 | O | 19 | 8 | 11 | 3 | 7.566494 | 7.591781 | -2.5E-02 | 17697.629 | 17697.144 | 4.9E-01 | 17701.719 | 17701.234 | 4.9E-01 | 3.333 | 2.847 | 4.9E-01 |
| 85 | F | 19 | 9 | 10 | 1 | 7.779018 | 7.773302 | 5.7E-03 | 17692.297 | 17692.400 | -1.0E-01 | 17696.899 | 17697.002 | -1.0E-01 | -1.487 | -1.385 | -1.0E-01 |
| 86 | Ne | 19 | 10 | 9 | -1 | 7.567342 | 7.615234 | -4.8E-02 | 17695.025 | 17694.109 | 9.2E-01 | 17700.138 | 17699.222 | 9.2E-01 | 1.752 | 0.835 | 9.2E-01 |
| 87 | Na | 19 | 11 | 8 | -3 | 6.937885 | 7.002619 | -6.5E-02 | 17705.690 | 17704.453 | 1.2E+00 | 17711.316 | 17710.079 | 1.2E+00 | 12.929 | 11.691 | 1.2E+00 |
| 88 | Mg | 19 | 12 | 7 | -5 | 5.902025 | 5.872830 | 2.9E-02 | 17724.077 | 17724.624 | -5.5E-01 | 17730.215 | 17730.762 | -5.5E-01 | 31.828 | 32.374 | -5.5E-01 |
| 89 | C | 20 | 6 | 14 | 8 | 5.958733 | 5.986947 | -2.8E-02 | 18664.371 | 18663.803 | 5.7E-01 | 18667.438 | 18666.870 | 5.7E-01 | 37.558 | 36.989 | 5.7E-01 |
| 90 | N | 20 | 7 | 13 | 6 | 6.709240 | 6.738383 | -2.9E-02 | 18648.067 | 18647.480 | 5.9E-01 | 18651.645 | 18651.058 | 5.9E-01 | 21.765 | 21.177 | 5.9E-01 |
| 91 | O | 20 | 8 | 12 | 4 | 7.568570 | 7.559814 | 8.8E-03 | 18629.586 | 18629.757 | -1.7E-01 | 18633.676 | 18633.847 | -1.7E-01 | 3.796 | 3.965 | -1.7E-01 |
| 92 | F | 20 | 9 | 11 | 2 | 7.720134 | 7.703366 | 1.7E-02 | 18625.261 | 18625.591 | -3.3E-01 | 18629.863 | 18630.193 | -3.3E-01 | -0.017 | 0.311 | -3.3E-01 |
| 93 | Ne | 20 | 10 | 10 | 0 | 8.032240 | 7.999092 | 3.3E-02 | 18617.725 | 18618.382 | -6.6E-01 | 18622.838 | 18623.495 | -6.6E-01 | -7.042 | -6.386 | -6.6E-01 |
| 94 | Na | 20 | 11 | 9 | -2 | 7.298496 | 7.287781 | 1.1E-02 | 18631.105 | 18631.313 | -2.1E-01 | 18636.731 | 18636.938 | -2.1E-01 | 6.851 | 7.057 | -2.1E-01 |
| 95 | Mg | 20 | 12 | 8 | -4 | 6.723976 | 6.724618 | -6.4E-04 | 18641.302 | 18641.281 | 2.1E-02 | 18647.439 | 18647.418 | 2.1E-02 | 17.559 | 17.537 | 2.2E-02 |
| 96 | N | 21 | 7 | 14 | 7 | 6.608099 | 6.609126 | -1.0E-03 | 19583.047 | 19583.021 | 2.6E-02 | 19586.625 | 19586.600 | 2.6E-02 | 25.251 | 25.225 | 2.7E-02 |
| 97 | O | 21 | 8 | 13 | 5 | 7.389380 | 7.395009 | -5.6E-03 | 19565.346 | 19565.223 | 1.2E-01 | 19569.436 | 19569.313 | 1.2E-01 | 8.062 | 7.938 | 1.2E-01 |
| 98 | F | 21 | 9 | 12 | 3 | 7.738293 | 7.765882 | -2.8E-02 | 19556.725 | 19556.140 | 5.8E-01 | 19561.327 | 19560.742 | 5.8E-01 | -0.048 | -0.634 | 5.9E-01 |
| 99 | Ne | 21 | 10 | 11 | 1 | 7.971713 | 7.940934 | 3.1E-02 | 19550.529 | 19551.169 | -6.4E-01 | 19555.643 | 19556.283 | -6.4E-01 | -5.732 | -5.093 | -6.4E-01 |
| 100 | Na | 21 | 11 | 10 | -1 | 7.765547 | 7.759564 | 6.0E-03 | 19553.564 | 19553.683 | -1.2E-01 | 19559.190 | 19559.308 | -1.2E-01 | -2.185 | -2.067 | -1.2E-01 |
| 101 | Mg | 21 | 12 | 9 | -3 | 7.104571 | 7.152806 | -4.8E-02 | 19566.150 | 19565.130 | 1.0E+00 | 19572.288 | 19571.267 | 1.0E+00 | 10.914 | 9.892 | 1.0E+00 |
| 102 | C | 22 | 6 | 16 | 10 | 5.422030 | 5.419570 | 2.5E-03 | 20543.392 | 20543.442 | -5.1E-02 | 20546.459 | 20546.509 | -5.1E-02 | 53.590 | 53.640 | -5.0E-02 |
| 103 | N | 22 | 7 | 15 | 8 | 6.366085 | 6.390681 | -2.5E-02 | 20521.329 | 20520.783 | 5.5E-01 | 20524.907 | 20524.362 | 5.5E-01 | 32.039 | 31.492 | 5.5E-01 |
| 104 | O | 22 | 8 | 14 | 6 | 7.364858 | 7.335639 | 2.9E-02 | 20498.062 | 20498.700 | -6.4E-01 | 20502.152 | 20502.790 | -6.4E-01 | 9.283 | 9.920 | -6.4E-01 |
| 105 | F | 22 | 9 | 13 | 4 | 7.624295 | 7.639929 | -1.6E-02 | 20491.060 | 20490.711 | 3.5E-01 | 20495.662 | 20495.312 | 3.5E-01 | 2.793 | 2.443 | 3.5E-01 |
| 106 | Ne | 22 | 10 | 12 | 2 | 8.080465 | 8.040962 | 4.0E-02 | 20479.730 | 20480.593 | -8.6E-01 | 20484.844 | 20485.706 | -8.6E-01 | -8.025 | -7.163 | -8.6E-01 |
| 107 | Na | 22 | 11 | 11 | 0 | 7.915667 | 7.875152 | 4.1E-02 | 20482.061 | 20482.946 | -8.8E-01 | 20487.687 | 20488.571 | -8.8E-01 | -5.182 | -4.298 | -8.8E-01 |
| 108 | Mg | 22 | 12 | 10 | -2 | 7.662762 | 7.683458 | -2.1E-02 | 20486.331 | 20485.868 | 4.6E-01 | 20492.468 | 20492.005 | 4.6E-01 | -0.400 | -0.864 | 4.6E-01 |
| 109 | O | 23 | 8 | 15 | 7 | 7.163516 | 7.137057 | 2.6E-02 | 21434.893 | 21435.497 | -6.0E-01 | 21438.983 | 21439.587 | -6.0E-01 | 14.621 | 15.223 | -6.0E-01 |
| 110 | F | 23 | 9 | 14 | 5 | 7.621136 | 7.616902 | 4.2E-03 | 21423.074 | 21423.166 | -9.2E-02 | 21427.675 | 21427.767 | -9.2E-02 | 3.313 | 3.404 | -9.1E-02 |
| 111 | Ne | 23 | 10 | 13 | 3 | 7.955255 | 7.947677 | 7.6E-03 | 21414.095 | 21414.263 | -1.7E-01 | 21419.208 | 21419.376 | -1.7E-01 | -5.154 | -4.987 | -1.7E-01 |
| 112 | Na | 23 | 11 | 12 | 1 | 8.111493 | 8.093587 | 1.8E-02 | 21409.207 | 21409.612 | -4.0E-01 | 21414.832 | 21415.237 | -4.0E-01 | -9.530 | -9.126 | -4.0E-01 |
| 113 | Mg | 23 | 12 | 11 | -1 | 7.901104 | 7.891284 | 9.8E-03 | 21412.752 | 21412.970 | -2.2E-01 | 21418.889 | 21419.107 | -2.2E-01 | -5.473 | -5.256 | -2.2E-01 |
| 114 | Al | 23 | 13 | 10 | -3 | 7.335727 | 7.385819 | -5.0E-02 | 21424.461 | 21423.300 | 1.2E+00 | 21431.110 | 21429.950 | 1.2E+00 | 6.748 | 5.587 | 1.2E+00 |
| 115 | O | 24 | 8 | 16 | 8 | 7.039685 | 7.022191 | 1.7E-02 | 22370.267 | 22370.682 | -4.2E-01 | 22374.357 | 22374.772 | -4.2E-01 | 18.500 | 18.914 | -4.1E-01 |
| 116 | F | 24 | 9 | 15 | 6 | 7.462957 | 7.459623 | 3.3E-03 | 22358.814 | 22358.889 | -7.5E-02 | 22363.416 | 22363.491 | -7.5E-02 | 7.560 | 7.633 | -7.3E-02 |
| 117 | Ne | 24 | 10 | 14 | 4 | 7.993324 | 8.000515 | -7.2E-03 | 22344.791 | 22344.613 | 1.8E-01 | 22349.905 | 22349.726 | 1.8E-01 | -5.952 | -6.131 | 1.8E-01 |
| 118 | Na | 24 | 11 | 13 | 2 | 8.063490 | 8.041738 | 2.2E-02 | 22341.813 | 22342.328 | -5.2E-01 | 22347.438 | 22347.954 | -5.2E-01 | -8.418 | -7.904 | -5.1E-01 |
| 119 | Mg | 24 | 12 | 12 | 0 | 8.260709 | 8.228893 | 3.2E-02 | 22335.785 | 22336.541 | -7.6E-01 | 22341.923 | 22342.679 | -7.6E-01 | -13.934 | -13.179 | -7.5E-01 |
| 120 | Al | 24 | 13 | 11 | -2 | 7.649530 | 7.675560 | -2.6E-02 | 22349.159 | 22348.526 | 6.3E-01 | 22355.809 | 22355.176 | 6.3E-01 | -0.048 | -0.682 | 6.3E-01 |
| 121 | Si | 24 | 14 | 10 | -4 | 7.167267 | 7.182479 | -1.5E-02 | 22359.439 | 22359.065 | 3.7E-01 | 22366.601 | 22366.227 | 3.7E-01 | 10.744 | 10.369 | 3.8E-01 |
| 122 | O | 25 | 8 | 17 | 9 | 6.727057 | 6.722141 | 4.9E-03 | 23310.608 | 23310.726 | -1.2E-01 | 23314.698 | 23314.816 | -1.2E-01 | 27.348 | 27.465 | -1.2E-01 |
| 123 | F | 25 | 9 | 16 | 7 | 7.335132 | 7.330151 | 5.0E-03 | 23294.112 | 23294.231 | -1.2E-01 | 23298.714 | 23298.833 | -1.2E-01 | 11.364 | 11.482 | -1.2E-01 |
| 124 | Ne | 25 | 10 | 15 | 5 | 7.840771 | 7.852208 | -1.1E-02 | 23280.177 | 23279.885 | 2.9E-01 | 23285.291 | 23284.999 | 2.9E-01 | -2.060 | -2.353 | 2.9E-01 |
| 125 | Na | 25 | 11 | 14 | 3 | 8.101397 | 8.088370 | 1.3E-02 | 23272.367 | 23272.686 | -3.2E-01 | 23277.993 | 23278.311 | -3.2E-01 | -9.358 | -9.040 | -3.2E-01 |
| 126 | Mg | 25 | 12 | 13 | 1 | 8.223502 | 8.228803 | -5.3E-03 | 23268.020 | 23267.880 | 1.4E-01 | 23274.158 | 23274.018 | 1.4E-01 | -13.193 | -13.334 | 1.4E-01 |
| 127 | Al | 25 | 13 | 12 | -1 | 8.021144 | 8.044475 | -2.3E-02 | 23271.785 | 23271.193 | 5.9E-01 | 23278.434 | 23277.843 | 5.9E-01 | -8.916 | -9.509 | 5.9E-01 |
| 128 | Si | 25 | 14 | 11 | -3 | 7.480110 | 7.503489 | -2.3E-02 | 23284.016 | 23283.422 | 5.9E-01 | 23291.178 | 23290.584 | 5.9E-01 | 3.827 | 3.233 | 5.9E-01 |
| 129 | O | 26 | 8 | 18 | 10 | 6.494709 | 6.516314 | -2.2E-02 | 24249.487 | 24248.921 | 5.7E-01 | 24253.577 | 24253.011 | 5.7E-01 | 34.733 | 34.166 | 5.7E-01 |
| 130 | F | 26 | 9 | 17 | 8 | 7.082618 | 7.078588 | 4.0E-03 | 24232.908 | 24233.007 | -9.9E-02 | 24237.509 | 24237.609 | -1.0E-01 | 18.665 | 18.763 | -9.8E-02 |
| 131 | Ne | 26 | 10 | 16 | 6 | 7.751974 | 7.773953 | -2.2E-02 | 24214.210 | 24213.633 | 5.8E-01 | 24219.324 | 24218.746 | 5.8E-01 | 0.479 | -0.099 | 5.8E-01 |
| 132 | Na | 26 | 11 | 15 | 4 | 8.004201 | 7.992514 | 1.2E-02 | 24206.358 | 24206.655 | -3.0E-01 | 24211.984 | 24212.281 | -3.0E-01 | -6.861 | -6.565 | -3.0E-01 |
| 133 | Mg | 26 | 12 | 14 | 2 | 8.333870 | 8.383823 | -5.0E-02 | 24196.492 | 24195.186 | 1.3E+00 | 24202.630 | 24201.324 | 1.3E+00 | -16.215 | -17.522 | 1.3E+00 |
| 134 | Al | 26 | 13 | 13 | 0 | 8.149764 | 8.154243 | -4.5E-03 | 24199.985 | 24199.860 | 1.2E-01 | 24206.634 | 24206.510 | 1.2E-01 | -12.210 | -12.336 | 1.3E-01 |
| 135 | Si | 26 | 14 | 12 | -2 | 7.924707 | 7.962739 | -3.8E-02 | 24204.542 | 24203.544 | 1.0E+00 | 24211.703 | 24210.706 | 1.0E+00 | -7.141 | -8.140 | 1.0E+00 |

| | | | | | | | | | | | | | | | | | | |
|---|---|---|---|---|---|---|---|---|---|---|---|---|---|---|---|---|---|---|
| 136 | F | 27 | 9 | 18 | 9 | 6.898326 | 6.885379 | 1.3E-02 | 25170.366 | 25170.711 | -3.4E-01 | 25174.968 | 25175.312 | -3.4E-01 | 24.630 | 24.973 | -3.4E-01 |
| 137 | Ne | 27 | 10 | 17 | 7 | 7.520973 | 7.542352 | -2.1E-02 | 25152.261 | 25151.678 | 5.8E-01 | 25157.374 | 25156.791 | 5.8E-01 | 7.036 | 6.451 | 5.8E-01 |
| 138 | Na | 27 | 11 | 16 | 5 | 7.956942 | 7.954339 | 2.6E-03 | 25139.195 | 25139.259 | -6.4E-02 | 25144.821 | 25144.884 | -6.4E-02 | -5.518 | -5.455 | -6.2E-02 |
| 139 | Mg | 27 | 12 | 15 | 3 | 8.263853 | 8.276897 | -1.3E-02 | 25129.614 | 25129.255 | 3.6E-01 | 25135.752 | 25135.392 | 3.6E-01 | -14.587 | -14.948 | 3.6E-01 |
| 140 | Al | 27 | 13 | 14 | 1 | 8.331548 | 8.338559 | -7.0E-03 | 25126.492 | 25126.295 | 2.0E-01 | 25133.142 | 25132.944 | 2.0E-01 | -17.197 | -17.395 | 2.0E-01 |
| 141 | Si | 27 | 14 | 13 | -1 | 8.124337 | 8.162001 | -3.8E-02 | 25130.792 | 25129.766 | 1.0E+00 | 25137.954 | 25136.928 | 1.0E+00 | -12.384 | -13.412 | 1.0E+00 |
| 142 | P | 27 | 15 | 12 | -3 | 7.663438 | 7.646325 | 1.7E-02 | 25141.942 | 25142.394 | -4.5E-01 | 25149.616 | 25150.068 | -4.5E-01 | -0.722 | -0.271 | -4.5E-01 |
| 143 | F | 28 | 9 | 19 | 10 | 6.644100 | 6.629039 | 1.5E-02 | 26110.152 | 26110.568 | -4.2E-01 | 26114.753 | 26115.170 | -4.2E-01 | 32.921 | 33.336 | -4.2E-01 |
| 144 | Ne | 28 | 10 | 18 | 8 | 7.388637 | 7.384117 | 4.5E-03 | 26088.010 | 26088.131 | -1.2E-01 | 26093.124 | 26093.245 | -1.2E-01 | 11.292 | 11.411 | -1.2E-01 |
| 145 | Na | 28 | 11 | 17 | 6 | 7.799264 | 7.800567 | -1.3E-03 | 26075.219 | 26075.176 | 4.3E-02 | 26080.844 | 26080.801 | 4.3E-02 | -0.988 | -1.033 | 4.4E-02 |
| 146 | Mg | 28 | 12 | 16 | 4 | 8.272409 | 8.296679 | -2.4E-02 | 26060.676 | 26059.989 | 6.9E-01 | 26066.814 | 26066.127 | 6.9E-01 | -15.019 | -15.707 | 6.9E-01 |
| 147 | Al | 28 | 13 | 15 | 2 | 8.309889 | 8.324335 | -1.4E-02 | 26058.332 | 26057.920 | 4.1E-01 | 26064.982 | 26064.569 | 4.1E-01 | -16.851 | -17.264 | 4.1E-01 |
| 148 | Si | 28 | 14 | 14 | 0 | 8.447744 | 8.427705 | 2.0E-02 | 26053.178 | 26053.730 | -5.5E-01 | 26060.340 | 26060.892 | -5.5E-01 | -21.493 | -20.942 | -5.5E-01 |
| 149 | P | 28 | 15 | 13 | -2 | 7.907479 | 7.925664 | -1.8E-02 | 26067.010 | 26066.492 | 5.2E-01 | 26074.685 | 26074.166 | 5.2E-01 | -7.148 | -7.668 | 5.2E-01 |
| 150 | S | 28 | 16 | 12 | -4 | 7.478790 | 7.486041 | -7.3E-03 | 26077.719 | 26077.505 | 2.1E-01 | 26085.906 | 26085.692 | 2.1E-01 | 4.073 | 3.859 | 2.1E-01 |
| 151 | Ne | 29 | 10 | 19 | 9 | 7.167067 | 7.154101 | 1.3E-02 | 27026.613 | 27026.983 | -3.7E-01 | 27031.726 | 27032.096 | -3.7E-01 | 18.400 | 18.769 | -3.7E-01 |
| 152 | Na | 29 | 11 | 18 | 7 | 7.682152 | 7.694278 | -1.2E-02 | 27010.381 | 27010.023 | 3.6E-01 | 27016.006 | 27015.648 | 3.6E-01 | 2.680 | 2.320 | 3.6E-01 |
| 153 | Mg | 29 | 12 | 17 | 5 | 8.113202 | 8.137074 | -2.4E-02 | 26996.586 | 26995.887 | 7.0E-01 | 27002.724 | 27002.024 | 7.0E-01 | -10.603 | -11.304 | 7.0E-01 |
| 154 | Al | 29 | 13 | 16 | 3 | 8.348357 | 8.364850 | -1.6E-02 | 26988.472 | 26987.986 | 4.9E-01 | 26995.122 | 26994.635 | 4.9E-01 | -18.205 | -18.692 | 4.9E-01 |
| 155 | Si | 29 | 14 | 15 | 1 | 8.448635 | 8.421079 | 2.8E-02 | 26984.269 | 26985.060 | -7.9E-01 | 26991.431 | 26992.222 | -7.9E-01 | -21.895 | -21.106 | -7.9E-01 |
| 156 | P | 29 | 15 | 14 | -1 | 8.251222 | 8.240294 | 1.1E-02 | 26988.700 | 26989.007 | -3.1E-01 | 26996.374 | 26996.681 | -3.1E-01 | -16.952 | -16.646 | -3.1E-01 |
| 157 | S | 29 | 16 | 13 | -3 | 7.748519 | 7.717029 | 3.1E-02 | 27001.983 | 27002.886 | -9.0E-01 | 27010.170 | 27011.073 | -9.0E-01 | -3.156 | -2.255 | -9.0E-01 |
| 158 | Ne | 30 | 10 | 20 | 10 | 7.042549 | 7.033300 | 9.2E-03 | 27962.746 | 27963.018 | -2.7E-01 | 27967.860 | 27968.132 | -2.7E-01 | 23.040 | 23.310 | -2.7E-01 |
| 159 | Na | 30 | 11 | 19 | 8 | 7.501968 | 7.514740 | -1.3E-02 | 27947.670 | 27947.280 | 3.9E-01 | 27953.295 | 27952.905 | 3.9E-01 | 8.475 | 8.084 | 3.9E-01 |
| 160 | Mg | 30 | 12 | 18 | 6 | 8.054503 | 8.077918 | -2.3E-02 | 27929.799 | 27929.090 | 7.1E-01 | 27935.937 | 27935.227 | 7.1E-01 | -8.884 | -9.595 | 7.1E-01 |
| 161 | Al | 30 | 13 | 17 | 4 | 8.261382 | 8.272374 | -1.1E-02 | 27922.298 | 27921.961 | 3.4E-01 | 27928.948 | 27928.610 | 3.4E-01 | -15.872 | -16.212 | 3.4E-01 |
| 162 | Si | 30 | 14 | 16 | 2 | 8.520654 | 8.539490 | -1.9E-02 | 27913.226 | 27912.652 | 5.7E-01 | 27920.387 | 27919.814 | 5.7E-01 | -24.433 | -25.008 | 5.8E-01 |
| 163 | P | 30 | 15 | 15 | 0 | 8.353497 | 8.338199 | 1.5E-02 | 27916.946 | 27917.395 | -4.5E-01 | 27924.620 | 27925.069 | -4.5E-01 | -20.201 | -19.753 | -4.5E-01 |
| 164 | S | 30 | 16 | 14 | -2 | 8.122699 | 8.115766 | 6.9E-03 | 27922.575 | 27922.772 | -2.0E-01 | 27930.761 | 27930.959 | -2.0E-01 | -14.059 | -13.863 | -2.0E-01 |
| 165 | Ne | 31 | 10 | 21 | 11 | 6.824743 | 6.824659 | 8.4E-05 | 28902.021 | 28902.018 | 3.2E-03 | 28907.135 | 28907.132 | 3.1E-03 | 30.820 | 30.816 | 4.7E-03 |
| 166 | Na | 31 | 11 | 20 | 9 | 7.398196 | 7.391858 | 6.3E-03 | 28882.950 | 28883.140 | -1.9E-01 | 28888.575 | 28888.765 | -1.9E-01 | 12.261 | 12.449 | -1.9E-01 |
| 167 | Mg | 31 | 12 | 19 | 7 | 7.869194 | 7.899105 | -3.0E-02 | 28867.055 | 28866.120 | 9.3E-01 | 28873.192 | 28872.258 | 9.3E-01 | -3.122 | -4.058 | 9.4E-01 |
| 168 | Al | 31 | 13 | 18 | 5 | 8.225655 | 8.227250 | -1.6E-03 | 28854.710 | 28854.652 | 5.7E-02 | 28861.359 | 28861.302 | 5.7E-02 | -14.955 | -15.014 | 5.9E-02 |
| 169 | Si | 31 | 14 | 17 | 3 | 8.458291 | 8.443006 | 1.5E-02 | 28846.204 | 28846.669 | -4.7E-01 | 28853.365 | 28853.831 | -4.7E-01 | -22.949 | -22.485 | -4.6E-01 |
| 170 | P | 31 | 15 | 16 | 1 | 8.481167 | 8.475004 | 6.2E-03 | 28844.200 | 28844.381 | -1.8E-01 | 28851.874 | 28852.056 | -1.8E-01 | -24.441 | -24.260 | -1.8E-01 |
| 171 | S | 31 | 16 | 15 | -1 | 8.281800 | 8.280847 | 9.5E-04 | 28849.085 | 28849.104 | -1.9E-02 | 28857.272 | 28857.291 | -1.9E-02 | -19.043 | -19.025 | -1.8E-02 |
| 172 | Cl | 31 | 17 | 14 | -3 | 7.870228 | 7.824000 | 4.6E-02 | 28860.549 | 28861.971 | -1.4E+00 | 28869.248 | 28870.671 | -1.4E+00 | -7.066 | -5.645 | -1.4E+00 |
| 173 | Na | 32 | 11 | 21 | 10 | 7.214584 | 7.196884 | 1.8E-02 | 29820.993 | 29821.553 | -5.6E-01 | 29826.618 | 29827.178 | -5.6E-01 | 18.810 | 19.368 | -5.6E-01 |
| 174 | Mg | 32 | 12 | 20 | 8 | 7.803837 | 7.798067 | 5.8E-03 | 29800.842 | 29801.020 | -1.8E-01 | 29806.980 | 29807.157 | -1.8E-01 | -0.829 | -0.653 | -1.8E-01 |
| 175 | Al | 32 | 13 | 19 | 6 | 8.100318 | 8.100196 | 1.2E-04 | 29790.060 | 29790.056 | 4.0E-03 | 29796.710 | 29796.706 | 4.0E-03 | -11.099 | -11.104 | 5.5E-03 |
| 176 | Si | 32 | 14 | 18 | 4 | 8.481468 | 8.459634 | 2.2E-02 | 29776.569 | 29777.259 | -6.9E-01 | 29783.731 | 29784.421 | -6.9E-01 | -24.078 | -23.389 | -6.9E-01 |
| 177 | P | 32 | 15 | 17 | 2 | 8.464120 | 8.455088 | 9.0E-03 | 29775.829 | 29776.109 | -2.8E-01 | 29783.504 | 29783.783 | -2.8E-01 | -24.305 | -24.027 | -2.8E-01 |
| 178 | S | 32 | 16 | 16 | 0 | 8.493129 | 8.505387 | -1.2E-02 | 29773.606 | 29773.204 | 4.0E-01 | 29781.793 | 29781.391 | 4.0E-01 | -26.016 | -26.419 | 4.0E-01 |
| 179 | Cl | 32 | 17 | 15 | -2 | 8.072404 | 8.055422 | 1.7E-02 | 29785.774 | 29786.307 | -5.3E-01 | 29794.474 | 29795.006 | -5.3E-01 | -13.335 | -12.804 | -5.3E-01 |
| 180 | Ar | 32 | 18 | 14 | -4 | 7.700008 | 7.693214 | 6.8E-03 | 29796.396 | 29796.602 | -2.1E-01 | 29805.608 | 29805.814 | -2.1E-01 | -2.200 | -1.996 | -2.0E-01 |
| 181 | Mg | 33 | 12 | 21 | 9 | 7.636458 | 7.625601 | 1.1E-02 | 30738.127 | 30738.478 | -3.5E-01 | 30744.265 | 30744.616 | -3.5E-01 | 4.962 | 5.312 | -3.5E-01 |
| 182 | Al | 33 | 13 | 20 | 7 | 8.019733 | 8.021757 | -2.0E-03 | 30724.185 | 30724.110 | 7.5E-02 | 30730.834 | 30730.760 | 7.5E-02 | -8.468 | -8.544 | 7.6E-02 |
| 183 | Si | 33 | 14 | 19 | 5 | 8.361059 | 8.328593 | 3.2E-02 | 30711.626 | 30712.689 | -1.1E+00 | 30718.788 | 30719.851 | -1.1E+00 | -20.514 | -19.453 | -1.1E+00 |
| 184 | P | 33 | 15 | 18 | 3 | 8.513806 | 8.484682 | 2.9E-02 | 30705.291 | 30706.243 | -9.5E-01 | 30712.965 | 30713.917 | -9.5E-01 | -26.337 | -25.387 | -9.5E-01 |
| 185 | S | 33 | 16 | 17 | 1 | 8.497630 | 8.501013 | -3.4E-03 | 30704.530 | 30704.408 | 1.2E-01 | 30712.717 | 30712.595 | 1.2E-01 | -26.586 | -26.709 | 1.2E-01 |
| 186 | Cl | 33 | 17 | 16 | -1 | 8.304755 | 8.318378 | -1.4E-02 | 30709.600 | 30709.139 | 4.6E-01 | 30718.299 | 30717.839 | 4.6E-01 | -21.003 | -21.465 | 4.6E-01 |
| 187 | Ar | 33 | 18 | 15 | -3 | 7.928955 | 7.885968 | 4.3E-02 | 30720.706 | 30722.113 | -1.4E+00 | 30729.918 | 30731.325 | -1.4E+00 | -9.384 | -7.979 | -1.4E+00 |
| 188 | Mg | 34 | 12 | 22 | 10 | 7.550390 | 7.530298 | 2.0E-02 | 31672.982 | 31673.658 | -6.8E-01 | 31679.120 | 31679.796 | -6.8E-01 | 8.323 | 8.998 | -6.7E-01 |
| 189 | Al | 34 | 13 | 21 | 8 | 7.862446 | 7.885628 | -2.3E-02 | 31661.078 | 31660.282 | 8.0E-01 | 31667.728 | 31666.932 | 8.0E-01 | -3.069 | -3.867 | 8.0E-01 |
| 190 | Si | 34 | 14 | 20 | 6 | 8.336137 | 8.292973 | 4.3E-02 | 31643.678 | 31645.137 | -1.5E+00 | 31650.840 | 31652.299 | -1.5E+00 | -19.957 | -18.499 | -1.5E+00 |
| 191 | P | 34 | 15 | 19 | 4 | 8.448185 | 8.413391 | 3.5E-02 | 31638.573 | 31639.747 | -1.2E+00 | 31646.248 | 31647.422 | -1.2E+00 | -24.549 | -23.377 | -1.2E+00 |
| 192 | S | 34 | 16 | 18 | 2 | 8.583498 | 8.594767 | -1.1E-02 | 31632.678 | 31632.285 | 3.9E-01 | 31640.865 | 31640.472 | 3.9E-01 | -29.932 | -30.326 | 3.9E-01 |
| 193 | Cl | 34 | 17 | 17 | 0 | 8.398970 | 8.408327 | -9.4E-03 | 31637.657 | 31637.328 | 3.3E-01 | 31646.356 | 31646.028 | 3.3E-01 | -24.440 | -24.771 | 3.3E-01 |
| 194 | Ar | 34 | 18 | 16 | -2 | 8.197672 | 8.210405 | -1.3E-02 | 31643.206 | 31642.761 | 4.4E-01 | 31652.418 | 31651.974 | 4.4E-01 | -18.378 | -18.824 | 4.5E-01 |
| 195 | Mg | 35 | 12 | 23 | 11 | 7.356233 | 7.340699 | 1.6E-02 | 32611.793 | 32612.329 | -5.4E-01 | 32617.930 | 32618.467 | -5.4E-01 | 15.640 | 16.175 | -5.3E-01 |
| 196 | Al | 35 | 13 | 22 | 9 | 7.787012 | 7.793636 | -6.6E-03 | 32595.421 | 32595.181 | 2.4E-01 | 32602.071 | 32601.831 | 2.4E-01 | -0.220 | -0.461 | 2.4E-01 |
| 197 | Si | 35 | 14 | 21 | 7 | 8.168676 | 8.156456 | 1.2E-02 | 32580.768 | 32581.187 | -4.2E-01 | 32587.930 | 32588.349 | -4.2E-01 | -14.360 | -13.943 | -4.2E-01 |
| 198 | P | 35 | 15 | 20 | 5 | 8.446248 | 8.387539 | 5.9E-02 | 32569.758 | 32571.804 | -2.0E+00 | 32577.433 | 32579.478 | -2.0E+00 | -24.858 | -22.814 | -2.0E+00 |
| 199 | S | 35 | 16 | 19 | 3 | 8.537851 | 8.523188 | 1.5E-02 | 32565.257 | 32565.761 | -5.0E-01 | 32573.444 | 32573.948 | -5.0E-01 | -28.846 | -28.345 | -5.0E-01 |
| 200 | Cl | 35 | 17 | 18 | 1 | 8.520279 | 8.520366 | -8.7E-05 | 32564.577 | 32564.564 | 1.4E-02 | 32573.277 | 32573.263 | 1.4E-02 | -29.014 | -29.029 | 1.5E-02 |
| 201 | Ar | 35 | 18 | 17 | -1 | 8.327465 | 8.337478 | -1.0E-02 | 32570.031 | 32569.669 | 3.6E-01 | 32579.243 | 32578.881 | 3.6E-01 | -23.047 | -23.411 | 3.6E-01 |
| 202 | K | 35 | 19 | 16 | -3 | 7.965840 | 7.959278 | 6.6E-03 | 32581.392 | 32581.610 | -2.2E-01 | 32591.118 | 32591.335 | -2.2E-01 | -11.173 | -10.957 | -2.2E-01 |
| 203 | Mg | 36 | 12 | 24 | 12 | 7.244419 | 7.231152 | 1.3E-02 | 33548.027 | 33548.498 | -4.7E-01 | 33554.165 | 33554.635 | -4.7E-01 | 20.380 | 20.849 | -4.7E-01 |
| 204 | Al | 36 | 13 | 23 | 10 | 7.623515 | 7.636194 | -1.3E-02 | 33533.085 | 33532.621 | 4.6E-01 | 33539.735 | 33539.271 | 4.6E-01 | 5.950 | 5.484 | 4.7E-01 |
| 205 | Si | 36 | 14 | 22 | 8 | 8.111330 | 8.094633 | 1.7E-02 | 33514.229 | 33514.822 | -5.9E-01 | 33521.391 | 33521.984 | -5.9E-01 | -12.393 | -11.802 | -5.9E-01 |
| 206 | P | 36 | 15 | 21 | 6 | 8.307868 | 8.295165 | 1.3E-02 | 33505.859 | 33506.307 | -4.5E-01 | 33513.533 | 33513.982 | -4.5E-01 | -20.251 | -19.805 | -4.5E-01 |
| 207 | S | 36 | 16 | 20 | 4 | 8.575389 | 8.549134 | 2.6E-02 | 33494.934 | 33495.869 | -9.4E-01 | 33503.120 | 33504.056 | -9.4E-01 | -30.664 | -29.731 | -9.3E-01 |
| 208 | Cl | 36 | 17 | 19 | 2 | 8.521932 | 8.513795 | 8.1E-03 | 33495.563 | 33495.845 | -2.8E-01 | 33504.263 | 33504.545 | -2.8E-01 | -29.522 | -29.241 | -2.8E-01 |
| 209 | Ar | 36 | 18 | 18 | 0 | 8.519909 | 8.514593 | 5.3E-03 | 33494.341 | 33494.521 | -1.8E-01 | 33503.553 | 33503.733 | -1.8E-01 | -30.232 | -30.053 | -1.8E-01 |
| 210 | K | 36 | 19 | 17 | -2 | 8.142219 | 8.155512 | -1.3E-02 | 33506.642 | 33506.151 | 4.9E-01 | 33516.367 | 33515.877 | 4.9E-01 | -17.417 | -17.909 | 4.9E-01 |
| 211 | Ca | 36 | 20 | 16 | -4 | 7.815879 | 7.821612 | -5.7E-03 | 33517.095 | 33516.876 | 2.2E-01 | 33527.333 | 33527.114 | 2.2E-01 | -6.451 | -6.672 | 2.2E-01 |
| 212 | Al | 37 | 13 | 24 | 11 | 7.531315 | 7.551095 | -2.0E-02 | 34468.438 | 34467.699 | 7.4E-01 | 34475.088 | 34474.349 | 7.4E-01 | 9.810 | 9.068 | 7.4E-01 |
| 213 | Si | 37 | 14 | 23 | 9 | 7.953518 | 7.961661 | -8.1E-03 | 34451.522 | 34451.213 | 3.1E-01 | 34458.684 | 34458.375 | 3.1E-01 | -6.594 | -6.906 | 3.1E-01 |
| 214 | P | 37 | 15 | 22 | 7 | 8.267558 | 8.245161 | 2.2E-02 | 34438.608 | 34439.428 | -8.2E-01 | 34446.282 | 34447.102 | -8.2E-01 | -18.996 | -18.178 | -8.2E-01 |
| 215 | S | 37 | 16 | 21 | 5 | 8.459935 | 8.453733 | 6.2E-03 | 34430.195 | 34430.415 | -2.2E-01 | 34438.382 | 34438.602 | -2.2E-01 | -26.896 | -26.679 | -2.2E-01 |
| 216 | Cl | 37 | 17 | 20 | 3 | 8.570280 | 8.551065 | 1.9E-02 | 34424.817 | 34425.518 | -7.0E-01 | 34433.517 | 34434.217 | -7.0E-01 | -31.762 | -31.063 | -7.0E-01 |
| 217 | Ar | 37 | 18 | 19 | 1 | 8.527139 | 8.527442 | -3.0E-04 | 34425.118 | 34425.096 | 2.3E-02 | 34434.331 | 34434.308 | 2.2E-02 | -30.948 | -30.972 | 2.4E-02 |
| 218 | K | 37 | 19 | 18 | -1 | 8.339847 | 8.354622 | -1.5E-02 | 34430.753 | 34430.194 | 5.6E-01 | 34440.478 | 34439.920 | 5.6E-01 | -24.800 | -25.361 | 5.6E-01 |
| 219 | Ca | 37 | 20 | 17 | -3 | 8.003456 | 7.994865 | 8.6E-03 | 34441.904 | 34442.209 | -3.1E-01 | 34452.142 | 34452.448 | -3.1E-01 | -13.136 | -12.833 | -3.0E-01 |
| 220 | Al | 38 | 13 | 25 | 12 | 7.377097 | 7.381892 | -4.8E-03 | 35406.333 | 35406.143 | 1.9E-01 | 35412.982 | 35412.793 | 1.9E-01 | 16.210 | 16.018 | 1.9E-01 |
| 221 | Si | 38 | 14 | 24 | 10 | 7.892829 | 7.893774 | -9.5E-04 | 35385.440 | 35385.396 | 4.4E-02 | 35392.602 | 35392.558 | 4.4E-02 | -4.170 | -4.216 | 4.6E-02 |
| 222 | P | 38 | 15 | 23 | 8 | 8.148537 | 8.146313 | 2.2E-03 | 35374.429 | 35374.504 | -7.5E-02 | 35382.103 | 35382.178 | -7.5E-02 | -14.670 | -14.596 | -7.4E-02 |
| 223 | S | 38 | 16 | 22 | 6 | 8.448781 | 8.443350 | 5.4E-03 | 35361.724 | 35361.921 | -2.0E-01 | 35369.911 | 35370.108 | -2.0E-01 | -26.861 | -26.666 | -1.9E-01 |
| 224 | Cl | 38 | 17 | 21 | 4 | 8.505480 | 8.506658 | -1.2E-03 | 35358.275 | 35358.220 | 5.5E-02 | 35366.974 | 35366.919 | 5.5E-02 | -29.798 | -29.855 | 5.7E-02 |
| 225 | Ar | 38 | 18 | 20 | 2 | 8.614280 | 8.615035 | -7.5E-04 | 35352.845 | 35352.805 | 4.0E-02 | 35362.058 | 35362.018 | 4.0E-02 | -34.715 | -34.757 | 4.2E-02 |
| 226 | K | 38 | 19 | 19 | 0 | 8.438058 | 8.434552 | 3.5E-03 | 35358.246 | 35358.368 | -1.2E-01 | 35367.972 | 35368.093 | -1.2E-01 | -28.801 | -28.681 | -1.2E-01 |
| 227 | Ca | 38 | 20 | 18 | -2 | 8.240043 | 8.253356 | -1.3E-02 | 35364.475 | 35363.957 | 5.2E-01 | 35374.714 | 35374.195 | 5.2E-01 | -22.059 | -22.579 | 5.2E-01 |
| 228 | Si | 39 | 14 | 25 | 11 | 7.730979 | 7.748589 | -1.8E-02 | 36323.425 | 36322.730 | 7.0E-01 | 36330.587 | 36329.892 | 7.0E-01 | 2.320 | 1.623 | 7.0E-01 |
| 229 | P | 39 | 15 | 24 | 9 | 8.099370 | 8.087879 | 1.1E-02 | 36307.763 | 36308.202 | -4.4E-01 | 36315.437 | 36315.876 | -4.4E-01 | -12.829 | -12.392 | -4.4E-01 |
| 230 | S | 39 | 16 | 23 | 7 | 8.344261 | 8.339897 | 4.4E-03 | 36296.917 | 36297.078 | -1.6E-01 | 36305.104 | 36305.265 | -1.6E-01 | -23.162 | -23.004 | -1.6E-01 |
| 231 | Cl | 39 | 17 | 22 | 5 | 8.494402 | 8.502244 | -7.8E-03 | 36289.767 | 36289.450 | 3.2E-01 | 36298.466 | 36298.150 | 3.2E-01 | -29.800 | -30.118 | 3.2E-01 |
| 232 | Ar | 39 | 18 | 21 | 3 | 8.562598 | 8.577102 | -1.5E-02 | 36285.812 | 36285.235 | 5.8E-01 | 36295.024 | 36294.448 | 5.8E-01 | -33.242 | -33.821 | 5.8E-01 |
| 233 | K | 39 | 19 | 20 | 1 | 8.557025 | 8.536074 | 2.1E-02 | 36284.734 | 36285.539 | -8.1E-01 | 36294.459 | 36295.265 | -8.1E-01 | -33.807 | -33.004 | -8.0E-01 |
| 234 | Ca | 39 | 20 | 19 | -1 | 8.369670 | 8.359807 | 9.9E-03 | 36290.745 | 36291.117 | -3.7E-01 | 36300.984 | 36301.356 | -3.7E-01 | -27.283 | -26.913 | -3.7E-01 |
| 235 | Sc | 39 | 21 | 18 | -3 | 8.013456 | 8.023538 | -1.0E-02 | 36303.342 | 36302.935 | 4.1E-01 | 36314.094 | 36313.687 | 4.1E-01 | -14.173 | -14.581 | 4.1E-01 |
| 236 | Si | 40 | 14 | 26 | 12 | 7.661754 | 7.670004 | -8.3E-03 | 37258.028 | 37257.690 | 3.4E-01 | 37265.190 | 37264.852 | 3.4E-01 | 5.430 | 5.090 | 3.4E-01 |
| 237 | P | 40 | 15 | 25 | 10 | 7.979798 | 7.978499 | 1.3E-03 | 37244.012 | 37244.055 | -4.3E-02 | 37251.686 | 37251.729 | -4.3E-02 | -8.074 | -8.033 | -4.1E-02 |
| 238 | S | 40 | 16 | 24 | 8 | 8.329325 | 8.314024 | 1.5E-02 | 37228.736 | 37229.338 | -6.0E-01 | 37236.923 | 37237.525 | -6.0E-01 | -22.838 | -22.237 | -6.0E-01 |
| 239 | Cl | 40 | 17 | 23 | 6 | 8.427765 | 8.439028 | -1.1E-02 | 37223.503 | 37223.042 | 4.6E-01 | 37232.203 | 37231.742 | 4.6E-01 | -27.558 | -28.021 | 4.6E-01 |
| 240 | Ar | 40 | 18 | 22 | 4 | 8.595259 | 8.616770 | -2.2E-02 | 37215.508 | 37214.637 | 8.7E-01 | 37224.721 | 37223.849 | 8.7E-01 | -35.040 | -35.913 | 8.7E-01 |
| 241 | K | 40 | 19 | 21 | 2 | 8.538090 | 8.549121 | -1.1E-02 | 37216.500 | 37216.047 | 4.5E-01 | 37226.225 | 37225.772 | 4.5E-01 | -33.535 | -33.991 | 4.6E-01 |
| 242 | Ca | 40 | 20 | 20 | 0 | 8.551304 | 8.506306 | 4.5E-02 | 37214.676 | 37216.463 | -1.8E+00 | 37224.914 | 37226.701 | -1.8E+00 | -34.846 | -33.061 | -1.8E+00 |
| 243 | Sc | 40 | 21 | 19 | -2 | 8.173669 | 8.191855 | -1.8E-02 | 37228.485 | 37227.745 | 7.4E-01 | 37239.237 | 37238.496 | 7.4E-01 | -20.523 | -21.266 | 7.4E-01 |
| 244 | Ti | 40 | 22 | 18 | -4 | 7.862286 | 7.884232 | -2.2E-02 | 37239.645 | 37238.753 | 8.9E-01 | 37250.910 | 37250.018 | 8.9E-01 | -8.850 | -9.744 | 8.9E-01 |
| 245 | Si | 41 | 14 | 27 | 13 | 7.508573 | 7.511676 | -3.1E-03 | 38196.212 | 38196.077 | 1.4E-01 | 38203.374 | 38203.239 | 1.4E-01 | 12.120 | 11.982 | 1.4E-01 |
| 246 | P | 41 | 15 | 26 | 11 | 7.906551 | 7.913674 | -7.1E-03 | 38178.600 | 38178.300 | 3.0E-01 | 38186.275 | 38185.974 | 3.0E-01 | -4.980 | -5.283 | 3.0E-01 |
| 247 | S | 41 | 16 | 25 | 9 | 8.229635 | 8.214564 | 1.5E-02 | 38164.059 | 38164.667 | -6.1E-01 | 38172.246 | 38172.854 | -6.1E-01 | -19.009 | -18.402 | -6.1E-01 |
| 248 | Cl | 41 | 17 | 24 | 7 | 8.412959 | 8.413987 | -1.0E-03 | 38155.248 | 38155.195 | 5.3E-02 | 38163.947 | 38163.895 | 5.3E-02 | -27.307 | -27.362 | 5.5E-02 |
| 249 | Ar | 41 | 18 | 23 | 5 | 8.534372 | 8.554419 | -2.0E-02 | 38148.975 | 38148.142 | 8.3E-01 | 38158.187 | 38157.354 | 8.3E-01 | -33.068 | -33.902 | 8.3E-01 |
| 250 | K | 41 | 19 | 22 | 3 | 8.576072 | 8.598744 | -2.3E-02 | 38145.970 | 38145.028 | 9.4E-01 | 38155.695 | 38154.754 | 9.4E-01 | -35.560 | -36.503 | 9.4E-01 |
| 251 | Ca | 41 | 20 | 21 | 1 | 8.546706 | 8.538247 | 8.5E-03 | 38145.878 | 38146.212 | -3.3E-01 | 38156.117 | 38156.451 | -3.3E-01 | -35.138 | -34.806 | -3.3E-01 |
| 252 | Sc | 41 | 21 | 20 | -1 | 8.369198 | 8.364346 | 4.9E-03 | 38151.860 | 38152.046 | -1.9E-01 | 38162.612 | 38162.798 | -1.9E-01 | -28.642 | -28.459 | -1.8E-01 |
| 253 | Ti | 41 | 22 | 19 | -3 | 8.034388 | 8.031423 | 3.0E-03 | 38164.292 | 38164.399 | -1.1E-01 | 38175.557 | 38175.665 | -1.1E-01 | -15.698 | -15.592 | -1.1E-01 |
| 254 | P | 42 | 15 | 27 | 12 | 7.767866 | 7.773303 | -5.4E-03 | 39116.084 | 39115.847 | 2.4E-01 | 39123.758 | 39123.521 | 2.4E-01 | 1.010 | 0.770 | 2.4E-01 |
| 255 | S | 42 | 16 | 26 | 10 | 8.193227 | 8.184879 | 8.3E-03 | 39096.924 | 39097.265 | -3.4E-01 | 39105.111 | 39105.452 | -3.4E-01 | -17.638 | -17.299 | -3.4E-01 |
| 256 | Cl | 42 | 17 | 25 | 8 | 8.347819 | 8.347529 | 2.9E-04 | 39089.136 | 39089.138 | -1.9E-03 | 39097.836 | 39097.837 | -1.8E-03 | -24.913 | -24.913 | 1.2E-04 |
| 257 | Ar | 42 | 18 | 24 | 6 | 8.555613 | 8.566565 | -1.1E-02 | 39079.113 | 39078.642 | 4.7E-01 | 39088.326 | 39087.855 | 4.7E-01 | -34.423 | -34.896 | 4.7E-01 |
| 258 | K | 42 | 19 | 23 | 4 | 8.551256 | 8.578826 | -2.8E-02 | 39078.001 | 39076.831 | 1.2E+00 | 39087.727 | 39086.557 | 1.2E+00 | -35.022 | -36.194 | 1.2E+00 |
| 259 | Ca | 42 | 20 | 22 | 2 | 8.616563 | 8.628769 | -1.2E-02 | 39073.963 | 39073.438 | 5.3E-01 | 39084.201 | 39083.676 | 5.3E-01 | -38.547 | -39.074 | 5.3E-01 |
| 260 | Sc | 42 | 21 | 21 | 0 | 8.444933 | 8.444364 | 5.7E-04 | 39079.876 | 39079.886 | -1.1E-02 | 39090.627 | 39090.638 | -1.1E-02 | -32.121 | -32.113 | -8.6E-03 |
| 261 | Ti | 42 | 22 | 20 | -2 | 8.259247 | 8.264501 | -5.3E-03 | 39086.379 | 39086.144 | 2.3E-01 | 39097.644 | 39097.409 | 2.3E-01 | -25.105 | -25.341 | 2.4E-01 |
| 262 | P | 43 | 15 | 28 | 13 | 7.689572 | 7.698882 | -9.3E-03 | 40051.248 | 40050.839 | 4.1E-01 | 40058.922 | 40058.513 | 4.1E-01 | 4.680 | 4.269 | 4.1E-01 |
| 263 | S | 43 | 16 | 27 | 11 | 8.063827 | 8.075469 | -1.2E-02 | 40033.860 | 40033.350 | 5.1E-01 | 40042.047 | 40041.537 | 5.1E-01 | -12.195 | -12.708 | 5.1E-01 |
| 264 | Cl | 43 | 17 | 26 | 9 | 8.327660 | 8.317007 | 1.1E-02 | 40021.220 | 40021.668 | -4.5E-01 | 40029.920 | 40030.368 | -4.5E-01 | -24.323 | -23.877 | -4.5E-01 |
| 265 | Ar | 43 | 18 | 25 | 7 | 8.488237 | 8.495368 | -7.1E-03 | 40013.020 | 40012.703 | 3.2E-01 | 40022.233 | 40021.915 | 3.2E-01 | -32.010 | -32.330 | 3.2E-01 |
| 266 | K | 43 | 19 | 24 | 5 | 8.576220 | 8.594870 | -1.9E-02 | 40007.942 | 40007.128 | 8.1E-01 | 40017.667 | 40016.853 | 8.1E-01 | -36.575 | -37.391 | 8.2E-01 |
| 267 | Ca | 43 | 20 | 23 | 3 | 8.600663 | 8.617615 | -1.7E-02 | 40005.595 | 40004.854 | 7.4E-01 | 40015.834 | 40015.092 | 7.4E-01 | -38.409 | -39.152 | 7.4E-01 |
| 268 | Sc | 43 | 21 | 22 | 1 | 8.530825 | 8.547106 | -1.6E-02 | 40007.303 | 40006.589 | 7.1E-01 | 40018.055 | 40017.341 | 7.1E-01 | -36.188 | -36.903 | 7.2E-01 |
| 269 | Ti | 43 | 22 | 21 | -1 | 8.352932 | 8.368977 | -1.6E-02 | 40013.656 | 40012.952 | 7.0E-01 | 40024.922 | 40024.218 | 7.0E-01 | -29.321 | -30.027 | 7.1E-01 |
| 270 | V | 43 | 23 | 20 | -3 | 8.069512 | 8.056978 | 1.3E-02 | 40024.547 | 40025.072 | -5.2E-01 | 40036.326 | 40036.851 | -5.2E-01 | -17.916 | -17.394 | -5.2E-01 |
| 271 | S | 44 | 16 | 28 | 12 | 7.996015 | 8.030791 | -3.5E-02 | 40968.346 | 40966.806 | 1.5E+00 | 40976.532 | 40974.993 | 1.5E+00 | -9.204 | -10.746 | 1.5E+00 |
| 272 | Cl | 44 | 17 | 27 | 10 | 8.237450 | 8.242834 | -5.4E-03 | 40956.427 | 40956.180 | 2.5E-01 | 40965.127 | 40964.880 | 2.5E-01 | -20.610 | -20.859 | 2.5E-01 |
| 273 | Ar | 44 | 18 | 26 | 8 | 8.493840 | 8.497326 | -3.5E-03 | 40943.851 | 40943.687 | 1.6E-01 | 40953.063 | 40952.899 | 1.6E-01 | -32.673 | -32.840 | 1.7E-01 |
| 274 | K | 44 | 19 | 25 | 6 | 8.546701 | 8.558731 | -1.2E-02 | 40940.230 | 40939.689 | 5.4E-01 | 40949.955 | 40949.414 | 5.4E-01 | -35.781 | -36.325 | 5.4E-01 |
| 275 | Ca | 44 | 20 | 24 | 4 | 8.658175 | 8.671464 | -1.3E-02 | 40934.029 | 40933.432 | 6.0E-01 | 40944.268 | 40943.671 | 6.0E-01 | -41.469 | -42.068 | 6.0E-01 |
| 276 | Sc | 44 | 21 | 23 | 2 | 8.557379 | 8.577089 | -2.0E-02 | 40937.169 | 40936.288 | 8.8E-01 | 40947.921 | 40947.040 | 8.8E-01 | -37.816 | -38.698 | 8.8E-01 |
| 277 | Ti | 44 | 22 | 22 | 0 | 8.533520 | 8.510661 | 2.3E-02 | 40936.923 | 40937.915 | -9.9E-01 | 40948.188 | 40949.180 | -9.9E-01 | -37.549 | -36.559 | -9.9E-01 |
| 278 | V | 44 | 23 | 21 | -2 | 8.210463 | 8.217080 | -6.6E-03 | 40949.841 | 40949.535 | 3.1E-01 | 40961.620 | 40961.314 | 3.1E-01 | -24.116 | -24.424 | 3.1E-01 |
| 279 | S | 45 | 16 | 29 | 13 | 7.881807 | 7.882191 | -3.8E-04 | 41905.054 | 41905.027 | 2.7E-02 | 41913.241 | 41913.214 | 2.7E-02 | -3.990 | -4.018 | 2.9E-02 |
| 280 | Cl | 45 | 17 | 28 | 11 | 8.183758 | 8.203262 | -2.0E-02 | 41890.171 | 41889.283 | 8.9E-01 | 41898.871 | 41897.983 | 8.9E-01 | -18.360 | -19.250 | 8.9E-01 |
| 281 | Ar | 45 | 18 | 27 | 9 | 8.419952 | 8.422333 | -2.4E-03 | 41878.247 | 41878.129 | 1.2E-01 | 41887.460 | 41887.342 | 1.2E-01 | -29.771 | -29.891 | 1.2E-01 |
| 282 | K | 45 | 19 | 26 | 7 | 8.554674 | 8.557550 | -2.9E-03 | 41870.890 | 41870.748 | 1.4E-01 | 41880.615 | 41880.474 | 1.4E-01 | -36.616 | -36.759 | 1.4E-01 |
| 283 | Ca | 45 | 20 | 25 | 5 | 8.630544 | 8.635052 | -4.5E-03 | 41866.180 | 41865.965 | 2.2E-01 | 41876.418 | 41876.203 | 2.2E-01 | -40.812 | -41.030 | 2.2E-01 |
| 284 | Sc | 45 | 21 | 24 | 3 | 8.618915 | 8.638259 | -1.9E-02 | 41865.408 | 41864.524 | 8.8E-01 | 41876.159 | 41875.276 | 8.8E-01 | -41.071 | -41.957 | 8.9E-01 |
| 285 | Ti | 45 | 22 | 23 | 1 | 8.555706 | 8.554323 | 1.4E-03 | 41866.956 | 41867.005 | -4.8E-02 | 41878.222 | 41878.270 | -4.8E-02 | -39.009 | -38.963 | -4.6E-02 |
| 286 | V | 45 | 23 | 22 | -1 | 8.379908 | 8.388015 | -8.1E-03 | 41873.571 | 41873.192 | 3.8E-01 | 41885.350 | 41884.971 | 3.8E-01 | -31.881 | -32.262 | 3.8E-01 |
| 287 | Cr | 45 | 24 | 21 | -3 | 8.087728 | 8.081299 | 6.4E-03 | 41885.423 | 41885.697 | -2.7E-01 | 41897.716 | 41897.990 | -2.7E-01 | -19.515 | -19.243 | -2.7E-01 |
| 288 | Cl | 46 | 17 | 29 | 12 | 8.082414 | 8.102278 | -2.0E-02 | 42826.215 | 42825.291 | 9.2E-01 | 42834.914 | 42833.990 | 9.2E-01 | -13.810 | -14.736 | 9.3E-01 |
| 289 | Ar | 46 | 18 | 28 | 10 | 8.411502 | 8.410737 | 7.7E-04 | 42809.781 | 42809.806 | -2.4E-02 | 42818.994 | 42819.018 | -2.4E-02 | -29.731 | -29.709 | -2.2E-02 |
| 290 | K | 46 | 19 | 27 | 8 | 8.518042 | 8.514454 | 3.6E-03 | 42803.585 | 42803.739 | -1.5E-01 | 42813.311 | 42813.464 | -1.5E-01 | -35.414 | -35.263 | -1.5E-01 |
| 291 | Ca | 46 | 20 | 26 | 6 | 8.668958 | 8.666181 | 2.8E-03 | 42795.348 | 42795.463 | -1.2E-01 | 42805.586 | 42805.702 | -1.2E-01 | -43.138 | -43.025 | -1.1E-01 |
| 292 | Sc | 46 | 21 | 25 | 4 | 8.621996 | 8.637285 | -1.5E-02 | 42796.212 | 42795.496 | 7.2E-01 | 42806.964 | 42806.248 | 7.2E-01 | -41.760 | -42.479 | 7.2E-01 |
| 293 | Ti | 46 | 22 | 24 | 2 | 8.656434 | 8.650765 | 5.7E-03 | 42793.332 | 42793.579 | -2.5E-01 | 42804.598 | 42804.845 | -2.5E-01 | -44.127 | -43.882 | -2.4E-01 |
| 294 | V | 46 | 23 | 23 | 0 | 8.486113 | 8.471083 | 1.5E-02 | 42799.871 | 42800.548 | -6.8E-01 | 42811.650 | 42812.327 | -6.8E-01 | -37.075 | -36.400 | -6.7E-01 |
| 295 | Cr | 46 | 24 | 22 | -2 | 8.303865 | 8.300151 | 3.7E-03 | 42806.958 | 42807.114 | -1.6E-01 | 42819.251 | 42819.407 | -1.6E-01 | -29.474 | -29.320 | -1.5E-01 |
| 296 | Ar | 47 | 18 | 29 | 11 | 8.308100 | 8.316818 | -8.7E-03 | 43745.795 | 43745.375 | 4.2E-01 | 43755.008 | 43754.587 | 4.2E-01 | -25.211 | -25.634 | 4.2E-01 |
| 297 | K | 47 | 19 | 28 | 9 | 8.514879 | 8.496335 | 1.9E-02 | 43734.781 | 43735.641 | -8.6E-01 | 43744.507 | 43745.367 | -8.6E-01 | -35.712 | -34.854 | -8.6E-01 |
| 298 | Ca | 47 | 20 | 27 | 7 | 8.639328 | 8.617527 | 2.2E-02 | 43727.637 | 43728.649 | -1.0E+00 | 43737.875 | 43738.888 | -1.0E+00 | -42.343 | -41.333 | -1.0E+00 |
| 299 | Sc | 47 | 21 | 26 | 5 | 8.665069 | 8.668454 | -3.4E-03 | 43725.131 | 43724.959 | 1.7E-01 | 43735.883 | 43735.711 | 1.7E-01 | -44.336 | -44.510 | 1.7E-01 |
| 300 | Ti | 47 | 22 | 25 | 3 | 8.661206 | 8.656601 | 4.6E-03 | 43724.017 | 43724.220 | -2.0E-01 | 43735.282 | 43735.485 | -2.0E-01 | -44.936 | -44.736 | -2.0E-01 |
| 301 | V | 47 | 23 | 24 | 1 | 8.582207 | 8.574251 | 8.0E-03 | 43726.434 | 43726.793 | -3.6E-01 | 43738.213 | 43738.572 | -3.6E-01 | -42.006 | -41.649 | -3.6E-01 |
| 302 | Cr | 47 | 24 | 23 | -1 | 8.407159 | 8.409837 | -2.7E-03 | 43733.365 | 43733.224 | 1.4E-01 | 43745.658 | 43745.517 | 1.4E-01 | -34.561 | -34.704 | 1.4E-01 |
| 303 | Mn | 47 | 25 | 22 | -3 | 8.135291 | 8.125468 | 9.8E-03 | 43744.847 | 43745.292 | -4.5E-01 | 43757.653 | 43758.099 | -4.5E-01 | -22.565 | -22.122 | -4.4E-01 |
| 304 | K | 48 | 19 | 29 | 10 | 8.434232 | 8.428731 | 5.5E-03 | 44669.703 | 44669.955 | -2.5E-01 | 44679.428 | 44679.681 | -2.5E-01 | -32.284 | -32.034 | -2.5E-01 |
| 305 | Ca | 48 | 20 | 28 | 8 | 8.666689 | 8.630687 | 3.6E-02 | 44657.249 | 44658.965 | -1.7E+00 | 44667.488 | 44669.204 | -1.7E+00 | -44.225 | -42.511 | -1.7E+00 |
| 306 | Sc | 48 | 21 | 27 | 6 | 8.656195 | 8.649879 | 6.3E-03 | 44656.457 | 44656.748 | -2.9E-01 | 44667.209 | 44667.499 | -2.9E-01 | -44.503 | -44.216 | -2.9E-01 |
| 307 | Ti | 48 | 22 | 26 | 4 | 8.722986 | 8.720244 | 2.7E-03 | 44651.956 | 44652.074 | -1.2E-01 | 44663.221 | 44663.339 | -1.2E-01 | -48.492 | -48.376 | -1.2E-01 |
| 308 | V | 48 | 23 | 25 | 2 | 8.623042 | 8.614491 | 8.6E-03 | 44655.457 | 44655.853 | -4.0E-01 | 44667.236 | 44667.632 | -4.0E-01 | -44.477 | -44.083 | -3.9E-01 |
| 309 | Cr | 48 | 24 | 24 | 0 | 8.572262 | 8.549135 | 2.3E-02 | 44656.598 | 44657.693 | -1.1E+00 | 44668.891 | 44669.986 | -1.1E+00 | -42.822 | -41.729 | -1.1E+00 |
| 310 | Mn | 48 | 25 | 23 | -2 | 8.274750 | 8.275709 | -9.6E-04 | 44669.583 | 44669.521 | 6.2E-02 | 44682.389 | 44682.327 | 6.2E-02 | -29.323 | -29.388 | 6.4E-02 |
| 311 | K | 49 | 19 | 30 | 11 | 8.372274 | 8.387814 | -1.6E-02 | 45603.870 | 45603.097 | 7.7E-01 | 45613.595 | 45612.822 | 7.7E-01 | -29.611 | -30.387 | 7.8E-01 |
| 312 | Ca | 49 | 20 | 29 | 9 | 8.594847 | 8.559920 | 3.5E-02 | 45591.668 | 45593.367 | -1.7E+00 | 45601.907 | 45603.606 | -1.7E+00 | -41.300 | -39.603 | -1.7E+00 |
| 313 | Sc | 49 | 21 | 28 | 7 | 8.686252 | 8.658876 | 2.7E-02 | 45585.894 | 45587.222 | -1.3E+00 | 45596.646 | 45597.974 | -1.3E+00 | -46.561 | -45.235 | -1.3E+00 |
| 314 | Ti | 49 | 22 | 27 | 5 | 8.711137 | 8.699621 | 1.2E-02 | 45583.379 | 45583.929 | -5.5E-01 | 45594.644 | 45595.195 | -5.5E-01 | -48.563 | -48.015 | -5.5E-01 |
| 315 | V | 49 | 23 | 26 | 3 | 8.682888 | 8.679798 | 3.1E-03 | 45583.467 | 45583.604 | -1.4E-01 | 45595.246 | 45595.383 | -1.4E-01 | -47.961 | -47.826 | -1.3E-01 |
| 316 | Cr | 49 | 24 | 25 | 1 | 8.613284 | 8.603545 | 9.7E-03 | 45585.581 | 45586.043 | -4.6E-01 | 45597.874 | 45598.336 | -4.6E-01 | -45.333 | -44.873 | -4.6E-01 |
| 317 | Mn | 49 | 25 | 24 | -1 | 8.440257 | 8.435798 | 4.5E-03 | 45592.763 | 45592.966 | -2.0E-01 | 45605.570 | 45605.773 | -2.0E-01 | -37.637 | -37.436 | -2.0E-01 |
| 318 | Fe | 49 | 26 | 23 | -3 | 8.161311 | 8.157395 | 3.9E-03 | 45605.135 | 45605.311 | -1.8E-01 | 45618.456 | 45618.631 | -1.8E-01 | -24.751 | -24.578 | -1.7E-01 |
| 319 | K | 50 | 19 | 31 | 12 | 8.288582 | 8.290243 | -1.7E-03 | 46539.247 | 46539.153 | 9.4E-02 | 46548.973 | 46548.878 | 9.4E-02 | -25.728 | -25.825 | 9.7E-02 |
| 320 | Ca | 50 | 20 | 30 | 10 | 8.550163 | 8.544026 | 6.1E-03 | 46524.873 | 46525.168 | -2.9E-01 | 46535.111 | 46535.406 | -2.9E-01 | -39.589 | -39.297 | -2.9E-01 |
| 321 | Sc | 50 | 21 | 29 | 8 | 8.633682 | 8.618066 | 1.6E-02 | 46519.401 | 46520.169 | -7.7E-01 | 46530.153 | 46530.921 | -7.7E-01 | -44.548 | -43.782 | -7.7E-01 |
| 322 | Ti | 50 | 22 | 28 | 6 | 8.755698 | 8.738521 | 1.7E-02 | 46512.005 | 46512.850 | -8.5E-01 | 46523.270 | 46524.115 | -8.5E-01 | -51.431 | -50.588 | -8.4E-01 |
| 323 | V | 50 | 23 | 27 | 4 | 8.695915 | 8.688954 | 7.0E-03 | 46513.698 | 46514.032 | -3.3E-01 | 46525.477 | 46525.811 | -3.3E-01 | -49.224 | -48.893 | -3.3E-01 |
| 324 | Cr | 50 | 24 | 26 | 2 | 8.701025 | 8.697851 | 3.2E-03 | 46512.146 | 46512.290 | -1.4E-01 | 46524.439 | 46524.583 | -1.4E-01 | -50.262 | -50.120 | -1.4E-01 |
| 325 | Mn | 50 | 25 | 25 | 0 | 8.532689 | 8.533355 | -6.7E-04 | 46519.267 | 46519.218 | 4.9E-02 | 46532.073 | 46532.024 | 4.9E-02 | -42.627 | -42.679 | 5.1E-02 |
| 326 | Fe | 50 | 26 | 24 | -2 | 8.354270 | 8.345909 | 8.4E-03 | 46526.891 | 46527.293 | -4.0E-01 | 46540.212 | 46540.614 | -4.0E-01 | -34.489 | -34.089 | -4.0E-01 |
| 327 | K | 51 | 19 | 32 | 13 | 8.221346 | 8.239036 | -1.8E-02 | 47473.953 | 47473.040 | 9.1E-01 | 47483.679 | 47482.765 | 9.1E-01 | -22.516 | -23.432 | 9.2E-01 |
| 328 | Ca | 51 | 20 | 31 | 11 | 8.477035 | 8.445531 | 3.2E-02 | 47459.618 | 47461.212 | -1.6E+00 | 47469.856 | 47471.451 | -1.6E+00 | -36.339 | -34.746 | -1.6E+00 |
| 329 | Sc | 51 | 21 | 30 | 9 | 8.596798 | 8.595425 | 1.4E-03 | 47452.214 | 47452.271 | -5.7E-02 | 47462.966 | 47463.023 | -5.7E-02 | -43.229 | -43.174 | -5.5E-02 |
| 330 | Ti | 51 | 22 | 29 | 7 | 8.708969 | 8.693812 | 1.5E-02 | 47445.198 | 47445.957 | -7.6E-01 | 47456.463 | 47457.222 | -7.6E-01 | -49.732 | -48.975 | -7.6E-01 |
| 331 | V | 51 | 23 | 28 | 5 | 8.742096 | 8.725239 | 1.7E-02 | 47442.212 | 47443.058 | -8.5E-01 | 47453.991 | 47454.836 | -8.5E-01 | -52.204 | -51.361 | -8.4E-01 |
| 332 | Cr | 51 | 24 | 27 | 3 | 8.711998 | 8.709802 | 2.2E-03 | 47442.451 | 47442.548 | -9.7E-02 | 47454.744 | 47454.841 | -9.7E-02 | -51.451 | -51.356 | -9.5E-02 |
| 333 | Mn | 51 | 25 | 26 | 1 | 8.633765 | 8.634945 | -1.2E-03 | 47445.145 | 47445.069 | 7.6E-02 | 47457.951 | 47457.875 | 7.6E-02 | -48.243 | -48.322 | 7.8E-02 |
| 334 | Fe | 51 | 26 | 25 | -1 | 8.460755 | 8.467446 | -6.7E-03 | 47452.672 | 47452.314 | 3.6E-01 | 47465.992 | 47465.635 | 3.6E-01 | -40.202 | -40.562 | 3.6E-01 |
| 335 | Co | 51 | 27 | 24 | -3 | 8.193254 | 8.187364 | 5.9E-03 | 47465.017 | 47465.301 | -2.8E-01 | 47478.853 | 47479.136 | -2.8E-01 | -27.342 | -27.061 | -2.8E-01 |
| 336 | Ca | 52 | 20 | 32 | 12 | 8.429319 | 8.410395 | 1.9E-02 | 48393.187 | 48394.159 | -9.7E-01 | 48403.426 | 48404.398 | -9.7E-01 | -34.263 | -33.293 | -9.7E-01 |
| 337 | Sc | 52 | 21 | 31 | 10 | 8.527802 | 8.521912 | 5.9E-03 | 48386.770 | 48387.064 | -2.9E-01 | 48397.522 | 48397.816 | -2.9E-01 | -40.167 | -39.875 | -2.9E-01 |
| 338 | Ti | 52 | 22 | 30 | 8 | 8.691648 | 8.701453 | -9.8E-03 | 48376.955 | 48376.431 | 5.2E-01 | 48388.220 | 48387.697 | 5.2E-01 | -49.469 | -49.995 | 5.3E-01 |
| 339 | V | 52 | 23 | 29 | 6 | 8.714579 | 8.709185 | 5.4E-03 | 48374.466 | 48374.732 | -2.7E-01 | 48386.245 | 48386.511 | -2.7E-01 | -51.444 | -51.180 | -2.6E-01 |
| 340 | Cr | 52 | 24 | 28 | 4 | 8.775967 | 8.772137 | 3.8E-03 | 48369.978 | 48370.162 | -1.8E-01 | 48382.271 | 48382.455 | -1.8E-01 | -55.418 | -55.236 | -1.8E-01 |
| 341 | Mn | 52 | 25 | 27 | 2 | 8.670321 | 8.673612 | -3.3E-03 | 48374.175 | 48373.988 | 1.9E-01 | 48386.982 | 48386.795 | 1.9E-01 | -50.707 | -50.896 | 1.9E-01 |
| 342 | Fe | 52 | 26 | 26 | 0 | 8.609611 | 8.606865 | 2.7E-03 | 48376.036 | 48376.162 | -1.3E-01 | 48389.356 | 48389.483 | -1.3E-01 | -48.332 | -48.208 | -1.2E-01 |
| 343 | Sc | 53 | 21 | 32 | 11 | 8.480339 | 8.475942 | 4.4E-03 | 49320.323 | 49320.544 | -2.2E-01 | 49331.075 | 49331.296 | -2.2E-01 | -38.107 | -37.890 | -2.2E-01 |
| 344 | Ti | 53 | 22 | 31 | 9 | 8.630154 | 8.624767 | 5.4E-03 | 49311.087 | 49311.360 | -2.7E-01 | 49322.353 | 49322.625 | -2.7E-01 | -46.830 | -46.560 | -2.7E-01 |
| 345 | V | 53 | 23 | 30 | 7 | 8.710110 | 8.714185 | -4.1E-03 | 49305.554 | 49305.324 | 2.3E-01 | 49317.333 | 49317.103 | 2.3E-01 | -51.850 | -52.083 | 2.3E-01 |
| 346 | Cr | 53 | 24 | 29 | 5 | 8.760177 | 8.754984 | 5.2E-03 | 49301.604 | 49301.864 | -2.6E-01 | 49313.897 | 49314.157 | -2.6E-01 | -55.286 | -55.028 | -2.6E-01 |
| 347 | Mn | 53 | 25 | 28 | 3 | 8.734155 | 8.732781 | 1.4E-03 | 49301.687 | 49301.744 | -5.7E-02 | 49314.494 | 49314.551 | -5.7E-02 | -54.689 | -54.634 | -5.5E-02 |
| 348 | Fe | 53 | 26 | 27 | 1 | 8.648784 | 8.656685 | -7.9E-03 | 49304.915 | 49304.480 | 4.4E-01 | 49318.236 | 49317.801 | 4.4E-01 | -50.947 | -51.384 | 4.4E-01 |
| 349 | Co | 53 | 27 | 26 | -1 | 8.477643 | 8.494190 | -1.7E-02 | 49312.689 | 49311.795 | 8.9E-01 | 49326.524 | 49325.630 | 8.9E-01 | -42.659 | -43.555 | 9.0E-01 |
| 350 | Ni | 53 | 28 | 25 | -3 | 8.217074 | 8.218006 | -9.3E-04 | 49325.202 | 49325.135 | 6.7E-02 | 49339.552 | 49339.485 | 6.7E-02 | -29.631 | -29.701 | 7.0E-02 |
| 351 | Sc | 54 | 21 | 33 | 12 | 8.389275 | 8.381316 | 8.0E-03 | 50256.326 | 50256.743 | -4.2E-01 | 50267.078 | 50267.495 | -4.2E-01 | -33.599 | -33.184 | -4.1E-01 |
| 352 | Ti | 54 | 22 | 32 | 10 | 8.596973 | 8.603176 | -6.2E-03 | 50243.814 | 50243.466 | 3.5E-01 | 50255.080 | 50254.731 | 3.5E-01 | -45.597 | -45.948 | 3.5E-01 |
| 353 | V | 54 | 23 | 31 | 8 | 8.662023 | 8.666477 | -4.5E-03 | 50239.006 | 50238.751 | 2.5E-01 | 50250.785 | 50250.530 | 2.5E-01 | -49.892 | -50.149 | 2.6E-01 |
| 354 | Cr | 54 | 24 | 30 | 6 | 8.777935 | 8.787535 | -9.6E-03 | 50231.450 | 50230.917 | 5.3E-01 | 50243.743 | 50243.210 | 5.3E-01 | -56.934 | -57.470 | 5.4E-01 |
| 355 | Mn | 54 | 25 | 29 | 4 | 8.737944 | 8.739089 | -1.1E-03 | 50232.314 | 50232.236 | 7.7E-02 | 50245.120 | 50245.043 | 7.7E-02 | -55.557 | -55.636 | 8.0E-02 |
| 356 | Fe | 54 | 26 | 28 | 2 | 8.736370 | 8.738806 | -2.4E-03 | 50231.102 | 50230.954 | 1.5E-01 | 50244.423 | 50244.275 | 1.5E-01 | -56.254 | -56.404 | 1.5E-01 |
| 357 | Co | 54 | 27 | 27 | 0 | 8.569205 | 8.581584 | -1.2E-02 | 50238.832 | 50238.147 | 6.9E-01 | 50252.667 | 50251.982 | 6.9E-01 | -48.009 | -48.697 | 6.9E-01 |
| 358 | Ni | 54 | 28 | 26 | -2 | 8.392006 | 8.401896 | -9.9E-03 | 50247.104 | 50246.552 | 5.5E-01 | 50261.454 | 50260.902 | 5.5E-01 | -39.223 | -39.777 | 5.5E-01 |
| 359 | Sc | 55 | 21 | 34 | 13 | 8.317630 | 8.331444 | -1.4E-02 | 51191.442 | 51190.670 | 7.7E-01 | 51202.194 | 51201.422 | 7.7E-01 | -29.977 | -30.751 | 7.7E-01 |
| 360 | Ti | 55 | 22 | 33 | 11 | 8.515970 | 8.506868 | 9.1E-03 | 51179.238 | 51179.725 | -4.9E-01 | 51190.503 | 51190.991 | -4.9E-01 | -41.668 | -41.183 | -4.8E-01 |
| 361 | V | 55 | 23 | 32 | 9 | 8.637674 | 8.639358 | -1.7E-03 | 51171.248 | 51171.142 | 1.1E-01 | 51183.027 | 51182.921 | 1.1E-01 | -49.144 | -49.253 | 1.1E-01 |
| 362 | Cr | 55 | 24 | 31 | 7 | 8.731905 | 8.739952 | -8.0E-03 | 51164.769 | 51164.312 | 4.6E-01 | 51177.062 | 51176.605 | 4.6E-01 | -55.109 | -55.569 | 4.6E-01 |
| 363 | Mn | 55 | 25 | 30 | 5 | 8.765009 | 8.769841 | -4.8E-03 | 51161.652 | 51161.371 | 2.8E-01 | 51174.459 | 51174.178 | 2.8E-01 | -57.712 | -57.996 | 2.8E-01 |
| 364 | Fe | 55 | 26 | 29 | 3 | 8.746583 | 8.745025 | 1.6E-03 | 51161.369 | 51161.439 | -6.9E-02 | 51174.690 | 51174.760 | -6.9E-02 | -57.481 | -57.414 | -6.7E-02 |
| 365 | Co | 55 | 27 | 28 | 1 | 8.669606 | 8.670319 | -7.1E-04 | 51164.306 | 51164.250 | 5.6E-02 | 51178.142 | 51178.085 | 5.6E-02 | -54.029 | -54.088 | 5.9E-02 |
| 366 | Ni | 55 | 28 | 27 | -1 | 8.497308 | 8.508439 | -1.1E-02 | 51172.486 | 51171.856 | 6.3E-01 | 51186.836 | 51186.206 | 6.3E-01 | -45.335 | -45.968 | 6.3E-01 |
| 367 | Cu | 55 | 29 | 26 | -3 | 8.233996 | 8.234588 | -5.9E-04 | 51185.671 | 51185.620 | 5.1E-02 | 51200.535 | 51200.484 | 5.1E-02 | -31.635 | -31.689 | 5.4E-02 |
| 368 | Ti | 56 | 22 | 34 | 12 | 8.464062 | 8.475624 | -1.2E-02 | 52113.194 | 52112.533 | 6.6E-01 | 52124.459 | 52123.799 | 6.6E-01 | -39.205 | -39.869 | 6.6E-01 |
| 369 | V | 56 | 23 | 33 | 10 | 8.573623 | 8.567238 | 6.4E-03 | 52105.763 | 52106.106 | -3.4E-01 | 52117.542 | 52117.885 | -3.4E-01 | -46.123 | -45.782 | -3.4E-01 |
| 370 | Cr | 56 | 24 | 32 | 8 | 8.723191 | 8.740135 | -1.7E-02 | 52096.091 | 52095.127 | 9.6E-01 | 52108.384 | 52107.420 | 9.6E-01 | -55.281 | -56.248 | 9.7E-01 |
| 371 | Mn | 56 | 25 | 31 | 6 | 8.738320 | 8.747778 | -9.5E-03 | 52093.947 | 52093.402 | 5.5E-01 | 52106.754 | 52106.209 | 5.5E-01 | -56.911 | -57.459 | 5.5E-01 |
| 372 | Fe | 56 | 26 | 30 | 4 | 8.790342 | 8.796540 | -6.2E-03 | 52089.738 | 52089.374 | 3.6E-01 | 52103.058 | 52102.695 | 3.6E-01 | -60.606 | -60.972 | 3.7E-01 |
| 373 | Co | 56 | 27 | 29 | 2 | 8.694825 | 8.699464 | -4.6E-03 | 52093.790 | 52093.513 | 2.8E-01 | 52107.625 | 52107.348 | 2.8E-01 | -56.040 | -56.319 | 2.8E-01 |
| 374 | Ni | 56 | 28 | 28 | 0 | 8.642767 | 8.629747 | 1.3E-02 | 52095.408 | 52096.120 | -7.1E-01 | 52109.758 | 52110.469 | -7.1E-01 | -53.907 | -53.198 | -7.1E-01 |
| 375 | Ti | 57 | 22 | 35 | 13 | 8.363527 | 8.373263 | -9.7E-03 | 53050.026 | 53049.458 | 5.7E-01 | 53061.291 | 53060.723 | 5.7E-01 | -33.868 | -34.438 | 5.7E-01 |
| 376 | V | 57 | 23 | 34 | 11 | 8.531570 | 8.527818 | 3.8E-03 | 53039.151 | 53039.351 | -2.0E-01 | 53050.930 | 53051.130 | -2.0E-01 | -44.228 | -44.031 | -2.0E-01 |
| 377 | Cr | 57 | 24 | 33 | 9 | 8.663384 | 8.667271 | -3.9E-03 | 53030.342 | 53030.106 | 2.4E-01 | 53042.635 | 53042.398 | 2.4E-01 | -52.524 | -52.763 | 2.4E-01 |
| 378 | Mn | 57 | 25 | 32 | 7 | 8.736711 | 8.749361 | -1.3E-02 | 53024.866 | 53024.130 | 7.4E-01 | 53037.673 | 53036.936 | 7.4E-01 | -57.486 | -58.225 | 7.4E-01 |
| 379 | Fe | 57 | 26 | 31 | 5 | 8.770267 | 8.775027 | -4.8E-03 | 53021.657 | 53021.369 | 2.9E-01 | 53034.978 | 53034.690 | 2.9E-01 | -60.181 | -60.471 | 2.9E-01 |
| 380 | Co | 57 | 27 | 30 | 3 | 8.741871 | 8.747356 | -5.5E-03 | 53021.979 | 53021.649 | 3.3E-01 | 53035.814 | 53035.484 | 3.3E-01 | -59.345 | -59.677 | 3.3E-01 |
| 381 | Ni | 57 | 28 | 29 | 1 | 8.670923 | 8.666987 | 3.9E-03 | 53024.726 | 53024.933 | -2.1E-01 | 53039.076 | 53039.282 | -2.1E-01 | -56.083 | -55.879 | -2.0E-01 |
| 382 | Cu | 57 | 29 | 28 | -1 | 8.503251 | 8.505699 | -2.4E-03 | 53032.986 | 53032.828 | 1.6E-01 | 53047.851 | 53047.693 | 1.6E-01 | -47.308 | -47.469 | 1.6E-01 |
| 383 | V | 58 | 23 | 35 | 12 | 8.456240 | 8.446905 | 9.3E-03 | 53974.554 | 53975.082 | -5.3E-01 | 53986.333 | 53986.861 | -5.3E-01 | -40.319 | -39.795 | -5.2E-01 |
| 384 | Cr | 58 | 24 | 34 | 10 | 8.641290 | 8.649777 | -8.5E-03 | 53962.525 | 53962.018 | 5.1E-01 | 53974.818 | 53974.311 | 5.1E-01 | -51.835 | -52.344 | 5.1E-01 |
| 385 | Mn | 58 | 25 | 33 | 8 | 8.696643 | 8.701442 | -4.8E-03 | 53958.019 | 53957.725 | 2.9E-01 | 53970.825 | 53970.532 | 2.9E-01 | -55.828 | -56.124 | 3.0E-01 |
| 386 | Fe | 58 | 26 | 32 | 6 | 8.792239 | 8.799574 | -7.3E-03 | 53951.178 | 53950.736 | 4.4E-01 | 53964.498 | 53964.057 | 4.4E-01 | -62.154 | -62.599 | 4.4E-01 |
| 387 | Co | 58 | 27 | 31 | 4 | 8.738959 | 8.745811 | -6.9E-03 | 53952.971 | 53952.557 | 4.1E-01 | 53966.806 | 53966.392 | 4.1E-01 | -59.847 | -60.264 | 4.2E-01 |
| 388 | Ni | 58 | 28 | 30 | 2 | 8.732049 | 8.736100 | -4.1E-03 | 53952.075 | 53951.823 | 2.5E-01 | 53966.425 | 53966.172 | 2.5E-01 | -60.228 | -60.483 | 2.6E-01 |
| 389 | Cu | 58 | 29 | 29 | 0 | 8.570957 | 8.575587 | -4.6E-03 | 53960.121 | 53959.835 | 2.9E-01 | 53974.986 | 53974.699 | 2.9E-01 | -51.667 | -51.957 | 2.9E-01 |
| 390 | Zn | 58 | 30 | 28 | -2 | 8.395934 | 8.402863 | -6.9E-03 | 53968.976 | 53968.555 | 4.2E-01 | 53984.355 | 53983.934 | 4.2E-01 | -42.298 | -42.722 | 4.2E-01 |
| 391 | V | 59 | 23 | 36 | 13 | 8.407555 | 8.405712 | 1.8E-03 | 54908.536 | 54908.631 | -9.5E-02 | 54920.315 | 54920.410 | -9.5E-02 | -37.832 | -37.740 | -9.2E-02 |
| 392 | Cr | 59 | 24 | 35 | 11 | 8.564795 | 8.568337 | -3.5E-03 | 54897.963 | 54897.739 | 2.2E-01 | 54910.255 | 54910.032 | 2.2E-01 | -47.891 | -48.118 | 2.3E-01 |
| 393 | Mn | 59 | 25 | 34 | 9 | 8.680921 | 8.683962 | -3.0E-03 | 54889.815 | 54889.620 | 1.9E-01 | 54902.621 | 54902.427 | 1.9E-01 | -55.525 | -55.723 | 2.0E-01 |
| 394 | Fe | 59 | 26 | 33 | 7 | 8.754760 | 8.754729 | 3.1E-05 | 54884.162 | 54884.148 | 1.4E-02 | 54897.483 | 54897.468 | 1.4E-02 | -60.664 | -60.681 | 1.7E-02 |
| 395 | Co | 59 | 27 | 32 | 5 | 8.768025 | 8.770011 | -2.0E-03 | 54882.083 | 54881.949 | 1.3E-01 | 54895.918 | 54895.784 | 1.3E-01 | -62.229 | -62.366 | 1.4E-01 |
| 396 | Ni | 59 | 28 | 31 | 3 | 8.736578 | 8.735562 | 1.0E-03 | 54882.641 | 54882.684 | -4.2E-02 | 54896.991 | 54897.033 | -4.2E-02 | -61.156 | -61.117 | -4.0E-02 |
| 397 | Cu | 59 | 29 | 30 | 1 | 8.641990 | 8.650485 | -8.5E-03 | 54886.925 | 54886.405 | 5.2E-01 | 54901.789 | 54901.270 | 5.2E-01 | -56.358 | -56.880 | 5.2E-01 |
| 398 | Zn | 59 | 30 | 29 | -1 | 8.473767 | 8.484372 | -1.1E-02 | 54895.553 | 54894.908 | 6.4E-01 | 54910.932 | 54910.287 | 6.4E-01 | -47.215 | -47.862 | 6.5E-01 |
| 399 | V | 60 | 23 | 37 | 14 | 8.325450 | 8.320125 | 5.3E-03 | 55844.620 | 55844.926 | -3.1E-01 | 55856.399 | 55856.705 | -3.1E-01 | -33.242 | -32.939 | -3.0E-01 |
| 400 | Cr | 60 | 24 | 36 | 12 | 8.533443 | 8.545237 | -1.2E-02 | 55830.844 | 55830.122 | 7.2E-01 | 55843.137 | 55842.415 | 7.2E-01 | -46.504 | -47.229 | 7.3E-01 |
| 401 | Mn | 60 | 25 | 35 | 10 | 8.628138 | 8.623930 | 4.2E-03 | 55823.866 | 55824.103 | -2.4E-01 | 55836.673 | 55836.910 | -2.4E-01 | -52.968 | -52.734 | -2.3E-01 |
| 402 | Fe | 60 | 26 | 34 | 8 | 8.755840 | 8.758556 | -2.7E-03 | 55814.908 | 55814.729 | 1.8E-01 | 55828.228 | 55828.049 | 1.8E-01 | -61.412 | -61.594 | 1.8E-01 |
| 403 | Co | 60 | 27 | 33 | 6 | 8.746757 | 8.745954 | 8.0E-04 | 55814.156 | 55814.187 | -3.1E-02 | 55827.991 | 55828.023 | -3.1E-02 | -61.650 | -61.621 | -2.9E-02 |
| 404 | Ni | 60 | 28 | 32 | 4 | 8.780764 | 8.779024 | 1.7E-03 | 55810.819 | 55810.906 | -8.7E-02 | 55825.168 | 55825.255 | -8.7E-02 | -64.473 | -64.388 | -8.4E-02 |
| 405 | Cu | 60 | 29 | 31 | 2 | 8.665592 | 8.672980 | -7.4E-03 | 55816.432 | 55815.971 | 4.6E-01 | 55831.296 | 55830.835 | 4.6E-01 | -58.345 | -58.809 | 4.6E-01 |
| 406 | Zn | 60 | 30 | 30 | 0 | 8.583039 | 8.589725 | -6.7E-03 | 55820.088 | 55819.668 | 4.2E-01 | 55835.467 | 55835.047 | 4.2E-01 | -54.174 | -54.597 | 4.2E-01 |
| 407 | V | 61 | 23 | 38 | 15 | 8.276439 | 8.273331 | 3.1E-03 | 56778.849 | 56779.025 | -1.8E-01 | 56790.628 | 56790.804 | -1.8E-01 | -30.506 | -30.334 | -1.7E-01 |
| 408 | Cr | 61 | 24 | 37 | 13 | 8.459494 | 8.458575 | 9.2E-04 | 56766.387 | 56766.429 | -4.2E-02 | 56778.680 | 56778.721 | -4.2E-02 | -42.455 | -42.416 | -3.9E-02 |
| 409 | Mn | 61 | 25 | 36 | 11 | 8.598915 | 8.600630 | -1.7E-03 | 56756.586 | 56756.466 | 1.2E-01 | 56769.393 | 56769.273 | 1.2E-01 | -51.742 | -51.865 | 1.2E-01 |
| 410 | Fe | 61 | 26 | 35 | 9 | 8.703768 | 8.702061 | 1.7E-03 | 56748.894 | 56748.982 | -8.8E-02 | 56762.214 | 56762.302 | -8.8E-02 | -58.920 | -58.835 | -8.5E-02 |
| 411 | Co | 61 | 27 | 34 | 7 | 8.756141 | 8.752281 | 3.9E-03 | 56744.402 | 56744.621 | -2.2E-01 | 56758.237 | 56758.456 | -2.2E-01 | -62.898 | -62.682 | -2.2E-01 |
| 412 | Ni | 61 | 28 | 33 | 5 | 8.765016 | 8.757811 | 7.2E-03 | 56742.564 | 56742.986 | -4.2E-01 | 56756.914 | 56757.336 | -4.2E-01 | -64.221 | -63.802 | -4.2E-01 |
| 413 | Cu | 61 | 29 | 32 | 3 | 8.715510 | 8.715997 | -4.9E-04 | 56744.287 | 56744.239 | 4.8E-02 | 56759.151 | 56759.103 | 4.8E-02 | -61.984 | -62.035 | 5.1E-02 |
| 414 | Zn | 61 | 30 | 31 | 1 | 8.610306 | 8.618438 | -8.1E-03 | 56749.407 | 56748.892 | 5.1E-01 | 56764.786 | 56764.271 | 5.1E-01 | -56.349 | -56.867 | 5.2E-01 |
| 415 | Ga | 61 | 31 | 30 | -1 | 8.446429 | 8.458190 | -1.2E-02 | 56758.106 | 56757.369 | 7.4E-01 | 56774.000 | 56773.263 | 7.4E-01 | -47.135 | -47.875 | 7.4E-01 |
| 416 | Cr | 62 | 24 | 38 | 14 | 8.428069 | 8.428484 | -4.2E-04 | 57699.441 | 57699.401 | 4.0E-02 | 57711.734 | 57711.694 | 4.0E-02 | -40.895 | -40.938 | 4.3E-02 |
| 417 | Fe | 62 | 26 | 36 | 10 | 8.692882 | 8.696683 | -3.8E-03 | 57680.430 | 57680.178 | 2.5E-01 | 57693.751 | 57693.499 | 2.5E-01 | -58.878 | -59.133 | 2.5E-01 |
| 418 | Co | 62 | 27 | 35 | 8 | 8.721325 | 8.714221 | 7.1E-03 | 57677.370 | 57677.794 | -4.2E-01 | 57691.205 | 57691.629 | -4.2E-01 | -61.424 | -61.003 | -4.2E-01 |
| 419 | Ni | 62 | 28 | 34 | 6 | 8.794546 | 8.783001 | 1.2E-02 | 57671.533 | 57672.232 | -7.0E-01 | 57685.883 | 57686.581 | -7.0E-01 | -66.746 | -66.050 | -7.0E-01 |
| 420 | Cu | 62 | 29 | 33 | 4 | 8.718074 | 8.714754 | 3.3E-03 | 57674.978 | 57675.165 | -1.9E-01 | 57689.842 | 57690.030 | -1.9E-01 | -62.787 | -62.602 | -1.8E-01 |
| 421 | Zn | 62 | 30 | 32 | 2 | 8.679335 | 8.685169 | -5.8E-03 | 57676.082 | 57675.702 | 3.8E-01 | 57691.461 | 57691.081 | 3.8E-01 | -61.168 | -61.551 | 3.8E-01 |
| 422 | Ga | 62 | 31 | 31 | 0 | 8.518635 | 8.515658 | 3.0E-03 | 57684.748 | 57684.913 | -1.7E-01 | 57700.642 | 57700.807 | -1.6E-01 | -51.986 | -51.824 | -1.6E-01 |
| 423 | Cr | 63 | 24 | 39 | 15 | 8.340298 | 8.336076 | 4.2E-03 | 58636.108 | 58636.360 | -2.5E-01 | 58648.401 | 58648.652 | -2.5E-01 | -35.722 | -35.474 | -2.5E-01 |
| 424 | Mn | 63 | 25 | 38 | 13 | 8.505101 | 8.504483 | 6.2E-04 | 58624.429 | 58624.453 | -2.4E-02 | 58637.236 | 58637.260 | -2.4E-02 | -46.887 | -46.866 | -2.1E-02 |
| 425 | Fe | 63 | 26 | 37 | 11 | 8.631549 | 8.634763 | -3.2E-03 | 58615.166 | 58614.948 | 2.2E-01 | 58628.487 | 58628.269 | 2.2E-01 | -55.636 | -55.857 | 2.2E-01 |
| 426 | Co | 63 | 27 | 36 | 9 | 8.717788 | 8.711820 | 6.0E-03 | 58608.437 | 58608.796 | -3.6E-01 | 58622.272 | 58622.631 | -3.6E-01 | -61.851 | -61.495 | -3.6E-01 |
| 427 | Ni | 63 | 28 | 35 | 7 | 8.763486 | 8.748855 | 1.5E-02 | 58604.261 | 58605.165 | -9.0E-01 | 58618.611 | 58619.515 | -9.0E-01 | -65.512 | -64.611 | -9.0E-01 |
| 428 | Cu | 63 | 29 | 34 | 5 | 8.752131 | 8.741804 | 1.0E-02 | 58603.679 | 58604.312 | -6.3E-01 | 58618.544 | 58619.176 | -6.3E-01 | -65.579 | -64.950 | -6.3E-01 |
| 429 | Zn | 63 | 30 | 33 | 3 | 8.686281 | 8.688244 | -2.0E-03 | 58606.531 | 58606.388 | 1.4E-01 | 58621.910 | 58621.767 | 1.4E-01 | -62.213 | -62.359 | 1.5E-01 |
| 430 | Ga | 63 | 31 | 32 | 1 | 8.583926 | 8.586730 | -2.8E-03 | 58611.682 | 58611.486 | 2.0E-01 | 58627.576 | 58627.380 | 2.0E-01 | -56.547 | -56.746 | 2.0E-01 |
| 431 | Ge | 63 | 32 | 31 | -1 | 8.418716 | 8.419710 | -9.9E-04 | 58620.792 | 58620.710 | 8.3E-02 | 58637.202 | 58637.119 | 8.3E-02 | -46.921 | -47.007 | 8.6E-02 |
| 432 | Mn | 64 | 25 | 39 | 14 | 8.437417 | 8.430074 | 7.3E-03 | 59559.821 | 59560.276 | -4.5E-01 | 59572.628 | 59573.083 | -4.5E-01 | -42.989 | -42.537 | -4.5E-01 |
| 433 | Fe | 64 | 26 | 38 | 12 | 8.612388 | 8.621208 | -8.8E-03 | 59547.327 | 59546.746 | 5.8E-01 | 59560.647 | 59560.067 | 5.8E-01 | -54.970 | -55.553 | 5.8E-01 |
| 434 | Co | 64 | 27 | 37 | 10 | 8.675513 | 8.666080 | 9.4E-03 | 59541.990 | 59542.577 | -5.9E-01 | 59555.825 | 59556.412 | -5.9E-01 | -59.792 | -59.208 | -5.8E-01 |
| 435 | Ni | 64 | 28 | 36 | 8 | 8.777454 | 8.762134 | 1.5E-02 | 59534.169 | 59535.132 | -9.6E-01 | 59548.518 | 59549.482 | -9.6E-01 | -67.098 | -66.138 | -9.6E-01 |
| 436 | Cu | 64 | 29 | 35 | 6 | 8.739068 | 8.725309 | 1.4E-02 | 59535.329 | 59536.191 | -8.6E-01 | 59550.193 | 59551.055 | -8.6E-01 | -65.424 | -64.565 | -8.6E-01 |
| 437 | Zn | 64 | 30 | 34 | 4 | 8.735901 | 8.735270 | 6.3E-04 | 59534.234 | 59534.256 | -2.2E-02 | 59549.613 | 59549.635 | -2.2E-02 | -66.004 | -65.985 | -1.9E-02 |
| 438 | Ga | 64 | 31 | 33 | 2 | 8.611630 | 8.614654 | -3.0E-03 | 59540.890 | 59540.677 | 2.1E-01 | 59556.784 | 59556.571 | 2.1E-01 | -58.833 | -59.049 | 2.2E-01 |
| 439 | Ge | 64 | 32 | 32 | 0 | 8.528823 | 8.518283 | 1.1E-02 | 59544.892 | 59545.547 | -6.5E-01 | 59561.301 | 59561.956 | -6.5E-01 | -54.315 | -53.664 | -6.5E-01 |
| 440 | Mn | 65 | 25 | 40 | 15 | 8.400681 | 8.393767 | 6.9E-03 | 60493.337 | 60493.771 | -4.3E-01 | 60506.144 | 60506.578 | -4.3E-01 | -40.967 | -40.536 | -4.3E-01 |
| 441 | Fe | 65 | 26 | 39 | 13 | 8.546401 | 8.552207 | -5.8E-03 | 60482.569 | 60482.175 | 3.9E-01 | 60495.889 | 60495.496 | 3.9E-01 | -51.221 | -51.618 | 4.0E-01 |
| 442 | Co | 65 | 27 | 38 | 11 | 8.656884 | 8.657280 | -4.0E-04 | 60474.091 | 60474.048 | 4.2E-02 | 60487.926 | 60487.884 | 4.2E-02 | -59.185 | -59.231 | 4.5E-02 |
| 443 | Ni | 65 | 28 | 37 | 9 | 8.736233 | 8.721186 | 1.5E-02 | 60467.636 | 60468.597 | -9.6E-01 | 60481.986 | 60482.946 | -9.6E-01 | -65.125 | -64.168 | -9.6E-01 |
| 444 | Cu | 65 | 29 | 36 | 7 | 8.757093 | 8.741685 | 1.5E-02 | 60464.983 | 60465.967 | -9.8E-01 | 60479.847 | 60480.831 | -9.8E-01 | -67.263 | -66.283 | -9.8E-01 |
| 445 | Zn | 65 | 30 | 35 | 5 | 8.724262 | 8.723421 | 8.4E-04 | 60465.820 | 60465.856 | -3.6E-02 | 60481.199 | 60481.235 | -3.6E-02 | -65.912 | -65.879 | -3.3E-02 |
| 446 | Ga | 65 | 31 | 34 | 3 | 8.662157 | 8.663542 | -1.4E-03 | 60468.560 | 60468.450 | 1.1E-01 | 60484.454 | 60484.344 | 1.1E-01 | -62.657 | -62.770 | 1.1E-01 |
| 447 | Ge | 65 | 32 | 33 | 1 | 8.555058 | 8.549752 | 5.3E-03 | 60474.223 | 60474.548 | -3.2E-01 | 60490.633 | 60490.957 | -3.2E-01 | -56.478 | -56.157 | -3.2E-01 |
| 448 | As | 65 | 33 | 32 | -1 | 8.396234 | 8.395355 | 8.8E-04 | 60483.249 | 60483.285 | -3.6E-02 | 60500.174 | 60500.210 | -3.6E-02 | -46.937 | -46.904 | -3.3E-02 |
| 449 | Mn | 66 | 25 | 41 | 16 | 8.331798 | 8.314250 | 1.8E-02 | 61429.048 | 61430.191 | -1.1E+00 | 61441.855 | 61442.998 | -1.1E+00 | -36.750 | -35.610 | -1.1E+00 |
| 450 | Fe | 66 | 26 | 40 | 14 | 8.521724 | 8.529400 | -7.7E-03 | 61415.216 | 61414.694 | 5.2E-01 | 61428.537 | 61428.015 | 5.2E-01 | -50.068 | -50.593 | 5.3E-01 |
| 451 | Co | 66 | 27 | 39 | 12 | 8.605941 | 8.602973 | 3.0E-03 | 61408.361 | 61408.541 | -1.8E-01 | 61422.196 | 61422.376 | -1.8E-01 | -56.409 | -56.232 | -1.8E-01 |
| 452 | Ni | 66 | 28 | 38 | 10 | 8.739508 | 8.725277 | 1.4E-02 | 61398.249 | 61399.171 | -9.2E-01 | 61412.599 | 61413.521 | -9.2E-01 | -66.006 | -65.087 | -9.2E-01 |
| 453 | Cu | 66 | 29 | 37 | 8 | 8.731469 | 8.715113 | 1.6E-02 | 61397.483 | 61398.544 | -1.1E+00 | 61412.347 | 61413.408 | -1.1E+00 | -66.258 | -65.200 | -1.1E+00 |
| 454 | Zn | 66 | 30 | 36 | 6 | 8.759630 | 8.756143 | 3.5E-03 | 61394.327 | 61394.538 | -2.1E-01 | 61409.706 | 61409.917 | -2.1E-01 | -68.899 | -68.691 | -2.1E-01 |
| 455 | Ga | 66 | 31 | 35 | 4 | 8.669367 | 8.671745 | -2.4E-03 | 61398.987 | 61398.810 | 1.8E-01 | 61414.881 | 61414.705 | 1.8E-01 | -63.724 | -63.904 | 1.8E-01 |
| 456 | Ge | 66 | 32 | 34 | 2 | 8.625437 | 8.623190 | 2.2E-03 | 61400.589 | 61400.717 | -1.3E-01 | 61416.998 | 61417.126 | -1.3E-01 | -61.607 | -61.482 | -1.3E-01 |
| 457 | As | 66 | 33 | 33 | 0 | 8.468403 | 8.451846 | 1.7E-02 | 61409.655 | 61410.727 | -1.1E+00 | 61426.580 | 61427.652 | -1.1E+00 | -52.025 | -50.956 | -1.1E+00 |
| 458 | Fe | 67 | 26 | 41 | 15 | 8.455310 | 8.453933 | 1.4E-03 | 62350.710 | 62350.786 | -7.7E-02 | 62364.030 | 62364.107 | -7.7E-02 | -46.069 | -45.995 | -7.3E-02 |
| 459 | Co | 67 | 27 | 40 | 13 | 8.581741 | 8.586557 | -4.8E-03 | 62340.942 | 62340.603 | 3.4E-01 | 62354.777 | 62354.438 | 3.4E-01 | -55.322 | -55.664 | 3.4E-01 |
| 460 | Ni | 67 | 28 | 39 | 11 | 8.695750 | 8.677624 | 1.8E-02 | 62332.007 | 62333.204 | -1.2E+00 | 62346.356 | 62347.554 | -1.2E+00 | -63.743 | -62.549 | -1.2E+00 |
| 461 | Cu | 67 | 29 | 38 | 9 | 8.737447 | 8.723041 | 1.4E-02 | 62327.916 | 62328.863 | -9.5E-01 | 62342.780 | 62343.728 | -9.5E-01 | -67.319 | -66.375 | -9.4E-01 |
| 462 | Zn | 67 | 30 | 37 | 7 | 8.734148 | 8.733891 | 2.6E-04 | 62326.840 | 62326.838 | 1.4E-03 | 62342.219 | 62342.217 | 1.4E-03 | -67.880 | -67.885 | 4.6E-03 |
| 463 | Ga | 67 | 31 | 36 | 5 | 8.707528 | 8.707127 | 4.0E-04 | 62327.326 | 62327.334 | -7.6E-03 | 62343.220 | 62343.228 | -7.6E-03 | -66.879 | -66.875 | -4.4E-03 |
| 464 | Ge | 67 | 32 | 35 | 3 | 8.632853 | 8.636209 | -3.4E-03 | 62331.032 | 62330.787 | 2.4E-01 | 62347.441 | 62347.196 | 2.4E-01 | -62.658 | -62.906 | 2.5E-01 |
| 465 | As | 67 | 33 | 34 | 1 | 8.530568 | 8.525643 | 4.9E-03 | 62336.587 | 62336.896 | -3.1E-01 | 62353.512 | 62353.821 | -3.1E-01 | -56.587 | -56.281 | -3.1E-01 |
| 466 | Se | 67 | 34 | 33 | -1 | 8.369534 | 8.363597 | 5.9E-03 | 62346.078 | 62346.455 | -3.8E-01 | 62363.519 | 62363.895 | -3.8E-01 | -46.580 | -46.207 | -3.7E-01 |
| 467 | Fe | 68 | 26 | 42 | 16 | 8.416675 | 8.423231 | -6.6E-03 | 63284.447 | 63283.985 | 4.6E-01 | 63297.768 | 63297.306 | 4.6E-01 | -43.825 | -44.290 | 4.6E-01 |
| 468 | Co | 68 | 27 | 41 | 14 | 8.524264 | 8.524150 | 1.1E-04 | 63275.834 | 63275.826 | 8.6E-03 | 63289.669 | 63289.661 | 8.6E-03 | -51.924 | -51.936 | 1.2E-02 |
| 469 | Ni | 68 | 28 | 40 | 12 | 8.682466 | 8.671818 | 1.1E-02 | 63263.780 | 63264.487 | -7.1E-01 | 63278.129 | 63278.836 | -7.1E-01 | -63.464 | -62.760 | -7.0E-01 |
| 470 | Cu | 68 | 29 | 39 | 10 | 8.701890 | 8.687586 | 1.4E-02 | 63261.162 | 63262.117 | -9.5E-01 | 63276.026 | 63276.981 | -9.5E-01 | -65.567 | -64.615 | -9.5E-01 |
| 471 | Zn | 68 | 30 | 38 | 8 | 8.755677 | 8.754605 | 1.1E-03 | 63256.207 | 63256.261 | -5.4E-02 | 63271.586 | 63271.640 | -5.4E-02 | -70.007 | -69.956 | -5.1E-02 |
| 472 | Ga | 68 | 31 | 37 | 6 | 8.701214 | 8.700732 | 4.8E-04 | 63258.613 | 63258.627 | -1.4E-02 | 63274.507 | 63274.521 | -1.4E-02 | -67.086 | -67.075 | -1.0E-02 |
| 473 | Ge | 68 | 32 | 36 | 4 | 8.688136 | 8.690840 | -2.7E-03 | 63258.205 | 63258.001 | 2.0E-01 | 63274.614 | 63274.410 | 2.0E-01 | -66.979 | -67.186 | 2.1E-01 |
| 474 | As | 68 | 33 | 35 | 2 | 8.557745 | 8.561923 | -4.2E-03 | 63265.774 | 63265.469 | 3.0E-01 | 63282.698 | 63282.394 | 3.0E-01 | -58.895 | -59.203 | 3.1E-01 |
| 475 | Se | 68 | 34 | 34 | 0 | 8.477047 | 8.461511 | 1.6E-02 | 63269.963 | 63270.998 | -1.0E+00 | 63287.403 | 63288.439 | -1.0E+00 | -54.189 | -53.158 | -1.0E+00 |
| 476 | Co | 69 | 27 | 42 | 15 | 8.492270 | 8.500402 | -8.1E-03 | 64209.083 | 64208.505 | 5.8E-01 | 64222.918 | 64222.340 | 5.8E-01 | -50.169 | -50.750 | 5.8E-01 |
| 477 | Ni | 69 | 28 | 41 | 13 | 8.623099 | 8.616806 | 6.3E-03 | 64198.759 | 64199.176 | -4.2E-01 | 64213.108 | 64213.526 | -4.2E-01 | -59.979 | -59.565 | -4.1E-01 |
| 478 | Cu | 69 | 29 | 40 | 11 | 8.695204 | 8.686860 | 8.3E-03 | 64192.487 | 64193.044 | -5.6E-01 | 64207.351 | 64207.909 | -5.6E-01 | -65.736 | -65.182 | -5.5E-01 |
| 479 | Zn | 69 | 30 | 39 | 9 | 8.722726 | 8.723946 | -1.2E-03 | 64189.290 | 64189.188 | 1.0E-01 | 64204.669 | 64204.567 | 1.0E-01 | -68.418 | -68.524 | 1.1E-01 |
| 480 | Ga | 69 | 31 | 38 | 7 | 8.724579 | 8.724950 | -3.7E-04 | 64187.865 | 64187.820 | 4.5E-02 | 64203.759 | 64203.714 | 4.5E-02 | -69.328 | -69.376 | 4.8E-02 |
| 481 | Ge | 69 | 32 | 37 | 5 | 8.680963 | 8.689067 | -8.1E-03 | 64189.577 | 64188.998 | 5.8E-01 | 64205.986 | 64205.407 | 5.8E-01 | -67.101 | -67.683 | 5.8E-01 |
| 482 | As | 69 | 33 | 36 | 3 | 8.611820 | 8.617042 | -5.2E-03 | 64193.050 | 64192.669 | 3.8E-01 | 64209.975 | 64209.594 | 3.8E-01 | -63.112 | -63.496 | 3.8E-01 |
| 483 | Se | 69 | 34 | 35 | 1 | 8.503707 | 8.497367 | 6.3E-03 | 64199.212 | 64199.628 | -4.2E-01 | 64216.652 | 64217.068 | -4.2E-01 | -56.435 | -56.022 | -4.1E-01 |
| 484 | Br | 69 | 35 | 34 | -1 | 8.342757 | 8.350207 | -7.5E-03 | 64209.019 | 64208.483 | 5.4E-01 | 64226.975 | 64226.439 | 5.4E-01 | -46.111 | -46.651 | 5.4E-01 |
| 485 | Co | 70 | 27 | 43 | 16 | 8.439831 | 8.432654 | 7.2E-03 | 65143.827 | 65144.313 | -4.9E-01 | 65157.662 | 65158.148 | -4.9E-01 | -46.919 | -46.437 | -4.8E-01 |
| 486 | Ni | 70 | 28 | 42 | 14 | 8.604291 | 8.601765 | 2.5E-03 | 65131.018 | 65131.177 | -1.6E-01 | 65145.367 | 65145.527 | -1.6E-01 | -59.214 | -59.057 | -1.6E-01 |
| 487 | Cu | 70 | 29 | 41 | 12 | 8.646865 | 8.642244 | 4.6E-03 | 65126.740 | 65127.046 | -3.1E-01 | 65141.605 | 65141.910 | -3.1E-01 | -62.976 | -62.674 | -3.0E-01 |
| 488 | Zn | 70 | 30 | 40 | 10 | 8.729807 | 8.733433 | -3.6E-03 | 65119.637 | 65119.365 | 2.7E-01 | 65135.016 | 65134.744 | 2.7E-01 | -69.565 | -69.840 | 2.8E-01 |
| 489 | Ga | 70 | 31 | 39 | 8 | 8.709280 | 8.707252 | 2.0E-03 | 65119.777 | 65119.900 | -1.2E-01 | 65135.671 | 65135.794 | -1.2E-01 | -68.910 | -68.791 | -1.2E-01 |
| 490 | Ge | 70 | 32 | 38 | 6 | 8.721699 | 8.728104 | -6.4E-03 | 65117.610 | 65117.142 | 4.7E-01 | 65134.019 | 65133.551 | 4.7E-01 | -70.562 | -71.033 | 4.7E-01 |
| 491 | As | 70 | 33 | 37 | 4 | 8.621666 | 8.633671 | -1.2E-02 | 65123.314 | 65122.453 | 8.6E-01 | 65140.239 | 65139.378 | 8.6E-01 | -64.342 | -65.206 | 8.6E-01 |
| 492 | Se | 70 | 34 | 36 | 2 | 8.576033 | 8.574523 | 1.5E-03 | 65125.211 | 65125.295 | -8.4E-02 | 65142.651 | 65142.735 | -8.4E-02 | -61.930 | -61.849 | -8.1E-02 |
| 493 | Br | 70 | 35 | 35 | 0 | 8.414796 | 8.408151 | 6.6E-03 | 65135.199 | 65135.642 | -4.4E-01 | 65153.155 | 65153.598 | -4.4E-01 | -51.426 | -50.986 | -4.4E-01 |
| 494 | Co | 71 | 27 | 44 | 17 | 8.398734 | 8.403614 | -4.9E-03 | 66077.870 | 66077.507 | 3.6E-01 | 66091.705 | 66091.342 | 3.6E-01 | -44.370 | -44.736 | 3.7E-01 |
| 495 | Ni | 71 | 28 | 43 | 15 | 8.543156 | 8.541271 | 1.9E-03 | 66066.319 | 66066.436 | -1.2E-01 | 66080.669 | 66080.786 | -1.2E-01 | -55.406 | -55.293 | -1.1E-01 |
| 496 | Cu | 71 | 29 | 42 | 13 | 8.635022 | 8.633088 | 1.9E-03 | 66058.500 | 66058.619 | -1.2E-01 | 66073.364 | 66073.484 | -1.2E-01 | -62.711 | -62.595 | -1.2E-01 |
| 497 | Zn | 71 | 30 | 41 | 11 | 8.689041 | 8.694319 | -5.3E-03 | 66053.367 | 66052.974 | 3.9E-01 | 66068.746 | 66068.353 | 3.9E-01 | -67.329 | -67.725 | 4.0E-01 |
| 498 | Ga | 71 | 31 | 40 | 9 | 8.717604 | 8.721046 | -3.4E-03 | 66050.042 | 66049.778 | 2.6E-01 | 66065.936 | 66065.672 | 2.6E-01 | -70.139 | -70.406 | 2.7E-01 |
| 499 | Ge | 71 | 32 | 39 | 7 | 8.703308 | 8.714731 | -1.1E-02 | 66049.759 | 66048.928 | 8.3E-01 | 66066.169 | 66065.338 | 8.3E-01 | -69.906 | -70.741 | 8.3E-01 |
| 500 | As | 71 | 33 | 38 | 5 | 8.663932 | 8.674806 | -1.1E-02 | 66051.257 | 66050.465 | 7.9E-01 | 66068.182 | 66067.389 | 7.9E-01 | -67.893 | -68.689 | 8.0E-01 |
| 501 | Se | 71 | 34 | 37 | 3 | 8.586060 | 8.594046 | -8.0E-03 | 66055.488 | 66054.900 | 5.9E-01 | 66072.928 | 66072.340 | 5.9E-01 | -63.147 | -63.738 | 5.9E-01 |
| 502 | Br | 71 | 35 | 36 | 1 | 8.481462 | 8.482126 | -6.6E-04 | 66061.617 | 66061.547 | 6.9E-02 | 66079.573 | 66079.503 | 6.9E-02 | -56.502 | -56.575 | 7.3E-02 |
| 503 | Kr | 71 | 36 | 35 | -1 | 8.327130 | 8.326545 | 5.9E-04 | 66071.276 | 66071.295 | -1.9E-02 | 66089.748 | 66089.767 | -1.9E-02 | -46.327 | -46.312 | -1.5E-02 |
| 504 | Ni | 72 | 28 | 44 | 16 | 8.520211 | 8.519754 | 4.6E-04 | 66998.993 | 66999.009 | -1.6E-02 | 67013.343 | 67013.359 | -1.6E-02 | -54.226 | -54.214 | -1.3E-02 |
| 505 | Cu | 72 | 29 | 43 | 14 | 8.586525 | 8.581485 | 5.0E-03 | 66992.922 | 66993.267 | -3.5E-01 | 67007.786 | 67008.131 | -3.5E-01 | -59.783 | -59.441 | -3.4E-01 |
| 506 | Zn | 72 | 30 | 42 | 12 | 8.691804 | 8.692992 | -1.2E-03 | 66984.044 | 66983.941 | 1.0E-01 | 66999.423 | 66999.320 | 1.0E-01 | -68.145 | -68.253 | 1.1E-01 |
| 507 | Ga | 72 | 31 | 41 | 10 | 8.687088 | 8.693037 | -5.9E-03 | 66983.087 | 66982.639 | 4.5E-01 | 66998.981 | 66998.533 | 4.5E-01 | -68.588 | -69.039 | 4.5E-01 |
| 508 | Ge | 72 | 32 | 40 | 8 | 8.731745 | 8.740508 | -8.8E-03 | 66978.574 | 66977.923 | 6.5E-01 | 66994.983 | 66994.332 | 6.5E-01 | -72.586 | -73.240 | 6.5E-01 |
| 509 | As | 72 | 33 | 39 | 6 | 8.660378 | 8.676033 | -1.6E-02 | 66982.415 | 66981.267 | 1.1E+00 | 66999.339 | 66998.192 | 1.1E+00 | -68.230 | -69.381 | 1.2E+00 |
| 510 | Se | 72 | 34 | 38 | 4 | 8.644489 | 8.652576 | -8.1E-03 | 66982.261 | 66981.657 | 6.0E-01 | 66999.701 | 66999.097 | 6.0E-01 | -67.868 | -68.475 | 6.1E-01 |
| 511 | Br | 72 | 35 | 37 | 2 | 8.511389 | 8.521900 | -1.1E-02 | 66990.546 | 66989.767 | 7.8E-01 | 67008.502 | 67007.723 | 7.8E-01 | -59.067 | -59.850 | 7.8E-01 |
| 512 | Kr | 72 | 36 | 36 | 0 | 8.429319 | 8.422184 | 7.1E-03 | 66995.156 | 66995.647 | -4.9E-01 | 67013.628 | 67014.119 | -4.9E-01 | -53.941 | -53.453 | -4.9E-01 |
| 513 | Ni | 73 | 28 | 45 | 17 | 8.457652 | 8.457312 | 3.4E-04 | 67934.605 | 67934.613 | -8.0E-03 | 67948.955 | 67948.963 | -8.0E-03 | -50.108 | -50.104 | -4.5E-03 |
| 514 | Cu | 73 | 29 | 44 | 15 | 8.568569 | 8.566208 | 2.4E-03 | 67925.211 | 67925.366 | -1.5E-01 | 67940.076 | 67940.230 | -1.5E-01 | -58.987 | -58.836 | -1.5E-01 |
| 515 | Zn | 73 | 30 | 43 | 13 | 8.648345 | 8.647096 | 1.2E-03 | 67918.091 | 67918.163 | -7.3E-02 | 67933.470 | 67933.542 | -7.3E-02 | -65.593 | -65.524 | -6.9E-02 |
| 516 | Ga | 73 | 31 | 42 | 11 | 8.693873 | 8.696306 | -2.4E-03 | 67913.470 | 67913.273 | 2.0E-01 | 67929.364 | 67929.167 | 2.0E-01 | -69.699 | -69.899 | 2.0E-01 |
| 517 | Ge | 73 | 32 | 41 | 9 | 8.705049 | 8.717083 | -1.2E-02 | 67911.356 | 67910.458 | 9.0E-01 | 67927.765 | 67926.867 | 9.0E-01 | -71.298 | -72.199 | 9.0E-01 |
| 518 | As | 73 | 33 | 40 | 7 | 8.689609 | 8.705463 | -1.6E-02 | 67911.186 | 67910.008 | 1.2E+00 | 67928.110 | 67926.932 | 1.2E+00 | -70.953 | -72.134 | 1.2E+00 |
| 519 | Se | 73 | 34 | 39 | 5 | 8.641558 | 8.657494 | -1.6E-02 | 67913.395 | 67912.211 | 1.2E+00 | 67930.836 | 67929.651 | 1.2E+00 | -68.227 | -69.415 | 1.2E+00 |
| 520 | Br | 73 | 35 | 38 | 3 | 8.568104 | 8.578882 | -1.1E-02 | 67917.459 | 67916.651 | 8.1E-01 | 67935.415 | 67934.607 | 8.1E-01 | -63.648 | -64.460 | 8.1E-01 |
| 521 | Kr | 73 | 36 | 37 | 1 | 8.460184 | 8.459101 | 1.1E-03 | 67924.039 | 67924.096 | -5.6E-02 | 67942.511 | 67942.568 | -5.6E-02 | -56.552 | -56.499 | -5.3E-02 |
| 522 | Cu | 74 | 29 | 45 | 16 | 8.521562 | 8.511498 | 1.0E-02 | 68859.687 | 68860.414 | -7.3E-01 | 68874.551 | 68875.278 | -7.3E-01 | -56.006 | -55.282 | -7.2E-01 |
| 523 | Zn | 74 | 30 | 44 | 14 | 8.642754 | 8.637867 | 4.9E-03 | 68849.421 | 68849.765 | -3.4E-01 | 68864.800 | 68865.144 | -3.4E-01 | -65.757 | -65.417 | -3.4E-01 |
| 524 | Ga | 74 | 31 | 43 | 12 | 8.663167 | 8.659985 | 3.2E-03 | 68846.613 | 68846.830 | -2.2E-01 | 68862.507 | 68862.724 | -2.2E-01 | -68.050 | -67.837 | -2.1E-01 |
| 525 | Ge | 74 | 32 | 42 | 10 | 8.725200 | 8.730208 | -5.0E-03 | 68840.725 | 68840.335 | 3.9E-01 | 68857.135 | 68856.744 | 3.9E-01 | -73.422 | -73.816 | 3.9E-01 |
| 526 | As | 74 | 33 | 41 | 8 | 8.680001 | 8.694544 | -1.5E-02 | 68842.772 | 68841.676 | 1.1E+00 | 68859.697 | 68858.600 | 1.1E+00 | -70.860 | -71.960 | 1.1E+00 |
| 527 | Se | 74 | 34 | 40 | 6 | 8.687715 | 8.700643 | -1.3E-02 | 68840.904 | 68839.926 | 9.8E-01 | 68858.344 | 68857.366 | 9.8E-01 | -72.213 | -73.195 | 9.8E-01 |
| 528 | Br | 74 | 35 | 39 | 4 | 8.583562 | 8.600128 | -1.7E-02 | 68847.313 | 68846.065 | 1.2E+00 | 68865.269 | 68864.021 | 1.2E+00 | -65.288 | -66.540 | 1.3E+00 |
| 529 | Kr | 74 | 36 | 38 | 2 | 8.533038 | 8.535500 | -2.5E-03 | 68849.753 | 68849.548 | 2.0E-01 | 68868.225 | 68868.020 | 2.0E-01 | -62.332 | -62.540 | 2.1E-01 |
| 530 | Rb | 74 | 37 | 37 | 0 | 8.381711 | 8.374862 | 6.8E-03 | 68859.653 | 68860.136 | -4.8E-01 | 68878.641 | 68879.125 | -4.8E-01 | -51.916 | -51.436 | -4.8E-01 |
| 531 | Cu | 75 | 29 | 46 | 17 | 8.495094 | 8.494537 | 5.6E-04 | 69792.715 | 69792.740 | -2.4E-02 | 69807.580 | 69807.604 | -2.4E-02 | -54.471 | -54.451 | -2.1E-02 |
| 532 | Zn | 75 | 30 | 45 | 15 | 8.592497 | 8.588663 | 3.8E-03 | 69784.113 | 69784.382 | -2.7E-01 | 69799.492 | 69799.762 | -2.7E-01 | -62.559 | -62.293 | -2.7E-01 |
| 533 | Ga | 75 | 31 | 44 | 13 | 8.660808 | 8.655290 | 5.5E-03 | 69777.692 | 69778.087 | -3.9E-01 | 69793.586 | 69793.981 | -3.9E-01 | -68.465 | -68.073 | -3.9E-01 |
| 534 | Ge | 75 | 32 | 43 | 11 | 8.695609 | 8.698508 | -2.9E-03 | 69773.785 | 69773.548 | 2.4E-01 | 69790.194 | 69789.957 | 2.4E-01 | -71.857 | -72.098 | 2.4E-01 |
| 535 | As | 75 | 33 | 42 | 9 | 8.700874 | 8.712093 | -1.1E-02 | 69772.092 | 69771.230 | 8.6E-01 | 69789.017 | 69788.155 | 8.6E-01 | -73.034 | -73.900 | 8.7E-01 |
| 536 | Se | 75 | 34 | 41 | 7 | 8.678913 | 8.693693 | -1.5E-02 | 69772.441 | 69771.312 | 1.1E+00 | 69789.882 | 69788.752 | 1.1E+00 | -72.169 | -73.303 | 1.1E+00 |
| 537 | Br | 75 | 35 | 40 | 5 | 8.627650 | 8.644575 | -1.7E-02 | 69774.988 | 69773.697 | 1.3E+00 | 69792.944 | 69791.653 | 1.3E+00 | -69.107 | -70.402 | 1.3E+00 |
| 538 | Kr | 75 | 36 | 39 | 3 | 8.553439 | 8.557927 | -4.5E-03 | 69779.255 | 69778.896 | 3.6E-01 | 69797.727 | 69797.368 | 3.6E-01 | -64.324 | -64.686 | 3.6E-01 |
| 539 | Rb | 75 | 37 | 38 | 1 | 8.448275 | 8.446595 | 1.7E-03 | 69785.844 | 69785.947 | -1.0E-01 | 69804.832 | 69804.935 | -1.0E-01 | -57.219 | -57.120 | -9.9E-02 |
| 540 | Sr | 75 | 38 | 37 | -1 | 8.296511 | 8.296192 | 3.2E-04 | 69795.928 | 69795.928 | 6.6E-05 | 69815.432 | 69815.432 | 1.3E-04 | -46.619 | -46.622 | 3.7E-03 |
| 541 | Cu | 76 | 29 | 47 | 18 | 8.443527 | 8.442435 | 1.1E-03 | 70727.705 | 70727.770 | -6.6E-02 | 70742.569 | 70742.635 | -6.6E-02 | -50.976 | -50.914 | -6.2E-02 |
| 542 | Zn | 76 | 30 | 46 | 16 | 8.582273 | 8.576473 | 5.8E-03 | 70715.863 | 70716.286 | -4.2E-01 | 70731.242 | 70731.665 | -4.2E-01 | -62.303 | -61.884 | -4.2E-01 |
| 543 | Ga | 76 | 31 | 45 | 14 | 8.624526 | 8.614442 | 1.0E-02 | 70711.354 | 70712.102 | -7.5E-01 | 70727.248 | 70727.996 | -7.5E-01 | -66.297 | -65.553 | -7.4E-01 |
| 544 | Ge | 76 | 32 | 44 | 12 | 8.705236 | 8.701960 | 3.3E-03 | 70703.923 | 70704.152 | -2.3E-01 | 70720.332 | 70720.561 | -2.3E-01 | -73.213 | -72.987 | -2.3E-01 |
| 545 | As | 76 | 33 | 43 | 10 | 8.682816 | 8.691427 | -8.6E-03 | 70704.329 | 70703.654 | 6.7E-01 | 70721.254 | 70720.579 | 6.7E-01 | -72.291 | -72.970 | 6.8E-01 |
| 546 | Se | 76 | 34 | 42 | 8 | 8.711477 | 8.722788 | -1.1E-02 | 70700.853 | 70699.972 | 8.8E-01 | 70718.293 | 70717.412 | 8.8E-01 | -75.252 | -76.136 | 8.8E-01 |
| 547 | Br | 76 | 35 | 41 | 6 | 8.635882 | 8.651169 | -1.5E-02 | 70705.300 | 70704.116 | 1.2E+00 | 70723.256 | 70722.072 | 1.2E+00 | -70.289 | -71.476 | 1.2E+00 |
| 548 | Kr | 76 | 36 | 40 | 4 | 8.608813 | 8.617894 | -9.1E-03 | 70706.059 | 70705.346 | 7.1E-01 | 70724.531 | 70723.818 | 7.1E-01 | -69.014 | -69.730 | 7.2E-01 |
| 549 | Rb | 76 | 37 | 39 | 2 | 8.486215 | 8.486917 | -7.0E-04 | 70714.078 | 70714.001 | 7.7E-02 | 70733.066 | 70732.989 | 7.7E-02 | -60.479 | -60.559 | 8.0E-02 |
| 550 | Sr | 76 | 38 | 38 | 0 | 8.393929 | 8.388009 | 5.9E-03 | 70719.793 | 70720.219 | -4.3E-01 | 70739.297 | 70739.723 | -4.3E-01 | -54.248 | -53.825 | -4.2E-01 |
| 551 | Zn | 77 | 30 | 47 | 17 | 8.530003 | 8.529472 | 5.3E-04 | 71650.871 | 71650.894 | -2.3E-02 | 71666.250 | 71666.273 | -2.3E-02 | -58.789 | -58.770 | -1.9E-02 |
| 552 | Ga | 77 | 31 | 46 | 15 | 8.613390 | 8.606206 | 7.2E-03 | 71643.153 | 71643.687 | -5.3E-01 | 71659.047 | 71659.581 | -5.3E-01 | -65.992 | -65.462 | -5.3E-01 |
| 553 | Ge | 77 | 32 | 45 | 13 | 8.671028 | 8.665367 | 5.7E-03 | 71637.417 | 71637.833 | -4.2E-01 | 71653.826 | 71654.243 | -4.2E-01 | -71.213 | -70.800 | -4.1E-01 |
| 554 | As | 77 | 33 | 44 | 11 | 8.695978 | 8.699352 | -3.4E-03 | 71634.198 | 71633.918 | 2.8E-01 | 71651.123 | 71650.843 | 2.8E-01 | -73.916 | -74.200 | 2.8E-01 |
| 555 | Se | 77 | 34 | 43 | 9 | 8.694690 | 8.706204 | -1.2E-02 | 71632.999 | 71632.092 | 9.1E-01 | 71650.440 | 71649.532 | 9.1E-01 | -74.599 | -75.511 | 9.1E-01 |
| 556 | Br | 77 | 35 | 42 | 7 | 8.666806 | 8.683727 | -1.7E-02 | 71633.848 | 71632.524 | 1.3E+00 | 71651.804 | 71650.480 | 1.3E+00 | -73.235 | -74.563 | 1.3E+00 |
| 557 | Kr | 77 | 36 | 41 | 5 | 8.616836 | 8.627231 | -1.0E-02 | 71636.398 | 71635.575 | 8.2E-01 | 71654.870 | 71654.047 | 8.2E-01 | -70.169 | -70.996 | 8.3E-01 |
| 558 | Rb | 77 | 37 | 40 | 3 | 8.537339 | 8.544106 | -6.8E-03 | 71641.220 | 71640.676 | 5.4E-01 | 71660.209 | 71659.664 | 5.4E-01 | -64.830 | -65.378 | 5.5E-01 |
| 559 | Sr | 77 | 38 | 39 | 1 | 8.435918 | 8.424652 | 1.1E-02 | 71647.731 | 71648.575 | -8.4E-01 | 71667.236 | 71668.079 | -8.4E-01 | -57.803 | -56.964 | -8.4E-01 |
| 560 | Cu | 78 | 29 | 49 | 20 | 8.350925 | 8.373891 | -2.3E-02 | 72597.171 | 72595.363 | 1.8E+00 | 72612.036 | 72610.227 | 1.8E+00 | -44.497 | -46.310 | 1.8E+00 |
| 561 | Zn | 78 | 30 | 48 | 18 | 8.507379 | 8.518618 | -1.1E-02 | 72583.671 | 72582.776 | 8.9E-01 | 72599.050 | 72598.155 | 8.9E-01 | -57.483 | -58.382 | 9.0E-01 |
| 562 | Ga | 78 | 31 | 47 | 16 | 8.577127 | 8.566552 | 1.1E-02 | 72576.933 | 72577.739 | -8.1E-01 | 72592.827 | 72593.633 | -8.1E-01 | -63.706 | -62.904 | -8.0E-01 |
| 563 | Ge | 78 | 32 | 46 | 14 | 8.671663 | 8.663797 | 7.9E-03 | 72568.262 | 72568.856 | -5.9E-01 | 72584.671 | 72585.265 | -5.9E-01 | -71.862 | -71.272 | -5.9E-01 |
| 564 | As | 78 | 33 | 45 | 12 | 8.673875 | 8.672551 | 1.3E-03 | 72566.791 | 72566.875 | -8.3E-02 | 72583.716 | 72583.799 | -8.3E-02 | -72.817 | -72.738 | -7.9E-02 |
| 565 | Se | 78 | 34 | 44 | 10 | 8.717806 | 8.724007 | -6.2E-03 | 72562.067 | 72561.562 | 5.0E-01 | 72579.507 | 72579.003 | 5.0E-01 | -77.026 | -77.534 | 5.1E-01 |
| 566 | Br | 78 | 35 | 43 | 8 | 8.661958 | 8.679043 | -1.7E-02 | 72565.125 | 72563.771 | 1.4E+00 | 72583.081 | 72581.727 | 1.4E+00 | -73.452 | -74.810 | 1.4E+00 |
| 567 | Kr | 78 | 36 | 42 | 6 | 8.661254 | 8.672613 | -1.1E-02 | 72563.882 | 72562.973 | 9.1E-01 | 72582.353 | 72581.445 | 9.1E-01 | -74.180 | -75.092 | 9.1E-01 |
| 568 | Rb | 78 | 37 | 41 | 4 | 8.558350 | 8.568044 | -9.7E-03 | 72570.609 | 72569.830 | 7.8E-01 | 72589.598 | 72588.818 | 7.8E-01 | -66.935 | -67.718 | 7.8E-01 |
| 569 | Sr | 78 | 38 | 40 | 2 | 8.500096 | 8.499334 | 7.6E-04 | 72573.855 | 72573.890 | -3.6E-02 | 72593.359 | 72593.395 | -3.6E-02 | -63.174 | -63.142 | -3.2E-02 |
| 570 | Zn | 79 | 30 | 49 | 19 | 8.450582 | 8.470335 | -2.0E-02 | 73519.216 | 73517.637 | 1.6E+00 | 73534.595 | 73533.016 | 1.6E+00 | -53.432 | -55.015 | 1.6E+00 |
| 571 | Ga | 79 | 31 | 48 | 17 | 8.556063 | 8.558908 | -2.8E-03 | 73509.585 | 73509.342 | 2.4E-01 | 73525.479 | 73525.236 | 2.4E-01 | -62.548 | -62.795 | 2.5E-01 |
| 572 | Ge | 79 | 32 | 47 | 15 | 8.634501 | 8.627486 | 7.0E-03 | 73502.091 | 73502.626 | -5.3E-01 | 73518.500 | 73519.035 | -5.3E-01 | -69.527 | -68.996 | -5.3E-01 |
| 573 | As | 79 | 33 | 46 | 13 | 8.676616 | 8.674810 | 1.8E-03 | 73497.466 | 73497.589 | -1.2E-01 | 73514.391 | 73514.514 | -1.2E-01 | -73.636 | -73.517 | -1.2E-01 |
| 574 | Se | 79 | 34 | 45 | 11 | 8.695591 | 8.700994 | -5.4E-03 | 73494.669 | 73494.222 | 4.5E-01 | 73512.110 | 73511.662 | 4.5E-01 | -75.917 | -76.369 | 4.5E-01 |
| 575 | Br | 79 | 35 | 44 | 9 | 8.687595 | 8.701036 | -1.3E-02 | 73494.003 | 73492.920 | 1.1E+00 | 73511.959 | 73510.876 | 1.1E+00 | -76.068 | -77.155 | 1.1E+00 |
| 576 | Kr | 79 | 36 | 43 | 7 | 8.657112 | 8.671256 | -1.4E-02 | 73495.113 | 73493.973 | 1.1E+00 | 73513.585 | 73512.445 | 1.1E+00 | -74.442 | -75.586 | 1.1E+00 |
| 577 | Rb | 79 | 37 | 42 | 5 | 8.601142 | 8.614281 | -1.3E-02 | 73498.236 | 73497.175 | 1.1E+00 | 73517.224 | 73516.163 | 1.1E+00 | -70.803 | -71.868 | 1.1E+00 |
| 578 | Sr | 79 | 38 | 41 | 3 | 8.523820 | 8.523485 | 3.3E-04 | 73503.046 | 73503.048 | -2.6E-03 | 73522.550 | 73522.553 | -2.7E-03 | -65.477 | -65.478 | 1.2E-03 |
| 579 | Y | 79 | 39 | 40 | 1 | 8.423790 | 8.412308 | 1.1E-02 | 73509.649 | 73510.532 | -8.8E-01 | 73529.670 | 73530.553 | -8.8E-01 | -58.357 | -57.478 | -8.8E-01 |
| 580 | Zn | 80 | 30 | 50 | 20 | 8.423545 | 8.444887 | -2.1E-02 | 74452.493 | 74450.768 | 1.7E+00 | 74467.872 | 74466.147 | 1.7E+00 | -51.649 | -53.378 | 1.7E+00 |
| 581 | Ga | 80 | 31 | 49 | 18 | 8.508454 | 8.518123 | -9.7E-03 | 74444.403 | 74443.611 | 7.9E-01 | 74460.297 | 74459.505 | 7.9E-01 | -59.224 | -60.020 | 8.0E-01 |
| 582 | Ge | 80 | 32 | 48 | 16 | 8.627570 | 8.625428 | 2.1E-03 | 74433.577 | 74433.728 | -1.5E-01 | 74449.986 | 74450.138 | -1.5E-01 | -69.535 | -69.387 | -1.5E-01 |
| 583 | As | 80 | 33 | 47 | 14 | 8.651280 | 8.646939 | 4.3E-03 | 74430.382 | 74430.709 | -3.3E-01 | 74447.307 | 74447.634 | -3.3E-01 | -72.214 | -71.891 | -3.2E-01 |
| 584 | Se | 80 | 34 | 46 | 12 | 8.710813 | 8.711553 | -7.4E-04 | 74424.321 | 74424.241 | 8.0E-02 | 74441.762 | 74441.682 | 8.0E-02 | -77.759 | -77.843 | 8.4E-02 |
| 585 | Br | 80 | 35 | 45 | 10 | 8.677653 | 8.688543 | -1.1E-02 | 74425.676 | 74424.783 | 8.9E-01 | 74443.632 | 74442.739 | 8.9E-01 | -75.889 | -76.786 | 9.0E-01 |
| 586 | Kr | 80 | 36 | 44 | 8 | 8.692928 | 8.704268 | -1.1E-02 | 74423.156 | 74422.226 | 9.3E-01 | 74441.628 | 74440.698 | 9.3E-01 | -77.893 | -78.827 | 9.3E-01 |
| 587 | Rb | 80 | 37 | 43 | 6 | 8.611675 | 8.625471 | -1.4E-02 | 74428.357 | 74427.231 | 1.1E+00 | 74447.346 | 74446.219 | 1.1E+00 | -72.175 | -73.306 | 1.1E+00 |
| 588 | Sr | 80 | 38 | 42 | 4 | 8.578596 | 8.583897 | -5.3E-03 | 74429.705 | 74429.257 | 4.5E-01 | 74449.210 | 74448.762 | 4.5E-01 | -70.311 | -70.763 | 4.5E-01 |
| 589 | Y | 80 | 39 | 41 | 2 | 8.454259 | 8.452750 | 1.5E-03 | 74438.353 | 74438.449 | -9.6E-02 | 74458.374 | 74458.470 | -9.6E-02 | -61.147 | -61.054 | -9.2E-02 |
| 590 | Zr | 80 | 40 | 40 | 0 | 8.374107 | 8.354065 | 2.0E-02 | 74443.466 | 74445.044 | -1.6E+00 | 74464.004 | 74465.582 | -1.6E+00 | -55.517 | -53.943 | -1.6E+00 |
| 591 | Zn | 81 | 30 | 51 | 21 | 8.351925 | 8.371590 | -2.0E-02 | 75389.436 | 75387.826 | 1.6E+00 | 75404.815 | 75403.205 | 1.6E+00 | -46.200 | -47.814 | 1.6E+00 |
| 592 | Ga | 81 | 31 | 50 | 19 | 8.483357 | 8.496126 | -1.3E-02 | 75377.493 | 75376.440 | 1.1E+00 | 75393.387 | 75392.334 | 1.1E+00 | -57.628 | -58.685 | 1.1E+00 |
| 593 | Ge | 81 | 32 | 49 | 17 | 8.580658 | 8.587682 | -7.0E-03 | 75368.314 | 75367.726 | 5.9E-01 | 75384.723 | 75384.135 | 5.9E-01 | -66.292 | -66.884 | 5.9E-01 |
| 594 | As | 81 | 33 | 48 | 15 | 8.648056 | 8.647530 | 5.3E-04 | 75361.557 | 75361.580 | -2.3E-02 | 75378.482 | 75378.504 | -2.3E-02 | -72.533 | -72.515 | -1.9E-02 |
| 595 | Se | 81 | 34 | 47 | 13 | 8.685999 | 8.686467 | -4.7E-04 | 75357.186 | 75357.127 | 5.9E-02 | 75374.626 | 75374.567 | 5.9E-02 | -76.389 | -76.452 | 6.3E-02 |
| 596 | Br | 81 | 35 | 46 | 11 | 8.695929 | 8.702963 | -7.0E-03 | 75355.083 | 75354.492 | 5.9E-01 | 75373.039 | 75372.448 | 5.9E-01 | -77.976 | -78.571 | 6.0E-01 |
| 597 | Kr | 81 | 36 | 45 | 9 | 8.682803 | 8.695105 | -1.2E-02 | 75354.848 | 75353.830 | 1.0E+00 | 75373.320 | 75372.302 | 1.0E+00 | -77.695 | -78.717 | 1.0E+00 |
| 598 | Rb | 81 | 37 | 44 | 7 | 8.645513 | 8.661581 | -1.6E-02 | 75356.570 | 75355.246 | 1.3E+00 | 75375.558 | 75374.234 | 1.3E+00 | -75.457 | -76.785 | 1.3E+00 |
| 599 | Sr | 81 | 38 | 43 | 5 | 8.587354 | 8.597048 | -9.7E-03 | 75359.982 | 75359.173 | 8.1E-01 | 75379.487 | 75378.678 | 8.1E-01 | -71.528 | -72.341 | 8.1E-01 |
| 600 | Y | 81 | 39 | 42 | 3 | 8.505888 | 8.510136 | -4.2E-03 | 75365.282 | 75364.914 | 3.7E-01 | 75385.303 | 75384.935 | 3.7E-01 | -65.712 | -66.084 | 3.7E-01 |
| 601 | Zr | 81 | 40 | 41 | 1 | 8.405916 | 8.390841 | 1.5E-02 | 75372.081 | 75373.277 | -1.2E+00 | 75392.619 | 75393.815 | -1.2E+00 | -58.396 | -57.204 | -1.2E+00 |
| 602 | Ga | 82 | 31 | 51 | 20 | 8.421049 | 8.430954 | -9.9E-03 | 76313.684 | 76312.854 | 8.3E-01 | 76329.578 | 76328.748 | 8.3E-01 | -52.931 | -53.765 | 8.3E-01 |
| 603 | Ge | 82 | 32 | 50 | 18 | 8.563756 | 8.572167 | -8.4E-03 | 76300.685 | 76299.976 | 7.1E-01 | 76317.094 | 76316.385 | 7.1E-01 | -65.415 | -66.128 | 7.1E-01 |
| 604 | As | 82 | 33 | 49 | 16 | 8.611386 | 8.617749 | -6.4E-03 | 76295.481 | 76294.940 | 5.4E-01 | 76312.406 | 76311.864 | 5.4E-01 | -70.103 | -70.649 | 5.5E-01 |
| 605 | Se | 82 | 34 | 48 | 14 | 8.693197 | 8.693772 | -5.8E-04 | 76287.475 | 76287.407 | 6.8E-02 | 76304.915 | 76304.847 | 6.8E-02 | -77.594 | -77.666 | 7.2E-02 |
| 606 | Br | 82 | 35 | 47 | 12 | 8.682477 | 8.686782 | -4.3E-03 | 76287.056 | 76286.681 | 3.7E-01 | 76305.012 | 76304.637 | 3.7E-01 | -77.497 | -77.876 | 3.8E-01 |
| 607 | Kr | 82 | 36 | 46 | 10 | 8.710657 | 8.718838 | -8.2E-03 | 76283.447 | 76282.754 | 6.9E-01 | 76301.919 | 76301.226 | 6.9E-01 | -80.590 | -81.287 | 7.0E-01 |
| 608 | Rb | 82 | 37 | 45 | 8 | 8.647427 | 8.663530 | -1.6E-02 | 76287.333 | 76285.990 | 1.3E+00 | 76306.321 | 76304.978 | 1.3E+00 | -76.188 | -77.535 | 1.3E+00 |
| 609 | Sr | 82 | 38 | 44 | 6 | 8.635718 | 8.645175 | -9.5E-03 | 76286.995 | 76286.195 | 8.0E-01 | 76306.499 | 76305.700 | 8.0E-01 | -76.010 | -76.813 | 8.0E-01 |
| 610 | Y | 82 | 39 | 43 | 4 | 8.529262 | 8.536652 | -7.4E-03 | 76294.425 | 76293.795 | 6.3E-01 | 76314.446 | 76313.816 | 6.3E-01 | -68.063 | -68.697 | 6.3E-01 |
| 611 | Ga | 83 | 31 | 52 | 21 | 8.372575 | 8.381510 | -8.9E-03 | 77248.852 | 77248.092 | 7.6E-01 | 77264.746 | 77263.986 | 7.6E-01 | -49.257 | -50.021 | 7.6E-01 |
| 612 | Ge | 83 | 32 | 51 | 19 | 8.504345 | 8.510833 | -6.5E-03 | 77236.617 | 77236.060 | 5.6E-01 | 77253.027 | 77252.469 | 5.6E-01 | -60.976 | -61.538 | 5.6E-01 |
| 613 | As | 83 | 33 | 50 | 17 | 8.599653 | 8.605292 | -5.6E-03 | 77227.409 | 77226.921 | 4.9E-01 | 77244.334 | 77243.846 | 4.9E-01 | -69.669 | -70.161 | 4.9E-01 |
| 614 | Se | 83 | 34 | 49 | 15 | 8.658556 | 8.666027 | -7.5E-03 | 77221.222 | 77220.582 | 6.4E-01 | 77238.663 | 77238.022 | 6.4E-01 | -75.341 | -75.985 | 6.4E-01 |
| 615 | Br | 83 | 35 | 48 | 13 | 8.693381 | 8.696769 | -3.4E-03 | 77217.034 | 77216.731 | 3.0E-01 | 77234.990 | 77234.687 | 3.0E-01 | -79.013 | -79.320 | 3.1E-01 |
| 616 | Kr | 83 | 36 | 47 | 11 | 8.695721 | 8.705182 | -9.5E-03 | 77215.541 | 77214.734 | 8.1E-01 | 77234.013 | 77233.206 | 8.1E-01 | -79.990 | -80.801 | 8.1E-01 |
| 617 | Rb | 83 | 37 | 46 | 9 | 8.675218 | 8.690820 | -1.6E-02 | 77215.944 | 77214.627 | 1.3E+00 | 77234.932 | 77233.615 | 1.3E+00 | -79.071 | -80.392 | 1.3E+00 |
| 618 | Sr | 83 | 38 | 45 | 7 | 8.638407 | 8.649754 | -1.1E-02 | 77217.701 | 77216.736 | 9.7E-01 | 77237.206 | 77236.240 | 9.7E-01 | -76.798 | -77.767 | 9.7E-01 |
| 619 | Y | 83 | 39 | 44 | 5 | 8.573643 | 8.585486 | -1.2E-02 | 77221.777 | 77220.770 | 1.0E+00 | 77241.799 | 77240.791 | 1.0E+00 | -72.205 | -73.216 | 1.0E+00 |
| 620 | Zr | 83 | 40 | 43 | 3 | 8.488385 | 8.491013 | -2.6E-03 | 77227.555 | 77227.312 | 2.4E-01 | 77248.093 | 77247.849 | 2.4E-01 | -65.911 | -66.158 | 2.5E-01 |
| 621 | Nb | 83 | 41 | 42 | 1 | 8.388598 | 8.378414 | 1.0E-02 | 77234.538 | 77235.357 | -8.2E-01 | 77255.593 | 77256.412 | -8.2E-01 | -58.411 | -57.595 | -8.2E-01 |
| 622 | Ge | 84 | 32 | 52 | 20 | 8.465524 | 8.468387 | -2.9E-03 | 78170.939 | 78170.680 | 2.6E-01 | 78187.349 | 78187.089 | 2.6E-01 | -58.148 | -58.412 | 2.6E-01 |
| 623 | As | 84 | 33 | 51 | 18 | 8.547938 | 8.553547 | -5.6E-03 | 78162.719 | 78162.228 | 4.9E-01 | 78179.644 | 78179.153 | 4.9E-01 | -65.854 | -66.349 | 5.0E-01 |
| 624 | Se | 84 | 34 | 50 | 16 | 8.658793 | 8.660876 | -2.1E-03 | 78152.109 | 78151.914 | 2.0E-01 | 78169.549 | 78169.354 | 2.0E-01 | -75.948 | -76.147 | 2.0E-01 |
| 625 | Br | 84 | 35 | 49 | 14 | 8.671329 | 8.676792 | -5.5E-03 | 78149.758 | 78149.278 | 4.8E-01 | 78167.714 | 78167.234 | 4.8E-01 | -77.783 | -78.267 | 4.8E-01 |
| 626 | Kr | 84 | 36 | 48 | 12 | 8.717446 | 8.722544 | -5.1E-03 | 78144.586 | 78144.136 | 4.5E-01 | 78163.058 | 78162.608 | 4.5E-01 | -82.439 | -82.894 | 4.5E-01 |
| 627 | Rb | 84 | 37 | 47 | 10 | 8.676224 | 8.686553 | -1.0E-02 | 78146.750 | 78145.859 | 8.9E-01 | 78165.738 | 78164.848 | 8.9E-01 | -79.759 | -80.654 | 8.9E-01 |
| 628 | Sr | 84 | 38 | 46 | 8 | 8.677512 | 8.687388 | -9.9E-03 | 78145.343 | 78144.490 | 8.5E-01 | 78164.848 | 78163.994 | 8.5E-01 | -80.650 | -81.507 | 8.6E-01 |
| 629 | Y | 84 | 39 | 45 | 6 | 8.587769 | 8.601783 | -1.4E-02 | 78151.583 | 78150.381 | 1.2E+00 | 78171.604 | 78170.402 | 1.2E+00 | -73.893 | -75.099 | 1.2E+00 |
| 630 | Zr | 84 | 40 | 44 | 4 | 8.549016 | 8.553121 | -4.1E-03 | 78153.539 | 78153.169 | 3.7E-01 | 78174.077 | 78173.707 | 3.7E-01 | -71.421 | -71.794 | 3.7E-01 |
| 631 | Ge | 85 | 32 | 53 | 21 | 8.401768 | 8.392197 | 9.6E-03 | 79107.458 | 79108.253 | -7.9E-01 | 79123.868 | 79124.662 | -7.9E-01 | -53.123 | -52.333 | -7.9E-01 |
| 632 | As | 85 | 33 | 52 | 19 | 8.510984 | 8.514297 | -3.3E-03 | 79096.877 | 79096.576 | 3.0E-01 | 79113.802 | 79113.501 | 3.0E-01 | -63.189 | -63.495 | 3.1E-01 |
| 633 | Se | 85 | 34 | 51 | 17 | 8.610304 | 8.612646 | -2.3E-03 | 79087.137 | 79086.918 | 2.2E-01 | 79104.578 | 79104.358 | 2.2E-01 | -72.414 | -72.637 | 2.2E-01 |
| 634 | Br | 85 | 35 | 50 | 15 | 8.673592 | 8.674653 | -1.1E-03 | 79080.460 | 79080.348 | 1.1E-01 | 79098.416 | 79098.304 | 1.1E-01 | -78.575 | -78.691 | 1.2E-01 |
| 635 | Kr | 85 | 36 | 49 | 13 | 8.698562 | 8.704157 | -5.6E-03 | 79077.039 | 79076.541 | 5.0E-01 | 79095.511 | 79095.013 | 5.0E-01 | -81.480 | -81.982 | 5.0E-01 |
| 636 | Rb | 85 | 37 | 48 | 11 | 8.697441 | 8.706621 | -9.2E-03 | 79075.836 | 79075.033 | 8.0E-01 | 79094.824 | 79094.021 | 8.0E-01 | -82.167 | -82.974 | 8.1E-01 |
| 637 | Sr | 85 | 38 | 47 | 9 | 8.675718 | 8.685433 | -9.7E-03 | 79076.383 | 79075.534 | 8.5E-01 | 79095.888 | 79095.039 | 8.5E-01 | -81.103 | -81.957 | 8.5E-01 |
| 638 | Y | 85 | 39 | 46 | 7 | 8.628148 | 8.642156 | -1.4E-02 | 79079.128 | 79077.913 | 1.2E+00 | 79099.149 | 79097.934 | 1.2E+00 | -77.842 | -79.061 | 1.2E+00 |
| 639 | Zr | 85 | 40 | 45 | 5 | 8.564025 | 8.571001 | -7.0E-03 | 79083.279 | 79082.661 | 6.2E-01 | 79103.817 | 79103.199 | 6.2E-01 | -73.174 | -73.796 | 6.2E-01 |
| 640 | Nb | 85 | 41 | 44 | 3 | 8.473711 | 8.479188 | -5.5E-03 | 79089.657 | 79089.165 | 4.9E-01 | 79110.711 | 79110.220 | 4.9E-01 | -66.280 | -66.775 | 5.0E-01 |
| 641 | Mo | 85 | 42 | 43 | 1 | 8.361331 | 8.358036 | 3.3E-03 | 79097.909 | 79098.163 | -2.5E-01 | 79119.481 | 79119.735 | -2.5E-01 | -57.510 | -57.260 | -2.5E-01 |
| 642 | As | 86 | 33 | 53 | 20 | 8.456721 | 8.447532 | 9.2E-03 | 80032.598 | 80033.369 | -7.7E-01 | 80049.523 | 80050.293 | -7.7E-01 | -58.962 | -58.196 | -7.7E-01 |
| 643 | Se | 86 | 34 | 52 | 18 | 8.581822 | 8.581970 | -1.5E-04 | 80020.542 | 80020.508 | 3.3E-02 | 80037.982 | 80037.949 | 3.3E-02 | -70.503 | -70.541 | 3.7E-02 |
| 644 | Br | 86 | 35 | 51 | 16 | 8.632365 | 8.635892 | -3.5E-03 | 80014.897 | 80014.572 | 3.2E-01 | 80032.853 | 80032.528 | 3.2E-01 | -75.632 | -75.961 | 3.3E-01 |
| 645 | Kr | 86 | 36 | 50 | 14 | 8.712029 | 8.709234 | 2.8E-03 | 80006.748 | 80006.966 | -2.2E-01 | 80025.220 | 80025.438 | -2.2E-01 | -83.266 | -83.051 | -2.1E-01 |
| 646 | Rb | 86 | 37 | 49 | 12 | 8.696901 | 8.695954 | 9.5E-04 | 80006.750 | 80006.809 | -5.9E-02 | 80025.738 | 80025.797 | -5.9E-02 | -82.747 | -82.692 | -5.5E-02 |
| 647 | Sr | 86 | 38 | 48 | 10 | 8.708457 | 8.713780 | -5.3E-03 | 80004.457 | 80003.976 | 4.8E-01 | 80023.962 | 80023.481 | 4.8E-01 | -84.523 | -85.009 | 4.9E-01 |
| 648 | Y | 86 | 39 | 47 | 8 | 8.638430 | 8.650353 | -1.2E-02 | 80009.181 | 80008.131 | 1.0E+00 | 80029.202 | 80028.152 | 1.0E+00 | -79.283 | -80.337 | 1.1E+00 |
| 649 | Zr | 86 | 40 | 46 | 6 | 8.614047 | 8.622812 | -8.8E-03 | 80009.979 | 80009.200 | 7.8E-01 | 80030.517 | 80029.738 | 7.8E-01 | -77.969 | -78.751 | 7.8E-01 |
| 650 | Nb | 86 | 41 | 45 | 4 | 8.502209 | 8.509488 | -7.3E-03 | 80018.297 | 80017.646 | 6.5E-01 | 80039.352 | 80038.701 | 6.5E-01 | -69.133 | -69.789 | 6.6E-01 |
| 651 | Mo | 86 | 42 | 44 | 2 | 8.434709 | 8.432418 | 2.3E-03 | 80022.803 | 80022.973 | -1.7E-01 | 80044.375 | 80044.546 | -1.7E-01 | -64.110 | -63.944 | -1.7E-01 |
| 652 | As | 87 | 33 | 54 | 21 | 8.413851 | 8.403424 | 1.0E-02 | 80967.437 | 80968.324 | -8.9E-01 | 80984.361 | 80985.249 | -8.9E-01 | -55.618 | -54.735 | -8.8E-01 |
| 653 | Se | 87 | 34 | 53 | 19 | 8.529091 | 8.518017 | 1.1E-02 | 80956.113 | 80957.056 | -9.4E-01 | 80973.553 | 80974.496 | -9.4E-01 | -66.426 | -65.487 | -9.4E-01 |
| 654 | Br | 87 | 35 | 52 | 17 | 8.605910 | 8.609357 | -3.4E-03 | 80948.131 | 80947.810 | 3.2E-01 | 80966.088 | 80965.766 | 3.2E-01 | -73.892 | -74.217 | 3.3E-01 |
| 655 | Kr | 87 | 36 | 51 | 15 | 8.675283 | 8.673673 | 1.6E-03 | 80940.798 | 80940.916 | -1.2E-01 | 80959.270 | 80959.388 | -1.2E-01 | -80.710 | -80.596 | -1.1E-01 |
| 656 | Rb | 87 | 37 | 50 | 13 | 8.710983 | 8.703715 | 7.3E-03 | 80936.393 | 80937.003 | -6.1E-01 | 80955.381 | 80955.991 | -6.1E-01 | -84.598 | -83.992 | -6.1E-01 |
| 657 | Sr | 87 | 38 | 49 | 11 | 8.705235 | 8.704492 | 7.4E-04 | 80935.595 | 80935.636 | -4.1E-02 | 80955.099 | 80955.140 | -4.1E-02 | -84.880 | -84.843 | -3.7E-02 |
| 658 | Y | 87 | 39 | 48 | 9 | 8.674844 | 8.681357 | -6.5E-03 | 80936.940 | 80936.349 | 5.9E-01 | 80956.961 | 80956.370 | 5.9E-01 | -83.018 | -83.613 | 5.9E-01 |
| 659 | Zr | 87 | 40 | 47 | 7 | 8.623647 | 8.633031 | -9.4E-03 | 80940.095 | 80939.253 | 8.4E-01 | 80960.633 | 80959.791 | 8.4E-01 | -79.347 | -80.192 | 8.5E-01 |
| 660 | Nb | 87 | 41 | 46 | 5 | 8.551743 | 8.562280 | -1.1E-02 | 80945.051 | 80944.109 | 9.4E-01 | 80966.106 | 80965.164 | 9.4E-01 | -73.873 | -74.820 | 9.5E-01 |
| 661 | Mo | 87 | 42 | 45 | 3 | 8.462423 | 8.462674 | -2.5E-04 | 80951.522 | 80951.474 | 4.8E-02 | 80973.094 | 80973.046 | 4.8E-02 | -66.885 | -66.937 | 5.2E-02 |
| 662 | Tc | 87 | 43 | 44 | 1 | 8.347744 | 8.347657 | 8.7E-05 | 80960.200 | 80960.180 | 1.9E-02 | 80982.289 | 80982.270 | 2.0E-02 | -57.690 | -57.714 | 2.4E-02 |
| 663 | Se | 88 | 34 | 54 | 20 | 8.495004 | 8.481846 | 1.3E-02 | 81890.149 | 81891.286 | -1.1E+00 | 81907.589 | 81908.726 | -1.1E+00 | -63.884 | -62.751 | -1.1E+00 |
| 664 | Br | 88 | 35 | 53 | 18 | 8.563747 | 8.554830 | 8.9E-03 | 81882.801 | 81883.565 | -7.6E-01 | 81900.757 | 81901.521 | -7.6E-01 | -70.716 | -69.957 | -7.6E-01 |
| 665 | Kr | 88 | 36 | 52 | 16 | 8.656849 | 8.655942 | 9.1E-04 | 81873.310 | 81873.368 | -5.8E-02 | 81891.782 | 81891.840 | -5.8E-02 | -79.691 | -79.638 | -5.4E-02 |
| 666 | Rb | 88 | 37 | 51 | 14 | 8.681115 | 8.676881 | 4.2E-03 | 81869.876 | 81870.226 | -3.5E-01 | 81888.864 | 81889.214 | -3.5E-01 | -82.609 | -82.263 | -3.5E-01 |
| 667 | Sr | 88 | 38 | 50 | 12 | 8.732592 | 8.719457 | 1.3E-02 | 81864.047 | 81865.180 | -1.1E+00 | 81883.552 | 81884.684 | -1.1E+00 | -87.921 | -86.793 | -1.1E+00 |
| 668 | Y | 88 | 39 | 49 | 10 | 8.682536 | 8.680249 | 2.3E-03 | 81867.153 | 81867.330 | -1.8E-01 | 81887.174 | 81887.352 | -1.8E-01 | -84.299 | -84.126 | -1.7E-01 |
| 669 | Zr | 88 | 40 | 48 | 8 | 8.666027 | 8.673538 | -7.5E-03 | 81867.307 | 81866.621 | 6.9E-01 | 81887.845 | 81887.159 | 6.9E-01 | -83.628 | -84.318 | 6.9E-01 |
| 670 | Nb | 88 | 41 | 47 | 6 | 8.572451 | 8.583439 | -1.1E-02 | 81874.242 | 81873.250 | 9.9E-01 | 81895.297 | 81894.305 | 9.9E-01 | -76.176 | -77.173 | 1.0E+00 |
| 671 | Mo | 88 | 42 | 46 | 4 | 8.523908 | 8.527977 | -4.1E-03 | 81877.215 | 81876.830 | 3.8E-01 | 81898.787 | 81898.402 | 3.8E-01 | -72.687 | -73.075 | 3.9E-01 |
| 672 | Tc | 88 | 43 | 45 | 2 | 8.389958 | 8.391526 | -1.6E-03 | 81887.702 | 81887.537 | 1.7E-01 | 81909.792 | 81909.627 | 1.6E-01 | -61.681 | -61.851 | 1.7E-01 |
| 673 | Se | 89 | 34 | 55 | 21 | 8.435279 | 8.420180 | 1.5E-02 | 82826.535 | 82827.858 | -1.3E+00 | 82843.975 | 82845.298 | -1.3E+00 | -58.992 | -57.673 | -1.3E+00 |
| 674 | Br | 89 | 35 | 54 | 19 | 8.530779 | 8.521232 | 9.5E-03 | 82816.737 | 82817.566 | -8.3E-01 | 82834.693 | 82835.522 | -8.3E-01 | -68.274 | -67.450 | -8.2E-01 |
| 675 | Kr | 89 | 36 | 53 | 17 | 8.614815 | 8.604639 | 1.0E-02 | 82807.959 | 82808.843 | -8.8E-01 | 82826.431 | 82827.315 | -8.8E-01 | -76.536 | -75.656 | -8.8E-01 |
| 676 | Rb | 89 | 37 | 52 | 15 | 8.664187 | 8.663503 | 6.8E-04 | 82802.267 | 82802.305 | -3.8E-02 | 82821.255 | 82821.293 | -3.8E-02 | -81.712 | -81.678 | -3.4E-02 |
| 677 | Sr | 89 | 38 | 51 | 13 | 8.705919 | 8.695034 | 1.1E-02 | 82797.254 | 82798.199 | -9.5E-01 | 82816.758 | 82817.704 | -9.5E-01 | -86.209 | -85.268 | -9.4E-01 |
| 678 | Y | 89 | 39 | 50 | 11 | 8.713987 | 8.697520 | 1.6E-02 | 82795.237 | 82796.678 | -1.4E+00 | 82815.258 | 82816.700 | -1.4E+00 | -87.709 | -86.272 | -1.4E+00 |
| 679 | Zr | 89 | 40 | 49 | 9 | 8.673368 | 8.673760 | -3.9E-04 | 82797.553 | 82797.493 | 6.0E-02 | 82818.091 | 82818.031 | 6.0E-02 | -84.876 | -84.940 | 6.4E-02 |
| 680 | Nb | 89 | 41 | 48 | 7 | 8.616814 | 8.626405 | -9.6E-03 | 82801.287 | 82800.408 | 8.8E-01 | 82822.342 | 82821.463 | 8.8E-01 | -80.625 | -81.509 | 8.8E-01 |
| 681 | Mo | 89 | 42 | 47 | 5 | 8.544984 | 8.550600 | -5.6E-03 | 82806.380 | 82805.854 | 5.3E-01 | 82827.952 | 82827.426 | 5.3E-01 | -75.015 | -75.545 | 5.3E-01 |
| 682 | Tc | 89 | 43 | 46 | 3 | 8.450575 | 8.454603 | -4.0E-03 | 82813.483 | 82813.097 | 3.9E-01 | 82835.572 | 82835.187 | 3.9E-01 | -67.395 | -67.785 | 3.9E-01 |
| 683 | Se | 90 | 34 | 56 | 22 | 8.395766 | 8.385778 | 1.0E-02 | 83761.221 | 83762.099 | -8.8E-01 | 83778.661 | 83779.540 | -8.8E-01 | -55.800 | -54.926 | -8.7E-01 |
| 684 | Br | 90 | 35 | 55 | 20 | 8.478186 | 8.468336 | 9.8E-03 | 83752.505 | 83753.370 | -8.7E-01 | 83770.461 | 83771.326 | -8.7E-01 | -64.000 | -63.139 | -8.6E-01 |
| 685 | Kr | 90 | 36 | 54 | 18 | 8.591259 | 8.578875 | 1.2E-02 | 83741.030 | 83742.123 | -1.1E+00 | 83759.502 | 83760.595 | -1.1E+00 | -74.959 | -73.871 | -1.1E+00 |
| 686 | Rb | 90 | 37 | 53 | 16 | 8.631516 | 8.621288 | 1.0E-02 | 83736.108 | 83737.006 | -9.0E-01 | 83755.096 | 83755.994 | -9.0E-01 | -79.365 | -78.471 | -8.9E-01 |
| 687 | Sr | 90 | 38 | 52 | 14 | 8.695981 | 8.690060 | 5.9E-03 | 83729.008 | 83729.517 | -5.1E-01 | 83748.512 | 83749.022 | -5.1E-01 | -85.949 | -85.444 | -5.1E-01 |
| 688 | Y | 90 | 39 | 51 | 12 | 8.693354 | 8.681107 | 1.2E-02 | 83727.945 | 83729.024 | -1.1E+00 | 83747.966 | 83749.045 | -1.1E+00 | -86.495 | -85.421 | -1.1E+00 |
| 689 | Zr | 90 | 40 | 50 | 10 | 8.709980 | 8.698731 | 1.1E-02 | 83725.150 | 83726.137 | -9.9E-01 | 83745.688 | 83746.675 | -9.9E-01 | -88.774 | -87.790 | -9.8E-01 |
| 690 | Nb | 90 | 41 | 49 | 8 | 8.633384 | 8.635649 | -2.3E-03 | 83730.744 | 83730.515 | 2.3E-01 | 83751.799 | 83751.570 | 2.3E-01 | -82.662 | -82.896 | 2.3E-01 |
| 691 | Mo | 90 | 42 | 48 | 6 | 8.597032 | 8.604253 | -7.2E-03 | 83732.716 | 83732.040 | 6.8E-01 | 83754.288 | 83753.612 | 6.8E-01 | -80.173 | -80.853 | 6.8E-01 |
| 692 | Tc | 90 | 43 | 47 | 4 | 8.483359 | 8.488817 | -5.5E-03 | 83741.647 | 83741.129 | 5.2E-01 | 83763.737 | 83763.218 | 5.2E-01 | -70.725 | -71.247 | 5.2E-01 |
| 693 | Ru | 90 | 44 | 46 | 2 | 8.409768 | 8.408613 | 1.2E-03 | 83746.970 | 83747.047 | -7.6E-02 | 83769.577 | 83769.654 | -7.6E-02 | -64.884 | -64.812 | -7.2E-02 |
| 694 | Br | 91 | 35 | 56 | 21 | 8.441923 | 8.435518 | 6.4E-03 | 84686.892 | 84687.454 | -5.6E-01 | 84704.848 | 84705.410 | -5.6E-01 | -61.107 | -60.550 | -5.6E-01 |
| 695 | Kr | 91 | 36 | 55 | 19 | 8.541751 | 8.527876 | 1.4E-02 | 84676.509 | 84677.750 | -1.2E+00 | 84694.981 | 84696.222 | -1.2E+00 | -70.974 | -69.737 | -1.2E+00 |
| 696 | Rb | 91 | 37 | 54 | 17 | 8.607562 | 8.598784 | 8.8E-03 | 84669.222 | 84669.998 | -7.8E-01 | 84688.210 | 84688.986 | -7.8E-01 | -77.745 | -76.973 | -7.7E-01 |
| 697 | Sr | 91 | 38 | 53 | 15 | 8.663880 | 8.651071 | 1.3E-02 | 84662.798 | 84663.941 | -1.1E+00 | 84682.303 | 84683.445 | -1.1E+00 | -83.652 | -82.514 | -1.1E+00 |
| 698 | Y | 91 | 39 | 52 | 13 | 8.684947 | 8.679823 | 5.1E-03 | 84659.582 | 84660.025 | -4.4E-01 | 84679.603 | 84680.046 | -4.4E-01 | -86.352 | -85.914 | -4.4E-01 |
| 699 | Zr | 91 | 40 | 51 | 11 | 8.693320 | 8.683849 | 9.5E-03 | 84657.521 | 84658.358 | -8.4E-01 | 84678.059 | 84678.896 | -8.4E-01 | -87.896 | -87.063 | -8.3E-01 |
| 700 | Nb | 91 | 41 | 50 | 9 | 8.670904 | 8.662866 | 8.0E-03 | 84658.262 | 84658.968 | -7.1E-01 | 84679.317 | 84680.023 | -7.1E-01 | -86.639 | -85.937 | -7.0E-01 |
| 701 | Mo | 91 | 42 | 49 | 7 | 8.613627 | 8.614891 | -1.3E-03 | 84662.174 | 84662.033 | 1.4E-01 | 84683.747 | 84683.605 | 1.4E-01 | -82.209 | -82.354 | 1.5E-01 |
| 702 | Tc | 91 | 43 | 48 | 5 | 8.536651 | 8.543683 | -7.0E-03 | 84667.879 | 84667.213 | 6.7E-01 | 84689.969 | 84689.302 | 6.7E-01 | -75.986 | -76.657 | 6.7E-01 |
| 703 | Ru | 91 | 44 | 47 | 3 | 8.442925 | 8.442939 | -1.4E-05 | 84675.109 | 84675.080 | 2.9E-02 | 84697.716 | 84697.687 | 2.9E-02 | -68.240 | -68.273 | 3.3E-02 |
| 704 | Br | 92 | 35 | 57 | 22 | 8.384911 | 8.383539 | 1.4E-03 | 85623.260 | 85623.366 | -1.1E-01 | 85641.216 | 85641.322 | -1.1E-01 | -56.233 | -56.132 | -1.0E-01 |
| 705 | Kr | 92 | 36 | 56 | 20 | 8.512674 | 8.502623 | 1.0E-02 | 85610.208 | 85611.111 | -9.0E-01 | 85628.680 | 85629.583 | -9.0E-01 | -68.769 | -67.871 | -9.0E-01 |
| 706 | Rb | 92 | 37 | 55 | 18 | 8.569423 | 8.555758 | 1.4E-02 | 85603.689 | 85604.923 | -1.2E+00 | 85622.677 | 85623.911 | -1.2E+00 | -74.773 | -73.542 | -1.2E+00 |
| 707 | Sr | 92 | 38 | 54 | 16 | 8.648907 | 8.636235 | 1.3E-02 | 85595.077 | 85596.220 | -1.1E+00 | 85614.582 | 85615.724 | -1.1E+00 | -82.867 | -81.729 | -1.1E+00 |
| 708 | Y | 92 | 39 | 53 | 14 | 8.661595 | 8.649354 | 1.2E-02 | 85592.611 | 85593.713 | -1.1E+00 | 85612.632 | 85613.734 | -1.1E+00 | -84.817 | -83.719 | -1.1E+00 |
| 709 | Zr | 92 | 40 | 52 | 12 | 8.692684 | 8.690430 | 2.3E-03 | 85588.452 | 85588.634 | -1.8E-01 | 85608.990 | 85609.172 | -1.8E-01 | -88.460 | -88.281 | -1.8E-01 |
| 710 | Nb | 92 | 41 | 51 | 10 | 8.662377 | 8.655808 | 6.6E-03 | 85589.941 | 85590.520 | -5.8E-01 | 85610.995 | 85611.574 | -5.8E-01 | -86.454 | -85.879 | -5.7E-01 |
| 711 | Mo | 92 | 42 | 50 | 8 | 8.657722 | 8.650695 | 7.0E-03 | 85589.069 | 85589.690 | -6.2E-01 | 85610.641 | 85611.262 | -6.2E-01 | -86.808 | -86.192 | -6.2E-01 |
| 712 | Tc | 92 | 43 | 49 | 6 | 8.563543 | 8.564137 | -5.9E-04 | 85596.434 | 85596.353 | 8.1E-02 | 85618.524 | 85618.442 | 8.1E-02 | -78.926 | -79.012 | 8.6E-02 |
| 713 | Ru | 92 | 44 | 48 | 4 | 8.504773 | 8.509432 | -4.7E-03 | 85600.541 | 85600.085 | 4.6E-01 | 85623.148 | 85622.692 | 4.6E-01 | -74.301 | -74.762 | 4.6E-01 |
| 714 | Rh | 92 | 45 | 47 | 2 | 8.373420 | 8.373606 | -1.9E-04 | 85611.325 | 85611.280 | 4.5E-02 | 85634.450 | 85634.405 | 4.5E-02 | -62.999 | -63.049 | 5.0E-02 |
| 715 | Br | 93 | 35 | 58 | 23 | 8.346459 | 8.347789 | -1.3E-03 | 86558.017 | 86557.872 | 1.4E-01 | 86575.973 | 86575.828 | 1.4E-01 | -52.970 | -53.119 | 1.5E-01 |
| 716 | Kr | 93 | 36 | 57 | 21 | 8.458108 | 8.452031 | 6.1E-03 | 86546.335 | 86546.879 | -5.4E-01 | 86564.807 | 86565.351 | -5.4E-01 | -64.136 | -63.597 | -5.4E-01 |
| 717 | Rb | 93 | 37 | 56 | 19 | 8.540921 | 8.532747 | 8.2E-03 | 86537.335 | 86538.073 | -7.4E-01 | 86556.323 | 86557.061 | -7.4E-01 | -72.620 | -71.887 | -7.3E-01 |
| 718 | Sr | 93 | 38 | 55 | 17 | 8.612788 | 8.595319 | 1.7E-02 | 86529.353 | 86530.954 | -1.6E+00 | 86548.857 | 86550.459 | -1.6E+00 | -80.086 | -78.489 | -1.6E+00 |
| 719 | Y | 93 | 39 | 54 | 15 | 8.648910 | 8.637976 | 1.1E-02 | 86524.694 | 86525.688 | -9.9E-01 | 86544.716 | 86545.709 | -9.9E-01 | -84.228 | -83.239 | -9.9E-01 |
| 720 | Zr | 93 | 40 | 53 | 13 | 8.671627 | 8.662483 | 9.1E-03 | 86521.283 | 86522.108 | -8.3E-01 | 86541.821 | 86542.646 | -8.3E-01 | -87.123 | -86.301 | -8.2E-01 |
| 721 | Nb | 93 | 41 | 52 | 11 | 8.664186 | 8.665376 | -1.2E-03 | 86520.675 | 86520.539 | 1.4E-01 | 86541.730 | 86541.594 | 1.4E-01 | -87.213 | -87.353 | 1.4E-01 |
| 722 | Mo | 93 | 42 | 51 | 9 | 8.651401 | 8.644773 | 6.6E-03 | 86520.565 | 86521.155 | -5.9E-01 | 86542.137 | 86542.727 | -5.9E-01 | -86.806 | -86.220 | -5.9E-01 |
| 723 | Tc | 93 | 43 | 50 | 7 | 8.608569 | 8.601892 | 6.7E-03 | 86523.248 | 86523.843 | -5.9E-01 | 86545.338 | 86545.932 | -5.9E-01 | -83.605 | -83.016 | -5.9E-01 |
| 724 | Ru | 93 | 44 | 49 | 5 | 8.531462 | 8.530852 | 6.1E-04 | 86529.119 | 86529.149 | -2.9E-02 | 86551.727 | 86551.756 | -2.9E-02 | -77.217 | -77.192 | -2.5E-02 |
| 725 | Rh | 93 | 45 | 48 | 3 | 8.434825 | 8.438590 | -3.8E-03 | 86536.806 | 86536.428 | 3.8E-01 | 86559.931 | 86559.553 | 3.8E-01 | -69.012 | -69.395 | 3.8E-01 |
| 726 | Kr | 94 | 36 | 58 | 22 | 8.424331 | 8.424265 | 6.6E-05 | 87480.618 | 87480.602 | 1.5E-02 | 87499.089 | 87499.074 | 1.5E-02 | -61.348 | -61.368 | 2.0E-02 |
| 727 | Rb | 94 | 37 | 57 | 20 | 8.492764 | 8.490087 | 2.7E-03 | 87472.886 | 87473.116 | -2.3E-01 | 87491.874 | 87492.104 | -2.3E-01 | -68.563 | -68.338 | -2.2E-01 |
| 728 | Sr | 94 | 38 | 56 | 18 | 8.593834 | 8.579235 | 1.5E-02 | 87462.087 | 87463.436 | -1.3E+00 | 87481.592 | 87482.941 | -1.3E+00 | -78.846 | -77.501 | -1.3E+00 |
| 729 | Y | 94 | 39 | 55 | 16 | 8.622821 | 8.604507 | 1.8E-02 | 87458.063 | 87459.761 | -1.7E+00 | 87478.084 | 87479.782 | -1.7E+00 | -82.353 | -80.660 | -1.7E+00 |
| 730 | Zr | 94 | 40 | 54 | 14 | 8.666818 | 8.658581 | 8.2E-03 | 87452.628 | 87453.378 | -7.5E-01 | 87473.166 | 87473.916 | -7.5E-01 | -87.271 | -86.526 | -7.5E-01 |
| 731 | Nb | 94 | 41 | 53 | 12 | 8.648902 | 8.645323 | 3.6E-03 | 87453.013 | 87453.324 | -3.1E-01 | 87474.068 | 87474.379 | -3.1E-01 | -86.369 | -86.063 | -3.1E-01 |
| 732 | Mo | 94 | 42 | 52 | 10 | 8.662320 | 8.662112 | 2.1E-04 | 87450.452 | 87450.446 | 6.5E-03 | 87472.024 | 87472.018 | 6.5E-03 | -88.413 | -88.424 | 1.1E-02 |
| 733 | Tc | 94 | 43 | 51 | 8 | 8.608723 | 8.604043 | 4.7E-03 | 87454.191 | 87454.604 | -4.1E-01 | 87476.280 | 87476.693 | -4.1E-01 | -84.157 | -83.748 | -4.1E-01 |
| 734 | Ru | 94 | 44 | 50 | 6 | 8.583661 | 8.577972 | 5.7E-03 | 87455.246 | 87455.754 | -5.1E-01 | 87477.854 | 87478.361 | -5.1E-01 | -82.584 | -82.081 | -5.0E-01 |
| 735 | Rh | 94 | 45 | 49 | 4 | 8.472402 | 8.470107 | 2.3E-03 | 87464.405 | 87464.592 | -1.9E-01 | 87487.530 | 87487.717 | -1.9E-01 | -72.908 | -72.725 | -1.8E-01 |
| 736 | Pd | 94 | 46 | 48 | 2 | 8.391669 | 8.395706 | -4.0E-03 | 87470.693 | 87470.285 | 4.1E-01 | 87494.336 | 87493.928 | 4.1E-01 | -66.101 | -66.514 | 4.1E-01 |
| 737 | Kr | 95 | 36 | 59 | 23 | 8.365995 | 8.370712 | -4.7E-03 | 88417.300 | 88416.831 | 4.7E-01 | 88435.772 | 88435.303 | 4.7E-01 | -56.159 | -56.633 | 4.7E-01 |
| 738 | Rb | 95 | 37 | 58 | 21 | 8.460234 | 8.464087 | -3.9E-03 | 88407.049 | 88406.661 | 3.9E-01 | 88426.037 | 88425.649 | 3.9E-01 | -65.894 | -66.287 | 3.9E-01 |
| 739 | Sr | 95 | 38 | 57 | 19 | 8.549139 | 8.538251 | 1.1E-02 | 88397.304 | 88398.316 | -1.0E+00 | 88416.809 | 88417.820 | -1.0E+00 | -75.122 | -74.116 | -1.0E+00 |
| 740 | Y | 95 | 39 | 56 | 17 | 8.604999 | 8.590988 | 1.4E-02 | 88390.699 | 88392.006 | -1.3E+00 | 88410.720 | 88412.027 | -1.3E+00 | -81.211 | -79.909 | -1.3E+00 |
| 741 | Zr | 95 | 40 | 55 | 15 | 8.643609 | 8.627104 | 1.7E-02 | 88385.732 | 88387.275 | -1.5E+00 | 88406.270 | 88407.813 | -1.5E+00 | -85.662 | -84.123 | -1.5E+00 |
| 742 | Nb | 95 | 41 | 54 | 13 | 8.647200 | 8.644767 | 2.4E-03 | 88384.091 | 88384.297 | -2.1E-01 | 88405.146 | 88405.352 | -2.1E-01 | -86.785 | -86.584 | -2.0E-01 |
| 743 | Mo | 95 | 42 | 53 | 11 | 8.648707 | 8.643933 | 4.8E-03 | 88382.649 | 88383.076 | -4.3E-01 | 88404.221 | 88404.648 | -4.3E-01 | -87.711 | -87.288 | -4.2E-01 |
| 744 | Tc | 95 | 43 | 52 | 9 | 8.622677 | 8.623800 | -1.1E-03 | 88383.822 | 88383.688 | 1.3E-01 | 88405.911 | 88405.778 | 1.3E-01 | -86.020 | -86.158 | 1.4E-01 |
| 745 | Ru | 95 | 44 | 51 | 7 | 8.587457 | 8.580809 | 6.6E-03 | 88385.868 | 88386.472 | -6.0E-01 | 88408.475 | 88409.079 | -6.0E-01 | -83.456 | -82.857 | -6.0E-01 |
| 746 | Rh | 95 | 45 | 50 | 5 | 8.525370 | 8.518040 | 7.3E-03 | 88390.466 | 88391.134 | -6.7E-01 | 88413.591 | 88414.259 | -6.7E-01 | -78.341 | -77.677 | -6.6E-01 |
| 747 | Pd | 95 | 46 | 49 | 3 | 8.428967 | 8.427057 | 1.9E-03 | 88398.323 | 88398.476 | -1.5E-01 | 88421.967 | 88422.119 | -1.5E-01 | -69.965 | -69.817 | -1.5E-01 |
| 748 | Kr | 96 | 36 | 60 | 24 | 8.330850 | 8.338643 | -7.8E-03 | 89351.874 | 89351.104 | 7.7E-01 | 89370.346 | 89369.576 | 7.7E-01 | -53.080 | -53.854 | 7.7E-01 |
| 749 | Rb | 96 | 37 | 59 | 22 | 8.408896 | 8.418812 | -9.9E-03 | 89343.083 | 89342.109 | 9.7E-01 | 89362.071 | 89361.097 | 9.7E-01 | -61.354 | -62.333 | 9.8E-01 |
| 750 | Sr | 96 | 38 | 58 | 20 | 8.521324 | 8.519326 | 2.0E-03 | 89330.991 | 89331.160 | -1.7E-01 | 89350.495 | 89350.664 | -1.7E-01 | -72.930 | -72.766 | -1.6E-01 |
| 751 | Y | 96 | 39 | 57 | 18 | 8.569547 | 8.557067 | 1.2E-02 | 89325.063 | 89326.237 | -1.2E+00 | 89345.084 | 89346.258 | -1.2E+00 | -78.342 | -77.172 | -1.2E+00 |
| 752 | Zr | 96 | 40 | 56 | 16 | 8.635387 | 8.620126 | 1.5E-02 | 89317.443 | 89318.883 | -1.4E+00 | 89337.981 | 89339.421 | -1.4E+00 | -85.445 | -84.009 | -1.4E+00 |
| 753 | Nb | 96 | 41 | 55 | 14 | 8.628928 | 8.620401 | 8.5E-03 | 89316.764 | 89317.557 | -7.9E-01 | 89337.818 | 89338.612 | -7.9E-01 | -85.607 | -84.818 | -7.9E-01 |
| 754 | Mo | 96 | 42 | 54 | 12 | 8.653974 | 8.650795 | 3.2E-03 | 89313.060 | 89313.339 | -2.8E-01 | 89334.632 | 89334.911 | -2.8E-01 | -88.794 | -88.519 | -2.7E-01 |
| 755 | Tc | 96 | 43 | 53 | 10 | 8.614853 | 8.613195 | 1.7E-03 | 89315.515 | 89315.648 | -1.3E-01 | 89337.605 | 89337.738 | -1.3E-01 | -85.820 | -85.692 | -1.3E-01 |
| 756 | Ru | 96 | 44 | 52 | 8 | 8.609399 | 8.608518 | 8.8E-04 | 89314.739 | 89314.796 | -5.7E-02 | 89337.346 | 89337.403 | -5.7E-02 | -86.079 | -86.027 | -5.3E-02 |
| 757 | Rh | 96 | 45 | 51 | 6 | 8.534660 | 8.528994 | 5.7E-03 | 89320.614 | 89321.130 | -5.2E-01 | 89343.739 | 89344.255 | -5.2E-01 | -79.686 | -79.175 | -5.1E-01 |
| 758 | Pd | 96 | 46 | 50 | 4 | 8.490007 | 8.484634 | 5.4E-03 | 89323.600 | 89324.087 | -4.9E-01 | 89347.243 | 89347.730 | -4.9E-01 | -76.182 | -75.700 | -4.8E-01 |
| 759 | Ag | 96 | 47 | 49 | 2 | 8.360290 | 8.360399 | -1.1E-04 | 89334.752 | 89334.712 | 4.0E-02 | 89358.914 | 89358.874 | 4.0E-02 | -64.512 | -64.556 | 4.5E-02 |
| 760 | Kr | 97 | 36 | 61 | 25 | 8.269864 | 8.281159 | -1.1E-02 | 90289.024 | 90287.907 | 1.1E+00 | 90307.496 | 90306.379 | 1.1E+00 | -47.423 | -48.545 | 1.1E+00 |
| 761 | Rb | 97 | 37 | 60 | 23 | 8.376186 | 8.387940 | -1.2E-02 | 90277.412 | 90276.250 | 1.2E+00 | 90296.400 | 90295.238 | 1.2E+00 | -58.519 | -59.686 | 1.2E+00 |
| 762 | Sr | 97 | 38 | 59 | 21 | 8.471864 | 8.475512 | -3.6E-03 | 90266.833 | 90266.456 | 3.8E-01 | 90286.337 | 90285.960 | 3.8E-01 | -68.582 | -68.964 | 3.8E-01 |
| 763 | Y | 97 | 39 | 58 | 19 | 8.541582 | 8.540391 | 1.2E-03 | 90258.771 | 90258.863 | -9.2E-02 | 90278.792 | 90278.884 | -9.2E-02 | -76.127 | -76.040 | -8.7E-02 |
| 764 | Zr | 97 | 40 | 57 | 17 | 8.603838 | 8.587819 | 1.6E-02 | 90251.433 | 90252.962 | -1.5E+00 | 90271.971 | 90273.500 | -1.5E+00 | -82.948 | -81.424 | -1.5E+00 |
| 765 | Nb | 97 | 41 | 56 | 15 | 8.623192 | 8.616205 | 7.0E-03 | 90248.256 | 90248.909 | -6.5E-01 | 90269.311 | 90269.964 | -6.5E-01 | -85.608 | -84.960 | -6.5E-01 |
| 766 | Mo | 97 | 42 | 55 | 13 | 8.635080 | 8.628343 | 6.7E-03 | 90245.804 | 90246.431 | -6.3E-01 | 90267.376 | 90268.003 | -6.3E-01 | -87.544 | -86.921 | -6.2E-01 |
| 767 | Tc | 97 | 43 | 54 | 11 | 8.623667 | 8.623128 | 5.4E-04 | 90245.611 | 90245.637 | -2.6E-02 | 90267.700 | 90267.726 | -2.6E-02 | -87.219 | -87.198 | -2.1E-02 |
| 768 | Ru | 97 | 44 | 53 | 9 | 8.604266 | 8.599155 | 5.1E-03 | 90246.193 | 90246.661 | -4.7E-01 | 90268.800 | 90269.268 | -4.7E-01 | -86.119 | -85.656 | -4.6E-01 |
| 769 | Rh | 97 | 45 | 52 | 7 | 8.559881 | 8.558392 | 1.5E-03 | 90249.198 | 90249.314 | -1.2E-01 | 90272.323 | 90272.439 | -1.2E-01 | -82.596 | -82.485 | -1.1E-01 |
| 770 | Pd | 97 | 46 | 51 | 5 | 8.502430 | 8.495416 | 7.0E-03 | 90253.470 | 90254.122 | -6.5E-01 | 90277.113 | 90277.765 | -6.5E-01 | -77.806 | -77.159 | -6.5E-01 |
| 771 | Ag | 97 | 47 | 50 | 3 | 8.422405 | 8.416523 | 5.9E-03 | 90259.932 | 90260.473 | -5.4E-01 | 90284.093 | 90284.635 | -5.4E-01 | -70.826 | -70.289 | -5.4E-01 |
| 772 | Rb | 98 | 37 | 61 | 24 | 8.330210 | 8.339158 | -8.9E-03 | 91213.107 | 91212.208 | 9.0E-01 | 91232.095 | 91231.196 | 9.0E-01 | -54.318 | -55.222 | 9.0E-01 |
| 773 | Sr | 98 | 38 | 60 | 22 | 8.445774 | 8.452077 | -6.3E-03 | 91200.483 | 91199.842 | 6.4E-01 | 91219.987 | 91219.347 | 6.4E-01 | -66.426 | -67.071 | 6.5E-01 |
| 774 | Y | 98 | 39 | 59 | 20 | 8.497736 | 8.503803 | -6.1E-03 | 91194.092 | 91193.473 | 6.2E-01 | 91214.113 | 91213.495 | 6.2E-01 | -72.301 | -72.923 | 6.2E-01 |
| 775 | Zr | 98 | 40 | 58 | 18 | 8.581507 | 8.577521 | 4.0E-03 | 91184.583 | 91184.949 | -3.7E-01 | 91205.121 | 91205.487 | -3.7E-01 | -81.293 | -80.931 | -3.6E-01 |
| 776 | Nb | 98 | 41 | 57 | 16 | 8.596360 | 8.590362 | 6.0E-03 | 91181.828 | 91182.391 | -5.6E-01 | 91202.883 | 91203.446 | -5.6E-01 | -83.530 | -82.972 | -5.6E-01 |
| 777 | Mo | 98 | 42 | 56 | 14 | 8.635157 | 8.630705 | 4.5E-03 | 91176.726 | 91177.137 | -4.1E-01 | 91198.298 | 91198.709 | -4.1E-01 | -88.115 | -87.709 | -4.1E-01 |
| 778 | Tc | 98 | 43 | 55 | 12 | 8.609993 | 8.607663 | 2.3E-03 | 91177.893 | 91178.094 | -2.0E-01 | 91199.982 | 91200.184 | -2.0E-01 | -86.431 | -86.234 | -2.0E-01 |
| 779 | Ru | 98 | 44 | 54 | 10 | 8.620312 | 8.616516 | 3.8E-03 | 91175.581 | 91175.926 | -3.4E-01 | 91198.189 | 91198.533 | -3.4E-01 | -88.225 | -87.885 | -3.4E-01 |
| 780 | Rh | 98 | 45 | 53 | 8 | 8.560802 | 8.556381 | 4.4E-03 | 91180.113 | 91180.518 | -4.1E-01 | 91203.238 | 91203.643 | -4.1E-01 | -83.175 | -82.775 | -4.0E-01 |
| 781 | Pd | 98 | 46 | 52 | 6 | 8.533899 | 8.532701 | 1.2E-03 | 91181.449 | 91181.538 | -8.9E-02 | 91205.092 | 91205.181 | -8.9E-02 | -81.321 | -81.237 | -8.4E-02 |
| 782 | Ag | 98 | 47 | 51 | 4 | 8.441686 | 8.434848 | 6.8E-03 | 91189.185 | 91189.826 | -6.4E-01 | 91213.347 | 91213.988 | -6.4E-01 | -73.066 | -72.430 | -6.4E-01 |
| 783 | Cd | 98 | 48 | 50 | 2 | 8.378294 | 8.375781 | 2.5E-03 | 91194.097 | 91194.313 | -2.2E-01 | 91218.777 | 91218.993 | -2.2E-01 | -67.636 | -67.425 | -2.1E-01 |
| 784 | Rb | 99 | 37 | 62 | 25 | 8.296151 | 8.303285 | -7.1E-03 | 92147.714 | 92146.985 | 7.3E-01 | 92166.702 | 92165.974 | 7.3E-01 | -51.205 | -51.939 | 7.3E-01 |
| 785 | Sr | 99 | 38 | 61 | 23 | 8.402456 | 8.404398 | -1.9E-03 | 92135.891 | 92135.676 | 2.2E-01 | 92155.395 | 92155.180 | 2.1E-01 | -62.512 | -62.732 | 2.2E-01 |
| 786 | Y | 99 | 39 | 60 | 21 | 8.476814 | 8.482303 | -5.5E-03 | 92127.231 | 92126.664 | 5.7E-01 | 92147.252 | 92146.685 | 5.7E-01 | -70.656 | -71.227 | 5.7E-01 |
| 787 | Zr | 99 | 40 | 59 | 19 | 8.539303 | 8.542509 | -3.2E-03 | 92119.745 | 92119.403 | 3.4E-01 | 92140.283 | 92139.941 | 3.4E-01 | -77.624 | -77.971 | 3.5E-01 |
| 788 | Nb | 99 | 41 | 58 | 17 | 8.578949 | 8.582578 | -3.6E-03 | 92114.521 | 92114.136 | 3.8E-01 | 92135.576 | 92135.191 | 3.8E-01 | -82.332 | -82.721 | 3.9E-01 |
| 789 | Mo | 99 | 42 | 57 | 15 | 8.607786 | 8.606513 | 1.3E-03 | 92110.366 | 92110.467 | -1.0E-01 | 92131.938 | 92132.039 | -1.0E-01 | -85.969 | -85.873 | -9.6E-02 |
| 790 | Tc | 99 | 43 | 56 | 13 | 8.613599 | 8.613054 | 5.4E-04 | 92108.491 | 92108.518 | -2.7E-02 | 92130.581 | 92130.608 | -2.7E-02 | -87.327 | -87.304 | -2.3E-02 |
| 791 | Ru | 99 | 44 | 55 | 11 | 8.608677 | 8.602745 | 5.9E-03 | 92107.678 | 92108.238 | -5.6E-01 | 92130.285 | 92130.846 | -5.6E-01 | -87.622 | -87.067 | -5.6E-01 |
| 792 | Rh | 99 | 45 | 54 | 9 | 8.580129 | 8.576327 | 3.8E-03 | 92109.204 | 92109.553 | -3.5E-01 | 92132.329 | 92132.678 | -3.5E-01 | -85.578 | -85.234 | -3.4E-01 |
| 793 | Pd | 99 | 46 | 53 | 7 | 8.537916 | 8.531144 | 6.8E-03 | 92112.083 | 92112.725 | -6.4E-01 | 92135.726 | 92136.368 | -6.4E-01 | -82.181 | -81.544 | -6.4E-01 |
| 794 | Ag | 99 | 47 | 52 | 5 | 8.474774 | 8.472481 | 2.3E-03 | 92117.033 | 92117.231 | -2.0E-01 | 92141.195 | 92141.392 | -2.0E-01 | -76.712 | -76.520 | -1.9E-01 |
| 795 | Cd | 99 | 48 | 51 | 3 | 8.398373 | 8.392974 | 5.4E-03 | 92123.296 | 92123.801 | -5.0E-01 | 92147.976 | 92148.481 | -5.0E-01 | -69.931 | -69.432 | -5.0E-01 |
| 796 | Sr | 100 | 38 | 62 | 24 | 8.372327 | 8.376418 | -4.1E-03 | 93070.067 | 93069.635 | 4.3E-01 | 93089.571 | 93089.139 | 4.3E-01 | -59.830 | -60.267 | 4.4E-01 |
| 797 | Y | 100 | 39 | 61 | 22 | 8.439535 | 8.442144 | -2.6E-03 | 93062.047 | 93061.763 | 2.8E-01 | 93082.068 | 93081.784 | 2.8E-01 | -67.333 | -67.622 | 2.9E-01 |
| 798 | Zr | 100 | 40 | 60 | 20 | 8.522198 | 8.527551 | -5.4E-03 | 93052.481 | 93051.922 | 5.6E-01 | 93073.019 | 93072.460 | 5.6E-01 | -76.382 | -76.946 | 5.6E-01 |
| 799 | Nb | 100 | 41 | 59 | 18 | 8.548588 | 8.553948 | -5.4E-03 | 93048.543 | 93047.982 | 5.6E-01 | 93069.598 | 93069.037 | 5.6E-01 | -79.803 | -80.369 | 5.7E-01 |
| 800 | Mo | 100 | 42 | 58 | 16 | 8.604626 | 8.604812 | -1.9E-04 | 93041.640 | 93041.595 | 4.4E-02 | 93063.212 | 93063.168 | 4.4E-02 | -86.189 | -86.239 | 4.9E-02 |
| 801 | Tc | 100 | 43 | 57 | 14 | 8.595107 | 8.595121 | -1.4E-05 | 93041.292 | 93041.264 | 2.8E-02 | 93063.381 | 93063.354 | 2.8E-02 | -86.020 | -86.053 | 3.3E-02 |
| 802 | Ru | 100 | 44 | 56 | 12 | 8.619323 | 8.614844 | 4.5E-03 | 93037.570 | 93037.991 | -4.2E-01 | 93060.178 | 93060.598 | -4.2E-01 | -89.224 | -88.808 | -4.2E-01 |
| 803 | Rh | 100 | 45 | 55 | 10 | 8.575143 | 8.569383 | 5.8E-03 | 93040.688 | 93041.236 | -5.5E-01 | 93063.813 | 93064.361 | -5.5E-01 | -85.588 | -85.045 | -5.4E-01 |
| 804 | Pd | 100 | 46 | 54 | 8 | 8.563710 | 8.558443 | 5.3E-03 | 93040.531 | 93041.029 | -5.0E-01 | 93064.174 | 93064.672 | -5.0E-01 | -85.227 | -84.734 | -4.9E-01 |
| 805 | Ag | 100 | 47 | 53 | 6 | 8.484994 | 8.477727 | 7.3E-03 | 93047.102 | 93047.799 | -7.0E-01 | 93071.263 | 93071.961 | -7.0E-01 | -78.138 | -77.445 | -6.9E-01 |
| 806 | Cd | 100 | 48 | 52 | 4 | 8.437737 | 8.437986 | -2.5E-04 | 93050.527 | 93050.472 | 5.5E-02 | 93075.207 | 93075.152 | 5.5E-02 | -74.195 | -74.254 | 6.0E-02 |
| 807 | In | 100 | 49 | 51 | 2 | 8.331097 | 8.324903 | 6.2E-03 | 93059.890 | 93060.478 | -5.9E-01 | 93085.088 | 93085.677 | -5.9E-01 | -64.313 | -63.729 | -5.8E-01 |
| 808 | Sn | 100 | 50 | 50 | 0 | 8.252974 | 8.246644 | 6.3E-03 | 93066.401 | 93067.002 | -6.0E-01 | 93092.118 | 93092.720 | -6.0E-01 | -57.283 | -56.686 | -6.0E-01 |
| 809 | Sr | 101 | 38 | 63 | 25 | 8.327088 | 8.324863 | 2.2E-03 | 94005.829 | 94006.031 | -2.0E-01 | 94025.333 | 94025.535 | -2.0E-01 | -55.562 | -55.365 | -2.0E-01 |
| 810 | Y | 101 | 39 | 62 | 23 | 8.413451 | 8.415644 | -2.2E-03 | 93995.807 | 93995.562 | 2.5E-01 | 94015.828 | 94015.583 | 2.5E-01 | -65.067 | -65.317 | 2.5E-01 |
| 811 | Zr | 101 | 40 | 61 | 21 | 8.485940 | 8.488848 | -2.9E-03 | 93987.187 | 93986.869 | 3.2E-01 | 94007.725 | 94007.407 | 3.2E-01 | -73.171 | -73.493 | 3.2E-01 |
| 812 | Nb | 101 | 41 | 60 | 19 | 8.534800 | 8.541331 | -6.5E-03 | 93980.952 | 93980.268 | 6.8E-01 | 94002.007 | 94001.323 | 6.8E-01 | -78.888 | -79.577 | 6.9E-01 |
| 813 | Mo | 101 | 42 | 59 | 17 | 8.572880 | 8.577807 | -4.9E-03 | 93975.807 | 93975.284 | 5.2E-01 | 93997.379 | 93996.856 | 5.2E-01 | -83.516 | -84.044 | 5.3E-01 |
| 814 | Tc | 101 | 43 | 58 | 15 | 8.593101 | 8.596321 | -3.2E-03 | 93972.465 | 93972.113 | 3.5E-01 | 93994.554 | 93994.203 | 3.5E-01 | -86.341 | -86.697 | 3.6E-01 |
| 815 | Ru | 101 | 44 | 57 | 13 | 8.601330 | 8.598673 | 2.7E-03 | 93970.334 | 93970.575 | -2.4E-01 | 93992.941 | 93993.182 | -2.4E-01 | -87.955 | -87.718 | -2.4E-01 |
| 816 | Rh | 101 | 45 | 56 | 11 | 8.588201 | 8.584458 | 3.7E-03 | 93970.359 | 93970.710 | -3.5E-01 | 93993.485 | 93993.835 | -3.5E-01 | -87.411 | -87.065 | -3.5E-01 |
| 817 | Pd | 101 | 46 | 55 | 9 | 8.560849 | 8.552697 | 8.2E-03 | 93971.822 | 93972.616 | -7.9E-01 | 93995.465 | 93996.259 | -7.9E-01 | -85.431 | -84.641 | -7.9E-01 |
| 818 | Ag | 101 | 47 | 54 | 7 | 8.512546 | 8.506870 | 5.7E-03 | 93975.400 | 93975.943 | -5.4E-01 | 93999.561 | 94000.105 | -5.4E-01 | -81.334 | -80.795 | -5.4E-01 |
| 819 | Cd | 101 | 48 | 53 | 5 | 8.450365 | 8.442614 | 7.8E-03 | 93980.379 | 93981.132 | -7.5E-01 | 94005.059 | 94005.812 | -7.5E-01 | -75.836 | -75.088 | -7.5E-01 |
| 820 | Sn | 101 | 50 | 51 | 1 | 8.281102 | 8.270724 | 1.0E-02 | 93994.872 | 93995.889 | -1.0E+00 | 94020.590 | 94021.606 | -1.0E+00 | -60.306 | -59.294 | -1.0E+00 |
| 821 | Sr | 102 | 38 | 64 | 26 | 8.293172 | 8.292739 | 4.3E-04 | 94940.526 | 94940.548 | -2.1E-02 | 94960.031 | 94960.052 | -2.1E-02 | -52.358 | -52.342 | -1.7E-02 |
| 822 | Y | 102 | 39 | 63 | 24 | 8.371924 | 8.372005 | -8.1E-05 | 94931.195 | 94931.163 | 3.2E-02 | 94951.216 | 94951.184 | 3.2E-02 | -61.173 | -61.210 | 3.7E-02 |
| 823 | Zr | 102 | 40 | 62 | 22 | 8.466414 | 8.469184 | -2.8E-03 | 94920.258 | 94919.951 | 3.1E-01 | 94940.796 | 94940.489 | 3.1E-01 | -71.594 | -71.905 | 3.1E-01 |
| 824 | Nb | 102 | 41 | 61 | 20 | 8.504988 | 8.509127 | -4.1E-03 | 94915.024 | 94914.577 | 4.5E-01 | 94936.079 | 94935.632 | 4.5E-01 | -76.311 | -76.763 | 4.5E-01 |
| 825 | Mo | 102 | 42 | 60 | 18 | 8.568493 | 8.571170 | -2.7E-03 | 94907.247 | 94906.948 | 3.0E-01 | 94928.819 | 94928.520 | 3.0E-01 | -83.570 | -83.874 | 3.0E-01 |
| 826 | Tc | 102 | 43 | 59 | 16 | 8.570628 | 8.575208 | -4.6E-03 | 94905.729 | 94905.236 | 4.9E-01 | 94927.819 | 94927.325 | 4.9E-01 | -84.571 | -85.069 | 5.0E-01 |
| 827 | Ru | 102 | 44 | 58 | 14 | 8.607392 | 8.605973 | 1.4E-03 | 94900.679 | 94900.797 | -1.2E-01 | 94923.286 | 94923.404 | -1.2E-01 | -89.103 | -88.990 | -1.1E-01 |
| 828 | Rh | 102 | 45 | 57 | 12 | 8.576953 | 8.574428 | 2.5E-03 | 94902.484 | 94902.714 | -2.3E-01 | 94925.609 | 94925.839 | -2.3E-01 | -86.780 | -86.556 | -2.2E-01 |
| 829 | Pd | 102 | 46 | 56 | 10 | 8.580563 | 8.574466 | 6.1E-03 | 94900.815 | 94901.409 | -5.9E-01 | 94924.458 | 94925.052 | -5.9E-01 | -87.931 | -87.343 | -5.9E-01 |
| 830 | Ag | 102 | 47 | 55 | 8 | 8.517163 | 8.507674 | 9.5E-03 | 94905.981 | 94906.920 | -9.4E-01 | 94930.143 | 94931.081 | -9.4E-01 | -82.247 | -81.313 | -9.3E-01 |
| 831 | Cd | 102 | 48 | 54 | 6 | 8.484130 | 8.478734 | 5.4E-03 | 94908.050 | 94908.570 | -5.2E-01 | 94932.730 | 94933.250 | -5.2E-01 | -79.660 | -79.144 | -5.2E-01 |
| 832 | In | 102 | 49 | 53 | 4 | 8.388560 | 8.379341 | 9.2E-03 | 94916.497 | 94917.406 | -9.1E-01 | 94941.696 | 94942.605 | -9.1E-01 | -70.694 | -69.789 | -9.0E-01 |
| 833 | Sn | 102 | 50 | 52 | 2 | 8.324419 | 8.322884 | 1.5E-03 | 94921.738 | 94921.863 | -1.3E-01 | 94947.456 | 94947.581 | -1.3E-01 | -64.934 | -64.813 | -1.2E-01 |
| 834 | Y | 103 | 39 | 64 | 25 | 8.342640 | 8.340781 | 1.9E-03 | 95865.405 | 95865.573 | -1.7E-01 | 95885.426 | 95885.594 | -1.7E-01 | -58.458 | -58.295 | -1.6E-01 |
| 835 | Zr | 103 | 40 | 63 | 23 | 8.425954 | 8.426757 | -8.0E-04 | 95855.524 | 95855.417 | 1.1E-01 | 95876.062 | 95875.955 | 1.1E-01 | -67.821 | -67.933 | 1.1E-01 |
| 836 | Nb | 103 | 41 | 62 | 21 | 8.488297 | 8.491525 | -3.2E-03 | 95847.803 | 95847.446 | 3.6E-01 | 95868.858 | 95868.501 | 3.6E-01 | -75.025 | -75.387 | 3.6E-01 |
| 837 | Mo | 103 | 42 | 61 | 19 | 8.538387 | 8.540553 | -2.2E-03 | 95841.344 | 95841.096 | 2.5E-01 | 95862.917 | 95862.668 | 2.5E-01 | -80.967 | -81.220 | 2.5E-01 |
| 838 | Tc | 103 | 43 | 60 | 17 | 8.566084 | 8.571394 | -5.3E-03 | 95837.192 | 95836.619 | 5.7E-01 | 95859.281 | 95858.708 | 5.7E-01 | -84.602 | -85.180 | 5.8E-01 |
| 839 | Ru | 103 | 44 | 59 | 15 | 8.584331 | 8.586707 | -2.4E-03 | 95834.013 | 95833.741 | 2.7E-01 | 95856.620 | 95856.348 | 2.7E-01 | -87.264 | -87.540 | 2.8E-01 |
| 840 | Rh | 103 | 45 | 58 | 13 | 8.584157 | 8.584884 | -7.3E-04 | 95832.730 | 95832.628 | 1.0E-01 | 95855.855 | 95855.753 | 1.0E-01 | -88.028 | -88.136 | 1.1E-01 |
| 841 | Pd | 103 | 46 | 57 | 11 | 8.571289 | 8.566016 | 5.3E-03 | 95832.755 | 95833.270 | -5.1E-01 | 95856.398 | 95856.913 | -5.1E-01 | -87.485 | -86.975 | -5.1E-01 |
| 842 | Ag | 103 | 47 | 56 | 9 | 8.537628 | 8.531964 | 5.7E-03 | 95834.922 | 95835.476 | -5.5E-01 | 95859.083 | 95859.637 | -5.5E-01 | -84.800 | -84.251 | -5.5E-01 |
| 843 | Cd | 103 | 48 | 55 | 7 | 8.489758 | 8.480013 | 9.7E-03 | 95838.551 | 95839.525 | -9.7E-01 | 95863.231 | 95864.205 | -9.7E-01 | -80.652 | -79.683 | -9.7E-01 |
| 844 | In | 103 | 49 | 54 | 5 | 8.423691 | 8.416185 | 7.5E-03 | 95844.055 | 95844.798 | -7.4E-01 | 95869.254 | 95869.996 | -7.4E-01 | -74.630 | -73.892 | -7.4E-01 |
| 845 | Sn | 103 | 50 | 53 | 3 | 8.341727 | 8.333024 | 8.7E-03 | 95851.196 | 95852.061 | -8.6E-01 | 95876.914 | 95877.779 | -8.7E-01 | -66.970 | -66.109 | -8.6E-01 |
| 846 | Zr | 104 | 40 | 64 | 24 | 8.402436 | 8.402787 | -3.5E-04 | 96789.109 | 96789.049 | 6.1E-02 | 96809.647 | 96809.587 | 6.1E-02 | -65.730 | -65.796 | 6.6E-02 |
| 847 | Nb | 104 | 41 | 63 | 22 | 8.453519 | 8.455848 | -2.3E-03 | 96782.497 | 96782.230 | 2.7E-01 | 96803.552 | 96803.285 | 2.7E-01 | -71.825 | -72.097 | 2.7E-01 |
| 848 | Mo | 104 | 42 | 62 | 20 | 8.528023 | 8.528991 | -9.7E-04 | 96773.449 | 96773.323 | 1.3E-01 | 96795.021 | 96794.895 | 1.3E-01 | -80.356 | -80.487 | 1.3E-01 |
| 849 | Tc | 104 | 43 | 61 | 18 | 8.541185 | 8.546531 | -5.3E-03 | 96770.781 | 96770.198 | 5.8E-01 | 96792.870 | 96792.288 | 5.8E-01 | -82.507 | -83.094 | 5.9E-01 |
| 850 | Ru | 104 | 44 | 60 | 16 | 8.587380 | 8.588670 | -1.3E-03 | 96764.676 | 96764.515 | 1.6E-01 | 96787.284 | 96787.122 | 1.6E-01 | -88.094 | -88.260 | 1.7E-01 |
| 851 | Rh | 104 | 45 | 59 | 14 | 8.568914 | 8.571250 | -2.3E-03 | 96765.297 | 96765.026 | 2.7E-01 | 96788.422 | 96788.151 | 2.7E-01 | -86.956 | -87.231 | 2.8E-01 |
| 852 | Pd | 104 | 46 | 58 | 12 | 8.584846 | 8.582507 | 2.3E-03 | 96762.339 | 96762.554 | -2.1E-01 | 96785.982 | 96786.197 | -2.1E-01 | -89.395 | -89.185 | -2.1E-01 |
| 853 | Ag | 104 | 47 | 57 | 10 | 8.536183 | 8.529444 | 6.7E-03 | 96766.100 | 96766.771 | -6.7E-01 | 96790.261 | 96790.933 | -6.7E-01 | -85.116 | -84.450 | -6.7E-01 |
| 854 | Cd | 104 | 48 | 56 | 8 | 8.517621 | 8.510867 | 6.8E-03 | 96766.729 | 96767.402 | -6.7E-01 | 96791.409 | 96792.082 | -6.7E-01 | -83.968 | -83.301 | -6.7E-01 |
| 855 | In | 104 | 49 | 55 | 6 | 8.435237 | 8.423760 | 1.1E-02 | 96773.996 | 96775.159 | -1.2E+00 | 96799.195 | 96800.358 | -1.2E+00 | -76.183 | -75.025 | -1.2E+00 |
| 856 | Sn | 104 | 50 | 54 | 4 | 8.383911 | 8.376835 | 7.1E-03 | 96778.033 | 96778.737 | -7.0E-01 | 96803.750 | 96804.455 | -7.0E-01 | -71.627 | -70.928 | -7.0E-01 |
| 857 | Sb | 104 | 51 | 53 | 2 | 8.256614 | 8.257110 | -5.0E-04 | 96789.970 | 96789.886 | 8.4E-02 | 96816.207 | 96816.123 | 8.4E-02 | -59.171 | -59.259 | 8.9E-02 |
| 858 | Zr | 105 | 40 | 65 | 25 | 8.358718 | 8.356747 | 2.0E-03 | 97724.863 | 97725.045 | -1.8E-01 | 97745.400 | 97745.583 | -1.8E-01 | -61.471 | -61.293 | -1.8E-01 |
| 859 | Nb | 105 | 41 | 64 | 23 | 8.431657 | 8.433568 | -1.9E-03 | 97715.905 | 97715.679 | 2.3E-01 | 97736.960 | 97736.734 | 2.3E-01 | -69.912 | -70.142 | 2.3E-01 |
| 860 | Mo | 105 | 42 | 63 | 21 | 8.494981 | 8.494781 | 2.0E-04 | 97707.956 | 97707.952 | 4.5E-03 | 97729.528 | 97729.524 | 4.4E-03 | -77.343 | -77.353 | 9.5E-03 |
| 861 | Tc | 105 | 43 | 62 | 19 | 8.534670 | 8.537608 | -2.9E-03 | 97702.489 | 97702.154 | 3.3E-01 | 97724.578 | 97724.244 | 3.3E-01 | -82.293 | -82.633 | 3.4E-01 |
| 862 | Ru | 105 | 44 | 61 | 17 | 8.561882 | 8.565704 | -3.8E-03 | 97698.332 | 97697.903 | 4.3E-01 | 97720.939 | 97720.511 | 4.3E-01 | -85.933 | -86.366 | 4.3E-01 |
| 863 | Rh | 105 | 45 | 60 | 15 | 8.572698 | 8.576541 | -3.8E-03 | 97695.896 | 97695.465 | 4.3E-01 | 97719.021 | 97718.590 | 4.3E-01 | -87.851 | -88.287 | 4.4E-01 |
| 864 | Pd | 105 | 46 | 59 | 13 | 8.570649 | 8.570785 | -1.4E-04 | 97694.810 | 97694.768 | 4.3E-02 | 97718.454 | 97718.411 | 4.3E-02 | -88.418 | -88.466 | 4.8E-02 |
| 865 | Ag | 105 | 47 | 58 | 11 | 8.550370 | 8.548800 | 1.6E-03 | 97695.639 | 97695.775 | -1.4E-01 | 97719.801 | 97719.936 | -1.4E-01 | -87.071 | -86.940 | -1.3E-01 |
| 866 | Cd | 105 | 48 | 57 | 9 | 8.516852 | 8.509375 | 7.5E-03 | 97697.858 | 97698.613 | -7.6E-01 | 97722.538 | 97723.293 | -7.6E-01 | -84.334 | -83.584 | -7.5E-01 |
| 867 | In | 105 | 49 | 56 | 7 | 8.464704 | 8.456516 | 8.2E-03 | 97702.032 | 97702.861 | -8.3E-01 | 97727.231 | 97728.060 | -8.3E-01 | -79.641 | -78.816 | -8.2E-01 |
| 868 | Sn | 105 | 50 | 55 | 5 | 8.397228 | 8.384388 | 1.3E-02 | 97707.816 | 97709.133 | -1.3E+00 | 97733.533 | 97734.850 | -1.3E+00 | -73.338 | -72.026 | -1.3E+00 |
| 869 | Sb | 105 | 51 | 54 | 3 | 8.300997 | 8.300553 | 4.4E-04 | 97716.618 | 97716.633 | -1.5E-02 | 97742.855 | 97742.870 | -1.5E-02 | -64.016 | -64.007 | -9.6E-03 |
| 870 | Te | 105 | 52 | 53 | 1 | 8.186836 | 8.192080 | -5.2E-03 | 97727.303 | 97726.720 | 5.8E-01 | 97754.060 | 97753.477 | 5.8E-01 | -52.812 | -53.400 | 5.9E-01 |
| 871 | Nb | 106 | 41 | 65 | 24 | 8.393237 | 8.394661 | -1.4E-03 | 98651.111 | 98650.935 | 1.8E-01 | 98672.166 | 98671.990 | 1.8E-01 | -66.200 | -66.380 | 1.8E-01 |
| 872 | Mo | 106 | 42 | 64 | 22 | 8.479640 | 8.478759 | 8.8E-04 | 98640.652 | 98640.720 | -6.8E-02 | 98662.225 | 98662.293 | -6.8E-02 | -76.141 | -76.078 | -6.3E-02 |
| 873 | Tc | 106 | 43 | 63 | 20 | 8.506548 | 8.509158 | -2.6E-03 | 98636.500 | 98636.198 | 3.0E-01 | 98658.590 | 98658.287 | 3.0E-01 | -79.775 | -80.083 | 3.1E-01 |
| 874 | Ru | 106 | 44 | 62 | 18 | 8.560932 | 8.562387 | -1.5E-03 | 98629.436 | 98629.255 | 1.8E-01 | 98652.043 | 98651.862 | 1.8E-01 | -86.322 | -86.509 | 1.9E-01 |
| 875 | Rh | 106 | 45 | 61 | 16 | 8.553923 | 8.558879 | -5.0E-03 | 98628.879 | 98628.326 | 5.5E-01 | 98652.004 | 98651.451 | 5.5E-01 | -86.362 | -86.920 | 5.6E-01 |
| 876 | Pd | 106 | 46 | 60 | 14 | 8.579991 | 8.581668 | -1.7E-03 | 98624.815 | 98624.609 | 2.1E-01 | 98648.458 | 98648.252 | 2.1E-01 | -89.907 | -90.119 | 2.1E-01 |
| 877 | Ag | 106 | 47 | 59 | 12 | 8.544638 | 8.542430 | 2.2E-03 | 98627.262 | 98627.467 | -2.0E-01 | 98651.423 | 98651.628 | -2.0E-01 | -86.942 | -86.742 | -2.0E-01 |
| 878 | Cd | 106 | 48 | 58 | 10 | 8.539047 | 8.534582 | 4.5E-03 | 98626.553 | 98626.997 | -4.4E-01 | 98651.233 | 98651.677 | -4.4E-01 | -87.132 | -86.694 | -4.4E-01 |
| 879 | In | 106 | 49 | 57 | 8 | 8.470119 | 8.460946 | 9.2E-03 | 98632.559 | 98633.501 | -9.4E-01 | 98657.757 | 98658.699 | -9.4E-01 | -80.608 | -79.671 | -9.4E-01 |
| 880 | Sn | 106 | 50 | 56 | 6 | 8.432037 | 8.423891 | 8.1E-03 | 98635.294 | 98636.126 | -8.3E-01 | 98661.012 | 98661.844 | -8.3E-01 | -77.354 | -76.526 | -8.3E-01 |
| 881 | Sb | 106 | 51 | 55 | 4 | 8.322012 | 8.315380 | 6.6E-03 | 98645.655 | 98646.326 | -6.7E-01 | 98671.892 | 98672.563 | -6.7E-01 | -66.473 | -65.807 | -6.7E-01 |
| 882 | Te | 106 | 52 | 54 | 2 | 8.236757 | 8.245127 | -8.4E-03 | 98653.390 | 98652.470 | 9.2E-01 | 98680.147 | 98679.227 | 9.2E-01 | -58.219 | -59.144 | 9.3E-01 |
| 883 | Nb | 107 | 41 | 66 | 25 | 8.367054 | 8.367964 | -9.1E-04 | 99585.084 | 99584.962 | 1.2E-01 | 99606.139 | 99606.017 | 1.2E-01 | -63.720 | -63.847 | 1.3E-01 |
| 884 | Mo | 107 | 42 | 65 | 23 | 8.442339 | 8.441128 | 1.2E-03 | 99575.729 | 99575.834 | -1.0E-01 | 99597.302 | 99597.406 | -1.0E-01 | -72.558 | -72.459 | -9.9E-02 |
| 885 | Tc | 107 | 43 | 64 | 21 | 8.492878 | 8.495499 | -2.6E-03 | 99569.022 | 99568.715 | 3.1E-01 | 99591.112 | 99590.805 | 3.1E-01 | -78.748 | -79.060 | 3.1E-01 |
| 886 | Ru | 107 | 44 | 63 | 19 | 8.533348 | 8.535770 | -2.4E-03 | 99563.392 | 99563.106 | 2.9E-01 | 99585.999 | 99585.713 | 2.9E-01 | -83.861 | -84.152 | 2.9E-01 |
| 887 | Rh | 107 | 45 | 62 | 17 | 8.554105 | 8.558950 | -4.8E-03 | 99559.871 | 99559.325 | 5.5E-01 | 99582.996 | 99582.450 | 5.5E-01 | -86.864 | -87.415 | 5.5E-01 |
| 888 | Pd | 107 | 46 | 61 | 15 | 8.560893 | 8.566231 | -5.3E-03 | 99557.844 | 99557.244 | 6.0E-01 | 99581.487 | 99580.887 | 6.0E-01 | -88.373 | -88.977 | 6.0E-01 |
| 889 | Ag | 107 | 47 | 60 | 13 | 8.553899 | 8.556547 | -2.6E-03 | 99557.291 | 99556.979 | 3.1E-01 | 99581.453 | 99581.141 | 3.1E-01 | -88.407 | -88.724 | 3.2E-01 |
| 890 | Cd | 107 | 48 | 59 | 11 | 8.533351 | 8.529678 | 3.7E-03 | 99558.189 | 99558.552 | -3.6E-01 | 99582.869 | 99583.232 | -3.6E-01 | -86.990 | -86.632 | -3.6E-01 |
| 891 | In | 107 | 49 | 58 | 9 | 8.494021 | 8.488377 | 5.6E-03 | 99561.096 | 99561.670 | -5.7E-01 | 99586.295 | 99586.869 | -5.7E-01 | -83.564 | -82.996 | -5.7E-01 |
| 892 | Sn | 107 | 50 | 57 | 7 | 8.439494 | 8.428916 | 1.1E-02 | 99565.629 | 99566.730 | -1.1E+00 | 99591.347 | 99592.448 | -1.1E+00 | -78.512 | -77.417 | -1.1E+00 |
| 893 | Sb | 107 | 51 | 56 | 5 | 8.358733 | 8.356171 | 2.6E-03 | 99572.969 | 99573.211 | -2.4E-01 | 99599.206 | 99599.449 | -2.4E-01 | -70.653 | -70.416 | -2.4E-01 |
| 894 | Te | 107 | 52 | 55 | 3 | 8.256871 | 8.260519 | -3.6E-03 | 99582.567 | 99582.144 | 4.2E-01 | 99609.323 | 99608.900 | 4.2E-01 | -60.536 | -60.964 | 4.3E-01 |
| 895 | Nb | 108 | 41 | 67 | 26 | 8.325667 | 8.325974 | -3.1E-04 | 100520.753 | 100520.695 | 5.8E-02 | 100541.807 | 100541.750 | 5.7E-02 | -59.546 | -59.609 | 6.3E-02 |
| 896 | Mo | 108 | 42 | 66 | 24 | 8.422278 | 8.421062 | 1.2E-03 | 100509.019 | 100509.125 | -1.1E-01 | 100530.591 | 100530.697 | -1.1E-01 | -70.762 | -70.661 | -1.0E-01 |
| 897 | Tc | 108 | 43 | 65 | 22 | 8.462797 | 8.463800 | -1.0E-03 | 100503.343 | 100503.209 | 1.3E-01 | 100525.433 | 100525.298 | 1.3E-01 | -75.921 | -76.060 | 1.4E-01 |
| 898 | Ru | 108 | 44 | 64 | 20 | 8.527207 | 8.527685 | -4.8E-04 | 100495.087 | 100495.009 | 7.8E-02 | 100517.694 | 100517.616 | 7.8E-02 | -83.659 | -83.743 | 8.4E-02 |
| 899 | Rh | 108 | 45 | 63 | 18 | 8.532672 | 8.537412 | -4.7E-03 | 100493.196 | 100492.657 | 5.4E-01 | 100516.321 | 100515.782 | 5.4E-01 | -85.032 | -85.577 | 5.4E-01 |
| 900 | Pd | 108 | 46 | 62 | 16 | 8.567025 | 8.571600 | -4.6E-03 | 100488.186 | 100487.663 | 5.2E-01 | 100511.829 | 100511.307 | 5.2E-01 | -89.524 | -90.052 | 5.3E-01 |
| 901 | Ag | 108 | 47 | 61 | 14 | 8.542025 | 8.546034 | -4.0E-03 | 100489.585 | 100489.123 | 4.6E-01 | 100513.747 | 100513.285 | 4.6E-01 | -87.607 | -88.074 | 4.7E-01 |
| 902 | Cd | 108 | 48 | 60 | 12 | 8.550020 | 8.549056 | 9.6E-04 | 100487.421 | 100487.495 | -7.4E-02 | 100512.101 | 100512.175 | -7.4E-02 | -89.253 | -89.183 | -6.9E-02 |
| 903 | In | 108 | 49 | 59 | 10 | 8.495252 | 8.488754 | 6.5E-03 | 100492.035 | 100492.706 | -6.7E-01 | 100517.233 | 100517.905 | -6.7E-01 | -84.120 | -83.454 | -6.7E-01 |
| 904 | Sn | 108 | 50 | 58 | 8 | 8.469027 | 8.462415 | 6.6E-03 | 100493.566 | 100494.249 | -6.8E-01 | 100519.283 | 100519.966 | -6.8E-01 | -82.070 | -81.392 | -6.8E-01 |
| 905 | Sb | 108 | 51 | 57 | 6 | 8.372666 | 8.368090 | 4.6E-03 | 100502.671 | 100503.133 | -4.6E-01 | 100528.908 | 100529.370 | -4.6E-01 | -72.445 | -71.988 | -4.6E-01 |
| 906 | Te | 108 | 52 | 56 | 4 | 8.303721 | 8.309446 | -5.7E-03 | 100508.815 | 100508.164 | 6.5E-01 | 100535.572 | 100534.921 | 6.5E-01 | -65.782 | -66.438 | 6.6E-01 |
| 907 | I | 108 | 53 | 55 | 2 | 8.174856 | 8.177737 | -2.9E-03 | 100521.431 | 100521.086 | 3.4E-01 | 100548.707 | 100548.362 | 3.4E-01 | -52.647 | -52.996 | 3.5E-01 |
| 908 | Nb | 109 | 41 | 68 | 27 | 8.296489 | 8.295004 | 1.5E-03 | 101455.173 | 101455.310 | -1.4E-01 | 101476.227 | 101476.365 | -1.4E-01 | -56.620 | -56.488 | -1.3E-01 |
| 909 | Mo | 109 | 42 | 67 | 25 | 8.381536 | 8.380076 | 1.5E-03 | 101444.603 | 101444.737 | -1.3E-01 | 101466.175 | 101466.309 | -1.3E-01 | -66.672 | -66.544 | -1.3E-01 |
| 910 | Tc | 109 | 43 | 66 | 23 | 8.444160 | 8.445759 | -1.6E-03 | 101436.477 | 101436.277 | 2.0E-01 | 101458.567 | 101458.366 | 2.0E-01 | -74.281 | -74.486 | 2.1E-01 |
| 911 | Ru | 109 | 44 | 65 | 21 | 8.496208 | 8.497678 | -1.5E-03 | 101429.504 | 101429.317 | 1.9E-01 | 101452.111 | 101451.924 | 1.9E-01 | -80.736 | -80.929 | 1.9E-01 |
| 912 | Rh | 109 | 45 | 64 | 19 | 8.528146 | 8.532626 | -4.5E-03 | 101424.722 | 101424.207 | 5.2E-01 | 101447.847 | 101447.332 | 5.2E-01 | -85.000 | -85.521 | 5.2E-01 |
| 913 | Pd | 109 | 46 | 63 | 17 | 8.544883 | 8.552475 | -7.6E-03 | 101421.598 | 101420.742 | 8.6E-01 | 101445.241 | 101444.385 | 8.6E-01 | -87.607 | -88.468 | 8.6E-01 |
| 914 | Ag | 109 | 47 | 62 | 15 | 8.547919 | 8.554992 | -7.1E-03 | 101419.966 | 101419.166 | 8.0E-01 | 101444.128 | 101443.328 | 8.0E-01 | -88.720 | -89.525 | 8.1E-01 |
| 915 | Cd | 109 | 48 | 61 | 13 | 8.538764 | 8.540502 | -1.7E-03 | 101419.663 | 101419.444 | 2.2E-01 | 101444.343 | 101444.124 | 2.2E-01 | -88.504 | -88.729 | 2.2E-01 |
| 916 | In | 109 | 49 | 60 | 11 | 8.513087 | 8.510646 | 2.4E-03 | 101421.161 | 101421.396 | -2.4E-01 | 101446.359 | 101446.595 | -2.4E-01 | -86.488 | -86.257 | -2.3E-01 |
| 917 | Sn | 109 | 50 | 59 | 9 | 8.470524 | 8.463654 | 6.9E-03 | 101424.499 | 101425.216 | -7.2E-01 | 101450.217 | 101450.934 | -7.2E-01 | -82.631 | -81.918 | -7.1E-01 |
| 918 | Sb | 109 | 51 | 58 | 7 | 8.404816 | 8.403349 | 1.5E-03 | 101430.359 | 101430.487 | -1.3E-01 | 101456.596 | 101456.724 | -1.3E-01 | -76.251 | -76.128 | -1.2E-01 |
| 919 | Te | 109 | 52 | 57 | 5 | 8.319330 | 8.322098 | -2.8E-03 | 101438.375 | 101438.041 | 3.3E-01 | 101465.132 | 101464.798 | 3.3E-01 | -67.715 | -68.055 | 3.4E-01 |
| 920 | I | 109 | 53 | 56 | 3 | 8.220022 | 8.226326 | -6.3E-03 | 101447.898 | 101447.177 | 7.2E-01 | 101475.174 | 101474.454 | 7.2E-01 | -57.673 | -58.399 | 7.3E-01 |
| 921 | Xe | 109 | 54 | 55 | 1 | 8.107307 | 8.106639 | 6.7E-04 | 101458.881 | 101458.920 | -3.9E-02 | 101486.678 | 101486.717 | -3.9E-02 | -46.170 | -46.136 | -3.4E-02 |
| 922 | Mo | 110 | 42 | 68 | 26 | 8.359413 | 8.356229 | 3.2E-03 | 102378.220 | 102378.545 | -3.3E-01 | 102399.792 | 102400.117 | -3.3E-01 | -64.549 | -64.229 | -3.2E-01 |
| 923 | Tc | 110 | 43 | 67 | 24 | 8.411240 | 8.411035 | 2.0E-04 | 102371.219 | 102371.216 | 3.0E-03 | 102393.309 | 102393.306 | 3.4E-03 | -71.032 | -71.041 | 8.7E-03 |
| 924 | Ru | 110 | 44 | 66 | 22 | 8.486293 | 8.485342 | 9.5E-04 | 102361.664 | 102361.742 | -7.8E-02 | 102384.271 | 102384.349 | -7.8E-02 | -80.071 | -79.998 | -7.3E-02 |
| 925 | Rh | 110 | 45 | 65 | 20 | 8.504257 | 8.507597 | -3.3E-03 | 102358.387 | 102357.993 | 3.9E-01 | 102381.512 | 102381.118 | 3.9E-01 | -82.829 | -83.229 | 4.0E-01 |
| 926 | Pd | 110 | 46 | 64 | 18 | 8.547168 | 8.552796 | -5.6E-03 | 102352.367 | 102351.720 | 6.5E-01 | 102376.010 | 102375.363 | 6.5E-01 | -88.332 | -88.984 | 6.5E-01 |
| 927 | Ag | 110 | 47 | 63 | 16 | 8.532112 | 8.540449 | -8.3E-03 | 102352.722 | 102351.776 | 9.5E-01 | 102376.884 | 102375.938 | 9.5E-01 | -87.458 | -88.409 | 9.5E-01 |
| 928 | Cd | 110 | 48 | 62 | 14 | 8.551282 | 8.554276 | -3.0E-03 | 102349.313 | 102348.954 | 3.6E-01 | 102373.993 | 102373.634 | 3.6E-01 | -90.349 | -90.713 | 3.6E-01 |
| 929 | In | 110 | 49 | 61 | 12 | 8.508915 | 8.506835 | 2.1E-03 | 102352.672 | 102352.870 | -2.0E-01 | 102377.871 | 102378.069 | -2.0E-01 | -86.471 | -86.278 | -1.9E-01 |
| 930 | Sn | 110 | 50 | 60 | 10 | 8.496087 | 8.490831 | 5.3E-03 | 102352.782 | 102353.329 | -5.5E-01 | 102378.499 | 102379.046 | -5.5E-01 | -85.842 | -85.300 | -5.4E-01 |
| 931 | Sb | 110 | 51 | 59 | 8 | 8.412681 | 8.410659 | 2.0E-03 | 102360.655 | 102360.845 | -1.9E-01 | 102386.892 | 102387.082 | -1.9E-01 | -77.450 | -77.264 | -1.9E-01 |
| 932 | Te | 110 | 52 | 58 | 6 | 8.358115 | 8.364738 | -6.6E-03 | 102365.355 | 102364.594 | 7.6E-01 | 102392.112 | 102391.351 | 7.6E-01 | -72.230 | -72.996 | 7.7E-01 |
| 933 | I | 110 | 53 | 57 | 4 | 8.244043 | 8.248003 | -4.0E-03 | 102376.601 | 102376.132 | 4.7E-01 | 102403.877 | 102403.408 | 4.7E-01 | -60.464 | -60.938 | 4.7E-01 |
| 934 | Xe | 110 | 54 | 56 | 2 | 8.159243 | 8.165577 | -6.3E-03 | 102384.627 | 102383.896 | 7.3E-01 | 102412.423 | 102411.692 | 7.3E-01 | -51.919 | -52.655 | 7.4E-01 |
| 935 | Mo | 111 | 42 | 69 | 27 | 8.315273 | 8.311842 | 3.4E-03 | 103314.326 | 103314.681 | -3.6E-01 | 103335.898 | 103336.253 | -3.6E-01 | -59.938 | -59.587 | -3.5E-01 |
| 936 | Tc | 111 | 43 | 68 | 25 | 8.390071 | 8.388817 | 1.3E-03 | 103304.723 | 103304.837 | -1.1E-01 | 103326.813 | 103326.926 | -1.1E-01 | -69.023 | -68.915 | -1.1E-01 |
| 937 | Ru | 111 | 44 | 67 | 23 | 8.452938 | 8.452121 | 8.2E-04 | 103296.445 | 103296.509 | -6.4E-02 | 103319.052 | 103319.116 | -6.4E-02 | -76.783 | -76.724 | -5.9E-02 |
| 938 | Rh | 111 | 45 | 66 | 21 | 8.495633 | 8.498362 | -2.7E-03 | 103290.406 | 103290.075 | 3.3E-01 | 103313.531 | 103313.201 | 3.3E-01 | -82.305 | -82.640 | 3.4E-01 |
| 939 | Pd | 111 | 46 | 65 | 19 | 8.521755 | 8.530230 | -8.5E-03 | 103286.206 | 103285.237 | 9.7E-01 | 103309.849 | 103308.880 | 9.7E-01 | -85.987 | -86.961 | 9.7E-01 |
| 940 | Ag | 111 | 47 | 64 | 17 | 8.534795 | 8.544615 | -9.8E-03 | 103283.458 | 103282.339 | 1.1E+00 | 103307.619 | 103306.500 | 1.1E+00 | -88.216 | -89.340 | 1.1E+00 |
| 941 | Cd | 111 | 48 | 63 | 15 | 8.537087 | 8.542151 | -5.1E-03 | 103281.902 | 103281.311 | 5.9E-01 | 103306.582 | 103305.991 | 5.9E-01 | -89.253 | -89.850 | 6.0E-01 |
| 942 | In | 111 | 49 | 62 | 13 | 8.522271 | 8.523494 | -1.2E-03 | 103282.246 | 103282.080 | 1.7E-01 | 103307.445 | 103307.279 | 1.7E-01 | -88.391 | -88.562 | 1.7E-01 |
| 943 | Sn | 111 | 50 | 61 | 11 | 8.493139 | 8.488269 | 4.9E-03 | 103284.178 | 103284.688 | -5.1E-01 | 103309.896 | 103310.405 | -5.1E-01 | -85.940 | -85.435 | -5.0E-01 |
| 944 | Sb | 111 | 51 | 60 | 9 | 8.440119 | 8.439594 | 5.3E-04 | 103288.762 | 103288.788 | -2.6E-02 | 103314.999 | 103315.025 | -2.6E-02 | -80.837 | -80.815 | -2.1E-02 |
| 945 | Te | 111 | 52 | 59 | 7 | 8.367763 | 8.372762 | -5.0E-03 | 103295.491 | 103294.904 | 5.9E-01 | 103322.248 | 103321.661 | 5.9E-01 | -73.587 | -74.180 | 5.9E-01 |
| 946 | I | 111 | 53 | 58 | 5 | 8.282934 | 8.291772 | -8.8E-03 | 103303.605 | 103302.591 | 1.0E+00 | 103330.882 | 103329.867 | 1.0E+00 | -64.954 | -65.973 | 1.0E+00 |
| 947 | Xe | 111 | 54 | 57 | 3 | 8.180739 | 8.187315 | -6.6E-03 | 103313.647 | 103312.883 | 7.6E-01 | 103341.443 | 103340.679 | 7.6E-01 | -54.393 | -55.162 | 7.7E-01 |
| 948 | Tc | 112 | 43 | 69 | 26 | 8.353586 | 8.351156 | 2.4E-03 | 104239.985 | 104240.231 | -2.5E-01 | 104262.074 | 104262.321 | -2.5E-01 | -65.255 | -65.014 | -2.4E-01 |
| 949 | Ru | 112 | 44 | 68 | 24 | 8.439223 | 8.435904 | 3.3E-03 | 104229.094 | 104229.439 | -3.4E-01 | 104251.701 | 104252.046 | -3.4E-01 | -75.629 | -75.289 | -3.4E-01 |
| 950 | Rh | 112 | 45 | 67 | 22 | 8.468882 | 8.470173 | -1.3E-03 | 104224.472 | 104224.300 | 1.7E-01 | 104247.597 | 104247.425 | 1.7E-01 | -79.733 | -79.910 | 1.8E-01 |
| 951 | Pd | 112 | 46 | 66 | 20 | 8.520727 | 8.526030 | -5.3E-03 | 104217.364 | 104216.743 | 6.2E-01 | 104241.008 | 104240.386 | 6.2E-01 | -86.322 | -86.949 | 6.3E-01 |
| 952 | Ag | 112 | 47 | 65 | 18 | 8.516080 | 8.526363 | -1.0E-02 | 104216.584 | 104215.404 | 1.2E+00 | 104240.746 | 104239.565 | 1.2E+00 | -86.584 | -87.770 | 1.2E+00 |
| 953 | Cd | 112 | 48 | 64 | 16 | 8.544738 | 8.550785 | -6.0E-03 | 104212.074 | 104211.367 | 7.1E-01 | 104236.754 | 104236.047 | 7.1E-01 | -90.576 | -91.288 | 7.1E-01 |
| 954 | In | 112 | 49 | 63 | 14 | 8.514675 | 8.515653 | -9.8E-04 | 104214.140 | 104214.000 | 1.4E-01 | 104239.338 | 104239.199 | 1.4E-01 | -87.991 | -88.136 | 1.5E-01 |
| 955 | Sn | 112 | 50 | 62 | 12 | 8.513627 | 8.509584 | 4.0E-03 | 104212.956 | 104213.378 | -4.2E-01 | 104238.673 | 104239.095 | -4.2E-01 | -88.656 | -88.239 | -4.2E-01 |
| 956 | Sb | 112 | 51 | 61 | 10 | 8.443632 | 8.442320 | 1.3E-03 | 104219.493 | 104219.609 | -1.2E-01 | 104245.730 | 104245.846 | -1.2E-01 | -81.599 | -81.489 | -1.1E-01 |
| 957 | Te | 112 | 52 | 60 | 8 | 8.400652 | 8.407979 | -7.3E-03 | 104223.005 | 104222.152 | 8.5E-01 | 104249.762 | 104248.909 | 8.5E-01 | -77.568 | -78.426 | 8.6E-01 |
| 958 | I | 112 | 53 | 59 | 6 | 8.299879 | 8.307862 | -8.0E-03 | 104232.990 | 104232.063 | 9.3E-01 | 104260.266 | 104259.339 | 9.3E-01 | -67.063 | -67.996 | 9.3E-01 |
| 959 | Xe | 112 | 54 | 58 | 4 | 8.230065 | 8.240240 | -1.0E-02 | 104239.507 | 104238.333 | 1.2E+00 | 104267.303 | 104266.130 | 1.2E+00 | -60.026 | -61.205 | 1.2E+00 |
| 960 | Cs | 112 | 55 | 57 | 2 | 8.100408 | 8.104171 | -3.8E-03 | 104252.726 | 104252.270 | 4.6E-01 | 104281.042 | 104280.586 | 4.6E-01 | -46.287 | -46.749 | 4.6E-01 |
| 961 | Tc | 113 | 43 | 70 | 27 | 8.329464 | 8.324838 | 4.6E-03 | 105173.922 | 105174.419 | -5.0E-01 | 105196.012 | 105196.509 | -5.0E-01 | -62.812 | -62.320 | -4.9E-01 |
| 962 | Ru | 113 | 44 | 69 | 25 | 8.402707 | 8.399506 | 3.2E-03 | 105164.346 | 105164.681 | -3.4E-01 | 105186.953 | 105187.288 | -3.4E-01 | -71.870 | -71.541 | -3.3E-01 |
| 963 | Rh | 113 | 45 | 68 | 23 | 8.456823 | 8.456781 | 4.2E-05 | 105156.931 | 105156.908 | 2.3E-02 | 105180.056 | 105180.033 | 2.3E-02 | -78.768 | -78.796 | 2.8E-02 |
| 964 | Pd | 113 | 46 | 67 | 21 | 8.492586 | 8.500243 | -7.7E-03 | 105151.589 | 105150.696 | 8.9E-01 | 105175.232 | 105174.339 | 8.9E-01 | -83.591 | -84.490 | 9.0E-01 |
| 965 | Ag | 113 | 47 | 66 | 19 | 8.516065 | 8.526154 | -1.0E-02 | 105147.635 | 105146.466 | 1.2E+00 | 105171.797 | 105170.628 | 1.2E+00 | -87.027 | -88.201 | 1.2E+00 |
| 966 | Cd | 113 | 48 | 65 | 17 | 8.526986 | 8.535318 | -8.3E-03 | 105145.100 | 105144.129 | 9.7E-01 | 105169.780 | 105168.809 | 9.7E-01 | -89.043 | -90.020 | 9.8E-01 |
| 967 | In | 113 | 49 | 64 | 15 | 8.522917 | 8.527533 | -4.6E-03 | 105144.259 | 105143.707 | 5.5E-01 | 105169.458 | 105168.906 | 5.5E-01 | -89.366 | -89.923 | 5.6E-01 |
| 968 | Sn | 113 | 50 | 63 | 13 | 8.506812 | 8.503460 | 3.4E-03 | 105144.777 | 105145.125 | -3.5E-01 | 105170.495 | 105170.843 | -3.5E-01 | -88.328 | -87.986 | -3.4E-01 |
| 969 | Sb | 113 | 51 | 62 | 11 | 8.465276 | 8.465556 | -2.8E-04 | 105148.169 | 105148.106 | 6.3E-02 | 105174.406 | 105174.343 | 6.3E-02 | -84.417 | -84.486 | 6.9E-02 |
| 970 | Te | 113 | 52 | 61 | 9 | 8.404636 | 8.411368 | -6.7E-03 | 105153.720 | 105152.927 | 7.9E-01 | 105180.476 | 105179.683 | 7.9E-01 | -78.347 | -79.145 | 8.0E-01 |
| 971 | I | 113 | 53 | 60 | 7 | 8.333752 | 8.344334 | -1.1E-02 | 105160.428 | 105159.199 | 1.2E+00 | 105187.704 | 105186.475 | 1.2E+00 | -71.120 | -72.354 | 1.2E+00 |
| 972 | Xe | 113 | 54 | 59 | 5 | 8.247927 | 8.256786 | -8.9E-03 | 105168.824 | 105167.789 | 1.0E+00 | 105196.620 | 105195.585 | 1.0E+00 | -62.204 | -63.244 | 1.0E+00 |
| 973 | Cs | 113 | 55 | 58 | 3 | 8.148617 | 8.155462 | -6.8E-03 | 105178.743 | 105177.935 | 8.1E-01 | 105207.060 | 105206.251 | 8.1E-01 | -51.764 | -52.577 | 8.1E-01 |
| 974 | Ru | 114 | 44 | 70 | 26 | 8.385341 | 8.379629 | 5.7E-03 | 106097.488 | 106098.113 | -6.3E-01 | 106120.095 | 106120.720 | -6.3E-01 | -70.222 | -69.603 | -6.2E-01 |
| 975 | Rh | 114 | 45 | 69 | 24 | 8.426649 | 8.425632 | 1.0E-03 | 106091.479 | 106091.568 | -8.9E-02 | 106114.604 | 106114.693 | -8.9E-02 | -75.713 | -75.630 | -8.3E-02 |
| 976 | Pd | 114 | 46 | 68 | 22 | 8.488012 | 8.491955 | -3.9E-03 | 106083.183 | 106082.706 | 4.8E-01 | 106106.826 | 106106.349 | 4.8E-01 | -83.491 | -83.974 | 4.8E-01 |
| 977 | Ag | 114 | 47 | 67 | 20 | 8.493778 | 8.504512 | -1.1E-02 | 106081.225 | 106079.973 | 1.3E+00 | 106105.387 | 106104.134 | 1.3E+00 | -84.931 | -86.189 | 1.3E+00 |
| 978 | Cd | 114 | 48 | 66 | 18 | 8.531512 | 8.539333 | -7.8E-03 | 106075.623 | 106074.702 | 9.2E-01 | 106100.303 | 106099.382 | 9.2E-01 | -90.015 | -90.941 | 9.3E-01 |
| 979 | In | 114 | 49 | 65 | 16 | 8.511961 | 8.515936 | -4.0E-03 | 106076.550 | 106076.067 | 4.8E-01 | 106101.749 | 106101.266 | 4.8E-01 | -88.568 | -89.057 | 4.9E-01 |
| 980 | Sn | 114 | 50 | 64 | 14 | 8.522545 | 8.519439 | 3.1E-03 | 106074.042 | 106074.366 | -3.2E-01 | 106099.760 | 106100.083 | -3.2E-01 | -90.557 | -90.239 | -3.2E-01 |
| 981 | Sb | 114 | 51 | 63 | 12 | 8.462510 | 8.464120 | -1.6E-03 | 106079.585 | 106079.370 | 2.2E-01 | 106105.822 | 106105.607 | 2.2E-01 | -84.496 | -84.716 | 2.2E-01 |
| 982 | Te | 114 | 52 | 62 | 10 | 8.432778 | 8.439960 | -7.2E-03 | 106081.672 | 106080.821 | 8.5E-01 | 106108.429 | 106107.578 | 8.5E-01 | -81.889 | -82.745 | 8.6E-01 |
| 983 | Xe | 114 | 54 | 60 | 6 | 8.289205 | 8.301424 | -1.2E-02 | 106095.435 | 106094.009 | 1.4E+00 | 106123.232 | 106121.805 | 1.4E+00 | -67.086 | -68.518 | 1.4E+00 |
| 984 | Cs | 114 | 55 | 59 | 4 | 8.173538 | 8.181732 | -8.2E-03 | 106107.319 | 106106.350 | 9.7E-01 | 106135.635 | 106134.667 | 9.7E-01 | -54.682 | -55.656 | 9.7E-01 |
| 985 | Ba | 114 | 56 | 58 | 2 | 8.090160 | 8.095315 | -5.2E-03 | 106115.521 | 106114.898 | 6.2E-01 | 106144.358 | 106143.735 | 6.2E-01 | -45.960 | -46.588 | 6.3E-01 |
| 986 | Ru | 115 | 44 | 71 | 27 | 8.348540 | 8.339990 | 8.5E-03 | 107032.900 | 107033.857 | -9.6E-01 | 107055.508 | 107056.464 | -9.6E-01 | -66.304 | -65.353 | -9.5E-01 |
| 987 | Rh | 115 | 45 | 70 | 25 | 8.410654 | 8.408244 | 2.4E-03 | 107024.457 | 107024.707 | -2.5E-01 | 107047.582 | 107047.832 | -2.5E-01 | -74.229 | -73.985 | -2.4E-01 |
| 988 | Pd | 115 | 46 | 69 | 23 | 8.457740 | 8.463065 | -5.3E-03 | 107017.742 | 107017.102 | 6.4E-01 | 107041.385 | 107040.745 | 6.4E-01 | -80.427 | -81.072 | 6.5E-01 |
| 989 | Ag | 115 | 47 | 68 | 21 | 8.490555 | 8.500250 | -9.7E-03 | 107012.667 | 107011.524 | 1.1E+00 | 107036.829 | 107035.685 | 1.1E+00 | -84.983 | -86.132 | 1.1E+00 |
| 990 | Cd | 115 | 48 | 67 | 19 | 8.510724 | 8.520744 | -1.0E-02 | 107009.047 | 107007.865 | 1.2E+00 | 107033.727 | 107032.545 | 1.2E+00 | -88.084 | -89.272 | 1.2E+00 |
| 991 | In | 115 | 49 | 66 | 17 | 8.516546 | 8.523511 | -7.0E-03 | 107007.076 | 107006.245 | 8.3E-01 | 107032.275 | 107031.444 | 8.3E-01 | -89.536 | -90.373 | 8.4E-01 |
| 992 | Sn | 115 | 50 | 65 | 15 | 8.514069 | 8.510003 | 4.1E-03 | 107006.060 | 107006.497 | -4.4E-01 | 107031.778 | 107032.215 | -4.4E-01 | -90.034 | -89.602 | -4.3E-01 |
| 993 | Sb | 115 | 51 | 64 | 13 | 8.480915 | 8.482273 | -1.4E-03 | 107008.571 | 107008.383 | 1.9E-01 | 107034.808 | 107034.620 | 1.9E-01 | -87.003 | -87.197 | 1.9E-01 |
| 994 | Te | 115 | 52 | 63 | 11 | 8.431150 | 8.439430 | -8.3E-03 | 107012.992 | 107012.008 | 9.8E-01 | 107039.749 | 107038.764 | 9.8E-01 | -82.063 | -83.053 | 9.9E-01 |
| 995 | I | 115 | 53 | 62 | 9 | 8.374564 | 8.384244 | -9.7E-03 | 107018.197 | 107017.051 | 1.1E+00 | 107045.474 | 107044.328 | 1.1E+00 | -76.338 | -77.490 | 1.2E+00 |
| 996 | Xe | 115 | 54 | 61 | 7 | 8.300970 | 8.311926 | -1.1E-02 | 107025.358 | 107024.065 | 1.3E+00 | 107053.155 | 107051.861 | 1.3E+00 | -68.657 | -69.956 | 1.3E+00 |
| 997 | Ru | 116 | 44 | 72 | 28 | 8.326883 | 8.316609 | 1.0E-02 | 107966.629 | 107967.795 | -1.2E+00 | 107989.237 | 107990.402 | -1.2E+00 | -64.069 | -62.909 | -1.2E+00 |
| 998 | Rh | 116 | 45 | 71 | 26 | 8.377640 | 8.374226 | 3.4E-03 | 107959.441 | 107959.810 | -3.7E-01 | 107982.566 | 107982.935 | -3.7E-01 | -70.739 | -70.376 | -3.6E-01 |
| 999 | Pd | 116 | 46 | 70 | 24 | 8.449282 | 8.450999 | -1.7E-03 | 107949.830 | 107949.603 | 2.3E-01 | 107973.474 | 107973.247 | 2.3E-01 | -79.832 | -80.064 | 2.3E-01 |
| 1000 | Ag | 116 | 47 | 69 | 22 | 8.465906 | 8.475458 | -9.6E-03 | 107946.601 | 107945.465 | 1.1E+00 | 107970.763 | 107969.626 | 1.1E+00 | -82.543 | -83.685 | 1.1E+00 |
| 1001 | Cd | 116 | 48 | 68 | 20 | 8.512351 | 8.520588 | -8.2E-03 | 107939.913 | 107938.928 | 9.8E-01 | 107964.593 | 107963.608 | 9.8E-01 | -88.713 | -89.703 | 9.9E-01 |
| 1002 | In | 116 | 49 | 67 | 18 | 8.501617 | 8.508452 | -6.8E-03 | 107939.857 | 107939.034 | 8.2E-01 | 107965.056 | 107964.233 | 8.2E-01 | -88.250 | -89.078 | 8.3E-01 |
| 1003 | Sn | 116 | 50 | 66 | 16 | 8.523116 | 8.521170 | 1.9E-03 | 107936.062 | 107936.257 | -1.9E-01 | 107961.780 | 107961.974 | -1.9E-01 | -91.526 | -91.337 | -1.9E-01 |
| 1004 | Sb | 116 | 51 | 65 | 14 | 8.475817 | 8.476985 | -1.2E-03 | 107940.247 | 107940.080 | 1.7E-01 | 107966.484 | 107966.317 | 1.7E-01 | -86.822 | -86.994 | 1.7E-01 |
| 1005 | Te | 116 | 52 | 64 | 12 | 8.455687 | 8.462234 | -6.5E-03 | 107941.280 | 107940.488 | 7.9E-01 | 107968.037 | 107967.245 | 7.9E-01 | -85.269 | -86.066 | 8.0E-01 |
| 1006 | I | 116 | 53 | 63 | 10 | 8.381902 | 8.389574 | -7.7E-03 | 107948.537 | 107947.614 | 9.2E-01 | 107975.813 | 107974.890 | 9.2E-01 | -77.492 | -78.421 | 9.3E-01 |
| 1007 | Xe | 116 | 54 | 62 | 8 | 8.336834 | 8.348490 | -1.2E-02 | 107952.462 | 107951.077 | 1.4E+00 | 107980.259 | 107978.873 | 1.4E+00 | -73.047 | -74.438 | 1.4E+00 |
| 1008 | Ru | 117 | 44 | 73 | 29 | 8.285816 | 8.273743 | 1.2E-02 | 108902.673 | 108904.059 | -1.4E+00 | 108925.280 | 108926.666 | -1.4E+00 | -59.520 | -58.139 | -1.4E+00 |
| 1009 | Rh | 117 | 45 | 72 | 27 | 8.359283 | 8.352941 | 6.3E-03 | 108892.777 | 108893.492 | -7.1E-01 | 108915.902 | 108916.617 | -7.1E-01 | -68.898 | -68.188 | -7.1E-01 |
| 1010 | Pd | 117 | 46 | 71 | 25 | 8.416930 | 8.419033 | -2.1E-03 | 108884.732 | 108884.458 | 2.7E-01 | 108908.375 | 108908.101 | 2.7E-01 | -76.425 | -76.704 | 2.8E-01 |
| 1011 | Ag | 117 | 47 | 70 | 23 | 8.459452 | 8.467356 | -7.9E-03 | 108878.456 | 108877.503 | 9.5E-01 | 108902.617 | 108901.664 | 9.5E-01 | -82.182 | -83.141 | 9.6E-01 |
| 1012 | Cd | 117 | 48 | 69 | 21 | 8.488974 | 8.499015 | -1.0E-02 | 108873.701 | 108872.497 | 1.2E+00 | 108898.381 | 108897.177 | 1.2E+00 | -86.418 | -87.628 | 1.2E+00 |
| 1013 | In | 117 | 49 | 68 | 19 | 8.503865 | 8.512126 | -8.3E-03 | 108870.658 | 108869.661 | 1.0E+00 | 108895.856 | 108894.860 | 1.0E+00 | -88.943 | -89.945 | 1.0E+00 |
| 1014 | Sn | 117 | 50 | 67 | 17 | 8.509612 | 8.508663 | 9.5E-04 | 108868.684 | 108868.764 | -8.0E-02 | 108894.402 | 108894.482 | -8.0E-02 | -90.398 | -90.323 | -7.5E-02 |
| 1015 | Sb | 117 | 51 | 66 | 15 | 8.487897 | 8.490556 | -2.7E-03 | 108869.923 | 108869.580 | 3.4E-01 | 108896.160 | 108895.818 | 3.4E-01 | -88.640 | -88.988 | 3.5E-01 |
| 1016 | Te | 117 | 52 | 65 | 13 | 8.450919 | 8.458207 | -7.3E-03 | 108872.948 | 108872.063 | 8.9E-01 | 108899.704 | 108898.819 | 8.8E-01 | -85.095 | -85.986 | 8.9E-01 |
| 1017 | I | 117 | 53 | 64 | 11 | 8.404409 | 8.413616 | -9.2E-03 | 108877.087 | 108875.977 | 1.1E+00 | 108904.363 | 108903.253 | 1.1E+00 | -80.436 | -81.552 | 1.1E+00 |
| 1018 | Xe | 117 | 54 | 63 | 9 | 8.344297 | 8.354004 | -9.7E-03 | 108882.818 | 108881.649 | 1.2E+00 | 108910.614 | 108909.445 | 1.2E+00 | -74.185 | -75.360 | 1.2E+00 |
| 1019 | Cs | 117 | 55 | 62 | 7 | 8.271864 | 8.282592 | -1.1E-02 | 108889.990 | 108888.700 | 1.3E+00 | 108918.306 | 108917.017 | 1.3E+00 | -66.493 | -67.788 | 1.3E+00 |
| 1020 | Ba | 117 | 56 | 61 | 5 | 8.189354 | 8.192839 | -3.5E-03 | 108898.341 | 108897.898 | 4.4E-01 | 108927.178 | 108926.735 | 4.4E-01 | -57.622 | -58.070 | 4.5E-01 |
| 1021 | Rh | 118 | 45 | 73 | 28 | 8.322860 | 8.316178 | 6.7E-03 | 109828.281 | 109829.042 | -7.6E-01 | 109851.406 | 109852.167 | -7.6E-01 | -64.888 | -64.132 | -7.6E-01 |
| 1022 | Pd | 118 | 46 | 72 | 26 | 8.405223 | 8.403431 | 1.8E-03 | 109817.261 | 109817.445 | -1.8E-01 | 109840.905 | 109841.088 | -1.8E-01 | -75.389 | -75.211 | -1.8E-01 |
| 1023 | Ag | 118 | 47 | 71 | 24 | 8.433889 | 8.439576 | -5.7E-03 | 109812.578 | 109811.879 | 7.0E-01 | 109836.740 | 109836.040 | 7.0E-01 | -79.554 | -80.259 | 7.1E-01 |
| 1024 | Cd | 118 | 48 | 70 | 22 | 8.487835 | 8.495035 | -7.2E-03 | 109804.912 | 109804.033 | 8.8E-01 | 109829.592 | 109828.713 | 8.8E-01 | -86.702 | -87.586 | 8.8E-01 |
| 1025 | In | 118 | 49 | 69 | 20 | 8.485667 | 8.493846 | -8.2E-03 | 109803.866 | 109802.871 | 9.9E-01 | 109829.065 | 109828.070 | 9.9E-01 | -87.228 | -88.229 | 1.0E+00 |
| 1026 | Sn | 118 | 50 | 68 | 18 | 8.516534 | 8.515511 | 1.0E-03 | 109798.923 | 109799.013 | -9.0E-02 | 109824.641 | 109824.731 | -9.0E-02 | -91.653 | -91.568 | -8.4E-02 |
| 1027 | Sb | 118 | 51 | 67 | 16 | 8.478915 | 8.481694 | -2.8E-03 | 109802.060 | 109801.701 | 3.6E-01 | 109828.297 | 109827.938 | 3.6E-01 | -87.996 | -88.361 | 3.6E-01 |
| 1028 | Te | 118 | 52 | 66 | 14 | 8.469747 | 8.475791 | -6.0E-03 | 109801.840 | 109801.095 | 7.5E-01 | 109828.597 | 109827.852 | 7.5E-01 | -87.697 | -88.448 | 7.5E-01 |
| 1029 | I | 118 | 53 | 65 | 12 | 8.406120 | 8.414776 | -8.7E-03 | 109808.046 | 109806.992 | 1.1E+00 | 109835.322 | 109834.268 | 1.1E+00 | -80.971 | -82.031 | 1.1E+00 |
| 1030 | Xe | 118 | 54 | 64 | 10 | 8.374981 | 8.383916 | -8.9E-03 | 109810.418 | 109809.330 | 1.1E+00 | 109838.214 | 109837.127 | 1.1E+00 | -78.079 | -79.173 | 1.1E+00 |
| 1031 | Cs | 118 | 55 | 63 | 8 | 8.286404 | 8.295500 | -9.1E-03 | 109819.568 | 109818.460 | 1.1E+00 | 109847.884 | 109846.777 | 1.1E+00 | -68.409 | -69.523 | 1.1E+00 |
| 1032 | Rh | 119 | 45 | 74 | 29 | 8.303394 | 8.291314 | 1.2E-02 | 110761.840 | 110763.250 | -1.4E+00 | 110784.965 | 110786.375 | -1.4E+00 | -62.823 | -61.418 | -1.4E+00 |
| 1033 | Pd | 119 | 46 | 73 | 27 | 8.368966 | 8.368452 | 5.1E-04 | 110752.736 | 110752.770 | -3.4E-02 | 110776.379 | 110776.413 | -3.4E-02 | -71.408 | -71.380 | -2.8E-02 |
| 1034 | Ag | 119 | 47 | 72 | 25 | 8.423212 | 8.427787 | -4.6E-03 | 110744.980 | 110744.407 | 5.7E-01 | 110769.142 | 110768.569 | 5.7E-01 | -78.646 | -79.224 | 5.8E-01 |
| 1035 | Cd | 119 | 48 | 71 | 23 | 8.461439 | 8.470545 | -9.1E-03 | 110739.130 | 110738.018 | 1.1E+00 | 110763.810 | 110762.698 | 1.1E+00 | -83.977 | -85.096 | 1.1E+00 |
| 1036 | In | 119 | 49 | 70 | 21 | 8.486145 | 8.493917 | -7.8E-03 | 110734.889 | 110733.934 | 9.5E-01 | 110760.088 | 110759.133 | 9.5E-01 | -87.700 | -88.660 | 9.6E-01 |
| 1037 | Sn | 119 | 50 | 69 | 19 | 8.499449 | 8.500128 | -6.8E-04 | 110732.005 | 110731.893 | 1.1E-01 | 110757.722 | 110757.611 | 1.1E-01 | -90.065 | -90.182 | 1.2E-01 |
| 1038 | Sb | 119 | 51 | 68 | 17 | 8.487910 | 8.491127 | -3.2E-03 | 110732.076 | 110731.662 | 4.1E-01 | 110758.313 | 110757.899 | 4.1E-01 | -89.474 | -89.894 | 4.2E-01 |
| 1039 | Te | 119 | 52 | 67 | 15 | 8.462067 | 8.468522 | -6.5E-03 | 110733.850 | 110733.049 | 8.0E-01 | 110760.606 | 110759.806 | 8.0E-01 | -87.181 | -87.987 | 8.1E-01 |
| 1040 | I | 119 | 53 | 66 | 13 | 8.426789 | 8.433793 | -7.0E-03 | 110736.746 | 110735.879 | 8.7E-01 | 110764.022 | 110763.156 | 8.7E-01 | -83.766 | -84.638 | 8.7E-01 |
| 1041 | Xe | 119 | 54 | 65 | 11 | 8.378441 | 8.385465 | -7.0E-03 | 110741.197 | 110740.327 | 8.7E-01 | 110768.993 | 110768.124 | 8.7E-01 | -78.794 | -79.669 | 8.8E-01 |
| 1042 | Cs | 119 | 55 | 64 | 9 | 8.317334 | 8.325743 | -8.4E-03 | 110747.166 | 110746.131 | 1.0E+00 | 110775.482 | 110774.448 | 1.0E+00 | -72.305 | -73.346 | 1.0E+00 |
| 1043 | Ba | 119 | 56 | 63 | 7 | 8.245928 | 8.250550 | -4.6E-03 | 110754.360 | 110753.775 | 5.8E-01 | 110783.197 | 110782.613 | 5.8E-01 | -64.590 | -65.181 | 5.9E-01 |
| 1044 | Pd | 120 | 46 | 74 | 28 | 8.357086 | 8.349740 | 7.3E-03 | 111685.358 | 111686.212 | -8.5E-01 | 111709.001 | 111709.855 | -8.5E-01 | -70.280 | -69.432 | -8.5E-01 |
| 1045 | Ag | 120 | 47 | 73 | 26 | 8.395327 | 8.397224 | -1.9E-03 | 111679.469 | 111679.213 | 2.6E-01 | 111703.630 | 111703.374 | 2.6E-01 | -75.652 | -75.913 | 2.6E-01 |
| 1046 | Cd | 120 | 48 | 72 | 24 | 8.458023 | 8.463035 | -5.0E-03 | 111670.644 | 111670.014 | 6.3E-01 | 111695.324 | 111694.694 | 6.3E-01 | -83.957 | -84.594 | 6.4E-01 |
| 1047 | In | 120 | 49 | 71 | 22 | 8.466264 | 8.472607 | -6.3E-03 | 111668.354 | 111667.563 | 7.9E-01 | 111693.553 | 111692.762 | 7.9E-01 | -85.729 | -86.525 | 8.0E-01 |
| 1048 | Sn | 120 | 50 | 70 | 20 | 8.504494 | 8.503083 | 1.4E-03 | 111662.465 | 111662.604 | -1.4E-01 | 111688.183 | 111688.322 | -1.4E-01 | -91.099 | -90.966 | -1.3E-01 |
| 1049 | Sb | 120 | 51 | 69 | 18 | 8.475636 | 8.478960 | -3.3E-03 | 111664.626 | 111664.196 | 4.3E-01 | 111690.864 | 111690.433 | 4.3E-01 | -88.418 | -88.854 | 4.4E-01 |
| 1050 | Te | 120 | 52 | 68 | 16 | 8.477035 | 8.481369 | -4.3E-03 | 111663.157 | 111662.605 | 5.5E-01 | 111689.913 | 111689.361 | 5.5E-01 | -89.368 | -89.926 | 5.6E-01 |
| 1051 | I | 120 | 53 | 67 | 14 | 8.423724 | 8.431141 | -7.4E-03 | 111668.252 | 111667.329 | 9.2E-01 | 111695.528 | 111694.606 | 9.2E-01 | -83.753 | -84.682 | 9.3E-01 |
| 1052 | Xe | 120 | 54 | 66 | 12 | 8.404031 | 8.409643 | -5.6E-03 | 111669.313 | 111668.606 | 7.1E-01 | 111697.109 | 111696.402 | 7.1E-01 | -82.172 | -82.885 | 7.1E-01 |
| 1053 | Cs | 120 | 55 | 65 | 10 | 8.328480 | 8.333736 | -5.3E-03 | 111677.076 | 111676.412 | 6.6E-01 | 111705.393 | 111704.728 | 6.6E-01 | -73.889 | -74.559 | 6.7E-01 |
| 1054 | Ba | 120 | 56 | 64 | 8 | 8.280294 | 8.288057 | -7.8E-03 | 111681.556 | 111680.589 | 9.7E-01 | 111710.393 | 111709.426 | 9.7E-01 | -68.889 | -69.861 | 9.7E-01 |
| 1055 | Pd | 121 | 46 | 75 | 29 | 8.320858 | 8.312316 | 8.5E-03 | 112620.950 | 112621.956 | -1.0E+00 | 112644.593 | 112645.599 | -1.0E+00 | -66.182 | -65.182 | -1.0E+00 |
| 1056 | Ag | 121 | 47 | 74 | 27 | 8.382330 | 8.382091 | 2.4E-04 | 112612.211 | 112612.212 | -8.6E-04 | 112636.373 | 112636.373 | -3.5E-04 | -74.403 | -74.408 | 5.2E-03 |
| 1057 | Cd | 121 | 48 | 73 | 25 | 8.430996 | 8.435740 | -4.7E-03 | 112605.022 | 112604.419 | 6.0E-01 | 112629.702 | 112629.099 | 6.0E-01 | -81.074 | -81.683 | 6.1E-01 |
| 1058 | In | 121 | 49 | 72 | 23 | 8.463889 | 8.469314 | -5.4E-03 | 112599.741 | 112599.054 | 6.9E-01 | 112624.939 | 112624.253 | 6.9E-01 | -85.836 | -86.528 | 6.9E-01 |
| 1059 | Sn | 121 | 50 | 71 | 21 | 8.485203 | 8.484974 | 2.3E-04 | 112595.860 | 112595.857 | 2.6E-03 | 112621.578 | 112621.575 | 2.8E-03 | -89.197 | -89.206 | 8.7E-03 |
| 1060 | Sb | 121 | 51 | 70 | 19 | 8.482052 | 8.484652 | -2.6E-03 | 112594.940 | 112594.594 | 3.5E-01 | 112621.177 | 112620.831 | 3.5E-01 | -89.599 | -89.950 | 3.5E-01 |
| 1061 | Te | 121 | 52 | 69 | 17 | 8.466873 | 8.471085 | -4.2E-03 | 112595.475 | 112594.933 | 5.4E-01 | 112622.231 | 112621.690 | 5.4E-01 | -88.544 | -89.092 | 5.5E-01 |
| 1062 | I | 121 | 53 | 68 | 15 | 8.441460 | 8.445574 | -4.1E-03 | 112597.248 | 112596.717 | 5.3E-01 | 112624.524 | 112623.993 | 5.3E-01 | -86.252 | -86.788 | 5.4E-01 |
| 1063 | Xe | 121 | 54 | 67 | 13 | 8.403832 | 8.407679 | -3.8E-03 | 112600.498 | 112599.999 | 5.0E-01 | 112628.295 | 112627.796 | 5.0E-01 | -82.481 | -82.986 | 5.0E-01 |
| 1064 | Cs | 121 | 55 | 66 | 11 | 8.352914 | 8.358464 | -5.6E-03 | 112605.357 | 112604.651 | 7.1E-01 | 112633.673 | 112632.968 | 7.1E-01 | -77.102 | -77.814 | 7.1E-01 |
| 1065 | Ba | 121 | 56 | 65 | 9 | 8.293907 | 8.295484 | -1.6E-03 | 112611.194 | 112610.968 | 2.3E-01 | 112640.031 | 112639.805 | 2.3E-01 | -70.745 | -70.976 | 2.3E-01 |
| 1066 | Pd | 122 | 46 | 76 | 30 | 8.305975 | 8.291826 | 1.4E-02 | 113554.010 | 113555.709 | -1.7E+00 | 113577.653 | 113579.352 | -1.7E+00 | -64.616 | -62.923 | -1.7E+00 |
| 1067 | Ag | 122 | 47 | 75 | 28 | 8.352758 | 8.349418 | 3.3E-03 | 113547.002 | 113547.381 | -3.8E-01 | 113571.163 | 113571.543 | -3.8E-01 | -71.106 | -70.733 | -3.7E-01 |
| 1068 | Cd | 122 | 48 | 74 | 26 | 8.424266 | 8.425171 | -9.0E-04 | 113536.977 | 113536.838 | 1.4E-01 | 113561.657 | 113561.518 | 1.4E-01 | -80.612 | -80.758 | 1.5E-01 |
| 1069 | In | 122 | 49 | 73 | 24 | 8.442120 | 8.445216 | -3.1E-03 | 113533.498 | 113533.090 | 4.1E-01 | 113558.697 | 113558.289 | 4.1E-01 | -83.573 | -83.986 | 4.1E-01 |
| 1070 | Sn | 122 | 50 | 72 | 22 | 8.487909 | 8.484427 | 3.5E-03 | 113526.610 | 113527.005 | -3.9E-01 | 113552.328 | 113552.722 | -3.9E-01 | -89.942 | -89.553 | -3.9E-01 |
| 1071 | Sb | 122 | 51 | 71 | 20 | 8.468317 | 8.469447 | -1.1E-03 | 113527.699 | 113527.530 | 1.7E-01 | 113553.936 | 113553.767 | 1.7E-01 | -88.334 | -88.509 | 1.7E-01 |
| 1072 | Te | 122 | 52 | 70 | 18 | 8.478140 | 8.479668 | -1.5E-03 | 113525.199 | 113524.980 | 2.2E-01 | 113551.955 | 113551.737 | 2.2E-01 | -90.314 | -90.539 | 2.2E-01 |
| 1073 | I | 122 | 53 | 69 | 16 | 8.437023 | 8.439395 | -2.4E-03 | 113528.913 | 113528.591 | 3.2E-01 | 113556.189 | 113555.867 | 3.2E-01 | -86.080 | -86.408 | 3.3E-01 |
| 1074 | Xe | 122 | 54 | 68 | 14 | 8.424664 | 8.426658 | -2.0E-03 | 113529.118 | 113528.842 | 2.8E-01 | 113556.915 | 113556.638 | 2.8E-01 | -85.355 | -85.637 | 2.8E-01 |
| 1075 | Cs | 122 | 55 | 67 | 12 | 8.359151 | 8.362286 | -3.1E-03 | 113535.808 | 113535.392 | 4.2E-01 | 113564.125 | 113563.708 | 4.2E-01 | -78.145 | -78.567 | 4.2E-01 |
| 1076 | Ba | 122 | 56 | 66 | 10 | 8.323756 | 8.326580 | -2.8E-03 | 113538.824 | 113538.444 | 3.8E-01 | 113567.661 | 113567.281 | 3.8E-01 | -74.609 | -74.994 | 3.9E-01 |
| 1077 | Ag | 123 | 47 | 76 | 29 | 8.337803 | 8.332141 | 5.7E-03 | 114480.054 | 114480.722 | -6.7E-01 | 114504.216 | 114504.884 | -6.7E-01 | -69.548 | -68.886 | -6.6E-01 |
| 1078 | Cd | 123 | 48 | 75 | 27 | 8.395395 | 8.395640 | -2.5E-04 | 114471.669 | 114471.610 | 5.9E-02 | 114496.349 | 114496.290 | 5.9E-02 | -77.414 | -77.479 | 6.5E-02 |
| 1079 | In | 123 | 49 | 74 | 25 | 8.437947 | 8.438960 | -1.0E-03 | 114465.134 | 114464.980 | 1.5E-01 | 114490.333 | 114490.179 | 1.5E-01 | -83.430 | -83.591 | 1.6E-01 |
| 1080 | Sn | 123 | 50 | 73 | 23 | 8.467244 | 8.463789 | 3.5E-03 | 114460.229 | 114460.624 | -4.0E-01 | 114485.947 | 114486.342 | -3.9E-01 | -87.816 | -87.428 | -3.9E-01 |
| 1081 | Sb | 123 | 51 | 72 | 21 | 8.472335 | 8.471773 | 5.6E-04 | 114458.302 | 114458.340 | -3.8E-02 | 114484.539 | 114484.577 | -3.8E-02 | -89.225 | -89.193 | -3.2E-02 |
| 1082 | Te | 123 | 52 | 71 | 19 | 8.465546 | 8.466603 | -1.1E-03 | 114457.835 | 114457.673 | 1.6E-01 | 114484.591 | 114484.430 | 1.6E-01 | -89.172 | -89.340 | 1.7E-01 |
| 1083 | I | 123 | 53 | 70 | 17 | 8.449198 | 8.449656 | -4.6E-04 | 114458.544 | 114458.455 | 8.9E-02 | 114485.820 | 114485.731 | 8.9E-02 | -87.944 | -88.039 | 9.5E-02 |
| 1084 | Xe | 123 | 54 | 69 | 15 | 8.420927 | 8.421435 | -5.1E-04 | 114460.719 | 114460.623 | 9.6E-02 | 114488.515 | 114488.419 | 9.6E-02 | -85.249 | -85.350 | 1.0E-01 |
| 1085 | Cs | 123 | 55 | 68 | 13 | 8.380379 | 8.382064 | -1.7E-03 | 114464.403 | 114464.162 | 2.4E-01 | 114492.720 | 114492.479 | 2.4E-01 | -81.044 | -81.291 | 2.5E-01 |
| 1086 | Ba | 123 | 56 | 67 | 11 | 8.330208 | 8.330139 | 6.9E-05 | 114469.272 | 114469.245 | 2.7E-02 | 114498.109 | 114498.082 | 2.7E-02 | -75.655 | -75.687 | 3.2E-02 |
| 1087 | Ag | 124 | 47 | 77 | 30 | 8.308655 | 8.299317 | 9.3E-03 | 115414.896 | 115416.026 | -1.1E+00 | 115439.057 | 115440.187 | -1.1E+00 | -66.200 | -65.076 | -1.1E+00 |
| 1088 | Cd | 124 | 48 | 76 | 28 | 8.387035 | 8.383290 | 3.7E-03 | 115403.876 | 115404.311 | -4.4E-01 | 115428.556 | 115428.991 | -4.4E-01 | -76.702 | -76.272 | -4.3E-01 |
| 1089 | In | 124 | 49 | 75 | 26 | 8.414343 | 8.412761 | 1.6E-03 | 115399.189 | 115399.355 | -1.7E-01 | 115424.387 | 115424.554 | -1.7E-01 | -80.870 | -80.710 | -1.6E-01 |
| 1090 | Sn | 124 | 50 | 74 | 24 | 8.467421 | 8.460296 | 7.1E-03 | 115391.306 | 115392.159 | -8.5E-01 | 115417.023 | 115417.877 | -8.5E-01 | -88.234 | -87.387 | -8.5E-01 |
| 1091 | Sb | 124 | 51 | 73 | 22 | 8.456167 | 8.453873 | 2.3E-03 | 115391.400 | 115391.653 | -2.5E-01 | 115417.637 | 115417.890 | -2.5E-01 | -87.621 | -87.374 | -2.5E-01 |
| 1092 | Te | 124 | 52 | 72 | 20 | 8.473279 | 8.471421 | 1.9E-03 | 115387.976 | 115388.174 | -2.0E-01 | 115414.732 | 115414.931 | -2.0E-01 | -90.525 | -90.333 | -1.9E-01 |
| 1093 | I | 124 | 53 | 71 | 18 | 8.441489 | 8.440276 | 1.2E-03 | 115390.616 | 115390.733 | -1.2E-01 | 115417.892 | 115418.010 | -1.2E-01 | -87.366 | -87.254 | -1.1E-01 |
| 1094 | Xe | 124 | 54 | 70 | 16 | 8.437562 | 8.435689 | 1.9E-03 | 115389.800 | 115389.999 | -2.0E-01 | 115417.597 | 115417.796 | -2.0E-01 | -87.661 | -87.468 | -1.9E-01 |
| 1095 | Cs | 124 | 55 | 69 | 14 | 8.383432 | 8.382083 | 1.3E-03 | 115395.210 | 115395.343 | -1.3E-01 | 115423.526 | 115423.660 | -1.3E-01 | -81.731 | -81.604 | -1.3E-01 |
| 1096 | Ba | 124 | 56 | 68 | 12 | 8.355820 | 8.355641 | 1.8E-04 | 115397.331 | 115397.318 | 1.3E-02 | 115426.168 | 115426.155 | 1.3E-02 | -79.090 | -79.108 | 1.8E-02 |
| 1097 | La | 124 | 57 | 67 | 10 | 8.278292 | 8.277826 | 4.7E-04 | 115405.641 | 115405.664 | -2.3E-02 | 115434.999 | 115435.021 | -2.2E-02 | -70.259 | -70.242 | -1.6E-02 |
| 1098 | Ag | 125 | 47 | 78 | 31 | 8.290997 | 8.282510 | 8.5E-03 | 116348.360 | 116349.393 | -1.0E+00 | 116372.521 | 116373.554 | -1.0E+00 | -64.230 | -63.204 | -1.0E+00 |
| 1099 | Cd | 125 | 48 | 77 | 29 | 8.357681 | 8.353317 | 4.4E-03 | 116338.723 | 116339.240 | -5.2E-01 | 116363.404 | 116363.920 | -5.2E-01 | -73.348 | -72.837 | -5.1E-01 |
| 1100 | In | 125 | 49 | 76 | 27 | 8.408452 | 8.404707 | 3.7E-03 | 116331.076 | 116331.515 | -4.4E-01 | 116356.275 | 116356.713 | -4.4E-01 | -80.477 | -80.044 | -4.3E-01 |
| 1101 | Sn | 125 | 50 | 75 | 25 | 8.445550 | 8.437742 | 7.8E-03 | 116325.137 | 116326.083 | -9.5E-01 | 116350.855 | 116351.801 | -9.5E-01 | -85.896 | -84.957 | -9.4E-01 |

| | | | | | | | | | | | | | | | | | | |
|---|---|---|---|---|---|---|---|---|---|---|---|---|---|---|---|---|---|---|
| 1102 | Sb | 125 | 51 | 74 | 23 | 8.458170 | 8.453374 | 4.8E-03 | 116322.258 | 116322.827 | -5.7E-01 | 116348.495 | 116349.064 | -5.7E-01 | -88.256 | -87.694 | -5.6E-01 |
| 1103 | Te | 125 | 52 | 73 | 21 | 8.458045 | 8.455905 | 2.1E-03 | 116320.972 | 116321.208 | -2.4E-01 | 116347.729 | 116347.964 | -2.4E-01 | -89.023 | -88.793 | -2.3E-01 |
| 1104 | I | 125 | 53 | 72 | 19 | 8.450300 | 8.446823 | 3.5E-03 | 116320.638 | 116321.040 | -4.0E-01 | 116347.914 | 116348.317 | -4.0E-01 | -88.837 | -88.441 | -4.0E-01 |
| 1105 | Xe | 125 | 54 | 71 | 17 | 8.430888 | 8.427487 | 3.4E-03 | 116321.762 | 116322.154 | -3.9E-01 | 116349.559 | 116349.950 | -3.9E-01 | -87.193 | -86.807 | -3.9E-01 |
| 1106 | Cs | 125 | 55 | 70 | 15 | 8.399787 | 8.397320 | 2.5E-03 | 116324.347 | 116324.622 | -2.7E-01 | 116352.664 | 116352.938 | -2.7E-01 | -84.088 | -83.819 | -2.7E-01 |
| 1107 | Ba | 125 | 56 | 69 | 13 | 8.358178 | 8.355730 | 2.4E-03 | 116328.246 | 116328.517 | -2.7E-01 | 116357.083 | 116357.354 | -2.7E-01 | -79.669 | -79.404 | -2.7E-01 |
| 1108 | La | 125 | 57 | 68 | 11 | 8.304643 | 8.303252 | 1.4E-03 | 116333.634 | 116333.773 | -1.4E-01 | 116362.992 | 116363.131 | -1.4E-01 | -73.759 | -73.627 | -1.3E-01 |
| 1109 | Cd | 126 | 48 | 78 | 30 | 8.346747 | 8.341751 | 5.0E-03 | 117271.309 | 117271.910 | -6.0E-01 | 117295.989 | 117296.590 | -6.0E-01 | -72.257 | -71.662 | -5.9E-01 |
| 1110 | In | 126 | 49 | 77 | 28 | 8.384317 | 8.378246 | 6.1E-03 | 117265.274 | 117266.009 | -7.4E-01 | 117290.473 | 117291.208 | -7.4E-01 | -77.773 | -77.044 | -7.3E-01 |
| 1111 | Sn | 126 | 50 | 76 | 26 | 8.443523 | 8.432582 | 1.1E-02 | 117256.513 | 117257.861 | -1.3E+00 | 117282.230 | 117283.579 | -1.3E+00 | -86.015 | -84.673 | -1.3E+00 |
| 1112 | Sb | 126 | 51 | 75 | 24 | 8.440314 | 8.433532 | 6.8E-03 | 117255.615 | 117256.439 | -8.2E-01 | 117281.852 | 117282.676 | -8.2E-01 | -86.393 | -85.576 | -8.2E-01 |
| 1113 | Te | 126 | 52 | 74 | 22 | 8.463248 | 8.457646 | 5.6E-03 | 117251.424 | 117252.098 | -6.7E-01 | 117278.180 | 117278.855 | -6.7E-01 | -90.065 | -89.397 | -6.7E-01 |
| 1114 | I | 126 | 53 | 73 | 20 | 8.439944 | 8.434698 | 5.2E-03 | 117253.058 | 117253.687 | -6.3E-01 | 117280.334 | 117280.963 | -6.3E-01 | -87.911 | -87.289 | -6.2E-01 |
| 1115 | Xe | 126 | 54 | 72 | 18 | 8.443530 | 8.437565 | 6.0E-03 | 117251.304 | 117252.022 | -7.2E-01 | 117279.100 | 117279.819 | -7.2E-01 | -89.146 | -88.433 | -7.1E-01 |
| 1116 | Cs | 126 | 55 | 71 | 16 | 8.399265 | 8.393896 | 5.4E-03 | 117255.579 | 117256.221 | -6.4E-01 | 117283.895 | 117284.538 | -6.4E-01 | -84.350 | -83.714 | -6.4E-01 |
| 1117 | Ba | 126 | 56 | 70 | 14 | 8.379718 | 8.376147 | 3.6E-03 | 117256.739 | 117257.154 | -4.2E-01 | 117285.576 | 117285.991 | -4.2E-01 | -82.670 | -82.261 | -4.1E-01 |
| 1118 | La | 126 | 57 | 69 | 12 | 8.312426 | 8.309495 | 2.9E-03 | 117263.914 | 117264.248 | -3.3E-01 | 117293.272 | 117293.606 | -3.3E-01 | -74.973 | -74.646 | -3.3E-01 |
| 1119 | Ce | 126 | 58 | 68 | 10 | 8.273257 | 8.270142 | 3.1E-03 | 117267.546 | 117267.903 | -3.6E-01 | 117297.425 | 117297.781 | -3.6E-01 | -70.821 | -70.470 | -3.5E-01 |
| 1120 | Cd | 127 | 48 | 79 | 31 | 8.314922 | 8.313540 | 1.4E-03 | 118206.569 | 118206.716 | -1.5E-01 | 118231.249 | 118231.396 | -1.5E-01 | -68.491 | -68.350 | -1.4E-01 |
| 1121 | In | 127 | 49 | 78 | 29 | 8.374965 | 8.370763 | 4.2E-03 | 118197.643 | 118198.147 | -5.0E-01 | 118222.841 | 118223.346 | -5.0E-01 | -76.898 | -76.400 | -5.0E-01 |
| 1122 | Sn | 127 | 50 | 77 | 27 | 8.420560 | 8.409804 | 1.1E-02 | 118190.551 | 118191.886 | -1.3E+00 | 118216.268 | 118217.604 | -1.3E+00 | -83.471 | -82.141 | -1.3E+00 |
| 1123 | Sb | 127 | 51 | 76 | 25 | 8.439820 | 8.431421 | 8.4E-03 | 118186.803 | 118187.839 | -1.0E+00 | 118213.040 | 118214.076 | -1.0E+00 | -86.699 | -85.670 | -1.0E+00 |
| 1124 | Te | 127 | 52 | 75 | 23 | 8.446117 | 8.440400 | 5.7E-03 | 118184.701 | 118185.396 | -6.9E-01 | 118211.458 | 118212.152 | -6.9E-01 | -88.282 | -87.593 | -6.9E-01 |
| 1125 | I | 127 | 53 | 74 | 21 | 8.445487 | 8.438191 | 7.3E-03 | 118183.479 | 118184.374 | -8.9E-01 | 118210.756 | 118211.650 | -8.9E-01 | -88.984 | -88.096 | -8.9E-01 |
| 1126 | Xe | 127 | 54 | 73 | 19 | 8.434111 | 8.426803 | 7.3E-03 | 118183.622 | 118184.517 | -8.9E-01 | 118211.418 | 118212.313 | -9.0E-01 | -88.322 | -87.433 | -8.9E-01 |
| 1127 | Cs | 127 | 55 | 72 | 17 | 8.411562 | 8.405072 | 6.5E-03 | 118185.183 | 118185.973 | -7.9E-01 | 118213.499 | 118214.290 | -7.9E-01 | -86.240 | -85.456 | -7.8E-01 |
| 1128 | Ba | 127 | 56 | 71 | 15 | 8.378455 | 8.373074 | 5.4E-03 | 118188.085 | 118188.734 | -6.5E-01 | 118216.922 | 118217.571 | -6.5E-01 | -82.818 | -82.175 | -6.4E-01 |
| 1129 | La | 127 | 57 | 70 | 13 | 8.333540 | 8.330179 | 3.4E-03 | 118192.486 | 118192.877 | -3.9E-01 | 118221.844 | 118222.235 | -3.9E-01 | -77.896 | -77.511 | -3.9E-01 |
| 1130 | Ce | 127 | 58 | 69 | 11 | 8.280791 | 8.275496 | 5.3E-03 | 118197.882 | 118198.518 | -6.4E-01 | 118227.760 | 118228.397 | -6.4E-01 | -71.979 | -71.349 | -6.3E-01 |
| 1131 | Cd | 128 | 48 | 80 | 32 | 8.303264 | 8.302926 | 3.4E-04 | 119139.312 | 119139.326 | -1.4E-02 | 119163.992 | 119164.006 | -1.4E-02 | -67.242 | -67.233 | -8.5E-03 |
| 1132 | In | 128 | 49 | 79 | 30 | 8.351090 | 8.346207 | 4.9E-03 | 119131.889 | 119132.485 | -6.0E-01 | 119157.088 | 119157.683 | -6.0E-01 | -74.146 | -73.556 | -5.9E-01 |
| 1133 | Sn | 128 | 50 | 78 | 28 | 8.416979 | 8.405367 | 1.2E-02 | 119122.154 | 119123.610 | -1.5E+00 | 119147.872 | 119149.328 | -1.5E+00 | -83.362 | -81.912 | -1.5E+00 |
| 1134 | Sb | 128 | 51 | 77 | 26 | 8.420775 | 8.411399 | 9.4E-03 | 119120.366 | 119121.536 | -1.2E+00 | 119146.603 | 119147.773 | -1.2E+00 | -84.630 | -83.467 | -1.2E+00 |
| 1135 | Te | 128 | 52 | 76 | 24 | 8.448752 | 8.440398 | 8.4E-03 | 119115.483 | 119116.521 | -1.0E+00 | 119142.240 | 119143.278 | -1.0E+00 | -88.994 | -87.962 | -1.0E+00 |
| 1136 | I | 128 | 53 | 75 | 22 | 8.432835 | 8.424174 | 8.7E-03 | 119116.219 | 119117.295 | -1.1E+00 | 119143.495 | 119144.571 | -1.1E+00 | -87.739 | -86.669 | -1.1E+00 |
| 1137 | Xe | 128 | 54 | 74 | 20 | 8.443298 | 8.433483 | 9.8E-03 | 119113.577 | 119114.800 | -1.2E+00 | 119141.373 | 119142.597 | -1.2E+00 | -89.860 | -88.643 | -1.2E+00 |
| 1138 | Cs | 128 | 55 | 73 | 18 | 8.406493 | 8.398723 | 7.8E-03 | 119116.986 | 119117.946 | -9.6E-01 | 119145.302 | 119146.263 | -9.6E-01 | -85.932 | -84.977 | -9.5E-01 |
| 1139 | Ba | 128 | 56 | 72 | 16 | 8.396063 | 8.388959 | 7.1E-03 | 119117.018 | 119117.893 | -8.7E-01 | 119145.855 | 119146.730 | -8.7E-01 | -85.379 | -84.510 | -8.7E-01 |
| 1140 | La | 128 | 57 | 71 | 14 | 8.337190 | 8.332769 | 4.4E-03 | 119123.250 | 119123.781 | -5.3E-01 | 119152.608 | 119153.139 | -5.3E-01 | -78.625 | -78.101 | -5.2E-01 |
| 1141 | Ce | 128 | 58 | 70 | 12 | 8.306925 | 8.302263 | 4.7E-03 | 119125.821 | 119126.382 | -5.6E-01 | 119155.700 | 119156.260 | -5.6E-01 | -75.534 | -74.979 | -5.5E-01 |
| 1142 | Pr | 128 | 59 | 69 | 10 | 8.228913 | 8.222892 | 6.0E-03 | 119134.503 | 119135.237 | -7.3E-01 | 119164.903 | 119165.637 | -7.3E-01 | -66.331 | -65.603 | -7.3E-01 |
| 1143 | In | 129 | 49 | 80 | 31 | 8.338782 | 8.339219 | -4.4E-04 | 120064.691 | 120064.605 | 8.6E-02 | 120089.890 | 120089.804 | 8.6E-02 | -72.838 | -72.930 | 9.2E-02 |
| 1144 | Sn | 129 | 50 | 79 | 29 | 8.392943 | 8.384327 | 8.6E-03 | 120056.403 | 120057.484 | -1.1E+00 | 120082.121 | 120083.202 | -1.1E+00 | -80.607 | -79.532 | -1.1E+00 |
| 1145 | Sb | 129 | 51 | 78 | 27 | 8.418059 | 8.409904 | 8.2E-03 | 120051.861 | 120052.882 | -1.0E+00 | 120078.098 | 120079.119 | -1.0E+00 | -84.629 | -83.614 | -1.0E+00 |
| 1146 | Te | 129 | 52 | 77 | 25 | 8.430409 | 8.423068 | 7.3E-03 | 120048.966 | 120049.882 | -9.2E-01 | 120075.723 | 120076.638 | -9.2E-01 | -87.005 | -86.096 | -9.1E-01 |
| 1147 | I | 129 | 53 | 76 | 23 | 8.435990 | 8.425880 | 1.0E-02 | 120046.944 | 120048.216 | -1.3E+00 | 120074.221 | 120075.492 | -1.3E+00 | -88.507 | -87.241 | -1.3E+00 |

| | | | | | | | | | | | | | | | | | | |
|---|---|---|---|---|---|---|---|---|---|---|---|---|---|---|---|---|---|---|
| 1148 | Xe | 129 | 54 | 75 | 21 | 8.431390 | 8.420967 | 1.0E-02 | 120046.235 | 120047.547 | -1.3E+00 | 120074.032 | 120075.343 | -1.3E+00 | -88.696 | -87.391 | -1.3E+00 |
| 1149 | Cs | 129 | 55 | 74 | 19 | 8.416047 | 8.406553 | 9.5E-03 | 120046.912 | 120048.103 | -1.2E+00 | 120075.228 | 120076.419 | -1.2E+00 | -87.499 | -86.314 | -1.2E+00 |
| 1150 | Ba | 129 | 56 | 73 | 17 | 8.391099 | 8.383178 | 7.9E-03 | 120048.827 | 120049.815 | -9.9E-01 | 120077.664 | 120078.652 | -9.9E-01 | -85.063 | -84.082 | -9.8E-01 |
| 1151 | La | 129 | 57 | 72 | 15 | 8.356052 | 8.349179 | 6.9E-03 | 120052.045 | 120052.897 | -8.5E-01 | 120081.403 | 120082.255 | -8.5E-01 | -81.325 | -80.479 | -8.5E-01 |
| 1152 | Ce | 129 | 58 | 71 | 13 | 8.310940 | 8.304372 | 6.6E-03 | 120056.561 | 120057.373 | -8.1E-01 | 120086.440 | 120087.252 | -8.1E-01 | -76.287 | -75.482 | -8.1E-01 |
| 1153 | Pr | 129 | 59 | 70 | 11 | 8.254380 | 8.249086 | 5.3E-03 | 120062.554 | 120063.200 | -6.5E-01 | 120092.954 | 120093.600 | -6.5E-01 | -69.774 | -69.133 | -6.4E-01 |
| 1154 | Cd | 130 | 48 | 82 | 34 | 8.255788 | 8.251244 | 4.5E-03 | 121008.008 | 121008.570 | -5.6E-01 | 121032.688 | 121033.250 | -5.6E-01 | -61.534 | -60.978 | -5.6E-01 |
| 1155 | In | 130 | 49 | 81 | 32 | 8.314001 | 8.311534 | 2.5E-03 | 120999.139 | 120999.430 | -2.9E-01 | 121024.338 | 121024.629 | -2.9E-01 | -69.884 | -69.599 | -2.9E-01 |
| 1156 | Sn | 130 | 50 | 80 | 30 | 8.386821 | 8.380586 | 6.2E-03 | 120988.371 | 120989.151 | -7.8E-01 | 121014.089 | 121014.869 | -7.8E-01 | -80.133 | -79.359 | -7.7E-01 |
| 1157 | Sb | 130 | 51 | 79 | 28 | 8.397368 | 8.391668 | 5.7E-03 | 120985.698 | 120986.408 | -7.1E-01 | 121011.935 | 121012.646 | -7.1E-01 | -82.286 | -81.582 | -7.0E-01 |
| 1158 | Te | 130 | 52 | 78 | 26 | 8.430324 | 8.423570 | 6.8E-03 | 120980.112 | 120980.959 | -8.5E-01 | 121006.869 | 121007.715 | -8.5E-01 | -87.353 | -86.513 | -8.4E-01 |
| 1159 | I | 130 | 53 | 77 | 24 | 8.421100 | 8.411671 | 9.4E-03 | 120980.009 | 120981.203 | -1.2E+00 | 121007.286 | 121008.479 | -1.2E+00 | -86.936 | -85.749 | -1.2E+00 |
| 1160 | Xe | 130 | 54 | 76 | 22 | 8.437731 | 8.425602 | 1.2E-02 | 120976.545 | 120978.089 | -1.5E+00 | 121004.341 | 121005.885 | -1.5E+00 | -89.880 | -88.343 | -1.5E+00 |
| 1161 | Cs | 130 | 55 | 75 | 20 | 8.408784 | 8.398178 | 1.1E-02 | 120979.005 | 120980.350 | -1.3E+00 | 121007.322 | 121008.667 | -1.3E+00 | -86.900 | -85.561 | -1.3E+00 |
| 1162 | Ba | 130 | 56 | 74 | 18 | 8.405550 | 8.395332 | 1.0E-02 | 120978.123 | 120979.417 | -1.3E+00 | 121006.960 | 121008.254 | -1.3E+00 | -87.262 | -85.974 | -1.3E+00 |
| 1163 | La | 130 | 57 | 73 | 16 | 8.356191 | 8.348644 | 7.5E-03 | 120983.237 | 120984.182 | -9.5E-01 | 121012.594 | 121013.540 | -9.5E-01 | -81.627 | -80.688 | -9.4E-01 |
| 1164 | Ce | 130 | 58 | 72 | 14 | 8.333216 | 8.326387 | 6.8E-03 | 120984.920 | 120985.772 | -8.5E-01 | 121014.799 | 121015.651 | -8.5E-01 | -79.423 | -78.577 | -8.5E-01 |
| 1165 | Pr | 130 | 59 | 71 | 12 | 8.263756 | 8.257482 | 6.3E-03 | 120992.646 | 120993.425 | -7.8E-01 | 121023.046 | 121023.825 | -7.8E-01 | -71.175 | -70.403 | -7.7E-01 |
| 1166 | Nd | 130 | 60 | 70 | 10 | 8.222513 | 8.214938 | 7.6E-03 | 120996.704 | 120997.651 | -9.5E-01 | 121027.625 | 121028.573 | -9.5E-01 | -66.596 | -65.655 | -9.4E-01 |
| 1167 | In | 131 | 49 | 82 | 33 | 8.297963 | 8.293917 | 4.0E-03 | 121932.491 | 121932.992 | -5.0E-01 | 121957.690 | 121958.191 | -5.0E-01 | -68.026 | -67.531 | -4.9E-01 |
| 1168 | Sn | 131 | 50 | 81 | 31 | 8.362575 | 8.356220 | 6.4E-03 | 121922.726 | 121923.528 | -8.0E-01 | 121948.444 | 121949.246 | -8.0E-01 | -77.272 | -76.476 | -8.0E-01 |
| 1169 | Sb | 131 | 51 | 80 | 29 | 8.392556 | 8.390670 | 1.9E-03 | 121917.497 | 121917.713 | -2.2E-01 | 121943.734 | 121943.950 | -2.2E-01 | -81.982 | -81.772 | -2.1E-01 |
| 1170 | Te | 131 | 52 | 79 | 27 | 8.411233 | 8.407934 | 3.3E-03 | 121913.748 | 121914.149 | -4.0E-01 | 121940.505 | 121940.905 | -4.0E-01 | -85.211 | -84.817 | -3.9E-01 |
| 1171 | I | 131 | 53 | 78 | 25 | 8.422297 | 8.413659 | 8.6E-03 | 121910.997 | 121912.096 | -1.1E+00 | 121938.273 | 121939.372 | -1.1E+00 | -87.443 | -86.350 | -1.1E+00 |
| 1172 | Xe | 131 | 54 | 77 | 23 | 8.423736 | 8.412895 | 1.1E-02 | 121909.506 | 121910.893 | -1.4E+00 | 121937.302 | 121938.689 | -1.4E+00 | -88.414 | -87.033 | -1.4E+00 |
| 1173 | Cs | 131 | 55 | 76 | 21 | 8.415056 | 8.403912 | 1.1E-02 | 121909.340 | 121910.766 | -1.4E+00 | 121937.657 | 121939.083 | -1.4E+00 | -88.059 | -86.639 | -1.4E+00 |
| 1174 | Ba | 131 | 56 | 75 | 19 | 8.398588 | 8.387662 | 1.1E-02 | 121910.195 | 121911.592 | -1.4E+00 | 121939.032 | 121940.429 | -1.4E+00 | -86.684 | -85.293 | -1.4E+00 |
| 1175 | La | 131 | 57 | 74 | 17 | 8.370367 | 8.361485 | 8.9E-03 | 121912.589 | 121913.717 | -1.1E+00 | 121941.946 | 121943.075 | -1.1E+00 | -83.769 | -82.647 | -1.1E+00 |
| 1176 | Ce | 131 | 58 | 73 | 15 | 8.333396 | 8.325695 | 7.7E-03 | 121916.128 | 121917.101 | -9.7E-01 | 121946.007 | 121946.980 | -9.7E-01 | -79.708 | -78.742 | -9.7E-01 |
| 1177 | Pr | 131 | 59 | 72 | 13 | 8.286143 | 8.279348 | 6.8E-03 | 121921.015 | 121921.869 | -8.5E-01 | 121951.415 | 121952.269 | -8.5E-01 | -74.301 | -73.453 | -8.5E-01 |
| 1178 | Nd | 131 | 60 | 71 | 11 | 8.230304 | 8.221970 | 8.3E-03 | 121927.026 | 121928.080 | -1.1E+00 | 121957.948 | 121959.002 | -1.1E+00 | -67.768 | -66.720 | -1.0E+00 |
| 1179 | In | 132 | 49 | 83 | 34 | 8.253695 | 8.250475 | 3.2E-03 | 122869.602 | 122869.998 | -4.0E-01 | 122894.801 | 122895.197 | -4.0E-01 | -62.409 | -62.019 | -3.9E-01 |
| 1180 | Sn | 132 | 50 | 82 | 32 | 8.354852 | 8.342250 | 1.3E-02 | 122854.948 | 122856.582 | -1.6E+00 | 122880.666 | 122882.299 | -1.6E+00 | -76.544 | -74.917 | -1.6E+00 |
| 1181 | Sb | 132 | 51 | 81 | 30 | 8.372347 | 8.369438 | 2.9E-03 | 122851.337 | 122851.690 | -3.5E-01 | 122877.574 | 122877.927 | -3.5E-01 | -79.636 | -79.289 | -3.5E-01 |
| 1182 | Te | 132 | 52 | 80 | 28 | 8.408485 | 8.408925 | -4.4E-04 | 122845.265 | 122845.175 | 9.0E-02 | 122872.022 | 122871.932 | 9.0E-02 | -85.188 | -85.284 | 9.6E-02 |
| 1183 | I | 132 | 53 | 79 | 26 | 8.406462 | 8.401015 | 5.4E-03 | 122844.230 | 122844.917 | -6.9E-01 | 122871.506 | 122872.193 | -6.9E-01 | -85.703 | -85.023 | -6.8E-01 |
| 1184 | Xe | 132 | 54 | 78 | 24 | 8.427622 | 8.417530 | 1.0E-02 | 122840.134 | 122841.434 | -1.3E+00 | 122867.931 | 122869.230 | -1.3E+00 | -89.279 | -87.986 | -1.3E+00 |
| 1185 | Cs | 132 | 55 | 77 | 22 | 8.405614 | 8.395085 | 1.1E-02 | 122841.737 | 122843.093 | -1.4E+00 | 122870.053 | 122871.410 | -1.4E+00 | -87.156 | -85.806 | -1.3E+00 |
| 1186 | Ba | 132 | 56 | 76 | 20 | 8.409375 | 8.397375 | 1.2E-02 | 122839.938 | 122841.487 | -1.5E+00 | 122868.775 | 122870.324 | -1.5E+00 | -88.435 | -86.892 | -1.5E+00 |
| 1187 | La | 132 | 57 | 75 | 18 | 8.367756 | 8.358698 | 9.1E-03 | 122844.128 | 122845.289 | -1.2E+00 | 122873.486 | 122874.647 | -1.2E+00 | -83.724 | -82.569 | -1.2E+00 |
| 1188 | Ce | 132 | 58 | 74 | 16 | 8.352339 | 8.343718 | 8.6E-03 | 122844.860 | 122845.962 | -1.1E+00 | 122874.739 | 122875.841 | -1.1E+00 | -82.471 | -81.375 | -1.1E+00 |
| 1189 | Pr | 132 | 59 | 73 | 14 | 8.291431 | 8.284506 | 6.9E-03 | 122851.596 | 122852.474 | -8.8E-01 | 122881.996 | 122882.874 | -8.8E-01 | -75.213 | -74.342 | -8.7E-01 |
| 1190 | Nd | 132 | 60 | 72 | 12 | 8.256810 | 8.250080 | 6.7E-03 | 122854.862 | 122855.713 | -8.5E-01 | 122885.784 | 122886.635 | -8.5E-01 | -71.426 | -70.581 | -8.4E-01 |
| 1191 | Sn | 133 | 50 | 83 | 33 | 8.310091 | 8.301792 | 8.3E-03 | 123792.112 | 123793.186 | -1.1E+00 | 123817.830 | 123818.903 | -1.1E+00 | -70.874 | -69.807 | -1.1E+00 |
| 1192 | Sb | 133 | 51 | 82 | 31 | 8.364722 | 8.358306 | 6.4E-03 | 123783.544 | 123784.367 | -8.2E-01 | 123809.781 | 123810.604 | -8.2E-01 | -78.923 | -78.106 | -8.2E-01 |
| 1193 | Te | 133 | 52 | 81 | 29 | 8.388987 | 8.390371 | -1.4E-03 | 123779.015 | 123778.800 | 2.2E-01 | 123805.772 | 123805.556 | 2.2E-01 | -82.932 | -83.154 | 2.2E-01 |

| | | | | | | | | | | | | | | | | | | |
|---|---|---|---|---|---|---|---|---|---|---|---|---|---|---|---|---|---|---|
| 1194 | I | 133 | 53 | 80 | 27 | 8.405319 | 8.403192 | 2.1E-03 | 123775.541 | 123775.792 | -2.5E-01 | 123802.817 | 123803.068 | -2.5E-01 | -85.887 | -85.642 | -2.4E-01 |
| 1195 | Xe | 133 | 54 | 79 | 25 | 8.412647 | 8.406171 | 6.5E-03 | 123773.264 | 123774.092 | -8.3E-01 | 123801.060 | 123801.889 | -8.3E-01 | -87.644 | -86.821 | -8.2E-01 |
| 1196 | Cs | 133 | 55 | 78 | 23 | 8.409978 | 8.400545 | 9.4E-03 | 123772.316 | 123773.537 | -1.2E+00 | 123800.633 | 123801.854 | -1.2E+00 | -88.071 | -86.856 | -1.2E+00 |
| 1197 | Ba | 133 | 56 | 77 | 21 | 8.400206 | 8.389218 | 1.1E-02 | 123772.313 | 123773.740 | -1.4E+00 | 123801.150 | 123802.577 | -1.4E+00 | -87.554 | -86.133 | -1.4E+00 |
| 1198 | La | 133 | 57 | 76 | 19 | 8.378841 | 8.369119 | 9.7E-03 | 123773.852 | 123775.109 | -1.3E+00 | 123803.209 | 123804.467 | -1.3E+00 | -85.494 | -84.243 | -1.3E+00 |
| 1199 | Ce | 133 | 58 | 75 | 17 | 8.349829 | 8.340988 | 8.8E-03 | 123776.407 | 123777.547 | -1.1E+00 | 123806.286 | 123807.426 | -1.1E+00 | -82.418 | -81.284 | -1.1E+00 |
| 1200 | Pr | 133 | 59 | 74 | 15 | 8.310258 | 8.302690 | 7.6E-03 | 123780.366 | 123781.336 | -9.7E-01 | 123810.766 | 123811.736 | -9.7E-01 | -77.938 | -76.974 | -9.6E-01 |
| 1201 | Nd | 133 | 60 | 73 | 13 | 8.262231 | 8.254343 | 7.9E-03 | 123785.450 | 123786.462 | -1.0E+00 | 123816.371 | 123817.383 | -1.0E+00 | -72.332 | -71.327 | -1.0E+00 |
| 1202 | Pm | 133 | 61 | 72 | 11 | 8.204283 | 8.196130 | 8.2E-03 | 123791.853 | 123792.899 | -1.0E+00 | 123823.296 | 123824.343 | -1.0E+00 | -65.408 | -64.368 | -1.0E+00 |
| 1203 | Sn | 134 | 50 | 84 | 34 | 8.275160 | 8.271920 | 3.2E-03 | 124728.048 | 124728.452 | -4.0E-01 | 124753.766 | 124754.170 | -4.0E-01 | -66.432 | -66.034 | -4.0E-01 |
| 1204 | Sb | 134 | 51 | 83 | 32 | 8.325950 | 8.321662 | 4.3E-03 | 124719.940 | 124720.484 | -5.4E-01 | 124746.177 | 124746.721 | -5.4E-01 | -74.021 | -73.483 | -5.4E-01 |
| 1205 | Te | 134 | 52 | 82 | 30 | 8.383660 | 8.381858 | 1.8E-03 | 124710.905 | 124711.115 | -2.1E-01 | 124737.662 | 124737.872 | -2.1E-01 | -82.536 | -82.332 | -2.0E-01 |
| 1206 | I | 134 | 53 | 81 | 28 | 8.389188 | 8.387895 | 1.3E-03 | 124708.862 | 124709.004 | -1.4E-01 | 124736.139 | 124736.280 | -1.4E-01 | -84.059 | -83.924 | -1.3E-01 |
| 1207 | Xe | 134 | 54 | 80 | 26 | 8.413687 | 8.410790 | 2.9E-03 | 124704.277 | 124704.633 | -3.6E-01 | 124732.073 | 124732.429 | -3.6E-01 | -88.124 | -87.775 | -3.5E-01 |
| 1208 | Cs | 134 | 55 | 79 | 24 | 8.398646 | 8.392726 | 5.9E-03 | 124704.990 | 124705.750 | -7.6E-01 | 124733.307 | 124734.066 | -7.6E-01 | -86.891 | -86.138 | -7.5E-01 |
| 1209 | Ba | 134 | 56 | 78 | 22 | 8.408173 | 8.398236 | 9.9E-03 | 124702.411 | 124703.708 | -1.3E+00 | 124731.248 | 124732.545 | -1.3E+00 | -88.950 | -87.659 | -1.3E+00 |
| 1210 | La | 134 | 57 | 77 | 20 | 8.374488 | 8.365446 | 9.0E-03 | 124705.621 | 124706.798 | -1.2E+00 | 124734.979 | 124736.156 | -1.2E+00 | -85.219 | -84.048 | -1.2E+00 |
| 1211 | Ce | 134 | 58 | 76 | 18 | 8.365771 | 8.356171 | 9.6E-03 | 124705.486 | 124706.737 | -1.3E+00 | 124735.365 | 124736.615 | -1.3E+00 | -84.833 | -83.589 | -1.2E+00 |
| 1212 | Pr | 134 | 59 | 75 | 16 | 8.312881 | 8.305393 | 7.5E-03 | 124711.270 | 124712.237 | -9.7E-01 | 124741.670 | 124742.637 | -9.7E-01 | -78.528 | -77.567 | -9.6E-01 |
| 1213 | Nd | 134 | 60 | 74 | 14 | 8.285538 | 8.278334 | 7.2E-03 | 124713.630 | 124714.558 | -9.3E-01 | 124744.551 | 124745.480 | -9.3E-01 | -75.646 | -74.725 | -9.2E-01 |
| 1214 | Pm | 134 | 61 | 73 | 12 | 8.213225 | 8.206691 | 6.5E-03 | 124722.016 | 124722.853 | -8.4E-01 | 124753.459 | 124754.297 | -8.4E-01 | -66.739 | -65.908 | -8.3E-01 |
| 1215 | Sn | 135 | 50 | 85 | 35 | 8.230687 | 8.221141 | 9.5E-03 | 125665.342 | 125666.601 | -1.3E+00 | 125691.060 | 125692.318 | -1.3E+00 | -60.632 | -59.380 | -1.3E+00 |
| 1216 | Sb | 135 | 51 | 84 | 33 | 8.291983 | 8.294512 | -2.5E-03 | 125655.765 | 125655.393 | 3.7E-01 | 125682.002 | 125681.630 | 3.7E-01 | -69.690 | -70.068 | 3.8E-01 |
| 1217 | Te | 135 | 52 | 83 | 31 | 8.345731 | 8.348369 | -2.6E-03 | 125647.207 | 125646.820 | 3.9E-01 | 125673.964 | 125673.577 | 3.9E-01 | -77.728 | -78.122 | 3.9E-01 |
| 1218 | I | 135 | 53 | 82 | 29 | 8.384833 | 8.380856 | 4.0E-03 | 125640.626 | 125641.131 | -5.1E-01 | 125667.903 | 125668.408 | -5.0E-01 | -83.789 | -83.291 | -5.0E-01 |
| 1219 | Xe | 135 | 54 | 81 | 27 | 8.398503 | 8.396803 | 1.7E-03 | 125637.479 | 125637.675 | -2.0E-01 | 125665.275 | 125665.472 | -2.0E-01 | -86.417 | -86.227 | -1.9E-01 |
| 1220 | Cs | 135 | 55 | 80 | 25 | 8.401338 | 8.397782 | 3.6E-03 | 125635.793 | 125636.240 | -4.5E-01 | 125664.110 | 125664.557 | -4.5E-01 | -87.582 | -87.142 | -4.4E-01 |
| 1221 | Ba | 135 | 56 | 79 | 23 | 8.397534 | 8.390789 | 6.7E-03 | 125635.004 | 125635.880 | -8.8E-01 | 125663.841 | 125664.717 | -8.8E-01 | -87.851 | -86.981 | -8.7E-01 |
| 1222 | La | 135 | 57 | 78 | 21 | 8.382797 | 8.374931 | 7.9E-03 | 125635.690 | 125636.717 | -1.0E+00 | 125665.048 | 125666.075 | -1.0E+00 | -86.644 | -85.623 | -1.0E+00 |
| 1223 | Ce | 135 | 58 | 77 | 19 | 8.361986 | 8.352565 | 9.4E-03 | 125637.197 | 125638.433 | -1.2E+00 | 125667.075 | 125668.312 | -1.2E+00 | -84.616 | -83.387 | -1.2E+00 |
| 1224 | Pr | 135 | 59 | 76 | 17 | 8.328928 | 8.320877 | 8.1E-03 | 125640.356 | 125641.406 | -1.1E+00 | 125670.756 | 125671.806 | -1.1E+00 | -80.936 | -79.892 | -1.0E+00 |
| 1225 | Nd | 135 | 60 | 75 | 15 | 8.288153 | 8.280481 | 7.7E-03 | 125644.557 | 125645.555 | -1.0E+00 | 125675.478 | 125676.477 | -1.0E+00 | -76.214 | -75.221 | -9.9E-01 |
| 1226 | Pm | 135 | 61 | 74 | 13 | 8.236530 | 8.230234 | 6.3E-03 | 125650.221 | 125651.034 | -8.1E-01 | 125681.665 | 125682.477 | -8.1E-01 | -70.027 | -69.221 | -8.1E-01 |
| 1227 | Sm | 135 | 62 | 73 | 11 | 8.177626 | 8.169790 | 7.8E-03 | 125656.869 | 125657.888 | -1.0E+00 | 125688.835 | 125689.854 | -1.0E+00 | -62.857 | -61.844 | -1.0E+00 |
| 1228 | Sb | 136 | 51 | 85 | 34 | 8.252274 | 8.247895 | 4.4E-03 | 126592.439 | 126593.004 | -5.6E-01 | 126618.676 | 126619.241 | -5.6E-01 | -64.510 | -63.951 | -5.6E-01 |
| 1229 | Te | 136 | 52 | 84 | 32 | 8.319433 | 8.324531 | -5.1E-03 | 126582.003 | 126581.279 | 7.2E-01 | 126608.760 | 126608.035 | 7.2E-01 | -74.426 | -75.157 | 7.3E-01 |
| 1230 | I | 136 | 53 | 83 | 30 | 8.351325 | 8.351629 | -3.0E-04 | 126576.364 | 126576.291 | 7.3E-02 | 126603.640 | 126603.567 | 7.3E-02 | -79.545 | -79.625 | 8.0E-02 |
| 1231 | Xe | 136 | 54 | 82 | 28 | 8.396188 | 8.392844 | 3.3E-03 | 126568.960 | 126569.382 | -4.2E-01 | 126596.757 | 126597.179 | -4.2E-01 | -86.429 | -86.014 | -4.2E-01 |
| 1232 | Cs | 136 | 55 | 81 | 26 | 8.389772 | 8.387411 | 2.4E-03 | 126568.530 | 126568.818 | -2.9E-01 | 126596.847 | 126597.135 | -2.9E-01 | -86.339 | -86.058 | -2.8E-01 |
| 1233 | Ba | 136 | 56 | 80 | 24 | 8.402756 | 8.399012 | 3.7E-03 | 126565.462 | 126565.937 | -4.7E-01 | 126594.299 | 126594.774 | -4.7E-01 | -88.887 | -88.419 | -4.7E-01 |
| 1234 | La | 136 | 57 | 79 | 22 | 8.376050 | 8.371447 | 4.6E-03 | 126567.790 | 126568.382 | -5.9E-01 | 126597.148 | 126597.740 | -5.9E-01 | -86.037 | -85.453 | -5.8E-01 |
| 1235 | Ce | 136 | 58 | 78 | 20 | 8.373762 | 8.366280 | 7.5E-03 | 126566.798 | 126567.780 | -9.8E-01 | 126596.677 | 126597.659 | -9.8E-01 | -86.509 | -85.533 | -9.8E-01 |
| 1236 | Pr | 136 | 59 | 77 | 18 | 8.330008 | 8.322219 | 7.8E-03 | 126571.445 | 126572.468 | -1.0E+00 | 126601.845 | 126602.868 | -1.0E+00 | -81.340 | -80.324 | -1.0E+00 |
| 1237 | Nd | 136 | 60 | 76 | 16 | 8.308512 | 8.301310 | 7.2E-03 | 126573.065 | 126574.007 | -9.4E-01 | 126603.986 | 126604.929 | -9.4E-01 | -79.199 | -78.263 | -9.4E-01 |
| 1238 | Pm | 136 | 61 | 75 | 14 | 8.243798 | 8.238255 | 5.5E-03 | 126580.562 | 126581.278 | -7.2E-01 | 126612.005 | 126612.721 | -7.2E-01 | -71.181 | -70.471 | -7.1E-01 |
| 1239 | Sm | 136 | 62 | 74 | 12 | 8.205916 | 8.199620 | 6.3E-03 | 126584.409 | 126585.227 | -8.2E-01 | 126616.375 | 126617.193 | -8.2E-01 | -66.811 | -66.000 | -8.1E-01 |
| 1240 | Sb | 137 | 51 | 86 | 35 | 8.218255 | 8.216816 | 1.4E-03 | 127528.413 | 127528.579 | -1.7E-01 | 127554.650 | 127554.816 | -1.7E-01 | -60.030 | -59.870 | -1.6E-01 |
| 1241 | Te | 137 | 52 | 85 | 33 | 8.280238 | 8.280959 | -7.2E-04 | 127518.619 | 127518.489 | 1.3E-01 | 127545.376 | 127545.246 | 1.3E-01 | -69.304 | -69.441 | 1.4E-01 |
| 1242 | I | 137 | 53 | 84 | 31 | 8.326002 | 8.329571 | -3.6E-03 | 127511.047 | 127510.526 | 5.2E-01 | 127538.324 | 127537.803 | 5.2E-01 | -76.356 | -76.884 | 5.3E-01 |
| 1243 | Xe | 137 | 54 | 83 | 29 | 8.364286 | 8.365672 | -1.4E-03 | 127504.500 | 127504.278 | 2.2E-01 | 127532.296 | 127532.074 | 2.2E-01 | -82.383 | -82.613 | 2.3E-01 |
| 1244 | Cs | 137 | 55 | 82 | 27 | 8.388958 | 8.384233 | 4.7E-03 | 127499.817 | 127500.431 | -6.1E-01 | 127528.134 | 127528.748 | -6.1E-01 | -86.546 | -85.938 | -6.1E-01 |
| 1245 | Ba | 137 | 56 | 81 | 25 | 8.391828 | 8.388958 | 2.9E-03 | 127498.121 | 127498.480 | -3.6E-01 | 127526.958 | 127527.317 | -3.6E-01 | -87.721 | -87.369 | -3.5E-01 |
| 1246 | La | 137 | 57 | 80 | 23 | 8.381880 | 8.379740 | 2.1E-03 | 127498.181 | 127498.440 | -2.6E-01 | 127527.539 | 127527.797 | -2.6E-01 | -87.141 | -86.889 | -2.5E-01 |
| 1247 | Ce | 137 | 58 | 79 | 21 | 8.367249 | 8.362587 | 4.7E-03 | 127498.882 | 127499.485 | -6.0E-01 | 127528.761 | 127529.364 | -6.0E-01 | -85.919 | -85.322 | -6.0E-01 |
| 1248 | Pr | 137 | 59 | 78 | 19 | 8.341707 | 8.336093 | 5.6E-03 | 127501.078 | 127501.811 | -7.3E-01 | 127531.478 | 127532.211 | -7.3E-01 | -83.202 | -82.476 | -7.3E-01 |
| 1249 | Nd | 137 | 60 | 77 | 17 | 8.309593 | 8.302215 | 7.4E-03 | 127504.174 | 127505.147 | -9.7E-01 | 127535.095 | 127536.069 | -9.7E-01 | -79.585 | -78.617 | -9.7E-01 |
| 1250 | Pm | 137 | 61 | 76 | 15 | 8.263651 | 8.258922 | 4.7E-03 | 127509.164 | 127509.774 | -6.1E-01 | 127540.607 | 127541.217 | -6.1E-01 | -74.073 | -73.469 | -6.0E-01 |
| 1251 | Sm | 137 | 62 | 75 | 13 | 8.213806 | 8.206467 | 7.3E-03 | 127514.688 | 127515.655 | -9.7E-01 | 127546.653 | 127547.620 | -9.7E-01 | -68.027 | -67.066 | -9.6E-01 |
| 1252 | Te | 138 | 52 | 86 | 34 | 8.252579 | 8.253479 | -9.0E-04 | 128453.721 | 128453.566 | 1.6E-01 | 128480.478 | 128480.322 | 1.6E-01 | -65.696 | -65.858 | 1.6E-01 |
| 1253 | I | 138 | 53 | 85 | 32 | 8.292444 | 8.290774 | 1.7E-03 | 128446.918 | 128447.116 | -2.0E-01 | 128474.194 | 128474.393 | -2.0E-01 | -71.980 | -71.788 | -1.9E-01 |
| 1254 | Xe | 138 | 54 | 84 | 30 | 8.344690 | 8.347593 | -2.9E-03 | 128438.405 | 128437.972 | 4.3E-01 | 128466.202 | 128465.768 | 4.3E-01 | -79.972 | -80.412 | 4.4E-01 |
| 1255 | Cs | 138 | 55 | 83 | 28 | 8.360143 | 8.361735 | -1.6E-03 | 128434.970 | 128434.717 | 2.5E-01 | 128463.287 | 128463.034 | 2.5E-01 | -82.887 | -83.147 | 2.6E-01 |
| 1256 | Ba | 138 | 56 | 82 | 26 | 8.393422 | 8.389468 | 4.0E-03 | 128429.075 | 128429.586 | -5.1E-01 | 128457.912 | 128458.424 | -5.1E-01 | -88.262 | -87.757 | -5.0E-01 |
| 1257 | La | 138 | 57 | 81 | 24 | 8.375144 | 8.373500 | 1.6E-03 | 128430.294 | 128430.486 | -1.9E-01 | 128459.652 | 128459.844 | -1.9E-01 | -86.522 | -86.336 | -1.9E-01 |
| 1258 | Ce | 138 | 58 | 80 | 22 | 8.377061 | 8.374560 | 2.5E-03 | 128428.726 | 128429.036 | -3.1E-01 | 128458.605 | 128458.915 | -3.1E-01 | -87.569 | -87.266 | -3.0E-01 |
| 1259 | Pr | 138 | 59 | 79 | 20 | 8.339240 | 8.336716 | 2.5E-03 | 128432.642 | 128432.954 | -3.1E-01 | 128463.042 | 128463.354 | -3.1E-01 | -83.132 | -82.826 | -3.1E-01 |
| 1260 | Nd | 138 | 60 | 78 | 18 | 8.325502 | 8.320839 | 4.7E-03 | 128433.234 | 128433.840 | -6.1E-01 | 128464.155 | 128464.762 | -6.1E-01 | -82.018 | -81.418 | -6.0E-01 |
| 1261 | Pm | 138 | 61 | 77 | 16 | 8.268544 | 8.265188 | 3.4E-03 | 128439.790 | 128440.215 | -4.3E-01 | 128471.233 | 128471.659 | -4.3E-01 | -74.940 | -74.522 | -4.2E-01 |
| 1262 | Sm | 138 | 62 | 76 | 14 | 8.237928 | 8.232958 | 5.0E-03 | 128442.711 | 128443.358 | -6.5E-01 | 128474.676 | 128475.323 | -6.5E-01 | -71.498 | -70.857 | -6.4E-01 |
| 1263 | Eu | 138 | 63 | 75 | 12 | 8.161620 | 8.158003 | 3.6E-03 | 128451.936 | 128452.396 | -4.6E-01 | 128484.424 | 128484.884 | -4.6E-01 | -61.750 | -61.296 | -4.5E-01 |
| 1264 | Te | 139 | 52 | 87 | 35 | 8.211771 | 8.210024 | 1.7E-03 | 129390.706 | 129390.918 | -2.1E-01 | 129417.463 | 129417.674 | -2.1E-01 | -60.205 | -60.000 | -2.1E-01 |
| 1265 | I | 139 | 53 | 86 | 33 | 8.265523 | 8.264673 | 8.5E-04 | 129381.932 | 129382.019 | -8.7E-02 | 129409.209 | 129409.295 | -8.6E-02 | -68.459 | -68.379 | -8.0E-02 |
| 1266 | Xe | 139 | 54 | 85 | 31 | 8.311590 | 8.310917 | 6.7E-04 | 129374.227 | 129374.288 | -6.1E-02 | 129402.023 | 129402.084 | -6.1E-02 | -75.645 | -75.590 | -5.4E-02 |
| 1267 | Cs | 139 | 55 | 84 | 29 | 8.342339 | 8.344989 | -2.7E-03 | 129368.650 | 129368.248 | 4.0E-01 | 129396.967 | 129396.565 | 4.0E-01 | -80.701 | -81.109 | 4.1E-01 |
| 1268 | Ba | 139 | 56 | 83 | 27 | 8.367019 | 8.368139 | -1.1E-03 | 129363.917 | 129363.727 | 1.9E-01 | 129392.754 | 129392.564 | 1.9E-01 | -84.914 | -85.110 | 2.0E-01 |
| 1269 | La | 139 | 57 | 82 | 25 | 8.378043 | 8.374425 | 3.6E-03 | 129361.081 | 129361.550 | -4.7E-01 | 129390.439 | 129390.907 | -4.7E-01 | -87.229 | -86.767 | -4.6E-01 |
| 1270 | Ce | 139 | 58 | 81 | 23 | 8.370411 | 8.368026 | 2.4E-03 | 129360.839 | 129361.135 | -3.0E-01 | 129390.718 | 129391.014 | -3.0E-01 | -86.950 | -86.661 | -2.9E-01 |
| 1271 | Pr | 139 | 59 | 80 | 21 | 8.349466 | 8.348518 | 9.5E-04 | 129362.447 | 129362.542 | -9.5E-02 | 129392.847 | 129392.942 | -9.5E-02 | -84.821 | -84.732 | -8.9E-02 |
| 1272 | Nd | 139 | 60 | 79 | 19 | 8.323647 | 8.320850 | 2.8E-03 | 129364.732 | 129365.083 | -3.5E-01 | 129395.653 | 129396.005 | -3.5E-01 | -82.015 | -81.669 | -3.5E-01 |
| 1273 | Pm | 139 | 61 | 78 | 17 | 8.285544 | 8.283647 | 1.9E-03 | 129368.724 | 129368.950 | -2.3E-01 | 129400.167 | 129400.393 | -2.3E-01 | -77.501 | -77.281 | -2.2E-01 |
| 1274 | Sm | 139 | 62 | 77 | 15 | 8.243078 | 8.238309 | 4.8E-03 | 129373.322 | 129373.947 | -6.2E-01 | 129405.288 | 129405.912 | -6.2E-01 | -72.380 | -71.762 | -6.2E-01 |
| 1275 | Eu | 139 | 63 | 76 | 13 | 8.187218 | 8.183816 | 3.4E-03 | 129379.782 | 129380.216 | -4.3E-01 | 129412.270 | 129412.704 | -4.3E-01 | -65.398 | -64.971 | -4.3E-01 |
| 1276 | Te | 140 | 52 | 88 | 36 | 8.183279 | 8.183656 | -3.8E-04 | 130326.049 | 130325.965 | 8.4E-02 | 130352.805 | 130352.721 | 8.4E-02 | -56.357 | -56.447 | 9.0E-02 |
| 1277 | I | 140 | 53 | 87 | 34 | 8.229399 | 8.226313 | 3.1E-03 | 130318.290 | 130318.690 | -4.0E-01 | 130345.566 | 130345.966 | -4.0E-01 | -63.596 | -63.202 | -3.9E-01 |
| 1278 | Xe | 140 | 54 | 86 | 32 | 8.290887 | 8.288994 | 1.9E-03 | 130308.379 | 130308.612 | -2.3E-01 | 130336.175 | 130336.408 | -2.3E-01 | -72.986 | -72.761 | -2.3E-01 |
| 1279 | Cs | 140 | 55 | 85 | 30 | 8.314326 | 8.313505 | 8.2E-04 | 130303.795 | 130303.877 | -8.2E-02 | 130332.111 | 130332.193 | -8.2E-02 | -77.050 | -76.975 | -7.5E-02 |
| 1280 | Ba | 140 | 56 | 84 | 28 | 8.353166 | 8.356043 | -2.9E-03 | 130297.055 | 130296.618 | 4.4E-01 | 130325.892 | 130325.455 | 4.4E-01 | -83.270 | -83.714 | 4.4E-01 |
| 1281 | La | 140 | 57 | 83 | 26 | 8.355064 | 8.357847 | -2.8E-03 | 130295.486 | 130295.061 | 4.2E-01 | 130324.844 | 130324.419 | 4.2E-01 | -84.318 | -84.749 | 4.3E-01 |
| 1282 | Ce | 140 | 58 | 82 | 24 | 8.376339 | 8.372932 | 3.4E-03 | 130291.204 | 130291.645 | -4.4E-01 | 130321.083 | 130321.524 | -4.4E-01 | -88.079 | -87.644 | -4.3E-01 |
| 1283 | Pr | 140 | 59 | 81 | 22 | 8.346551 | 8.345961 | 5.9E-04 | 130294.071 | 130294.117 | -4.6E-02 | 130324.471 | 130324.517 | -4.6E-02 | -84.691 | -84.651 | -4.0E-02 |
| 1284 | Nd | 140 | 60 | 80 | 20 | 8.337838 | 8.336759 | 1.1E-03 | 130293.986 | 130294.101 | -1.1E-01 | 130324.908 | 130325.022 | -1.1E-01 | -84.254 | -84.146 | -1.1E-01 |
| 1285 | Pm | 140 | 61 | 79 | 18 | 8.289070 | 8.288349 | 7.2E-04 | 130299.510 | 130299.573 | -6.3E-02 | 130330.953 | 130331.017 | -6.4E-02 | -78.209 | -78.152 | -5.7E-02 |
| 1286 | Sm | 140 | 62 | 78 | 16 | 8.263820 | 8.261989 | 1.8E-03 | 130301.740 | 130301.958 | -2.2E-01 | 130333.706 | 130333.924 | -2.2E-01 | -75.456 | -75.245 | -2.1E-01 |
| 1287 | Eu | 140 | 63 | 77 | 14 | 8.197732 | 8.194929 | 2.8E-03 | 130309.688 | 130310.041 | -3.5E-01 | 130342.176 | 130342.529 | -3.5E-01 | -66.986 | -66.639 | -3.5E-01 |
| 1288 | Gd | 140 | 64 | 76 | 12 | 8.154975 | 8.151924 | 3.1E-03 | 130314.369 | 130314.756 | -3.9E-01 | 130347.380 | 130347.767 | -3.9E-01 | -61.782 | -61.402 | -3.8E-01 |
| 1289 | Tb | 140 | 65 | 75 | 10 | 8.068672 | 8.065722 | 2.9E-03 | 130325.146 | 130325.519 | -3.7E-01 | 130358.680 | 130359.052 | -3.7E-01 | -50.482 | -50.116 | -3.7E-01 |
| 1290 | Xe | 141 | 54 | 87 | 33 | 8.255364 | 8.252392 | 3.0E-03 | 131244.662 | 131245.049 | -3.9E-01 | 131272.459 | 131272.845 | -3.9E-01 | -68.197 | -67.817 | -3.8E-01 |
| 1291 | Cs | 141 | 55 | 86 | 31 | 8.294353 | 8.292523 | 1.8E-03 | 131237.862 | 131238.087 | -2.3E-01 | 131266.179 | 131266.404 | -2.2E-01 | -74.477 | -74.259 | -2.2E-01 |
| 1292 | Ba | 141 | 56 | 85 | 29 | 8.326079 | 8.325919 | 1.6E-04 | 131232.086 | 131232.075 | 1.1E-02 | 131260.923 | 131260.912 | 1.1E-02 | -79.733 | -79.751 | 1.8E-02 |
| 1293 | La | 141 | 57 | 84 | 27 | 8.343238 | 8.346887 | -3.6E-03 | 131228.363 | 131227.814 | 5.5E-01 | 131257.721 | 131257.172 | 5.5E-01 | -82.935 | -83.491 | 5.6E-01 |
| 1294 | Ce | 141 | 58 | 83 | 25 | 8.355430 | 8.356906 | -1.5E-03 | 131225.341 | 131225.098 | 2.4E-01 | 131255.220 | 131254.976 | 2.4E-01 | -85.436 | -85.686 | 2.5E-01 |
| 1295 | Pr | 141 | 59 | 82 | 23 | 8.353998 | 8.350994 | 3.0E-03 | 131224.239 | 131224.627 | -3.9E-01 | 131254.639 | 131255.027 | -3.9E-01 | -86.016 | -85.636 | -3.8E-01 |
| 1296 | Nd | 141 | 60 | 81 | 21 | 8.335520 | 8.333522 | 2.0E-03 | 131225.541 | 131225.786 | -2.4E-01 | 131256.462 | 131256.707 | -2.5E-01 | -84.193 | -83.955 | -2.4E-01 |
| 1297 | Pm | 141 | 61 | 80 | 19 | 8.303940 | 8.303881 | 5.9E-05 | 131228.689 | 131228.660 | 2.9E-02 | 131260.133 | 131260.104 | 2.9E-02 | -80.523 | -80.559 | 3.6E-02 |
| 1298 | Sm | 141 | 62 | 79 | 17 | 8.265845 | 8.265741 | 1.0E-04 | 131232.756 | 131232.733 | 2.3E-02 | 131264.722 | 131264.698 | 2.4E-02 | -75.934 | -75.964 | 3.0E-02 |
| 1299 | Eu | 141 | 63 | 78 | 15 | 8.217684 | 8.218088 | -4.0E-04 | 131238.242 | 131238.146 | 9.6E-02 | 131270.730 | 131270.634 | 9.6E-02 | -69.926 | -70.028 | 1.0E-01 |
| 1300 | Gd | 141 | 64 | 77 | 13 | 8.164608 | 8.161637 | 3.0E-03 | 131244.421 | 131244.800 | -3.8E-01 | 131277.432 | 131277.811 | -3.8E-01 | -63.224 | -62.852 | -3.7E-01 |
| 1301 | Tb | 141 | 65 | 76 | 11 | 8.097475 | 8.096543 | 9.3E-04 | 131252.582 | 131252.673 | -9.1E-02 | 131286.115 | 131286.206 | -9.1E-02 | -54.541 | -54.457 | -8.4E-02 |
| 1302 | I | 142 | 53 | 89 | 36 | 8.165019 | 8.163242 | 1.8E-03 | 132190.104 | 132190.324 | -2.2E-01 | 132217.380 | 132217.601 | -2.2E-01 | -54.770 | -54.556 | -2.1E-01 |
| 1303 | Xe | 142 | 54 | 88 | 34 | 8.233169 | 8.231532 | 1.6E-03 | 132179.124 | 132179.324 | -2.0E-01 | 132206.920 | 132207.120 | -2.0E-01 | -65.230 | -65.036 | -1.9E-01 |
| 1304 | Cs | 142 | 55 | 87 | 32 | 8.264900 | 8.261237 | 3.7E-03 | 132173.315 | 132173.803 | -4.9E-01 | 132201.632 | 132202.119 | -4.9E-01 | -70.518 | -70.038 | -4.8E-01 |
| 1305 | Ba | 142 | 56 | 86 | 30 | 8.310974 | 8.309663 | 1.3E-03 | 132165.470 | 132165.622 | -1.5E-01 | 132194.307 | 132194.460 | -1.5E-01 | -77.843 | -77.697 | -1.5E-01 |
| 1306 | La | 142 | 57 | 85 | 28 | 8.320825 | 8.321991 | -1.2E-03 | 132162.768 | 132162.568 | 2.0E-01 | 132192.126 | 132191.926 | 2.0E-01 | -80.024 | -80.231 | 2.1E-01 |
| 1307 | Ce | 142 | 58 | 84 | 26 | 8.347068 | 8.350809 | -3.7E-03 | 132157.738 | 132157.172 | 5.7E-01 | 132187.617 | 132187.051 | 5.7E-01 | -84.533 | -85.106 | 5.7E-01 |
| 1308 | Pr | 142 | 59 | 83 | 24 | 8.336316 | 8.339629 | -3.3E-03 | 132157.962 | 132157.455 | 5.1E-01 | 132188.362 | 132187.855 | 5.1E-01 | -83.788 | -84.302 | 5.1E-01 |
| 1309 | Nd | 142 | 60 | 82 | 22 | 8.346029 | 8.342704 | 3.3E-03 | 132155.278 | 132155.714 | -4.4E-01 | 132186.200 | 132186.635 | -4.4E-01 | -85.950 | -85.521 | -4.3E-01 |
| 1310 | Pm | 142 | 61 | 81 | 20 | 8.306662 | 8.304860 | 1.8E-03 | 132159.564 | 132159.783 | -2.2E-01 | 132191.008 | 132191.226 | -2.2E-01 | -81.142 | -80.931 | -2.1E-01 |
| 1311 | Sm | 142 | 62 | 80 | 18 | 8.285973 | 8.285810 | 1.6E-04 | 132161.198 | 132161.183 | 1.5E-02 | 132193.163 | 132193.148 | 1.5E-02 | -78.987 | -79.009 | 2.2E-02 |
| 1312 | Eu | 142 | 63 | 79 | 16 | 8.226428 | 8.226940 | -5.1E-04 | 132168.348 | 132168.237 | 1.1E-01 | 132200.836 | 132200.725 | 1.1E-01 | -71.314 | -71.432 | 1.2E-01 |
| 1313 | Gd | 142 | 64 | 78 | 14 | 8.190256 | 8.190471 | -2.2E-04 | 132172.180 | 132172.110 | 7.0E-02 | 132205.190 | 132205.120 | 7.0E-02 | -66.960 | -67.037 | 7.7E-02 |
| 1314 | Tb | 142 | 65 | 77 | 12 | 8.111507 | 8.112412 | -9.1E-04 | 132182.057 | 132181.888 | 1.7E-01 | 132215.590 | 132215.421 | 1.7E-01 | -56.560 | -56.735 | 1.8E-01 |
| 1315 | Xe | 143 | 54 | 89 | 35 | 8.196885 | 8.195351 | 1.5E-03 | 133115.645 | 133115.832 | -1.9E-01 | 133143.441 | 133143.628 | -1.9E-01 | -60.203 | -60.023 | -1.8E-01 |
| 1316 | Cs | 143 | 55 | 88 | 33 | 8.243657 | 8.240923 | 2.7E-03 | 133107.654 | 133108.012 | -3.6E-01 | 133135.970 | 133136.328 | -3.6E-01 | -67.674 | -67.323 | -3.5E-01 |
| 1317 | Ba | 143 | 56 | 87 | 31 | 8.281986 | 8.279311 | 2.7E-03 | 133100.870 | 133101.218 | -3.5E-01 | 133129.707 | 133130.056 | -3.5E-01 | -73.937 | -73.595 | -3.4E-01 |
| 1318 | La | 143 | 57 | 86 | 29 | 8.306126 | 8.306543 | -4.2E-04 | 133096.115 | 133096.020 | 9.5E-02 | 133125.472 | 133125.378 | 9.4E-02 | -78.171 | -78.272 | 1.0E-01 |
| 1319 | Ce | 143 | 58 | 85 | 27 | 8.324675 | 8.326805 | -2.1E-03 | 133092.159 | 133091.819 | 3.4E-01 | 133122.038 | 133121.698 | 3.4E-01 | -81.606 | -81.953 | 3.5E-01 |
| 1320 | Pr | 143 | 59 | 84 | 25 | 8.329426 | 8.334436 | -5.0E-03 | 133090.176 | 133089.423 | 7.5E-01 | 133120.576 | 133119.823 | 7.5E-01 | -83.068 | -83.827 | 7.6E-01 |
| 1321 | Nd | 143 | 60 | 83 | 23 | 8.330487 | 8.331397 | -9.1E-04 | 133088.720 | 133088.553 | 1.7E-01 | 133119.642 | 133119.475 | 1.7E-01 | -84.002 | -84.176 | 1.7E-01 |
| 1322 | Pm | 143 | 61 | 82 | 21 | 8.317732 | 8.313901 | 3.8E-03 | 133089.240 | 133089.750 | -5.1E-01 | 133120.683 | 133121.194 | -5.1E-01 | -82.960 | -82.457 | -5.0E-01 |
| 1323 | Sm | 143 | 62 | 81 | 19 | 8.288179 | 8.285821 | 2.4E-03 | 133092.162 | 133092.461 | -3.0E-01 | 133124.127 | 133124.426 | -3.0E-01 | -79.517 | -79.225 | -2.9E-01 |
| 1324 | Eu | 143 | 63 | 80 | 17 | 8.245817 | 8.246410 | -5.9E-04 | 133096.915 | 133096.791 | 1.2E-01 | 133129.403 | 133129.279 | 1.2E-01 | -74.241 | -74.372 | 1.3E-01 |
| 1325 | Gd | 143 | 64 | 79 | 15 | 8.198318 | 8.198039 | 2.8E-04 | 133102.402 | 133102.402 | -2.9E-04 | 133135.413 | 133135.413 | 2.0E-04 | -68.231 | -68.238 | 6.6E-03 |
| 1326 | Tb | 143 | 65 | 78 | 13 | 8.138217 | 8.140325 | -2.1E-03 | 133109.691 | 133109.349 | 3.4E-01 | 133143.225 | 133142.883 | 3.4E-01 | -60.419 | -60.768 | 3.5E-01 |
| 1327 | Dy | 143 | 66 | 77 | 11 | 8.075052 | 8.073198 | 1.9E-03 | 133117.418 | 133117.642 | -2.2E-01 | 133151.475 | 133151.699 | -2.2E-01 | -52.169 | -51.952 | -2.2E-01 |
| 1328 | Xe | 144 | 54 | 90 | 36 | 8.172884 | 8.173778 | -8.9E-04 | 134050.469 | 134050.308 | 1.6E-01 | 134078.266 | 134078.104 | 1.6E-01 | -56.872 | -57.040 | 1.7E-01 |
| 1329 | Cs | 144 | 55 | 89 | 34 | 8.211883 | 8.210473 | 1.4E-03 | 134043.551 | 134043.721 | -1.7E-01 | 134071.867 | 134072.037 | -1.7E-01 | -63.271 | -63.107 | -1.6E-01 |
| 1330 | Ba | 144 | 56 | 88 | 32 | 8.265454 | 8.263876 | 1.6E-03 | 134034.534 | 134034.727 | -1.9E-01 | 134063.371 | 134063.564 | -1.9E-01 | -71.767 | -71.581 | -1.9E-01 |
| 1331 | La | 144 | 57 | 87 | 30 | 8.281428 | 8.281322 | 1.1E-04 | 134030.930 | 134030.911 | 1.9E-02 | 134060.288 | 134060.269 | 1.9E-02 | -74.850 | -74.876 | 2.6E-02 |
| 1332 | Ce | 144 | 58 | 86 | 28 | 8.314759 | 8.316203 | -1.4E-03 | 134024.827 | 134024.584 | 2.4E-01 | 134054.706 | 134054.463 | 2.4E-01 | -80.432 | -80.682 | 2.5E-01 |
| 1333 | Pr | 144 | 59 | 85 | 26 | 8.311539 | 8.315517 | -4.0E-03 | 134023.987 | 134023.379 | 6.1E-01 | 134054.387 | 134053.779 | 6.1E-01 | -80.750 | -81.366 | 6.2E-01 |
| 1334 | Nd | 144 | 60 | 84 | 24 | 8.326922 | 8.331032 | -4.1E-03 | 134020.468 | 134019.840 | 6.3E-01 | 134051.390 | 134050.762 | 6.3E-01 | -83.748 | -84.383 | 6.4E-01 |
| 1335 | Pm | 144 | 61 | 83 | 22 | 8.305295 | 8.307174 | -1.9E-03 | 134022.279 | 134021.971 | 3.1E-01 | 134053.722 | 134053.414 | 3.1E-01 | -81.416 | -81.731 | 3.1E-01 |
| 1336 | Sm | 144 | 62 | 82 | 20 | 8.303678 | 8.299206 | 4.5E-03 | 134021.207 | 134021.813 | -6.1E-01 | 134053.172 | 134053.778 | -6.1E-01 | -81.965 | -81.366 | -6.0E-01 |
| 1337 | Eu | 144 | 63 | 81 | 18 | 8.254173 | 8.250956 | 3.2E-03 | 134027.031 | 134027.455 | -4.2E-01 | 134059.519 | 134059.943 | -4.2E-01 | -75.619 | -75.202 | -4.2E-01 |
| 1338 | Gd | 144 | 64 | 80 | 16 | 8.221937 | 8.222491 | -5.5E-04 | 134030.368 | 134030.249 | 1.2E-01 | 134063.378 | 134063.259 | 1.2E-01 | -71.760 | -71.886 | 1.3E-01 |
| 1339 | Tb | 144 | 65 | 79 | 14 | 8.151287 | 8.153421 | -2.1E-03 | 134039.236 | 134038.889 | 3.5E-01 | 134072.770 | 134072.422 | 3.5E-01 | -62.368 | -62.723 | 3.5E-01 |
| 1340 | Dy | 144 | 66 | 78 | 12 | 8.105589 | 8.107217 | -1.6E-03 | 134044.511 | 134044.236 | 2.8E-01 | 134078.568 | 134078.292 | 2.8E-01 | -56.570 | -56.852 | 2.8E-01 |
| 1341 | Ho | 144 | 67 | 77 | 10 | 8.017097 | 8.018756 | -1.7E-03 | 134055.948 | 134055.668 | 2.8E-01 | 134090.528 | 134090.248 | 2.8E-01 | -44.610 | -44.897 | 2.9E-01 |
| 1342 | Xe | 145 | 54 | 91 | 37 | 8.135087 | 8.135737 | -6.5E-04 | 134987.342 | 134987.216 | 1.3E-01 | 135015.139 | 135015.012 | 1.3E-01 | -51.493 | -51.627 | 1.3E-01 |
| 1343 | Cs | 145 | 55 | 90 | 35 | 8.188743 | 8.189192 | -4.5E-04 | 134978.260 | 134978.161 | 9.9E-02 | 135006.576 | 135006.478 | 9.8E-02 | -60.056 | -60.161 | 1.0E-01 |
| 1344 | Ba | 145 | 56 | 89 | 33 | 8.234798 | 8.234145 | 6.5E-04 | 134970.279 | 134970.340 | -6.1E-02 | 134999.116 | 134999.177 | -6.1E-02 | -67.516 | -67.462 | -5.4E-02 |
| 1345 | La | 145 | 57 | 88 | 31 | 8.266087 | 8.266270 | -1.8E-04 | 134964.439 | 134964.378 | 6.1E-02 | 134993.797 | 134993.736 | 6.1E-02 | -72.835 | -72.903 | 6.8E-02 |
| 1346 | Ce | 145 | 58 | 87 | 29 | 8.289875 | 8.291450 | -1.6E-03 | 134959.686 | 134959.423 | 2.6E-01 | 134989.565 | 134989.301 | 2.6E-01 | -77.067 | -77.337 | 2.7E-01 |
| 1347 | Pr | 145 | 59 | 86 | 27 | 8.302127 | 8.305632 | -3.5E-03 | 134956.606 | 134956.062 | 5.4E-01 | 134987.006 | 134986.462 | 5.4E-01 | -79.626 | -80.177 | 5.5E-01 |
| 1348 | Nd | 145 | 60 | 85 | 25 | 8.309186 | 8.312634 | -3.4E-03 | 134954.278 | 134953.742 | 5.4E-01 | 134985.200 | 134984.663 | 5.4E-01 | -81.432 | -81.975 | 5.4E-01 |
| 1349 | Pm | 145 | 61 | 84 | 23 | 8.302656 | 8.307384 | -4.7E-03 | 134953.921 | 134953.198 | 7.2E-01 | 134985.365 | 134984.642 | 7.2E-01 | -81.267 | -81.997 | 7.3E-01 |
| 1350 | Sm | 145 | 62 | 83 | 21 | 8.293012 | 8.292093 | 9.2E-04 | 134954.015 | 134954.110 | -9.5E-02 | 134985.981 | 134986.076 | -9.5E-02 | -80.651 | -80.563 | -8.8E-02 |
| 1351 | Eu | 145 | 63 | 82 | 19 | 8.269274 | 8.263930 | 5.3E-03 | 134956.152 | 134956.889 | -7.4E-01 | 134988.640 | 134989.376 | -7.4E-01 | -77.992 | -77.262 | -7.3E-01 |
| 1352 | Gd | 145 | 64 | 81 | 17 | 8.228930 | 8.225800 | 3.1E-03 | 134960.697 | 134961.112 | -4.1E-01 | 134993.708 | 134994.122 | -4.1E-01 | -72.924 | -72.517 | -4.1E-01 |
| 1353 | Tb | 145 | 65 | 80 | 15 | 8.177865 | 8.177009 | 8.6E-04 | 134966.797 | 134966.880 | -8.3E-02 | 135000.330 | 135000.414 | -8.4E-02 | -66.302 | -66.225 | -7.7E-02 |
| 1354 | Dy | 145 | 66 | 79 | 13 | 8.116888 | 8.118670 | -1.8E-03 | 134974.333 | 134974.033 | 3.0E-01 | 135008.389 | 135008.090 | 3.0E-01 | -58.243 | -58.549 | 3.1E-01 |
| 1355 | Ho | 145 | 67 | 78 | 11 | 8.048578 | 8.051413 | -2.8E-03 | 134982.932 | 134982.479 | 4.5E-01 | 135017.512 | 135017.059 | 4.5E-01 | -49.120 | -49.580 | 4.6E-01 |
| 1356 | Xe | 146 | 54 | 92 | 38 | 8.110415 | 8.111694 | -1.3E-03 | 135922.375 | 135922.156 | 2.2E-01 | 135950.171 | 135949.952 | 2.2E-01 | -47.955 | -48.181 | 2.3E-01 |
| 1357 | Cs | 146 | 55 | 91 | 36 | 8.157207 | 8.157405 | -2.0E-04 | 135914.240 | 135914.179 | 6.1E-02 | 135942.557 | 135942.495 | 6.2E-02 | -55.569 | -55.638 | 6.9E-02 |
| 1358 | Ba | 146 | 56 | 90 | 34 | 8.216036 | 8.218194 | -2.2E-03 | 135904.349 | 135904.000 | 3.5E-01 | 135933.186 | 135932.837 | 3.5E-01 | -64.940 | -65.296 | 3.6E-01 |
| 1359 | La | 146 | 57 | 89 | 32 | 8.238804 | 8.241908 | -3.1E-03 | 135899.721 | 135899.234 | 4.9E-01 | 135929.079 | 135928.591 | 4.9E-01 | -69.047 | -69.541 | 4.9E-01 |
| 1360 | Ce | 146 | 58 | 88 | 30 | 8.278570 | 8.281180 | -2.6E-03 | 135892.612 | 135892.196 | 4.2E-01 | 135922.491 | 135922.075 | 4.2E-01 | -75.635 | -76.058 | 4.2E-01 |
| 1361 | Pr | 146 | 59 | 87 | 28 | 8.280374 | 8.285717 | -5.3E-03 | 135891.045 | 135890.229 | 8.2E-01 | 135921.445 | 135920.629 | 8.2E-01 | -76.681 | -77.504 | 8.2E-01 |
| 1362 | Nd | 146 | 60 | 86 | 26 | 8.304091 | 8.307530 | -3.4E-03 | 135886.279 | 135885.740 | 5.4E-01 | 135917.200 | 135916.661 | 5.4E-01 | -80.926 | -81.471 | 5.5E-01 |
| 1363 | Pm | 146 | 61 | 85 | 24 | 8.288653 | 8.293898 | -5.2E-03 | 135887.228 | 135886.425 | 8.0E-01 | 135918.672 | 135917.869 | 8.0E-01 | -79.454 | -80.264 | 8.1E-01 |
| 1364 | Sm | 146 | 62 | 84 | 22 | 8.293856 | 8.297041 | -3.2E-03 | 135885.164 | 135884.661 | 5.0E-01 | 135917.130 | 135916.627 | 5.0E-01 | -80.996 | -81.506 | 5.1E-01 |
| 1365 | Eu | 146 | 63 | 83 | 20 | 8.261931 | 8.261406 | 5.3E-04 | 135888.521 | 135888.558 | -3.7E-02 | 135921.008 | 135921.046 | -3.8E-02 | -77.117 | -77.087 | -3.1E-02 |
| 1366 | Gd | 146 | 64 | 82 | 18 | 8.249504 | 8.243376 | 6.1E-03 | 135889.030 | 135889.885 | -8.6E-01 | 135922.040 | 135922.896 | -8.6E-01 | -76.086 | -75.237 | -8.5E-01 |
| 1367 | Tb | 146 | 65 | 81 | 16 | 8.187145 | 8.185223 | 1.9E-03 | 135896.829 | 135897.069 | -2.4E-01 | 135930.362 | 135930.603 | -2.4E-01 | -67.763 | -67.530 | -2.3E-01 |
| 1368 | Dy | 146 | 66 | 80 | 14 | 8.146112 | 8.147686 | -1.6E-03 | 135901.514 | 135901.244 | 2.7E-01 | 135935.571 | 135935.300 | 2.7E-01 | -62.555 | -62.833 | 2.8E-01 |
| 1369 | Ho | 146 | 67 | 79 | 12 | 8.063242 | 8.068820 | -5.6E-03 | 135912.308 | 135911.451 | 8.6E-01 | 135946.888 | 135946.032 | 8.6E-01 | -51.238 | -52.101 | 8.6E-01 |
| 1370 | Er | 146 | 68 | 78 | 10 | 8.010512 | 8.013448 | -2.9E-03 | 135918.700 | 135918.229 | 4.7E-01 | 135953.804 | 135953.333 | 4.7E-01 | -44.322 | -44.800 | 4.8E-01 |
| 1371 | Cs | 147 | 55 | 92 | 37 | 8.132468 | 8.133334 | -8.7E-04 | 136849.285 | 136849.125 | 1.6E-01 | 136877.602 | 136877.441 | 1.6E-01 | -52.018 | -52.186 | 1.7E-01 |
| 1372 | Ba | 147 | 56 | 91 | 35 | 8.183240 | 8.186978 | -3.7E-03 | 136840.519 | 136839.936 | 5.8E-01 | 136869.356 | 136868.773 | 5.8E-01 | -60.264 | -60.854 | 5.9E-01 |
| 1373 | La | 147 | 57 | 90 | 33 | 8.221553 | 8.226156 | -4.6E-03 | 136833.584 | 136832.873 | 7.1E-01 | 136862.942 | 136862.231 | 7.1E-01 | -66.678 | -67.396 | 7.2E-01 |
| 1374 | Ce | 147 | 58 | 89 | 31 | 8.252527 | 8.257066 | -4.5E-03 | 136827.727 | 136827.025 | 7.0E-01 | 136857.606 | 136856.904 | 7.0E-01 | -72.014 | -72.723 | 7.1E-01 |
| 1375 | Pr | 147 | 59 | 88 | 29 | 8.270539 | 8.275736 | -5.2E-03 | 136823.776 | 136822.976 | 8.0E-01 | 136854.176 | 136853.376 | 8.0E-01 | -75.444 | -76.251 | 8.1E-01 |
| 1376 | Nd | 147 | 60 | 87 | 27 | 8.283602 | 8.287810 | -4.2E-03 | 136820.552 | 136819.896 | 6.6E-01 | 136851.473 | 136850.818 | 6.5E-01 | -78.147 | -78.809 | 6.6E-01 |
| 1377 | Pm | 147 | 61 | 86 | 25 | 8.284370 | 8.289365 | -5.0E-03 | 136819.135 | 136818.363 | 7.7E-01 | 136850.578 | 136849.807 | 7.7E-01 | -79.042 | -79.820 | 7.8E-01 |
| 1378 | Sm | 147 | 62 | 85 | 23 | 8.280572 | 8.283682 | -3.1E-03 | 136818.388 | 136817.893 | 4.9E-01 | 136850.354 | 136849.859 | 5.0E-01 | -79.266 | -79.768 | 5.0E-01 |
| 1379 | Eu | 147 | 63 | 84 | 21 | 8.263539 | 8.266533 | -3.0E-03 | 136819.588 | 136819.109 | 4.8E-01 | 136852.076 | 136851.597 | 4.8E-01 | -77.544 | -78.030 | 4.9E-01 |
| 1380 | Gd | 147 | 64 | 83 | 19 | 8.243333 | 8.240054 | 3.3E-03 | 136821.253 | 136821.695 | -4.4E-01 | 136854.263 | 136854.706 | -4.4E-01 | -75.357 | -74.921 | -4.4E-01 |
| 1381 | Tb | 147 | 65 | 82 | 17 | 8.206623 | 8.202100 | 4.5E-03 | 136825.344 | 136825.969 | -6.2E-01 | 136858.877 | 136859.502 | -6.3E-01 | -70.743 | -70.125 | -6.2E-01 |
| 1382 | Dy | 147 | 66 | 81 | 15 | 8.156767 | 8.154374 | 2.4E-03 | 136831.367 | 136831.678 | -3.1E-01 | 136865.424 | 136865.735 | -3.1E-01 | -64.196 | -63.892 | -3.0E-01 |
| 1383 | Ho | 147 | 67 | 80 | 13 | 8.094037 | 8.096663 | -2.6E-03 | 136839.283 | 136838.855 | 4.3E-01 | 136873.863 | 136873.435 | 4.3E-01 | -55.757 | -56.192 | 4.3E-01 |
| 1384 | Er | 147 | 68 | 79 | 11 | 8.026475 | 8.028820 | -2.3E-03 | 136847.908 | 136847.521 | 3.9E-01 | 136883.012 | 136882.625 | 3.9E-01 | -46.608 | -47.002 | 3.9E-01 |
| 1385 | Tm | 147 | 69 | 78 | 9 | 7.948817 | 7.952412 | -3.6E-03 | 136858.017 | 136857.446 | 5.7E-01 | 136893.646 | 136893.074 | 5.7E-01 | -35.974 | -36.553 | 5.8E-01 |
| 1386 | Cs | 148 | 55 | 93 | 38 | 8.100151 | 8.098971 | 1.2E-03 | 137785.501 | 137785.643 | -1.4E-01 | 137813.817 | 137813.959 | -1.4E-01 | -47.296 | -47.162 | -1.3E-01 |
| 1387 | Ba | 148 | 56 | 92 | 36 | 8.164441 | 8.168749 | -4.3E-03 | 137774.683 | 137774.012 | 6.7E-01 | 137803.520 | 137802.849 | 6.7E-01 | -57.594 | -58.272 | 6.8E-01 |
| 1388 | La | 148 | 57 | 91 | 34 | 8.193716 | 8.200768 | -7.1E-03 | 137769.047 | 137767.969 | 1.1E+00 | 137798.405 | 137797.327 | 1.1E+00 | -62.709 | -63.794 | 1.1E+00 |
| 1389 | Ce | 148 | 58 | 90 | 32 | 8.240387 | 8.246406 | -6.0E-03 | 137760.837 | 137759.911 | 9.3E-01 | 137790.716 | 137789.790 | 9.3E-01 | -70.398 | -71.331 | 9.3E-01 |
| 1390 | Pr | 148 | 59 | 89 | 30 | 8.249540 | 8.256494 | -7.0E-03 | 137758.178 | 137757.113 | 1.1E+00 | 137788.579 | 137787.514 | 1.1E+00 | -72.535 | -73.607 | 1.1E+00 |
| 1391 | Nd | 148 | 60 | 88 | 28 | 8.277175 | 8.282374 | -5.2E-03 | 137752.784 | 137751.979 | 8.1E-01 | 137783.706 | 137782.900 | 8.1E-01 | -77.408 | -78.221 | 8.1E-01 |
| 1392 | Pm | 148 | 61 | 87 | 26 | 8.268222 | 8.274265 | -6.0E-03 | 137752.805 | 137751.874 | 9.3E-01 | 137784.249 | 137783.317 | 9.3E-01 | -76.865 | -77.804 | 9.4E-01 |
| 1393 | Sm | 148 | 62 | 86 | 24 | 8.279632 | 8.283833 | -4.2E-03 | 137749.812 | 137749.153 | 6.6E-01 | 137781.778 | 137781.118 | 6.6E-01 | -79.336 | -80.003 | 6.7E-01 |
| 1394 | Eu | 148 | 63 | 85 | 22 | 8.253828 | 8.257940 | -4.1E-03 | 137752.327 | 137751.679 | 6.5E-01 | 137784.815 | 137784.167 | 6.5E-01 | -76.299 | -76.954 | 6.5E-01 |
| 1395 | Gd | 148 | 64 | 84 | 20 | 8.248338 | 8.249883 | -1.5E-03 | 137751.834 | 137751.566 | 2.7E-01 | 137784.845 | 137784.577 | 2.7E-01 | -76.269 | -76.544 | 2.8E-01 |
| 1396 | Tb | 148 | 65 | 83 | 18 | 8.204282 | 8.203483 | 8.0E-04 | 137757.049 | 137757.127 | -7.8E-02 | 137790.583 | 137790.661 | -7.8E-02 | -70.531 | -70.460 | -7.1E-02 |
| 1397 | Dy | 148 | 66 | 82 | 16 | 8.180901 | 8.176166 | 4.7E-03 | 137759.204 | 137759.864 | -6.6E-01 | 137793.261 | 137793.921 | -6.6E-01 | -67.853 | -67.200 | -6.5E-01 |
| 1398 | Ho | 148 | 67 | 81 | 14 | 8.108979 | 8.108649 | 3.3E-04 | 137768.543 | 137768.550 | -6.8E-03 | 137803.123 | 137803.130 | -7.0E-03 | -57.991 | -57.991 | -1.0E-04 |
| 1399 | Er | 148 | 68 | 80 | 12 | 8.059692 | 8.062538 | -2.8E-03 | 137774.531 | 137774.067 | 4.6E-01 | 137809.635 | 137809.171 | 4.6E-01 | -51.479 | -51.950 | 4.7E-01 |
| 1400 | Tm | 148 | 69 | 79 | 10 | 7.968500 | 7.974130 | -5.6E-03 | 137786.721 | 137785.845 | 8.8E-01 | 137822.349 | 137821.473 | 8.8E-01 | -38.765 | -39.648 | 8.8E-01 |
| 1401 | La | 149 | 57 | 92 | 35 | 8.176191 | 8.182570 | -6.4E-03 | 138703.030 | 138702.045 | 9.8E-01 | 138732.388 | 138731.403 | 9.8E-01 | -60.220 | -61.212 | 9.9E-01 |
| 1402 | Ce | 149 | 58 | 91 | 33 | 8.214229 | 8.221203 | -7.0E-03 | 138696.059 | 138694.985 | 1.1E+00 | 138725.938 | 138724.864 | 1.1E+00 | -66.670 | -67.751 | 1.1E+00 |
| 1403 | Pr | 149 | 59 | 90 | 31 | 8.238303 | 8.245962 | -7.7E-03 | 138691.168 | 138689.992 | 1.2E+00 | 138721.569 | 138720.392 | 1.2E+00 | -71.039 | -72.223 | 1.2E+00 |
| 1404 | Nd | 149 | 60 | 89 | 29 | 8.255441 | 8.263104 | -7.7E-03 | 138687.311 | 138686.133 | 1.2E+00 | 138718.233 | 138717.055 | 1.2E+00 | -74.375 | -75.561 | 1.2E+00 |
| 1405 | Pm | 149 | 61 | 88 | 27 | 8.261522 | 8.269033 | -7.5E-03 | 138685.101 | 138683.945 | 1.2E+00 | 138716.544 | 138715.388 | 1.2E+00 | -76.064 | -77.227 | 1.2E+00 |
| 1406 | Sm | 149 | 62 | 87 | 25 | 8.263462 | 8.268705 | -5.2E-03 | 138683.507 | 138682.688 | 8.2E-01 | 138715.473 | 138714.654 | 8.2E-01 | -77.135 | -77.961 | 8.3E-01 |
| 1407 | Eu | 149 | 63 | 86 | 23 | 8.253550 | 8.258421 | -4.9E-03 | 138683.680 | 138682.915 | 7.6E-01 | 138716.167 | 138715.403 | 7.6E-01 | -76.440 | -77.212 | 7.7E-01 |
| 1408 | Gd | 149 | 64 | 85 | 21 | 8.239482 | 8.240979 | -1.5E-03 | 138684.471 | 138684.208 | 2.6E-01 | 138717.481 | 138717.219 | 2.6E-01 | -75.127 | -75.396 | 2.7E-01 |
| 1409 | Tb | 149 | 65 | 84 | 19 | 8.209816 | 8.213073 | -3.3E-03 | 138687.586 | 138687.060 | 5.3E-01 | 138721.119 | 138720.594 | 5.3E-01 | -71.489 | -72.021 | 5.3E-01 |
| 1410 | Dy | 149 | 66 | 83 | 17 | 8.179133 | 8.176361 | 2.8E-03 | 138690.852 | 138691.224 | -3.7E-01 | 138724.909 | 138725.281 | -3.7E-01 | -67.699 | -67.334 | -3.6E-01 |
| 1411 | Ho | 149 | 67 | 82 | 15 | 8.133366 | 8.129435 | 3.9E-03 | 138696.365 | 138696.910 | -5.4E-01 | 138730.945 | 138731.490 | -5.4E-01 | -61.662 | -61.125 | -5.4E-01 |
| 1412 | Er | 149 | 68 | 81 | 13 | 8.074955 | 8.072683 | 2.3E-03 | 138703.762 | 138704.059 | -3.0E-01 | 138738.866 | 138739.163 | -3.0E-01 | -53.742 | -53.452 | -2.9E-01 |
| 1413 | Ce | 150 | 58 | 92 | 34 | 8.201122 | 8.208552 | -7.4E-03 | 139629.376 | 139628.227 | 1.1E+00 | 139659.255 | 139658.106 | 1.1E+00 | -64.847 | -66.003 | 1.2E+00 |
| 1414 | Pr | 150 | 59 | 91 | 32 | 8.218930 | 8.225964 | -7.0E-03 | 139625.401 | 139624.311 | 1.1E+00 | 139655.802 | 139654.711 | 1.1E+00 | -68.300 | -69.398 | 1.1E+00 |
| 1415 | Nd | 150 | 60 | 90 | 30 | 8.249572 | 8.257249 | -7.7E-03 | 139619.501 | 139618.313 | 1.2E+00 | 139650.423 | 139649.235 | 1.2E+00 | -73.679 | -74.874 | 1.2E+00 |
| 1416 | Pm | 150 | 61 | 89 | 28 | 8.243806 | 8.254244 | -1.0E-02 | 139619.062 | 139617.459 | 1.6E+00 | 139650.505 | 139648.903 | 1.6E+00 | -73.596 | -75.206 | 1.6E+00 |
| 1417 | Sm | 150 | 62 | 88 | 26 | 8.261617 | 8.267841 | -6.2E-03 | 139615.086 | 139614.115 | 9.7E-01 | 139647.051 | 139646.080 | 9.7E-01 | -77.050 | -78.029 | 9.8E-01 |
| 1418 | Eu | 150 | 63 | 87 | 24 | 8.241344 | 8.247788 | -6.4E-03 | 139616.822 | 139615.817 | 1.0E+00 | 139649.310 | 139648.305 | 1.0E+00 | -74.792 | -75.804 | 1.0E+00 |
| 1419 | Gd | 150 | 64 | 86 | 22 | 8.242607 | 8.246069 | -3.5E-03 | 139615.328 | 139614.769 | 5.6E-01 | 139648.338 | 139647.780 | 5.6E-01 | -75.764 | -76.329 | 5.7E-01 |
| 1420 | Tb | 150 | 65 | 85 | 20 | 8.206338 | 8.208865 | -2.5E-03 | 139619.463 | 139619.044 | 4.2E-01 | 139652.996 | 139652.577 | 4.2E-01 | -71.106 | -71.532 | 4.3E-01 |
| 1421 | Dy | 150 | 66 | 84 | 18 | 8.189148 | 8.190715 | -1.6E-03 | 139620.736 | 139620.460 | 2.8E-01 | 139654.793 | 139654.517 | 2.8E-01 | -69.309 | -69.593 | 2.8E-01 |
| 1422 | Ho | 150 | 67 | 83 | 16 | 8.134841 | 8.134535 | 3.1E-04 | 139627.576 | 139627.581 | -4.5E-03 | 139662.156 | 139662.161 | -4.7E-03 | -61.946 | -61.948 | 2.7E-03 |
| 1423 | Er | 150 | 68 | 82 | 14 | 8.102195 | 8.098735 | 3.5E-03 | 139631.167 | 139631.644 | -4.8E-01 | 139666.271 | 139666.747 | -4.8E-01 | -57.831 | -57.362 | -4.7E-01 |
| 1424 | Ce | 151 | 58 | 93 | 35 | 8.176277 | 8.180922 | -4.6E-03 | 140564.492 | 140563.756 | 7.4E-01 | 140594.371 | 140593.635 | 7.4E-01 | -61.225 | -61.969 | 7.4E-01 |
| 1425 | Pr | 151 | 59 | 92 | 33 | 8.207880 | 8.213361 | -5.5E-03 | 140558.416 | 140557.553 | 8.6E-01 | 140588.816 | 140587.953 | 8.6E-01 | -66.779 | -67.650 | 8.7E-01 |
| 1426 | Nd | 151 | 60 | 91 | 31 | 8.230268 | 8.237202 | -6.9E-03 | 140553.732 | 140552.649 | 1.1E+00 | 140584.654 | 140583.570 | 1.1E+00 | -70.942 | -72.033 | 1.1E+00 |
| 1427 | Pm | 151 | 61 | 90 | 29 | 8.241266 | 8.248442 | -7.2E-03 | 140550.767 | 140549.646 | 1.1E+00 | 140582.211 | 140581.090 | 1.1E+00 | -73.385 | -74.513 | 1.1E+00 |
| 1428 | Sm | 151 | 62 | 89 | 27 | 8.243967 | 8.252835 | -8.9E-03 | 140549.055 | 140547.678 | 1.4E+00 | 140581.020 | 140579.643 | 1.4E+00 | -74.576 | -75.960 | 1.4E+00 |
| 1429 | Eu | 151 | 63 | 88 | 25 | 8.239292 | 8.247023 | -7.7E-03 | 140548.456 | 140547.250 | 1.2E+00 | 140580.944 | 140579.738 | 1.2E+00 | -74.652 | -75.865 | 1.2E+00 |
| 1430 | Gd | 151 | 64 | 87 | 23 | 8.231038 | 8.235149 | -4.1E-03 | 140548.398 | 140547.738 | 6.6E-01 | 140581.408 | 140580.748 | 6.6E-01 | -74.188 | -74.855 | 6.7E-01 |
| 1431 | Tb | 151 | 65 | 86 | 21 | 8.208870 | 8.213957 | -5.1E-03 | 140550.440 | 140549.631 | 8.1E-01 | 140583.973 | 140583.165 | 8.1E-01 | -71.623 | -72.438 | 8.2E-01 |
| 1432 | Dy | 151 | 66 | 85 | 19 | 8.184676 | 8.185751 | -1.1E-03 | 140552.787 | 140552.584 | 2.0E-01 | 140586.844 | 140586.641 | 2.0E-01 | -68.752 | -68.962 | 2.1E-01 |
| 1433 | Ho | 151 | 67 | 84 | 17 | 8.145525 | 8.148244 | -2.7E-03 | 140557.393 | 140556.941 | 4.5E-01 | 140591.973 | 140591.521 | 4.5E-01 | -63.623 | -64.082 | 4.6E-01 |
| 1434 | Er | 151 | 68 | 83 | 15 | 8.104872 | 8.102276 | 2.6E-03 | 140562.226 | 140562.575 | -3.5E-01 | 140597.330 | 140597.679 | -3.5E-01 | -58.266 | -57.924 | -3.4E-01 |
| 1435 | Tm | 151 | 69 | 82 | 13 | 8.050097 | 8.046857 | 3.2E-03 | 140569.190 | 140569.637 | -4.5E-01 | 140604.818 | 140605.265 | -4.5E-01 | -50.778 | -50.339 | -4.4E-01 |
| 1436 | Yb | 151 | 70 | 81 | 11 | 7.983755 | 7.981340 | 2.4E-03 | 140577.901 | 140578.222 | -3.2E-01 | 140614.054 | 140614.375 | -3.2E-01 | -41.542 | -41.228 | -3.1E-01 |
| 1437 | Pr | 152 | 59 | 93 | 34 | 8.187104 | 8.191356 | -4.3E-03 | 141492.932 | 141492.250 | 6.8E-01 | 141523.332 | 141522.650 | 6.8E-01 | -63.758 | -64.447 | 6.9E-01 |
| 1438 | Nd | 152 | 60 | 92 | 32 | 8.224001 | 8.229612 | -5.6E-03 | 141486.020 | 141485.130 | 8.9E-01 | 141516.941 | 141516.052 | 8.9E-01 | -70.149 | -71.045 | 9.0E-01 |
| 1439 | Pm | 152 | 61 | 91 | 30 | 8.226122 | 8.233055 | -6.9E-03 | 141484.393 | 141483.302 | 1.1E+00 | 141515.836 | 141514.746 | 1.1E+00 | -71.254 | -72.352 | 1.1E+00 |
| 1440 | Sm | 152 | 62 | 90 | 28 | 8.244057 | 8.251366 | -7.3E-03 | 141480.362 | 141479.214 | 1.1E+00 | 141512.328 | 141511.179 | 1.1E+00 | -74.762 | -75.918 | 1.2E+00 |
| 1441 | Eu | 152 | 63 | 89 | 26 | 8.226577 | 8.236270 | -9.7E-03 | 141481.715 | 141480.203 | 1.5E+00 | 141514.203 | 141512.691 | 1.5E+00 | -72.887 | -74.406 | 1.5E+00 |
| 1442 | Gd | 152 | 64 | 88 | 24 | 8.233397 | 8.238668 | -5.3E-03 | 141479.373 | 141478.533 | 8.4E-01 | 141512.384 | 141511.543 | 8.4E-01 | -74.706 | -75.554 | 8.5E-01 |
| 1443 | Tb | 152 | 65 | 87 | 22 | 8.202000 | 8.207475 | -5.5E-03 | 141482.840 | 141481.968 | 8.7E-01 | 141516.374 | 141515.502 | 8.7E-01 | -70.716 | -71.596 | 8.8E-01 |
| 1444 | Dy | 152 | 66 | 86 | 20 | 8.192916 | 8.195433 | -2.5E-03 | 141482.915 | 141482.492 | 4.2E-01 | 141516.972 | 141516.549 | 4.2E-01 | -70.118 | -70.548 | 4.3E-01 |
| 1445 | Ho | 152 | 67 | 85 | 18 | 8.144882 | 8.147995 | -3.1E-03 | 141488.911 | 141488.396 | 5.1E-01 | 141523.491 | 141522.976 | 5.1E-01 | -63.599 | -64.121 | 5.2E-01 |
| 1446 | Er | 152 | 68 | 84 | 16 | 8.119306 | 8.120909 | -1.6E-03 | 141491.492 | 141491.206 | 2.9E-01 | 141526.596 | 141526.310 | 2.9E-01 | -60.494 | -60.787 | 2.9E-01 |
| 1447 | Tm | 152 | 69 | 83 | 14 | 8.056769 | 8.055565 | 1.2E-03 | 141499.691 | 141499.832 | -1.4E-01 | 141535.319 | 141535.460 | -1.4E-01 | -51.771 | -51.638 | -1.3E-01 |
| 1448 | Yb | 152 | 70 | 82 | 12 | 8.015767 | 8.011678 | 4.1E-03 | 141504.617 | 141505.195 | -5.8E-01 | 141540.769 | 141541.347 | -5.8E-01 | -46.321 | -45.750 | -5.7E-01 |
| 1449 | Pr | 153 | 59 | 94 | 35 | 8.172036 | 8.175814 | -3.8E-03 | 142426.615 | 142426.002 | 6.1E-01 | 142457.016 | 142456.402 | 6.1E-01 | -61.568 | -62.189 | 6.2E-01 |
| 1450 | Nd | 153 | 60 | 93 | 33 | 8.204582 | 8.207550 | -3.0E-03 | 142420.332 | 142419.842 | 4.9E-01 | 142451.254 | 142450.763 | 4.9E-01 | -67.330 | -67.828 | 5.0E-01 |
| 1451 | Pm | 153 | 61 | 92 | 31 | 8.221150 | 8.225484 | -4.3E-03 | 142416.493 | 142415.793 | 7.0E-01 | 142447.936 | 142447.236 | 7.0E-01 | -70.648 | -71.355 | 7.1E-01 |
| 1452 | Sm | 153 | 62 | 91 | 29 | 8.228530 | 8.235761 | -7.2E-03 | 142414.059 | 142412.915 | 1.1E+00 | 142446.025 | 142444.881 | 1.1E+00 | -72.559 | -73.710 | 1.2E+00 |
| 1453 | Eu | 153 | 63 | 90 | 27 | 8.228693 | 8.234773 | -6.1E-03 | 142412.730 | 142411.761 | 9.7E-01 | 142445.218 | 142444.249 | 9.7E-01 | -73.366 | -74.342 | 9.8E-01 |
| 1454 | Gd | 153 | 64 | 89 | 25 | 8.220414 | 8.227528 | -7.1E-03 | 142412.692 | 142411.564 | 1.1E+00 | 142445.702 | 142444.574 | 1.1E+00 | -72.882 | -74.017 | 1.1E+00 |
| 1455 | Tb | 153 | 65 | 88 | 23 | 8.205045 | 8.210926 | -5.9E-03 | 142413.738 | 142412.798 | 9.4E-01 | 142447.271 | 142446.331 | 9.4E-01 | -71.313 | -72.260 | 9.5E-01 |
| 1456 | Dy | 153 | 66 | 87 | 21 | 8.185747 | 8.188350 | -2.6E-03 | 142415.385 | 142414.946 | 4.4E-01 | 142449.441 | 142449.003 | 4.4E-01 | -69.143 | -69.589 | 4.5E-01 |
| 1457 | Ho | 153 | 67 | 86 | 19 | 8.153638 | 8.157306 | -3.7E-03 | 142418.992 | 142418.389 | 6.0E-01 | 142453.572 | 142452.969 | 6.0E-01 | -65.012 | -65.622 | 6.1E-01 |
| 1458 | Er | 153 | 68 | 85 | 17 | 8.118851 | 8.119475 | -6.2E-04 | 142423.008 | 142422.870 | 1.4E-01 | 142458.112 | 142457.974 | 1.4E-01 | -60.472 | -60.617 | 1.4E-01 |
| 1459 | Tm | 153 | 69 | 84 | 15 | 8.071365 | 8.073180 | -1.8E-03 | 142428.967 | 142428.646 | 3.2E-01 | 142464.595 | 142464.274 | 3.2E-01 | -53.989 | -54.317 | 3.3E-01 |
| 1460 | Lu | 153 | 71 | 82 | 11 | 7.959399 | 7.955100 | 4.3E-03 | 142443.484 | 142444.097 | -6.1E-01 | 142480.161 | 142480.774 | -6.1E-01 | -38.423 | -37.817 | -6.1E-01 |
| 1461 | Pr | 154 | 59 | 95 | 36 | 8.149473 | 8.151232 | -1.8E-03 | 143361.484 | 143361.177 | 3.1E-01 | 143391.884 | 143391.577 | 3.1E-01 | -58.194 | -58.508 | 3.1E-01 |
| 1462 | Nd | 154 | 60 | 94 | 34 | 8.193029 | 8.197417 | -4.4E-03 | 143353.472 | 143352.760 | 7.1E-01 | 143384.394 | 143383.682 | 7.1E-01 | -65.684 | -66.404 | 7.2E-01 |
| 1463 | Pm | 154 | 61 | 93 | 32 | 8.206176 | 8.208393 | -2.2E-03 | 143350.143 | 143349.765 | 3.8E-01 | 143381.587 | 143381.208 | 3.8E-01 | -68.491 | -68.877 | 3.9E-01 |
| 1464 | Sm | 154 | 62 | 92 | 30 | 8.226830 | 8.232719 | -5.9E-03 | 143345.658 | 143344.713 | 9.4E-01 | 143377.623 | 143376.679 | 9.4E-01 | -72.455 | -73.406 | 9.5E-01 |
| 1465 | Eu | 154 | 63 | 91 | 28 | 8.217092 | 8.223452 | -6.4E-03 | 143345.853 | 143344.835 | 1.0E+00 | 143378.341 | 143377.323 | 1.0E+00 | -71.737 | -72.762 | 1.0E+00 |
| 1466 | Gd | 154 | 64 | 90 | 26 | 8.224792 | 8.230160 | -5.4E-03 | 143343.362 | 143342.496 | 8.7E-01 | 143376.373 | 143375.507 | 8.7E-01 | -73.705 | -74.579 | 8.7E-01 |
| 1467 | Tb | 154 | 65 | 89 | 24 | 8.196662 | 8.203943 | -7.3E-03 | 143346.389 | 143345.228 | 1.2E+00 | 143379.922 | 143378.761 | 1.2E+00 | -70.156 | -71.324 | 1.2E+00 |
| 1468 | Dy | 154 | 66 | 88 | 22 | 8.193128 | 8.196074 | -2.9E-03 | 143345.628 | 143345.133 | 4.9E-01 | 143379.684 | 143379.190 | 4.9E-01 | -70.394 | -70.895 | 5.0E-01 |
| 1469 | Ho | 154 | 67 | 87 | 20 | 8.150681 | 8.154660 | -4.0E-03 | 143350.859 | 143350.205 | 6.5E-01 | 143385.439 | 143384.785 | 6.5E-01 | -64.639 | -65.301 | 6.6E-01 |

| | | | | | | | | | | | | | | | | | | |
|---|---|---|---|---|---|---|---|---|---|---|---|---|---|---|---|---|---|---|
| 1470 | Er | 154 | 68 | 86 | 18 | 8.132391 | 8.133434 | -1.0E-03 | 143352.369 | 143352.167 | 2.0E-01 | 143387.473 | 143387.270 | 2.0E-01 | -62.605 | -62.815 | 2.1E-01 |
| 1471 | Tm | 154 | 69 | 85 | 16 | 8.074208 | 8.076583 | -2.4E-03 | 143360.023 | 143359.614 | 4.1E-01 | 143395.651 | 143395.242 | 4.1E-01 | -54.427 | -54.843 | 4.2E-01 |
| 1472 | Yb | 154 | 70 | 84 | 14 | 8.039940 | 8.041236 | -1.3E-03 | 143363.993 | 143363.750 | 2.4E-01 | 143400.146 | 143399.903 | 2.4E-01 | -49.932 | -50.183 | 2.5E-01 |
| 1473 | Pr | 155 | 59 | 96 | 37 | 8.131038 | 8.132402 | -1.4E-03 | 144295.757 | 144295.510 | 2.5E-01 | 144326.157 | 144325.910 | 2.5E-01 | -55.415 | -55.669 | 2.5E-01 |
| 1474 | Nd | 155 | 60 | 95 | 35 | 8.170304 | 8.172710 | -2.4E-03 | 144288.367 | 144287.958 | 4.1E-01 | 144319.288 | 144318.879 | 4.1E-01 | -62.284 | -62.700 | 4.2E-01 |
| 1475 | Pm | 155 | 61 | 94 | 33 | 8.195296 | 8.198225 | -2.9E-03 | 144283.189 | 144282.698 | 4.9E-01 | 144314.632 | 144314.141 | 4.9E-01 | -66.940 | -67.438 | 5.0E-01 |
| 1476 | Sm | 155 | 62 | 93 | 31 | 8.211218 | 8.215453 | -4.2E-03 | 144279.416 | 144278.722 | 6.9E-01 | 144311.382 | 144310.688 | 6.9E-01 | -70.190 | -70.892 | 7.0E-01 |
| 1477 | Eu | 155 | 63 | 92 | 29 | 8.216668 | 8.220376 | -3.7E-03 | 144277.267 | 144276.654 | 6.1E-01 | 144309.755 | 144309.142 | 6.1E-01 | -71.817 | -72.438 | 6.2E-01 |
| 1478 | Gd | 155 | 64 | 91 | 27 | 8.213247 | 8.218474 | -5.2E-03 | 144276.492 | 144275.643 | 8.5E-01 | 144309.503 | 144308.654 | 8.5E-01 | -72.069 | -72.926 | 8.6E-01 |
| 1479 | Tb | 155 | 65 | 90 | 25 | 8.202910 | 8.206446 | -3.5E-03 | 144276.789 | 144276.201 | 5.9E-01 | 144310.323 | 144309.735 | 5.9E-01 | -71.249 | -71.845 | 6.0E-01 |
| 1480 | Dy | 155 | 66 | 89 | 23 | 8.184350 | 8.188507 | -4.2E-03 | 144278.360 | 144277.676 | 6.8E-01 | 144312.417 | 144311.732 | 6.8E-01 | -69.155 | -69.847 | 6.9E-01 |
| 1481 | Ho | 155 | 67 | 88 | 21 | 8.159201 | 8.162078 | -2.9E-03 | 144280.953 | 144280.466 | 4.9E-01 | 144315.533 | 144315.046 | 4.9E-01 | -66.039 | -66.534 | 4.9E-01 |
| 1482 | Er | 155 | 68 | 87 | 19 | 8.129442 | 8.129832 | -3.9E-04 | 144284.259 | 144284.157 | 1.0E-01 | 144319.363 | 144319.261 | 1.0E-01 | -62.209 | -62.319 | 1.1E-01 |
| 1483 | Tm | 155 | 69 | 86 | 17 | 8.088375 | 8.089817 | -1.4E-03 | 144289.318 | 144289.052 | 2.7E-01 | 144324.946 | 144324.680 | 2.7E-01 | -56.626 | -56.899 | 2.7E-01 |
| 1484 | Yb | 155 | 70 | 85 | 15 | 8.043823 | 8.043075 | 7.5E-04 | 144294.917 | 144294.989 | -7.2E-02 | 144331.069 | 144331.142 | -7.3E-02 | -50.503 | -50.438 | -6.5E-02 |
| 1485 | Lu | 155 | 71 | 84 | 13 | 7.987469 | 7.988724 | -1.3E-03 | 144302.345 | 144302.106 | 2.4E-01 | 144339.022 | 144338.783 | 2.4E-01 | -42.550 | -42.796 | 2.5E-01 |
| 1486 | Nd | 156 | 60 | 96 | 36 | 8.158066 | 8.159705 | -1.6E-03 | 145221.671 | 145221.379 | 2.9E-01 | 145252.592 | 145252.301 | 2.9E-01 | -60.474 | -60.773 | 3.0E-01 |
| 1487 | Pm | 156 | 61 | 95 | 34 | 8.176705 | 8.178848 | -2.1E-03 | 145217.459 | 145217.088 | 3.7E-01 | 145248.902 | 145248.531 | 3.7E-01 | -64.164 | -64.542 | 3.8E-01 |
| 1488 | Sm | 156 | 62 | 94 | 32 | 8.205017 | 8.210101 | -5.1E-03 | 145211.738 | 145210.907 | 8.3E-01 | 145243.703 | 145242.873 | 8.3E-01 | -69.363 | -70.201 | 8.4E-01 |
| 1489 | Eu | 156 | 63 | 93 | 30 | 8.204633 | 8.207580 | -2.9E-03 | 145210.493 | 145209.995 | 5.0E-01 | 145242.981 | 145242.483 | 5.0E-01 | -70.085 | -70.591 | 5.1E-01 |
| 1490 | Gd | 156 | 64 | 92 | 28 | 8.215318 | 8.219588 | -4.3E-03 | 145207.521 | 145206.816 | 7.0E-01 | 145240.532 | 145239.827 | 7.1E-01 | -72.534 | -73.247 | 7.1E-01 |
| 1491 | Tb | 156 | 65 | 91 | 26 | 8.194635 | 8.198830 | -4.2E-03 | 145209.443 | 145208.748 | 6.9E-01 | 145242.976 | 145242.282 | 6.9E-01 | -70.090 | -70.792 | 7.0E-01 |
| 1492 | Dy | 156 | 66 | 90 | 24 | 8.192429 | 8.195060 | -2.6E-03 | 145208.481 | 145208.030 | 4.5E-01 | 145242.538 | 145242.087 | 4.5E-01 | -70.528 | -70.987 | 4.6E-01 |
| 1493 | Ho | 156 | 67 | 89 | 22 | 8.155042 | 8.158648 | -3.6E-03 | 145213.008 | 145212.404 | 6.0E-01 | 145247.588 | 145246.984 | 6.0E-01 | -65.478 | -66.090 | 6.1E-01 |
| 1494 | Er | 156 | 68 | 88 | 20 | 8.141899 | 8.141565 | 3.3E-04 | 145213.752 | 145213.762 | -9.9E-03 | 145248.856 | 145248.866 | -9.8E-03 | -64.210 | -64.208 | -2.7E-03 |
| 1495 | Tm | 156 | 69 | 87 | 18 | 8.089568 | 8.090723 | -1.2E-03 | 145220.609 | 145220.386 | 2.2E-01 | 145256.237 | 145256.014 | 2.2E-01 | -56.829 | -57.059 | 2.3E-01 |
| 1496 | Yb | 156 | 70 | 86 | 16 | 8.061665 | 8.061098 | 5.7E-04 | 145223.655 | 145223.700 | -4.5E-02 | 145259.808 | 145259.853 | -4.5E-02 | -53.258 | -53.221 | -3.8E-02 |
| 1497 | Lu | 156 | 71 | 85 | 14 | 7.995697 | 7.995579 | 1.2E-04 | 145232.639 | 145232.613 | 2.6E-02 | 145269.316 | 145269.291 | 2.5E-02 | -43.750 | -43.783 | 3.3E-02 |
| 1498 | Hf | 156 | 72 | 84 | 12 | 7.952973 | 7.952655 | 3.2E-04 | 145237.997 | 145238.001 | -4.4E-03 | 145275.199 | 145275.204 | -4.5E-03 | -37.867 | -37.870 | 2.7E-03 |
| 1499 | Nd | 157 | 60 | 97 | 37 | 8.131959 | 8.132062 | -1.0E-04 | 146157.177 | 146157.125 | 5.2E-02 | 146188.099 | 146188.046 | 5.3E-02 | -56.462 | -56.521 | 6.0E-02 |
| 1500 | Pm | 157 | 61 | 96 | 35 | 8.164144 | 8.165706 | -1.6E-03 | 146150.820 | 146150.538 | 2.8E-01 | 146182.263 | 146181.981 | 2.8E-01 | -62.297 | -62.586 | 2.9E-01 |
| 1501 | Sm | 157 | 62 | 95 | 33 | 8.187063 | 8.190552 | -3.5E-03 | 146145.917 | 146145.332 | 5.9E-01 | 146177.882 | 146177.297 | 5.8E-01 | -66.678 | -67.270 | 5.9E-01 |
| 1502 | Eu | 157 | 63 | 94 | 31 | 8.199791 | 8.202187 | -2.4E-03 | 146142.614 | 146142.200 | 4.1E-01 | 146175.102 | 146174.687 | 4.1E-01 | -69.458 | -69.880 | 4.2E-01 |
| 1503 | Gd | 157 | 64 | 93 | 29 | 8.203500 | 8.206503 | -3.0E-03 | 146140.727 | 146140.216 | 5.1E-01 | 146173.737 | 146173.227 | 5.1E-01 | -70.823 | -71.341 | 5.2E-01 |
| 1504 | Tb | 157 | 65 | 92 | 27 | 8.198134 | 8.199838 | -1.7E-03 | 146140.264 | 146139.957 | 3.1E-01 | 146173.797 | 146173.490 | 3.1E-01 | -70.763 | -71.077 | 3.1E-01 |
| 1505 | Dy | 157 | 66 | 91 | 25 | 8.184621 | 8.186925 | -2.3E-03 | 146141.080 | 146140.678 | 4.0E-01 | 146175.136 | 146174.734 | 4.0E-01 | -69.424 | -69.833 | 4.1E-01 |
| 1506 | Ho | 157 | 67 | 90 | 23 | 8.163122 | 8.164921 | -1.8E-03 | 146143.149 | 146142.826 | 3.2E-01 | 146177.729 | 146177.406 | 3.2E-01 | -66.831 | -67.162 | 3.3E-01 |
| 1507 | Er | 157 | 68 | 89 | 21 | 8.136221 | 8.137298 | -1.1E-03 | 146146.067 | 146145.856 | 2.1E-01 | 146181.171 | 146180.960 | 2.1E-01 | -63.389 | -63.608 | 2.2E-01 |
| 1508 | Tm | 157 | 69 | 88 | 19 | 8.101599 | 8.101891 | -2.9E-04 | 146150.196 | 146150.107 | 8.9E-02 | 146185.824 | 146185.735 | 8.9E-02 | -58.736 | -58.832 | 9.6E-02 |
| 1509 | Yb | 157 | 70 | 87 | 17 | 8.062790 | 8.060719 | 2.1E-03 | 146154.982 | 146155.264 | -2.8E-01 | 146191.135 | 146191.416 | -2.8E-01 | -53.426 | -53.151 | -2.7E-01 |
| 1510 | Lu | 157 | 71 | 86 | 15 | 8.013420 | 8.012548 | 8.7E-04 | 146161.426 | 146161.519 | -9.3E-02 | 146198.103 | 146198.196 | -9.3E-02 | -46.457 | -46.371 | -8.5E-02 |
| 1511 | Ta | 157 | 73 | 84 | 11 | 7.896364 | 7.895898 | 4.7E-04 | 146177.189 | 146177.217 | -2.8E-02 | 146214.916 | 146214.944 | -2.8E-02 | -29.644 | -29.624 | -2.0E-02 |
| 1512 | Pm | 158 | 61 | 97 | 36 | 8.143254 | 8.143781 | -5.3E-04 | 147085.521 | 147085.401 | 1.2E-01 | 147116.965 | 147116.845 | 1.2E-01 | -59.089 | -59.217 | 1.3E-01 |
| 1513 | Sm | 158 | 62 | 96 | 34 | 8.177296 | 8.182560 | -5.3E-03 | 147078.838 | 147077.969 | 8.7E-01 | 147110.804 | 147109.935 | 8.7E-01 | -65.250 | -66.127 | 8.8E-01 |
| 1514 | Eu | 158 | 63 | 95 | 32 | 8.185034 | 8.187363 | -2.3E-03 | 147076.311 | 147075.905 | 4.1E-01 | 147108.799 | 147108.393 | 4.1E-01 | -67.255 | -67.669 | 4.1E-01 |
| 1515 | Gd | 158 | 64 | 94 | 30 | 8.201815 | 8.205470 | -3.7E-03 | 147072.355 | 147071.738 | 6.2E-01 | 147105.365 | 147104.749 | 6.2E-01 | -70.689 | -71.313 | 6.2E-01 |

| | | | | | | | | | | | | | | | | | | |
|---|---|---|---|---|---|---|---|---|---|---|---|---|---|---|---|---|---|---|
| 1516 | Tb | 158 | 65 | 93 | 28 | 8.189149 | 8.190892 | -1.7E-03 | 147073.051 | 147072.736 | 3.2E-01 | 147106.584 | 147106.269 | 3.1E-01 | -69.470 | -69.793 | 3.2E-01 |
| 1517 | Dy | 158 | 66 | 92 | 26 | 8.190123 | 8.191932 | -1.8E-03 | 147071.591 | 147071.265 | 3.3E-01 | 147105.648 | 147105.322 | 3.3E-01 | -70.406 | -70.740 | 3.3E-01 |
| 1518 | Ho | 158 | 67 | 91 | 24 | 8.158464 | 8.160759 | -2.3E-03 | 147075.288 | 147074.884 | 4.0E-01 | 147109.868 | 147109.464 | 4.0E-01 | -66.186 | -66.598 | 4.1E-01 |
| 1519 | Er | 158 | 68 | 90 | 22 | 8.147926 | 8.147617 | 3.1E-04 | 147075.646 | 147075.653 | -7.4E-03 | 147110.750 | 147110.757 | -7.4E-03 | -65.304 | -65.304 | 4.7E-04 |
| 1520 | Tm | 158 | 69 | 89 | 20 | 8.101199 | 8.101809 | -6.1E-04 | 147081.723 | 147081.584 | 1.4E-01 | 147117.351 | 147117.212 | 1.4E-01 | -58.703 | -58.850 | 1.5E-01 |
| 1521 | Yb | 158 | 70 | 88 | 18 | 8.079203 | 8.076312 | 2.9E-03 | 147083.891 | 147084.305 | -4.1E-01 | 147120.044 | 147120.457 | -4.1E-01 | -56.010 | -55.604 | -4.1E-01 |
| 1522 | Lu | 158 | 71 | 87 | 16 | 8.018568 | 8.016809 | 1.8E-03 | 147092.165 | 147092.399 | -2.3E-01 | 147128.842 | 147129.076 | -2.3E-01 | -47.212 | -46.986 | -2.3E-01 |
| 1523 | Hf | 158 | 72 | 86 | 14 | 7.981276 | 7.979539 | 1.7E-03 | 147096.749 | 147096.979 | -2.3E-01 | 147133.952 | 147134.181 | -2.3E-01 | -42.103 | -41.880 | -2.2E-01 |
| 1524 | Pm | 159 | 61 | 98 | 37 | 8.126859 | 8.127483 | -6.2E-04 | 148019.550 | 148019.414 | 1.4E-01 | 148050.994 | 148050.858 | 1.4E-01 | -56.554 | -56.698 | 1.4E-01 |
| 1525 | Sm | 159 | 62 | 97 | 35 | 8.157495 | 8.160415 | -2.9E-03 | 148013.375 | 148012.873 | 5.0E-01 | 148045.340 | 148044.839 | 5.0E-01 | -62.208 | -62.717 | 5.1E-01 |
| 1526 | Eu | 159 | 63 | 96 | 33 | 8.176695 | 8.179283 | -2.6E-03 | 148009.017 | 148008.568 | 4.5E-01 | 148041.505 | 148041.056 | 4.5E-01 | -66.043 | -66.500 | 4.6E-01 |
| 1527 | Gd | 159 | 64 | 95 | 31 | 8.187610 | 8.190404 | -2.8E-03 | 148005.977 | 148005.494 | 4.8E-01 | 148038.987 | 148038.504 | 4.8E-01 | -68.561 | -69.051 | 4.9E-01 |
| 1528 | Tb | 159 | 65 | 94 | 29 | 8.188796 | 8.189782 | -9.9E-04 | 148004.483 | 148004.287 | 2.0E-01 | 148038.016 | 148037.820 | 2.0E-01 | -69.532 | -69.736 | 2.0E-01 |
| 1529 | Dy | 159 | 66 | 93 | 27 | 8.181577 | 8.182575 | -1.0E-03 | 148004.325 | 148004.126 | 2.0E-01 | 148038.382 | 148038.183 | 2.0E-01 | -69.166 | -69.373 | 2.1E-01 |
| 1530 | Ho | 159 | 67 | 92 | 25 | 8.165100 | 8.165551 | -4.5E-04 | 148005.639 | 148005.527 | 1.1E-01 | 148040.219 | 148040.107 | 1.1E-01 | -67.329 | -67.449 | 1.2E-01 |
| 1531 | Er | 159 | 68 | 91 | 23 | 8.142767 | 8.142740 | 2.7E-05 | 148007.884 | 148007.847 | 3.7E-02 | 148042.988 | 148042.950 | 3.8E-02 | -64.560 | -64.605 | 4.5E-02 |
| 1532 | Tm | 159 | 69 | 90 | 21 | 8.112754 | 8.111667 | 1.1E-03 | 148011.350 | 148011.480 | -1.3E-01 | 148046.978 | 148047.108 | -1.3E-01 | -60.570 | -60.448 | -1.2E-01 |
| 1533 | Yb | 159 | 70 | 89 | 19 | 8.078075 | 8.075137 | 2.9E-03 | 148015.557 | 148015.981 | -4.2E-01 | 148051.709 | 148052.133 | -4.2E-01 | -55.839 | -55.422 | -4.2E-01 |
| 1534 | Lu | 159 | 71 | 88 | 17 | 8.034600 | 8.031582 | 3.0E-03 | 148021.162 | 148021.598 | -4.4E-01 | 148057.839 | 148058.275 | -4.4E-01 | -49.709 | -49.280 | -4.3E-01 |
| 1535 | Hf | 159 | 72 | 87 | 15 | 7.986561 | 7.982208 | 4.4E-03 | 148027.493 | 148028.141 | -6.5E-01 | 148064.695 | 148065.343 | -6.5E-01 | -42.853 | -42.213 | -6.4E-01 |
| 1536 | Ta | 159 | 73 | 86 | 13 | 7.928757 | 7.926618 | 2.1E-03 | 148035.376 | 148035.671 | -3.0E-01 | 148073.104 | 148073.399 | -2.9E-01 | -34.444 | -34.157 | -2.9E-01 |
| 1537 | Sm | 160 | 62 | 98 | 36 | 8.144625 | 8.149637 | -5.0E-03 | 148946.842 | 148946.003 | 8.4E-01 | 148978.807 | 148977.968 | 8.4E-01 | -60.235 | -61.082 | 8.5E-01 |
| 1538 | Eu | 160 | 63 | 97 | 34 | 8.160021 | 8.162169 | -2.1E-03 | 148943.074 | 148942.692 | 3.8E-01 | 148975.562 | 148975.180 | 3.8E-01 | -63.480 | -63.870 | 3.9E-01 |
| 1539 | Gd | 160 | 64 | 96 | 32 | 8.183009 | 8.186912 | -3.9E-03 | 148938.091 | 148937.427 | 6.6E-01 | 148971.101 | 148970.438 | 6.6E-01 | -67.941 | -68.612 | 6.7E-01 |
| 1540 | Tb | 160 | 65 | 95 | 30 | 8.177461 | 8.179003 | -1.5E-03 | 148937.673 | 148937.387 | 2.9E-01 | 148971.207 | 148970.920 | 2.9E-01 | -67.836 | -68.129 | 2.9E-01 |
| 1541 | Dy | 160 | 66 | 94 | 28 | 8.184046 | 8.185524 | -1.5E-03 | 148935.314 | 148935.037 | 2.8E-01 | 148969.371 | 148969.094 | 2.8E-01 | -69.671 | -69.956 | 2.8E-01 |
| 1542 | Ho | 160 | 67 | 93 | 26 | 8.158593 | 8.160137 | -1.5E-03 | 148938.081 | 148937.793 | 2.9E-01 | 148972.661 | 148972.373 | 2.9E-01 | -66.381 | -66.677 | 3.0E-01 |
| 1543 | Er | 160 | 68 | 92 | 24 | 8.151721 | 8.151443 | 2.8E-04 | 148937.874 | 148937.877 | -2.7E-03 | 148972.978 | 148972.981 | -2.7E-03 | -66.064 | -66.069 | 4.9E-03 |
| 1544 | Tm | 160 | 69 | 91 | 22 | 8.110818 | 8.110750 | 6.8E-05 | 148943.112 | 148943.080 | 3.2E-02 | 148978.740 | 148978.708 | 3.2E-02 | -60.302 | -60.341 | 3.9E-02 |
| 1545 | Yb | 160 | 70 | 90 | 20 | 8.092571 | 8.089110 | 3.5E-03 | 148944.725 | 148945.235 | -5.1E-01 | 148980.877 | 148981.388 | -5.1E-01 | -58.165 | -57.662 | -5.0E-01 |
| 1546 | Lu | 160 | 71 | 89 | 18 | 8.038338 | 8.034689 | 3.6E-03 | 148952.095 | 148952.635 | -5.4E-01 | 148988.772 | 148989.312 | -5.4E-01 | -50.270 | -49.738 | -5.3E-01 |
| 1547 | Hf | 160 | 72 | 88 | 16 | 8.006332 | 8.001569 | 4.8E-03 | 148955.909 | 148956.626 | -7.2E-01 | 148993.111 | 148993.828 | -7.2E-01 | -45.931 | -45.222 | -7.1E-01 |
| 1548 | Ta | 160 | 73 | 87 | 14 | 7.938585 | 7.934072 | 4.5E-03 | 148965.440 | 148966.117 | -6.8E-01 | 149003.168 | 149003.845 | -6.8E-01 | -35.874 | -35.205 | -6.7E-01 |
| 1549 | W | 160 | 74 | 86 | 12 | 7.893088 | 7.889932 | 3.2E-03 | 148971.412 | 148971.871 | -4.6E-01 | 149009.665 | 149010.124 | -4.6E-01 | -29.377 | -28.926 | -4.5E-01 |
| 1550 | Sm | 161 | 62 | 99 | 37 | 8.122040 | 8.124702 | -2.7E-03 | 149881.899 | 149881.433 | 4.7E-01 | 149913.864 | 149913.398 | 4.7E-01 | -56.672 | -57.145 | 4.7E-01 |
| 1551 | Eu | 161 | 63 | 98 | 35 | 8.148980 | 8.151218 | -2.2E-03 | 149876.257 | 149875.858 | 4.0E-01 | 149908.745 | 149908.346 | 4.0E-01 | -61.792 | -62.198 | 4.1E-01 |
| 1552 | Gd | 161 | 64 | 97 | 33 | 8.167186 | 8.169562 | -2.4E-03 | 149872.021 | 149871.599 | 4.2E-01 | 149905.031 | 149904.610 | 4.2E-01 | -65.505 | -65.934 | 4.3E-01 |
| 1553 | Tb | 161 | 65 | 96 | 31 | 8.174474 | 8.175426 | -9.5E-04 | 149869.542 | 149869.349 | 1.9E-01 | 149903.075 | 149902.883 | 1.9E-01 | -67.461 | -67.661 | 2.0E-01 |
| 1554 | Dy | 161 | 66 | 95 | 29 | 8.173302 | 8.174418 | -1.1E-03 | 149868.425 | 149868.205 | 2.2E-01 | 149902.482 | 149902.262 | 2.2E-01 | -68.054 | -68.282 | 2.3E-01 |
| 1555 | Ho | 161 | 67 | 94 | 27 | 8.163114 | 8.162940 | 1.7E-04 | 149868.759 | 149868.747 | 1.2E-02 | 149903.340 | 149903.327 | 1.3E-02 | -67.197 | -67.217 | 2.1E-02 |
| 1556 | Er | 161 | 68 | 93 | 25 | 8.145856 | 8.145459 | 4.0E-04 | 149870.232 | 149870.254 | -2.2E-02 | 149905.336 | 149905.358 | -2.2E-02 | -65.200 | -65.186 | -1.5E-02 |
| 1557 | Tm | 161 | 69 | 92 | 23 | 8.120490 | 8.119106 | 1.4E-03 | 149873.009 | 149873.190 | -1.8E-01 | 149908.637 | 149908.818 | -1.8E-01 | -61.899 | -61.726 | -1.7E-01 |
| 1558 | Yb | 161 | 70 | 91 | 21 | 8.090416 | 8.087274 | 3.1E-03 | 149876.545 | 149877.007 | -4.6E-01 | 149912.697 | 149913.160 | -4.6E-01 | -57.839 | -57.384 | -4.6E-01 |
| 1559 | Lu | 161 | 71 | 90 | 19 | 8.052781 | 8.048014 | 4.8E-03 | 149881.297 | 149882.020 | -7.2E-01 | 149917.974 | 149918.697 | -7.2E-01 | -52.562 | -51.846 | -7.2E-01 |
| 1560 | Hf | 161 | 72 | 89 | 17 | 8.009121 | 8.003337 | 5.8E-03 | 149887.019 | 149887.905 | -8.9E-01 | 149924.221 | 149925.107 | -8.9E-01 | -46.315 | -45.436 | -8.8E-01 |
| 1561 | Ta | 161 | 73 | 88 | 15 | 7.956970 | 7.952382 | 4.6E-03 | 149894.107 | 149894.801 | -6.9E-01 | 149931.835 | 149932.528 | -6.9E-01 | -38.701 | -38.016 | -6.9E-01 |
| 1562 | Re | 161 | 75 | 86 | 11 | 7.836626 | 7.833369 | 3.3E-03 | 149910.866 | 149911.344 | -4.8E-01 | 149949.645 | 149950.123 | -4.8E-01 | -20.891 | -20.421 | -4.7E-01 |
| 1563 | Eu | 162 | 63 | 99 | 36 | 8.129384 | 8.131659 | -2.3E-03 | 150810.848 | 150810.441 | 4.1E-01 | 150843.336 | 150842.929 | 4.1E-01 | -58.695 | -59.109 | 4.1E-01 |
| 1564 | Gd | 162 | 64 | 98 | 34 | 8.159030 | 8.163481 | -4.5E-03 | 150804.740 | 150803.980 | 7.6E-01 | 150837.751 | 150836.991 | 7.6E-01 | -64.280 | -65.047 | 7.7E-01 |
| 1565 | Tb | 162 | 65 | 97 | 32 | 8.162812 | 8.162566 | 2.5E-04 | 150802.822 | 150802.822 | -4.6E-04 | 150836.355 | 150836.356 | -8.9E-04 | -65.675 | -65.682 | 7.3E-03 |
| 1566 | Dy | 162 | 66 | 96 | 30 | 8.173449 | 8.175024 | -1.6E-03 | 150799.793 | 150799.498 | 3.0E-01 | 150833.850 | 150833.555 | 3.0E-01 | -68.180 | -68.483 | 3.0E-01 |
| 1567 | Ho | 162 | 67 | 95 | 28 | 8.155413 | 8.155827 | -4.1E-04 | 150801.409 | 150801.301 | 1.1E-01 | 150835.989 | 150835.881 | 1.1E-01 | -66.041 | -66.156 | 1.2E-01 |
| 1568 | Er | 162 | 68 | 94 | 26 | 8.152389 | 8.152140 | 2.5E-04 | 150800.593 | 150800.592 | 1.3E-03 | 150835.697 | 150835.696 | 1.3E-03 | -66.333 | -66.342 | 9.0E-03 |
| 1569 | Tm | 162 | 69 | 93 | 24 | 8.117580 | 8.116969 | 6.1E-04 | 150804.926 | 150804.982 | -5.6E-02 | 150840.554 | 150840.610 | -5.6E-02 | -61.476 | -61.428 | -4.9E-02 |
| 1570 | Yb | 162 | 70 | 92 | 22 | 8.102566 | 8.099537 | 3.0E-03 | 150806.051 | 150806.499 | -4.5E-01 | 150842.204 | 150842.651 | -4.5E-01 | -59.827 | -59.387 | -4.4E-01 |
| 1571 | Lu | 162 | 71 | 91 | 20 | 8.054559 | 8.050179 | 4.4E-03 | 150812.521 | 150813.187 | -6.7E-01 | 150849.198 | 150849.864 | -6.7E-01 | -52.832 | -52.174 | -6.6E-01 |
| 1572 | Hf | 162 | 72 | 90 | 18 | 8.027120 | 8.020899 | 6.2E-03 | 150815.659 | 150816.622 | -9.6E-01 | 150852.861 | 150853.824 | -9.6E-01 | -49.169 | -48.213 | -9.6E-01 |
| 1573 | Ta | 162 | 73 | 89 | 16 | 7.964336 | 7.958551 | 5.8E-03 | 150824.522 | 150825.414 | -8.9E-01 | 150862.250 | 150863.142 | -8.9E-01 | -39.780 | -38.896 | -8.8E-01 |
| 1574 | W | 162 | 74 | 88 | 14 | 7.923828 | 7.918562 | 5.3E-03 | 150829.777 | 150830.584 | -8.1E-01 | 150868.030 | 150868.837 | -8.1E-01 | -34.000 | -33.201 | -8.0E-01 |
| 1575 | Eu | 163 | 63 | 100 | 37 | 8.116415 | 8.117709 | -1.3E-03 | 151744.397 | 151744.149 | 2.5E-01 | 151776.885 | 151776.636 | 2.5E-01 | -56.639 | -56.895 | 2.6E-01 |
| 1576 | Gd | 163 | 64 | 99 | 35 | 8.140297 | 8.143656 | -3.4E-03 | 151739.200 | 151738.614 | 5.9E-01 | 151772.210 | 151771.624 | 5.9E-01 | -61.314 | -61.908 | 5.9E-01 |
| 1577 | Tb | 163 | 65 | 98 | 33 | 8.155625 | 8.156359 | -7.3E-04 | 151735.396 | 151735.237 | 1.6E-01 | 151768.929 | 151768.770 | 1.6E-01 | -64.595 | -64.762 | 1.7E-01 |
| 1578 | Dy | 163 | 66 | 97 | 31 | 8.161777 | 8.161880 | -1.0E-04 | 151733.088 | 151733.031 | 5.7E-02 | 151767.144 | 151767.087 | 5.7E-02 | -66.380 | -66.444 | 6.5E-02 |
| 1579 | Ho | 163 | 67 | 96 | 29 | 8.156962 | 8.156320 | 6.4E-04 | 151732.567 | 151732.631 | -6.4E-02 | 151767.147 | 151767.211 | -6.4E-02 | -66.377 | -66.321 | -5.6E-02 |
| 1580 | Er | 163 | 68 | 95 | 27 | 8.144734 | 8.144582 | 1.5E-04 | 151733.254 | 151733.237 | 1.7E-02 | 151768.358 | 151768.341 | 1.7E-02 | -65.167 | -65.191 | 2.4E-02 |
| 1581 | Tm | 163 | 69 | 94 | 25 | 8.124972 | 8.123412 | 1.6E-03 | 151735.169 | 151735.380 | -2.1E-01 | 151770.797 | 151771.009 | -2.1E-01 | -62.728 | -62.523 | -2.0E-01 |
| 1582 | Yb | 163 | 70 | 93 | 23 | 8.099139 | 8.096659 | 2.5E-03 | 151738.072 | 151738.434 | -3.6E-01 | 151774.225 | 151774.586 | -3.6E-01 | -59.299 | -58.946 | -3.5E-01 |
| 1583 | Lu | 163 | 71 | 92 | 21 | 8.066684 | 8.061953 | 4.7E-03 | 151742.056 | 151742.783 | -7.3E-01 | 151778.733 | 151779.460 | -7.3E-01 | -54.791 | -54.072 | -7.2E-01 |
| 1584 | Hf | 163 | 72 | 91 | 19 | 8.027974 | 8.021942 | 6.0E-03 | 151747.058 | 151747.997 | -9.4E-01 | 151784.260 | 151785.199 | -9.4E-01 | -49.264 | -48.333 | -9.3E-01 |
| 1585 | Ta | 163 | 73 | 90 | 17 | 7.981890 | 7.975297 | 6.6E-03 | 151753.262 | 151754.291 | -1.0E+00 | 151790.989 | 151792.019 | -1.0E+00 | -42.535 | -41.513 | -1.0E+00 |
| 1586 | W | 163 | 74 | 89 | 15 | 7.930303 | 7.923158 | 7.1E-03 | 151760.363 | 151761.481 | -1.1E+00 | 151798.616 | 151799.734 | -1.1E+00 | -34.908 | -33.797 | -1.1E+00 |
| 1587 | Re | 163 | 75 | 88 | 13 | 7.870897 | 7.865593 | 5.3E-03 | 151768.738 | 151769.555 | -8.2E-01 | 151807.517 | 151808.334 | -8.2E-01 | -26.007 | -25.197 | -8.1E-01 |
| 1588 | Tb | 164 | 65 | 99 | 34 | 8.139757 | 8.141282 | -1.5E-03 | 152669.408 | 152669.119 | 2.9E-01 | 152702.941 | 152702.652 | 2.9E-01 | -62.077 | -62.374 | 3.0E-01 |
| 1589 | Dy | 164 | 66 | 98 | 32 | 8.158706 | 8.160036 | -1.3E-03 | 152664.995 | 152664.737 | 2.6E-01 | 152699.051 | 152698.793 | 2.6E-01 | -65.967 | -66.233 | 2.7E-01 |
| 1590 | Ho | 164 | 67 | 97 | 30 | 8.147924 | 8.147275 | 6.5E-04 | 152665.457 | 152665.523 | -6.6E-02 | 152700.037 | 152700.103 | -6.6E-02 | -64.981 | -64.923 | -5.8E-02 |
| 1591 | Er | 164 | 68 | 96 | 28 | 8.149012 | 8.148984 | 2.8E-05 | 152663.973 | 152663.936 | 3.7E-02 | 152699.077 | 152699.040 | 3.7E-02 | -65.942 | -65.986 | 4.5E-02 |
| 1592 | Tm | 164 | 69 | 95 | 26 | 8.119621 | 8.119668 | -4.7E-05 | 152667.486 | 152667.437 | 4.9E-02 | 152703.114 | 152703.065 | 4.9E-02 | -61.904 | -61.961 | 5.8E-02 |
| 1593 | Yb | 164 | 70 | 94 | 24 | 8.109447 | 8.106896 | 2.6E-03 | 152667.848 | 152668.224 | -3.8E-01 | 152704.001 | 152704.376 | -3.7E-01 | -61.018 | -60.650 | -3.7E-01 |
| 1594 | Lu | 164 | 71 | 93 | 22 | 8.065804 | 8.062890 | 2.9E-03 | 152673.699 | 152674.133 | -4.3E-01 | 152710.376 | 152710.810 | -4.3E-01 | -54.642 | -54.216 | -4.3E-01 |
| 1595 | Hf | 164 | 72 | 92 | 20 | 8.043813 | 8.037681 | 6.1E-03 | 152675.998 | 152676.959 | -9.6E-01 | 152713.200 | 152714.161 | -9.6E-01 | -51.818 | -50.865 | -9.5E-01 |
| 1596 | Ta | 164 | 73 | 91 | 18 | 7.986997 | 7.980410 | 6.6E-03 | 152684.008 | 152685.043 | -1.0E+00 | 152721.735 | 152722.770 | -1.0E+00 | -43.283 | -42.256 | -1.0E+00 |
| 1597 | W | 164 | 74 | 90 | 16 | 7.951404 | 7.944281 | 7.1E-03 | 152688.537 | 152689.660 | -1.1E+00 | 152726.790 | 152727.913 | -1.1E+00 | -38.228 | -37.113 | -1.1E+00 |
| 1598 | Re | 164 | 75 | 89 | 14 | 7.881360 | 7.874658 | 6.7E-03 | 152698.716 | 152699.769 | -1.1E+00 | 152737.495 | 152738.548 | -1.1E+00 | -27.523 | -26.478 | -1.0E+00 |
| 1599 | Os | 164 | 76 | 88 | 12 | 7.833599 | 7.828732 | 4.9E-03 | 152705.240 | 152705.991 | -7.5E-01 | 152744.545 | 152745.296 | -7.5E-01 | -20.473 | -19.730 | -7.4E-01 |
| 1600 | Dy | 165 | 66 | 99 | 33 | 8.143902 | 8.144686 | -7.8E-04 | 153598.844 | 153598.675 | 1.7E-01 | 153632.901 | 153632.731 | 1.7E-01 | -63.611 | -63.789 | 1.8E-01 |
| 1601 | Ho | 165 | 67 | 98 | 31 | 8.146960 | 8.145312 | 1.6E-03 | 153597.034 | 153597.265 | -2.3E-01 | 153631.614 | 153631.845 | -2.3E-01 | -64.898 | -64.675 | -2.2E-01 |
| 1602 | Er | 165 | 68 | 97 | 29 | 8.139928 | 8.139575 | 3.5E-04 | 153596.888 | 153596.905 | -1.7E-02 | 153631.992 | 153632.009 | -1.7E-02 | -64.520 | -64.512 | -8.9E-03 |
| 1603 | Tm | 165 | 69 | 96 | 27 | 8.125541 | 8.123905 | 1.6E-03 | 153597.955 | 153598.183 | -2.3E-01 | 153633.583 | 153633.811 | -2.3E-01 | -62.929 | -62.709 | -2.2E-01 |
| 1604 | Yb | 165 | 70 | 95 | 25 | 8.104839 | 8.102572 | 2.3E-03 | 153600.064 | 153600.396 | -3.3E-01 | 153636.217 | 153636.548 | -3.3E-01 | -60.295 | -59.972 | -3.2E-01 |
| 1605 | Lu | 165 | 71 | 94 | 23 | 8.076745 | 8.072785 | 4.0E-03 | 153603.393 | 153604.003 | -6.1E-01 | 153640.070 | 153640.680 | -6.1E-01 | -56.442 | -55.840 | -6.0E-01 |
| 1606 | Hf | 165 | 72 | 93 | 21 | 8.042872 | 8.037705 | 5.2E-03 | 153607.675 | 153608.483 | -8.1E-01 | 153644.877 | 153645.685 | -8.1E-01 | -51.636 | -50.835 | -8.0E-01 |
| 1607 | Ta | 165 | 73 | 92 | 19 | 8.003052 | 7.995536 | 7.5E-03 | 153612.937 | 153614.132 | -1.2E+00 | 153650.665 | 153651.860 | -1.2E+00 | -45.848 | -44.660 | -1.2E+00 |
| 1608 | W | 165 | 74 | 91 | 17 | 7.955969 | 7.948086 | 7.9E-03 | 153619.398 | 153620.653 | -1.3E+00 | 153657.651 | 153658.906 | -1.3E+00 | -38.861 | -37.614 | -1.2E+00 |
| 1609 | Re | 165 | 75 | 90 | 15 | 7.901424 | 7.894823 | 6.6E-03 | 153627.089 | 153628.132 | -1.0E+00 | 153665.868 | 153666.911 | -1.0E+00 | -30.644 | -29.609 | -1.0E+00 |
| 1610 | Tb | 166 | 65 | 101 | 36 | 8.113672 | 8.114929 | -1.3E-03 | 154536.589 | 154536.341 | 2.5E-01 | 154570.123 | 154569.875 | 2.5E-01 | -57.883 | -58.139 | 2.6E-01 |
| 1611 | Dy | 166 | 66 | 100 | 34 | 8.137273 | 8.140340 | -3.1E-03 | 154531.366 | 154530.817 | 5.5E-01 | 154565.423 | 154564.874 | 5.5E-01 | -62.583 | -63.141 | 5.6E-01 |
| 1612 | Ho | 166 | 67 | 99 | 32 | 8.135494 | 8.134223 | 1.3E-03 | 154530.355 | 154530.526 | -1.7E-01 | 154564.936 | 154565.106 | -1.7E-01 | -63.071 | -62.908 | -1.6E-01 |
| 1613 | Er | 166 | 68 | 98 | 30 | 8.141956 | 8.141607 | 3.5E-04 | 154527.977 | 154527.993 | -1.6E-02 | 154563.081 | 154563.097 | -1.6E-02 | -64.926 | -64.917 | -8.5E-03 |
| 1614 | Tm | 166 | 69 | 97 | 28 | 8.118944 | 8.118331 | 6.1E-04 | 154530.490 | 154530.550 | -6.0E-02 | 154566.118 | 154566.178 | -6.0E-02 | -61.888 | -61.836 | -5.2E-02 |
| 1615 | Yb | 166 | 70 | 96 | 26 | 8.112468 | 8.110556 | 1.9E-03 | 154530.258 | 154530.533 | -2.7E-01 | 154566.411 | 154566.685 | -2.7E-01 | -61.595 | -61.329 | -2.7E-01 |
| 1616 | Lu | 166 | 71 | 95 | 24 | 8.074175 | 8.072168 | 2.0E-03 | 154535.308 | 154535.598 | -2.9E-01 | 154571.985 | 154572.275 | -2.9E-01 | -56.021 | -55.739 | -2.8E-01 |
| 1617 | Hf | 166 | 72 | 94 | 22 | 8.056438 | 8.051381 | 5.1E-03 | 154536.945 | 154537.740 | -8.0E-01 | 154574.147 | 154574.942 | -8.0E-01 | -53.859 | -53.072 | -7.9E-01 |
| 1618 | Ta | 166 | 73 | 93 | 20 | 8.004971 | 7.999376 | 5.6E-03 | 154544.181 | 154545.065 | -8.8E-01 | 154581.908 | 154582.792 | -8.8E-01 | -46.098 | -45.222 | -8.8E-01 |
| 1619 | W | 166 | 74 | 92 | 18 | 7.974899 | 7.967264 | 7.6E-03 | 154547.865 | 154549.086 | -1.2E+00 | 154586.118 | 154587.340 | -1.2E+00 | -41.888 | -40.675 | -1.2E+00 |
| 1620 | Re | 166 | 75 | 91 | 16 | 7.909977 | 7.902738 | 7.2E-03 | 154557.334 | 154558.489 | -1.2E+00 | 154596.113 | 154597.268 | -1.2E+00 | -31.894 | -30.746 | -1.1E+00 |
| 1621 | Os | 166 | 76 | 90 | 14 | 7.866368 | 7.860594 | 5.8E-03 | 154563.264 | 154564.175 | -9.1E-01 | 154602.569 | 154603.481 | -9.1E-01 | -25.437 | -24.534 | -9.0E-01 |
| 1622 | Dy | 167 | 66 | 101 | 35 | 8.120995 | 8.122642 | -1.6E-03 | 155465.512 | 155465.198 | 3.1E-01 | 155499.569 | 155499.254 | 3.1E-01 | -59.931 | -60.254 | 3.2E-01 |
| 1623 | Ho | 167 | 67 | 100 | 33 | 8.130382 | 8.129725 | 6.6E-04 | 155462.639 | 155462.708 | -6.9E-02 | 155497.219 | 155497.288 | -6.9E-02 | -62.281 | -62.220 | -6.1E-02 |
| 1624 | Er | 167 | 68 | 99 | 31 | 8.131743 | 8.130199 | 1.5E-03 | 155461.106 | 155461.322 | -2.2E-01 | 155496.209 | 155496.426 | -2.2E-01 | -63.291 | -63.082 | -2.1E-01 |
| 1625 | Tm | 167 | 69 | 98 | 29 | 8.122587 | 8.120228 | 2.4E-03 | 155461.328 | 155461.680 | -3.5E-01 | 155496.956 | 155497.308 | -3.5E-01 | -62.544 | -62.200 | -3.4E-01 |
| 1626 | Yb | 167 | 70 | 97 | 27 | 8.106207 | 8.104522 | 1.7E-03 | 155462.757 | 155462.996 | -2.4E-01 | 155498.909 | 155499.148 | -2.4E-01 | -60.591 | -60.360 | -2.3E-01 |
| 1627 | Lu | 167 | 71 | 96 | 25 | 8.083021 | 8.079925 | 3.1E-03 | 155465.322 | 155465.795 | -4.7E-01 | 155501.999 | 155502.472 | -4.7E-01 | -57.501 | -57.036 | -4.7E-01 |
| 1628 | Hf | 167 | 72 | 95 | 23 | 8.054184 | 8.050042 | 4.1E-03 | 155468.830 | 155469.478 | -6.5E-01 | 155506.032 | 155506.680 | -6.5E-01 | -53.468 | -52.828 | -6.4E-01 |
| 1629 | Ta | 167 | 73 | 94 | 21 | 8.018861 | 8.012622 | 6.2E-03 | 155473.422 | 155474.419 | -1.0E+00 | 155511.149 | 155512.146 | -1.0E+00 | -48.351 | -47.362 | -9.9E-01 |
| 1630 | W | 167 | 74 | 93 | 19 | 7.976740 | 7.970037 | 6.7E-03 | 155479.148 | 155480.222 | -1.1E+00 | 155517.401 | 155518.475 | -1.1E+00 | -42.099 | -41.034 | -1.1E+00 |
| 1631 | Os | 167 | 76 | 91 | 15 | 7.873975 | 7.867014 | 7.0E-03 | 155493.693 | 155494.808 | -1.1E+00 | 155532.998 | 155534.113 | -1.1E+00 | -26.502 | -25.395 | -1.1E+00 |
| 1632 | Ir | 167 | 77 | 90 | 13 | 7.812856 | 7.808010 | 4.8E-03 | 155502.590 | 155503.352 | -7.6E-01 | 155542.422 | 155543.184 | -7.6E-01 | -17.078 | -16.324 | -7.5E-01 |
| 1633 | Dy | 168 | 66 | 102 | 36 | 8.112538 | 8.115740 | -3.2E-03 | 156398.378 | 156397.800 | 5.8E-01 | 156432.434 | 156431.856 | 5.8E-01 | -58.560 | -59.146 | 5.9E-01 |
| 1634 | Ho | 168 | 67 | 101 | 34 | 8.116814 | 8.116509 | 3.1E-04 | 156396.353 | 156396.364 | -1.1E-02 | 156430.933 | 156430.944 | -1.1E-02 | -60.061 | -60.058 | -2.7E-03 |
| 1635 | Er | 168 | 68 | 100 | 32 | 8.129598 | 8.129840 | -2.4E-04 | 156392.900 | 156392.818 | 8.2E-02 | 156428.003 | 156427.921 | 8.2E-02 | -62.991 | -63.081 | 9.0E-02 |
| 1636 | Tm | 168 | 69 | 99 | 30 | 8.114957 | 8.112728 | 2.2E-03 | 156394.053 | 156394.385 | -3.3E-01 | 156429.681 | 156430.013 | -3.3E-01 | -61.313 | -60.989 | -3.2E-01 |
| 1637 | Yb | 168 | 70 | 98 | 28 | 8.111896 | 8.110168 | 1.7E-03 | 156393.260 | 156393.508 | -2.5E-01 | 156429.413 | 156429.660 | -2.5E-01 | -61.581 | -61.342 | -2.4E-01 |
| 1638 | Lu | 168 | 71 | 97 | 26 | 8.080369 | 8.077545 | 2.8E-03 | 156397.250 | 156397.681 | -4.3E-01 | 156433.927 | 156434.358 | -4.3E-01 | -57.067 | -56.645 | -4.2E-01 |
| 1639 | Hf | 168 | 72 | 96 | 24 | 8.065553 | 8.061462 | 4.1E-03 | 156398.432 | 156399.075 | -6.4E-01 | 156435.634 | 156436.277 | -6.4E-01 | -55.361 | -54.726 | -6.3E-01 |
| 1640 | Ta | 168 | 73 | 95 | 22 | 8.019428 | 8.014923 | 4.5E-03 | 156404.873 | 156405.585 | -7.1E-01 | 156442.600 | 156443.312 | -7.1E-01 | -48.394 | -47.690 | -7.0E-01 |
| 1641 | W | 168 | 74 | 94 | 20 | 7.993931 | 7.987085 | 6.8E-03 | 156407.848 | 156408.953 | -1.1E+00 | 156446.101 | 156447.206 | -1.1E+00 | -44.893 | -43.796 | -1.1E+00 |
| 1642 | Re | 168 | 75 | 93 | 18 | 7.935120 | 7.927786 | 7.3E-03 | 156416.420 | 156417.606 | -1.2E+00 | 156455.199 | 156456.385 | -1.2E+00 | -35.795 | -34.617 | -1.2E+00 |
| 1643 | Os | 168 | 76 | 92 | 16 | 7.895892 | 7.889587 | 6.3E-03 | 156421.702 | 156422.714 | -1.0E+00 | 156461.007 | 156462.019 | -1.0E+00 | -29.987 | -28.983 | -1.0E+00 |
| 1644 | Ir | 168 | 77 | 91 | 14 | 7.824151 | 7.818452 | 5.7E-03 | 156432.445 | 156433.355 | -9.1E-01 | 156472.277 | 156473.187 | -9.1E-01 | -18.717 | -17.815 | -9.0E-01 |
| 1645 | Pt | 168 | 78 | 90 | 12 | 7.773907 | 7.771304 | 2.6E-03 | 156439.577 | 156439.966 | -3.9E-01 | 156479.936 | 156480.325 | -3.9E-01 | -11.058 | -10.677 | -3.8E-01 |
| 1646 | Dy | 169 | 66 | 103 | 37 | 8.094765 | 8.095554 | -7.9E-04 | 157332.834 | 157332.661 | 1.7E-01 | 157366.891 | 157366.718 | 1.7E-01 | -55.598 | -55.779 | 1.8E-01 |
| 1647 | Ho | 169 | 67 | 102 | 35 | 8.109071 | 8.109400 | -3.3E-04 | 157329.110 | 157329.014 | 9.6E-02 | 157363.691 | 157363.594 | 9.7E-02 | -58.798 | -58.902 | 1.0E-01 |
| 1648 | Er | 169 | 68 | 101 | 33 | 8.117016 | 8.116322 | 6.9E-04 | 157326.462 | 157326.538 | -7.6E-02 | 157361.566 | 157361.642 | -7.6E-02 | -60.923 | -60.855 | -6.8E-02 |
| 1649 | Tm | 169 | 69 | 100 | 31 | 8.114475 | 8.112231 | 2.2E-03 | 157325.585 | 157325.922 | -3.4E-01 | 157361.213 | 157361.550 | -3.4E-01 | -61.276 | -60.946 | -3.3E-01 |
| 1650 | Yb | 169 | 70 | 99 | 29 | 8.104529 | 8.102284 | 2.2E-03 | 157325.959 | 157326.295 | -3.4E-01 | 157362.111 | 157362.448 | -3.4E-01 | -60.377 | -60.049 | -3.3E-01 |
| 1651 | Lu | 169 | 71 | 98 | 27 | 8.086332 | 8.083034 | 3.3E-03 | 157327.727 | 157328.241 | -5.1E-01 | 157364.404 | 157364.918 | -5.1E-01 | -58.084 | -57.578 | -5.1E-01 |
| 1652 | Hf | 169 | 72 | 97 | 25 | 8.061778 | 8.058516 | 3.3E-03 | 157330.569 | 157331.076 | -5.1E-01 | 157367.771 | 157368.278 | -5.1E-01 | -54.717 | -54.218 | -5.0E-01 |
| 1653 | Ta | 169 | 73 | 96 | 23 | 8.030957 | 8.026067 | 4.9E-03 | 157334.470 | 157335.252 | -7.8E-01 | 157372.198 | 157372.979 | -7.8E-01 | -50.290 | -49.517 | -7.7E-01 |
| 1654 | W | 169 | 74 | 95 | 21 | 7.994537 | 7.988538 | 6.0E-03 | 157339.317 | 157340.286 | -9.7E-01 | 157377.570 | 157378.539 | -9.7E-01 | -44.918 | -43.958 | -9.6E-01 |
| 1655 | Re | 169 | 75 | 94 | 19 | 7.951395 | 7.944343 | 7.1E-03 | 157345.300 | 157346.445 | -1.1E+00 | 157384.079 | 157385.224 | -1.1E+00 | -38.409 | -37.272 | -1.1E+00 |
| 1656 | Os | 169 | 76 | 93 | 17 | 7.901285 | 7.894953 | 6.3E-03 | 157352.460 | 157353.483 | -1.0E+00 | 157391.765 | 157392.788 | -1.0E+00 | -30.723 | -29.708 | -1.0E+00 |
| 1657 | Ir | 169 | 77 | 92 | 15 | 7.845501 | 7.840233 | 5.3E-03 | 157360.578 | 157361.421 | -8.4E-01 | 157400.410 | 157401.253 | -8.4E-01 | -22.078 | -21.243 | -8.3E-01 |
| 1658 | Ho | 170 | 67 | 103 | 36 | 8.093799 | 8.093966 | -1.7E-04 | 158263.163 | 158263.094 | 6.9E-02 | 158297.743 | 158297.674 | 6.9E-02 | -56.239 | -56.316 | 7.7E-02 |
| 1659 | Er | 170 | 68 | 102 | 34 | 8.111961 | 8.113553 | -1.6E-03 | 158258.769 | 158258.457 | 3.1E-01 | 158293.873 | 158293.561 | 3.1E-01 | -60.109 | -60.429 | 3.2E-01 |
| 1660 | Tm | 170 | 69 | 101 | 32 | 8.105519 | 8.102750 | 2.8E-03 | 158258.558 | 158258.987 | -4.3E-01 | 158294.186 | 158294.615 | -4.3E-01 | -59.796 | -59.376 | -4.2E-01 |
| 1661 | Yb | 170 | 70 | 100 | 30 | 8.106614 | 8.105593 | 1.0E-03 | 158257.065 | 158257.196 | -1.3E-01 | 158293.217 | 158293.348 | -1.3E-01 | -60.765 | -60.642 | -1.2E-01 |
| 1662 | Lu | 170 | 71 | 99 | 28 | 8.081673 | 8.078801 | 2.9E-03 | 158259.998 | 158260.443 | -4.4E-01 | 158296.675 | 158297.120 | -4.5E-01 | -57.307 | -56.870 | -4.4E-01 |
| 1663 | Hf | 170 | 72 | 98 | 26 | 8.070875 | 8.067601 | 3.3E-03 | 158260.526 | 158261.039 | -5.1E-01 | 158297.728 | 158298.241 | -5.1E-01 | -56.254 | -55.750 | -5.0E-01 |
| 1664 | Ta | 170 | 73 | 97 | 24 | 8.030296 | 8.026644 | 3.7E-03 | 158266.117 | 158266.693 | -5.8E-01 | 158303.845 | 158304.420 | -5.8E-01 | -50.138 | -49.570 | -5.7E-01 |
| 1665 | W | 170 | 74 | 96 | 22 | 8.008946 | 8.003304 | 5.6E-03 | 158268.439 | 158269.352 | -9.1E-01 | 158306.692 | 158307.605 | -9.1E-01 | -47.290 | -46.385 | -9.1E-01 |
| 1666 | Re | 170 | 75 | 95 | 20 | 7.955092 | 7.949368 | 5.7E-03 | 158276.285 | 158277.212 | -9.3E-01 | 158315.064 | 158315.991 | -9.3E-01 | -38.918 | -37.999 | -9.2E-01 |
| 1667 | Os | 170 | 76 | 94 | 18 | 7.921130 | 7.915318 | 5.8E-03 | 158280.750 | 158281.691 | -9.4E-01 | 158320.056 | 158320.997 | -9.4E-01 | -33.926 | -32.994 | -9.3E-01 |
| 1668 | Pt | 170 | 78 | 92 | 14 | 7.808267 | 7.805822 | 2.4E-03 | 158297.319 | 158297.686 | -3.7E-01 | 158337.678 | 158338.045 | -3.7E-01 | -16.305 | -15.946 | -3.6E-01 |
| 1669 | Ho | 171 | 67 | 104 | 37 | 8.083611 | 8.084158 | -5.5E-04 | 159196.377 | 159196.243 | 1.3E-01 | 159230.957 | 159230.823 | 1.3E-01 | -54.519 | -54.662 | 1.4E-01 |
| 1670 | Er | 171 | 68 | 103 | 35 | 8.097749 | 8.097810 | -6.1E-05 | 159192.653 | 159192.601 | 5.2E-02 | 159227.757 | 159227.705 | 5.2E-02 | -57.719 | -57.779 | 6.0E-02 |
| 1671 | Tm | 171 | 69 | 102 | 33 | 8.101899 | 8.099819 | 2.1E-03 | 159190.637 | 159190.951 | -3.1E-01 | 159226.265 | 159226.579 | -3.1E-01 | -59.211 | -58.906 | -3.1E-01 |
| 1672 | Yb | 171 | 70 | 101 | 31 | 8.097889 | 8.095775 | 2.1E-03 | 159190.016 | 159190.335 | -3.2E-01 | 159226.168 | 159226.487 | -3.2E-01 | -59.308 | -58.997 | -3.1E-01 |
| 1673 | Lu | 171 | 71 | 100 | 29 | 8.084670 | 8.081980 | 2.7E-03 | 159190.969 | 159191.386 | -4.2E-01 | 159227.646 | 159228.063 | -4.2E-01 | -57.830 | -57.421 | -4.1E-01 |
| 1674 | Hf | 171 | 72 | 99 | 27 | 8.066068 | 8.062914 | 3.2E-03 | 159192.843 | 159193.338 | -5.0E-01 | 159230.045 | 159230.540 | -5.0E-01 | -55.431 | -54.944 | -4.9E-01 |
| 1675 | Ta | 171 | 73 | 98 | 25 | 8.039791 | 8.035566 | 4.2E-03 | 159196.028 | 159196.706 | -6.8E-01 | 159233.756 | 159234.434 | -6.8E-01 | -51.720 | -51.051 | -6.7E-01 |
| 1676 | W | 171 | 74 | 97 | 23 | 8.008115 | 8.003225 | 4.9E-03 | 159200.137 | 159200.928 | -7.9E-01 | 159238.390 | 159239.181 | -7.9E-01 | -47.086 | -46.304 | -7.8E-01 |
| 1677 | Re | 171 | 75 | 96 | 21 | 7.969412 | 7.963836 | 5.6E-03 | 159205.447 | 159206.354 | -9.1E-01 | 159244.226 | 159245.133 | -9.1E-01 | -41.250 | -40.351 | -9.0E-01 |
| 1678 | Os | 171 | 76 | 95 | 19 | 7.924211 | 7.919384 | 4.8E-03 | 159211.868 | 159212.646 | -7.8E-01 | 159251.173 | 159251.951 | -7.8E-01 | -34.303 | -33.533 | -7.7E-01 |
| 1679 | Ir | 171 | 77 | 94 | 17 | 7.873517 | 7.869139 | 4.4E-03 | 159219.227 | 159219.928 | -7.0E-01 | 159259.059 | 159259.760 | -7.0E-01 | -26.417 | -25.724 | -6.9E-01 |
| 1680 | Pt | 171 | 78 | 93 | 15 | 7.816621 | 7.813521 | 3.1E-03 | 159227.647 | 159228.129 | -4.8E-01 | 159268.006 | 159268.488 | -4.8E-01 | -17.470 | -16.997 | -4.7E-01 |
| 1681 | Au | 171 | 79 | 92 | 13 | 7.754137 | 7.753560 | 5.8E-04 | 159237.022 | 159237.072 | -5.0E-02 | 159277.909 | 159277.958 | -4.9E-02 | -7.568 | -7.526 | -4.1E-02 |
| 1682 | Er | 172 | 68 | 104 | 36 | 8.090413 | 8.092595 | -2.2E-03 | 160125.382 | 160124.966 | 4.2E-01 | 160160.486 | 160160.070 | 4.2E-01 | -56.484 | -56.909 | 4.2E-01 |
| 1683 | Tm | 172 | 69 | 103 | 34 | 8.091044 | 8.088301 | 2.7E-03 | 160123.967 | 160124.397 | -4.3E-01 | 160159.595 | 160160.025 | -4.3E-01 | -57.375 | -56.953 | -4.2E-01 |
| 1684 | Yb | 172 | 70 | 102 | 32 | 8.097433 | 8.096762 | 6.7E-04 | 160121.562 | 160121.635 | -7.3E-02 | 160157.714 | 160157.787 | -7.3E-02 | -59.256 | -59.192 | -6.5E-02 |
| 1685 | Lu | 172 | 71 | 101 | 30 | 8.078245 | 8.075860 | 2.4E-03 | 160123.555 | 160123.922 | -3.7E-01 | 160160.232 | 160160.599 | -3.7E-01 | -56.738 | -56.380 | -3.6E-01 |
| 1686 | Hf | 172 | 72 | 100 | 28 | 8.071743 | 8.069670 | 2.1E-03 | 160123.366 | 160123.679 | -3.1E-01 | 160160.568 | 160160.881 | -3.1E-01 | -56.402 | -56.098 | -3.0E-01 |
| 1687 | Ta | 172 | 73 | 99 | 26 | 8.037705 | 8.034332 | 3.4E-03 | 160127.913 | 160128.448 | -5.4E-01 | 160165.640 | 160166.176 | -5.4E-01 | -51.330 | -50.803 | -5.3E-01 |
| 1688 | W | 172 | 74 | 98 | 24 | 8.020175 | 8.015644 | 4.5E-03 | 160129.620 | 160130.354 | -7.3E-01 | 160167.873 | 160168.607 | -7.3E-01 | -49.097 | -48.372 | -7.3E-01 |
| 1689 | Re | 172 | 75 | 97 | 22 | 7.971610 | 7.967154 | 4.5E-03 | 160136.665 | 160137.385 | -7.2E-01 | 160175.444 | 160176.164 | -7.2E-01 | -41.526 | -40.814 | -7.1E-01 |
| 1690 | Os | 172 | 76 | 96 | 20 | 7.942162 | 7.937426 | 4.7E-03 | 160140.421 | 160141.189 | -7.7E-01 | 160179.727 | 160180.494 | -7.7E-01 | -37.244 | -36.484 | -7.6E-01 |
| 1691 | Ir | 172 | 77 | 95 | 18 | 7.880263 | 7.876652 | 3.6E-03 | 160149.759 | 160150.332 | -5.7E-01 | 160189.591 | 160190.164 | -5.7E-01 | -27.379 | -26.814 | -5.7E-01 |
| 1692 | Pt | 172 | 78 | 94 | 16 | 7.839192 | 7.837070 | 2.1E-03 | 160155.514 | 160155.830 | -3.2E-01 | 160195.873 | 160196.189 | -3.2E-01 | -21.097 | -20.789 | -3.1E-01 |
| 1693 | Au | 172 | 79 | 93 | 14 | 7.766453 | 7.764737 | 1.7E-03 | 160166.715 | 160166.961 | -2.5E-01 | 160207.602 | 160207.848 | -2.5E-01 | -9.369 | -9.131 | -2.4E-01 |
| 1694 | Hg | 172 | 80 | 92 | 12 | 7.713885 | 7.716361 | -2.5E-03 | 160174.447 | 160173.971 | 4.8E-01 | 160215.861 | 160215.385 | 4.8E-01 | -1.109 | -1.593 | 4.8E-01 |
| 1695 | Tm | 173 | 69 | 104 | 35 | 8.084453 | 8.082877 | 1.6E-03 | 161056.582 | 161056.813 | -2.3E-01 | 161092.210 | 161092.441 | -2.3E-01 | -56.254 | -56.032 | -2.2E-01 |
| 1696 | Yb | 173 | 70 | 103 | 33 | 8.087433 | 8.084928 | 2.5E-03 | 161054.760 | 161055.150 | -3.9E-01 | 161090.912 | 161091.303 | -3.9E-01 | -57.552 | -57.170 | -3.8E-01 |
| 1697 | Lu | 173 | 71 | 102 | 31 | 8.079040 | 8.076717 | 2.3E-03 | 161054.905 | 161055.263 | -3.6E-01 | 161091.582 | 161091.940 | -3.6E-01 | -56.883 | -56.532 | -3.5E-01 |
| 1698 | Hf | 173 | 72 | 101 | 29 | 8.066016 | 8.063171 | 2.8E-03 | 161055.850 | 161056.298 | -4.5E-01 | 161093.052 | 161093.501 | -4.5E-01 | -55.412 | -54.972 | -4.4E-01 |
| 1699 | Ta | 173 | 73 | 100 | 27 | 8.044064 | 8.040993 | 3.1E-03 | 161058.340 | 161058.827 | -4.9E-01 | 161096.068 | 161096.554 | -4.9E-01 | -52.397 | -51.918 | -4.8E-01 |
| 1700 | W | 173 | 74 | 99 | 25 | 8.018333 | 8.013905 | 4.4E-03 | 161061.484 | 161062.204 | -7.2E-01 | 161099.737 | 161100.458 | -7.2E-01 | -48.727 | -48.015 | -7.1E-01 |
| 1701 | Re | 173 | 75 | 98 | 23 | 7.983906 | 7.979435 | 4.5E-03 | 161066.131 | 161066.859 | -7.3E-01 | 161104.910 | 161105.638 | -7.3E-01 | -43.554 | -42.835 | -7.2E-01 |
| 1702 | Os | 173 | 76 | 97 | 21 | 7.944033 | 7.940014 | 4.0E-03 | 161071.721 | 161072.369 | -6.5E-01 | 161111.026 | 161111.675 | -6.5E-01 | -37.438 | -36.798 | -6.4E-01 |
| 1703 | Ir | 173 | 77 | 96 | 19 | 7.898066 | 7.894384 | 3.7E-03 | 161078.364 | 161078.953 | -5.9E-01 | 161118.196 | 161118.785 | -5.9E-01 | -30.268 | -29.687 | -5.8E-01 |
| 1704 | Pt | 173 | 78 | 95 | 17 | 7.845422 | 7.843506 | 1.9E-03 | 161086.162 | 161086.445 | -2.8E-01 | 161126.521 | 161126.804 | -2.8E-01 | -21.943 | -21.668 | -2.7E-01 |
| 1705 | Au | 173 | 79 | 94 | 15 | 7.788144 | 7.787683 | 4.6E-04 | 161094.761 | 161094.792 | -3.1E-02 | 161135.648 | 161135.678 | -3.0E-02 | -12.816 | -12.794 | -2.2E-02 |
| 1706 | Tm | 174 | 69 | 105 | 36 | 8.070648 | 8.069248 | 1.4E-03 | 161990.465 | 161990.667 | -2.0E-01 | 162026.093 | 162026.295 | -2.0E-01 | -53.866 | -53.672 | -1.9E-01 |
| 1707 | Yb | 174 | 70 | 104 | 34 | 8.083853 | 8.083593 | 2.6E-04 | 161986.860 | 161986.863 | -3.1E-03 | 162023.013 | 162023.016 | -2.5E-03 | -56.946 | -56.951 | 5.6E-03 |
| 1708 | Lu | 174 | 71 | 103 | 32 | 8.071464 | 8.068687 | 2.8E-03 | 161987.709 | 161988.149 | -4.4E-01 | 162024.386 | 162024.826 | -4.4E-01 | -55.572 | -55.140 | -4.3E-01 |
| 1709 | Hf | 174 | 72 | 102 | 30 | 8.068545 | 8.067638 | 9.1E-04 | 161986.909 | 161987.024 | -1.1E-01 | 162024.112 | 162024.226 | -1.1E-01 | -55.847 | -55.741 | -1.1E-01 |
| 1710 | Ta | 174 | 73 | 101 | 28 | 8.040452 | 8.037925 | 2.5E-03 | 161990.490 | 161990.885 | -4.0E-01 | 162028.217 | 162028.612 | -4.0E-01 | -51.741 | -51.354 | -3.9E-01 |
| 1711 | W | 174 | 74 | 100 | 26 | 8.027256 | 8.023984 | 3.3E-03 | 161991.478 | 161992.002 | -5.2E-01 | 162029.731 | 162030.255 | -5.2E-01 | -50.227 | -49.711 | -5.2E-01 |
| 1712 | Re | 174 | 75 | 99 | 24 | 7.985094 | 7.980970 | 4.1E-03 | 161997.506 | 161998.178 | -6.7E-01 | 162036.285 | 162036.957 | -6.7E-01 | -43.673 | -43.010 | -6.6E-01 |
| 1713 | Os | 174 | 76 | 98 | 22 | 7.959459 | 7.955693 | 3.8E-03 | 162000.658 | 162001.266 | -6.1E-01 | 162039.963 | 162040.572 | -6.1E-01 | -39.995 | -39.395 | -6.0E-01 |
| 1714 | Ir | 174 | 77 | 97 | 20 | 7.902513 | 7.900205 | 2.3E-03 | 162009.258 | 162009.612 | -3.5E-01 | 162049.090 | 162049.443 | -3.5E-01 | -30.869 | -30.523 | -3.5E-01 |
| 1715 | Pt | 174 | 78 | 96 | 18 | 7.866118 | 7.864695 | 1.4E-03 | 162014.281 | 162014.480 | -2.0E-01 | 162054.640 | 162054.839 | -2.0E-01 | -25.318 | -25.128 | -1.9E-01 |
| 1716 | Hg | 174 | 80 | 94 | 14 | 7.749816 | 7.752449 | -2.6E-03 | 162031.898 | 162031.390 | 5.1E-01 | 162073.312 | 162072.804 | 5.1E-01 | -6.646 | -7.163 | 5.2E-01 |
| 1717 | Tm | 175 | 69 | 106 | 37 | 8.061772 | 8.061258 | 5.1E-04 | 162923.513 | 162923.561 | -4.8E-02 | 162959.141 | 162959.189 | -4.8E-02 | -52.312 | -52.272 | -4.0E-02 |
| 1718 | Yb | 175 | 70 | 105 | 35 | 8.070930 | 8.069648 | 1.3E-03 | 162920.603 | 162920.785 | -1.8E-01 | 162956.756 | 162956.938 | -1.8E-01 | -54.697 | -54.523 | -1.7E-01 |
| 1719 | Lu | 175 | 71 | 104 | 33 | 8.069151 | 8.067201 | 2.0E-03 | 162919.608 | 162919.906 | -3.0E-01 | 162956.285 | 162956.583 | -3.0E-01 | -55.168 | -54.878 | -2.9E-01 |
| 1720 | Hf | 175 | 72 | 103 | 31 | 8.060774 | 8.059273 | 1.5E-03 | 162919.766 | 162919.985 | -2.2E-01 | 162956.968 | 162957.187 | -2.2E-01 | -54.484 | -54.274 | -2.1E-01 |
| 1721 | Ta | 175 | 73 | 102 | 29 | 8.044445 | 8.042327 | 2.1E-03 | 162921.316 | 162921.642 | -3.3E-01 | 162959.044 | 162959.370 | -3.3E-01 | -52.409 | -52.091 | -3.2E-01 |
| 1722 | W | 175 | 74 | 101 | 27 | 8.024112 | 8.020521 | 3.6E-03 | 162923.566 | 162924.150 | -5.8E-01 | 162961.819 | 162962.403 | -5.8E-01 | -49.633 | -49.058 | -5.7E-01 |
| 1723 | Re | 175 | 75 | 100 | 25 | 7.994816 | 7.991029 | 3.8E-03 | 162927.385 | 162928.002 | -6.2E-01 | 162966.164 | 162966.781 | -6.2E-01 | -45.288 | -44.680 | -6.1E-01 |
| 1724 | Os | 175 | 76 | 99 | 23 | 7.960728 | 7.956694 | 4.0E-03 | 162932.042 | 162932.701 | -6.6E-01 | 162971.347 | 162972.006 | -6.6E-01 | -40.105 | -39.454 | -6.5E-01 |
| 1725 | Ir | 175 | 77 | 98 | 21 | 7.917910 | 7.915775 | 2.1E-03 | 162938.226 | 162938.552 | -3.3E-01 | 162978.058 | 162978.384 | -3.3E-01 | -33.394 | -33.077 | -3.2E-01 |
| 1726 | Pt | 175 | 78 | 97 | 19 | 7.869472 | 7.869710 | -2.4E-04 | 162945.393 | 162945.303 | 9.0E-02 | 162985.752 | 162985.662 | 9.0E-02 | -25.700 | -25.798 | 9.8E-02 |
| 1727 | Au | 175 | 79 | 96 | 17 | 7.817661 | 7.818198 | -5.4E-04 | 162953.151 | 162953.007 | 1.4E-01 | 162994.037 | 162993.894 | 1.4E-01 | -17.415 | -17.567 | 1.5E-01 |
| 1728 | Hg | 175 | 80 | 95 | 15 | 7.759232 | 7.760934 | -1.7E-03 | 162962.065 | 162961.718 | 3.5E-01 | 163003.479 | 163003.132 | 3.5E-01 | -7.973 | -8.329 | 3.6E-01 |
| 1729 | Tm | 176 | 69 | 107 | 38 | 8.045111 | 8.045436 | -3.2E-04 | 163857.949 | 163857.850 | 9.9E-02 | 163893.577 | 163893.478 | 9.9E-02 | -49.370 | -49.477 | 1.1E-01 |
| 1730 | Yb | 176 | 70 | 106 | 36 | 8.064075 | 8.065968 | -1.9E-03 | 163853.304 | 163852.929 | 3.8E-01 | 163889.457 | 163889.081 | 3.8E-01 | -53.490 | -53.874 | 3.8E-01 |
| 1731 | Lu | 176 | 71 | 105 | 34 | 8.059031 | 8.057218 | 1.8E-03 | 163852.885 | 163853.161 | -2.8E-01 | 163889.562 | 163889.838 | -2.8E-01 | -53.384 | -53.117 | -2.7E-01 |
| 1732 | Hf | 176 | 72 | 104 | 32 | 8.061371 | 8.061485 | -1.1E-04 | 163851.166 | 163851.102 | 6.4E-02 | 163888.368 | 163888.304 | 6.4E-02 | -54.578 | -54.651 | 7.2E-02 |
| 1733 | Ta | 176 | 73 | 103 | 30 | 8.038670 | 8.037422 | 1.2E-03 | 163853.853 | 163854.028 | -1.8E-01 | 163891.581 | 163891.756 | -1.7E-01 | -51.365 | -51.199 | -1.7E-01 |
| 1734 | W | 176 | 74 | 102 | 28 | 8.030112 | 8.028310 | 1.8E-03 | 163854.052 | 163854.324 | -2.7E-01 | 163892.305 | 163892.577 | -2.7E-01 | -50.642 | -50.378 | -2.6E-01 |
| 1735 | Re | 176 | 75 | 101 | 26 | 7.993970 | 7.990762 | 3.2E-03 | 163859.104 | 163859.623 | -5.2E-01 | 163897.883 | 163898.402 | -5.2E-01 | -45.063 | -44.553 | -5.1E-01 |
| 1736 | Os | 176 | 76 | 100 | 24 | 7.972679 | 7.970028 | 2.7E-03 | 163861.543 | 163861.963 | -4.2E-01 | 163900.848 | 163901.268 | -4.2E-01 | -42.098 | -41.687 | -4.1E-01 |
| 1737 | Ir | 176 | 77 | 99 | 22 | 7.921424 | 7.919845 | 1.6E-03 | 163869.255 | 163869.485 | -2.3E-01 | 163909.087 | 163909.317 | -2.3E-01 | -33.859 | -33.638 | -2.2E-01 |
| 1738 | Pt | 176 | 78 | 98 | 20 | 7.888992 | 7.888523 | 4.7E-04 | 163873.654 | 163873.688 | -3.4E-02 | 163914.013 | 163914.047 | -3.4E-02 | -28.934 | -28.908 | -2.6E-02 |
| 1739 | Au | 176 | 79 | 97 | 18 | 7.824676 | 7.826235 | -1.6E-03 | 163883.663 | 163883.340 | 3.2E-01 | 163924.550 | 163924.226 | 3.2E-01 | -18.397 | -18.728 | 3.3E-01 |
| 1740 | Hg | 176 | 80 | 96 | 16 | 7.782599 | 7.785035 | -2.4E-03 | 163889.759 | 163889.281 | 4.8E-01 | 163931.173 | 163930.695 | 4.8E-01 | -11.773 | -12.260 | 4.9E-01 |
| 1741 | Tl | 176 | 81 | 95 | 14 | 7.707957 | 7.710767 | -2.8E-03 | 163901.586 | 163901.041 | 5.5E-01 | 163943.528 | 163942.983 | 5.5E-01 | 0.581 | 0.028 | 5.5E-01 |
| 1742 | Yb | 177 | 70 | 107 | 37 | 8.049964 | 8.049804 | 1.6E-04 | 164787.303 | 164787.289 | 1.4E-02 | 164823.456 | 164823.442 | 1.4E-02 | -50.985 | -51.007 | 2.2E-02 |
| 1743 | Lu | 177 | 71 | 106 | 35 | 8.053459 | 8.053348 | 1.1E-04 | 164785.377 | 164785.354 | 2.3E-02 | 164822.054 | 164822.031 | 2.3E-02 | -52.386 | -52.418 | 3.2E-02 |
| 1744 | Hf | 177 | 72 | 105 | 33 | 8.051849 | 8.051193 | 6.6E-04 | 164784.355 | 164784.427 | -7.2E-02 | 164821.557 | 164821.630 | -7.3E-02 | -52.883 | -52.819 | -6.4E-02 |
| 1745 | Ta | 177 | 73 | 104 | 31 | 8.040841 | 8.039572 | 1.3E-03 | 164784.996 | 164785.176 | -1.8E-01 | 164822.723 | 164822.903 | -1.8E-01 | -51.717 | -51.545 | -1.7E-01 |
| 1746 | W | 177 | 74 | 103 | 29 | 8.025035 | 8.023082 | 2.0E-03 | 164786.485 | 164786.786 | -3.0E-01 | 164824.739 | 164825.039 | -3.0E-01 | -49.702 | -49.410 | -2.9E-01 |
| 1747 | Re | 177 | 75 | 102 | 27 | 8.001222 | 7.998607 | 2.6E-03 | 164789.392 | 164789.809 | -4.2E-01 | 164828.171 | 164828.588 | -4.2E-01 | -46.269 | -45.861 | -4.1E-01 |
| 1748 | Os | 177 | 76 | 101 | 25 | 7.972396 | 7.969383 | 3.0E-03 | 164793.186 | 164793.672 | -4.9E-01 | 164832.491 | 164832.978 | -4.9E-01 | -41.949 | -41.471 | -4.8E-01 |
| 1749 | Ir | 177 | 77 | 100 | 23 | 7.934632 | 7.933239 | 1.4E-03 | 164798.561 | 164798.760 | -2.0E-01 | 164838.393 | 164838.592 | -2.0E-01 | -36.047 | -35.857 | -1.9E-01 |
| 1750 | Pt | 177 | 78 | 99 | 21 | 7.892489 | 7.892028 | 4.6E-04 | 164804.711 | 164804.744 | -3.3E-02 | 164845.070 | 164845.103 | -3.3E-02 | -29.370 | -29.346 | -2.5E-02 |
| 1751 | Au | 177 | 79 | 98 | 19 | 7.843858 | 7.844923 | -1.1E-03 | 164812.009 | 164811.772 | 2.4E-01 | 164852.895 | 164852.658 | 2.4E-01 | -21.545 | -21.791 | 2.5E-01 |
| 1752 | Hg | 177 | 80 | 97 | 17 | 7.789932 | 7.792190 | -2.3E-03 | 164820.244 | 164819.795 | 4.5E-01 | 164861.658 | 164861.209 | 4.5E-01 | -12.783 | -13.240 | 4.6E-01 |
| 1753 | Tl | 177 | 81 | 96 | 15 | 7.732078 | 7.734446 | -2.4E-03 | 164829.174 | 164828.704 | 4.7E-01 | 164871.116 | 164870.646 | 4.7E-01 | -3.325 | -3.803 | 4.8E-01 |
| 1754 | Yb | 178 | 70 | 108 | 38 | 8.042832 | 8.043740 | -9.1E-04 | 165720.088 | 165719.884 | 2.0E-01 | 165756.240 | 165756.037 | 2.0E-01 | -49.694 | -49.906 | 2.1E-01 |
| 1755 | Lu | 178 | 71 | 107 | 36 | 8.042065 | 8.041354 | 7.1E-04 | 165718.917 | 165719.001 | -8.4E-02 | 165755.595 | 165755.678 | -8.3E-02 | -50.340 | -50.265 | -7.5E-02 |
| 1756 | Hf | 178 | 72 | 106 | 34 | 8.049456 | 8.051163 | -1.7E-03 | 165716.295 | 165715.947 | 3.5E-01 | 165753.497 | 165753.149 | 3.5E-01 | -52.438 | -52.794 | 3.6E-01 |
| 1757 | W | 178 | 74 | 104 | 30 | 8.029271 | 8.028637 | 6.3E-04 | 165717.272 | 165717.339 | -6.7E-02 | 165755.525 | 165755.593 | -6.8E-02 | -50.409 | -50.350 | -5.9E-02 |
| 1758 | Re | 178 | 75 | 103 | 28 | 7.998157 | 7.996541 | 1.6E-03 | 165721.502 | 165721.744 | -2.4E-01 | 165760.281 | 165760.523 | -2.4E-01 | -45.653 | -45.420 | -2.3E-01 |
| 1759 | Os | 178 | 76 | 102 | 26 | 7.981911 | 7.980424 | 1.5E-03 | 165723.085 | 165723.303 | -2.2E-01 | 165762.390 | 165762.608 | -2.2E-01 | -43.544 | -43.335 | -2.1E-01 |
| 1760 | Ir | 178 | 77 | 101 | 24 | 7.936549 | 7.935544 | 1.0E-03 | 165729.850 | 165729.982 | -1.3E-01 | 165769.682 | 165769.814 | -1.3E-01 | -36.252 | -36.129 | -1.2E-01 |
| 1761 | Pt | 178 | 78 | 100 | 22 | 7.908251 | 7.908501 | -2.5E-04 | 165733.578 | 165733.486 | 9.2E-02 | 165773.937 | 165773.845 | 9.2E-02 | -31.997 | -32.098 | 1.0E-01 |
| 1762 | Au | 178 | 79 | 99 | 20 | 7.849524 | 7.851265 | -1.7E-03 | 165742.722 | 165742.363 | 3.6E-01 | 165783.608 | 165783.249 | 3.6E-01 | -22.326 | -22.693 | 3.7E-01 |
| 1763 | Hg | 178 | 80 | 98 | 18 | 7.811365 | 7.813907 | -2.5E-03 | 165748.204 | 165747.702 | 5.0E-01 | 165789.618 | 165789.116 | 5.0E-01 | -16.316 | -16.827 | 5.1E-01 |
| 1764 | Pb | 178 | 82 | 96 | 14 | 7.690861 | 7.696697 | -5.8E-03 | 165767.033 | 165765.943 | 1.1E+00 | 165809.503 | 165808.413 | 1.1E+00 | 3.569 | 2.470 | 1.1E+00 |
| 1765 | Lu | 179 | 71 | 108 | 37 | 8.035084 | 8.035040 | 4.4E-05 | 166651.690 | 166651.655 | 3.5E-02 | 166688.367 | 166688.332 | 3.5E-02 | -49.061 | -49.105 | 4.4E-02 |
| 1766 | Hf | 179 | 72 | 107 | 35 | 8.038560 | 8.038862 | -3.0E-04 | 166649.761 | 166649.663 | 9.8E-02 | 166686.963 | 166686.865 | 9.8E-02 | -50.465 | -50.572 | 1.1E-01 |
| 1767 | Ta | 179 | 73 | 106 | 33 | 8.033599 | 8.032709 | 8.9E-04 | 166649.341 | 166649.456 | -1.2E-01 | 166687.069 | 166687.184 | -1.1E-01 | -50.360 | -50.253 | -1.1E-01 |
| 1768 | W | 179 | 74 | 105 | 31 | 8.023294 | 8.021611 | 1.7E-03 | 166649.878 | 166650.134 | -2.6E-01 | 166688.131 | 166688.387 | -2.6E-01 | -49.297 | -49.050 | -2.5E-01 |
| 1769 | Re | 179 | 75 | 104 | 29 | 8.003769 | 8.002194 | 1.6E-03 | 166652.064 | 166652.300 | -2.4E-01 | 166690.844 | 166691.080 | -2.4E-01 | -46.585 | -46.357 | -2.3E-01 |
| 1770 | Os | 179 | 76 | 103 | 27 | 7.979479 | 7.978098 | 1.4E-03 | 166655.104 | 166655.304 | -2.0E-01 | 166694.409 | 166694.610 | -2.0E-01 | -43.019 | -42.827 | -1.9E-01 |
| 1771 | Ir | 179 | 77 | 102 | 25 | 7.947511 | 7.946779 | 7.3E-04 | 166659.517 | 166659.601 | -8.4E-02 | 166699.349 | 166699.433 | -8.4E-02 | -38.079 | -38.004 | -7.5E-02 |
| 1772 | Pt | 179 | 78 | 101 | 23 | 7.910674 | 7.910451 | 2.2E-04 | 166664.801 | 166664.793 | 7.6E-03 | 166705.160 | 166705.152 | 7.7E-03 | -32.268 | -32.285 | 1.7E-02 |
| 1773 | Au | 179 | 79 | 100 | 21 | 7.865637 | 7.867814 | -2.2E-03 | 166671.553 | 166671.115 | 4.4E-01 | 166712.440 | 166712.001 | 4.4E-01 | -24.989 | -25.436 | 4.5E-01 |
| 1774 | Hg | 179 | 80 | 99 | 19 | 7.816213 | 7.819640 | -3.4E-03 | 166679.090 | 166678.427 | 6.6E-01 | 166720.504 | 166719.841 | 6.6E-01 | -16.924 | -17.596 | 6.7E-01 |
| 1775 | Tl | 179 | 81 | 98 | 17 | 7.763553 | 7.766014 | -2.5E-03 | 166687.206 | 166686.715 | 4.9E-01 | 166729.148 | 166728.657 | 4.9E-01 | -8.280 | -8.780 | 5.0E-01 |
| 1776 | Pb | 179 | 82 | 97 | 15 | 7.701469 | 7.705890 | -4.4E-03 | 166697.008 | 166696.166 | 8.4E-01 | 166739.479 | 166738.636 | 8.4E-01 | 2.050 | 1.199 | 8.5E-01 |
| 1777 | Lu | 180 | 71 | 109 | 38 | 8.022052 | 8.020968 | 1.1E-03 | 167585.566 | 167585.719 | -1.5E-01 | 167622.244 | 167622.396 | -1.5E-01 | -46.679 | -46.535 | -1.4E-01 |
| 1778 | Hf | 180 | 72 | 108 | 36 | 8.034944 | 8.036579 | -1.6E-03 | 167581.938 | 167581.600 | 3.4E-01 | 167619.141 | 167618.803 | 3.4E-01 | -49.782 | -50.128 | 3.5E-01 |
| 1779 | Ta | 180 | 73 | 107 | 34 | 8.025900 | 8.024084 | 1.8E-03 | 167582.259 | 167582.541 | -2.8E-01 | 167619.986 | 167620.269 | -2.8E-01 | -48.936 | -48.662 | -2.7E-01 |
| 1780 | W | 180 | 74 | 106 | 32 | 8.025456 | 8.024972 | 4.8E-04 | 167581.031 | 167581.073 | -4.2E-02 | 167619.284 | 167619.326 | -4.2E-02 | -49.639 | -49.605 | -3.3E-02 |
| 1781 | Re | 180 | 75 | 105 | 30 | 7.999991 | 7.998344 | 1.6E-03 | 167584.306 | 167584.557 | -2.5E-01 | 167623.085 | 167623.336 | -2.5E-01 | -45.837 | -45.595 | -2.4E-01 |
| 1782 | Os | 180 | 76 | 104 | 28 | 7.987448 | 7.986914 | 5.3E-04 | 167585.255 | 167585.305 | -5.0E-02 | 167624.560 | 167624.610 | -5.0E-02 | -44.362 | -44.321 | -4.1E-02 |
| 1783 | Ir | 180 | 77 | 103 | 26 | 7.947633 | 7.947321 | 3.1E-04 | 167591.113 | 167591.122 | -8.7E-03 | 167630.945 | 167630.954 | -8.7E-03 | -37.978 | -37.977 | -2.7E-04 |
| 1784 | Pt | 180 | 78 | 102 | 24 | 7.923609 | 7.924646 | -1.0E-03 | 167594.128 | 167593.893 | 2.3E-01 | 167634.487 | 167634.252 | 2.3E-01 | -34.436 | -34.679 | 2.4E-01 |
| 1785 | Au | 180 | 79 | 101 | 22 | 7.870144 | 7.872452 | -2.3E-03 | 167602.442 | 167601.978 | 4.6E-01 | 167643.328 | 167642.864 | 4.6E-01 | -25.594 | -26.067 | 4.7E-01 |
| 1786 | Hg | 180 | 80 | 100 | 20 | 7.836110 | 7.839020 | -2.9E-03 | 167607.258 | 167606.685 | 5.7E-01 | 167648.672 | 167648.099 | 5.7E-01 | -20.250 | -20.832 | 5.8E-01 |
| 1787 | Tl | 180 | 81 | 99 | 18 | 7.770718 | 7.774447 | -3.7E-03 | 167617.718 | 167616.997 | 7.2E-01 | 167659.660 | 167658.939 | 7.2E-01 | -9.262 | -9.992 | 7.3E-01 |
| 1788 | Pb | 180 | 82 | 98 | 16 | 7.725636 | 7.730329 | -4.7E-03 | 167624.522 | 167623.627 | 9.0E-01 | 167666.993 | 167666.097 | 9.0E-01 | -1.930 | -2.834 | 9.0E-01 |
| 1789 | Lu | 181 | 71 | 110 | 39 | 8.011929 | 8.012152 | -2.2E-04 | 168518.942 | 168518.859 | 8.3E-02 | 168555.619 | 168555.536 | 8.3E-02 | -44.797 | -44.889 | 9.2E-02 |
| 1790 | Hf | 181 | 72 | 109 | 37 | 8.022015 | 8.022181 | -1.7E-04 | 168515.809 | 168515.735 | 7.4E-02 | 168553.011 | 168552.937 | 7.4E-02 | -47.405 | -47.488 | 8.2E-02 |
| 1791 | Ta | 181 | 73 | 108 | 35 | 8.023418 | 8.021682 | 1.7E-03 | 168514.247 | 168514.517 | -2.7E-01 | 168551.975 | 168552.245 | -2.7E-01 | -48.442 | -48.180 | -2.6E-01 |
| 1792 | W | 181 | 74 | 107 | 33 | 8.018056 | 8.016094 | 2.0E-03 | 168513.910 | 168514.220 | -3.1E-01 | 168552.163 | 168552.473 | -3.1E-01 | -48.253 | -47.952 | -3.0E-01 |
| 1793 | Re | 181 | 75 | 106 | 31 | 8.004162 | 8.001815 | 2.3E-03 | 168515.116 | 168515.495 | -3.8E-01 | 168553.895 | 168554.275 | -3.8E-01 | -46.521 | -46.151 | -3.7E-01 |
| 1794 | Os | 181 | 76 | 105 | 29 | 7.983426 | 7.982892 | 5.3E-04 | 168517.561 | 168517.611 | -5.0E-02 | 168556.866 | 168556.917 | -5.1E-02 | -43.550 | -43.509 | -4.1E-02 |
| 1795 | Ir | 181 | 77 | 104 | 27 | 7.956571 | 7.956435 | 1.4E-04 | 168521.113 | 168521.090 | 2.3E-02 | 168560.945 | 168560.922 | 2.3E-02 | -39.472 | -39.503 | 3.1E-02 |
| 1796 | Pt | 181 | 78 | 103 | 25 | 7.924087 | 7.925005 | -9.2E-04 | 168525.683 | 168525.469 | 2.1E-01 | 168566.042 | 168565.828 | 2.1E-01 | -34.374 | -34.597 | 2.2E-01 |
| 1797 | Au | 181 | 79 | 102 | 23 | 7.883835 | 7.886901 | -3.1E-03 | 168531.659 | 168531.055 | 6.0E-01 | 168572.545 | 168571.942 | 6.0E-01 | -27.871 | -28.483 | 6.1E-01 |
| 1798 | Hg | 181 | 80 | 101 | 21 | 7.839678 | 7.843293 | -3.6E-03 | 168538.341 | 168537.638 | 7.0E-01 | 168579.755 | 168579.052 | 7.0E-01 | -20.661 | -21.374 | 7.1E-01 |
| 1799 | Tl | 181 | 81 | 100 | 19 | 7.791918 | 7.793825 | -1.9E-03 | 168545.676 | 168545.280 | 4.0E-01 | 168587.618 | 168587.222 | 4.0E-01 | -12.799 | -13.203 | 4.0E-01 |
| 1800 | Pb | 181 | 82 | 99 | 17 | 7.734107 | 7.738157 | -4.1E-03 | 168554.829 | 168554.045 | 7.8E-01 | 168597.299 | 168596.515 | 7.8E-01 | -3.117 | -3.910 | 7.9E-01 |
| 1801 | Hf | 182 | 72 | 110 | 38 | 8.014850 | 8.017628 | -2.8E-03 | 169448.656 | 169448.107 | 5.5E-01 | 169485.858 | 169485.309 | 5.5E-01 | -46.052 | -46.610 | 5.6E-01 |
| 1802 | Ta | 182 | 73 | 109 | 36 | 8.012647 | 8.011143 | 1.5E-03 | 169447.750 | 169447.979 | -2.3E-01 | 169485.477 | 169485.707 | -2.3E-01 | -46.433 | -46.213 | -2.2E-01 |
| 1803 | W | 182 | 74 | 108 | 34 | 8.018318 | 8.017285 | 1.0E-03 | 169445.410 | 169445.552 | -1.4E-01 | 169483.663 | 169483.806 | -1.4E-01 | -48.248 | -48.114 | -1.3E-01 |
| 1804 | Re | 182 | 75 | 107 | 32 | 7.998634 | 7.996177 | 2.5E-03 | 169447.684 | 169448.085 | -4.0E-01 | 169486.463 | 169486.864 | -4.0E-01 | -45.448 | -45.055 | -3.9E-01 |
| 1805 | Os | 182 | 76 | 106 | 30 | 7.989728 | 7.989535 | 1.9E-04 | 169447.996 | 169447.984 | 1.2E-02 | 169487.301 | 169487.290 | 1.1E-02 | -44.609 | -44.629 | 2.0E-02 |
| 1806 | Ir | 182 | 77 | 105 | 28 | 7.954894 | 7.955248 | -3.5E-04 | 169453.027 | 169452.915 | 1.1E-01 | 169492.859 | 169492.747 | 1.1E-01 | -39.052 | -39.172 | 1.2E-01 |
| 1807 | Pt | 182 | 78 | 104 | 26 | 7.934752 | 7.937010 | -2.3E-03 | 169455.383 | 169454.924 | 4.6E-01 | 169495.742 | 169495.283 | 4.6E-01 | -36.168 | -36.636 | 4.7E-01 |
| 1808 | Au | 182 | 79 | 103 | 24 | 7.887226 | 7.889831 | -2.6E-03 | 169462.723 | 169462.201 | 5.2E-01 | 169503.610 | 169503.087 | 5.2E-01 | -28.301 | -28.832 | 5.3E-01 |
| 1809 | Hg | 182 | 80 | 102 | 22 | 7.856971 | 7.860413 | -3.4E-03 | 169466.920 | 169466.244 | 6.8E-01 | 169508.334 | 169507.658 | 6.8E-01 | -23.577 | -24.261 | 6.8E-01 |
| 1810 | Tl | 182 | 81 | 101 | 20 | 7.796251 | 7.800621 | -4.4E-03 | 169476.660 | 169475.815 | 8.5E-01 | 169518.602 | 169517.757 | 8.5E-01 | -13.308 | -14.162 | 8.5E-01 |
| 1811 | Pb | 182 | 82 | 100 | 18 | 7.756334 | 7.760219 | -3.9E-03 | 169482.615 | 169481.857 | 7.6E-01 | 169525.085 | 169524.327 | 7.6E-01 | -6.826 | -7.592 | 7.7E-01 |
| 1812 | Lu | 183 | 71 | 112 | 41 | 7.984812 | 7.984612 | 2.0E-04 | 170387.011 | 170387.005 | 6.0E-03 | 170423.688 | 170423.682 | 5.9E-03 | -39.716 | -39.731 | 1.5E-02 |
| 1813 | Hf | 183 | 72 | 111 | 39 | 8.000045 | 8.001056 | -1.0E-03 | 170382.916 | 170382.688 | 2.3E-01 | 170420.118 | 170419.890 | 2.3E-01 | -43.286 | -43.523 | 2.4E-01 |
| 1814 | Ta | 183 | 73 | 110 | 37 | 8.006753 | 8.006417 | 3.4E-04 | 170380.381 | 170380.398 | -1.7E-02 | 170418.108 | 170418.126 | -1.8E-02 | -45.296 | -45.288 | -8.5E-03 |
| 1815 | W | 183 | 74 | 109 | 35 | 8.008331 | 8.006496 | 1.8E-03 | 170378.784 | 170379.075 | -2.9E-01 | 170417.037 | 170417.328 | -2.9E-01 | -46.367 | -46.085 | -2.8E-01 |
| 1816 | Re | 183 | 75 | 108 | 33 | 8.001018 | 7.997466 | 3.6E-03 | 170378.814 | 170379.418 | -6.0E-01 | 170417.593 | 170418.197 | -6.0E-01 | -45.811 | -45.216 | -6.0E-01 |
| 1817 | Os | 183 | 76 | 107 | 31 | 7.985010 | 7.983779 | 1.2E-03 | 170380.435 | 170380.614 | -1.8E-01 | 170419.740 | 170419.919 | -1.8E-01 | -43.664 | -43.494 | -1.7E-01 |
| 1818 | Ir | 183 | 77 | 106 | 29 | 7.961823 | 7.962255 | -4.3E-04 | 170383.369 | 170383.243 | 1.3E-01 | 170423.201 | 170423.075 | 1.3E-01 | -40.203 | -40.338 | 1.3E-01 |
| 1819 | Pt | 183 | 78 | 105 | 27 | 7.933335 | 7.935791 | -2.5E-03 | 170387.273 | 170386.776 | 5.0E-01 | 170427.632 | 170427.135 | 5.0E-01 | -35.772 | -36.278 | 5.1E-01 |
| 1820 | Au | 183 | 79 | 104 | 25 | 7.898551 | 7.902250 | -3.7E-03 | 170392.329 | 170391.603 | 7.3E-01 | 170433.215 | 170432.490 | 7.3E-01 | -30.189 | -30.924 | 7.3E-01 |
| 1821 | Hg | 183 | 80 | 103 | 23 | 7.859387 | 7.863190 | -3.8E-03 | 170398.186 | 170397.441 | 7.5E-01 | 170439.600 | 170438.855 | 7.5E-01 | -23.805 | -24.559 | 7.5E-01 |
| 1822 | Tl | 183 | 81 | 102 | 21 | 7.815673 | 7.817919 | -2.2E-03 | 170404.875 | 170404.414 | 4.6E-01 | 170446.817 | 170446.356 | 4.6E-01 | -16.587 | -17.057 | 4.7E-01 |
| 1823 | Pb | 183 | 82 | 101 | 19 | 7.762130 | 7.766641 | -4.5E-03 | 170413.363 | 170412.487 | 8.8E-01 | 170455.833 | 170454.957 | 8.8E-01 | -7.571 | -8.456 | 8.9E-01 |
| 1824 | Hf | 184 | 72 | 112 | 40 | 7.990733 | 7.994243 | -3.5E-03 | 171316.195 | 171315.506 | 6.9E-01 | 171353.397 | 171352.708 | 6.9E-01 | -41.502 | -42.200 | 7.0E-01 |
| 1825 | Ta | 184 | 73 | 111 | 38 | 7.993764 | 7.993929 | -1.6E-04 | 171314.329 | 171314.255 | 7.4E-02 | 171352.057 | 171351.982 | 7.5E-02 | -42.842 | -42.925 | 8.3E-02 |
| 1826 | W | 184 | 74 | 110 | 36 | 8.005088 | 8.005532 | -4.4E-04 | 171310.938 | 171310.811 | 1.3E-01 | 171349.191 | 171349.064 | 1.3E-01 | -45.708 | -45.843 | 1.4E-01 |
| 1827 | Re | 184 | 75 | 109 | 34 | 7.992778 | 7.990030 | 2.7E-03 | 171311.895 | 171312.355 | -4.6E-01 | 171350.674 | 171351.134 | -4.6E-01 | -44.225 | -43.774 | -4.5E-01 |
| 1828 | Os | 184 | 76 | 108 | 32 | 7.988699 | 7.988301 | 4.0E-04 | 171311.336 | 171311.363 | -2.7E-02 | 171350.642 | 171350.669 | -2.7E-02 | -44.257 | -44.238 | -1.8E-02 |
| 1829 | Ir | 184 | 77 | 107 | 30 | 7.959198 | 7.959339 | -1.4E-04 | 171315.456 | 171315.383 | 7.3E-02 | 171355.288 | 171355.215 | 7.3E-02 | -39.611 | -39.693 | 8.2E-02 |
| 1830 | Pt | 184 | 78 | 106 | 28 | 7.942599 | 7.945656 | -3.1E-03 | 171317.201 | 171316.590 | 6.1E-01 | 171357.560 | 171356.949 | 6.1E-01 | -37.339 | -37.958 | 6.2E-01 |
| 1831 | Au | 184 | 79 | 105 | 26 | 7.900194 | 7.903537 | -3.3E-03 | 171323.693 | 171323.030 | 6.6E-01 | 171364.580 | 171363.916 | 6.6E-01 | -30.319 | -30.991 | 6.7E-01 |
| 1832 | Hg | 184 | 80 | 104 | 24 | 7.874366 | 7.878173 | -3.8E-03 | 171327.136 | 171326.386 | 7.5E-01 | 171368.550 | 171367.800 | 7.5E-01 | -26.349 | -27.107 | 7.6E-01 |
| 1833 | Tl | 184 | 81 | 103 | 22 | 7.818617 | 7.823064 | -4.4E-03 | 171336.083 | 171335.215 | 8.7E-01 | 171378.025 | 171377.157 | 8.7E-01 | -16.873 | -17.750 | 8.8E-01 |
| 1834 | Pb | 184 | 82 | 102 | 20 | 7.782725 | 7.786413 | -3.7E-03 | 171341.377 | 171340.647 | 7.3E-01 | 171383.847 | 171383.118 | 7.3E-01 | -11.052 | -11.790 | 7.4E-01 |
| 1835 | Bi | 184 | 83 | 101 | 18 | 7.711957 | 7.718844 | -6.9E-03 | 171353.087 | 171351.768 | 1.3E+00 | 171396.086 | 171394.767 | 1.3E+00 | 1.187 | -0.140 | 1.3E+00 |
| 1836 | Hf | 185 | 72 | 113 | 41 | 7.973970 | 7.975503 | -1.5E-03 | 172250.870 | 172250.543 | 3.3E-01 | 172288.073 | 172287.746 | 3.3E-01 | -38.320 | -38.656 | 3.4E-01 |
| 1837 | Ta | 185 | 73 | 112 | 39 | 7.986372 | 7.986873 | -5.0E-04 | 172247.269 | 172247.132 | 1.4E-01 | 172284.996 | 172284.859 | 1.4E-01 | -41.396 | -41.542 | 1.5E-01 |
| 1838 | W | 185 | 74 | 111 | 37 | 7.992919 | 7.992780 | 1.4E-04 | 172244.749 | 172244.730 | 1.9E-02 | 172283.002 | 172282.983 | 1.9E-02 | -43.390 | -43.418 | 2.8E-02 |
| 1839 | Re | 185 | 75 | 110 | 35 | 7.991029 | 7.989134 | 1.9E-03 | 172243.791 | 172244.096 | -3.0E-01 | 172282.570 | 172282.875 | -3.0E-01 | -43.823 | -43.527 | -3.0E-01 |
| 1840 | Os | 185 | 76 | 109 | 33 | 7.981325 | 7.980776 | 5.5E-04 | 172244.277 | 172244.332 | -5.5E-02 | 172283.583 | 172283.638 | -5.5E-02 | -42.810 | -42.763 | -4.6E-02 |
| 1841 | Ir | 185 | 77 | 108 | 31 | 7.963722 | 7.964264 | -5.4E-04 | 172246.225 | 172246.078 | 1.5E-01 | 172286.057 | 172285.910 | 1.5E-01 | -40.336 | -40.492 | 1.6E-01 |
| 1842 | Pt | 185 | 78 | 107 | 29 | 7.939777 | 7.942817 | -3.0E-03 | 172249.345 | 172248.735 | 6.1E-01 | 172289.704 | 172289.094 | 6.1E-01 | -36.688 | -37.307 | 6.2E-01 |
| 1843 | Au | 185 | 79 | 106 | 27 | 7.909487 | 7.913940 | -4.5E-03 | 172253.639 | 172252.767 | 8.7E-01 | 172294.526 | 172293.653 | 8.7E-01 | -31.867 | -32.748 | 8.8E-01 |
| 1844 | Hg | 185 | 80 | 105 | 25 | 7.874495 | 7.879513 | -5.0E-03 | 172258.803 | 172257.825 | 9.8E-01 | 172300.217 | 172299.239 | 9.8E-01 | -26.176 | -27.162 | 9.9E-01 |
| 1845 | Tl | 185 | 81 | 104 | 23 | 7.835576 | 7.838404 | -2.8E-03 | 172264.693 | 172264.119 | 5.7E-01 | 172306.635 | 172306.061 | 5.7E-01 | -19.758 | -20.340 | 5.8E-01 |
| 1846 | Pb | 185 | 82 | 103 | 21 | 7.786932 | 7.791388 | -4.5E-03 | 172272.381 | 172271.506 | 8.8E-01 | 172314.851 | 172313.976 | 8.7E-01 | -11.541 | -12.425 | 8.8E-01 |
| 1847 | Hf | 186 | 72 | 114 | 42 | 7.964302 | 7.966597 | -2.3E-03 | 173184.260 | 173183.790 | 4.7E-01 | 173221.462 | 173220.992 | 4.7E-01 | -36.424 | -36.903 | 4.8E-01 |
| 1848 | Ta | 186 | 73 | 113 | 40 | 7.971847 | 7.972474 | -6.3E-04 | 173181.549 | 173181.389 | 1.6E-01 | 173219.277 | 173219.116 | 1.6E-01 | -38.610 | -38.779 | 1.7E-01 |
| 1849 | W | 186 | 74 | 112 | 38 | 7.988614 | 7.989697 | -1.1E-03 | 173177.123 | 173176.876 | 2.5E-01 | 173215.376 | 173215.129 | 2.5E-01 | -42.511 | -42.766 | 2.6E-01 |
| 1850 | Re | 186 | 75 | 111 | 36 | 7.981288 | 7.979898 | 1.4E-03 | 173177.177 | 173177.390 | -2.1E-01 | 173215.956 | 173216.169 | -2.1E-01 | -41.931 | -41.726 | -2.0E-01 |
| 1851 | Os | 186 | 76 | 110 | 34 | 7.982844 | 7.983222 | -3.8E-04 | 173175.579 | 173175.462 | 1.2E-01 | 173214.884 | 173214.768 | 1.2E-01 | -43.002 | -43.128 | 1.3E-01 |
| 1852 | Ir | 186 | 77 | 109 | 32 | 7.958060 | 7.959633 | -1.6E-03 | 173178.880 | 173178.540 | 3.4E-01 | 173218.712 | 173218.372 | 3.4E-01 | -39.175 | -39.523 | 3.5E-01 |
| 1853 | Pt | 186 | 78 | 108 | 30 | 7.946809 | 7.950615 | -3.8E-03 | 173179.663 | 173178.907 | 7.6E-01 | 173220.022 | 173219.266 | 7.6E-01 | -37.864 | -38.629 | 7.6E-01 |
| 1854 | Au | 186 | 79 | 107 | 28 | 7.909540 | 7.913570 | -4.0E-03 | 173185.285 | 173184.487 | 8.0E-01 | 173226.172 | 173225.374 | 8.0E-01 | -31.715 | -32.522 | 8.1E-01 |
| 1855 | Hg | 186 | 80 | 106 | 26 | 7.888259 | 7.892400 | -4.1E-03 | 173187.934 | 173187.114 | 8.2E-01 | 173229.348 | 173228.528 | 8.2E-01 | -28.539 | -29.367 | 8.3E-01 |
| 1856 | Tl | 186 | 81 | 105 | 24 | 7.837535 | 7.842016 | -4.5E-03 | 173196.058 | 173195.175 | 8.8E-01 | 173238.000 | 173237.116 | 8.8E-01 | -19.887 | -20.779 | 8.9E-01 |
| 1857 | Pb | 186 | 82 | 104 | 22 | 7.805347 | 7.809044 | -3.7E-03 | 173200.734 | 173199.996 | 7.4E-01 | 173243.205 | 173242.466 | 7.4E-01 | -14.682 | -15.429 | 7.5E-01 |
| 1858 | Bi | 186 | 83 | 103 | 20 | 7.739014 | 7.746058 | -7.0E-03 | 173211.761 | 173210.400 | 1.4E+00 | 173254.760 | 173253.399 | 1.4E+00 | -3.126 | -4.497 | 1.4E+00 |
| 1859 | Po | 186 | 84 | 102 | 18 | 7.695998 | 7.701982 | -6.0E-03 | 173218.451 | 173217.285 | 1.2E+00 | 173261.979 | 173260.814 | 1.2E+00 | 4.092 | 2.918 | 1.2E+00 |
| 1860 | Ta | 187 | 73 | 114 | 41 | 7.963236 | 7.963226 | 1.0E-05 | 174114.753 | 174114.711 | 4.2E-02 | 174152.480 | 174152.438 | 4.2E-02 | -36.900 | -36.951 | 5.1E-02 |
| 1861 | W | 187 | 74 | 113 | 39 | 7.975128 | 7.975003 | 1.2E-04 | 174111.221 | 174111.200 | 2.1E-02 | 174149.474 | 174149.453 | 2.1E-02 | -39.906 | -39.937 | 3.0E-02 |
| 1862 | Re | 187 | 75 | 112 | 37 | 7.977962 | 7.976842 | 1.1E-03 | 174109.383 | 174109.547 | -1.6E-01 | 174148.162 | 174148.326 | -1.6E-01 | -41.219 | -41.064 | -1.5E-01 |
| 1863 | Os | 187 | 76 | 111 | 35 | 7.973791 | 7.973910 | -1.2E-04 | 174108.854 | 174108.786 | 6.8E-02 | 174148.159 | 174148.091 | 6.8E-02 | -41.221 | -41.298 | 7.7E-02 |
| 1864 | Ir | 187 | 77 | 110 | 33 | 7.960668 | 7.962498 | -1.8E-03 | 174109.999 | 174109.610 | 3.9E-01 | 174149.831 | 174149.442 | 3.9E-01 | -39.549 | -39.947 | 4.0E-01 |
| 1865 | Pt | 187 | 78 | 109 | 31 | 7.941168 | 7.946143 | -5.0E-03 | 174112.336 | 174111.358 | 9.8E-01 | 174152.695 | 174151.717 | 9.8E-01 | -36.685 | -37.672 | 9.9E-01 |
| 1866 | Au | 187 | 79 | 108 | 29 | 7.917427 | 7.922008 | -4.6E-03 | 174115.466 | 174114.561 | 9.0E-01 | 174156.353 | 174155.448 | 9.1E-01 | -33.028 | -33.942 | 9.1E-01 |
| 1867 | Hg | 187 | 80 | 107 | 27 | 7.886985 | 7.892247 | -5.3E-03 | 174119.849 | 174118.816 | 1.0E+00 | 174161.263 | 174160.230 | 1.0E+00 | -28.118 | -29.160 | 1.0E+00 |
| 1868 | Tl | 187 | 81 | 106 | 25 | 7.852456 | 7.855406 | -2.9E-03 | 174124.995 | 174124.394 | 6.0E-01 | 174166.937 | 174166.336 | 6.0E-01 | -22.443 | -23.053 | 6.1E-01 |
| 1869 | Pb | 187 | 82 | 105 | 23 | 7.808400 | 7.812709 | -4.3E-03 | 174131.923 | 174131.067 | 8.6E-01 | 174174.394 | 174173.537 | 8.6E-01 | -14.987 | -15.852 | 8.7E-01 |
| 1870 | Bi | 187 | 83 | 104 | 21 | 7.758208 | 7.763899 | -5.7E-03 | 174139.998 | 174138.883 | 1.1E+00 | 174182.997 | 174181.881 | 1.1E+00 | -6.383 | -7.508 | 1.1E+00 |
| 1871 | Po | 187 | 84 | 103 | 19 | 7.704740 | 7.709207 | -4.5E-03 | 174148.685 | 174147.798 | 8.9E-01 | 174192.213 | 174191.326 | 8.9E-01 | 2.833 | 1.937 | 9.0E-01 |
| 1872 | Ta | 188 | 73 | 115 | 42 | 7.946321 | 7.947224 | -9.0E-04 | 175049.535 | 175049.321 | 2.1E-01 | 175087.262 | 175087.049 | 2.1E-01 | -33.612 | -33.835 | 2.2E-01 |
| 1873 | W | 188 | 74 | 114 | 40 | 7.969063 | 7.969968 | -9.1E-04 | 175043.951 | 175043.737 | 2.1E-01 | 175082.205 | 175081.990 | 2.2E-01 | -38.670 | -38.894 | 2.2E-01 |
| 1874 | Re | 188 | 75 | 113 | 38 | 7.966758 | 7.965869 | 8.9E-04 | 175043.077 | 175043.198 | -1.2E-01 | 175081.856 | 175081.977 | -1.2E-01 | -39.019 | -38.906 | -1.1E-01 |
| 1875 | Os | 188 | 76 | 112 | 36 | 7.973875 | 7.974351 | -4.8E-04 | 175040.430 | 175040.294 | 1.4E-01 | 175079.735 | 175079.600 | 1.4E-01 | -41.139 | -41.284 | 1.4E-01 |
| 1876 | Ir | 188 | 77 | 111 | 34 | 7.954885 | 7.956180 | -1.3E-03 | 175042.691 | 175042.401 | 2.9E-01 | 175082.523 | 175082.233 | 2.9E-01 | -38.351 | -38.651 | 3.0E-01 |
| 1877 | Pt | 188 | 78 | 110 | 32 | 7.947945 | 7.951944 | -4.0E-03 | 175042.686 | 175041.887 | 8.0E-01 | 175083.045 | 175082.246 | 8.0E-01 | -37.829 | -38.638 | 8.1E-01 |
| 1878 | Au | 188 | 79 | 109 | 30 | 7.914250 | 7.920010 | -5.8E-03 | 175047.711 | 175046.580 | 1.1E+00 | 175088.598 | 175087.466 | 1.1E+00 | -32.277 | -33.417 | 1.1E+00 |
| 1879 | Hg | 188 | 80 | 108 | 28 | 7.899102 | 7.903135 | -4.0E-03 | 175049.249 | 175048.442 | 8.1E-01 | 175090.663 | 175089.856 | 8.1E-01 | -30.211 | -31.028 | 8.2E-01 |
| 1880 | Tl | 188 | 81 | 107 | 26 | 7.853053 | 7.857445 | -4.4E-03 | 175056.596 | 175055.721 | 8.8E-01 | 175098.538 | 175097.662 | 8.8E-01 | -22.336 | -23.221 | 8.8E-01 |
| 1881 | Pb | 188 | 82 | 106 | 24 | 7.824841 | 7.828277 | -3.4E-03 | 175060.589 | 175059.893 | 7.0E-01 | 175103.059 | 175102.363 | 7.0E-01 | -17.815 | -18.520 | 7.1E-01 |
| 1882 | Bi | 188 | 83 | 105 | 22 | 7.764136 | 7.769858 | -5.7E-03 | 175070.691 | 175069.564 | 1.1E+00 | 175113.690 | 175112.563 | 1.1E+00 | -7.185 | -8.321 | 1.1E+00 |
| 1883 | Po | 188 | 84 | 104 | 20 | 7.724654 | 7.729400 | -4.7E-03 | 175076.802 | 175075.858 | 9.4E-01 | 175120.330 | 175119.386 | 9.4E-01 | -0.544 | -1.498 | 9.5E-01 |
| 1884 | W | 189 | 74 | 115 | 41 | 7.953454 | 7.953594 | -1.4E-04 | 175978.498 | 175978.427 | 7.1E-02 | 176016.751 | 176016.680 | 7.1E-02 | -35.618 | -35.698 | 8.0E-02 |
| 1885 | Re | 189 | 75 | 114 | 39 | 7.961818 | 7.960802 | 1.0E-03 | 175975.609 | 175975.755 | -1.5E-01 | 176014.388 | 176014.535 | -1.5E-01 | -37.981 | -37.843 | -1.4E-01 |
| 1886 | Os | 189 | 76 | 113 | 37 | 7.963011 | 7.963300 | -2.9E-04 | 175974.075 | 175973.974 | 1.0E-01 | 176013.380 | 176013.279 | 1.0E-01 | -38.988 | -39.098 | 1.1E-01 |
| 1887 | Ir | 189 | 77 | 112 | 35 | 7.956058 | 7.957042 | -9.8E-04 | 175974.080 | 175973.847 | 2.3E-01 | 176013.912 | 176013.679 | 2.3E-01 | -38.457 | -38.699 | 2.4E-01 |
| 1888 | Pt | 189 | 78 | 111 | 33 | 7.941489 | 7.945845 | -4.4E-03 | 175975.524 | 175974.653 | 8.7E-01 | 176015.883 | 176015.012 | 8.7E-01 | -36.485 | -37.365 | 8.8E-01 |
| 1889 | Au | 189 | 79 | 110 | 31 | 7.921987 | 7.926533 | -4.5E-03 | 175977.900 | 175976.993 | 9.1E-01 | 176018.787 | 176017.879 | 9.1E-01 | -33.582 | -34.499 | 9.2E-01 |
| 1890 | Hg | 189 | 80 | 109 | 29 | 7.896918 | 7.901498 | -4.6E-03 | 175981.328 | 175980.414 | 9.1E-01 | 176022.742 | 176021.828 | 9.1E-01 | -29.626 | -30.550 | 9.2E-01 |
| 1891 | Tl | 189 | 81 | 108 | 27 | 7.866196 | 7.868968 | -2.8E-03 | 175985.824 | 175985.251 | 5.7E-01 | 176027.766 | 176027.193 | 5.7E-01 | -24.602 | -25.185 | 5.8E-01 |
| 1892 | Pb | 189 | 82 | 107 | 25 | 7.826480 | 7.830549 | -4.1E-03 | 175992.020 | 175991.200 | 8.2E-01 | 176034.490 | 176033.671 | 8.2E-01 | -17.878 | -18.707 | 8.3E-01 |
| 1893 | Bi | 189 | 83 | 106 | 23 | 7.781000 | 7.785745 | -4.7E-03 | 175999.305 | 175998.357 | 9.5E-01 | 176042.304 | 176041.356 | 9.5E-01 | -10.065 | -11.022 | 9.6E-01 |
| 1894 | Po | 189 | 84 | 105 | 21 | 7.731131 | 7.735361 | -4.2E-03 | 176007.418 | 176006.567 | 8.5E-01 | 176050.946 | 176050.095 | 8.5E-01 | -1.422 | -2.282 | 8.6E-01 |
| 1895 | W | 190 | 74 | 116 | 42 | 7.947566 | 7.947169 | 4.0E-04 | 176911.228 | 176911.259 | -3.1E-02 | 176949.482 | 176949.512 | -3.0E-02 | -34.381 | -34.359 | -2.2E-02 |
| 1896 | Re | 190 | 75 | 115 | 40 | 7.950049 | 7.948371 | 1.7E-03 | 176909.448 | 176909.722 | -2.7E-01 | 176948.227 | 176948.501 | -2.7E-01 | -35.635 | -35.371 | -2.6E-01 |
| 1897 | Os | 190 | 76 | 114 | 38 | 7.962112 | 7.961923 | 1.9E-04 | 176905.848 | 176905.838 | 1.0E-02 | 176945.153 | 176945.143 | 1.0E-02 | -38.709 | -38.729 | 1.9E-02 |
| 1898 | Ir | 190 | 77 | 113 | 36 | 7.947712 | 7.949128 | -1.4E-03 | 176907.275 | 176906.959 | 3.2E-01 | 176947.107 | 176946.791 | 3.2E-01 | -36.756 | -37.081 | 3.3E-01 |
| 1899 | Pt | 190 | 78 | 112 | 34 | 7.946592 | 7.949751 | -3.2E-03 | 176906.178 | 176905.530 | 6.5E-01 | 176946.537 | 176945.889 | 6.5E-01 | -37.325 | -37.982 | 6.6E-01 |
| 1900 | Au | 190 | 79 | 111 | 32 | 7.919095 | 7.922956 | -3.9E-03 | 176910.093 | 176909.311 | 7.8E-01 | 176950.979 | 176950.197 | 7.8E-01 | -32.883 | -33.674 | 7.9E-01 |
| 1901 | Hg | 190 | 80 | 110 | 30 | 7.907014 | 7.910484 | -3.5E-03 | 176911.078 | 176910.370 | 7.1E-01 | 176952.492 | 176951.784 | 7.1E-01 | -31.370 | -32.087 | 7.2E-01 |
| 1902 | Pb | 190 | 82 | 108 | 26 | 7.841128 | 7.844155 | -3.0E-03 | 176920.976 | 176920.350 | 6.3E-01 | 176963.446 | 176962.820 | 6.3E-01 | -20.416 | -21.051 | 6.3E-01 |
| 1903 | Bi | 190 | 83 | 107 | 24 | 7.785339 | 7.790177 | -4.8E-03 | 176930.265 | 176929.294 | 9.7E-01 | 176973.264 | 176972.293 | 9.7E-01 | -10.599 | -11.578 | 9.8E-01 |
| 1904 | Po | 190 | 84 | 106 | 22 | 7.749458 | 7.753399 | -3.9E-03 | 176935.771 | 176934.970 | 8.0E-01 | 176979.299 | 176978.498 | 8.0E-01 | -4.564 | -5.374 | 8.1E-01 |
| 1905 | W | 191 | 74 | 117 | 43 | 7.931435 | 7.929994 | 1.4E-03 | 177845.927 | 177846.158 | -2.3E-01 | 177884.180 | 177884.411 | -2.3E-01 | -31.176 | -30.955 | -2.2E-01 |
| 1906 | Re | 191 | 75 | 116 | 41 | 7.943967 | 7.941781 | 2.2E-03 | 177842.225 | 177842.598 | -3.7E-01 | 177881.004 | 177881.377 | -3.7E-01 | -34.352 | -33.989 | -3.6E-01 |
| 1907 | Os | 191 | 76 | 115 | 39 | 7.950576 | 7.949374 | 1.2E-03 | 177839.654 | 177839.838 | -1.8E-01 | 177878.960 | 177879.143 | -1.8E-01 | -36.397 | -36.222 | -1.7E-01 |
| 1908 | Ir | 191 | 77 | 114 | 37 | 7.948124 | 7.948152 | -2.8E-05 | 177838.814 | 177838.762 | 5.2E-02 | 177878.646 | 177878.594 | 5.2E-02 | -36.711 | -36.772 | 6.1E-02 |
| 1909 | Pt | 191 | 78 | 113 | 35 | 7.938743 | 7.942095 | -3.4E-03 | 177839.296 | 177838.609 | 6.9E-01 | 177879.655 | 177878.967 | 6.9E-01 | -35.701 | -36.398 | 7.0E-01 |
| 1910 | Au | 191 | 79 | 112 | 33 | 7.924750 | 7.927648 | -2.9E-03 | 177840.659 | 177840.057 | 6.0E-01 | 177881.545 | 177880.944 | 6.0E-01 | -33.811 | -34.422 | 6.1E-01 |
| 1911 | Hg | 191 | 80 | 111 | 31 | 7.903805 | 7.907387 | -3.6E-03 | 177843.350 | 177842.617 | 7.3E-01 | 177884.764 | 177884.030 | 7.3E-01 | -30.593 | -31.335 | 7.4E-01 |
| 1912 | Tl | 191 | 81 | 110 | 29 | 7.877143 | 7.879221 | -2.1E-03 | 177847.132 | 177846.685 | 4.5E-01 | 177889.074 | 177888.627 | 4.5E-01 | -26.283 | -26.738 | 4.6E-01 |
| 1913 | Pb | 191 | 82 | 109 | 27 | 7.841387 | 7.845069 | -3.7E-03 | 177852.650 | 177851.897 | 7.5E-01 | 177895.121 | 177894.367 | 7.5E-01 | -20.236 | -20.999 | 7.6E-01 |
| 1914 | Bi | 191 | 83 | 108 | 25 | 7.800663 | 7.804225 | -3.6E-03 | 177859.118 | 177858.386 | 7.3E-01 | 177902.117 | 177901.385 | 7.3E-01 | -13.240 | -13.980 | 7.4E-01 |
| 1915 | Po | 191 | 84 | 107 | 23 | 7.753787 | 7.757994 | -4.2E-03 | 177866.760 | 177865.904 | 8.6E-01 | 177910.288 | 177909.432 | 8.6E-01 | -5.069 | -5.933 | 8.6E-01 |
| 1916 | At | 191 | 85 | 106 | 21 | 7.702923 | 7.706212 | -3.3E-03 | 177875.163 | 177874.482 | 6.8E-01 | 177919.220 | 177918.540 | 6.8E-01 | 3.864 | 3.174 | 6.9E-01 |
| 1917 | Re | 192 | 75 | 117 | 42 | 7.930238 | 7.928678 | 1.6E-03 | 178776.483 | 178776.737 | -2.5E-01 | 178815.262 | 178815.516 | -2.5E-01 | -31.589 | -31.344 | -2.5E-01 |
| 1918 | Os | 192 | 76 | 116 | 40 | 7.948534 | 7.946657 | 1.9E-03 | 178771.661 | 178771.976 | -3.1E-01 | 178810.967 | 178811.281 | -3.1E-01 | -35.884 | -35.579 | -3.1E-01 |
| 1919 | Ir | 192 | 77 | 115 | 38 | 7.939010 | 7.938910 | 1.0E-04 | 178772.181 | 178772.153 | 2.8E-02 | 178812.013 | 178811.985 | 2.8E-02 | -34.838 | -34.874 | 3.7E-02 |
| 1920 | Pt | 192 | 78 | 114 | 36 | 7.942511 | 7.944307 | -1.8E-03 | 178770.199 | 178769.807 | 3.9E-01 | 178810.558 | 178810.166 | 3.9E-01 | -36.292 | -36.694 | 4.0E-01 |
| 1921 | Au | 192 | 79 | 113 | 34 | 7.920122 | 7.922602 | -2.5E-03 | 178773.188 | 178772.664 | 5.2E-01 | 178814.075 | 178813.550 | 5.2E-01 | -32.776 | -33.309 | 5.3E-01 |
| 1922 | Hg | 192 | 80 | 112 | 32 | 7.912064 | 7.914598 | -2.5E-03 | 178773.426 | 178772.890 | 5.4E-01 | 178814.839 | 178814.304 | 5.4E-01 | -32.011 | -32.556 | 5.4E-01 |
| 1923 | Tl | 192 | 81 | 111 | 30 | 7.876016 | 7.878275 | -2.3E-03 | 178779.036 | 178778.553 | 4.8E-01 | 178820.978 | 178820.495 | 4.8E-01 | -25.872 | -26.365 | 4.9E-01 |
| 1924 | Pb | 192 | 82 | 110 | 28 | 7.854718 | 7.856837 | -2.1E-03 | 178781.815 | 178781.358 | 4.6E-01 | 178824.285 | 178823.828 | 4.6E-01 | -22.565 | -23.032 | 4.7E-01 |
| 1925 | Bi | 192 | 83 | 109 | 26 | 7.803613 | 7.807204 | -3.6E-03 | 178790.316 | 178789.575 | 7.4E-01 | 178833.315 | 178832.574 | 7.4E-01 | -13.535 | -14.285 | 7.5E-01 |
| 1926 | Po | 192 | 84 | 108 | 24 | 7.771075 | 7.774038 | -3.0E-03 | 178795.252 | 178794.631 | 6.2E-01 | 178838.780 | 178838.159 | 6.2E-01 | -8.071 | -8.701 | 6.3E-01 |
| 1927 | At | 192 | 85 | 107 | 22 | 7.709675 | 7.713192 | -3.5E-03 | 178805.729 | 178805.001 | 7.3E-01 | 178849.786 | 178849.059 | 7.3E-01 | 2.936 | 2.199 | 7.4E-01 |
| 1928 | Re | 193 | 75 | 118 | 43 | 7.923956 | 7.921723 | 2.2E-03 | 179709.330 | 179709.716 | -3.9E-01 | 179748.109 | 179748.495 | -3.9E-01 | -30.235 | -29.859 | -3.8E-01 |
| 1929 | Os | 193 | 76 | 117 | 41 | 7.936279 | 7.933276 | 3.0E-03 | 179705.643 | 179706.177 | -5.3E-01 | 179744.948 | 179745.482 | -5.3E-01 | -33.396 | -32.872 | -5.2E-01 |
| 1930 | Ir | 193 | 77 | 116 | 39 | 7.938144 | 7.936512 | 1.6E-03 | 179703.974 | 179704.243 | -2.7E-01 | 179743.806 | 179744.075 | -2.7E-01 | -34.538 | -34.279 | -2.6E-01 |
| 1931 | Pt | 193 | 78 | 115 | 37 | 7.933797 | 7.935321 | -1.5E-03 | 179703.504 | 179703.163 | 3.4E-01 | 179743.863 | 179743.521 | 3.4E-01 | -34.482 | -34.832 | 3.5E-01 |
| 1932 | Au | 193 | 79 | 114 | 35 | 7.924170 | 7.925637 | -1.5E-03 | 179704.052 | 179703.721 | 3.3E-01 | 179744.939 | 179744.607 | 3.3E-01 | -33.406 | -33.746 | 3.4E-01 |
| 1933 | Hg | 193 | 80 | 113 | 33 | 7.907976 | 7.910127 | -2.2E-03 | 179705.868 | 179705.404 | 4.6E-01 | 179747.282 | 179746.818 | 4.6E-01 | -31.063 | -31.536 | 4.7E-01 |
| 1934 | Tl | 193 | 81 | 112 | 31 | 7.885344 | 7.886339 | -9.9E-04 | 179708.925 | 179708.684 | 2.4E-01 | 179750.867 | 179750.626 | 2.4E-01 | -27.477 | -27.728 | 2.5E-01 |
| 1935 | Pb | 193 | 82 | 111 | 29 | 7.853919 | 7.856442 | -2.5E-03 | 179713.680 | 179713.142 | 5.4E-01 | 179756.150 | 179755.613 | 5.4E-01 | -22.194 | -22.741 | 5.5E-01 |
| 1936 | Bi | 193 | 83 | 110 | 27 | 7.817110 | 7.819527 | -2.4E-03 | 179719.473 | 179718.955 | 5.2E-01 | 179762.472 | 179761.954 | 5.2E-01 | -15.873 | -16.399 | 5.3E-01 |
| 1937 | Po | 193 | 84 | 109 | 25 | 7.774128 | 7.777337 | -3.2E-03 | 179726.457 | 179725.786 | 6.7E-01 | 179769.985 | 179769.314 | 6.7E-01 | -8.360 | -9.040 | 6.8E-01 |
| 1938 | At | 193 | 85 | 108 | 23 | 7.727111 | 7.729469 | -2.4E-03 | 179734.219 | 179733.712 | 5.1E-01 | 179778.277 | 179777.769 | 5.1E-01 | -0.068 | -0.584 | 5.2E-01 |
| 1939 | Rn | 193 | 86 | 107 | 21 | 7.675852 | 7.676913 | -1.1E-03 | 179742.800 | 179742.542 | 2.6E-01 | 179787.387 | 179787.130 | 2.6E-01 | 9.043 | 8.776 | 2.7E-01 |
| 1940 | Os | 194 | 76 | 118 | 42 | 7.932033 | 7.930214 | 1.8E-03 | 180638.096 | 180638.403 | -3.1E-01 | 180677.401 | 180677.708 | -3.1E-01 | -32.437 | -32.139 | -3.0E-01 |
| 1941 | Ir | 194 | 77 | 117 | 40 | 7.928498 | 7.926550 | 1.9E-03 | 180637.473 | 180637.804 | -3.3E-01 | 180677.305 | 180677.636 | -3.3E-01 | -32.534 | -32.212 | -3.2E-01 |
| 1942 | Pt | 194 | 78 | 116 | 38 | 7.935954 | 7.936251 | -3.0E-04 | 180634.717 | 180634.612 | 1.1E-01 | 180675.076 | 180674.971 | 1.1E-01 | -34.763 | -34.877 | 1.1E-01 |
| 1943 | Au | 194 | 79 | 115 | 36 | 7.918780 | 7.919367 | -5.9E-04 | 180636.739 | 180636.577 | 1.6E-01 | 180677.625 | 180677.463 | 1.6E-01 | -32.213 | -32.384 | 1.7E-01 |
| 1944 | Hg | 194 | 80 | 114 | 34 | 7.914597 | 7.915769 | -1.2E-03 | 180636.241 | 180635.965 | 2.8E-01 | 180677.655 | 180677.378 | 2.8E-01 | -32.184 | -32.469 | 2.9E-01 |
| 1945 | Tl | 194 | 81 | 113 | 32 | 7.883520 | 7.884051 | -5.3E-04 | 180640.959 | 180640.807 | 1.5E-01 | 180682.901 | 180682.749 | 1.5E-01 | -26.937 | -27.099 | 1.6E-01 |
| 1946 | Pb | 194 | 82 | 112 | 30 | 7.865415 | 7.866520 | -1.1E-03 | 180643.161 | 180642.896 | 2.6E-01 | 180685.631 | 180685.367 | 2.6E-01 | -24.207 | -24.481 | 2.7E-01 |
| 1947 | Po | 194 | 84 | 110 | 26 | 7.789294 | 7.791541 | -2.2E-03 | 180655.306 | 180654.818 | 4.9E-01 | 180698.834 | 180698.346 | 4.9E-01 | -11.005 | -11.502 | 5.0E-01 |
| 1948 | At | 194 | 85 | 109 | 24 | 7.732204 | 7.734997 | -2.8E-03 | 180665.069 | 180664.475 | 5.9E-01 | 180709.127 | 180708.533 | 5.9E-01 | -0.712 | -1.315 | 6.0E-01 |
| 1949 | Rn | 194 | 86 | 108 | 22 | 7.695001 | 7.695373 | -3.7E-04 | 180670.975 | 180670.849 | 1.3E-01 | 180715.562 | 180715.437 | 1.3E-01 | 5.723 | 5.589 | 1.3E-01 |
| 1950 | Os | 195 | 76 | 119 | 43 | 7.917744 | 7.917184 | 5.6E-04 | 181572.516 | 181572.579 | -6.3E-02 | 181611.821 | 181611.884 | -6.3E-02 | -29.512 | -29.458 | -5.4E-02 |
| 1951 | Ir | 195 | 77 | 118 | 41 | 7.924926 | 7.923548 | 1.4E-03 | 181569.806 | 181570.028 | -2.2E-01 | 181609.638 | 181609.860 | -2.2E-01 | -31.694 | -31.482 | -2.1E-01 |
| 1952 | Pt | 195 | 78 | 117 | 39 | 7.926565 | 7.926440 | 1.2E-04 | 181568.177 | 181568.154 | 2.3E-02 | 181608.536 | 181608.513 | 2.3E-02 | -32.796 | -32.829 | 3.2E-02 |
| 1953 | Au | 195 | 79 | 116 | 37 | 7.921389 | 7.921097 | 2.9E-04 | 181567.877 | 181567.886 | -8.9E-03 | 181608.763 | 181608.772 | -9.1E-03 | -32.569 | -32.570 | 3.0E-04 |
| 1954 | Hg | 195 | 80 | 115 | 35 | 7.909327 | 7.910124 | -8.0E-04 | 181568.919 | 181568.715 | 2.0E-01 | 181610.333 | 181610.129 | 2.0E-01 | -31.000 | -31.213 | 2.1E-01 |
| 1955 | Tl | 195 | 81 | 114 | 33 | 7.890727 | 7.890618 | 1.1E-04 | 181571.236 | 181571.207 | 2.9E-02 | 181613.177 | 181613.149 | 2.8E-02 | -28.155 | -28.193 | 3.7E-02 |
| 1956 | Pb | 195 | 82 | 113 | 31 | 7.863936 | 7.864907 | -9.7E-04 | 181575.149 | 181574.910 | 2.4E-01 | 181617.619 | 181617.380 | 2.4E-01 | -23.713 | -23.962 | 2.5E-01 |
| 1957 | Bi | 195 | 83 | 112 | 29 | 7.830757 | 7.831868 | -1.1E-03 | 181580.308 | 181580.040 | 2.7E-01 | 181623.307 | 181623.039 | 2.7E-01 | -18.026 | -18.302 | 2.8E-01 |
| 1958 | Po | 195 | 84 | 111 | 27 | 7.791029 | 7.793618 | -2.6E-03 | 181586.743 | 181586.187 | 5.6E-01 | 181630.272 | 181629.715 | 5.6E-01 | -11.061 | -11.627 | 5.7E-01 |
| 1959 | At | 195 | 85 | 110 | 25 | 7.748119 | 7.749541 | -1.4E-03 | 181593.799 | 181593.470 | 3.3E-01 | 181637.857 | 181637.527 | 3.3E-01 | -3.476 | -3.815 | 3.4E-01 |
| 1960 | Rn | 195 | 86 | 109 | 23 | 7.700384 | 7.701071 | -6.9E-04 | 181601.795 | 181601.608 | 1.9E-01 | 181646.383 | 181646.196 | 1.9E-01 | 5.050 | 4.854 | 2.0E-01 |
| 1961 | Os | 196 | 76 | 120 | 44 | 7.912238 | 7.915002 | -2.8E-03 | 182505.242 | 182504.655 | 5.9E-01 | 182544.548 | 182543.960 | 5.9E-01 | -28.279 | -28.876 | 6.0E-01 |
| 1962 | Ir | 196 | 77 | 119 | 42 | 7.914160 | 7.913833 | 3.3E-04 | 182503.557 | 182503.574 | -1.7E-02 | 182543.389 | 182543.406 | -1.7E-02 | -29.438 | -29.430 | -8.3E-03 |
| 1963 | Pt | 196 | 78 | 118 | 40 | 7.926541 | 7.926799 | -2.6E-04 | 182499.821 | 182499.723 | 9.8E-02 | 182540.180 | 182540.082 | 9.8E-02 | -32.647 | -32.754 | 1.1E-01 |
| 1964 | Au | 196 | 79 | 117 | 38 | 7.914861 | 7.914065 | 8.0E-04 | 182500.800 | 182500.908 | -1.1E-01 | 182541.687 | 182541.795 | -1.1E-01 | -31.140 | -31.041 | -9.9E-02 |
| 1965 | Hg | 196 | 80 | 116 | 36 | 7.914373 | 7.914544 | -1.7E-04 | 182499.586 | 182499.504 | 8.2E-02 | 182541.000 | 182540.918 | 8.2E-02 | -31.827 | -31.918 | 9.1E-02 |
| 1966 | Tl | 196 | 81 | 115 | 34 | 7.888289 | 7.887200 | 1.1E-03 | 182503.388 | 182503.552 | -1.6E-01 | 182545.330 | 182545.494 | -1.6E-01 | -27.497 | -27.342 | -1.5E-01 |
| 1967 | Pb | 196 | 82 | 114 | 32 | 7.873400 | 7.873497 | -9.7E-05 | 182504.996 | 182504.927 | 6.9E-02 | 182547.466 | 182547.397 | 6.9E-02 | -25.361 | -25.439 | 7.8E-02 |
| 1968 | Bi | 196 | 83 | 113 | 30 | 7.831900 | 7.832238 | -3.4E-04 | 182511.819 | 182511.701 | 1.2E-01 | 182554.818 | 182554.700 | 1.2E-01 | -18.009 | -18.136 | 1.3E-01 |
| 1969 | Po | 196 | 84 | 112 | 28 | 7.804814 | 7.806146 | -1.3E-03 | 182515.816 | 182515.503 | 3.1E-01 | 182559.344 | 182559.031 | 3.1E-01 | -13.483 | -13.805 | 3.2E-01 |
| 1970 | At | 196 | 85 | 111 | 26 | 7.751997 | 7.753736 | -1.7E-03 | 182524.856 | 182524.463 | 3.9E-01 | 182568.914 | 182568.521 | 3.9E-01 | -3.913 | -4.315 | 4.0E-01 |
| 1971 | Rn | 196 | 86 | 110 | 24 | 7.717986 | 7.717634 | 3.5E-04 | 182530.210 | 182530.226 | -1.6E-02 | 182574.798 | 182574.814 | -1.6E-02 | 1.971 | 1.978 | -6.5E-03 |
| 1972 | Ir | 197 | 77 | 120 | 43 | 7.909008 | 7.911125 | -2.1E-03 | 183436.223 | 183435.759 | 4.6E-01 | 183476.055 | 183475.591 | 4.6E-01 | -28.266 | -28.739 | 4.7E-01 |
| 1973 | Pt | 197 | 78 | 119 | 41 | 7.915982 | 7.916901 | -9.2E-04 | 183433.540 | 183433.311 | 2.3E-01 | 183473.899 | 183473.670 | 2.3E-01 | -30.422 | -30.660 | 2.4E-01 |
| 1974 | Au | 197 | 79 | 118 | 39 | 7.915660 | 7.915056 | 6.0E-04 | 183432.293 | 183432.365 | -7.2E-02 | 183473.180 | 183473.251 | -7.1E-02 | -31.141 | -31.079 | -6.2E-02 |
| 1975 | Hg | 197 | 80 | 117 | 37 | 7.908644 | 7.908092 | 5.5E-04 | 183432.366 | 183432.426 | -6.0E-02 | 183473.780 | 183473.839 | -5.9E-02 | -30.541 | -30.491 | -5.1E-02 |
| 1976 | Tl | 197 | 81 | 116 | 35 | 7.893499 | 7.892559 | 9.4E-04 | 183434.039 | 183434.175 | -1.4E-01 | 183475.981 | 183476.117 | -1.4E-01 | -28.340 | -28.214 | -1.3E-01 |
| 1977 | Pb | 197 | 82 | 115 | 33 | 7.871298 | 7.870833 | 4.6E-04 | 183437.102 | 183437.143 | -4.1E-02 | 183479.572 | 183479.613 | -4.1E-02 | -24.749 | -24.717 | -3.2E-02 |
| 1978 | Bi | 197 | 83 | 114 | 31 | 7.841633 | 7.841540 | 9.3E-05 | 183441.635 | 183441.602 | 3.3E-02 | 183484.634 | 183484.601 | 3.3E-02 | -19.687 | -19.729 | 4.2E-02 |
| 1979 | Po | 197 | 84 | 113 | 29 | 7.805535 | 7.807098 | -1.6E-03 | 183447.434 | 183447.075 | 3.6E-01 | 183490.962 | 183490.603 | 3.6E-01 | -13.358 | -13.727 | 3.7E-01 |
| 1980 | At | 197 | 85 | 112 | 27 | 7.765961 | 7.766692 | -7.3E-04 | 183453.919 | 183453.723 | 2.0E-01 | 183497.976 | 183497.780 | 2.0E-01 | -6.344 | -6.550 | 2.1E-01 |
| 1981 | Rn | 197 | 86 | 111 | 25 | 7.722292 | 7.722131 | 1.6E-04 | 183461.209 | 183461.188 | 2.1E-02 | 183505.797 | 183505.775 | 2.2E-02 | 1.476 | 1.445 | 3.1E-02 |
| 1982 | Pt | 198 | 78 | 120 | 42 | 7.914159 | 7.917355 | -3.2E-03 | 184365.550 | 184364.870 | 6.8E-01 | 184405.909 | 184405.229 | 6.8E-01 | -29.906 | -30.595 | 6.9E-01 |
| 1983 | Au | 198 | 79 | 119 | 40 | 7.908573 | 7.907853 | 7.2E-04 | 184365.346 | 184365.441 | -9.5E-02 | 184406.233 | 184406.327 | -9.4E-02 | -29.582 | -29.497 | -8.5E-02 |
| 1984 | Hg | 198 | 80 | 118 | 38 | 7.911555 | 7.911777 | -2.2E-04 | 184363.446 | 184363.353 | 9.3E-02 | 184404.860 | 184404.767 | 9.3E-02 | -30.955 | -31.057 | 1.0E-01 |
| 1985 | Tl | 198 | 81 | 117 | 36 | 7.890129 | 7.888343 | 1.8E-03 | 184366.378 | 184366.682 | -3.0E-01 | 184408.320 | 184408.624 | -3.0E-01 | -27.495 | -27.200 | -2.9E-01 |
| 1986 | Pb | 198 | 82 | 116 | 34 | 7.878881 | 7.878217 | 6.6E-04 | 184367.294 | 184367.376 | -8.2E-02 | 184409.764 | 184409.846 | -8.2E-02 | -26.050 | -25.978 | -7.2E-02 |
| 1987 | Bi | 198 | 83 | 115 | 32 | 7.841189 | 7.840843 | 3.5E-04 | 184373.446 | 184373.464 | -1.8E-02 | 184416.445 | 184416.463 | -1.8E-02 | -19.369 | -19.361 | -8.4E-03 |
| 1988 | Po | 198 | 84 | 114 | 30 | 7.817559 | 7.818143 | -5.8E-04 | 184376.814 | 184376.646 | 1.7E-01 | 184420.342 | 184420.175 | 1.7E-01 | -15.473 | -15.650 | 1.8E-01 |
| 1989 | Rn | 198 | 86 | 112 | 26 | 7.737724 | 7.736980 | 7.4E-04 | 184389.997 | 184390.091 | -9.4E-02 | 184434.584 | 184434.679 | -9.5E-02 | -1.230 | -1.145 | -8.5E-02 |
| 1990 | Ir | 199 | 77 | 122 | 45 | 7.891214 | 7.898763 | -7.5E-03 | 185303.076 | 185301.528 | 1.5E+00 | 185342.908 | 185341.360 | 1.5E+00 | -24.400 | -25.958 | 1.6E+00 |
| 1991 | Pt | 199 | 78 | 121 | 43 | 7.902309 | 7.907491 | -5.2E-03 | 185299.559 | 185298.481 | 1.1E+00 | 185339.918 | 185338.840 | 1.1E+00 | -27.390 | -28.478 | 1.1E+00 |
| 1992 | Au | 199 | 79 | 120 | 41 | 7.906943 | 7.908543 | -1.6E-03 | 185297.327 | 185296.961 | 3.7E-01 | 185338.214 | 185337.848 | 3.7E-01 | -29.095 | -29.471 | 3.8E-01 |
| 1993 | Hg | 199 | 80 | 119 | 39 | 7.905280 | 7.904934 | 3.5E-04 | 185296.348 | 185296.369 | -2.1E-02 | 185337.762 | 185337.783 | -2.1E-02 | -29.546 | -29.536 | -1.1E-02 |
| 1994 | Tl | 199 | 81 | 118 | 37 | 7.893877 | 7.892861 | 1.0E-03 | 185297.307 | 185297.460 | -1.5E-01 | 185339.249 | 185339.402 | -1.5E-01 | -28.059 | -27.916 | -1.4E-01 |
| 1995 | Pb | 199 | 82 | 117 | 35 | 7.875736 | 7.874749 | 9.9E-04 | 185299.607 | 185299.753 | -1.5E-01 | 185342.077 | 185342.223 | -1.5E-01 | -25.232 | -25.095 | -1.4E-01 |
| 1996 | Bi | 199 | 83 | 116 | 33 | 7.849522 | 7.848942 | 5.8E-04 | 185303.512 | 185303.577 | -6.5E-02 | 185346.511 | 185346.576 | -6.5E-02 | -20.797 | -20.742 | -5.5E-02 |
| 1997 | Po | 199 | 84 | 115 | 31 | 7.817533 | 7.818102 | -5.7E-04 | 185308.567 | 185308.402 | 1.7E-01 | 185352.095 | 185351.930 | 1.7E-01 | -15.214 | -15.388 | 1.7E-01 |
| 1998 | At | 199 | 85 | 114 | 29 | 7.781488 | 7.781206 | 2.8E-04 | 185314.428 | 185314.432 | -3.6E-03 | 185358.485 | 185358.489 | -4.1E-03 | -8.823 | -8.829 | 5.6E-03 |
| 1999 | Rn | 199 | 86 | 113 | 27 | 7.740755 | 7.740381 | 3.7E-04 | 185321.221 | 185321.243 | -2.2E-02 | 185365.809 | 185365.830 | -2.1E-02 | -1.500 | -1.488 | -1.2E-02 |
| 2000 | Fr | 199 | 87 | 112 | 25 | 7.695310 | 7.694397 | 9.1E-04 | 185328.952 | 185329.080 | -1.3E-01 | 185374.070 | 185374.198 | -1.3E-01 | 6.761 | 6.880 | -1.2E-01 |
| 2001 | Pt | 200 | 78 | 122 | 44 | 7.899206 | 7.907163 | -8.0E-03 | 186231.843 | 186230.204 | 1.6E+00 | 186272.202 | 186270.563 | 1.6E+00 | -26.601 | -28.249 | 1.6E+00 |
| 2002 | Au | 200 | 79 | 121 | 42 | 7.898491 | 7.901138 | -2.6E-03 | 186230.676 | 186230.099 | 5.8E-01 | 186271.562 | 186270.985 | 5.8E-01 | -27.240 | -27.827 | 5.9E-01 |
| 2003 | Hg | 200 | 80 | 120 | 40 | 7.905896 | 7.908178 | -2.3E-03 | 186227.885 | 186227.380 | 5.0E-01 | 186269.299 | 186268.794 | 5.0E-01 | -29.504 | -30.018 | 5.1E-01 |
| 2004 | Tl | 200 | 81 | 119 | 38 | 7.889705 | 7.888161 | 1.5E-03 | 186229.813 | 186230.073 | -2.6E-01 | 186271.755 | 186272.015 | -2.6E-01 | -27.048 | -26.798 | -2.5E-01 |
| 2005 | Pb | 200 | 82 | 118 | 36 | 7.881808 | 7.881239 | 5.7E-04 | 186230.082 | 186230.146 | -6.4E-02 | 186272.552 | 186272.616 | -6.4E-02 | -26.251 | -26.196 | -5.4E-02 |
| 2006 | Bi | 200 | 83 | 117 | 34 | 7.848497 | 7.847396 | 1.1E-03 | 186235.433 | 186235.602 | -1.7E-01 | 186278.432 | 186278.601 | -1.7E-01 | -20.371 | -20.211 | -1.6E-01 |
| 2007 | Po | 200 | 84 | 116 | 32 | 7.827503 | 7.827879 | -3.8E-04 | 186238.320 | 186238.194 | 1.3E-01 | 186281.848 | 186281.722 | 1.3E-01 | -16.954 | -17.090 | 1.4E-01 |
| 2008 | At | 200 | 85 | 115 | 30 | 7.783760 | 7.783134 | 6.3E-04 | 186245.757 | 186245.830 | -7.3E-02 | 186289.815 | 186289.888 | -7.3E-02 | -8.988 | -8.925 | -6.3E-02 |
| 2009 | Rn | 200 | 86 | 114 | 28 | 7.754980 | 7.753694 | 1.3E-03 | 186250.201 | 186250.405 | -2.0E-01 | 186294.788 | 186294.993 | -2.0E-01 | -4.014 | -3.820 | -1.9E-01 |
| 2010 | Fr | 200 | 87 | 113 | 26 | 7.700322 | 7.699973 | 3.5E-04 | 186259.820 | 186259.836 | -1.6E-02 | 186304.937 | 186304.954 | -1.7E-02 | 6.135 | 6.142 | -6.7E-03 |
| 2011 | Pt | 201 | 78 | 123 | 45 | 7.885834 | 7.894922 | -9.1E-03 | 187166.197 | 187164.323 | 1.9E+00 | 187206.556 | 187204.682 | 1.9E+00 | -23.741 | -25.624 | 1.9E+00 |
| 2012 | Au | 201 | 79 | 122 | 43 | 7.895175 | 7.900606 | -5.4E-03 | 187163.009 | 187161.870 | 1.1E+00 | 187203.896 | 187202.756 | 1.1E+00 | -26.401 | -27.550 | 1.1E+00 |
| 2013 | Hg | 201 | 80 | 121 | 41 | 7.897561 | 7.900755 | -3.2E-03 | 187161.220 | 187160.530 | 6.9E-01 | 187202.634 | 187201.943 | 6.9E-01 | -27.663 | -28.363 | 7.0E-01 |
| 2014 | Tl | 201 | 81 | 120 | 39 | 7.891262 | 7.891977 | -7.2E-04 | 187161.176 | 187160.983 | 1.9E-01 | 187203.118 | 187202.925 | 1.9E-01 | -27.179 | -27.382 | 2.0E-01 |
| 2015 | Pb | 201 | 82 | 119 | 37 | 7.877820 | 7.877149 | 6.7E-04 | 187162.567 | 187162.652 | -8.5E-02 | 187205.037 | 187205.122 | -8.5E-02 | -25.259 | -25.184 | -7.5E-02 |
| 2016 | Bi | 201 | 83 | 118 | 35 | 7.854799 | 7.854507 | 2.9E-04 | 187165.883 | 187165.891 | -8.1E-03 | 187208.882 | 187208.890 | -8.0E-03 | -21.415 | -21.416 | 1.6E-03 |
| 2017 | Po | 201 | 84 | 117 | 33 | 7.826580 | 7.826989 | -4.1E-04 | 187170.244 | 187170.110 | 1.3E-01 | 187213.772 | 187213.638 | 1.3E-01 | -16.525 | -16.668 | 1.4E-01 |
| 2018 | At | 201 | 85 | 116 | 31 | 7.794152 | 7.793387 | 7.6E-04 | 187175.450 | 187175.551 | -1.0E-01 | 187219.507 | 187219.609 | -1.0E-01 | -10.789 | -10.697 | -9.2E-02 |
| 2019 | Rn | 201 | 86 | 115 | 29 | 7.756843 | 7.756102 | 7.4E-04 | 187181.637 | 187181.733 | -9.6E-02 | 187226.224 | 187226.320 | -9.6E-02 | -4.072 | -3.986 | -8.6E-02 |
| 2020 | Fr | 201 | 87 | 114 | 27 | 7.714771 | 7.713612 | 1.2E-03 | 187188.781 | 187188.960 | -1.8E-01 | 187233.898 | 187234.078 | -1.8E-01 | 3.602 | 3.771 | -1.7E-01 |
| 2021 | Pt | 202 | 78 | 124 | 46 | 7.881560 | 7.890033 | -8.5E-03 | 188098.740 | 188096.981 | 1.8E+00 | 188139.099 | 188137.340 | 1.8E+00 | -22.692 | -24.460 | 1.8E+00 |
| 2022 | Au | 202 | 79 | 123 | 44 | 7.885909 | 7.890869 | -5.0E-03 | 188096.551 | 188095.502 | 1.0E+00 | 188137.438 | 188136.388 | 1.0E+00 | -24.353 | -25.412 | 1.1E+00 |
| 2023 | Hg | 202 | 80 | 122 | 42 | 7.896851 | 7.902664 | -5.8E-03 | 188093.031 | 188091.809 | 1.2E+00 | 188134.445 | 188133.222 | 1.2E+00 | -27.345 | -28.578 | 1.2E+00 |
| 2024 | Tl | 202 | 81 | 121 | 40 | 7.886250 | 7.886531 | -2.8E-04 | 188093.862 | 188093.756 | 1.1E-01 | 188135.804 | 188135.698 | 1.1E-01 | -25.986 | -26.102 | 1.2E-01 |
| 2025 | Pb | 202 | 82 | 120 | 38 | 7.882148 | 7.882801 | -6.5E-04 | 188093.380 | 188093.198 | 1.8E-01 | 188135.850 | 188135.669 | 1.8E-01 | -25.940 | -26.132 | 1.9E-01 |
| 2026 | Bi | 202 | 83 | 119 | 36 | 7.852535 | 7.852231 | 3.0E-04 | 188098.051 | 188098.062 | -1.1E-02 | 188141.050 | 188141.061 | -1.1E-02 | -20.741 | -20.740 | -1.2E-03 |
| 2027 | Po | 202 | 84 | 118 | 34 | 7.834718 | 7.835674 | -9.6E-04 | 188100.338 | 188100.094 | 2.4E-01 | 188143.866 | 188143.622 | 2.4E-01 | -17.924 | -18.178 | 2.5E-01 |
| 2028 | At | 202 | 85 | 117 | 32 | 7.794542 | 7.794383 | 1.6E-04 | 188107.142 | 188107.122 | 2.0E-02 | 188151.200 | 188151.180 | 2.0E-02 | -10.591 | -10.621 | 3.0E-02 |
| 2029 | Rn | 202 | 86 | 116 | 30 | 7.769300 | 7.768040 | 1.3E-03 | 188110.929 | 188111.131 | -2.0E-01 | 188155.516 | 188155.718 | -2.0E-01 | -6.274 | -6.082 | -1.9E-01 |
| 2030 | Ra | 202 | 88 | 114 | 26 | 7.685487 | 7.683338 | 2.1E-03 | 188125.234 | 188125.613 | -3.8E-01 | 188170.882 | 188171.262 | -3.8E-01 | 9.091 | 9.461 | -3.7E-01 |
| 2031 | Au | 203 | 79 | 124 | 45 | 7.880864 | 7.885887 | -5.0E-03 | 189029.255 | 189028.188 | 1.1E+00 | 189070.141 | 189069.074 | 1.1E+00 | -23.143 | -24.220 | 1.1E+00 |
| 2032 | Hg | 203 | 80 | 123 | 43 | 7.887480 | 7.892742 | -5.3E-03 | 189026.602 | 189025.485 | 1.1E+00 | 189068.016 | 189066.899 | 1.1E+00 | -25.269 | -26.395 | 1.1E+00 |
| 2033 | Tl | 203 | 81 | 122 | 41 | 7.886050 | 7.888753 | -2.7E-03 | 189025.582 | 189024.984 | 6.0E-01 | 189067.524 | 189066.926 | 6.0E-01 | -25.761 | -26.368 | 6.1E-01 |
| 2034 | Pb | 203 | 82 | 121 | 39 | 7.877394 | 7.877736 | -3.4E-04 | 189026.028 | 189025.909 | 1.2E-01 | 189068.499 | 189068.379 | 1.2E-01 | -24.786 | -24.915 | 1.3E-01 |
| 2035 | Bi | 203 | 83 | 120 | 37 | 7.857472 | 7.858305 | -8.3E-04 | 189028.761 | 189028.542 | 2.2E-01 | 189071.760 | 189071.541 | 2.2E-01 | -21.524 | -21.754 | 2.3E-01 |
| 2036 | Po | 203 | 84 | 119 | 35 | 7.832862 | 7.833958 | -1.1E-03 | 189032.446 | 189032.172 | 2.7E-01 | 189075.974 | 189075.700 | 2.7E-01 | -17.311 | -17.594 | 2.8E-01 |
| 2037 | At | 203 | 85 | 118 | 33 | 7.803648 | 7.803462 | 1.9E-04 | 189037.065 | 189037.050 | 1.5E-02 | 189081.122 | 189081.108 | 1.4E-02 | -12.163 | -12.186 | 2.4E-02 |
| 2038 | Rn | 203 | 86 | 117 | 31 | 7.770221 | 7.769521 | 7.0E-04 | 189042.538 | 189042.627 | -8.9E-02 | 189087.125 | 189087.215 | -9.0E-02 | -6.159 | -6.080 | -8.0E-02 |
| 2039 | Fr | 203 | 87 | 116 | 29 | 7.731709 | 7.730356 | 1.4E-03 | 189049.043 | 189049.265 | -2.2E-01 | 189094.161 | 189094.382 | -2.2E-01 | 0.876 | 1.088 | -2.1E-01 |
| 2040 | Ra | 203 | 88 | 115 | 27 | 7.689478 | 7.688095 | 1.4E-03 | 189056.303 | 189056.530 | -2.3E-01 | 189101.952 | 189102.178 | -2.3E-01 | 8.667 | 8.884 | -2.2E-01 |
| 2041 | Hg | 204 | 80 | 124 | 44 | 7.885545 | 7.890591 | -5.0E-03 | 189958.675 | 189957.597 | 1.1E+00 | 190000.088 | 189999.011 | 1.1E+00 | -24.690 | -25.778 | 1.1E+00 |
| 2042 | Tl | 204 | 81 | 123 | 42 | 7.880021 | 7.880882 | -8.6E-04 | 189958.491 | 189958.266 | 2.2E-01 | 190000.433 | 190000.208 | 2.2E-01 | -24.346 | -24.580 | 2.3E-01 |
| 2043 | Pb | 204 | 82 | 122 | 40 | 7.879930 | 7.881721 | -1.8E-03 | 189957.199 | 189956.784 | 4.2E-01 | 189999.669 | 189999.254 | 4.1E-01 | -25.109 | -25.534 | 4.2E-01 |
| 2044 | Bi | 204 | 83 | 121 | 38 | 7.854213 | 7.854931 | -7.2E-04 | 189961.134 | 189960.937 | 2.0E-01 | 190004.133 | 190003.936 | 2.0E-01 | -20.646 | -20.852 | 2.1E-01 |
| 2045 | Po | 204 | 84 | 120 | 36 | 7.839080 | 7.841467 | -2.4E-03 | 189962.910 | 189962.372 | 5.4E-01 | 190006.438 | 190005.900 | 5.4E-01 | -18.341 | -18.889 | 5.5E-01 |
| 2046 | At | 204 | 85 | 119 | 34 | 7.803552 | 7.803517 | 3.5E-05 | 189968.846 | 189968.801 | 4.5E-02 | 190012.903 | 190012.859 | 4.4E-02 | -11.875 | -11.930 | 5.4E-02 |
| 2047 | Rn | 204 | 86 | 118 | 32 | 7.780637 | 7.780163 | 4.7E-04 | 189972.208 | 189972.252 | -4.4E-02 | 190016.796 | 190016.840 | -4.4E-02 | -7.983 | -7.949 | -3.4E-02 |
| 2048 | Fr | 204 | 87 | 117 | 30 | 7.734692 | 7.733781 | 9.1E-04 | 189980.268 | 189980.401 | -1.3E-01 | 190025.386 | 190025.519 | -1.3E-01 | 0.607 | 0.730 | -1.2E-01 |
| 2049 | Ra | 204 | 88 | 116 | 28 | 7.704191 | 7.702163 | 2.0E-03 | 189985.178 | 189985.537 | -3.6E-01 | 190030.826 | 190031.186 | -3.6E-01 | 6.047 | 6.397 | -3.5E-01 |
| 2050 | Hg | 205 | 80 | 125 | 45 | 7.874731 | 7.875249 | -5.2E-04 | 190892.571 | 190892.417 | 1.5E-01 | 190933.985 | 190933.831 | 1.5E-01 | -22.287 | -22.452 | 1.6E-01 |
| 2051 | Tl | 205 | 81 | 124 | 43 | 7.878392 | 7.879182 | -7.9E-04 | 190890.510 | 190890.300 | 2.1E-01 | 190932.452 | 190932.241 | 2.1E-01 | -23.820 | -24.041 | 2.2E-01 |
| 2052 | Pb | 205 | 82 | 123 | 41 | 7.874328 | 7.874179 | 1.5E-04 | 190890.033 | 190890.014 | 1.9E-02 | 190932.503 | 190932.484 | 1.9E-02 | -23.770 | -23.799 | 2.9E-02 |
| 2053 | Bi | 205 | 83 | 122 | 39 | 7.857314 | 7.859170 | -1.9E-03 | 190892.210 | 190891.779 | 4.3E-01 | 190935.209 | 190934.778 | 4.3E-01 | -21.064 | -21.505 | 4.4E-01 |
| 2054 | Po | 205 | 84 | 121 | 37 | 7.836157 | 7.838510 | -2.4E-03 | 190895.235 | 190894.702 | 5.3E-01 | 190938.763 | 190938.230 | 5.3E-01 | -17.509 | -18.053 | 5.4E-01 |
| 2055 | At | 205 | 85 | 120 | 35 | 7.810199 | 7.811266 | -1.1E-03 | 190899.245 | 190898.974 | 2.7E-01 | 190943.302 | 190943.032 | 2.7E-01 | -12.970 | -13.251 | 2.8E-01 |
| 2056 | Rn | 205 | 86 | 119 | 33 | 7.780742 | 7.780640 | 1.0E-04 | 190903.971 | 190903.940 | 3.1E-02 | 190948.559 | 190948.527 | 3.2E-02 | -7.714 | -7.756 | 4.2E-02 |
| 2057 | Fr | 205 | 87 | 118 | 31 | 7.745686 | 7.744712 | 9.7E-04 | 190909.845 | 190909.992 | -1.5E-01 | 190954.963 | 190955.109 | -1.5E-01 | -1.310 | -1.173 | -1.4E-01 |
| 2058 | Ra | 205 | 88 | 117 | 29 | 7.706998 | 7.705901 | 1.1E-03 | 190916.463 | 190916.634 | -1.7E-01 | 190962.111 | 190962.282 | -1.7E-01 | 5.839 | 6.000 | -1.6E-01 |
| 2059 | Hg | 206 | 80 | 126 | 46 | 7.869169 | 7.867052 | 2.1E-03 | 191825.408 | 191825.795 | -3.9E-01 | 191866.821 | 191867.209 | -3.9E-01 | -20.945 | -20.567 | -3.8E-01 |
| 2060 | Tl | 206 | 81 | 125 | 44 | 7.871719 | 7.866470 | 5.2E-03 | 191823.572 | 191824.604 | -1.0E+00 | 191865.514 | 191866.546 | -1.0E+00 | -22.253 | -21.230 | -1.0E+00 |
| 2061 | Pb | 206 | 82 | 124 | 42 | 7.875359 | 7.874550 | 8.1E-04 | 191821.511 | 191821.628 | -1.2E-01 | 191863.982 | 191864.099 | -1.2E-01 | -23.785 | -23.678 | -1.1E-01 |
| 2062 | Bi | 206 | 83 | 123 | 40 | 7.853322 | 7.853402 | -8.0E-05 | 191824.740 | 191824.673 | 6.7E-02 | 191867.739 | 191867.672 | 6.7E-02 | -20.028 | -20.104 | 7.7E-02 |
| 2063 | Po | 206 | 84 | 122 | 38 | 7.840595 | 7.844125 | -3.5E-03 | 191826.050 | 191825.272 | 7.8E-01 | 191869.578 | 191868.800 | 7.8E-01 | -18.188 | -18.977 | 7.9E-01 |
| 2064 | At | 206 | 85 | 121 | 36 | 7.808839 | 7.809987 | -1.1E-03 | 191831.280 | 191830.992 | 2.9E-01 | 191875.338 | 191875.049 | 2.9E-01 | -12.429 | -12.727 | 3.0E-01 |
| 2065 | Rn | 206 | 86 | 120 | 34 | 7.788956 | 7.789831 | -8.7E-04 | 191834.064 | 191833.831 | 2.3E-01 | 191878.651 | 191878.418 | 2.3E-01 | -9.115 | -9.358 | 2.4E-01 |
| 2066 | Fr | 206 | 87 | 119 | 32 | 7.746940 | 7.747015 | -7.5E-05 | 191841.407 | 191841.338 | 6.9E-02 | 191886.524 | 191886.456 | 6.8E-02 | -1.242 | -1.321 | 7.9E-02 |
| 2067 | Ra | 206 | 88 | 118 | 30 | 7.719801 | 7.718493 | 1.3E-03 | 191845.684 | 191845.900 | -2.2E-01 | 191891.332 | 191891.548 | -2.2E-01 | 3.566 | 3.771 | -2.1E-01 |
| 2068 | Hg | 207 | 80 | 127 | 47 | 7.848610 | 7.846013 | 2.6E-03 | 192761.359 | 192761.849 | -4.9E-01 | 192802.773 | 192803.263 | -4.9E-01 | -16.487 | -16.008 | -4.8E-01 |
| 2069 | Tl | 207 | 81 | 126 | 45 | 7.866792 | 7.859176 | 7.6E-03 | 192756.285 | 192757.813 | -1.5E+00 | 192798.227 | 192799.755 | -1.5E+00 | -21.033 | -19.516 | -1.5E+00 |
| 2070 | Pb | 207 | 82 | 125 | 43 | 7.869864 | 7.862503 | 7.4E-03 | 192754.339 | 192755.813 | -1.5E+00 | 192796.809 | 192798.283 | -1.5E+00 | -22.452 | -20.987 | -1.5E+00 |
| 2071 | Bi | 207 | 83 | 124 | 41 | 7.854502 | 7.854149 | 3.5E-04 | 192756.208 | 192756.230 | -2.2E-02 | 192799.207 | 192799.229 | -2.2E-02 | -20.054 | -20.041 | -1.3E-02 |
| 2072 | Po | 207 | 84 | 123 | 39 | 7.836671 | 7.838776 | -2.1E-03 | 192758.587 | 192758.101 | 4.9E-01 | 192802.115 | 192801.629 | 4.9E-01 | -17.145 | -17.642 | 5.0E-01 |
| 2073 | At | 207 | 85 | 122 | 37 | 7.813962 | 7.815731 | -1.8E-03 | 192761.976 | 192761.558 | 4.2E-01 | 192806.034 | 192805.616 | 4.2E-01 | -13.227 | -13.655 | 4.3E-01 |
| 2074 | Rn | 207 | 86 | 121 | 35 | 7.787998 | 7.788891 | -8.9E-04 | 192766.039 | 192765.801 | 2.4E-01 | 192810.626 | 192810.389 | 2.4E-01 | -8.635 | -8.882 | 2.5E-01 |
| 2075 | Fr | 207 | 87 | 120 | 33 | 7.756246 | 7.756402 | -1.6E-04 | 192771.299 | 192771.213 | 8.6E-02 | 192816.416 | 192816.331 | 8.5E-02 | -2.844 | -2.940 | 9.6E-02 |
| 2076 | Ra | 207 | 88 | 119 | 31 | 7.721629 | 7.721085 | 5.4E-04 | 192777.151 | 192777.210 | -5.9E-02 | 192822.800 | 192822.858 | -5.8E-02 | 3.539 | 3.588 | -4.9E-02 |
| 2077 | Ac | 207 | 89 | 118 | 29 | 7.681100 | 7.680610 | 4.9E-04 | 192784.228 | 192784.274 | -4.6E-02 | 192830.407 | 192830.454 | -4.7E-02 | 11.146 | 11.183 | -3.7E-02 |
| 2078 | Hg | 208 | 80 | 128 | 48 | 7.834191 | 7.833306 | 8.8E-04 | 193696.075 | 193696.211 | -1.4E-01 | 193737.489 | 193737.625 | -1.4E-01 | -13.265 | -13.139 | -1.3E-01 |
| 2079 | Tl | 208 | 81 | 127 | 46 | 7.847178 | 7.841378 | 5.8E-03 | 193692.064 | 193693.221 | -1.2E+00 | 193734.006 | 193735.163 | -1.2E+00 | -16.749 | -15.601 | -1.1E+00 |
| 2080 | Pb | 208 | 82 | 126 | 44 | 7.867450 | 7.857889 | 9.6E-03 | 193686.536 | 193688.476 | -1.9E+00 | 193729.007 | 193730.946 | -1.9E+00 | -21.748 | -19.819 | -1.9E+00 |
| 2081 | Bi | 208 | 83 | 125 | 42 | 7.849851 | 7.844324 | 5.5E-03 | 193688.886 | 193689.985 | -1.1E+00 | 193731.885 | 193732.984 | -1.1E+00 | -18.870 | -17.780 | -1.1E+00 |
| 2082 | Po | 208 | 84 | 124 | 40 | 7.839356 | 7.841130 | -1.8E-03 | 193689.757 | 193689.337 | 4.2E-01 | 193733.286 | 193732.865 | 4.2E-01 | -17.469 | -17.899 | 4.3E-01 |
| 2083 | At | 208 | 85 | 123 | 38 | 7.811558 | 7.812141 | -5.8E-04 | 193694.228 | 193694.055 | 1.7E-01 | 193738.285 | 193738.112 | 1.7E-01 | -12.470 | -12.653 | 1.8E-01 |
| 2084 | Rn | 208 | 86 | 122 | 36 | 7.794266 | 7.796042 | -1.8E-03 | 193696.512 | 193696.090 | 4.2E-01 | 193741.099 | 193740.678 | 4.2E-01 | -9.655 | -10.087 | 4.3E-01 |
| 2085 | Fr | 208 | 87 | 121 | 34 | 7.756902 | 7.757212 | -3.1E-04 | 193702.971 | 193702.854 | 1.2E-01 | 193748.089 | 193747.971 | 1.2E-01 | -2.666 | -2.793 | 1.3E-01 |
| 2086 | Ra | 208 | 88 | 120 | 32 | 7.732079 | 7.732023 | 5.6E-05 | 193706.822 | 193706.779 | 4.3E-02 | 193752.470 | 193752.428 | 4.2E-02 | 1.715 | 1.663 | 5.2E-02 |
| 2087 | Ac | 208 | 89 | 119 | 30 | 7.684836 | 7.685033 | -2.0E-04 | 193715.335 | 193715.239 | 9.6E-02 | 193761.514 | 193761.418 | 9.6E-02 | 10.759 | 10.654 | 1.1E-01 |
| 2088 | Th | 208 | 90 | 118 | 28 | 7.652638 | 7.651858 | 7.8E-04 | 193720.718 | 193720.825 | -1.1E-01 | 193767.429 | 193767.536 | -1.1E-01 | 16.674 | 16.771 | -9.7E-02 |
| 2089 | Tl | 209 | 81 | 128 | 47 | 7.833363 | 7.829732 | 3.6E-03 | 194626.669 | 194627.379 | -7.1E-01 | 194668.611 | 194669.321 | -7.1E-01 | -13.638 | -12.938 | -7.0E-01 |
| 2090 | Pb | 209 | 82 | 127 | 45 | 7.848646 | 7.841083 | 7.6E-03 | 194622.164 | 194623.696 | -1.5E+00 | 194664.635 | 194666.166 | -1.5E+00 | -17.614 | -16.093 | -1.5E+00 |
| 2091 | Bi | 209 | 83 | 126 | 43 | 7.847984 | 7.840427 | 7.6E-03 | 194620.992 | 194622.521 | -1.5E+00 | 194663.991 | 194665.520 | -1.5E+00 | -18.258 | -16.739 | -1.5E+00 |
| 2092 | Po | 209 | 84 | 125 | 41 | 7.835185 | 7.831995 | 3.2E-03 | 194622.355 | 194622.971 | -6.2E-01 | 194665.883 | 194666.499 | -6.2E-01 | -16.366 | -15.760 | -6.1E-01 |
| 2093 | At | 209 | 85 | 124 | 39 | 7.814776 | 7.814718 | 5.8E-05 | 194625.309 | 194625.269 | 4.0E-02 | 194669.366 | 194669.327 | 3.9E-02 | -12.882 | -12.932 | 5.0E-02 |
| 2094 | Rn | 209 | 86 | 123 | 37 | 7.792116 | 7.792819 | -7.0E-04 | 194628.733 | 194628.533 | 2.0E-01 | 194673.320 | 194673.121 | 2.0E-01 | -8.929 | -9.138 | 2.1E-01 |
| 2095 | Fr | 209 | 87 | 122 | 35 | 7.763680 | 7.764506 | -8.3E-04 | 194633.363 | 194633.137 | 2.3E-01 | 194678.481 | 194678.255 | 2.3E-01 | -3.768 | -4.004 | 2.4E-01 |
| 2096 | Ra | 209 | 88 | 121 | 33 | 7.733037 | 7.733095 | -5.8E-05 | 194638.455 | 194638.389 | 6.6E-02 | 194684.103 | 194684.037 | 6.6E-02 | 1.854 | 1.778 | 7.6E-02 |
| 2097 | Ac | 209 | 89 | 120 | 31 | 7.695847 | 7.696180 | -3.3E-04 | 194644.914 | 194644.790 | 1.2E-01 | 194691.093 | 194690.969 | 1.2E-01 | 8.844 | 8.710 | 1.3E-01 |
| 2098 | Th | 209 | 90 | 119 | 29 | 7.655298 | 7.656468 | -1.2E-03 | 194652.075 | 194651.775 | 3.0E-01 | 194698.786 | 194698.486 | 3.0E-01 | 16.537 | 16.227 | 3.1E-01 |
| 2099 | Tl | 210 | 81 | 129 | 48 | 7.813584 | 7.809028 | 4.6E-03 | 195562.555 | 195563.463 | -9.1E-01 | 195604.497 | 195605.405 | -9.1E-01 | -9.246 | -8.348 | -9.0E-01 |
| 2100 | Pb | 210 | 82 | 128 | 46 | 7.835963 | 7.832502 | 3.5E-03 | 195556.544 | 195557.222 | -6.8E-01 | 195599.015 | 195599.692 | -6.8E-01 | -14.728 | -14.061 | -6.7E-01 |
| 2101 | Bi | 210 | 83 | 127 | 44 | 7.832540 | 7.826328 | 6.2E-03 | 195555.952 | 195557.207 | -1.3E+00 | 195598.951 | 195600.206 | -1.3E+00 | -14.791 | -13.547 | -1.2E+00 |
| 2102 | Po | 210 | 84 | 126 | 42 | 7.834344 | 7.830158 | 4.2E-03 | 195554.262 | 195555.090 | -8.3E-01 | 195597.790 | 195598.618 | -8.3E-01 | -15.953 | -15.135 | -8.2E-01 |
| 2103 | At | 210 | 85 | 125 | 40 | 7.811661 | 7.807689 | 4.0E-03 | 195557.714 | 195558.496 | -7.8E-01 | 195601.771 | 195602.553 | -7.8E-01 | -11.972 | -11.199 | -7.7E-01 |
| 2104 | Rn | 210 | 86 | 124 | 38 | 7.796663 | 7.796985 | -3.2E-04 | 195559.551 | 195559.431 | 1.2E-01 | 195604.138 | 195604.018 | 1.2E-01 | -9.605 | -9.734 | 1.3E-01 |
| 2105 | Fr | 210 | 87 | 123 | 36 | 7.763071 | 7.763101 | -3.0E-05 | 195565.293 | 195565.233 | 6.0E-02 | 195610.410 | 195610.351 | 5.9E-02 | -3.333 | -3.402 | 6.9E-02 |
| 2106 | Ra | 210 | 88 | 122 | 34 | 7.741285 | 7.741919 | -6.3E-04 | 195568.555 | 195568.368 | 1.9E-01 | 195614.203 | 195614.016 | 1.9E-01 | 0.460 | 0.263 | 2.0E-01 |
| 2107 | Ac | 210 | 89 | 121 | 32 | 7.697896 | 7.699010 | -1.1E-03 | 195576.353 | 195576.065 | 2.9E-01 | 195622.532 | 195622.244 | 2.9E-01 | 8.790 | 8.491 | 3.0E-01 |
| 2108 | Th | 210 | 90 | 120 | 30 | 7.669075 | 7.669211 | -1.4E-04 | 195581.092 | 195581.008 | 8.4E-02 | 195627.802 | 195627.718 | 8.4E-02 | 14.060 | 13.966 | 9.4E-02 |
| 2109 | Tl | 211 | 81 | 130 | 49 | 7.799791 | 7.795346 | 4.4E-03 | 196497.217 | 196498.106 | -8.9E-01 | 196539.159 | 196540.048 | -8.9E-01 | -6.078 | -5.199 | -8.8E-01 |
| 2110 | Pb | 211 | 82 | 129 | 47 | 7.816999 | 7.812752 | 4.2E-03 | 196492.275 | 196493.122 | -8.5E-01 | 196534.745 | 196535.592 | -8.5E-01 | -10.491 | -9.655 | -8.4E-01 |
| 2111 | Bi | 211 | 83 | 128 | 45 | 7.819769 | 7.818591 | 1.2E-03 | 196490.380 | 196490.578 | -2.0E-01 | 196533.379 | 196533.577 | -2.0E-01 | -11.858 | -11.670 | -1.9E-01 |
| 2112 | Po | 211 | 84 | 127 | 43 | 7.818782 | 7.817017 | 1.8E-03 | 196489.277 | 196489.598 | -3.2E-01 | 196532.805 | 196533.126 | -3.2E-01 | -12.432 | -12.121 | -3.1E-01 |
| 2113 | At | 211 | 85 | 126 | 41 | 7.811352 | 7.806319 | 5.0E-03 | 196489.532 | 196490.543 | -1.0E+00 | 196533.590 | 196534.600 | -1.0E+00 | -11.647 | -10.646 | -1.0E+00 |

| | | | | | | | | | | | | | | | | | | |
|---|---|---|---|---|---|---|---|---|---|---|---|---|---|---|---|---|---|---|
| 2114 | Rn | 211 | 86 | 125 | 39 | 7.793939 | 7.790531 | 3.4E-03 | 196491.894 | 196492.561 | -6.7E-01 | 196536.482 | 196537.149 | -6.7E-01 | -8.755 | -8.098 | -6.6E-01 |
| 2115 | Fr | 211 | 87 | 124 | 37 | 7.768358 | 7.767498 | 8.6E-04 | 196495.979 | 196496.108 | -1.3E-01 | 196541.097 | 196541.225 | -1.3E-01 | -4.140 | -4.021 | -1.2E-01 |
| 2116 | Ra | 211 | 88 | 123 | 35 | 7.741087 | 7.740831 | 2.6E-04 | 196500.421 | 196500.421 | 3.4E-05 | 196546.069 | 196546.069 | -1.4E-04 | 0.832 | 0.822 | 9.8E-03 |
| 2117 | Ac | 211 | 89 | 122 | 33 | 7.707189 | 7.708045 | -8.6E-04 | 196506.260 | 196506.025 | 2.4E-01 | 196552.439 | 196552.204 | 2.4E-01 | 7.202 | 6.957 | 2.5E-01 |
| 2118 | Th | 211 | 90 | 121 | 31 | 7.671706 | 7.672255 | -5.5E-04 | 196512.433 | 196512.262 | 1.7E-01 | 196559.143 | 196558.972 | 1.7E-01 | 13.907 | 13.726 | 1.8E-01 |
| 2119 | Pb | 212 | 82 | 130 | 48 | 7.804312 | 7.802233 | 2.1E-03 | 197426.713 | 197427.105 | -3.9E-01 | 197469.184 | 197469.575 | -3.9E-01 | -7.547 | -7.166 | -3.8E-01 |
| 2120 | Bi | 212 | 83 | 129 | 46 | 7.803309 | 7.801746 | 1.6E-03 | 197425.615 | 197425.896 | -2.8E-01 | 197468.614 | 197468.895 | -2.8E-01 | -8.117 | -7.846 | -2.7E-01 |
| 2121 | Po | 212 | 84 | 128 | 44 | 7.810241 | 7.811648 | -1.4E-03 | 197422.834 | 197422.485 | 3.5E-01 | 197466.362 | 197466.013 | 3.5E-01 | -10.369 | -10.728 | 3.6E-01 |
| 2122 | At | 212 | 85 | 127 | 42 | 7.798338 | 7.795675 | 2.7E-03 | 197424.045 | 197424.558 | -5.1E-01 | 197468.103 | 197468.616 | -5.1E-01 | -8.628 | -8.125 | -5.0E-01 |
| 2123 | Rn | 212 | 86 | 126 | 40 | 7.794796 | 7.791097 | 3.7E-03 | 197423.484 | 197424.216 | -7.3E-01 | 197468.072 | 197468.803 | -7.3E-01 | -8.659 | -7.938 | -7.2E-01 |
| 2124 | Fr | 212 | 87 | 125 | 38 | 7.766844 | 7.763124 | 3.7E-03 | 197428.097 | 197428.833 | -7.4E-01 | 197473.215 | 197473.951 | -7.4E-01 | -3.516 | -2.790 | -7.3E-01 |
| 2125 | Ra | 212 | 88 | 124 | 36 | 7.747507 | 7.746897 | 6.1E-04 | 197430.884 | 197430.959 | -7.5E-02 | 197476.532 | 197476.608 | -7.6E-02 | -0.199 | -0.133 | -6.5E-02 |
| 2126 | Ac | 212 | 89 | 123 | 34 | 7.708549 | 7.708760 | -2.1E-04 | 197437.830 | 197437.730 | 1.0E-01 | 197484.009 | 197483.910 | 9.9E-02 | 7.278 | 7.169 | 1.1E-01 |
| 2127 | Th | 212 | 90 | 122 | 32 | 7.682123 | 7.682859 | -7.4E-04 | 197442.118 | 197441.907 | 2.1E-01 | 197488.829 | 197488.618 | 2.1E-01 | 12.098 | 11.877 | 2.2E-01 |
| 2128 | Pa | 212 | 91 | 121 | 30 | 7.633548 | 7.635751 | -2.2E-03 | 197451.102 | 197450.579 | 5.2E-01 | 197498.344 | 197497.821 | 5.2E-01 | 21.613 | 21.080 | 5.3E-01 |
| 2129 | Tl | 213 | 81 | 132 | 51 | 7.765430 | 7.759053 | 6.4E-03 | 198368.067 | 198369.377 | -1.3E+00 | 198410.009 | 198411.318 | -1.3E+00 | 1.784 | 3.083 | -1.3E+00 |
| 2130 | Pb | 213 | 82 | 131 | 49 | 7.785163 | 7.781206 | 4.0E-03 | 198362.553 | 198363.346 | -7.9E-01 | 198405.023 | 198405.817 | -7.9E-01 | -3.202 | -2.418 | -7.8E-01 |
| 2131 | Bi | 213 | 83 | 130 | 47 | 7.791014 | 7.791911 | -9.0E-04 | 198359.996 | 198359.755 | 2.4E-01 | 198402.994 | 198402.754 | 2.4E-01 | -5.230 | -5.481 | 2.5E-01 |
| 2132 | Po | 213 | 84 | 129 | 45 | 7.794021 | 7.795751 | -1.7E-03 | 198358.044 | 198357.624 | 4.2E-01 | 198401.572 | 198401.152 | 4.2E-01 | -6.653 | -7.083 | 4.3E-01 |
| 2133 | At | 213 | 85 | 128 | 43 | 7.790001 | 7.790867 | -8.7E-04 | 198357.588 | 198357.352 | 2.4E-01 | 198401.646 | 198401.410 | 2.4E-01 | -6.579 | -6.825 | 2.5E-01 |
| 2134 | Rn | 213 | 86 | 127 | 41 | 7.782191 | 7.781238 | 9.5E-04 | 198357.940 | 198358.090 | -1.5E-01 | 198402.527 | 198402.678 | -1.5E-01 | -5.698 | -5.557 | -1.4E-01 |
| 2135 | Fr | 213 | 87 | 126 | 39 | 7.768446 | 7.764106 | 4.3E-03 | 198359.555 | 198360.426 | -8.7E-01 | 198404.672 | 198405.544 | -8.7E-01 | -3.553 | -2.691 | -8.6E-01 |
| 2136 | Ra | 213 | 88 | 125 | 37 | 7.746414 | 7.743009 | 3.4E-03 | 198362.935 | 198363.606 | -6.7E-01 | 198408.583 | 198409.254 | -6.7E-01 | 0.358 | 1.019 | -6.6E-01 |
| 2137 | Ac | 213 | 89 | 124 | 35 | 7.715519 | 7.715132 | 3.9E-04 | 198368.202 | 198368.230 | -2.8E-02 | 198414.381 | 198414.409 | -2.8E-02 | 6.156 | 6.174 | -1.8E-02 |
| 2138 | Th | 213 | 90 | 123 | 33 | 7.683856 | 7.683870 | -1.4E-05 | 198373.632 | 198373.574 | 5.8E-02 | 198420.343 | 198420.285 | 5.8E-02 | 12.118 | 12.050 | 6.8E-02 |
| 2139 | Pa | 213 | 91 | 122 | 31 | 7.644760 | 7.646633 | -1.9E-03 | 198380.646 | 198380.191 | 4.6E-01 | 198427.888 | 198427.433 | 4.5E-01 | 19.663 | 19.198 | 4.6E-01 |
| 2140 | Pb | 214 | 82 | 132 | 50 | 7.772384 | 7.769867 | 2.5E-03 | 199297.068 | 199297.557 | -4.9E-01 | 199339.538 | 199340.027 | -4.9E-01 | -0.181 | 0.298 | -4.8E-01 |
| 2141 | Bi | 214 | 83 | 131 | 48 | 7.773490 | 7.773836 | -3.5E-04 | 199295.520 | 199295.396 | 1.2E-01 | 199338.519 | 199338.395 | 1.2E-01 | -1.200 | -1.334 | 1.3E-01 |
| 2142 | Po | 214 | 84 | 130 | 46 | 7.785114 | 7.788365 | -3.3E-03 | 199291.721 | 199290.975 | 7.5E-01 | 199335.249 | 199334.503 | 7.5E-01 | -4.470 | -5.226 | 7.6E-01 |
| 2143 | At | 214 | 85 | 129 | 44 | 7.776364 | 7.777657 | -1.3E-03 | 199292.282 | 199291.954 | 3.3E-01 | 199336.339 | 199336.011 | 3.3E-01 | -3.379 | -3.718 | 3.4E-01 |
| 2144 | Rn | 214 | 86 | 128 | 42 | 7.777100 | 7.778630 | -1.5E-03 | 199290.812 | 199290.433 | 3.8E-01 | 199335.399 | 199335.020 | 3.8E-01 | -4.319 | -4.709 | 3.9E-01 |
| 2145 | Fr | 214 | 87 | 127 | 40 | 7.757738 | 7.756633 | 1.1E-03 | 199293.643 | 199293.827 | -1.8E-01 | 199338.761 | 199338.944 | -1.8E-01 | -0.958 | -0.785 | -1.7E-01 |
| 2146 | Ra | 214 | 88 | 126 | 38 | 7.749170 | 7.745924 | 3.2E-03 | 199294.164 | 199294.805 | -6.4E-01 | 199339.812 | 199340.453 | -6.4E-01 | 0.093 | 0.724 | -6.3E-01 |
| 2147 | Ac | 214 | 89 | 125 | 36 | 7.715834 | 7.713233 | 2.6E-03 | 199299.984 | 199300.486 | -5.0E-01 | 199346.163 | 199346.666 | -5.0E-01 | 6.445 | 6.937 | -4.9E-01 |
| 2148 | Th | 214 | 90 | 124 | 34 | 7.692235 | 7.691908 | 3.3E-04 | 199303.721 | 199303.736 | -1.5E-02 | 199350.431 | 199350.446 | -1.5E-02 | 10.712 | 10.717 | -4.8E-03 |
| 2149 | Pa | 214 | 91 | 123 | 32 | 7.647584 | 7.649361 | -1.8E-03 | 199311.962 | 199311.526 | 4.4E-01 | 199359.204 | 199358.768 | 4.4E-01 | 19.485 | 19.039 | 4.5E-01 |
| 2150 | Bi | 215 | 83 | 132 | 49 | 7.761627 | 7.762949 | -1.3E-03 | 200229.862 | 200229.528 | 3.3E-01 | 200272.861 | 200272.527 | 3.3E-01 | 1.649 | 1.304 | 3.4E-01 |
| 2151 | Po | 215 | 84 | 131 | 47 | 7.768168 | 7.771055 | -2.9E-03 | 200227.145 | 200226.473 | 6.7E-01 | 200270.673 | 200270.001 | 6.7E-01 | -0.540 | -1.222 | 6.8E-01 |
| 2152 | At | 215 | 85 | 130 | 45 | 7.767854 | 7.770694 | -2.8E-03 | 200225.900 | 200225.239 | 6.6E-01 | 200269.958 | 200269.296 | 6.6E-01 | -1.255 | -1.927 | 6.7E-01 |
| 2153 | Rn | 215 | 86 | 129 | 43 | 7.763812 | 7.766207 | -2.4E-03 | 200225.457 | 200224.890 | 5.7E-01 | 200270.045 | 200269.478 | 5.7E-01 | -1.168 | -1.746 | 5.8E-01 |
| 2154 | Fr | 215 | 87 | 128 | 41 | 7.753259 | 7.754523 | -1.3E-03 | 200226.414 | 200226.089 | 3.2E-01 | 200271.531 | 200271.207 | 3.2E-01 | 0.318 | -0.017 | 3.3E-01 |
| 2155 | Ra | 215 | 88 | 127 | 39 | 7.739315 | 7.739106 | 2.1E-04 | 200228.099 | 200228.090 | 9.0E-03 | 200273.747 | 200273.738 | 8.8E-03 | 2.534 | 2.515 | 1.9E-02 |
| 2156 | Ac | 215 | 89 | 126 | 37 | 7.719411 | 7.716607 | 2.8E-03 | 200231.065 | 200231.613 | -5.5E-01 | 200277.244 | 200277.792 | -5.5E-01 | 6.031 | 6.569 | -5.4E-01 |
| 2157 | Th | 215 | 90 | 125 | 35 | 7.693025 | 7.690459 | 2.6E-03 | 200235.424 | 200235.921 | -5.0E-01 | 200282.134 | 200282.631 | -5.0E-01 | 10.922 | 11.408 | -4.9E-01 |
| 2158 | Pa | 215 | 91 | 124 | 33 | 7.657074 | 7.657783 | -7.1E-04 | 200241.839 | 200241.631 | 2.1E-01 | 200289.082 | 200288.873 | 2.1E-01 | 17.869 | 17.650 | 2.2E-01 |
| 2159 | Bi | 216 | 83 | 133 | 50 | 7.743499 | 7.744208 | -7.1E-04 | 201165.582 | 201165.379 | 2.0E-01 | 201208.581 | 201208.378 | 2.0E-01 | 5.874 | 5.661 | 2.1E-01 |
| 2160 | Po | 216 | 84 | 132 | 48 | 7.758812 | 7.762654 | -3.8E-03 | 201160.963 | 201160.082 | 8.8E-01 | 201204.491 | 201203.610 | 8.8E-01 | 1.784 | 0.893 | 8.9E-01 |
| 2161 | At | 216 | 85 | 131 | 46 | 7.752998 | 7.756127 | -3.1E-03 | 201160.907 | 201160.180 | 7.3E-01 | 201204.964 | 201204.237 | 7.3E-01 | 2.258 | 1.520 | 7.4E-01 |
| 2162 | Rn | 216 | 86 | 130 | 44 | 7.758655 | 7.761534 | -2.9E-03 | 201158.373 | 201157.699 | 6.7E-01 | 201202.960 | 201202.286 | 6.7E-01 | 0.253 | -0.431 | 6.8E-01 |
| 2163 | Fr | 216 | 87 | 129 | 42 | 7.742449 | 7.744658 | -2.2E-03 | 201160.561 | 201160.031 | 5.3E-01 | 201205.678 | 201205.148 | 5.3E-01 | 2.971 | 2.431 | 5.4E-01 |
| 2164 | Ra | 216 | 88 | 128 | 40 | 7.737346 | 7.739148 | -1.8E-03 | 201160.350 | 201159.907 | 4.4E-01 | 201205.998 | 201205.555 | 4.4E-01 | 3.291 | 2.838 | 4.5E-01 |
| 2165 | Ac | 216 | 89 | 127 | 38 | 7.711255 | 7.712006 | -7.5E-04 | 201164.672 | 201164.456 | 2.2E-01 | 201210.851 | 201210.635 | 2.2E-01 | 8.145 | 7.918 | 2.3E-01 |
| 2166 | Th | 216 | 90 | 126 | 36 | 7.697661 | 7.695686 | 2.0E-03 | 201166.295 | 201166.666 | -3.7E-01 | 201213.005 | 201213.377 | -3.7E-01 | 10.299 | 10.660 | -3.6E-01 |
| 2167 | Pa | 216 | 91 | 125 | 34 | 7.659311 | 7.658172 | 1.1E-03 | 201173.264 | 201173.455 | -1.9E-01 | 201220.507 | 201220.697 | -1.9E-01 | 17.800 | 17.980 | -1.8E-01 |
| 2168 | Bi | 217 | 83 | 134 | 51 | 7.731848 | 7.732439 | -5.9E-04 | 202099.932 | 202099.754 | 1.8E-01 | 202142.931 | 202142.753 | 1.8E-01 | 8.730 | 8.542 | 1.9E-01 |
| 2169 | Po | 217 | 84 | 133 | 49 | 7.741352 | 7.744488 | -3.1E-03 | 202096.558 | 202095.827 | 7.3E-01 | 202140.086 | 202139.355 | 7.3E-01 | 5.885 | 5.144 | 7.4E-01 |
| 2170 | At | 217 | 85 | 132 | 47 | 7.744610 | 7.747936 | -3.3E-03 | 202094.539 | 202093.766 | 7.7E-01 | 202138.597 | 202137.824 | 7.7E-01 | 4.396 | 3.613 | 7.8E-01 |
| 2171 | Rn | 217 | 86 | 131 | 45 | 7.744401 | 7.747600 | -3.2E-03 | 202093.272 | 202092.526 | 7.5E-01 | 202137.860 | 202137.114 | 7.5E-01 | 3.659 | 2.902 | 7.6E-01 |
| 2172 | Fr | 217 | 87 | 130 | 43 | 7.737773 | 7.740380 | -2.6E-03 | 202093.398 | 202092.780 | 6.2E-01 | 202138.516 | 202137.897 | 6.2E-01 | 4.315 | 3.686 | 6.3E-01 |
| 2173 | Ra | 217 | 88 | 129 | 41 | 7.726920 | 7.729957 | -3.0E-03 | 202094.440 | 202093.728 | 7.1E-01 | 202140.089 | 202139.376 | 7.1E-01 | 5.888 | 5.165 | 7.2E-01 |
| 2174 | Ac | 217 | 89 | 128 | 39 | 7.710337 | 7.712590 | -2.3E-03 | 202096.725 | 202096.182 | 5.4E-01 | 202142.905 | 202142.362 | 5.4E-01 | 8.704 | 8.150 | 5.5E-01 |
| 2175 | Th | 217 | 90 | 127 | 37 | 7.690537 | 7.691684 | -1.1E-03 | 202099.708 | 202099.405 | 3.0E-01 | 202146.419 | 202146.115 | 3.0E-01 | 12.218 | 11.904 | 3.1E-01 |
| 2176 | Pa | 217 | 91 | 126 | 35 | 7.664573 | 7.663926 | 6.5E-04 | 202104.028 | 202104.113 | -8.5E-02 | 202151.271 | 202151.356 | -8.5E-02 | 17.070 | 17.144 | -7.4E-02 |
| 2177 | Bi | 218 | 83 | 135 | 52 | 7.712827 | 7.712960 | -1.3E-04 | 203035.912 | 203035.833 | 7.9E-02 | 203078.911 | 203078.832 | 7.9E-02 | 13.216 | 13.127 | 8.9E-02 |
| 2178 | Po | 218 | 84 | 134 | 50 | 7.731519 | 7.735296 | -3.8E-03 | 203030.526 | 203029.652 | 8.7E-01 | 203074.054 | 203073.180 | 8.7E-01 | 8.359 | 7.475 | 8.8E-01 |
| 2179 | At | 218 | 85 | 133 | 48 | 7.729122 | 7.732590 | -3.5E-03 | 203029.736 | 203028.929 | 8.1E-01 | 203073.794 | 203072.987 | 8.1E-01 | 8.099 | 7.281 | 8.2E-01 |
| 2180 | Rn | 218 | 86 | 132 | 46 | 7.738750 | 7.741739 | -3.0E-03 | 203026.325 | 203025.622 | 7.0E-01 | 203070.913 | 203070.209 | 7.0E-01 | 5.218 | 4.504 | 7.1E-01 |
| 2181 | Fr | 218 | 87 | 131 | 44 | 7.726713 | 7.729063 | -2.4E-03 | 203027.637 | 203027.072 | 5.7E-01 | 203072.754 | 203072.189 | 5.6E-01 | 7.059 | 6.484 | 5.8E-01 |
| 2182 | Ra | 218 | 88 | 130 | 42 | 7.724996 | 7.727923 | -2.9E-03 | 203026.698 | 203026.007 | 6.9E-01 | 203072.346 | 203071.655 | 6.9E-01 | 6.651 | 5.949 | 7.0E-01 |
| 2183 | Ac | 218 | 89 | 129 | 40 | 7.702175 | 7.705759 | -3.6E-03 | 203030.360 | 203029.524 | 8.4E-01 | 203076.539 | 203075.704 | 8.4E-01 | 10.844 | 9.998 | 8.5E-01 |
| 2184 | Th | 218 | 90 | 128 | 38 | 7.691601 | 7.694292 | -2.7E-03 | 203031.351 | 203030.710 | 6.4E-01 | 203078.062 | 203077.420 | 6.4E-01 | 12.367 | 11.715 | 6.5E-01 |
| 2185 | Pa | 218 | 91 | 127 | 36 | 7.659033 | 7.661922 | -2.9E-03 | 203037.137 | 203036.452 | 6.9E-01 | 203084.379 | 203083.694 | 6.9E-01 | 18.684 | 17.989 | 7.0E-01 |
| 2186 | U | 218 | 92 | 126 | 34 | 7.640639 | 7.640135 | 5.0E-04 | 203039.832 | 203039.886 | -5.4E-02 | 203087.607 | 203087.660 | -5.3E-02 | 21.912 | 21.955 | -4.3E-02 |
| 2187 | Po | 219 | 84 | 135 | 51 | 7.713333 | 7.716234 | -2.9E-03 | 203966.342 | 203965.656 | 6.9E-01 | 204009.870 | 204009.184 | 6.9E-01 | 12.681 | 11.985 | 7.0E-01 |
| 2188 | At | 219 | 85 | 134 | 49 | 7.720191 | 7.723424 | -3.2E-03 | 203963.529 | 203962.769 | 7.6E-01 | 204007.586 | 204006.827 | 7.6E-01 | 10.397 | 9.627 | 7.7E-01 |
| 2189 | Rn | 219 | 86 | 133 | 47 | 7.723771 | 7.726854 | -3.1E-03 | 203961.432 | 203960.705 | 7.3E-01 | 204006.020 | 204005.293 | 7.3E-01 | 8.831 | 8.093 | 7.4E-01 |
| 2190 | Fr | 219 | 87 | 132 | 45 | 7.721168 | 7.723411 | -2.2E-03 | 203960.690 | 203960.146 | 5.4E-01 | 204005.807 | 204005.264 | 5.4E-01 | 8.619 | 8.064 | 5.5E-01 |
| 2191 | Ra | 219 | 88 | 131 | 43 | 7.714052 | 7.717158 | -3.1E-03 | 203960.935 | 203960.202 | 7.3E-01 | 204006.583 | 204005.850 | 7.3E-01 | 9.395 | 8.650 | 7.4E-01 |
| 2192 | Ac | 219 | 89 | 130 | 41 | 7.700547 | 7.704204 | -3.7E-03 | 203962.580 | 203961.725 | 8.6E-01 | 204008.759 | 204007.904 | 8.6E-01 | 11.570 | 10.704 | 8.7E-01 |
| 2193 | Th | 219 | 90 | 129 | 39 | 7.683718 | 7.688099 | -4.4E-03 | 203964.951 | 203963.937 | 1.0E+00 | 204011.662 | 204010.648 | 1.0E+00 | 14.473 | 13.448 | 1.0E+00 |
| 2194 | Pa | 219 | 91 | 128 | 37 | 7.661573 | 7.665144 | -3.6E-03 | 203968.487 | 203967.649 | 8.4E-01 | 204015.729 | 204014.892 | 8.4E-01 | 18.540 | 17.692 | 8.5E-01 |
| 2195 | U | 219 | 92 | 127 | 35 | 7.636329 | 7.638727 | -2.4E-03 | 203972.701 | 203972.119 | 5.8E-01 | 204020.475 | 204019.894 | 5.8E-01 | 23.287 | 22.695 | 5.9E-01 |
| 2196 | Po | 220 | 84 | 136 | 52 | 7.703224 | 7.706094 | -2.9E-03 | 204900.418 | 204899.736 | 6.8E-01 | 204943.946 | 204943.264 | 6.8E-01 | 15.263 | 14.571 | 6.9E-01 |
| 2197 | At | 220 | 85 | 135 | 50 | 7.703703 | 7.707320 | -3.6E-03 | 204899.001 | 204898.154 | 8.5E-01 | 204943.059 | 204942.212 | 8.5E-01 | 14.376 | 13.518 | 8.6E-01 |
| 2198 | Rn | 220 | 86 | 134 | 48 | 7.717247 | 7.720108 | -2.9E-03 | 204894.709 | 204894.028 | 6.8E-01 | 204939.296 | 204938.615 | 6.8E-01 | 10.614 | 9.922 | 6.9E-01 |
| 2199 | Fr | 220 | 87 | 133 | 46 | 7.709738 | 7.711205 | -1.5E-03 | 204895.048 | 204894.673 | 3.7E-01 | 204940.166 | 204939.791 | 3.8E-01 | 11.483 | 11.097 | 3.9E-01 |
| 2200 | Ra | 220 | 88 | 132 | 44 | 7.711694 | 7.713790 | -2.1E-03 | 204893.305 | 204892.791 | 5.1E-01 | 204938.953 | 204938.439 | 5.1E-01 | 10.271 | 9.746 | 5.2E-01 |
| 2201 | Ac | 220 | 89 | 131 | 42 | 7.692349 | 7.695849 | -3.5E-03 | 204896.248 | 204895.424 | 8.2E-01 | 204942.427 | 204941.603 | 8.2E-01 | 13.744 | 12.909 | 8.3E-01 |
| 2202 | Th | 220 | 90 | 130 | 40 | 7.684588 | 7.688652 | -4.1E-03 | 204896.642 | 204895.693 | 9.5E-01 | 204943.352 | 204942.403 | 9.5E-01 | 14.669 | 13.710 | 9.6E-01 |
| 2203 | Po | 221 | 84 | 137 | 53 | 7.684481 | 7.685916 | -1.4E-03 | 205836.423 | 205836.055 | 3.7E-01 | 205879.951 | 205879.583 | 3.7E-01 | 19.774 | 19.396 | 3.8E-01 |
| 2204 | At | 221 | 85 | 136 | 51 | 7.694475 | 7.697052 | -2.6E-03 | 205832.902 | 205832.281 | 6.2E-01 | 205876.960 | 205876.339 | 6.2E-01 | 16.783 | 16.151 | 6.3E-01 |
| 2205 | Rn | 221 | 86 | 135 | 49 | 7.701387 | 7.704324 | -2.9E-03 | 205830.062 | 205829.361 | 7.0E-01 | 205874.650 | 205873.949 | 7.0E-01 | 14.473 | 13.761 | 7.1E-01 |
| 2206 | Fr | 221 | 87 | 134 | 47 | 7.703250 | 7.704498 | -1.2E-03 | 205828.338 | 205828.010 | 3.3E-01 | 205873.455 | 205873.127 | 3.3E-01 | 13.279 | 12.940 | 3.4E-01 |
| 2207 | Ra | 221 | 88 | 133 | 45 | 7.701133 | 7.701977 | -8.4E-04 | 205827.493 | 205827.253 | 2.4E-01 | 205873.141 | 205872.901 | 2.4E-01 | 12.964 | 12.714 | 2.5E-01 |
| 2208 | Ac | 221 | 89 | 132 | 43 | 7.690537 | 7.692808 | -2.3E-03 | 205828.521 | 205827.965 | 5.6E-01 | 205874.700 | 205874.145 | 5.6E-01 | 14.523 | 13.957 | 5.7E-01 |
| 2209 | Th | 221 | 90 | 131 | 41 | 7.676070 | 7.680853 | -4.8E-03 | 205830.405 | 205829.293 | 1.1E+00 | 205877.115 | 205876.003 | 1.1E+00 | 16.938 | 15.816 | 1.1E+00 |
| 2210 | Pa | 221 | 91 | 130 | 39 | 7.656974 | 7.662201 | -5.2E-03 | 205833.311 | 205832.100 | 1.2E+00 | 205880.553 | 205879.342 | 1.2E+00 | 20.376 | 19.155 | 1.2E+00 |
| 2211 | Po | 222 | 84 | 138 | 54 | 7.674005 | 7.675132 | -1.1E-03 | 206770.629 | 206770.328 | 3.0E-01 | 206814.157 | 206813.857 | 3.0E-01 | 22.486 | 22.175 | 3.1E-01 |
| 2212 | At | 222 | 85 | 137 | 52 | 7.677387 | 7.680003 | -2.6E-03 | 206768.566 | 206767.935 | 6.3E-01 | 206812.624 | 206811.992 | 6.3E-01 | 20.953 | 20.311 | 6.4E-01 |
| 2213 | Rn | 222 | 86 | 136 | 50 | 7.694489 | 7.696623 | -2.1E-03 | 206763.458 | 206762.932 | 5.3E-01 | 206808.045 | 206807.519 | 5.3E-01 | 16.374 | 15.838 | 5.4E-01 |
| 2214 | Fr | 222 | 87 | 135 | 48 | 7.691074 | 7.691516 | -4.4E-04 | 206762.903 | 206762.753 | 1.5E-01 | 206808.021 | 206807.870 | 1.5E-01 | 16.350 | 16.189 | 1.6E-01 |
| 2215 | Ra | 222 | 88 | 134 | 46 | 7.696686 | 7.697628 | -9.4E-04 | 206760.344 | 206760.082 | 2.6E-01 | 206805.992 | 206805.730 | 2.6E-01 | 14.322 | 14.049 | 2.7E-01 |
| 2216 | Ac | 222 | 89 | 133 | 44 | 7.682801 | 7.683446 | -6.4E-04 | 206762.113 | 206761.916 | 2.0E-01 | 206808.293 | 206808.096 | 2.0E-01 | 16.622 | 16.414 | 2.1E-01 |
| 2217 | Th | 222 | 90 | 132 | 42 | 7.676657 | 7.679948 | -3.3E-03 | 206762.164 | 206761.378 | 7.9E-01 | 206808.874 | 206808.089 | 7.9E-01 | 17.203 | 16.407 | 8.0E-01 |
| 2218 | At | 223 | 85 | 138 | 53 | 7.668055 | 7.668932 | -8.8E-04 | 207702.535 | 207702.289 | 2.5E-01 | 207746.593 | 207746.346 | 2.5E-01 | 23.428 | 23.171 | 2.6E-01 |
| 2219 | Rn | 223 | 86 | 137 | 51 | 7.678171 | 7.679773 | -1.6E-03 | 207698.967 | 207698.558 | 4.1E-01 | 207743.555 | 207743.146 | 4.1E-01 | 20.390 | 19.970 | 4.2E-01 |
| 2220 | Fr | 223 | 87 | 136 | 49 | 7.683657 | 7.683717 | -6.0E-05 | 207696.431 | 207696.366 | 6.5E-02 | 207741.549 | 207741.483 | 6.6E-02 | 18.384 | 18.308 | 7.6E-02 |
| 2221 | Ra | 223 | 88 | 135 | 47 | 7.685302 | 7.684905 | 4.0E-04 | 207694.752 | 207694.787 | -3.5E-02 | 207740.400 | 207740.435 | -3.5E-02 | 17.235 | 17.260 | -2.5E-02 |
| 2222 | Ac | 223 | 89 | 134 | 45 | 7.679140 | 7.679267 | -1.3E-04 | 207694.812 | 207694.730 | 8.2E-02 | 207740.992 | 207740.909 | 8.3E-02 | 17.827 | 17.734 | 9.3E-02 |
| 2223 | Th | 223 | 90 | 133 | 43 | 7.668639 | 7.671008 | -2.4E-03 | 207695.840 | 207695.257 | 5.8E-01 | 207742.551 | 207741.968 | 5.8E-01 | 19.386 | 18.792 | 5.9E-01 |
| 2224 | Pa | 223 | 91 | 132 | 41 | 7.651969 | 7.656124 | -4.2E-03 | 207698.244 | 207697.262 | 9.8E-01 | 207745.486 | 207744.504 | 9.8E-01 | 22.321 | 21.329 | 9.9E-01 |
| 2225 | U | 223 | 92 | 131 | 39 | 7.632686 | 7.638362 | -5.7E-03 | 207701.229 | 207699.907 | 1.3E+00 | 207749.004 | 207747.682 | 1.3E+00 | 25.839 | 24.506 | 1.3E+00 |
| 2226 | At | 224 | 85 | 139 | 54 | 7.650735 | 7.652064 | -1.3E-03 | 208638.312 | 208637.964 | 3.5E-01 | 208682.370 | 208682.021 | 3.5E-01 | 27.711 | 27.352 | 3.6E-01 |
| 2227 | Rn | 224 | 86 | 138 | 52 | 7.670751 | 7.671437 | -6.9E-04 | 208632.517 | 208632.311 | 2.1E-01 | 208677.104 | 208676.899 | 2.1E-01 | 22.445 | 22.229 | 2.2E-01 |
| 2228 | Fr | 224 | 87 | 137 | 50 | 7.670160 | 7.669838 | 3.2E-04 | 208631.336 | 208631.356 | -2.0E-02 | 208676.454 | 208676.474 | -2.0E-02 | 21.795 | 21.804 | -8.9E-03 |
| 2229 | Ra | 224 | 88 | 136 | 48 | 7.679916 | 7.679597 | 3.2E-04 | 208627.838 | 208627.857 | -1.9E-02 | 208673.486 | 208673.505 | -1.9E-02 | 18.827 | 18.835 | -7.7E-03 |
| 2230 | Ac | 224 | 89 | 135 | 46 | 7.670139 | 7.669083 | 1.1E-03 | 208628.715 | 208628.897 | -1.8E-01 | 208674.894 | 208675.077 | -1.8E-01 | 20.235 | 20.407 | -1.7E-01 |
| 2231 | Th | 224 | 90 | 134 | 44 | 7.667723 | 7.669012 | -1.3E-03 | 208627.942 | 208627.599 | 3.4E-01 | 208674.653 | 208674.309 | 3.4E-01 | 19.994 | 19.640 | 3.5E-01 |
| 2232 | Pa | 224 | 91 | 133 | 42 | 7.646959 | 7.649353 | -2.4E-03 | 208631.279 | 208630.688 | 5.9E-01 | 208678.522 | 208677.930 | 5.9E-01 | 23.863 | 23.260 | 6.0E-01 |
| 2233 | U | 224 | 92 | 132 | 40 | 7.635201 | 7.639848 | -4.6E-03 | 208632.598 | 208631.502 | 1.1E+00 | 208680.373 | 208679.276 | 1.1E+00 | 25.714 | 24.607 | 1.1E+00 |
| 2234 | Rn | 225 | 86 | 139 | 53 | 7.654357 | 7.654626 | -2.7E-04 | 209568.100 | 209567.988 | 1.1E-01 | 209612.687 | 209612.575 | 1.1E-01 | 26.534 | 26.411 | 1.2E-01 |
| 2235 | Fr | 225 | 87 | 138 | 51 | 7.662940 | 7.661279 | 1.7E-03 | 209564.856 | 209565.177 | -3.2E-01 | 209609.974 | 209610.295 | -3.2E-01 | 23.821 | 24.131 | -3.1E-01 |
| 2236 | Ra | 225 | 88 | 137 | 49 | 7.667580 | 7.665867 | 1.7E-03 | 209562.499 | 209562.832 | -3.3E-01 | 209608.147 | 209608.480 | -3.3E-01 | 21.994 | 22.316 | -3.2E-01 |
| 2237 | Ac | 225 | 89 | 136 | 47 | 7.665684 | 7.663818 | 1.9E-03 | 209561.612 | 209561.978 | -3.7E-01 | 209607.792 | 209608.158 | -3.7E-01 | 21.639 | 21.994 | -3.6E-01 |
| 2238 | Th | 225 | 90 | 135 | 45 | 7.659221 | 7.659124 | 9.7E-05 | 209561.753 | 209561.720 | 3.3E-02 | 209608.463 | 209608.431 | 3.2E-02 | 22.311 | 22.267 | 4.4E-02 |
| 2239 | Pa | 225 | 91 | 134 | 43 | 7.646720 | 7.647696 | -9.8E-04 | 209563.251 | 209562.977 | 2.7E-01 | 209610.494 | 209610.219 | 2.8E-01 | 24.341 | 24.055 | 2.9E-01 |
| 2240 | U | 225 | 92 | 133 | 41 | 7.629745 | 7.633596 | -3.9E-03 | 209565.756 | 209564.834 | 9.2E-01 | 209613.531 | 209612.608 | 9.2E-01 | 27.378 | 26.445 | 9.3E-01 |
| 2241 | Np | 225 | 93 | 132 | 39 | 7.607557 | 7.613027 | -5.5E-03 | 209569.434 | 209568.146 | 1.3E+00 | 209617.741 | 209616.453 | 1.3E+00 | 31.588 | 30.290 | 1.3E+00 |
| 2242 | Rn | 226 | 86 | 140 | 54 | 7.646410 | 7.646115 | 3.0E-04 | 210501.807 | 210501.822 | -1.5E-02 | 210546.394 | 210546.409 | -1.5E-02 | 28.747 | 28.751 | -4.3E-03 |
| 2243 | Fr | 226 | 87 | 139 | 52 | 7.648287 | 7.647631 | 6.6E-04 | 210500.070 | 210500.166 | -9.6E-02 | 210545.187 | 210545.284 | -9.7E-02 | 27.541 | 27.626 | -8.5E-02 |
| 2244 | Ra | 226 | 88 | 138 | 50 | 7.661954 | 7.659971 | 2.0E-03 | 210495.668 | 210496.063 | -4.0E-01 | 210541.317 | 210541.712 | -3.9E-01 | 23.670 | 24.054 | -3.8E-01 |
| 2245 | Ac | 226 | 89 | 137 | 48 | 7.655656 | 7.652769 | 2.9E-03 | 210495.778 | 210496.377 | -6.0E-01 | 210541.957 | 210542.556 | -6.0E-01 | 24.310 | 24.898 | -5.9E-01 |
| 2246 | Th | 226 | 90 | 136 | 46 | 7.657120 | 7.656141 | 9.8E-04 | 210494.134 | 210494.301 | -1.7E-01 | 210540.844 | 210541.011 | -1.7E-01 | 23.197 | 23.353 | -1.6E-01 |
| 2247 | Pa | 226 | 91 | 135 | 44 | 7.641109 | 7.640029 | 1.1E-03 | 210496.438 | 210496.627 | -1.9E-01 | 210543.680 | 210543.869 | -1.9E-01 | 26.033 | 26.212 | -1.8E-01 |
| 2248 | U | 226 | 92 | 134 | 42 | 7.631914 | 7.633874 | -2.0E-03 | 210497.202 | 210496.703 | 5.0E-01 | 210544.976 | 210544.477 | 5.0E-01 | 27.329 | 26.819 | 5.1E-01 |
| 2249 | Rn | 227 | 86 | 141 | 55 | 7.630050 | 7.628676 | 1.4E-03 | 211437.439 | 211437.700 | -2.6E-01 | 211482.027 | 211482.287 | -2.6E-01 | 32.886 | 33.135 | -2.5E-01 |
| 2250 | Fr | 227 | 87 | 140 | 53 | 7.640701 | 7.638736 | 2.0E-03 | 211433.709 | 211434.103 | -3.9E-01 | 211478.827 | 211479.220 | -3.9E-01 | 29.686 | 30.069 | -3.8E-01 |
| 2251 | Ra | 227 | 88 | 139 | 51 | 7.648295 | 7.646383 | 1.9E-03 | 211430.672 | 211431.053 | -3.8E-01 | 211476.320 | 211476.702 | -3.8E-01 | 27.179 | 27.550 | -3.7E-01 |
| 2252 | Ac | 227 | 89 | 138 | 49 | 7.650701 | 7.646814 | 3.9E-03 | 211428.813 | 211429.641 | -8.3E-01 | 211474.992 | 211475.821 | -8.3E-01 | 25.851 | 26.669 | -8.2E-01 |
| 2253 | Th | 227 | 90 | 137 | 47 | 7.647451 | 7.645285 | 2.2E-03 | 211428.237 | 211428.674 | -4.4E-01 | 211474.947 | 211475.385 | -4.4E-01 | 25.806 | 26.233 | -4.3E-01 |
| 2254 | Pa | 227 | 91 | 136 | 45 | 7.639487 | 7.637263 | 2.2E-03 | 211428.731 | 211429.180 | -4.5E-01 | 211475.973 | 211476.423 | -4.5E-01 | 26.832 | 27.271 | -4.4E-01 |
| 2255 | U | 227 | 92 | 135 | 43 | 7.626391 | 7.626615 | -2.2E-04 | 211430.389 | 211430.282 | 1.1E-01 | 211478.163 | 211478.057 | 1.1E-01 | 29.022 | 28.905 | 1.2E-01 |
| 2256 | Np | 227 | 93 | 134 | 41 | 7.607350 | 7.609458 | -2.1E-03 | 211433.396 | 211432.861 | 5.3E-01 | 211481.703 | 211481.168 | 5.3E-01 | 32.562 | 32.016 | 5.5E-01 |
| 2257 | Rn | 228 | 86 | 142 | 56 | 7.621645 | 7.619797 | 1.8E-03 | 212371.291 | 212371.661 | -3.7E-01 | 212415.878 | 212416.248 | -3.7E-01 | 35.243 | 35.602 | -3.6E-01 |
| 2258 | Fr | 228 | 87 | 141 | 54 | 7.626435 | 7.624572 | 1.9E-03 | 212368.886 | 212369.259 | -3.7E-01 | 212414.004 | 212414.376 | -3.7E-01 | 33.369 | 33.731 | -3.6E-01 |
| 2259 | Ra | 228 | 88 | 140 | 52 | 7.642419 | 7.640318 | 2.1E-03 | 212363.929 | 212364.355 | -4.3E-01 | 212409.577 | 212410.003 | -4.3E-01 | 28.942 | 29.357 | -4.2E-01 |
| 2260 | Ac | 228 | 89 | 139 | 50 | 7.639189 | 7.636107 | 3.1E-03 | 212363.352 | 212364.001 | -6.5E-01 | 212409.531 | 212410.180 | -6.5E-01 | 28.896 | 29.534 | -6.4E-01 |
| 2261 | Th | 228 | 90 | 138 | 48 | 7.645074 | 7.641760 | 3.3E-03 | 212360.697 | 212361.398 | -7.0E-01 | 212407.407 | 212408.108 | -7.0E-01 | 26.772 | 27.463 | -6.9E-01 |
| 2262 | Pa | 228 | 91 | 137 | 46 | 7.632203 | 7.628729 | 3.5E-03 | 212362.317 | 212363.054 | -7.4E-01 | 212409.559 | 212410.296 | -7.4E-01 | 28.924 | 29.650 | -7.3E-01 |
| 2263 | U | 228 | 92 | 136 | 44 | 7.627465 | 7.625844 | 1.6E-03 | 212362.083 | 212362.397 | -3.1E-01 | 212409.857 | 212410.171 | -3.1E-01 | 29.222 | 29.525 | -3.0E-01 |
| 2264 | Np | 228 | 93 | 135 | 42 | 7.604851 | 7.604105 | 7.5E-04 | 212365.924 | 212366.037 | -1.1E-01 | 212414.231 | 212414.344 | -1.1E-01 | 33.596 | 33.699 | -1.0E-01 |
| 2265 | Pu | 228 | 94 | 134 | 40 | 7.590529 | 7.592076 | -1.5E-03 | 212367.874 | 212367.464 | 4.1E-01 | 212416.714 | 212416.304 | 4.1E-01 | 36.079 | 35.658 | 4.2E-01 |
| 2266 | Rn | 229 | 86 | 143 | 57 | 7.605622 | 7.602052 | 3.6E-03 | 213306.904 | 213307.670 | -7.7E-01 | 213351.491 | 213352.258 | -7.7E-01 | 39.362 | 40.118 | -7.6E-01 |
| 2267 | Fr | 229 | 87 | 142 | 55 | 7.618311 | 7.615088 | 3.2E-03 | 213302.686 | 213303.372 | -6.9E-01 | 213347.803 | 213348.489 | -6.9E-01 | 35.674 | 36.349 | -6.7E-01 |
| 2268 | Ra | 229 | 88 | 141 | 53 | 7.628544 | 7.626019 | 2.5E-03 | 213299.029 | 213299.555 | -5.3E-01 | 213344.677 | 213345.203 | -5.3E-01 | 32.549 | 33.063 | -5.1E-01 |
| 2269 | Ac | 229 | 89 | 140 | 51 | 7.633208 | 7.629872 | 3.3E-03 | 213296.648 | 213297.358 | -7.1E-01 | 213342.827 | 213343.538 | -7.1E-01 | 30.698 | 31.398 | -7.0E-01 |
| 2270 | Th | 229 | 90 | 139 | 49 | 7.634645 | 7.631194 | 3.5E-03 | 213295.005 | 213295.741 | -7.4E-01 | 213341.716 | 213342.452 | -7.4E-01 | 29.587 | 30.312 | -7.3E-01 |
| 2271 | Pa | 229 | 91 | 138 | 47 | 7.629868 | 7.625327 | 4.5E-03 | 213294.785 | 213295.770 | -9.8E-01 | 213342.027 | 213343.012 | -9.9E-01 | 29.898 | 30.872 | -9.7E-01 |
| 2272 | U | 229 | 92 | 137 | 45 | 7.620720 | 7.617622 | 3.1E-03 | 213295.565 | 213296.219 | -6.5E-01 | 213343.340 | 213343.994 | -6.5E-01 | 31.211 | 31.854 | -6.4E-01 |
| 2273 | Np | 229 | 93 | 136 | 43 | 7.606085 | 7.603742 | 2.3E-03 | 213297.602 | 213298.082 | -4.8E-01 | 213345.909 | 213346.389 | -4.8E-01 | 33.780 | 34.249 | -4.7E-01 |
| 2274 | Pu | 229 | 94 | 135 | 41 | 7.586889 | 7.587283 | -3.9E-04 | 213300.682 | 213300.535 | 1.5E-01 | 213349.523 | 213349.375 | 1.5E-01 | 37.394 | 37.235 | 1.6E-01 |
| 2275 | Fr | 230 | 87 | 143 | 56 | 7.603601 | 7.600752 | 2.8E-03 | 214238.016 | 214238.619 | -6.0E-01 | 214283.134 | 214283.737 | -6.0E-01 | 39.511 | 40.103 | -5.9E-01 |
| 2276 | Ra | 230 | 88 | 142 | 54 | 7.621914 | 7.619462 | 2.5E-03 | 214232.491 | 214233.002 | -5.1E-01 | 214278.139 | 214278.650 | -5.1E-01 | 34.516 | 35.016 | -5.0E-01 |
| 2277 | Ac | 230 | 89 | 141 | 52 | 7.621460 | 7.618562 | 2.9E-03 | 214231.282 | 214231.895 | -6.1E-01 | 214277.461 | 214278.074 | -6.1E-01 | 33.838 | 34.440 | -6.0E-01 |
| 2278 | Th | 230 | 90 | 140 | 50 | 7.630989 | 7.627561 | 3.4E-03 | 214227.777 | 214228.511 | -7.3E-01 | 214274.487 | 214275.222 | -7.3E-01 | 30.864 | 31.587 | -7.2E-01 |
| 2279 | Pa | 230 | 91 | 139 | 48 | 7.621890 | 7.617267 | 4.6E-03 | 214228.555 | 214229.564 | -1.0E+00 | 214275.798 | 214276.806 | -1.0E+00 | 32.175 | 33.172 | -1.0E+00 |
| 2280 | U | 230 | 92 | 138 | 46 | 7.620922 | 7.616323 | 4.6E-03 | 214227.463 | 214228.465 | -1.0E+00 | 214275.238 | 214276.240 | -1.0E+00 | 31.615 | 32.606 | -9.9E-01 |
| 2281 | Np | 230 | 93 | 137 | 44 | 7.601775 | 7.597488 | 4.3E-03 | 214230.552 | 214231.482 | -9.3E-01 | 214278.859 | 214279.789 | -9.3E-01 | 35.236 | 36.155 | -9.2E-01 |
| 2282 | Pu | 230 | 94 | 136 | 42 | 7.590993 | 7.588618 | 2.4E-03 | 214231.717 | 214232.206 | -4.9E-01 | 214280.557 | 214281.046 | -4.9E-01 | 36.934 | 37.412 | -4.8E-01 |
| 2283 | Fr | 231 | 87 | 144 | 57 | 7.594570 | 7.590825 | 3.7E-03 | 215172.064 | 215172.877 | -8.1E-01 | 215217.181 | 215217.995 | -8.1E-01 | 42.064 | 42.866 | -8.0E-01 |
| 2284 | Ra | 231 | 88 | 143 | 55 | 7.607841 | 7.604774 | 3.1E-03 | 215167.685 | 215168.341 | -6.6E-01 | 215213.333 | 215213.989 | -6.6E-01 | 38.216 | 38.861 | -6.4E-01 |
| 2285 | Ac | 231 | 89 | 142 | 53 | 7.615076 | 7.611626 | 3.5E-03 | 215164.701 | 215165.444 | -7.4E-01 | 215210.880 | 215211.623 | -7.4E-01 | 35.763 | 36.495 | -7.3E-01 |
| 2286 | Th | 231 | 90 | 141 | 51 | 7.620111 | 7.616225 | 3.9E-03 | 215162.224 | 215163.067 | -8.4E-01 | 215208.935 | 215209.778 | -8.4E-01 | 33.818 | 34.650 | -8.3E-01 |
| 2287 | Pa | 231 | 91 | 140 | 49 | 7.618419 | 7.613677 | 4.7E-03 | 215161.301 | 215162.341 | -1.0E+00 | 215208.543 | 215209.583 | -1.0E+00 | 33.426 | 34.455 | -1.0E+00 |
| 2288 | U | 231 | 92 | 139 | 47 | 7.613380 | 7.608535 | 4.8E-03 | 215161.150 | 215162.214 | -1.1E+00 | 215208.925 | 215209.988 | -1.1E+00 | 33.808 | 34.860 | -1.1E+00 |
| 2289 | Np | 231 | 93 | 138 | 45 | 7.602124 | 7.596502 | 5.6E-03 | 215162.435 | 215163.678 | -1.2E+00 | 215210.742 | 215211.985 | -1.2E+00 | 35.625 | 36.857 | -1.2E+00 |
| 2290 | Pu | 231 | 94 | 137 | 43 | 7.587220 | 7.582836 | 4.4E-03 | 215164.563 | 215165.518 | -9.6E-01 | 215213.403 | 215214.358 | -9.6E-01 | 38.286 | 39.230 | -9.4E-01 |
| 2291 | Ra | 232 | 88 | 144 | 56 | 7.600009 | 7.597914 | 2.1E-03 | 216101.460 | 216101.893 | -4.3E-01 | 216147.108 | 216147.541 | -4.3E-01 | 40.497 | 40.919 | -4.2E-01 |
| 2292 | Ac | 232 | 89 | 143 | 54 | 7.602424 | 7.600013 | 2.4E-03 | 216099.586 | 216100.092 | -5.1E-01 | 216145.765 | 216146.271 | -5.1E-01 | 39.154 | 39.649 | -4.9E-01 |
| 2293 | Th | 232 | 90 | 142 | 52 | 7.615024 | 7.611980 | 3.0E-03 | 216095.349 | 216096.001 | -6.5E-01 | 216142.060 | 216142.712 | -6.5E-01 | 35.449 | 36.090 | -6.4E-01 |
| 2294 | Pa | 232 | 91 | 141 | 50 | 7.609500 | 7.604943 | 4.6E-03 | 216095.317 | 216096.319 | -1.0E+00 | 216142.559 | 216143.561 | -1.0E+00 | 35.948 | 36.939 | -9.9E-01 |
| 2295 | U | 232 | 92 | 140 | 48 | 7.611891 | 7.607204 | 4.7E-03 | 216093.447 | 216094.479 | -1.0E+00 | 216141.222 | 216142.254 | -1.0E+00 | 34.611 | 35.632 | -1.0E+00 |
| 2296 | Pu | 232 | 94 | 138 | 44 | 7.588973 | 7.583614 | 5.4E-03 | 216096.134 | 216097.320 | -1.2E+00 | 216144.974 | 216146.160 | -1.2E+00 | 38.363 | 39.538 | -1.2E+00 |
| 2297 | Ra | 233 | 88 | 145 | 57 | 7.585614 | 7.582744 | 2.9E-03 | 217036.779 | 217037.395 | -6.2E-01 | 217082.427 | 217083.043 | -6.2E-01 | 44.322 | 44.927 | -6.0E-01 |
| 2298 | Ac | 233 | 89 | 144 | 55 | 7.595193 | 7.592567 | 2.6E-03 | 217033.234 | 217033.792 | -5.6E-01 | 217079.413 | 217079.972 | -5.6E-01 | 41.308 | 41.855 | -5.5E-01 |
| 2299 | Th | 233 | 90 | 143 | 53 | 7.602884 | 7.600130 | 2.8E-03 | 217030.128 | 217030.716 | -5.9E-01 | 217076.839 | 217077.426 | -5.9E-01 | 38.734 | 39.310 | -5.8E-01 |
| 2300 | Pa | 233 | 91 | 142 | 51 | 7.604864 | 7.600560 | 4.3E-03 | 217028.353 | 217029.301 | -9.5E-01 | 217075.595 | 217076.543 | -9.5E-01 | 37.490 | 38.427 | -9.4E-01 |
| 2301 | U | 233 | 92 | 141 | 49 | 7.603952 | 7.598603 | 5.3E-03 | 217027.251 | 217028.441 | -1.2E+00 | 217075.025 | 217076.216 | -1.2E+00 | 36.920 | 38.100 | -1.2E+00 |
| 2302 | Np | 233 | 93 | 140 | 47 | 7.596175 | 7.589739 | 6.4E-03 | 217027.748 | 217029.191 | -1.4E+00 | 217076.055 | 217077.498 | -1.4E+00 | 37.950 | 39.382 | -1.4E+00 |
| 2303 | Pu | 233 | 94 | 139 | 45 | 7.583795 | 7.578365 | 5.4E-03 | 217029.317 | 217030.525 | -1.2E+00 | 217078.157 | 217079.365 | -1.2E+00 | 40.052 | 41.249 | -1.2E+00 |
| 2304 | Cm | 233 | 96 | 137 | 41 | 7.546007 | 7.541317 | 4.7E-03 | 217035.489 | 217036.524 | -1.0E+00 | 217085.397 | 217086.431 | -1.0E+00 | 47.292 | 48.315 | -1.0E+00 |
| 2305 | Ra | 234 | 88 | 146 | 58 | 7.576703 | 7.575089 | 1.6E-03 | 217970.844 | 217971.169 | -3.3E-01 | 218016.492 | 218016.817 | -3.3E-01 | 46.893 | 47.207 | -3.1E-01 |
| 2306 | Ac | 234 | 89 | 145 | 56 | 7.582129 | 7.580627 | 1.5E-03 | 217968.261 | 217968.559 | -3.0E-01 | 218014.440 | 218014.738 | -3.0E-01 | 44.841 | 45.128 | -2.9E-01 |
| 2307 | Th | 234 | 90 | 144 | 54 | 7.596849 | 7.595456 | 1.4E-03 | 217963.503 | 217963.775 | -2.7E-01 | 218010.213 | 218010.485 | -2.7E-01 | 40.614 | 40.875 | -2.6E-01 |
| 2308 | Pa | 234 | 91 | 143 | 52 | 7.594677 | 7.591380 | 3.3E-03 | 217962.697 | 217963.414 | -7.2E-01 | 218009.939 | 218010.656 | -7.2E-01 | 40.340 | 41.046 | -7.1E-01 |
| 2309 | U | 234 | 92 | 142 | 50 | 7.600707 | 7.596555 | 4.2E-03 | 217959.971 | 217960.887 | -9.2E-01 | 218007.746 | 218008.662 | -9.2E-01 | 38.147 | 39.052 | -9.0E-01 |
| 2310 | Np | 234 | 93 | 141 | 48 | 7.589630 | 7.583328 | 6.3E-03 | 217961.249 | 217962.667 | -1.4E+00 | 218009.556 | 218010.974 | -1.4E+00 | 39.957 | 41.364 | -1.4E+00 |
| 2311 | Pu | 234 | 94 | 140 | 46 | 7.584606 | 7.579162 | 5.4E-03 | 217961.109 | 217962.326 | -1.2E+00 | 218009.949 | 218011.166 | -1.2E+00 | 40.350 | 41.556 | -1.2E+00 |
| 2312 | Cm | 234 | 96 | 138 | 42 | 7.550679 | 7.544052 | 6.6E-03 | 217966.415 | 217967.908 | -1.5E+00 | 218016.323 | 218017.815 | -1.5E+00 | 46.724 | 48.205 | -1.5E+00 |
| 2313 | Ac | 235 | 89 | 146 | 57 | 7.573504 | 7.572312 | 1.2E-03 | 218902.271 | 218902.498 | -2.3E-01 | 218948.450 | 218948.677 | -2.3E-01 | 47.357 | 47.573 | -2.2E-01 |
| 2314 | Th | 235 | 90 | 145 | 55 | 7.584385 | 7.583090 | 1.3E-03 | 218898.400 | 218898.651 | -2.5E-01 | 218945.111 | 218945.361 | -2.5E-01 | 44.018 | 44.257 | -2.4E-01 |
| 2315 | Pa | 235 | 91 | 144 | 53 | 7.588413 | 7.586358 | 2.1E-03 | 218896.140 | 218896.568 | -4.3E-01 | 218943.382 | 218943.810 | -4.3E-01 | 42.289 | 42.706 | -4.2E-01 |
| 2316 | U | 235 | 92 | 143 | 51 | 7.590906 | 7.587322 | 3.6E-03 | 218894.239 | 218895.026 | -7.9E-01 | 218942.014 | 218942.801 | -7.9E-01 | 40.921 | 41.696 | -7.8E-01 |
| 2317 | Np | 235 | 93 | 142 | 49 | 7.587049 | 7.581368 | 5.7E-03 | 218893.831 | 218895.109 | -1.3E+00 | 218942.138 | 218943.417 | -1.3E+00 | 41.045 | 42.312 | -1.3E+00 |
| 2318 | Pu | 235 | 94 | 141 | 47 | 7.578873 | 7.573060 | 5.8E-03 | 218894.437 | 218895.746 | -1.3E+00 | 218943.277 | 218944.586 | -1.3E+00 | 42.184 | 43.482 | -1.3E+00 |
| 2319 | Am | 235 | 95 | 140 | 45 | 7.565150 | 7.558162 | 7.0E-03 | 218896.346 | 218897.930 | -1.6E+00 | 218945.719 | 218947.304 | -1.6E+00 | 44.626 | 46.200 | -1.6E+00 |
| 2320 | Ac | 236 | 89 | 147 | 58 | 7.559242 | 7.559265 | -2.3E-05 | 219837.629 | 219837.570 | 5.9E-02 | 219883.808 | 219883.749 | 5.9E-02 | 51.221 | 51.151 | 7.0E-02 |
| 2321 | Th | 236 | 90 | 146 | 56 | 7.576968 | 7.577727 | -7.6E-04 | 219832.132 | 219831.899 | 2.3E-01 | 219878.842 | 219878.609 | 2.3E-01 | 46.255 | 46.011 | 2.4E-01 |
| 2322 | Pa | 236 | 91 | 145 | 54 | 7.577557 | 7.576747 | 8.1E-04 | 219830.679 | 219830.815 | -1.4E-01 | 219877.921 | 219878.057 | -1.4E-01 | 45.334 | 45.459 | -1.2E-01 |
| 2323 | U | 236 | 92 | 144 | 52 | 7.586476 | 7.584696 | 1.8E-03 | 219827.259 | 219827.624 | -3.6E-01 | 219875.034 | 219875.398 | -3.6E-01 | 42.447 | 42.800 | -3.5E-01 |
| 2324 | Np | 236 | 93 | 143 | 50 | 7.579208 | 7.574383 | 4.8E-03 | 219827.659 | 219828.742 | -1.1E+00 | 219875.967 | 219877.049 | -1.1E+00 | 43.379 | 44.451 | -1.1E+00 |
| 2325 | Pu | 236 | 94 | 142 | 48 | 7.577913 | 7.573045 | 4.9E-03 | 219826.650 | 219827.742 | -1.1E+00 | 219875.490 | 219876.582 | -1.1E+00 | 42.903 | 43.983 | -1.1E+00 |
| 2326 | Cm | 236 | 96 | 140 | 44 | 7.550299 | 7.543880 | 6.4E-03 | 219830.534 | 219831.991 | -1.5E+00 | 219880.442 | 219881.899 | -1.5E+00 | 47.855 | 49.300 | -1.4E+00 |
| 2327 | Th | 237 | 90 | 147 | 57 | 7.563443 | 7.564219 | -7.8E-04 | 220767.326 | 220767.088 | 2.4E-01 | 220814.036 | 220813.798 | 2.4E-01 | 49.955 | 49.706 | 2.5E-01 |
| 2328 | Pa | 237 | 91 | 146 | 55 | 7.570384 | 7.570879 | -4.9E-04 | 220764.366 | 220764.194 | 1.7E-01 | 220811.609 | 220811.437 | 1.7E-01 | 47.528 | 47.344 | 1.8E-01 |
| 2329 | U | 237 | 92 | 145 | 53 | 7.576094 | 7.574834 | 1.3E-03 | 220761.699 | 220761.942 | -2.4E-01 | 220809.473 | 220809.716 | -2.4E-01 | 45.392 | 45.624 | -2.3E-01 |
| 2330 | Np | 237 | 93 | 144 | 51 | 7.574981 | 7.571654 | 3.3E-03 | 220760.647 | 220761.380 | -7.3E-01 | 220808.955 | 220809.687 | -7.3E-01 | 44.873 | 45.594 | -7.2E-01 |
| 2331 | Pu | 237 | 94 | 143 | 49 | 7.570752 | 7.566211 | 4.5E-03 | 220760.334 | 220761.354 | -1.0E+00 | 220809.175 | 220810.194 | -1.0E+00 | 45.093 | 46.101 | -1.0E+00 |
| 2332 | Cm | 237 | 96 | 141 | 45 | 7.546622 | 7.540061 | 6.6E-03 | 220763.421 | 220764.918 | -1.5E+00 | 220813.329 | 220814.825 | -1.5E+00 | 49.247 | 50.733 | -1.5E+00 |
| 2333 | Cf | 237 | 98 | 139 | 41 | 7.503356 | 7.497699 | 5.7E-03 | 220771.041 | 220772.322 | -1.3E+00 | 220822.018 | 220823.299 | -1.3E+00 | 57.937 | 59.206 | -1.3E+00 |
| 2334 | Pa | 238 | 91 | 147 | 56 | 7.558344 | 7.560338 | -2.0E-03 | 221699.227 | 221698.698 | 5.3E-01 | 221746.469 | 221745.940 | 5.3E-01 | 50.894 | 50.353 | 5.4E-01 |
| 2335 | U | 238 | 92 | 146 | 54 | 7.570120 | 7.571467 | -1.3E-03 | 221695.110 | 221694.734 | 3.8E-01 | 221742.884 | 221742.508 | 3.8E-01 | 47.309 | 46.922 | 3.9E-01 |
| 2336 | Np | 238 | 93 | 145 | 52 | 7.566213 | 7.564101 | 2.1E-03 | 221694.724 | 221695.171 | -4.5E-01 | 221743.032 | 221743.478 | -4.5E-01 | 47.456 | 47.892 | -4.4E-01 |
| 2337 | Pu | 238 | 94 | 144 | 50 | 7.568353 | 7.565476 | 2.9E-03 | 221692.900 | 221693.528 | -6.3E-01 | 221741.740 | 221742.368 | -6.3E-01 | 46.165 | 46.781 | -6.2E-01 |
| 2338 | Am | 238 | 95 | 143 | 48 | 7.555577 | 7.549167 | 6.4E-03 | 221694.625 | 221696.093 | -1.5E+00 | 221743.998 | 221745.466 | -1.5E+00 | 48.423 | 49.880 | -1.5E+00 |
| 2339 | Cm | 238 | 96 | 142 | 46 | 7.547997 | 7.541909 | 6.1E-03 | 221695.112 | 221696.503 | -1.4E+00 | 221745.020 | 221746.411 | -1.4E+00 | 49.445 | 50.824 | -1.4E+00 |
| 2340 | U | 239 | 92 | 147 | 55 | 7.558556 | 7.560561 | -2.0E-03 | 222629.869 | 222629.334 | 5.3E-01 | 222677.643 | 222677.109 | 5.3E-01 | 50.574 | 50.028 | 5.5E-01 |
| 2341 | Np | 239 | 93 | 146 | 53 | 7.560561 | 7.560458 | 1.0E-04 | 222628.074 | 222628.043 | 3.1E-02 | 222676.382 | 222676.350 | 3.2E-02 | 49.313 | 49.270 | 4.3E-02 |
| 2342 | Pu | 239 | 94 | 145 | 51 | 7.560310 | 7.557897 | 2.4E-03 | 222626.819 | 222627.339 | -5.2E-01 | 222675.659 | 222676.179 | -5.2E-01 | 48.590 | 49.099 | -5.1E-01 |
| 2343 | Am | 239 | 95 | 144 | 49 | 7.553681 | 7.548566 | 5.1E-03 | 222627.088 | 222628.252 | -1.2E+00 | 222676.461 | 222677.626 | -1.2E+00 | 49.392 | 50.546 | -1.2E+00 |

| | | | | | | | | | | | | | | | | |
|---|---|---|---|---|---|---|---|---|---|---|---|---|---|---|---|---|
| 2344 | Cm | 239 | 96 | 143 | 47 | 7.543059 | 7.537271 | 5.8E-03 | 222628.310 | 222629.635 | -1.3E+00 | 222678.218 | 222679.543 | -1.3E+00 | 51.148 | 52.462 | -1.3E+00 |
| 2345 | U | 240 | 92 | 148 | 56 | 7.551766 | 7.555772 | -4.0E-03 | 223563.505 | 223562.488 | 1.0E+00 | 223611.279 | 223610.263 | 1.0E+00 | 52.716 | 51.688 | 1.0E+00 |
| 2346 | Np | 240 | 93 | 147 | 54 | 7.550167 | 7.551995 | -1.8E-03 | 223562.574 | 223562.079 | 4.9E-01 | 223610.881 | 223610.386 | 4.9E-01 | 52.318 | 51.812 | 5.1E-01 |
| 2347 | Pu | 240 | 94 | 146 | 52 | 7.556035 | 7.556324 | -2.9E-04 | 223559.850 | 223559.724 | 1.3E-01 | 223608.690 | 223608.564 | 1.3E-01 | 50.127 | 49.990 | 1.4E-01 |
| 2348 | Am | 240 | 95 | 145 | 50 | 7.547005 | 7.542888 | 4.1E-03 | 223560.701 | 223561.632 | -9.3E-01 | 223610.075 | 223611.006 | -9.3E-01 | 51.512 | 52.431 | -9.2E-01 |
| 2349 | Cm | 240 | 96 | 144 | 48 | 7.542855 | 7.538276 | 4.6E-03 | 223560.381 | 223561.422 | -1.0E+00 | 223610.289 | 223611.330 | -1.0E+00 | 51.726 | 52.755 | -1.0E+00 |
| 2350 | Cf | 240 | 98 | 142 | 44 | 7.510230 | 7.504091 | 6.1E-03 | 223565.577 | 223566.991 | -1.4E+00 | 223616.554 | 223617.968 | -1.4E+00 | 57.991 | 59.393 | -1.4E+00 |
| 2351 | Np | 241 | 93 | 148 | 55 | 7.544262 | 7.546865 | -2.6E-03 | 224496.012 | 224495.329 | 6.8E-01 | 224544.319 | 224543.636 | 6.8E-01 | 54.262 | 53.567 | 6.9E-01 |
| 2352 | Pu | 241 | 94 | 147 | 53 | 7.546431 | 7.547701 | -1.3E-03 | 224494.174 | 224493.811 | 3.6E-01 | 224543.014 | 224542.651 | 3.6E-01 | 52.957 | 52.583 | 3.7E-01 |
| 2353 | Am | 241 | 95 | 146 | 51 | 7.543271 | 7.541289 | 2.0E-03 | 224493.620 | 224494.040 | -4.2E-01 | 224542.993 | 224543.414 | -4.2E-01 | 52.936 | 53.345 | -4.1E-01 |
| 2354 | Cm | 241 | 96 | 145 | 49 | 7.536841 | 7.532794 | 4.0E-03 | 224493.853 | 224494.770 | -9.2E-01 | 224543.761 | 224544.678 | -9.2E-01 | 53.704 | 54.609 | -9.1E-01 |
| 2355 | Np | 242 | 93 | 149 | 56 | 7.533396 | 7.536708 | -3.3E-03 | 225430.663 | 225429.805 | 8.6E-01 | 225478.970 | 225478.112 | 8.6E-01 | 57.419 | 56.550 | 8.7E-01 |
| 2356 | Pu | 242 | 94 | 148 | 54 | 7.541321 | 7.544818 | -3.5E-03 | 225427.429 | 225426.526 | 9.0E-01 | 225476.270 | 225475.367 | 9.0E-01 | 54.719 | 53.804 | 9.1E-01 |
| 2357 | Am | 242 | 95 | 147 | 52 | 7.534983 | 7.534668 | 3.2E-04 | 225427.647 | 225427.666 | -1.9E-02 | 225477.021 | 225477.040 | -1.9E-02 | 55.470 | 55.477 | -7.5E-03 |
| 2358 | Cm | 242 | 96 | 146 | 50 | 7.534496 | 7.532863 | 1.6E-03 | 225426.449 | 225426.786 | -3.4E-01 | 225476.357 | 225476.694 | -3.4E-01 | 54.805 | 55.131 | -3.3E-01 |
| 2359 | Cf | 242 | 98 | 144 | 46 | 7.509099 | 7.504075 | 5.0E-03 | 225429.961 | 225431.118 | -1.2E+00 | 225480.938 | 225482.094 | -1.2E+00 | 59.387 | 60.532 | -1.1E+00 |
| 2360 | Pu | 243 | 94 | 149 | 55 | 7.531002 | 7.534489 | -3.5E-03 | 226361.961 | 226361.057 | 9.0E-01 | 226410.801 | 226409.897 | 9.0E-01 | 57.756 | 56.840 | 9.2E-01 |
| 2361 | Am | 243 | 95 | 148 | 53 | 7.530168 | 7.531668 | -1.5E-03 | 226360.848 | 226360.426 | 4.2E-01 | 226410.221 | 226409.800 | 4.2E-01 | 57.176 | 56.743 | 4.3E-01 |
| 2362 | Cm | 243 | 96 | 147 | 51 | 7.526918 | 7.526315 | 6.0E-04 | 226360.321 | 226360.410 | -8.9E-02 | 226410.229 | 226410.317 | -8.8E-02 | 57.184 | 57.261 | -7.7E-02 |
| 2363 | Bk | 243 | 97 | 146 | 49 | 7.517494 | 7.514424 | 3.1E-03 | 226361.295 | 226361.982 | -6.9E-01 | 226411.737 | 226412.424 | -6.9E-01 | 58.691 | 59.367 | -6.8E-01 |
| 2364 | Pu | 244 | 94 | 150 | 56 | 7.524811 | 7.529524 | -4.7E-03 | 227295.506 | 227294.299 | 1.2E+00 | 227344.346 | 227343.140 | 1.2E+00 | 59.807 | 58.589 | 1.2E+00 |
| 2365 | Am | 244 | 95 | 149 | 54 | 7.521300 | 7.523558 | -2.3E-03 | 227295.047 | 227294.439 | 6.1E-01 | 227344.420 | 227343.812 | 6.1E-01 | 59.881 | 59.261 | 6.2E-01 |
| 2366 | Cm | 244 | 96 | 148 | 52 | 7.523944 | 7.525122 | -1.2E-03 | 227293.085 | 227292.740 | 3.4E-01 | 227342.993 | 227342.648 | 3.5E-01 | 58.454 | 58.097 | 3.6E-01 |
| 2367 | Bk | 244 | 97 | 147 | 50 | 7.511468 | 7.509459 | 2.0E-03 | 227294.813 | 227295.244 | -4.3E-01 | 227345.255 | 227345.686 | -4.3E-01 | 60.716 | 61.135 | -4.2E-01 |
| 2368 | Cf | 244 | 98 | 146 | 48 | 7.505131 | 7.502117 | 3.0E-03 | 227295.042 | 227295.718 | -6.8E-01 | 227346.019 | 227346.695 | -6.8E-01 | 61.479 | 62.144 | -6.6E-01 |
| 2369 | Pu | 245 | 94 | 151 | 57 | 7.513276 | 7.516689 | -3.4E-03 | 228230.373 | 228229.480 | 8.9E-01 | 228279.213 | 228278.320 | 8.9E-01 | 63.180 | 62.275 | 9.0E-01 |
| 2370 | Am | 245 | 95 | 150 | 55 | 7.515296 | 7.518507 | -3.2E-03 | 228228.562 | 228227.718 | 8.4E-01 | 228277.935 | 228277.092 | 8.4E-01 | 61.902 | 61.047 | 8.6E-01 |
| 2371 | Cm | 245 | 96 | 149 | 53 | 7.515765 | 7.517045 | -1.3E-03 | 228227.130 | 228226.759 | 3.7E-01 | 228277.038 | 228276.667 | 3.7E-01 | 61.005 | 60.622 | 3.8E-01 |
| 2372 | Bk | 245 | 97 | 148 | 51 | 7.509263 | 7.508404 | 8.6E-04 | 228227.407 | 228227.559 | -1.5E-01 | 228277.849 | 228278.001 | -1.5E-01 | 61.816 | 61.956 | -1.4E-01 |
| 2373 | Cf | 245 | 98 | 147 | 49 | 7.499656 | 7.497481 | 2.2E-03 | 228228.443 | 228228.917 | -4.7E-01 | 228279.420 | 228279.894 | -4.7E-01 | 63.387 | 63.849 | -4.6E-01 |
| 2374 | Pu | 246 | 94 | 152 | 58 | 7.506534 | 7.508973 | -2.4E-03 | 229164.083 | 229163.427 | 6.6E-01 | 229212.923 | 229212.267 | 6.6E-01 | 65.396 | 64.728 | 6.7E-01 |
| 2375 | Cm | 246 | 96 | 150 | 54 | 7.511464 | 7.514054 | -2.6E-03 | 229160.238 | 229159.543 | 6.9E-01 | 229210.146 | 229209.451 | 7.0E-01 | 62.619 | 61.912 | 7.1E-01 |
| 2376 | Bk | 246 | 97 | 149 | 52 | 7.502796 | 7.502085 | 7.1E-04 | 229161.054 | 229161.170 | -1.2E-01 | 229211.496 | 229211.612 | -1.2E-01 | 63.969 | 64.073 | -1.0E-01 |
| 2377 | Cf | 246 | 98 | 148 | 50 | 7.499114 | 7.497834 | 1.3E-03 | 229160.642 | 229160.898 | -2.6E-01 | 229211.619 | 229211.875 | -2.6E-01 | 64.092 | 64.336 | -2.4E-01 |
| 2378 | Fm | 246 | 100 | 146 | 46 | 7.467971 | 7.465075 | 2.9E-03 | 229165.668 | 229166.320 | -6.5E-01 | 229217.716 | 229218.367 | -6.5E-01 | 70.189 | 70.828 | -6.4E-01 |
| 2379 | Cm | 247 | 96 | 151 | 55 | 7.501926 | 7.503876 | -1.9E-03 | 230094.648 | 230094.109 | 5.4E-01 | 230144.556 | 230144.016 | 5.4E-01 | 65.534 | 64.983 | 5.5E-01 |
| 2380 | Bk | 247 | 97 | 150 | 53 | 7.498935 | 7.499227 | -2.9E-04 | 230094.070 | 230093.939 | 1.3E-01 | 230144.512 | 230144.381 | 1.3E-01 | 65.491 | 65.348 | 1.4E-01 |
| 2381 | Cf | 247 | 98 | 149 | 51 | 7.493285 | 7.491795 | 1.5E-03 | 230094.148 | 230094.457 | -3.1E-01 | 230145.125 | 230145.434 | -3.1E-01 | 66.104 | 66.401 | -3.0E-01 |
| 2382 | Es | 247 | 99 | 148 | 49 | 7.480099 | 7.477995 | 2.1E-03 | 230096.088 | 230096.548 | -4.6E-01 | 230147.600 | 230148.060 | -4.6E-01 | 68.578 | 69.027 | -4.5E-01 |
| 2383 | Cm | 248 | 96 | 152 | 56 | 7.496725 | 7.498585 | -1.9E-03 | 231028.001 | 231027.482 | 5.2E-01 | 231077.909 | 231077.390 | 5.2E-01 | 67.393 | 66.863 | 5.3E-01 |
| 2384 | Cf | 248 | 98 | 150 | 52 | 7.491035 | 7.490550 | 4.9E-04 | 231026.778 | 231026.840 | -6.2E-02 | 231077.755 | 231077.816 | -6.1E-02 | 67.240 | 67.289 | -4.9E-02 |
| 2385 | Fm | 248 | 100 | 148 | 48 | 7.465941 | 7.464054 | 1.9E-03 | 231030.366 | 231030.773 | -4.1E-01 | 231082.414 | 231082.821 | -4.1E-01 | 71.899 | 72.294 | -4.0E-01 |
| 2386 | Cm | 249 | 96 | 153 | 57 | 7.485547 | 7.485940 | -3.9E-04 | 231962.853 | 231962.698 | 1.6E-01 | 232012.761 | 232012.605 | 1.6E-01 | 70.751 | 70.584 | 1.7E-01 |
| 2387 | Bk | 249 | 97 | 152 | 55 | 7.486023 | 7.485994 | 2.9E-05 | 231961.418 | 231961.367 | 5.1E-02 | 232011.860 | 232011.809 | 5.1E-02 | 69.851 | 69.788 | 6.3E-02 |
| 2388 | Cf | 249 | 98 | 151 | 53 | 7.483382 | 7.482713 | 6.7E-04 | 231960.758 | 231960.866 | -1.1E-01 | 232011.735 | 232011.843 | -1.1E-01 | 69.726 | 69.821 | -9.5E-02 |
| 2389 | Fm | 249 | 100 | 149 | 49 | 7.461859 | 7.459904 | 2.0E-03 | 231963.482 | 231963.908 | -4.3E-01 | 232015.530 | 232015.956 | -4.3E-01 | 73.521 | 73.935 | -4.1E-01 |
| 2390 | Cm | 250 | 96 | 154 | 58 | 7.478935 | 7.478214 | 7.2E-04 | 232896.586 | 232896.709 | -1.2E-01 | 232946.494 | 232946.616 | -1.2E-01 | 72.990 | 73.101 | -1.1E-01 |
| 2391 | Bk | 250 | 97 | 153 | 56 | 7.475961 | 7.475707 | 2.5E-04 | 232896.013 | 232896.018 | -4.7E-03 | 232946.455 | 232946.460 | -4.7E-03 | 72.952 | 72.944 | 7.1E-03 |
| 2392 | Cf | 250 | 98 | 152 | 54 | 7.479949 | 7.479518 | 4.3E-04 | 232893.698 | 232893.747 | -4.9E-02 | 232944.675 | 232944.724 | -4.9E-02 | 71.172 | 71.209 | -3.7E-02 |
| 2393 | Fm | 250 | 100 | 150 | 50 | 7.462085 | 7.460266 | 1.8E-03 | 232895.529 | 232895.923 | -3.9E-01 | 232947.577 | 232947.971 | -3.9E-01 | 74.073 | 74.456 | -3.8E-01 |
| 2394 | Cm | 251 | 96 | 155 | 59 | 7.466717 | 7.463207 | 3.5E-03 | 233831.739 | 233832.562 | -8.2E-01 | 233881.647 | 233882.470 | -8.2E-01 | 76.649 | 77.461 | -8.1E-01 |
| 2395 | Bk | 251 | 97 | 154 | 57 | 7.469258 | 7.468241 | 1.0E-03 | 233829.785 | 233829.981 | -2.0E-01 | 233880.227 | 233880.423 | -2.0E-01 | 75.229 | 75.414 | -1.8E-01 |
| 2396 | Cf | 251 | 98 | 153 | 55 | 7.470495 | 7.469602 | 8.9E-04 | 233828.157 | 233828.322 | -1.6E-01 | 233879.134 | 233879.299 | -1.6E-01 | 74.136 | 74.289 | -1.5E-01 |
| 2397 | Es | 251 | 99 | 152 | 53 | 7.465875 | 7.463418 | 2.5E-03 | 233827.999 | 233828.556 | -5.6E-01 | 233879.511 | 233880.068 | -5.6E-01 | 74.514 | 75.058 | -5.4E-01 |
| 2398 | Fm | 251 | 100 | 151 | 51 | 7.457020 | 7.454545 | 2.5E-03 | 233828.903 | 233829.464 | -5.6E-01 | 233880.951 | 233881.512 | -5.6E-01 | 75.954 | 76.503 | -5.5E-01 |
| 2399 | Md | 251 | 101 | 150 | 49 | 7.441900 | 7.439572 | 2.3E-03 | 233831.380 | 233831.903 | -5.2E-01 | 233883.964 | 233884.487 | -5.2E-01 | 78.967 | 79.478 | -5.1E-01 |
| 2400 | Cf | 252 | 98 | 154 | 56 | 7.465344 | 7.464323 | 1.0E-03 | 234761.550 | 234761.748 | -2.0E-01 | 234812.527 | 234812.725 | -2.0E-01 | 76.035 | 76.221 | -1.9E-01 |
| 2401 | Es | 252 | 99 | 153 | 54 | 7.457240 | 7.455346 | 1.9E-03 | 234762.274 | 234762.692 | -4.2E-01 | 234813.787 | 234814.204 | -4.2E-01 | 77.295 | 77.701 | -4.1E-01 |
| 2402 | Fm | 252 | 100 | 152 | 52 | 7.456031 | 7.453219 | 2.8E-03 | 234761.261 | 234761.909 | -6.5E-01 | 234813.309 | 234813.957 | -6.5E-01 | 76.818 | 77.453 | -6.4E-01 |
| 2403 | No | 252 | 102 | 150 | 48 | 7.425795 | 7.424518 | 1.3E-03 | 234766.243 | 234766.503 | -2.6E-01 | 234819.363 | 234819.623 | -2.6E-01 | 82.872 | 83.120 | -2.5E-01 |
| 2404 | Cf | 253 | 98 | 155 | 57 | 7.454826 | 7.452345 | 2.5E-03 | 235696.311 | 235696.880 | -5.7E-01 | 235747.288 | 235747.856 | -5.7E-01 | 79.302 | 79.859 | -5.6E-01 |
| 2405 | Es | 253 | 99 | 154 | 55 | 7.452871 | 7.450536 | 2.3E-03 | 235695.488 | 235696.019 | -5.3E-01 | 235747.000 | 235747.531 | -5.3E-01 | 79.015 | 79.534 | -5.2E-01 |
| 2406 | Fm | 253 | 100 | 153 | 53 | 7.448457 | 7.445729 | 2.7E-03 | 235695.286 | 235695.916 | -6.3E-01 | 235747.334 | 235747.964 | -6.3E-01 | 79.349 | 79.967 | -6.2E-01 |
| 2407 | No | 253 | 102 | 151 | 49 | 7.422466 | 7.420751 | 1.7E-03 | 235699.225 | 235699.597 | -3.7E-01 | 235752.345 | 235752.717 | -3.7E-01 | 84.360 | 84.720 | -3.6E-01 |
| 2408 | Cf | 254 | 98 | 156 | 58 | 7.449223 | 7.445164 | 4.1E-03 | 236629.845 | 236630.817 | -9.7E-01 | 236680.821 | 236681.793 | -9.7E-01 | 81.342 | 82.302 | -9.6E-01 |
| 2409 | Es | 254 | 99 | 155 | 56 | 7.443583 | 7.440657 | 2.9E-03 | 236629.959 | 236630.643 | -6.8E-01 | 236681.471 | 236682.155 | -6.8E-01 | 81.992 | 82.663 | -6.7E-01 |
| 2410 | Fm | 254 | 100 | 154 | 54 | 7.444786 | 7.442611 | 2.2E-03 | 236628.336 | 236628.828 | -4.9E-01 | 236680.384 | 236680.876 | -4.9E-01 | 80.904 | 81.384 | -4.8E-01 |
| 2411 | No | 254 | 102 | 152 | 50 | 7.423585 | 7.421159 | 2.4E-03 | 236631.083 | 236631.638 | -5.5E-01 | 236684.204 | 236684.758 | -5.5E-01 | 84.725 | 85.267 | -5.4E-01 |
| 2412 | Es | 255 | 99 | 156 | 57 | 7.437816 | 7.433931 | 3.9E-03 | 237563.552 | 237564.483 | -9.3E-01 | 237615.064 | 237615.995 | -9.3E-01 | 84.091 | 85.009 | -9.2E-01 |
| 2413 | Fm | 255 | 100 | 155 | 55 | 7.435883 | 7.433332 | 2.6E-03 | 237562.726 | 237563.317 | -5.9E-01 | 237614.774 | 237615.365 | -5.9E-01 | 83.801 | 84.379 | -5.8E-01 |
| 2414 | Md | 255 | 101 | 154 | 53 | 7.428724 | 7.425838 | 2.9E-03 | 237563.234 | 237563.908 | -6.7E-01 | 237615.818 | 237616.492 | -6.7E-01 | 84.844 | 85.507 | -6.6E-01 |
| 2415 | No | 255 | 102 | 153 | 51 | 7.417958 | 7.415858 | 2.1E-03 | 237564.660 | 237565.134 | -4.7E-01 | 237617.781 | 237618.254 | -4.7E-01 | 86.807 | 87.269 | -4.6E-01 |
| 2416 | Lr | 255 | 103 | 152 | 49 | 7.402576 | 7.400092 | 2.5E-03 | 237567.263 | 237567.834 | -5.7E-01 | 237620.921 | 237621.492 | -5.7E-01 | 89.947 | 90.506 | -5.6E-01 |
| 2417 | Fm | 256 | 100 | 156 | 56 | 7.431778 | 7.428519 | 3.3E-03 | 238495.907 | 238496.681 | -7.7E-01 | 238547.955 | 238548.729 | -7.7E-01 | 85.487 | 86.249 | -7.6E-01 |
| 2418 | No | 256 | 102 | 154 | 52 | 7.416539 | 7.414701 | 1.8E-03 | 238497.171 | 238497.579 | -4.1E-01 | 238550.291 | 238550.700 | -4.1E-01 | 87.824 | 88.220 | -4.0E-01 |
| 2419 | Lr | 256 | 103 | 153 | 50 | 7.398160 | 7.395675 | 2.5E-03 | 238500.556 | 238501.130 | -5.7E-01 | 238554.214 | 238554.788 | -5.7E-01 | 91.747 | 92.308 | -5.6E-01 |
| 2420 | Rf | 256 | 104 | 152 | 48 | 7.385431 | 7.384186 | 1.2E-03 | 238502.495 | 238502.750 | -2.6E-01 | 238556.690 | 238556.946 | -2.6E-01 | 94.223 | 94.466 | -2.4E-01 |
| 2421 | Fm | 257 | 100 | 157 | 57 | 7.422191 | 7.417630 | 4.6E-03 | 239430.504 | 239431.616 | -1.1E+00 | 239482.552 | 239483.664 | -1.1E+00 | 88.591 | 89.690 | -1.1E+00 |
| 2422 | Md | 257 | 101 | 156 | 55 | 7.417566 | 7.414040 | 3.5E-03 | 239430.375 | 239431.220 | -8.4E-01 | 239482.959 | 239483.804 | -8.4E-01 | 88.997 | 89.830 | -8.3E-01 |
| 2423 | No | 257 | 102 | 155 | 53 | 7.409645 | 7.407843 | 1.8E-03 | 239431.091 | 239431.493 | -4.0E-01 | 239484.212 | 239484.613 | -4.0E-01 | 90.250 | 90.640 | -3.9E-01 |
| 2424 | Rf | 257 | 104 | 153 | 49 | 7.381700 | 7.380911 | 7.9E-04 | 239435.633 | 239435.773 | -1.4E-01 | 239489.829 | 239489.969 | -1.4E-01 | 95.868 | 95.995 | -1.3E-01 |
| 2425 | Md | 258 | 101 | 157 | 56 | 7.409668 | 7.404921 | 4.7E-03 | 240364.560 | 240365.724 | -1.2E+00 | 240417.144 | 240418.308 | -1.2E+00 | 91.688 | 92.840 | -1.2E+00 |
| 2426 | Rf | 258 | 104 | 154 | 50 | 7.382533 | 7.381592 | 9.4E-04 | 240367.602 | 240367.782 | -1.8E-01 | 240421.798 | 240421.977 | -1.8E-01 | 96.342 | 96.509 | -1.7E-01 |
| 2427 | Db | 259 | 105 | 154 | 49 | 7.360362 | 7.360050 | 3.1E-04 | 241304.207 | 241304.224 | -1.7E-02 | 241358.940 | 241358.957 | -1.7E-02 | 101.991 | 101.996 | -4.5E-03 |
| 2428 | Sg | 260 | 106 | 154 | 48 | 7.342560 | 7.342776 | -2.2E-04 | 242239.719 | 242239.599 | 1.2E-01 | 242294.991 | 242294.871 | 1.2E-01 | 106.548 | 106.415 | 1.3E-01 |
| 2429 | Rf | 261 | 104 | 157 | 53 | 7.371372 | 7.370313 | 1.1E-03 | 243167.064 | 243167.277 | -2.1E-01 | 243221.259 | 243221.473 | -2.1E-01 | 101.322 | 101.523 | -2.0E-01 |
| 2430 | Sg | 261 | 106 | 155 | 49 | 7.339766 | 7.340172 | -4.1E-04 | 243172.671 | 243172.501 | 1.7E-01 | 243227.944 | 243227.773 | 1.7E-01 | 108.006 | 107.823 | 1.8E-01 |
| 2431 | Sg | 262 | 106 | 156 | 50 | 7.341181 | 7.341337 | -1.6E-04 | 244104.526 | 244104.421 | 1.1E-01 | 244159.798 | 244159.693 | 1.0E-01 | 108.367 | 108.249 | 1.2E-01 |
| 2432 | Hs | 264 | 108 | 156 | 48 | 7.298373 | 7.298671 | -3.0E-04 | 245977.632 | 245977.488 | 1.4E-01 | 246033.983 | 246033.839 | 1.4E-01 | 119.564 | 119.407 | 1.6E-01 |
| 2433 | Hs | 265 | 108 | 157 | 49 | 7.296242 | 7.296859 | -6.2E-04 | 246910.464 | 246910.235 | 2.3E-01 | 246966.815 | 246966.586 | 2.3E-01 | 120.901 | 120.660 | 2.4E-01 |
| 2434 | Hs | 266 | 108 | 158 | 50 | 7.298269 | 7.298531 | -2.6E-04 | 247842.194 | 247842.058 | 1.4E-01 | 247898.545 | 247898.410 | 1.4E-01 | 121.137 | 120.989 | 1.5E-01 |
| 2435 | Ds | 269 | 110 | 159 | 49 | 7.250150 | 7.250239 | -8.9E-05 | 250649.293 | 250649.202 | 9.1E-02 | 250706.725 | 250706.634 | 9.1E-02 | 134.836 | 134.732 | 1.0E-01 |
| 2436 | Ds | 270 | 110 | 160 | 50 | 7.253771 | 7.252316 | 1.5E-03 | 251580.631 | 251580.957 | -3.3E-01 | 251638.063 | 251638.389 | -3.3E-01 | 134.679 | 134.992 | -3.1E-01 |
| 2437 | Eh | 294 | 117 | 177 | 60 | 7.093000 | 7.092966 | 3.4E-05 | 273994.050 | 273994.036 | 1.4E-02 | 274055.284 | 274055.270 | 1.4E-02 | 196.044 | 196.016 | 2.8E-02 |
| 2438 | Eh | 293 | 117 | 176 | 59 | 7.097000 | 7.095846 | 1.2E-03 | 273060.483 | 273060.720 | -2.4E-01 | 273121.716 | 273121.953 | -2.4E-01 | 193.970 | 194.194 | -2.2E-01 |
| 2439 | Eh | 292 | 117 | 175 | 58 | 7.096000 | 7.094275 | 1.7E-03 | 272128.269 | 272128.709 | -4.4E-01 | 272189.502 | 272189.943 | -4.4E-01 | 193.251 | 193.677 | -4.3E-01 |
| 2440 | Eh | 291 | 117 | 174 | 57 | 7.099000 | 7.096507 | 2.5E-03 | 271194.979 | 271195.588 | -6.1E-01 | 271256.212 | 271256.822 | -6.1E-01 | 191.454 | 192.050 | -6.0E-01 |
| 2441 | Lv | 293 | 116 | 177 | 61 | 7.111000 | 7.111951 | -9.5E-04 | 273057.536 | 273057.329 | 2.1E-01 | 273118.224 | 273118.018 | 2.1E-01 | 190.479 | 190.258 | 2.2E-01 |
| 2442 | Lv | 292 | 116 | 176 | 60 | 7.117000 | 7.114608 | 2.4E-03 | 272123.485 | 272124.100 | -6.1E-01 | 272184.173 | 272184.788 | -6.2E-01 | 187.921 | 188.523 | -6.0E-01 |
| 2443 | Lv | 291 | 116 | 175 | 59 | 7.116000 | 7.114835 | 1.2E-03 | 271191.371 | 271191.583 | -2.1E-01 | 271252.060 | 271252.272 | -2.1E-01 | 187.302 | 187.500 | -2.0E-01 |
| 2444 | Lv | 290 | 116 | 174 | 58 | 7.120000 | 7.116715 | 3.3E-03 | 270257.605 | 270258.587 | -9.8E-01 | 270318.294 | 270319.276 | -9.8E-01 | 185.030 | 185.998 | -9.7E-01 |
| 2445 | Lv | 289 | 116 | 173 | 57 | 7.119000 | 7.116225 | 2.8E-03 | 269325.668 | 269326.280 | -6.1E-01 | 269386.356 | 269386.969 | -6.1E-01 | 184.587 | 185.185 | -6.0E-01 |
| 2446 | Ef | 291 | 115 | 176 | 61 | 7.131000 | 7.129935 | 1.1E-03 | 271188.184 | 271188.516 | -3.3E-01 | 271248.328 | 271248.661 | -3.3E-01 | 183.570 | 183.889 | -3.2E-01 |
| 2447 | Ef | 290 | 115 | 175 | 60 | 7.132000 | 7.129986 | 2.0E-03 | 270255.669 | 270256.066 | -4.0E-01 | 270315.813 | 270316.210 | -4.0E-01 | 182.550 | 182.933 | -3.8E-01 |
| 2448 | Ef | 289 | 115 | 174 | 59 | 7.136000 | 7.133638 | 2.4E-03 | 269321.988 | 269322.575 | -5.9E-01 | 269382.132 | 269382.719 | -5.9E-01 | 180.363 | 180.936 | -5.7E-01 |
| 2449 | Ef | 288 | 115 | 173 | 58 | 7.136000 | 7.132842 | 3.2E-03 | 268389.667 | 268390.373 | -7.1E-01 | 268449.812 | 268450.517 | -7.1E-01 | 179.536 | 180.228 | -6.9E-01 |
| 2450 | Ef | 287 | 115 | 172 | 57 | 7.139000 | 7.135824 | 3.2E-03 | 267456.277 | 267457.084 | -8.1E-01 | 267516.421 | 267517.229 | -8.1E-01 | 177.639 | 178.433 | -7.9E-01 |
| 2451 | Fl | 289 | 114 | 175 | 61 | 7.149000 | 7.149624 | -6.2E-04 | 269319.543 | 269319.282 | 2.6E-01 | 269379.144 | 269378.883 | 2.6E-01 | 177.374 | 177.099 | 2.7E-01 |
| 2452 | Fl | 288 | 114 | 174 | 60 | 7.155000 | 7.153337 | 1.7E-03 | 268385.397 | 268385.797 | -4.0E-01 | 268444.998 | 268445.398 | -4.0E-01 | 174.722 | 175.108 | -3.9E-01 |
| 2453 | Fl | 287 | 114 | 173 | 59 | 7.154000 | 7.154128 | -1.3E-04 | 267453.168 | 267453.158 | 1.0E-02 | 267512.769 | 267512.759 | 1.0E-02 | 173.987 | 173.963 | 2.4E-02 |
| 2454 | Fl | 286 | 114 | 172 | 58 | 7.159000 | 7.157039 | 2.0E-03 | 266519.301 | 266519.914 | -6.1E-01 | 266578.902 | 266579.515 | -6.1E-01 | 171.613 | 172.213 | -6.0E-01 |
| 2455 | Fl | 285 | 114 | 171 | 57 | 7.158000 | 7.157089 | 9.1E-04 | 265587.255 | 265587.491 | -2.4E-01 | 265646.856 | 265647.092 | -2.4E-01 | 171.062 | 171.285 | -2.2E-01 |
| 2456 | Ed | 287 | 113 | 174 | 61 | 7.168000 | 7.167978 | 2.2E-05 | 267450.550 | 267450.509 | 4.1E-02 | 267509.607 | 267509.567 | 4.0E-02 | 170.825 | 170.771 | 5.4E-02 |
| 2457 | Ed | 286 | 113 | 173 | 60 | 7.169000 | 7.168760 | 2.4E-04 | 266517.956 | 266517.888 | 6.8E-02 | 266577.013 | 266576.946 | 6.7E-02 | 169.725 | 169.644 | 8.1E-02 |
| 2458 | Ed | 285 | 113 | 172 | 59 | 7.174000 | 7.173337 | 6.6E-04 | 265584.151 | 265584.187 | -3.6E-02 | 265643.208 | 265643.245 | -3.7E-02 | 167.415 | 167.437 | -2.2E-02 |
| 2459 | Ed | 284 | 113 | 171 | 58 | 7.174000 | 7.173245 | 7.5E-04 | 264651.725 | 264651.821 | -9.6E-02 | 264710.783 | 264710.879 | -9.6E-02 | 166.483 | 166.565 | -8.2E-02 |
| 2460 | Ed | 283 | 113 | 170 | 57 | 7.178000 | 7.177131 | 8.7E-04 | 263718.223 | 263718.329 | -1.1E-01 | 263777.281 | 263777.387 | -1.1E-01 | 164.475 | 164.567 | -9.2E-02 |
| 2461 | Ed | 282 | 113 | 169 | 56 | 7.178000 | 7.176212 | 1.8E-03 | 262785.891 | 262786.200 | -3.1E-01 | 262844.949 | 262845.258 | -3.1E-01 | 163.638 | 163.932 | -2.9E-01 |
| 2462 | Ed | 281 | 113 | 168 | 55 | 7.182000 | 7.179441 | 2.6E-03 | 261852.355 | 261852.903 | -5.5E-01 | 261911.413 | 261911.961 | -5.5E-01 | 161.596 | 162.130 | -5.3E-01 |
| 2463 | Ed | 280 | 113 | 167 | 54 | 7.180000 | 7.177718 | 2.3E-03 | 260920.349 | 260921.000 | -6.5E-01 | 260979.407 | 260980.058 | -6.5E-01 | 161.083 | 161.721 | -6.4E-01 |
| 2464 | Ed | 279 | 113 | 166 | 53 | 7.184000 | 7.180300 | 3.7E-03 | 259987.011 | 259987.892 | -8.8E-01 | 260046.069 | 260046.950 | -8.8E-01 | 159.239 | 160.107 | -8.7E-01 |
| 2465 | Ed | 278 | 113 | 165 | 52 | 7.182000 | 7.177774 | 4.2E-03 | 259055.170 | 259056.209 | -1.0E+00 | 259114.228 | 259115.267 | -1.0E+00 | 158.893 | 159.918 | -1.0E+00 |
| 2466 | Cn | 285 | 112 | 173 | 61 | 7.185000 | 7.188085 | -3.1E-03 | 265582.262 | 265581.309 | 9.5E-01 | 265640.777 | 265639.825 | 9.5E-01 | 164.983 | 164.017 | 9.7E-01 |
| 2467 | Cn | 284 | 112 | 172 | 60 | 7.192000 | 7.192910 | -9.1E-04 | 264648.010 | 264647.562 | 4.5E-01 | 264706.525 | 264706.077 | 4.5E-01 | 162.225 | 161.764 | 4.6E-01 |
| 2468 | Cn | 283 | 112 | 171 | 59 | 7.191000 | 7.194267 | -3.3E-03 | 263715.690 | 263714.805 | 8.9E-01 | 263774.206 | 263773.320 | 8.9E-01 | 161.400 | 160.501 | 9.0E-01 |
| 2469 | Cn | 282 | 112 | 170 | 58 | 7.197000 | 7.198268 | -1.3E-03 | 262781.612 | 262781.305 | 3.1E-01 | 262840.128 | 262839.821 | 3.1E-01 | 158.816 | 158.496 | 3.2E-01 |
| 2470 | Cn | 281 | 112 | 169 | 57 | 7.197000 | 7.198862 | -1.9E-03 | 261849.420 | 261848.772 | 6.5E-01 | 261907.935 | 261907.287 | 6.5E-01 | 158.117 | 157.456 | 6.6E-01 |
| 2471 | Cn | 280 | 112 | 168 | 56 | 7.202000 | 7.202086 | -8.6E-05 | 260915.505 | 260915.502 | 2.6E-03 | 260974.020 | 260974.018 | 2.2E-03 | 155.697 | 155.681 | 1.6E-02 |
| 2472 | Cn | 279 | 112 | 167 | 55 | 7.201000 | 7.201945 | -9.4E-04 | 259983.447 | 259983.178 | 2.7E-01 | 260041.963 | 260041.694 | 2.7E-01 | 155.133 | 154.851 | 2.8E-01 |
| 2473 | Cn | 278 | 112 | 166 | 54 | 7.206000 | 7.204415 | 1.6E-03 | 259049.731 | 259050.128 | -4.0E-01 | 259108.246 | 259108.644 | -4.0E-01 | 152.910 | 153.295 | -3.8E-01 |
| 2474 | Cn | 277 | 112 | 165 | 53 | 7.205000 | 7.203541 | 1.5E-03 | 258117.757 | 258118.009 | -2.5E-01 | 258176.272 | 258176.525 | -2.5E-01 | 152.430 | 152.670 | -2.4E-01 |
| 2475 | Cn | 276 | 112 | 164 | 52 | 7.209000 | 7.205258 | 3.7E-03 | 257184.185 | 257185.174 | -9.9E-01 | 257242.700 | 257243.689 | -9.9E-01 | 150.352 | 151.328 | -9.8E-01 |
| 2476 | Rg | 283 | 111 | 172 | 61 | 7.203000 | 7.206797 | -3.8E-03 | 263713.693 | 263712.584 | 1.1E+00 | 263771.666 | 263770.558 | 1.1E+00 | 158.861 | 157.738 | 1.1E+00 |
| 2477 | Rg | 282 | 111 | 171 | 60 | 7.205000 | 7.208262 | -3.3E-03 | 262780.872 | 262779.812 | 1.1E+00 | 262838.845 | 262837.786 | 1.1E+00 | 157.534 | 156.461 | 1.1E+00 |
| 2478 | Rg | 281 | 111 | 170 | 59 | 7.211000 | 7.213855 | -2.9E-03 | 261846.808 | 261845.884 | 9.2E-01 | 261904.781 | 261903.857 | 9.2E-01 | 154.963 | 154.026 | 9.4E-01 |
| 2479 | Rg | 280 | 111 | 169 | 58 | 7.212000 | 7.214427 | -2.4E-03 | 260914.175 | 260913.372 | 8.0E-01 | 260972.149 | 260971.345 | 8.0E-01 | 153.825 | 153.008 | 8.2E-01 |
| 2480 | Rg | 279 | 111 | 168 | 57 | 7.217000 | 7.219305 | -2.3E-03 | 259980.431 | 259979.660 | 7.7E-01 | 260038.404 | 260037.633 | 7.7E-01 | 151.574 | 150.790 | 7.8E-01 |
| 2481 | Rg | 278 | 111 | 167 | 56 | 7.218000 | 7.219027 | -1.0E-03 | 259047.792 | 259047.391 | 4.0E-01 | 259105.765 | 259105.365 | 4.0E-01 | 150.430 | 150.016 | 4.1E-01 |
| 2482 | Rg | 277 | 111 | 166 | 55 | 7.223000 | 7.223217 | -2.2E-04 | 258114.040 | 258113.884 | 1.6E-01 | 258172.013 | 258171.858 | 1.6E-01 | 148.172 | 148.003 | 1.7E-01 |
| 2483 | Rg | 276 | 111 | 165 | 54 | 7.222000 | 7.222103 | -1.0E-04 | 257181.861 | 257181.849 | 1.2E-02 | 257239.835 | 257239.823 | 1.2E-02 | 147.487 | 147.462 | 2.5E-02 |
| 2484 | Rg | 275 | 111 | 164 | 53 | 7.227000 | 7.225607 | 1.4E-03 | 256248.136 | 256248.543 | -4.1E-01 | 256306.110 | 256306.516 | -4.1E-01 | 145.257 | 145.649 | -3.9E-01 |
| 2485 | Rg | 274 | 111 | 163 | 52 | 7.227000 | 7.223646 | 3.4E-03 | 255316.003 | 255316.740 | -7.4E-01 | 255373.977 | 255374.714 | -7.4E-01 | 144.617 | 145.341 | -7.2E-01 |
| 2486 | Rg | 273 | 111 | 162 | 51 | 7.231000 | 7.226436 | 4.6E-03 | 254382.529 | 254383.637 | -1.1E+00 | 254440.502 | 254441.610 | -1.1E+00 | 142.636 | 143.732 | -1.1E+00 |
| 2487 | Rg | 272 | 111 | 161 | 50 | 7.227000 | 7.223589 | 3.4E-03 | 253451.171 | 253452.072 | -9.0E-01 | 253509.144 | 253510.046 | -9.0E-01 | 142.773 | 143.661 | -8.9E-01 |
| 2488 | Ds | 281 | 110 | 171 | 61 | 7.220000 | 7.226819 | -6.8E-03 | 261845.626 | 261843.565 | 2.1E+00 | 261903.059 | 261900.997 | 2.1E+00 | 153.241 | 151.166 | 2.1E+00 |
| 2489 | Ds | 280 | 110 | 170 | 60 | 7.227000 | 7.232801 | -5.8E-03 | 260911.152 | 260909.552 | 1.6E+00 | 260968.584 | 260966.984 | 1.6E+00 | 150.260 | 148.647 | 1.6E+00 |
| 2490 | Ds | 279 | 110 | 169 | 59 | 7.228000 | 7.234733 | -6.7E-03 | 259978.527 | 259976.680 | 1.8E+00 | 260035.959 | 260034.112 | 1.8E+00 | 149.130 | 147.269 | 1.9E+00 |
| 2491 | Ds | 278 | 110 | 168 | 58 | 7.236000 | 7.239870 | -3.9E-03 | 259044.185 | 259042.921 | 1.3E+00 | 259101.617 | 259100.353 | 1.3E+00 | 146.282 | 145.004 | 1.3E+00 |
| 2492 | Ds | 277 | 110 | 167 | 57 | 7.236000 | 7.241020 | -5.0E-03 | 258111.642 | 258110.277 | 1.4E+00 | 258169.075 | 258167.709 | 1.4E+00 | 145.233 | 143.854 | 1.4E+00 |
| 2493 | Ds | 276 | 110 | 166 | 56 | 7.243000 | 7.245352 | -2.4E-03 | 257177.457 | 257176.757 | 7.0E-01 | 257234.889 | 257234.189 | 7.0E-01 | 142.542 | 141.828 | 7.1E-01 |
| 2494 | Ds | 275 | 110 | 165 | 55 | 7.243000 | 7.245740 | -2.7E-03 | 256245.039 | 256244.330 | 7.1E-01 | 256302.471 | 256301.763 | 7.1E-01 | 141.618 | 140.896 | 7.2E-01 |
| 2495 | Ds | 274 | 110 | 164 | 54 | 7.249000 | 7.249281 | -2.8E-04 | 255311.103 | 255311.041 | 6.2E-02 | 255368.535 | 255368.473 | 6.2E-02 | 139.175 | 139.100 | 7.5E-02 |
| 2496 | Ds | 273 | 110 | 163 | 53 | 7.249000 | 7.248896 | 1.0E-04 | 254378.816 | 254378.830 | -1.4E-02 | 254436.248 | 254436.262 | -1.4E-02 | 138.383 | 138.383 | -1.3E-04 |
| 2497 | Ds | 272 | 110 | 162 | 52 | 7.255000 | 7.251629 | 3.4E-03 | 253444.956 | 253445.770 | -8.1E-01 | 253502.388 | 253503.202 | -8.1E-01 | 136.016 | 136.817 | -8.0E-01 |
| 2498 | Ds | 271 | 110 | 161 | 51 | 7.252000 | 7.250431 | 1.6E-03 | 252513.393 | 252513.781 | -3.9E-01 | 252570.825 | 252571.213 | -3.9E-01 | 135.947 | 136.322 | -3.8E-01 |
| 2499 | Ds | 270 | 110 | 160 | 50 | 7.253771 | 7.252316 | 1.5E-03 | 251580.631 | 251580.957 | -3.3E-01 | 251638.063 | 251638.389 | -3.3E-01 | 134.679 | 134.992 | -3.1E-01 |
| 2500 | Ds | 269 | 110 | 159 | 49 | 7.250150 | 7.250239 | -8.9E-05 | 250649.293 | 250649.202 | 9.1E-02 | 250706.725 | 250706.634 | 9.1E-02 | 134.836 | 134.732 | 1.0E-01 |
| 2501 | Ds | 268 | 110 | 158 | 48 | 7.252000 | 7.251217 | 7.8E-04 | 249716.612 | 249716.625 | -1.3E-02 | 249774.044 | 249774.057 | -1.3E-02 | 133.649 | 133.649 | 6.0E-05 |
| 2502 | Ds | 267 | 110 | 157 | 47 | 7.247000 | 7.248183 | -1.2E-03 | 248785.388 | 248785.121 | 2.7E-01 | 248842.821 | 248842.553 | 2.7E-01 | 133.919 | 133.639 | 2.8E-01 |
| 2503 | Mt | 279 | 109 | 170 | 61 | 7.238000 | 7.244798 | -6.8E-03 | 259977.184 | 259975.195 | 2.0E+00 | 260034.076 | 260032.087 | 2.0E+00 | 147.246 | 145.244 | 2.0E+00 |
| 2504 | Mt | 278 | 109 | 169 | 60 | 7.241000 | 7.247000 | -6.0E-03 | 259044.043 | 259042.263 | 1.8E+00 | 259100.935 | 259099.154 | 1.8E+00 | 145.599 | 143.805 | 1.8E+00 |
| 2505 | Mt | 277 | 109 | 168 | 59 | 7.248000 | 7.253660 | -5.7E-03 | 258109.718 | 258108.100 | 1.6E+00 | 258166.610 | 258164.991 | 1.6E+00 | 142.768 | 141.136 | 1.6E+00 |
| 2506 | Mt | 276 | 109 | 167 | 58 | 7.251000 | 7.254956 | -4.0E-03 | 257176.665 | 257175.430 | 1.2E+00 | 257233.556 | 257232.322 | 1.2E+00 | 141.209 | 139.961 | 1.2E+00 |
| 2507 | Mt | 275 | 109 | 166 | 57 | 7.257000 | 7.260877 | -3.9E-03 | 256242.587 | 256241.492 | 1.1E+00 | 256299.479 | 256298.383 | 1.1E+00 | 138.625 | 137.516 | 1.1E+00 |
| 2508 | Mt | 274 | 109 | 165 | 56 | 7.260000 | 7.261299 | -1.3E-03 | 255309.626 | 255309.071 | 5.5E-01 | 255366.517 | 255365.963 | 5.5E-01 | 137.158 | 136.590 | 5.7E-01 |
| 2509 | Mt | 273 | 109 | 164 | 55 | 7.266000 | 7.266496 | -5.0E-04 | 254375.481 | 254375.348 | 1.3E-01 | 254432.372 | 254432.240 | 1.3E-01 | 134.507 | 134.361 | 1.5E-01 |
| 2510 | Mt | 272 | 109 | 163 | 54 | 7.267000 | 7.266047 | 9.5E-04 | 253443.062 | 253443.172 | -1.1E-01 | 253499.953 | 253500.063 | -1.1E-01 | 133.582 | 133.679 | -9.7E-02 |
| 2511 | Mt | 271 | 109 | 162 | 53 | 7.273000 | 7.270504 | 2.5E-03 | 252509.088 | 252509.665 | -5.8E-01 | 252565.980 | 252566.556 | -5.8E-01 | 131.102 | 131.665 | -5.6E-01 |
| 2512 | Mt | 270 | 109 | 161 | 52 | 7.271000 | 7.269151 | 1.8E-03 | 251577.206 | 251577.735 | -5.3E-01 | 251634.097 | 251634.627 | -5.3E-01 | 130.714 | 131.230 | -5.2E-01 |
| 2513 | Mt | 269 | 109 | 160 | 51 | 7.274000 | 7.272821 | 1.2E-03 | 250644.310 | 250644.452 | -1.4E-01 | 250701.201 | 250701.343 | -1.4E-01 | 129.312 | 129.441 | -1.3E-01 |
| 2514 | Mt | 268 | 109 | 159 | 50 | 7.271000 | 7.270504 | 5.0E-04 | 249712.655 | 249712.780 | -1.3E-01 | 249769.546 | 249769.671 | -1.3E-01 | 129.151 | 129.263 | -1.1E-01 |
| 2515 | Mt | 267 | 109 | 158 | 49 | 7.273000 | 7.273318 | -3.2E-04 | 248779.801 | 248779.734 | 6.7E-02 | 248836.692 | 248836.625 | 6.7E-02 | 127.791 | 127.711 | 8.0E-02 |
| 2516 | Mt | 266 | 109 | 157 | 48 | 7.270000 | 7.269963 | 3.7E-05 | 247848.479 | 247848.334 | 1.4E-01 | 247905.371 | 247905.226 | 1.5E-01 | 127.963 | 127.805 | 1.6E-01 |
| 2517 | Mt | 265 | 109 | 156 | 47 | 7.271000 | 7.271845 | -8.4E-04 | 246915.702 | 246915.540 | 1.6E-01 | 246972.593 | 246972.431 | 1.6E-01 | 126.680 | 126.505 | 1.7E-01 |
| 2518 | Hs | 277 | 108 | 169 | 61 | 7.255000 | 7.263104 | -8.1E-03 | 258108.983 | 258106.807 | 2.2E+00 | 258165.335 | 258163.158 | 2.2E+00 | 141.493 | 139.303 | 2.2E+00 |
| 2519 | Hs | 276 | 108 | 168 | 60 | 7.264000 | 7.270317 | -6.3E-03 | 257174.281 | 257172.514 | 1.8E+00 | 257230.633 | 257228.865 | 1.8E+00 | 138.285 | 136.504 | 1.8E+00 |
| 2520 | Hs | 275 | 108 | 167 | 59 | 7.267000 | 7.272961 | -6.0E-03 | 256241.123 | 256239.492 | 1.6E+00 | 256297.474 | 256295.843 | 1.6E+00 | 136.621 | 134.976 | 1.6E+00 |
| 2521 | Hs | 274 | 108 | 166 | 58 | 7.276000 | 7.279311 | -3.3E-03 | 255306.495 | 255305.460 | 1.0E+00 | 255362.846 | 255361.811 | 1.0E+00 | 133.487 | 132.438 | 1.0E+00 |
| 2522 | Hs | 273 | 108 | 165 | 57 | 7.278000 | 7.281152 | -3.2E-03 | 254373.488 | 254372.671 | 8.2E-01 | 254429.839 | 254429.022 | 8.2E-01 | 131.973 | 131.143 | 8.3E-01 |
| 2523 | Hs | 272 | 108 | 164 | 56 | 7.286000 | 7.286666 | -6.7E-04 | 253439.028 | 253438.887 | 1.4E-01 | 253495.379 | 253495.238 | 1.4E-01 | 129.007 | 128.853 | 1.5E-01 |
| 2524 | Hs | 271 | 108 | 163 | 55 | 7.288000 | 7.287710 | 2.9E-04 | 252506.299 | 252506.325 | -2.6E-02 | 252562.651 | 252562.676 | -2.5E-02 | 127.773 | 127.786 | -1.3E-02 |
| 2525 | Hs | 270 | 108 | 162 | 54 | 7.295000 | 7.292381 | 2.6E-03 | 251572.123 | 251572.786 | -6.6E-01 | 251628.474 | 251629.138 | -6.6E-01 | 125.090 | 125.741 | -6.5E-01 |
| 2526 | Hs | 269 | 108 | 161 | 53 | 7.294000 | 7.292592 | 1.4E-03 | 250640.128 | 250640.456 | -3.3E-01 | 250696.480 | 250696.808 | -3.3E-01 | 124.590 | 124.905 | -3.2E-01 |
| 2527 | Hs | 268 | 108 | 160 | 52 | 7.298000 | 7.296380 | 1.6E-03 | 249706.876 | 249707.169 | -2.9E-01 | 249763.227 | 249763.520 | -2.9E-01 | 122.831 | 123.111 | -2.8E-01 |
| 2528 | Hs | 267 | 108 | 159 | 51 | 7.295000 | 7.295693 | -6.9E-04 | 248775.203 | 248775.083 | 1.2E-01 | 248831.554 | 248831.434 | 1.2E-01 | 122.653 | 122.520 | 1.3E-01 |
| 2529 | Hs | 266 | 108 | 158 | 50 | 7.298269 | 7.298531 | -2.6E-04 | 247842.194 | 247842.058 | 1.4E-01 | 247898.545 | 247898.410 | 1.4E-01 | 121.137 | 120.989 | 1.5E-01 |
| 2530 | Hs | 265 | 108 | 157 | 49 | 7.296242 | 7.296859 | -6.2E-04 | 246910.464 | 246910.235 | 2.3E-01 | 246966.815 | 246966.586 | 2.3E-01 | 120.901 | 120.660 | 2.4E-01 |
| 2531 | Hs | 264 | 108 | 156 | 48 | 7.298373 | 7.298671 | -3.0E-04 | 245977.632 | 245977.488 | 1.4E-01 | 246033.983 | 246033.839 | 1.4E-01 | 119.564 | 119.407 | 1.6E-01 |
| 2532 | Hs | 263 | 108 | 155 | 47 | 7.295000 | 7.295925 | -9.3E-04 | 245046.292 | 245045.943 | 3.5E-01 | 245102.643 | 245102.294 | 3.5E-01 | 119.718 | 119.356 | 3.6E-01 |
| 2533 | Bh | 275 | 107 | 168 | 61 | 7.273000 | 7.278588 | -5.6E-03 | 256240.733 | 256239.267 | 1.5E+00 | 256296.544 | 256295.078 | 1.5E+00 | 135.691 | 134.212 | 1.5E+00 |
| 2534 | Bh | 274 | 107 | 167 | 60 | 7.278000 | 7.281854 | -3.9E-03 | 255307.262 | 255306.085 | 1.2E+00 | 255363.074 | 255361.897 | 1.2E+00 | 133.714 | 132.524 | 1.2E+00 |
| 2535 | Bh | 273 | 107 | 166 | 59 | 7.286000 | 7.289595 | -3.6E-03 | 254372.687 | 254371.688 | 1.0E+00 | 254428.498 | 254427.500 | 1.0E+00 | 130.633 | 129.621 | 1.0E+00 |
| 2536 | Bh | 272 | 107 | 165 | 58 | 7.290000 | 7.291935 | -1.9E-03 | 253439.352 | 253438.776 | 5.8E-01 | 253495.164 | 253494.588 | 5.8E-01 | 128.792 | 128.203 | 5.9E-01 |
| 2537 | Bh | 271 | 107 | 164 | 57 | 7.298000 | 7.298915 | -9.1E-04 | 252505.056 | 252504.611 | 4.4E-01 | 252560.868 | 252560.423 | 4.5E-01 | 125.990 | 125.532 | 4.6E-01 |
| 2538 | Bh | 270 | 107 | 163 | 56 | 7.301000 | 7.300343 | 6.6E-04 | 251571.799 | 251571.959 | -1.6E-01 | 251627.610 | 251627.771 | -1.6E-01 | 124.227 | 124.374 | -1.5E-01 |
| 2539 | Bh | 269 | 107 | 162 | 55 | 7.309000 | 7.306554 | 2.4E-03 | 250637.560 | 250638.024 | -4.6E-01 | 250693.371 | 250693.835 | -4.6E-01 | 121.482 | 121.933 | -4.5E-01 |
| 2540 | Bh | 268 | 107 | 161 | 54 | 7.308000 | 7.307045 | 9.6E-04 | 249705.391 | 249705.633 | -2.4E-01 | 249761.203 | 249761.445 | -2.4E-01 | 120.807 | 121.036 | -2.3E-01 |
| 2541 | Bh | 267 | 107 | 160 | 53 | 7.313000 | 7.312440 | 5.6E-04 | 248771.856 | 248771.934 | -7.8E-02 | 248827.668 | 248827.746 | -7.8E-02 | 118.767 | 118.831 | -6.4E-02 |
| 2542 | Bh | 266 | 107 | 159 | 52 | 7.313000 | 7.311931 | 1.1E-03 | 247839.704 | 247839.817 | -1.1E-01 | 247895.515 | 247895.628 | -1.1E-01 | 118.108 | 118.208 | -1.0E-01 |
| 2543 | Bh | 265 | 107 | 158 | 51 | 7.316000 | 7.316431 | -4.3E-04 | 246906.459 | 246906.371 | 8.8E-02 | 246962.270 | 246962.182 | 8.8E-02 | 116.357 | 116.256 | 1.0E-01 |
| 2544 | Bh | 264 | 107 | 157 | 50 | 7.315000 | 7.314832 | 1.7E-04 | 245974.666 | 245974.544 | 1.2E-01 | 246030.477 | 246030.355 | 1.2E-01 | 116.058 | 115.923 | 1.3E-01 |
| 2545 | Bh | 263 | 107 | 156 | 49 | 7.318000 | 7.318346 | -3.5E-04 | 245041.609 | 245041.369 | 2.4E-01 | 245097.421 | 245097.181 | 2.4E-01 | 114.496 | 114.243 | 2.5E-01 |
| 2546 | Bh | 262 | 107 | 155 | 48 | 7.315000 | 7.315568 | -5.7E-04 | 244110.163 | 244109.850 | 3.1E-01 | 244165.974 | 244165.662 | 3.1E-01 | 114.543 | 114.218 | 3.3E-01 |
| 2547 | Bh | 261 | 107 | 154 | 47 | 7.317000 | 7.318013 | -1.0E-03 | 243177.261 | 243176.962 | 3.0E-01 | 243233.072 | 243232.774 | 3.0E-01 | 113.135 | 112.824 | 3.1E-01 |
| 2548 | Bh | 260 | 107 | 153 | 46 | 7.313000 | 7.313988 | -9.9E-04 | 242245.957 | 242245.761 | 2.0E-01 | 242301.768 | 242301.573 | 2.0E-01 | 113.324 | 113.117 | 2.1E-01 |
| 2549 | Sg | 273 | 106 | 167 | 61 | 7.291000 | 7.294263 | -3.3E-03 | 254372.611 | 254371.737 | 8.7E-01 | 254427.883 | 254427.009 | 8.7E-01 | 130.018 | 129.130 | 8.9E-01 |
| 2550 | Sg | 272 | 106 | 166 | 60 | 7.301000 | 7.302840 | -1.8E-03 | 253437.680 | 253437.132 | 5.5E-01 | 253492.952 | 253492.405 | 5.5E-01 | 126.580 | 126.020 | 5.6E-01 |
| 2551 | Sg | 271 | 106 | 165 | 59 | 7.305000 | 7.306487 | -1.5E-03 | 252504.363 | 252503.882 | 4.8E-01 | 252559.635 | 252559.154 | 4.8E-01 | 124.758 | 124.263 | 4.9E-01 |
| 2552 | Sg | 270 | 106 | 164 | 58 | 7.314000 | 7.314182 | -1.8E-04 | 251569.603 | 251569.545 | 5.8E-02 | 251624.875 | 251624.817 | 5.8E-02 | 121.492 | 121.421 | 7.1E-02 |
| 2553 | Sg | 269 | 106 | 163 | 57 | 7.318000 | 7.316995 | 1.0E-03 | 250636.432 | 250636.537 | -1.1E-01 | 250691.705 | 250691.809 | -1.0E-01 | 119.815 | 119.907 | -9.2E-02 |
| 2554 | Sg | 268 | 106 | 162 | 56 | 7.326000 | 7.323810 | 2.2E-03 | 249701.925 | 249702.462 | -5.4E-01 | 249757.197 | 249757.735 | -5.4E-01 | 116.802 | 117.326 | -5.2E-01 |
| 2555 | Sg | 267 | 106 | 161 | 55 | 7.327000 | 7.325762 | 1.2E-03 | 248769.467 | 248769.700 | -2.3E-01 | 248824.739 | 248824.972 | -2.3E-01 | 115.838 | 116.058 | -2.2E-01 |
| 2556 | Sg | 266 | 106 | 160 | 54 | 7.332000 | 7.331655 | 3.4E-04 | 247835.754 | 247835.892 | -1.4E-01 | 247891.026 | 247891.165 | -1.4E-01 | 113.619 | 113.744 | -1.3E-01 |
| 2557 | Sg | 265 | 106 | 159 | 53 | 7.333000 | 7.332675 | 3.3E-04 | 246903.439 | 246903.388 | 5.1E-02 | 246958.711 | 246958.661 | 5.0E-02 | 112.797 | 112.735 | 6.2E-02 |
| 2558 | Sg | 264 | 106 | 158 | 52 | 7.338000 | 7.337565 | 4.3E-04 | 245969.931 | 245969.865 | 6.6E-02 | 246025.203 | 246025.137 | 6.6E-02 | 110.784 | 110.705 | 7.9E-02 |
| 2559 | Sg | 263 | 106 | 157 | 51 | 7.337000 | 7.337547 | -5.5E-04 | 245037.843 | 245037.642 | 2.0E-01 | 245093.116 | 245092.914 | 2.0E-01 | 110.190 | 109.976 | 2.1E-01 |
| 2560 | Sg | 262 | 106 | 156 | 50 | 7.341181 | 7.341337 | -1.6E-04 | 244104.526 | 244104.421 | 1.1E-01 | 244159.798 | 244159.693 | 1.0E-01 | 108.367 | 108.249 | 1.2E-01 |
| 2561 | Sg | 261 | 106 | 155 | 49 | 7.339766 | 7.340172 | -4.1E-04 | 243172.671 | 243172.501 | 1.7E-01 | 243227.944 | 243227.773 | 1.7E-01 | 108.006 | 107.823 | 1.8E-01 |
| 2562 | Sg | 260 | 106 | 154 | 48 | 7.342560 | 7.342776 | -2.2E-04 | 242239.719 | 242239.599 | 1.2E-01 | 242294.991 | 242294.871 | 1.2E-01 | 106.548 | 106.415 | 1.3E-01 |
| 2563 | Sg | 259 | 106 | 153 | 47 | 7.340000 | 7.340381 | -3.8E-04 | 241308.236 | 241307.996 | 2.4E-01 | 241363.509 | 241363.269 | 2.4E-01 | 106.559 | 106.307 | 2.5E-01 |
| 2564 | Sg | 258 | 106 | 152 | 46 | 7.342000 | 7.341754 | 2.5E-04 | 240375.426 | 240375.417 | 8.9E-03 | 240430.698 | 240430.689 | 8.6E-03 | 105.243 | 105.222 | 2.1E-02 |

## Appendix C

### Table 4

**The prediction for the values of binding energy, nuclear mass, atomic mass and mass excess for some heavy nuclei (paper [37])**

| No | Element | A | Z | N | N-Z | $B_E$ | $Nu_{Mass}$ | $At_{Mass}$ | $Mas_{Exc}$ | [MeV] |
|---|---|---|---|---|---|---|---|---|---|---|
| 1 | 104266 | 266. | 104. | 162. | 58. | 0.7346076E+01 | 0.2478347E+06 | 0.2478889E+06 | 0.1114747E+03 | |
| 2 | 104267 | 267. | 104. | 163. | 59. | 0.7336931E+01 | 0.2487694E+06 | 0.2488236E+06 | 0.1146416E+03 | |
| 3 | 104265 | 266. | 105. | 161. | 57. | 0.7335649E+01 | 0.2478362E+06 | 0.2478909E+06 | 0.1134652E+03 | |
| 4 | 105270 | 270. | 105. | 165. | 60. | 0.7312202E+01 | 0.2515714E+06 | 0.2516261E+06 | 0.1227385E+03 | |
| 5 | 105268 | 268. | 105. | 163. | 58. | 0.7324809E+01 | 0.2497035E+06 | 0.2497582E+06 | 0.1178416E+03 | |
| 6 | 105267 | 267. | 105. | 162. | 57. | 0.7332857E+01 | 0.2487691E+06 | 0.2488239E+06 | 0.1149464E+03 | |
| 7 | 106271 | 271. | 106. | 165. | 59. | 0.7306487E+01 | 0.2525039E+06 | 0.2525592E+06 | 0.1242633E+03 | |
| 8 | 106269 | 269. | 106. | 163. | 57. | 0.7316995E+01 | 0.2506365E+06 | 0.2506918E+06 | 0.1199071E+03 | |
| 9 | 107274 | 274. | 107. | 167. | 60. | 0.7281854E+01 | 0.2553061E+06 | 0.2553619E+06 | 0.1325240E+03 | |
| 10 | 107272 | 272. | 107. | 165. | 58. | 0.7291935E+01 | 0.2534388E+06 | 0.2534946E+06 | 0.1282031E+03 | |
| 11 | 107271 | 271. | 107. | 164. | 57. | 0.7298915E+01 | 0.2525046E+06 | 0.2525604E+06 | 0.1255323E+03 | |
| 12 | 107270 | 270. | 107. | 163. | 56. | 0.7300343E+01 | 0.2515720E+06 | 0.2516278E+06 | 0.1243743E+03 | |
| 13 | 108278 | 278. | 108. | 170. | 62. | 0.7259620E+01 | 0.2590401E+06 | 0.2590964E+06 | 0.1410800E+03 | |
| 14 | 108277 | 277. | 108. | 169. | 61. | 0.7263104E+01 | 0.2581068E+06 | 0.2581632E+06 | 0.1393034E+03 | |
| 15 | 108275 | 275. | 108. | 167. | 59. | 0.7272961E+01 | 0.2562395E+06 | 0.2562958E+06 | 0.1349762E+03 | |
| 16 | 198273 | 273. | 108. | 165. | 57. | 0.7281152E+01 | 0.2543727E+06 | 0.2544290E+06 | 0.1311432E+03 | |
| 17 | 110275 | 275. | 110. | 165. | 55. | 0.7245740E+01 | 0.2562443E+06 | 0.2563018E+06 | 0.1408958E+03 | |
| 18 | 110282 | 282. | 110. | 172. | 62. | 0.7224057E+01 | 0.2627767E+06 | 0.2628341E+06 | 0.1527893E+03 | |
| 19 | 112279 | 279. | 112. | 167. | 55. | 0.7201945E+01 | 0.2599832E+06 | 0.2600417E+06 | 0.1548508E+03 | |
| 20 | 112286 | 286. | 112. | 174. | 62. | 0.7185912E+01 | 0.2665143E+06 | 0.2665728E+06 | 0.1655217E+03 | |
| 21 | 113285 | 285. | 113. | 173. | 60. | 0.7168760E+01 | 0.2665179E+06 | 0.2665769E+06 | 0.1696442E+03 | |
| 22 | 113284 | 284. | 113. | 172. | 59. | 0.7173337E+01 | 0.2655842E+06 | 0.2656432E+06 | 0.1674371E+03 | |
| 23 | 113283 | 283. | 113. | 171. | 58. | 0.7173245E+01 | 0.2646518E+06 | 0.2647109E+06 | 0.1665652E+03 | |
| 24 | 113282 | 282. | 113. | 170. | 57. | 0.7177131E+01 | 0.2637183E+06 | 0.2637774E+06 | 0.1645675E+03 | |
| 25 | 113281 | 281. | 113. | 169. | 56. | 0.7176212E+01 | 0.2627862E+06 | 0.2628453E+06 | 0.1639325E+03 | |
| 26 | 114283 | 283. | 114. | 169. | 55. | 0.7158593E+01 | 0.2637222E+06 | 0.2637819E+06 | 0.1690308E+03 | |
| 27 | 114290 | 290. | 114. | 176. | 62. | 0.7148038E+01 | 0.2702522E+06 | 0.2703118E+06 | 0.1784806E+03 | |

| | | | | | | | | | |
|---|---|---|---|---|---|---|---|---|---|
| 28 | 115290 | 290. | 115. | 175. | 60. | 0.7129986E+01 | 0.2702561E+06 | 0.2703162E+06 | 0.1829326E+03 |
| 29 | 115289 | 289. | 115. | 174. | 59. | 0.7133638E+01 | 0.2693226E+06 | 0.2693827E+06 | 0.1809359E+03 |
| 30 | 115288 | 288. | 115. | 173. | 58. | 0.7132842E+01 | 0.2683904E+06 | 0.2684505E+06 | 0.1802276E+03 |
| 31 | 115287 | 287. | 115. | 172. | 57. | 0.7135824E+01 | 0.2674571E+06 | 0.2675172E+06 | 0.1784331E+03 |
| 32 | 116294 | 294. | 116. | 178. | 62. | 0.7111165E+01 | 0.2739900E+06 | 0.2740507E+06 | 0.1914483E+03 |
| 33 | 116293 | 293. | 116. | 177. | 61. | 0.7111951E+01 | 0.2730573E+06 | 0.2731180E+06 | 0.1902580E+03 |
| 34 | 116292 | 292. | 116. | 176. | 60. | 0.7114608E+01 | 0.2721241E+06 | 0.2721848E+06 | 0.1885226E+03 |
| 35 | 116291 | 291. | 116. | 175. | 59. | 0.7114835E+01 | 0.2711916E+06 | 0.2712523E+06 | 0.1875000E+03 |
| 36 | 116290 | 290. | 116. | 174. | 58. | 0.7116715E+01 | 0.2702586E+06 | 0.2703193E+06 | 0.1859983E+03 |
| 37 | 118293 | 293. | 118. | 175. | 57. | 0.7078354E+01 | 0.2730645E+06 | 0.2731263E+06 | 0.1985357E+03 |
| 38 | 118294 | 294. | 118. | 176. | 58. | 0.7079403E+01 | 0.2739967E+06 | 0.2740585E+06 | 0.1992202E+03 |
| 39 | 118295 | 295. | 118. | 177. | 59. | 0.7078443E+01 | 0.2749295E+06 | 0.2749912E+06 | 0.2004954E+03 |
| 40 | 118296 | 296. | 118. | 178. | 60. | 0.7078998E+01 | 0.2758618E+06 | 0.2759236E+06 | 0.2013241E+03 |
| 41 | 118297 | 297. | 118. | 179. | 61. | 0.7077473E+01 | 0.2767947E+06 | 0.2768565E+06 | 0.2027690E+03 |
| 42 | 119295 | 295. | 119. | 176. | 57. | 0.7059040E+01 | 0.2749339E+06 | 0.2749962E+06 | 0.2054360E+03 |
| 43 | 119296 | 296. | 119. | 177. | 58. | 0.7057423E+01 | 0.2758668E+06 | 0.2759292E+06 | 0.2069270E+03 |
| 44 | 119297 | 297. | 119. | 178. | 59. | 0.7060021E+01 | 0.2767986E+06 | 0.2768609E+06 | 0.2071694E+03 |
| 45 | 120295 | 295. | 120. | 175. | 55. | 0.7038272E+01 | 0.2749387E+06 | 0.2750015E+06 | 0.2107797E+03 |
| 46 | 120296 | 296. | 120. | 176. | 56. | 0.7040502E+01 | 0.2758705E+06 | 0.2759334E+06 | 0.2111527E+03 |
| 47 | 120297 | 297. | 120. | 177. | 57. | 0.7041094E+01 | 0.2768029E+06 | 0.2768657E+06 | 0.2120076E+03 |
| 48 | 120298 | 298. | 120. | 178. | 58. | 0.7042875E+01 | 0.2777349E+06 | 0.2777977E+06 | 0.2125070E+03 |
| 49 | 120299 | 299. | 120. | 179. | 59. | 0.7042966E+01 | 0.2786674E+06 | 0.2787302E+06 | 0.2135083E+03 |

## APPENDIX D

IF WE DEFINE THE ALPHA DECAY AS FOLLOW

$$M(Z,N) -> M_d(Z-2,N-2) + M_a(2,2) ,$$

calculate the masses using formulae (1) and (4), then the total energy of the decay is

$$Q_t^{Th} = M(Z,N) - M_d(Z-2,N-2) - M_a(2,2)$$

and the kinetic energy of alpha particle is

$$E_k^{Th} = Q_{at}^{Th} \frac{M_d(Z-2,N-2)}{M_d(Z-2,N-2) + M_a(2,2)}$$

IN THE NEXT TABLE 5 ARE PRESENTES THE DATA FROM PAPER [37] AS FOLLOW: Mode, Nomep, Element, A, Z, N, M, $M_d$, $E_k^{Expt}$, $E_k^{Th}$, $Q_k^{Expt}$, $Q_k^{Th}$, $dE=\frac{E_k^{Expt}-E_k^{Th}}{E_k^{Th}}$, $dQ=\frac{Q_a^{Expt}-Q_a^{Th}}{Q_a^{Th}}$.

Table 5

| No | A | Z | N | $E_k^{Expt}$ | $E_k^{Th}$ | $ResE_k$ | $hi2E_k$ | $Q_t^{Expt}$ | $Q_t^{Th}$ | $ResQ_t$ | $hi2Q_t$ [Mev]] |
|---|---|---|---|---|---|---|---|---|---|---|---|
| 1 | 294 | 118 | 176 | 0.1054E+02 | 0.1166E+02 | -1.12 | -0.10 | 0.1068E+02 | 0.1182E+02 | -1.14 | -0.10 |
| 2 | 294 | 117 | 177 | 0.1040E+02 | 0.1081E+02 | -0.41 | -0.04 | 0.1055E+02 | 0.1118E+02 | -0.63 | -0.06 |
| 3 | 293 | 117 | 176 | 0.1057E+02 | 0.1060E+02 | -0.03 | -0.00 | 0.1072E+02 | 0.1132E+02 | -0.60 | -0.05 |
| 4 | 293 | 116 | 177 | 0.1048E+02 | 0.1056E+02 | -0.08 | -0.01 | 0.1062E+02 | 0.1071E+02 | -0.09 | -0.01 |
| 5 | 292 | 116 | 176 | 0.1073E+02 | 0.1063E+02 | 0.10 | 0.01 | 0.1088E+02 | 0.1078E+02 | 0.10 | 0.01 |
| 6 | 291 | 116 | 175 | 0.1085E+02 | 0.1074E+02 | 0.11 | 0.01 | 0.1100E+02 | 0.1089E+02 | 0.11 | 0.01 |
| 7 | 290. | 116. | 174. | 0.1109E+02 | 0.1085E+02 | 0.24 | 0.02 | 0.1125E+02 | 0.1100E+02 | 0.25 | 0.02 |
| 8 | 290. | 115. | 175. | 0.1061E+02 | 0.9780E+01 | 0.83 | 0.08 | 0.1075E+02 | 0.1041E+02 | 0.34 | 0.03 |
| 9 | 289. | 115. | 174. | 0.1081E+02 | 0.1015E+02 | 0.66 | 0.07 | 0.1096E+02 | 0.1049E+02 | 0.47 | 0.05 |
| 10 | 288. | 115. | 173. | 0.1097E+02 | 0.1029E+02 | 0.68 | 0.07 | 0.1113E+02 | 0.1049E+02 | 0.64 | 0.06 |
| 11 | 287. | 115. | 172. | 0.1117E+02 | 0.1061E+02 | 0.56 | 0.05 | 0.1133E+02 | 0.1076E+02 | 0.57 | 0.05 |
| 12 | 289. | 114. | 175. | 0.1040E+02 | 0.9840E+01 | 0.56 | 0.06 | 0.1055E+02 | 0.9980E+01 | 0.57 | 0.06 |
| 13 | 288. | 114. | 174. | 0.1066E+02 | 0.9930E+01 | 0.73 | 0.07 | 0.1081E+02 | 0.1007E+02 | 0.74 | 0.07 |
| 14 | 287. | 114. | 173. | 0.1078E+02 | 0.1003E+02 | 0.75 | 0.07 | 0.1093E+02 | 0.1017E+02 | 0.76 | 0.07 |
| 15 | 286. | 114. | 172. | 0.1103E+02 | 0.1021E+02 | 0.82 | 0.08 | 0.1118E+02 | 0.1035E+02 | 0.83 | 0.08 |
| 16 | 285. | 114. | 171. | 0.1114E+02 | 0.1114E+02 | 0.00 | 0.00 | 0.1130E+02 | 0.1130E+02 | 0.00 | 0.00 |
| 17 | 286. | 113. | 173. | 0.1050E+02 | 0.9610E+01 | 0.89 | 0.09 | 0.1065E+02 | 0.9790E+01 | 0.86 | 0.09 |
| 18 | 285. | 113. | 172. | 0.1073E+02 | 0.9470E+01 | 1.26 | 0.13 | 0.1088E+02 | 0.1001E+02 | 0.87 | 0.09 |
| 19 | 284. | 113. | 171. | 0.1087E+02 | 0.9100E+01 | 1.77 | 0.19 | 0.1102E+02 | 0.1012E+02 | 0.90 | 0.09 |
| 20 | 283. | 113. | 170. | 0.1109E+02 | 0.1023E+02 | 0.86 | 0.08 | 0.1124E+02 | 0.1038E+02 | 0.86 | 0.08 |
| 21 | 282. | 113. | 169. | 0.1122E+02 | 0.1063E+02 | 0.59 | 0.06 | 0.1138E+02 | 0.1078E+02 | 0.60 | 0.06 |
| 22 | 285. | 112. | 173. | 0.1018E+02 | 0.9190E+01 | 0.99 | 0.11 | 0.1032E+02 | 0.9320E+01 | 1.00 | 0.11 |
| 23 | 284. | 112. | 172. | 0.1044E+02 | 0.1044E+02 | 0.00 | 0.00 | 0.1059E+02 | 0.1059E+02 | 0.00 | 0.00 |
| 24 | 283. | 112. | 171. | 0.1055E+02 | 0.9530E+01 | 1.02 | 0.11 | 0.1070E+02 | 0.9660E+01 | 1.04 | 0.11 |
| 25 | 282. | 112. | 170. | 0.1080E+02 | 0.1080E+02 | 0.00 | 0.00 | 0.1096E+02 | 0.1096E+02 | 0.00 | 0.00 |
| 26 | 281. | 112. | 169. | 0.1091E+02 | 0.1031E+02 | 0.60 | 0.06 | 0.1107E+02 | 0.1046E+02 | 0.61 | 0.06 |
| 27 | 282. | 111. | 171. | 0.9982E+01 | 0.8860E+01 | 1.12 | 0.13 | 0.1013E+02 | 0.1053E+02 | -0.40 | -0.04 |
| 28 | 281. | 111. | 170. | 0.1021E+02 | 0.9280E+01 | 0.93 | 0.10 | 0.1036E+02 | 0.9410E+01 | 0.95 | 0.10 |
| 29 | 280. | 111. | 169. | 0.1037E+02 | 0.9090E+01 | 1.28 | 0.14 | 0.1052E+02 | 0.9910E+01 | 0.61 | 0.06 |
| 30 | 279. | 111. | 168. | 0.1059E+02 | 0.1038E+02 | 0.21 | 0.02 | 0.1074E+02 | 0.1053E+02 | 0.21 | 0.02 |
| 31 | 278. | 111. | 167. | 0.1074E+02 | 0.1069E+02 | 0.05 | 0.00 | 0.1090E+02 | 0.1085E+02 | 0.05 | 0.00 |
| 32 | 281. | 110. | 171. | 0.9201E+01 | 0.8730E+01 | 0.47 | 0.05 | 0.9334E+01 | 0.8850E+01 | 0.48 | 0.05 |
| 33 | 279. | 110. | 169. | 0.9624E+01 | 0.9710E+01 | -0.09 | -0.01 | 0.9764E+01 | 0.9850E+01 | -0.09 | -0.01 |
| 34 | 277. | 110. | 167. | 0.1004E+02 | 0.1057E+02 | -0.53 | -0.05 | 0.1018E+02 | 0.1072E+02 | -0.54 | -0.05 |
| 35 | 278. | 109. | 169. | 0.8627E+01 | 0.9380E+01 | -0.75 | -0.08 | 0.8753E+01 | 0.9580E+01 | -0.83 | -0.09 |
| 36 | 277. | 109. | 168. | 0.8857E+01 | 0.8857E+01 | 0.00 | 0.00 | 0.8987E+01 | 0.8987E+01 | 0.00 | 0.00 |
| 37 | 276. | 109. | 167. | 0.9096E+01 | 0.9170E+01 | -0.07 | -0.01 | 0.9230E+01 | 0.1003E+02 | -0.80 | -0.08 |
| 38 | 275. | 109. | 166. | 0.9319E+01 | 0.1033E+02 | -1.01 | -0.10 | 0.9456E+01 | 0.1048E+02 | -1.02 | -0.10 |
| 39 | 274. | 109. | 165. | 0.9546E+01 | 0.1000E+02 | -0.45 | -0.05 | 0.9688E+01 | 0.1010E+02 | -0.41 | -0.04 |
| 40 | 277. | 108. | 169. | 0.7536E+01 | 0.7536E+01 | 0.00 | 0.00 | 0.7647E+01 | 0.7647E+01 | 0.00 | 0.00 |
| 41 | 275. | 108. | 167. | 0.8067E+01 | 0.9310E+01 | -1.24 | -0.13 | 0.8186E+01 | 0.9450E+01 | -1.26 | -0.13 |
| 42 | 273. | 108. | 165. | 0.8582E+01 | 0.9590E+01 | -1.01 | -0.11 | 0.8709E+01 | 0.9730E+01 | -1.02 | -0.10 |
| 43 | 274. | 107. | 167. | 0.7154E+01 | 0.8730E+01 | -1.58 | -0.18 | 0.7260E+01 | 0.8840E+01 | -1.58 | -0.18 |
| 44 | 272. | 107. | 165. | 0.7721E+01 | 0.8550E+01 | -0.83 | -0.10 | 0.7836E+01 | 0.9180E+01 | -1.34 | -0.15 |
| 45 | 271. | 107. | 164. | 0.7941E+01 | 0.9280E+01 | -1.34 | -0.14 | 0.8060E+01 | 0.9420E+01 | -1.36 | -0.14 |
| 46 | 270. | 107. | 163. | 0.8259E+01 | 0.8930E+01 | -0.67 | -0.08 | 0.8383E+01 | 0.9060E+01 | -0.68 | -0.07 |
| 47 | 271. | 106. | 165. | 0.6992E+01 | 0.8540E+01 | -1.55 | -0.18 | 0.7097E+01 | 0.8670E+01 | -1.57 | -0.18 |
| 48 | 269. | 106. | 163. | 0.7572E+01 | 0.8570E+01 | -1.00 | -0.12 | 0.7686E+01 | 0.8700E+01 | -1.01 | -0.12 |
| 49 | 270. | 105. | 165. | 0.6541E+01 | 0.6541E+01 | 0.00 | 0.00 | 0.6639E+01 | 0.6639E+01 | 0.00 | 0.00 |
| 50 | 268. | 105. | 163. | 0.7142E+01 | 0.7142E+01 | 0.00 | 0.00 | 0.7250E+01 | 0.7250E+01 | 0.00 | 0.00 |
| 51 | 267. | 105. | 162. | 0.7364E+01 | 0.7364E+01 | 0.00 | 0.00 | 0.7476E+01 | 0.7476E+01 | 0.00 | 0.00 |
| 52 | 266. | 105. | 161. | 0.7698E+01 | 0.7698E+01 | 0.00 | 0.00 | 0.7816E+01 | 0.7816E+01 | 0.00 | 0.00 |
| 53 | 267. | 104. | 163. | 0.6777E+01 | 0.6777E+01 | 0.00 | 0.00 | 0.6880E+01 | 0.6880E+01 | 0.00 | 0.00 |
| 54 | 265. | 104. | 161. | 0.7356E+01 | 0.7356E+01 | 0.00 | 0.00 | 0.7469E+01 | 0.7469E+01 | 0.00 | 0.00 |

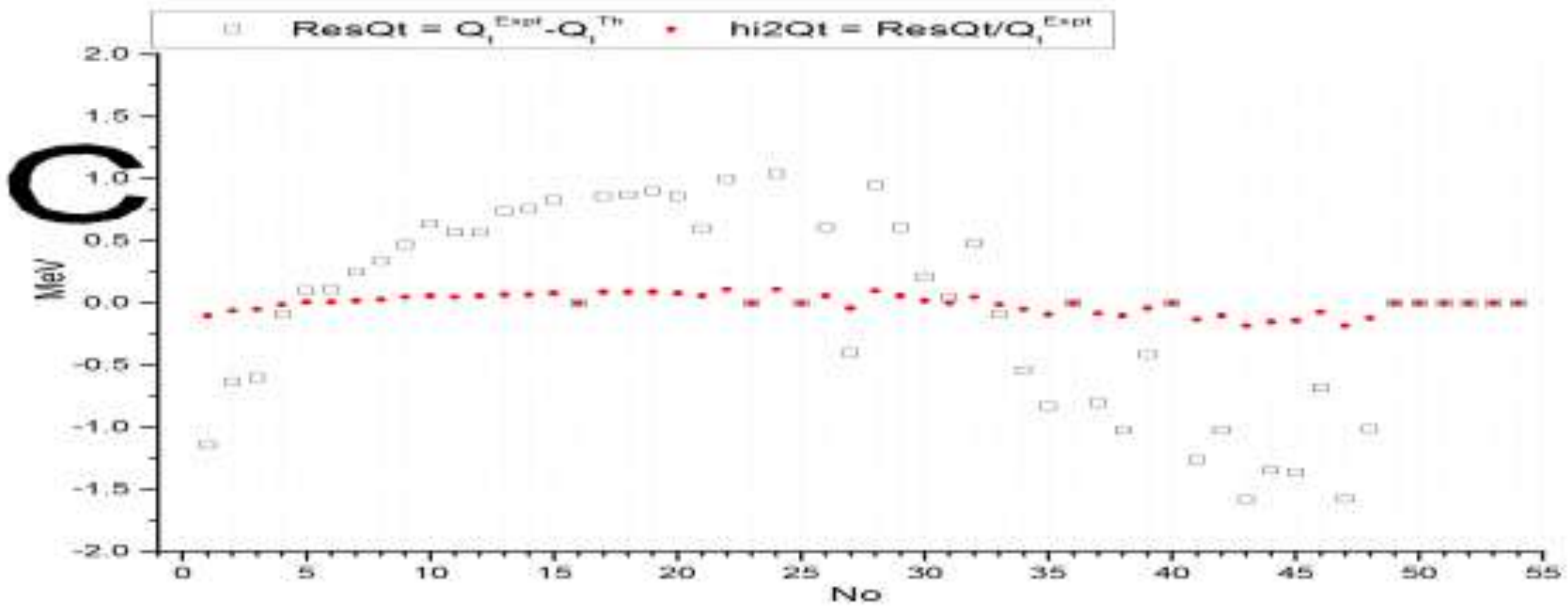

**FIG. 11. ILLUSTRATION OF DATA FROM TABLE 5 FOR VARIABLE E**

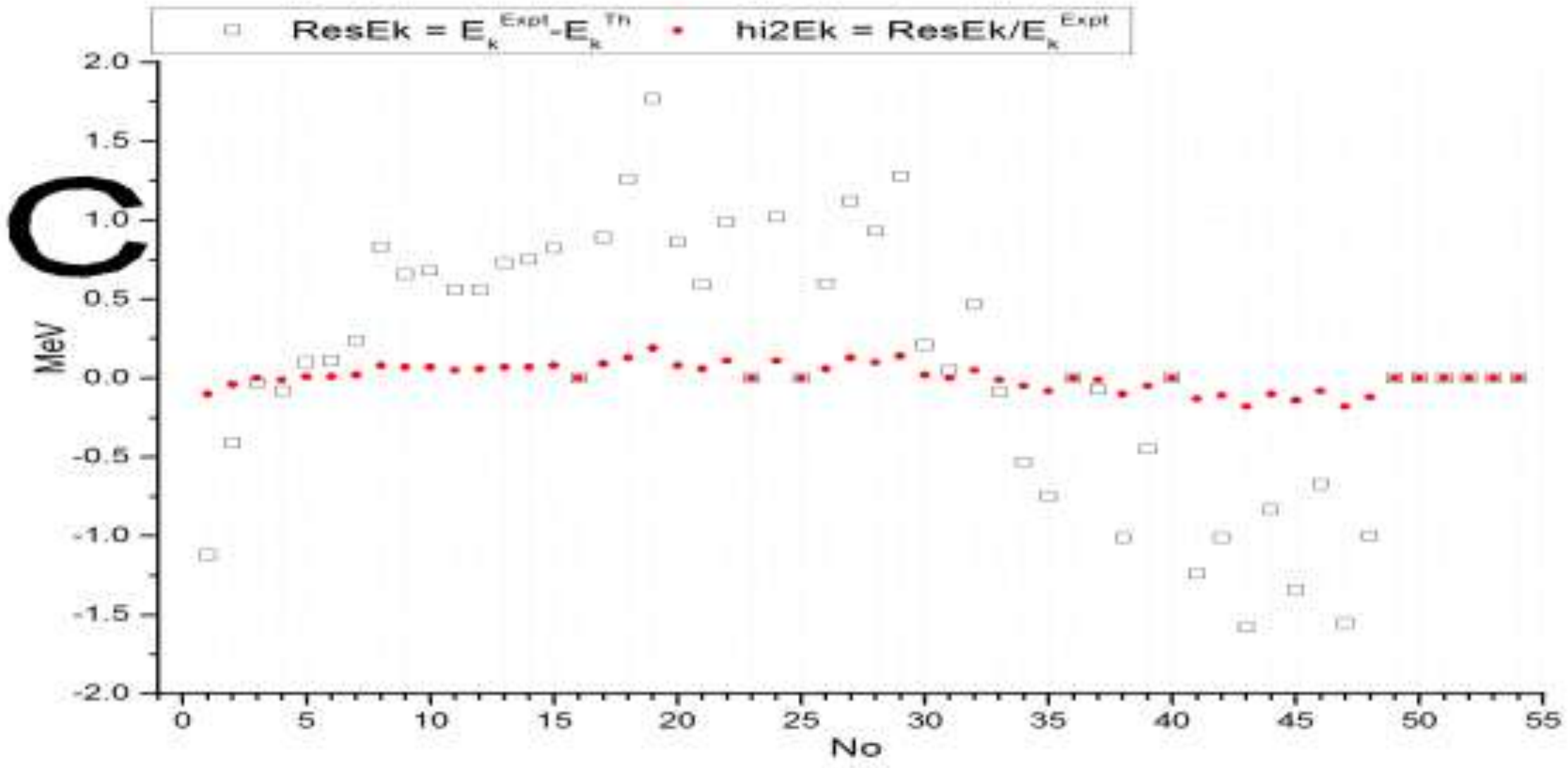

**FIG. 12. ILLUSTRATION OF DATA FROM TABLE 5 FOR VARIABLE Q**